\documentclass[12pt,twoside]{report}
\usepackage {graphicx}

\def\mnras{MNRAS}
\def\apjs{ApJS}
\def\apj{ApJ}
\def\apjl{ApJL}
\def\apjl{ApJL}

\def\aj{AJ}
\def\pasp{PASP}
\def\pasj{PASJ}
\def\aap{A\&A}
\def\nat{Nature}
\def\baas{BAAS}
\def\ha{H$\alpha$}
\def\hd{H$\delta$}
\def\hb{H$\beta$}
\def\oii{[O{\sc ii}]}
\def\nii{[N{\sc ii}]}
\def\simless{<}
\def\degree{\kern-.2em\r{}\kern-.3em}
{ }

\begin{document}

\title{Environmental Effects on Galaxy Evolution 
}
\bigskip

\author{Tomotsugu Goto  
\\ Department of Physics, Graduate School of Science, \\The University
 of Tokyo \\\\\\
 A dissertation submitted to The University of Tokyo\\ 
 for the degree of Doctor of Philosophy}
\date{June 6, 2003}
\maketitle

\chapter*{Ackowledgements}
\bigskip

 I am grateful to my advisers, Maki Sekiguchi, and  Sadanori Okamura
 for their continuous encouragement and support for
 not only this thesis, but also for my entire graduate school life. 
  I am indebted to Maki Sekiguchi for introducing me to an interesting
 field in astrophysics, galaxy evolution
 in clusters,  and for  his continuous support for my thesis project.
  The large amount of time I have spent with him during my graduate life
 has been extremely helpful to me. 
  I wish to express my deepest gratitude to Sadanori Okamura for his
 guidance on this work and for careful reading of  this
 manuscript. I owe it to  Sadanori and people in his group that I was able to
 concentrate on my thesis project and complete it during the last four
 months of my graduate life.

   I acknowledge Masafumi Yagi for his extensive help in solving
 statistical problems in this work.  His brain (often called
 a supercomputer) is indispensable to this project. 
 I wish to extend my sincere gratitude to Mamoru Doi for careful reading
   of the manuscript and many fruitful suggestions. 
 I would like to thank Naoki Yasuda, Shinichi Ichikawa, Takashi
 Ichikawa, Masaru Watanabe, Masaru Hamabe, Masataka Fukugita
   and Kazuhiro Shimasaku for instructing me how to use the SDSS data.
  The monthly SDSS meeting they used to have was a precious opportunity
 for a graduate student like myself to learn about the SDSS.
 I thank Masayuki Tanaka for always chatting with me whenever I wanted
 to. Many discussions with him often lead to a new idea in analysis.
   I thank Shuichi Aoki for sharing his sharp mind with me.
 Not only his expertise but also his attitude to research work set
 an excellent example for me to learn. 
  I thank Eiichiro and Midori Komatsu for their continuous friendship,
 and for teaching me  a spiritual aspect to be an astronomer. 
    I am grateful to Nell Hana Hoffman, Ricardo Colon, Emanuel Bowes,
 Shane Zabel, and Michael Crouch   
  for patiently correcting my English, which significantly  improved the quality of the
  manuscript.  
  I thank Masami Ouchi for showing an astronomical Samurai spirit 
 in studying 7 days without going home. The presence of such a brilliant
 graduate student has been a magnificent inspiration to me.
 I gratefully thank Takamitsu Miyaji not only for his continuous
 treatments to curry and 
 rice, but also helping me to settle down in Pittsburgh. It is a
 privilege to make friends with a world-famous X-ray astronomer.
 I thank Kentaro and Michiko Nagamine for their continuous friendship, especially in
 visiting us in Kashiwa. I hope to play basketball with them in the near
 future again.   
 Many thanks also to Yutaka Fujita and Tomonori Totani for showing me
 how fun it is to be an  astronomer.
 I thank Aya Bamba for kindly recommending me the best doughnut shop in
 Japan, where I sometimes hit on a good idea over a cup of coffee.
 I wish to thank Masatake Ohashi for his help in finishing
 this thesis.  
 I thank Koji Mori, Haruyoshi Katayama and Kenji Hamaguchi for offering me the
 opportunity to work on X-ray astronomy.
 I thank our secretary, Yuri Iinuma, for handling my many impossible requests for the
 grant.

  I wish to extend my many thanks to Christy Tremonti and Eric Peng for teaching me how to take a
 subway when I got lost in New York.
 I thank Neal A. Miller for sharing a ride in Pasadena. The discussion
 on radio astronomy we had in the car was quite influential to
 Appendix B of this work.
 I thank Marc
 Postman, Albert Conti, Sharon Busching, Alan Uomoto and Adrian
 C. Pope for showing me around Baltimore. 
  I thank James E. Gunn for using my plots in his review talk, and thus
  encouraging me to keep studying astronomy.
 I thank Wataru Kawasaki and Tadayuki Kodama for showing me a variety of
 wonderful science that a cluster finder can do.  
 I thank  Wolfgang Voges for being kind to every young astronomers
 including me. His presence gave me a great relief when I got nervous in
 visiting the States for the first time.
 I thank Jeff Peterson for saying hello to me every time we met at
 school. I thank Ravi Sheth for being always gentle to everyone, and 
 telling an interesting story to be a movie star.
       I  would like to express my sincere gratitude to Toru Yamada, Yutaka
  Komiyama,  Hisanori Furusawa,
 Francisco  Castander, Jeff Maki, Rupert Croft, Tiziana Di Matteo, Taotao Fang,
  Gulab Dewangan, Kavan Ratnatunga,  Douglas
 L. Tucker, Masanori Iye, Tadayuki Kodama, Shang-Shan Chong, Iskra
 Strateva, Brice Menard, Katsuko T. Nakahira, Connie Rockosi, Stefano
 Zibetti, Shiyin Shen, Simon White, Guinevere Kauffman and Stephane Charlot for their suggestions and
   discussions to improve the  work.     
   It has been a pleasure to work with them.
 I have been lucky to be blessed with a large number of wonderful
 coworkers and colleague graduate students. I thank the following for
 the fond memories, their kindness, 
 their enduring friendship and a plethora of experiences together which
 I will remember for good: Naohisa Inada, Kayo
 Issya, Chiaki Hikage,  Chris Miller, Percy Gomez, David Wake,  Makiko
 Yoshida, Tomoki Saito, Eri Yamanoi, Adam Knudson, 
  Shigeyuki Sako, Yoshihiko Yamada, Tomoaki Oyama, Fumiaki Nakata,
 Yasuhiro Shioya  and Ichi Tanaka.
 I thank Tim McKay and David Weinberg for a useful discussion in the
 cluster working group. 
 I thank Ching-Wa Yip and Sam Schmidt for their friendship at the
 University of Pittsburgh. 
 I thank Masayuki Ohama and Katsumi Kurasawa for performing a
 wonderful joke on top of Mt. Wakakusa, which lubricated the
 communication between the U.S. and Japanese astronomers. 
    I thank Yoshihiro Ueda for kindly teaching me how to ski. I thank
 Michael A. Strauss and Robert Lupton for kindly inviting me to a sushi
 restaurant. 
 I thank Chisato Yamauchi and Reiko Nakajima for constructing a useful software
 called AstroNomical IMage 
 Explore, or ANIME. I thank James Annis and Michael Vogeley for rescuing me
 at Fermilab. 
 I thank Ivan K. Baldry for showing me around a castle.  
 I thank J. Brinkmann for always appreciating my publications. 
 I thank Ani Thakar, Tamas Budavari, Istvan Szapudi, Marc SubbaRao,
 Mariangela Bernardi, Osamu Nakamura, Erika Kamikawa for
 kindly feeding deer in  my hometown, Nara. 
 I thank Yeong-Shang Loh, Lei Hao, Randall Rojas and Roy Gal for making
 a good friend with  me.       
 I thank Rita Kim and Neta Bahcall for their hospitality during my stay
 in Princeton. 
 I thank our secretary Masumi Nakaya for bringing a  wonderful smile to the office. 
 I thank David Hogg, Michael Blanton and Alex Quintero for a useful
 discussion on Appendix A \& B. 
  I am grateful to Sidney van den Bergh and Don York, for valuable
 comments on Chapter \ref{chap:PS} of this work.
  I appreciate Andrew and Tony Hopkins for their friendship during my
 stay in Pittsburgh.
  I would like to extend my appreciation to  A. Kathy Romer, Richard
 Griffiths, David A. Turnshek, Andrew Connolly, David Johnston, Erin 
 Scheldon, Sara Hansen, Satoru Ikeuchi, Yasushi Suto, Katsuhiko Sato,
 John Peoples, and Sandhya Rao
 for their kindness.  I thank Ian Smail for sending me the data that I needed
 before I asked. I thank Takashi Okamoto and Naoyuki Tamura for their
 hospitality during my visit in Durham. 
 I thank Dajana Dzanovic for useful scientific communication through
 e-mail.  I thank Chiaki Kobayashi for organizing the galaxy seminar.

   I thank the SDSS collaboration for creating such a wonderful data
 set. Working with such a great data set for my four years of graduate
 life has been quite an experience. 
   I thank Department of Physics of Carnegie Mellon University for its
 hospitality during my visit.  
   I acknowledge financial support from the Japan Society for the
 Promotion of Science (JSPS) through JSPS Research Fellowships for Young
 Scientists. This work owes its completion to this financial support.

 I thank my family, Hisako and Sayaka Goto for their continuous support for my
 entire life.

 Lastly I deeply thank my wife Miki Goto for her continuous support and
 genuine affection. 

\bigskip
 \begin{center}
 With pleasure, I dedicate this work to Miki.
 \end{center}
  
  \begin{flushright}
\bigskip
   
                                                    Tomotsugu Goto, June 6, 2003.
 \end{flushright}
\tableofcontents 

\begin{abstract}

 We investigated environmental effects on galaxy evolution using the
 Sloan Digital Sky Survey (SDSS) data. By developing a new, uniform galaxy
 cluster selection method (the Cut \& Enhance method), we have created
 one of the largest, most uniform galaxy cluster catalog with well
 determined selection function. Based on this
 cluster catalog, we derived extensive observational evidence of cluster
 galaxy evolution.

 Composite
 luminosity functions (LF) of these cluster galaxies show that cluster
 LFs have  a brighter characteristic magnitude ($M^*$) and a flatter 
       faint end slope than field LFs.    We also found that early-type
 galaxies always have flatter slopes 
       than  late-type galaxies. These results suggest that cluster
 galaxies have a quite different evolutionary history from that of field galaxies.

 We confirmed the existence of the Butcher-Oemler effect
 as an increase of fractions of blue cluster galaxies with 
 increasing redshift. This is direct evidence of $spectral$ evolution of
 cluster galaxies. Cluster galaxies evolve by changing their color from
 blue to red, perhaps reducing their star formation rate (SFR).
 We also found that fractions of morphologically spiral galaxies are
 larger in higher redshift. This ``morphological Butcher-Oemler effect'' is
 shown for the first time using an automated galaxy classification, and
 is direct evidence of $morphological$ cluster galaxy evolution.
 Cluster galaxies change their morphology from spiral to early-type galaxies.
  In
 addition to the redshift evolution, we found the slight dependence of
 blue/spiral fractions on cluster richness, in a sense that richer
 clusters have smaller fractions of blue/spiral galaxies. This result
 has significant implication for the underlying physical mechanism since
 it is  consistent with a theoretical prediction of a ram-pressure
 stripping model, where richer clusters have more effective ram-pressure. 

 While investigating the morphology-density relation in the SDSS, we
 found two characteristic environments where the morphology-density relation abruptly
 changes.
   In the sparsest regions (galaxy density below 2 galaxy Mpc$^{-2}$ or outside of 2 virial
 radius), the morphology-density relations  become less notable,
 suggesting that the
 responsible physical mechanisms require denser environment. 
    In the intermediate density regions, (galaxy density between 2 and 6
 galaxy 
 Mpc$^{-2}$ or virial radius between 0.3 and 2), S0 fractions increase
 toward denser regions, whereas late-spiral fractions
 decrease. Considering the median size of S0 galaxies are smaller than
 that of late-spiral galaxies and star formation rate radically declines in
 these regions, the mechanism
 that gradually reduces star formation might be responsible for
 morphological changes in these intermediate density regions
 (e.g., ram-pressure stripping, strangulation). 
    In the cluster core regions (above 6 galaxy Mpc$^{-2}$ or inside of
 0.3 virial radius), S0 fractions decreases radically and elliptical
 fractions increase. This is a contrasting results to that in
 intermediate regions and it suggests that yet another mechanism is
 responsible for morphological change in these regions.

   Finally, we found that passive spiral galaxies preferentially live in
 cluster infalling regions (galaxy density 1$\sim$2 Mpc$^{-2}$ and
 1$\sim$10 virial radius).  Thus the origins of passive spiral galaxies
 are likely to be cluster related.   The existence of passive spiral
 galaxies suggest that a physical mechanism that works calmly is
 preferred to dynamical origins such as major merger/interaction since such a
 mechanism can destroy spiral arm structures. 

   Considering all the observational results, we propose a new
   evolutionary scenario of cluster infalling galaxies. 
   Around 2 virial radius or galaxy density below 2 galaxy Mpc$^{-2}$, 
 infalling field spiral galaxies quiescently stop their star
 formation and are transformed into
 passive spiral galaxies calmly.  These  passive spiral galaxies later become S0
 galaxies. Possible responsible physical mechanisms in this region include ram-pressure
 stripping, strangulation, thermal evaporation and minor mergers,
 perhaps mainly happening in sub-clump regions around a cluster.
    In the cluster core regions, we speculate that S0 galaxies fade
    away to enhance the dominance of old bright elliptical galaxies.

 \end{abstract}
 


\chapter{General Introduction}

\begin{figure}[h]
\includegraphics[scale=0.7,angle=0]{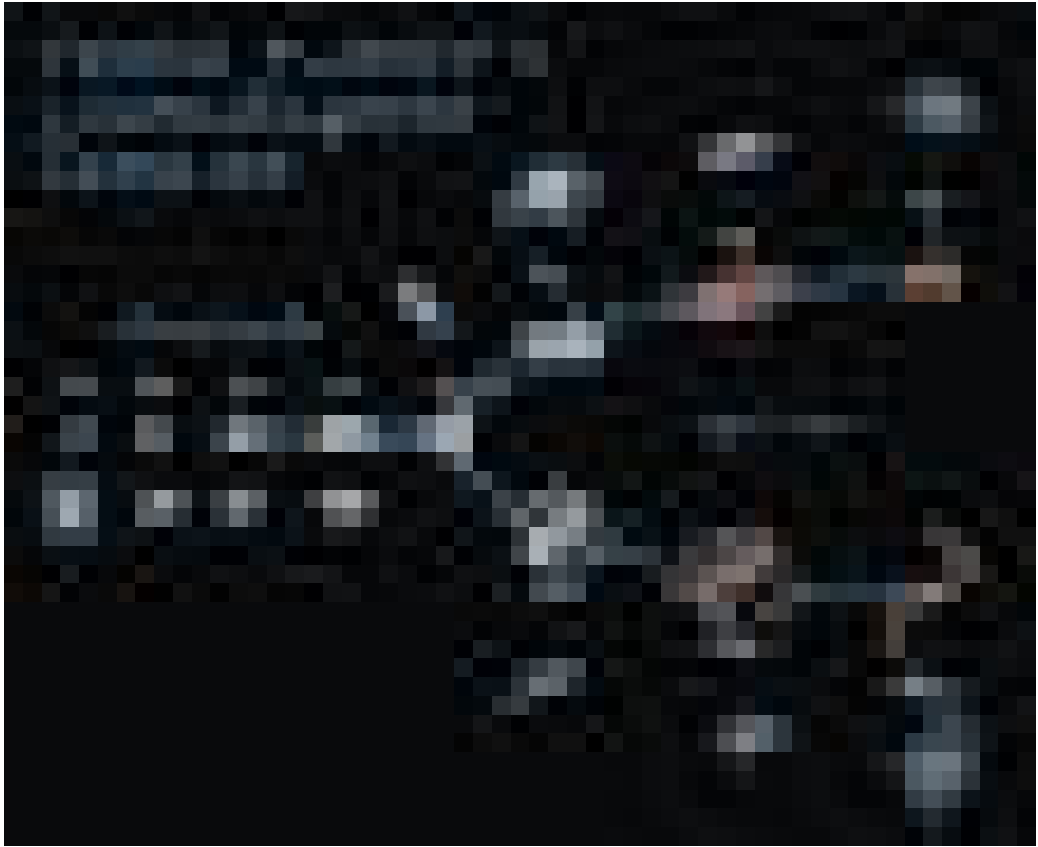}
\caption{Hubble's tuning fork diagram. 
 }\label{fig:hubble}
\end{figure}

 It is a remarkable feature that various properties of galaxies vary
 according to  their environments. In 1926, Hubble classified galaxies
 according to their morphology along the tuning fork diagram (Figure
 \ref{fig:hubble}). This is 
 the so-called Hubble's tuning fork diagram, and is still commonly used
 to understand variety in galaxy morphology. However, little is known
 on the origin of this variety in galaxy morphology.
 It has been known for a long time
 that galaxy 
 cluster regions are dominated by bright elliptical galaxies, whereas
 star-forming spiral galaxies are more numerous in the field. Many
 people observed that cluster
 elliptical galaxies have tight correlation between their color and magnitude
(so-called color-magnitude relation). This variety in galaxy properties
 and correlation with galaxy environments
 have driven many
 researchers to study galaxies and clusters. However, only little has been
 known on the origin of this variety in galaxies. 
   The purpose of this work is to clarify observationally 
 environment-dependent effects which affect galaxy properties causing the evolution of galaxies,
 in order to understand underlying physical mechanism governing
 this variety in properties of present day galaxies.


\section{Observational Evidence for Environmental Effects}

\subsection{The Morphology-Density Relation}

 In his pioneering work, 
 Dressler (1980) studied 55 nearby clusters and found that the fraction of
 elliptical galaxies are higher in the denser regions such as cluster
 cores. This correlation between galaxy morphology and its environment
 is called the morphology-density relation, and has been
 observed by so many people until today.
    Postman \& Geller (1984) extended morphology study
 to galaxy groups (Typically a group consists of  a few to dozens of galaxies.) using the
 data from the CfA Redshift Survey (Huchra et 
 al. 1983). The relation was completely consistent with Dressler
 (1980). At low densities below galaxy density $\sim$5 Mpc$^{-3}$,
 population fractions seems to be independent 
 of environment. At high density above $\sim$3000 galaxies
 Mpc$^{-3}$, the elliptical fraction increased steeply. 
    Whitmore et al. (1993) re-analyzed the 55 nearby clusters
 studied by Dressler (1980) and argued that the morphology-radius relation is more
 fundamental; the correlation between morphology and cluster centric radius seems tighter
 than the morphology-density relation. 
    Dressler et al. (1997) studied 10 high redshift clusters at $z\sim$0.5
 and found that the morphology-density relation is seen for centrally
 concentrated clusters. However, the relation was nearly absent for less
 concentrated or irregular clusters.    
     Fasano et al. (2000) studied nine clusters at intermediate redshift 
 (0.1$\leq z\leq$0.25) and found that
 morphology-density relation in clusters with high elliptical concentration,
 but not in those with low elliptical concentration.
    Hashimoto et al. (1999) used data from Las Campanas Redshift Survey
 (LCRS; Shectman et al. 1996) to study the concentration-density
 relation. They found that the ratio of high to low concentrated
 galaxies decreases smoothly with decreasing density.    
    Dominguez et al. (2001) analyzed nearby clusters with X-ray and found
 that mechanisms of global nature (X-ray mass density) dominate in high
 density environments, 
 namely the virialized regions of clusters, while local galaxy density
 is the relevant parameter in the outskirts where the influence of
 cluster as a whole is relatively small compared to local effects.  
    Dominguez et al. (2002) studied groups in 2dF Galaxy Group Catalog
 using PCA analysis of spectra as a galaxy classification and local galaxy
 density from redshift space as a measure of galaxy environment. 
 They found that both morphology-density relation and
 morphology-group-centric radius relation is clearly seen in high mass
 ($Mv\geq$10$^{13.5}M_{\odot}$) groups, but neither relation is observed for
 low mass ($Mv<$10$^{13.5}M_{\odot}$) groups. 


  Apparently previous studies have shown the fact that there is a
  correlation between galaxy morphology and environment. However, previous
  studies often had the following uncertainties; (i) two dimensional density
  estimation from the imaging data, (ii) subjective, eye-based galaxy
  classification, (iii) lack of the field data. Therefore, it is important to rectify these
  problems using a larger, more uniform data with three dimensional
  information, such as the SDSS data used in this work. Furthermore, to
  understand the underlying physical mechanism, it is of importance to
  specify the exact environment where galaxy morphology starts to change and the accurate
  amount of galaxies which undergo morphological change. 
  We try to  address these problems using the SDSS data in Chapter
  \ref{chap:MD}. 
 In addition to clarify various correlations between galaxy properties as
  have been claimed in previous work, 
 high quality and large quantity of the SDSS data offers us, for the
  first time, a chance to understand underlying physical mechanisms.

\subsection{Suppression of Star Formation in Cluster Regions}

   Similarly, there have been extensive evidence that cluster galaxies
   have smaller amount of star formation, with
   redder colors than field galaxies. It has been known for a long time that cluster
   regions lack emission line galaxies (Osterbrook 1960; Gisler 1978;
   Dressler, Thompson, \& Shectman 1985). 
    Recently, many studies reported that star formation in the cores of
 clusters is much lower than that in the surrounding field (e.g., Balogh
 et al. 1997,1998,1999; Poggianti et al. 1999; Martin, Lotz \& Ferguson 2000;
 Couch et al. 2001; Balogh et al. 2002).

   In addition, it is becoming possible to specify the environment where
    star formation in galaxies starts to change.
       Abraham et al.(1996) reported that cluster members
 become progressively bluer as a function of cluster-centric distance
 out to 5 Mpc in A2390 ($z=$0.23). 
     van Dokkum et al. (1998) found S0 galaxies in the outskirts of a
 cluster at $z=$0.33. These S0s show a much wider scatter in their colors
 and are bluer on average than those in cluster cores, providing
 possible evidence for recent infall of galaxies from the field. 
  Terlevich, Caldwell \& Bower (2001) reported that $U-V$
 colors of early-type galaxies are systematically bluer at outside the
 core of Coma cluster.  
   Pimbblet et al. (2002) studied 11 X-ray luminous clusters
 (0.07$<z<$0.16) and found that median galaxy color shifts bluewards
 with decreasing local  galaxy density. 
   At higher redshifts, Kodama et al. (2001) reported that colors of
 galaxies abruptly change at sub-clump regions surrounding a cluster
 at $z=$0.41.
      Recently, using statistically much larger sample, Lewis et
   al. (2002) and Gomez et al. (2003) showed that 
   star formation rate (SFR) of galaxies decreases toward cluster cores
   at around 1 virial radius.
      All of these results are often interpreted as the result of star-forming
    field disk galaxies infalling to a cluster, being transformed to
       passive cluster galaxies. 
    It is of importance to observationally
   specify the environment where galaxy properties start to change and how much
   they change 
   in order to understand
   the underlying physical mechanisms responsible for these differences.


\subsection{Color-Magnitude Relation}

  Cluster galaxies are known to have tight color-magnitude
  relation. 
      Visvanathan \& Sandage (1977) noted that bright early-type
  galaxies in Coma and eight other local clusters have a tight
  correlation between their colors and magnitude, in a sense that
  brighter galaxies have redder colors (the color-magnitude
  relation; CMR). Bower, Lucey \& Ellis (1992) later studied the CMR in
  detail using high precision photometry of early-type galaxies in Virgo
  and Coma clusters to find little scatter around the mean CMR.
    Various other studies also confirmed the existence of
  tight CMR in both low redshift cluster (Sandage \& Visvanathan 1978;
  van Dokkum et
  al. 1998; Pimbblet et al. 2002 ), and high redshift
  clusters(Aragon-Salamanca et al. 1993; Stanford  et al. 1995,1998; Dickinson 1996; Ellis et
  al. 1997; Kodama et al. 1998; van Dokkum et al. 2000). 
   These
  studies are in remarkable agreement: the slope and scatter of the CMR seem to be
  roughly constant with passive evolution of an old stellar population
  formed at high redshift. The rms of scatter about the mean CMR is
  typically $\sim$0.04 mag, a comparable size to observational errors,
  implying a virtually negligible intrinsic scatter. The universality 
  of the CMR also suggests the different star formation history between
  field galaxies and cluster galaxies.
   It is one of the purposes of this study to observationally collect the
  evidence for cluster galaxy evolution in order to understand the
  underlying physical mechanisms governing cluster galaxy evolution.

\subsection{Evolution of Cluster Galaxies}

\begin{figure}[h]
\includegraphics[scale=0.7,angle=0]{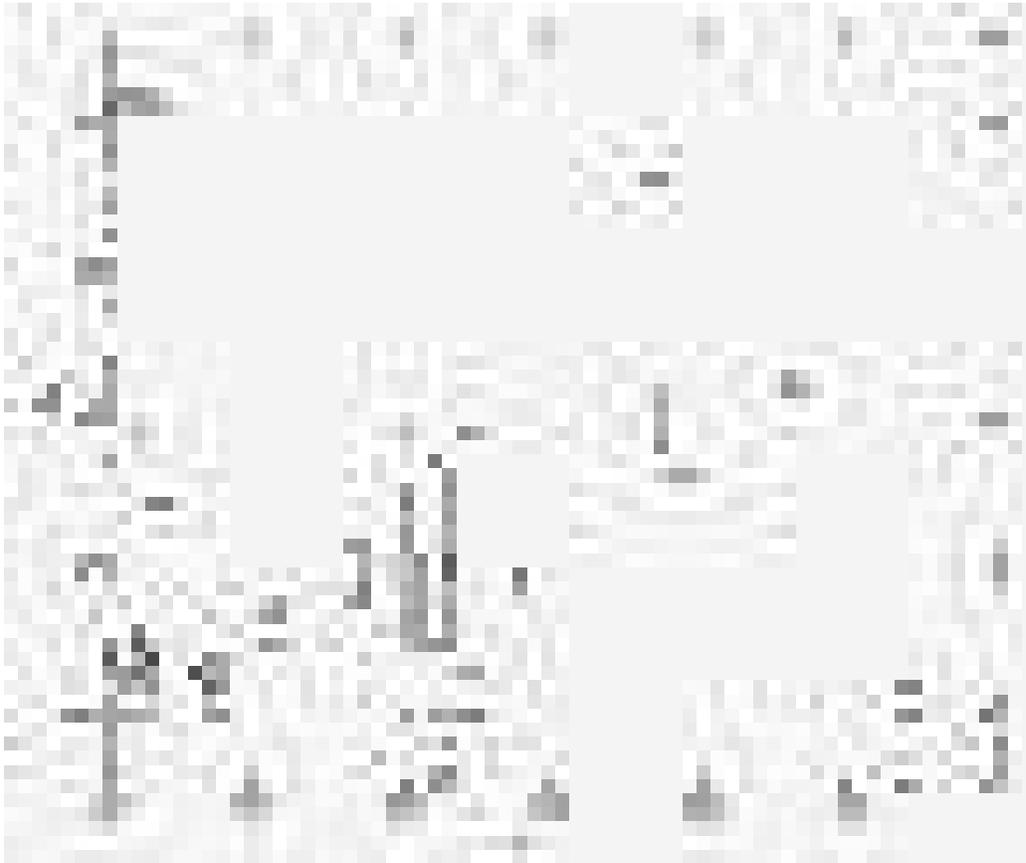}
\caption{Blue galaxy fraction vs. redshift. Filled circles, compact
 clusters ($C \geq 0.40$); open circles, irregular clusters ($C<0.35$);
 dotted circles, intermediate clusters ($0.35 \leq C < 0.40$).  
 }\label{fig:bo84}
\end{figure}

    There also have been direct evidence for cluster galaxy evolution.
   Butcher \& Oemler (1978, 1984) found that
 fractions of blue galaxies are larger in the past, showing that cluster
 galaxies evolve from blue to red (so-called the
 Butcher-Oemler effect; Figure \ref{fig:bo84}).  Butcher and Oemler's work made a strong impact
   since it showed direct evidence for the 
 evolution of cluster galaxies. Much work regarding the nature of
 these blue galaxies followed. Rakos \& Schombert (1995) found that the
 fraction of blue galaxies increases from 20\% at $z=$0.4 to 80\% at
 $z=$0.9, suggesting that the evolution in clusters is even stronger than
 previously thought. Margoniner \& De Carvalho (2000) studied 48
 clusters in the redshift range of 0.03$<z<$0.38 and detected a strong
 Butcher-Oemler effect consistent with  that of Rakos \& Schombert (1995). 
 Despite the global trend with redshift,  almost all previous work has reported 
 a wide range of blue fraction values at respective redshifts. 
 In particular, in a large sample of 295 Abell clusters, Margoniner et al. (2001)
 not only confirmed the existence of the Butcher-Oemler effect,
 but also found the blue fraction depends on cluster richness.
  Ellingson et al. (2001) studied 15 CNOC1 clusters (Yee, Ellingson, \&
 Carlberg 1996) between $z=$0.18 and $z=$0.55. Since they used spectroscopically observed
 galaxies, they do not suffer from the fore/background
 contamination (but see Diaferio et al. 2001). They used galaxies  brighter than $M_{r}=-19.0$ 
  and found a blue fraction of $\sim$0.15 at $z=$0.3.

\begin{figure}[h]
\includegraphics[scale=0.7,angle=0]{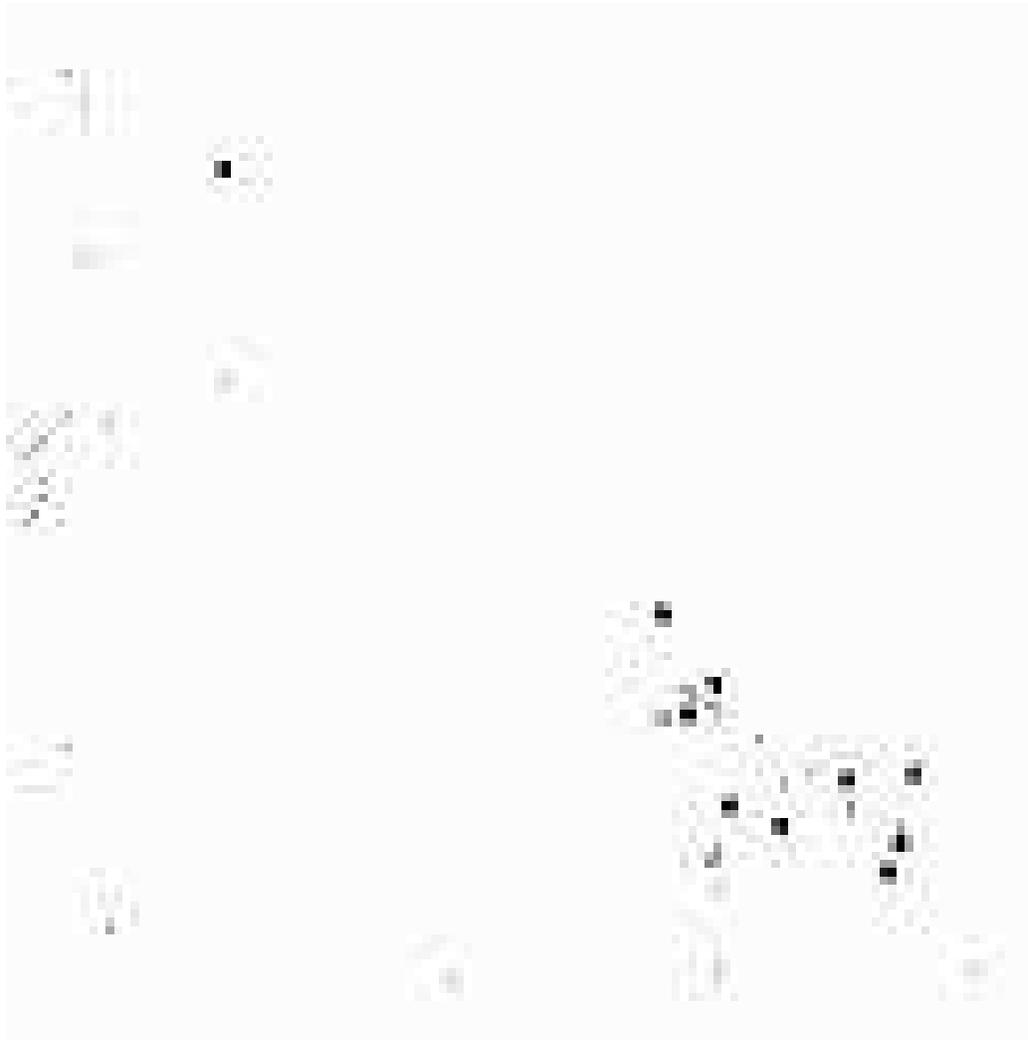}
\caption{The S0/E fraction for clusters in the sample as a function of
 redshift. The open circles are the outer fields in A370 and Cl 0939+47,
 which are not used in the least-squares fit, shown with its 1 $\sigma$
 errors as the solid and dotted lines. The extrapolation of this linear
 relation to zero redshift approximately matches the value S0/E $\sim$
 2, shown by the filled square, found for 11 clusters with 0.035 $<z<$
 0.044 of the D80 cluster sample (Dressler et al. 1997). 
 }\label{fig:D97}
\end{figure}

  In addition to the color evolution of cluster galaxies, morphological
   evolution of cluster galaxies have been observed.
 Dressler et al. (1997) studied 10 high redshift clusters at $z\sim$0.5
 and found that S0 fractions
 are much smaller than those in nearby clusters, suggesting that S0 galaxies
 are created fairly recently ($z\leq$0.5; Figure \ref{fig:D97}). 
  Fasano et al. (2000) studied 9 intermediate redshift clusters and plotted morphological
 fraction as a function of redshift to find that S0 fraction decreases
 with increasing redshift, whereas spiral fraction increases with
 redshift. 
     There also have been observations showing the deficit of S0
   galaxies in high redshift clusters (Andreon et al. 1997; Couch et al. 1998; Lubin et
   al. 1998; but also see Andreon et al. 1998; Fabricant et al. 2000).
  These observational results show us that cluster galaxies are likely
  to evolve,
  changing their color from red to blue, and their morphology from
  late-type to early-type. 

  However, almost all of previous work had a fundamental problem in
  using a heterogeneous sample of clusters. For example, one can never
  claim the evolution by comparing low redshift clusters observed by
  a ground based telescope with high redshift clusters observed by $the\ Hubble\
  Space\ Telescope$. In such a case, high redshift clusters are much
  richer than low redshift clusters, and thus subjective to richness
  related bias. There have been many people who doubt such an
  evolutionary study with heterogeneous sample (Newberry, Kirshner \&
  Boroson 1988; Allington-Smith et al. 1993;  Garilli et al. 1996;
 Smail et al. 1998; Andreon \& Ettori 1999; Balogh et al. 1999;  Fairley et
 al. 2002). 
 In Chapter \ref{chap:BO}, we try to rectify this problem using the
  largest sample of 514 clusters, uniformly selected from the same SDSS
  data, eliminating a richness related and redshift related bias.

\section{Theoretical Implications}

%

 Various physical mechanisms have been proposed to explain the
 observational results on cluster galaxy evolution. In this section, we aim to summarize
 them by classifying them into three broad categories; (1) interplay
 between  galaxies and intra-cluster medium (2) galaxy-galaxy gravitational
 interaction. (3) gravitational interaction between a galaxy and the cluster potential.

\begin{enumerate}
 \item  Interplay between  galaxies and intra-cluster medium.

	   Ram-pressure stripping
	of cold gas in the disk of a galaxy is a candidate since the
	star formation loses its source due to the removal of the cold gas.
              (Gunn \& Gott 1972; Farouki \&
	Shapiro 1980; Kent 1981; Fujita 1998; Fujita \& Nagashima 1999;
	Abadi, Moore \& Bower 1999; Quilis, 
	Moore \& Bower 2000; Toniazzo \& Schndler 2001; Fujita
	2001,2003). Ram-pressure stripping is likely to be effective in
	the central region of clusters where density of intra-cluster
	medium (ICM) is high. In fact, disk galaxies with a deficit of
	cold gas or a morphological sign of gas removal are often
	observed near cluster centers (e.g., Cayatte et al. 1990; Vollmer
	et al. 2000; Bravo-Alfaro et al. 2000, 2001; Solanes et
	al. 2001; Bureau \& Carignan 2002; Vollmer 2003).  Especially,
	Haynes \& Giovanelli (1986) showed that the frequency of HI
	deficient galaxies projected on the sky increases close to the
	center of the Virgo Cluster. 
        In addition,
	the ram-pressure stripping is predicted to be more effective in
	a richer cluster since the velocity of galaxies is larger
	(Bahcall 1977; Fujita et al. 1999,2001). This prediction can be
	observationally tested by studying fractions of blue galaxies as a
	function of cluster richness. We perform such an attempt in
	Chapter \ref{chap:MD}.  
          
         On the other hand, star formation rates of galaxies may
	decrease gradually by stripping of warm halo gas that
	would later infall on to the galactic disks, and  be the
	source of cold gas and stars in the galactic disks
            (strangulation; Larson, Tinsley \& Caldwell 1980; 
	Balogh et al. 2001; Bekki et al. 2002; Mo \& Mao 2002; Oh \&
	Benson 2002). If only warm halo gas is stripped, star formation may
	be allowed to continue by consuming remaining cold disk gas but,
	without infalling gas from halo to replenish this supply, star
	formation will die out on timescales of a few Gyr (Larson et
	al. 1980). Observationally, Finoguenov et al. (2003)
 found in Coma cluster the filamentary gas where strangulation is likely to be happening and
	they predicted quiescent star 
  formation in galaxy disks around the filament.  

          Another mechanism to slow down the star formation rate of
	galaxies is thermal evaporation of the cold gas in disk galaxies via
	heat conduction from the surrounding hot ICM (Cowie \& Songaila
	1977; Fujita 2003).
	 Pressure triggered star formation,  in which galactic gas clouds
	are compressed by the ICM pressure,
         can temporarily increase the star formation rate 
         (Dressler \& Gunn 1983; Evrard 1991;	Fujita 1998).

	All of the above mechanisms need relatively high density of hot
	intra-cluster gas, and thus likely to happen in the central region
	of clusters. Although several authors indicated that these
	mechanisms can not explain the the suppression of star formation
	as far as several Mpc from the center of a cluster (Balogh et
	al. 1997; Kodama et al. 2001; Lewis et al. 2002), 
	Fujita et al. (2003) pointed out that these
	mechanisms can happen in cluster sub-clump regions (small groups
	around a cluster). 
         The above
	mechanisms mainly affect the star formation rate of a galaxy,
	compared with the following dynamical mechanisms which directly
	affect the morphology of a galaxy. 

 \item  Galaxy-galaxy gravitational interaction. 

	This category includes interaction/merging of
 galaxies (Icke 1985; Lavery \& Henry 1988; Bekki 1998). Mergers between
	galaxies with compatible masses (major mergers) could create
	an elliptical galaxy (Toomre \& Toomre 1972). However, in the
	semi-analytic simulation, it is known that the observed
	fraction of intermediate bulge-to-disk ratio galaxies can not be
	explained solely by major mergers (Okamoto \& Nagashima 2001;
	Diaferio et al. 2001; Springel et al. 2001). Okamoto \&
	Nagashima (2003) indicated that mergers between galaxies with
	significantly different masses (minor mergers) might play an
	important role in creating intermediate  bulge-to-disk ratio
	galaxies. 
	Galaxy harassment via high speed 
	impulsive encounters also can affect morphology and star
	formation rate of cluster galaxies (Moore et al. 1996, 1998,
	1999).
  
 \item  Gravitational interaction between a galaxy and the cluster potential.

	Tidal compression of galactic gas via interaction with the
	cluster potential can  increase the star
	formation rate of a galaxy (Byrd \& Valtonen 1990, Valluri 1993;
	Fujita 1998). 
%

\end{enumerate}

    Unfortunately, there exists little evidence demonstrating that any one
    of these processes is actually responsible for driving galaxy
    evolution. Most of these processes act over an extended period of
    time, while observations at a certain redshift cannot easily provide
    the detailed information that is needed to elucidate subtle and
    complicated processes. To extract useful information in comparison
    with the observational data, it is of importance to
  specify the environment where galaxy properties change. For example,
  stripping mechanisms (ram-pressure stripping, strangulation) and
  evaporation require dense hot gas, and therefore, can not happen far
  away from a cluster where gas density is too low. They can happen only in cluster cores or in dense
  sub-clumps (Fujita et al. 2003). Merging/interaction is probably difficult to
  happen in cluster cores since relative velocity of galaxies are too
  high in such regions in the present universe  (Ostriker 1980; Binney
  \& Tremaine 1987; Mamon 1992; Makino \& Hut 1997).  It is also
  important to investigate the morphology of galaxies which are currently
  undertaking transformation. If major merging/interaction is the responsible
  mechanism, galaxies should have disturbed signature (e.g., tidal tails,
  multiple cores) in their
  morphology. If minor merger is dominant, galactic bulges may become
  larger and larger during the evolution (Ohama 2003).
  If stripping or evaporation is more effective, galaxies should
  be transformed calmly, reducing their luminosity and sizes. 
   In this work, we observationally try to clarify these points by
    examining morphology of galaxies in Sections \ref{chap:MD} and \ref{chap:PS}.

\section{Need for Larger, More Uniform Samples}
 
 The last two sections described various observational and theoretical
  implications on the evolution of cluster galaxies.
  However, almost all observational researches in the literature suffered from the
 lack of a large, uniform cluster catalog and galaxy catalog. The most
 commonly used cluster catalog, Abell cluster catalog (Abell 1958, Abell, Corwin
 and Olowin 1989) was constructed by the visual inspection of
 photographic plates. Although human eyes are an excellent tool to find
 galaxy clusters, it suffers from subjectivity. A computer based
 automated method is needed to create 
 an objective cluster catalog with well understood selection
 function. The effects could be more severe for evolutionary study. For
 example, Butcher \& Oemler (1984) used 10 clusters selected from the
 single color photographic plates to find the spectral evolution of
 cluster galaxies. Since higher redshift clusters are fainter due to
 long distance and redshifting of the flux, higher redshift sample could
 be richer clusters than the nearby sample. Thus it potentially has a
 malmquist type bias. As another example, Fasano et al. (2000) found
 morphological evolution of cluster galaxies, but their cluster sample
 consist of three different cluster samples observed with three different
 telescopes; local clusters of Dressler (1980), intermittent sample of
 their own and high redshift sample of Dressler (1997). It is necessary to
 use a large \& homogeneous sample of clusters when studying evolution
 of cluster galaxies. 

  Furthermore, in many studies, the morphological classifications of galaxies have also been performed
 by human eyes (Dressler et al. 1980,1997; Fasano et al. 2000), and thus
 could have been suffered from the subjectivity. By its nature, human
 based classification change from person to person (Lahav et al. 1995). Even the
 classification by the same person can change according to the 
 condition of the person, such as how tired he is, brightness of the
 display, specifications of the softwares used. 
 Computer based
 automated galaxy classification is preferred since it is easier to
 compute completeness and contamination rate of the classification, and
 to apply the same method to future observational data or computer based simulations. 

 With the advent of the Sloan Digital Sky Survey (SDSS; see Fukugita et
  al. 1996; Gunn et al. 1998;  Lupton 
  et al. 1999,2001; York et al. 2000; Eisenstein et al. 2001;
  Hogg et al. 2001; Blanton et al. 2003a; Pier et
  al. 2002; Richards et al. 2002; Stoughton et al. 2002; Strauss et
  al. 2002; Smith et al. 2002 and Abazajian et al. 2003 for more detail
  of the SDSS data), which is both an imaging and spectroscopic survey of 100,000
 deg$^2$ of the sky, it is now becoming possible to create a large,
 uniform cluster catalog and to study the evolution of cluster galaxies
 using a large, uniform data. The SDSS takes high quality CCD image of a quarter of
 the sky in 5 optical bands ($u,g,r,i$ and $z$), and produce the data
  that enable us to classify
 galaxies using an automated method, without depending on human eyes.
  In this thesis, we take a full advantage
 of the SDSS data and study the environmental effects on cluster galaxy
 evolution using one of the largest and most uniform samples to date. The sample
 allows us to reveal the environmental effects on galaxy evolution in much
 more detail than previous study, providing us with the first opportunity
  to understand the underlying physical mechanisms.

 In Chapter \ref{chap:SDSS}, we briefly describe the SDSS data. 
 In Chapter \ref{chap:CE}, we create a new, uniform cluster
 catalog using the SDSS data. 
 In Chapter \ref{chap:LF}, we study composite luminosity
 functions of clusters to reveal a statistical difference between
 cluster galaxies and  field galaxies. 
 In Chapter \ref{chap:BO}, we study the evolution of cluster
 galaxies and reveal that cluster galaxies evolve both spectrally and
 morphologically.
 In Chapter \ref{chap:MD}, we study the morphology-density
 relation and reveal the correlation between galaxy morphology and local
 galaxy density. 
 In Chapter \ref{chap:PS}, we study passive spiral galaxies,
 which turned out to be a galaxy in transition.
  In Chapter \ref{chap:fate}, we discuss  theoretical scenarios
 consistent with all the observational results.
 In Chapter \ref{chap:summary}, 
 we summarize this work. 
  Over the last decade, the origin of post-starburst (E+A) galaxies has
 been actively debated, often in connection to cluster related phenomena. We present an
 analysis of this post-starburst phenomena in Appendix \ref{EA1} and \ref{EA2},
 together with the reason why these chapters are not included in the
 main text.   
  The cosmological parameters
 adopted are $H_0$=75 km s$^{-1}$ Mpc$^{-1}$, and
 ($\Omega_m$,$\Omega_{\Lambda}$,$\Omega_k$)=(0.3,0.7,0.0), unless
 otherwise stated. 

\newpage

\newpage

\newpage
\chapter{The Sloan Digital Sky Survey}
\label{chap:SDSS}

%

\section{The Sloan Digital Sky Survey}
\label{Mar 12 12:23:11 2003}

\begin{figure}[h]
\includegraphics[scale=0.7,angle=0]{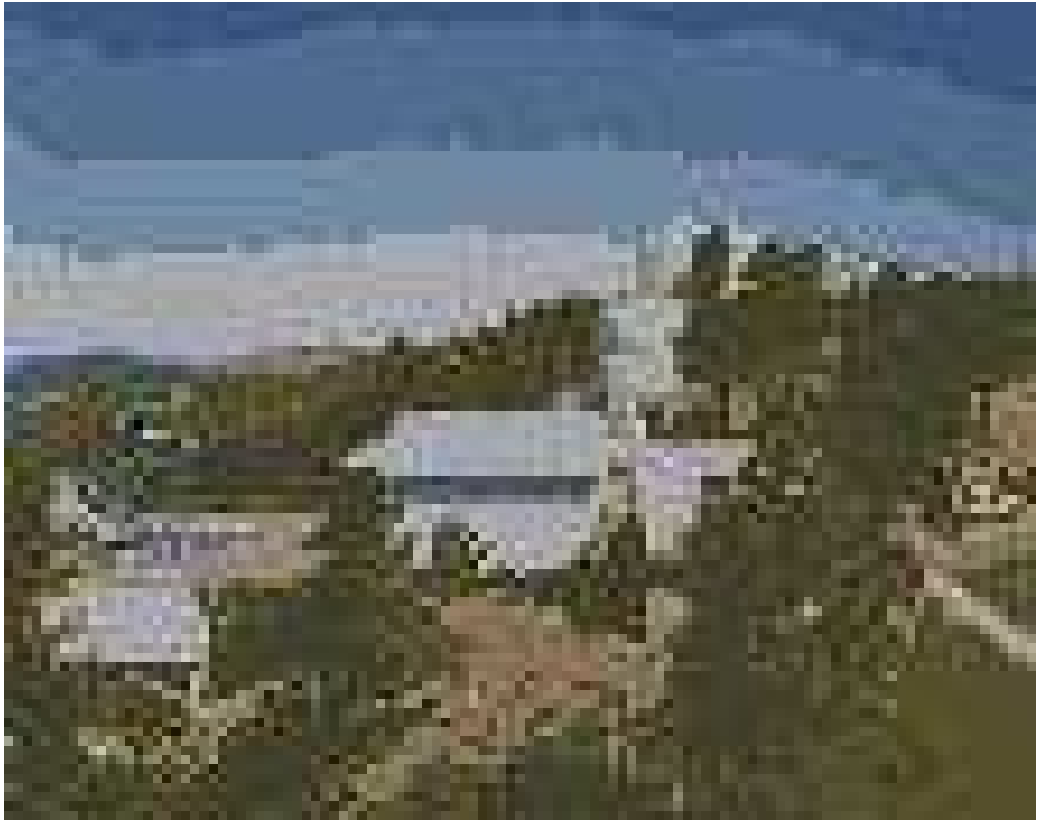}
\caption{The Apache Point Observatory in New Mexico, U.S.A. 
 }\label{fig:apo_small}
\end{figure}

\begin{figure}[h]
\includegraphics[scale=0.7]{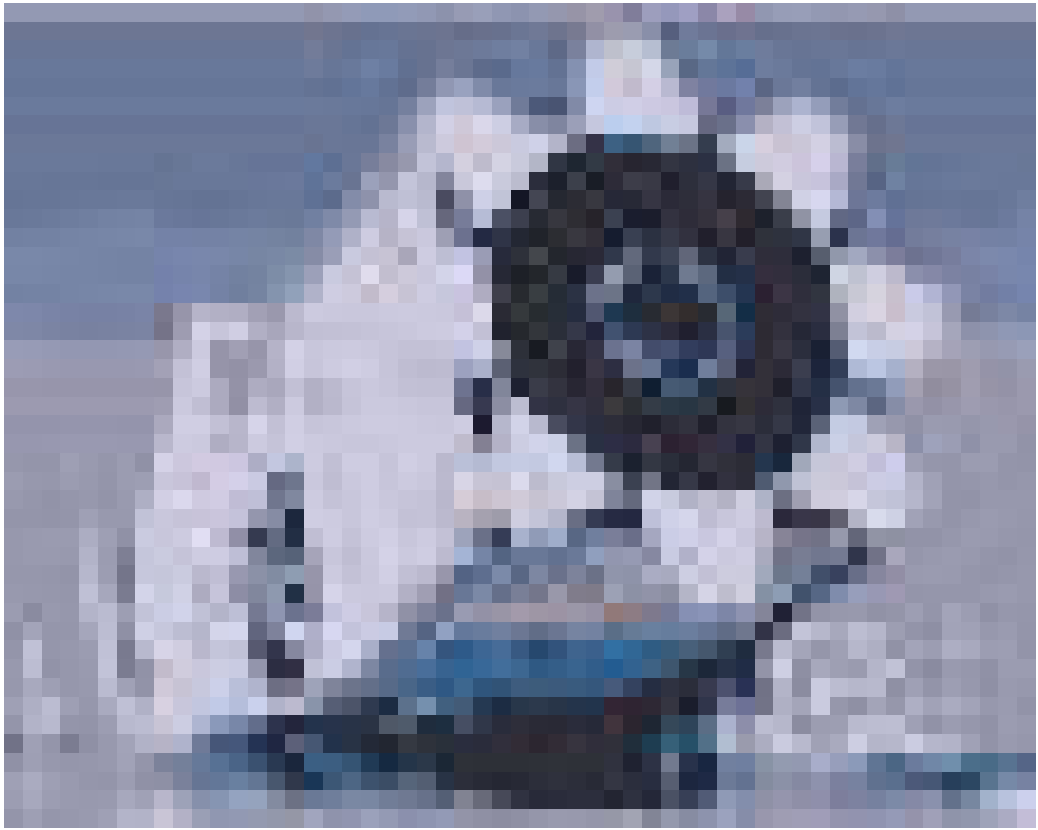}
\caption{The SDSS 2.5-meter telescope located at Apache Point Observatory
 }\label{fig:telescope}
\end{figure}

 The Sloan Digital Sky Survey (SDSS; see Fukugita et al. 1996; Gunn et al. 1998;  Lupton
  et al. 1999,2001; York et al. 2000; Eisenstein et al. 2001;
  Hogg et al. 2001; Blanton et al. 2003a; Pier et
  al. 2002; Richards et al. 2002; Stoughton et al. 2002; Strauss et
  al. 2002; Smith et al. 2002 and Abazajian et al. 2003 for more detail
  of the SDSS data.) is both an
 imaging and spectroscopic survey of a quarter of the
 sky. The SDSS began its operation in November 2000.
 For both imaging and spectroscopy, the SDSS uses a dedicated 2.5m
 telescope located at Apache Point Observatory in New Mexico, U.S.A
  (Figure \ref{fig:apo_small}). 
 The SDSS telescope has a modified f/5.0 Ritchy-Chretien optical design
 with a large secondary and two corrector lenses below the primary
 mirror.  The telescope has a wide ($3^{\circ}$ diameter),   essentially
  distortion-free field of view. 
 The SDSS telescope is shown in Figure \ref{fig:telescope}. 
 It is a unique feature of the SDSS that the telescope  will carry
  out both imaging and spectroscopic surveys of the region 
 of about  $\pi$ steradians centered on the north Galactic pole in five
 years.   Imaging part of the survey obtains CCD images of 10,000 deg$^2$
 of the sky in five
 optical bands ($u,g,r,i$ and $z$; Fukugita et al. 1996). The
 spectroscopic part of the survey observes one million galaxies
  uniformly selected
 from the imaging part of the survey. 
 We use this excellent data set to tackle the long standing problems on
 environmental effects on galaxy evolution. In this chapter, we briefly
 describe the design and products of the SDSS.

\subsection{The SDSS Imaging Survey}

\begin{figure}[h]
\includegraphics[scale=0.8]{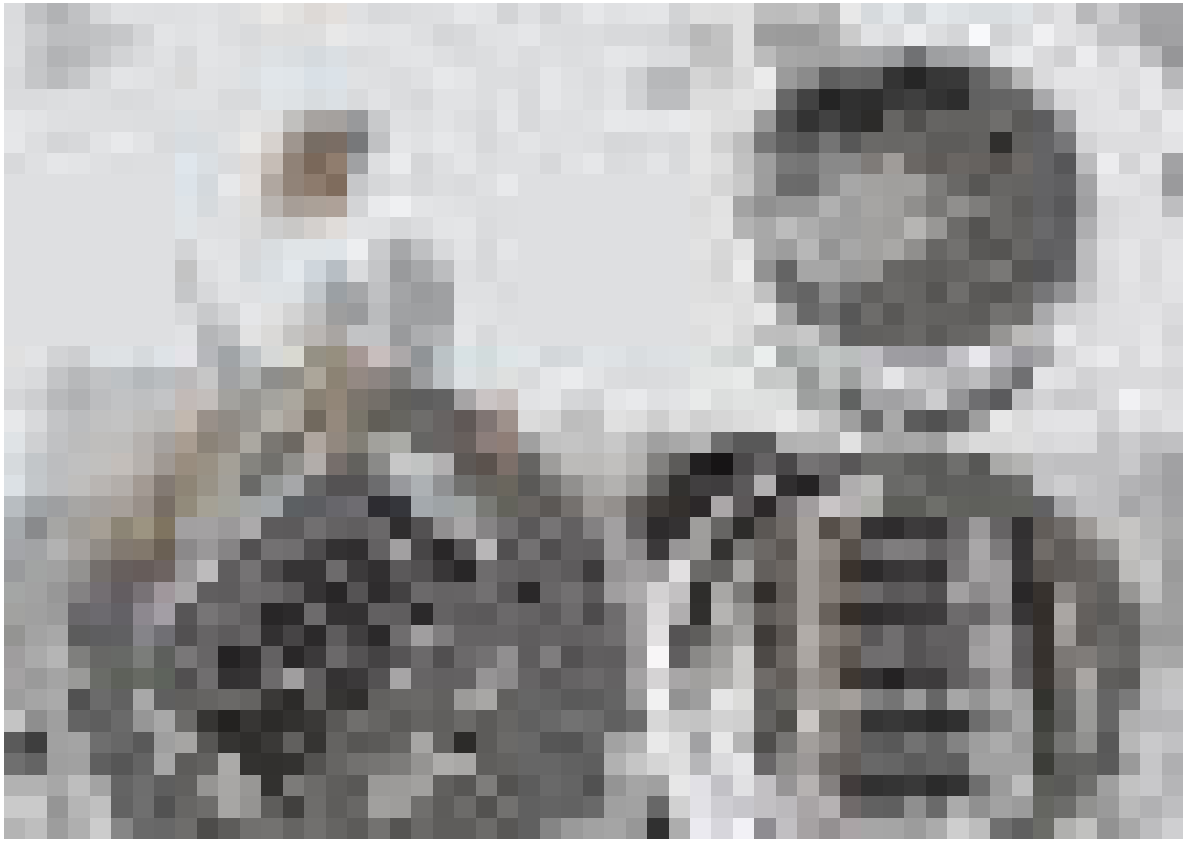}
\caption{The imaging camera of the SDSS. The five broadband filters are
 aligned in an order of $riuzg$. The narrow
 dark rectangles are astrometric CCDs.  
 }\label{fig:camera}
\end{figure}

\begin{figure}[h]
\includegraphics[scale=0.7]{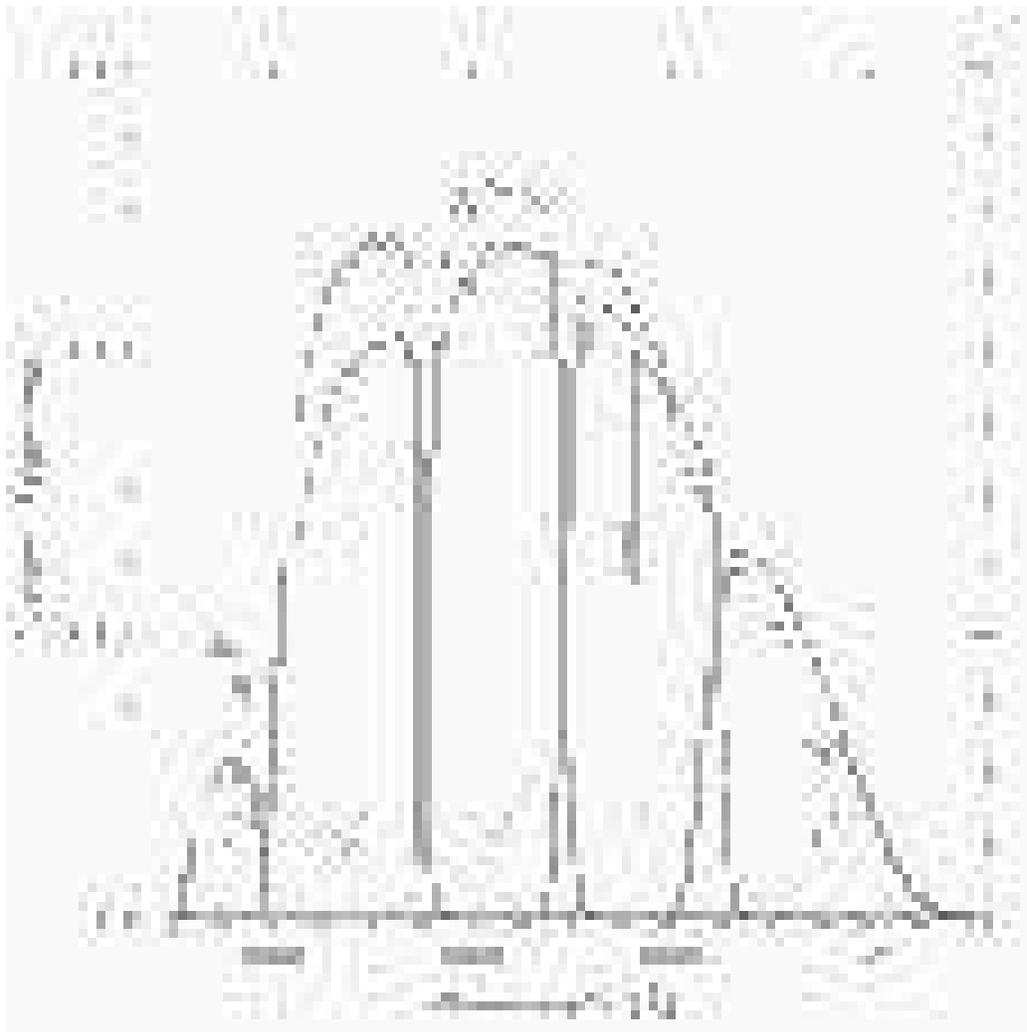}
\caption{The SDSS photometric system response as a function of
 wavelength. The upper curves is without atmospheric
 extinction. The lower curve is with atmospheric extinction when
 observed at a typical altitude of 56\degree.  
 }\label{fig:filter}
\end{figure}

\begin{figure}[h]
\includegraphics[scale=0.7,angle=0]{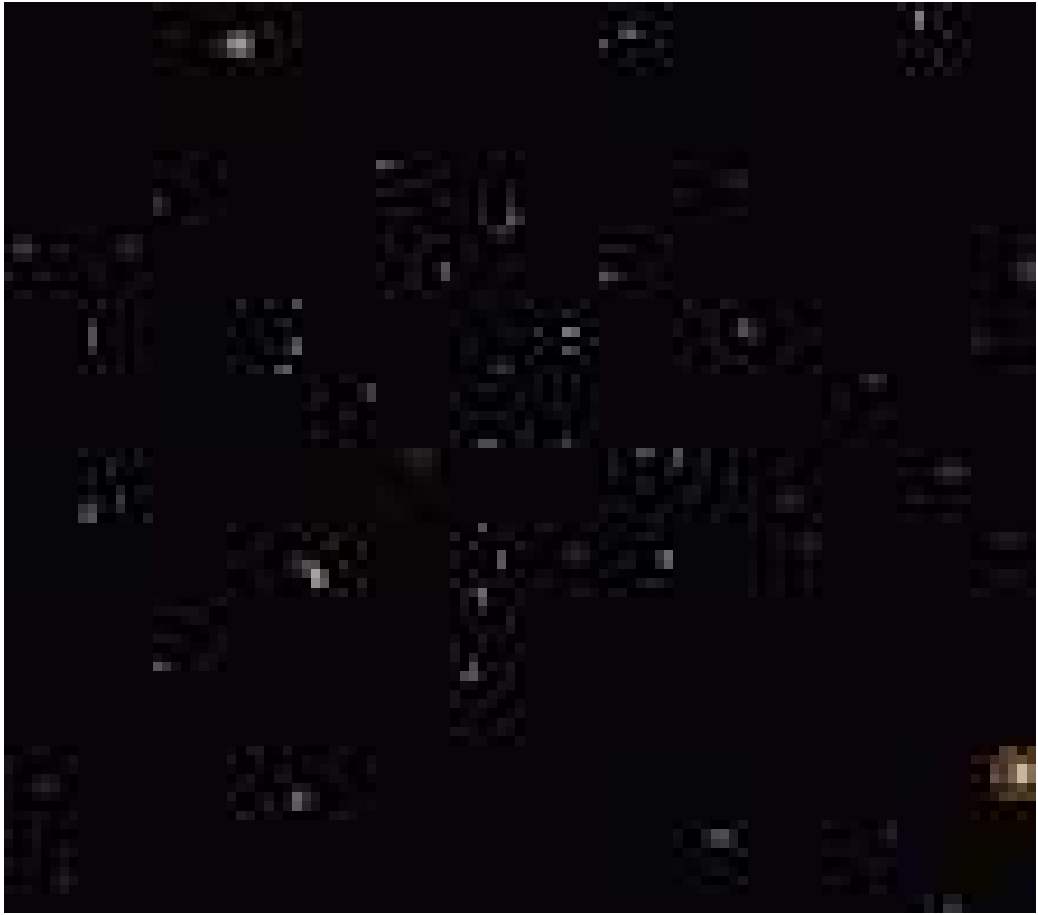}
\caption{A galaxy cluster Abell 168, observed with the SDSS imaging camera.
 }\label{fig:a168}
\end{figure}

\begin{figure}[h]
\includegraphics[scale=0.7,angle=0]{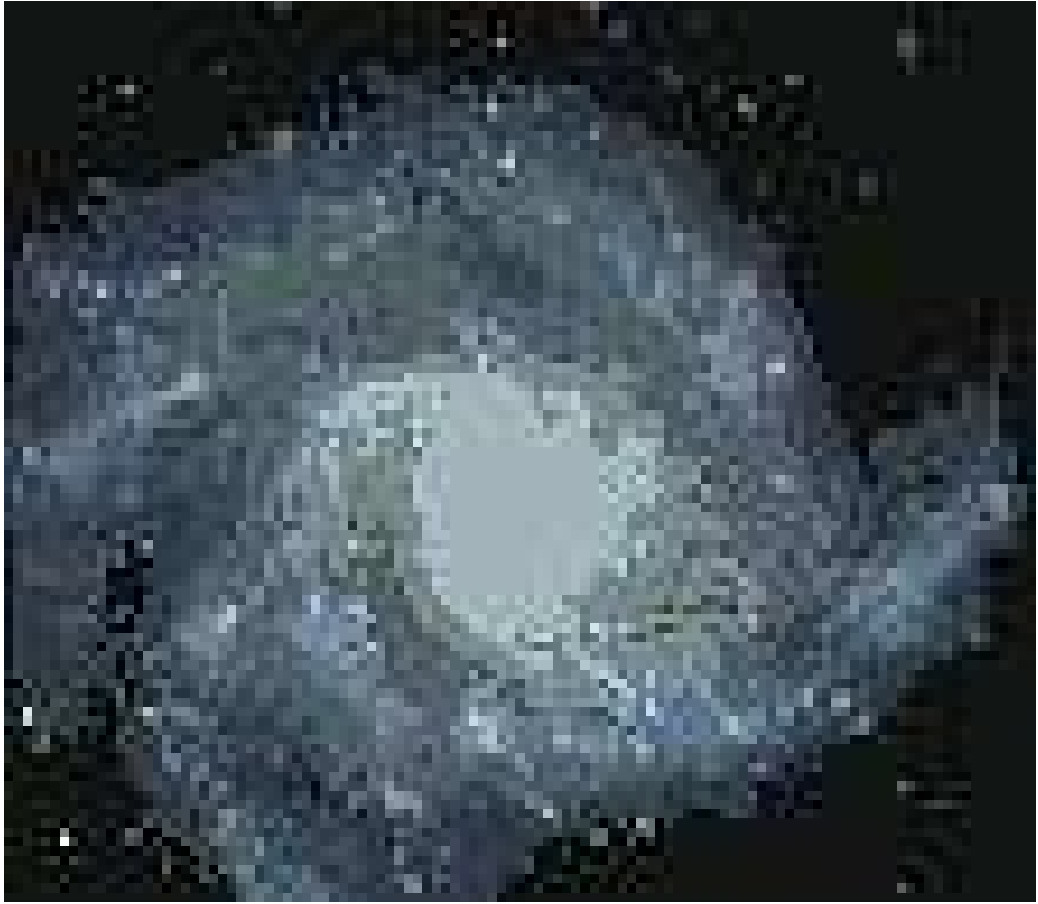}
\caption{A bright spiral galaxy M101, observed with the SDSS imaging camera.
 }\label{fig:m101}
\end{figure}

\begin{figure}[h]
\includegraphics[scale=0.7,angle=0]{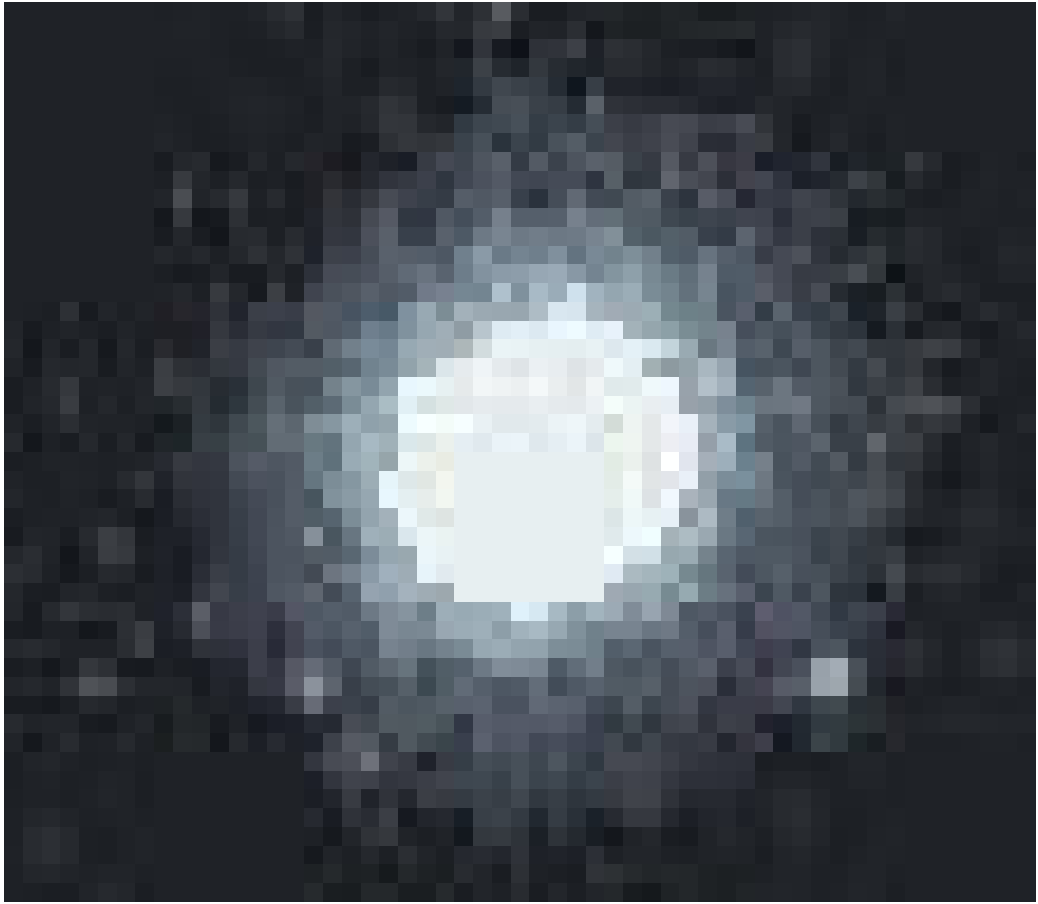}
\caption{A globular cluster M2, observed with the SDSS imaging camera.
 }\label{fig:m2}
\end{figure}

 The imaging part of the SDSS will image 10,000 square degrees in five
bandpasses to a depth of $g \sim 23$ mag with effective wavelengths of
[3561 \AA{}, 4676 \AA{}, 6176 \AA{}, 7494 \AA{} and 8873 \AA{}]
($u,g,r,i$ and $z$)
spanning from the atmospheric cutoff at 300nm to the limit of CCD
sensitivity at 1100nm (Fukugita et al. 1996). 
 The imaging camera (Gunn et al. 1998) consists of a mosaic of 30
  2048 $\times$ 2048 SITe CCDs with 24 $\mu$m pixels as shown in
 Figure \ref{fig:camera}. 
 A pixel in the imaging camera subtends $0.396''$ on the sky.
 The CCDs are
 arranged in six dewars (six columns) containing 5 CCDs each. The camera
 is mounted
 to the corrector plate  and observes the sky through five broad band filters ($u,g,r,i$ and $z$).
 The imaging CCDs saturate at about 14 mag, so to calibrate these data
 with existing astrometric catalogs, the camera contains an additional
 22 CCDs, each having 2048 $\times$ 400 pixels, with neutral density filters
 that saturate at only 8 mag. The photometric calibration is carried out
 with a separate 20'' photometric telescope equipped with a single-CCD
 camera and the SDSS filters (Uomoto et al. 2003). The photometric
 telescope is located near the SDSS telescope at the same site.
 However, at the time
 of writing, the calibration is still in preliminary stage. Thus we
 denote preliminary SDSS magnitudes as $u^*,g^*,r^*,i^*$ and $z^*$,
 rather than the notation $u,g,r,i$ and $z$ that will be used for the
 final SDSS photometric system (and is used in this paper to refer to
 the SDSS filters themselves). The response functions are presented in
 Figure \ref{fig:filter}. The limiting magnitude, i.e., 
 5$\sigma$ detection in $1''$ seeing limits, 
 for point sources are 22.3, 23.3, 23.1, 22.3 and 20.8
 in the $u,g,r,i$ and $z$ filters, respectively, at an airmass of 1.4. 
In general, the telescope is
driven along a great circle on the sky in such a way that objects pass directly
along a column of CCDs. This allows essentially simultaneous observations to be
obtained in each of the five passbands and provides very efficient survey
observing (the shutter never closes). Each object spends 5.7 mins to pass
 the entire CCD array.  Net integration time in each filter is 54.1
 sec.  We show three representative bright astronomical objects observed with the SDSS
 camera in Figures \ref{fig:a168}-\ref{fig:m2}. 
 
\begin{figure}[h]
\includegraphics[scale=0.5,angle=270]{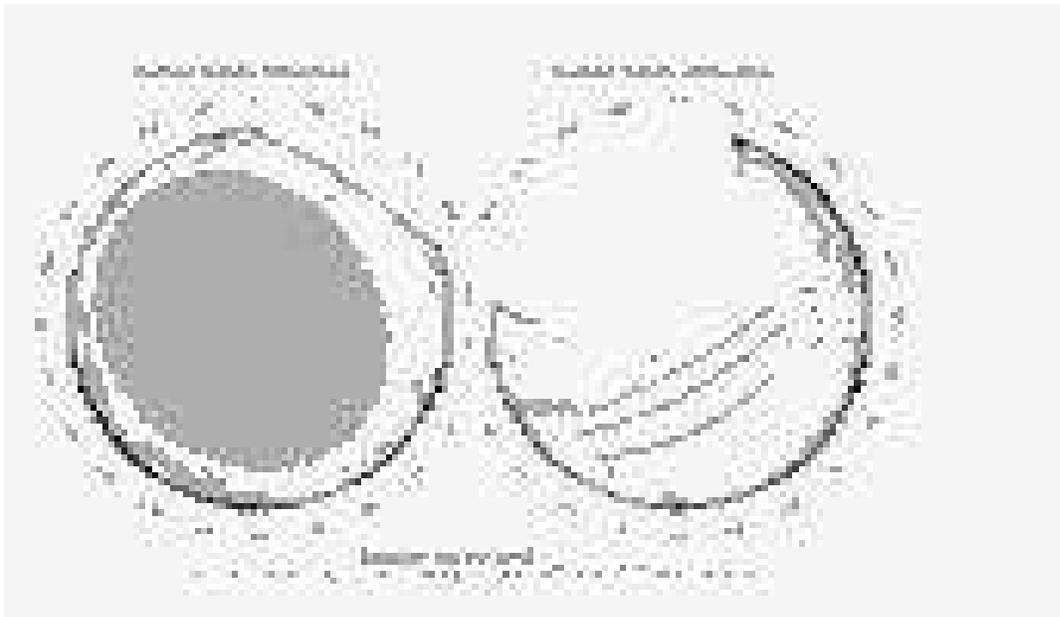}
\caption{The location of the survey imaging strips plotted in a polar
 projection for the North (left) and South (right) Galactic
 hemispheres. The contours indicate the amount of reddening due to
 dust in our own Galaxy. 
 Each solid line shows a 2.5 deg wide rectangular region ($stripe$) that the SDSS camera
 observes in two nights.  
 }\label{fig:area}
\end{figure}

The location of the survey imaging area is shown in Figure \ref{fig:area}. The
northern survey area is centered near the North Galactic Pole and it
lies within a nearly elliptical shape 130\degree\ E-W by 110\degree \
N-S chosen to minimize Galactic foreground extinction. All scans are
conducted along great circles in order to minimize the transit time
differences across the camera array. There are 45 great circles
($stripes$) in the northern survey regions separated by 2.5\degree.   The
 SDSS  observes three non-contiguous stripes in the Southern Galactic
 Hemisphere, at declinations of 0\degree, 15\degree and
 $-$10\degree, during the fall season when the northern sky is
 unobservable. Each stripe is scanned twice, with an offset
 perpendicular to the scan direction in order to interlace the
 photometric columns. A completed stripe slightly exceeds 2.5\degree\  in
 width and thus there is a small amount of overlap to allow for
 telescope mis-tracking and to provide multiple observations of some
 fraction of the sky for quality assurance purposes. The total stripe
 length for the 45 northern stripes will require a minimum of 650 hours
 of pristine photometric and seeing conditions to scan at a sidereal
 rate.  Based upon the current experience of observing at Apache Point Observatory, it seems
 likely that the SDSS will only complete about 75\% of the imaging after
 5 years of survey operations. 


\subsection{The SDSS Spectroscopic Survey}

\begin{figure}[h]
\includegraphics[scale=0.7,angle=0]{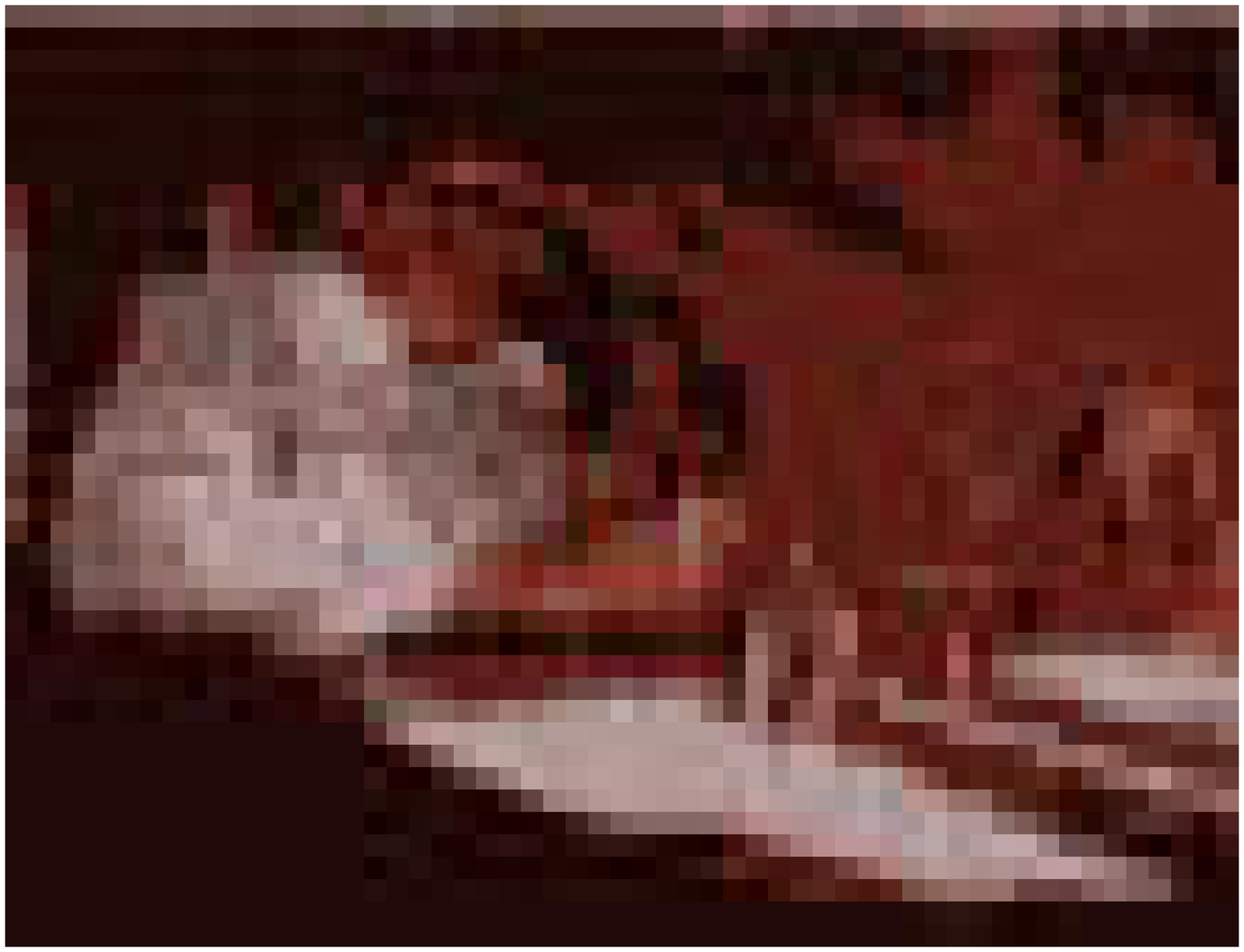}
\includegraphics[scale=0.7,angle=0]{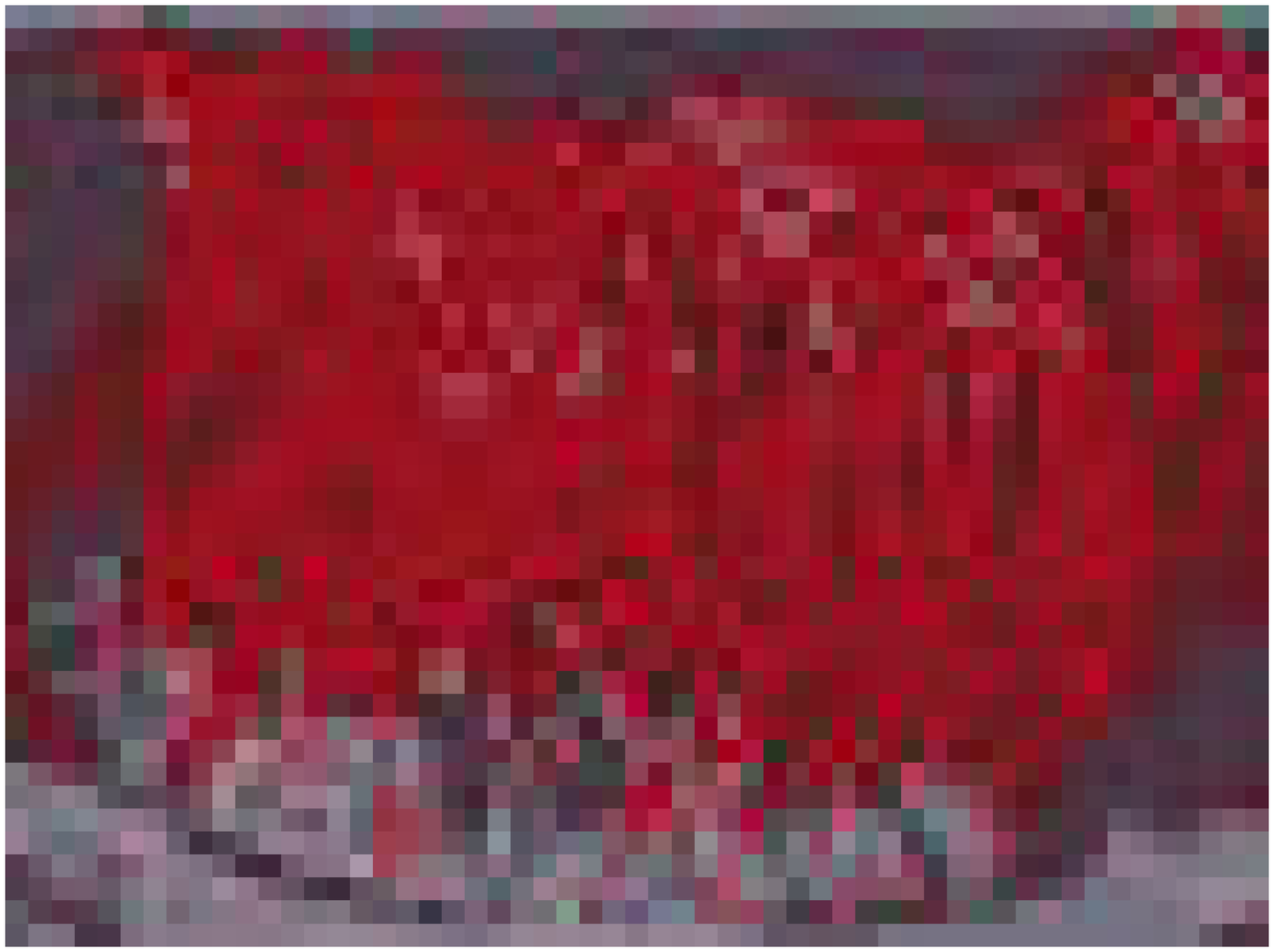}
\caption{
 Astrophysicist Rich Kron, of the University of Chicago and Fermilab, inserts optical fibers into a pre-drilled "plug-plate," part of the Sloan Digital Sky Survey's unique spectrographic system.
 }\label{fig:spectrograph}
\end{figure}

\begin{figure}[h]
\includegraphics[scale=0.7,angle=0]{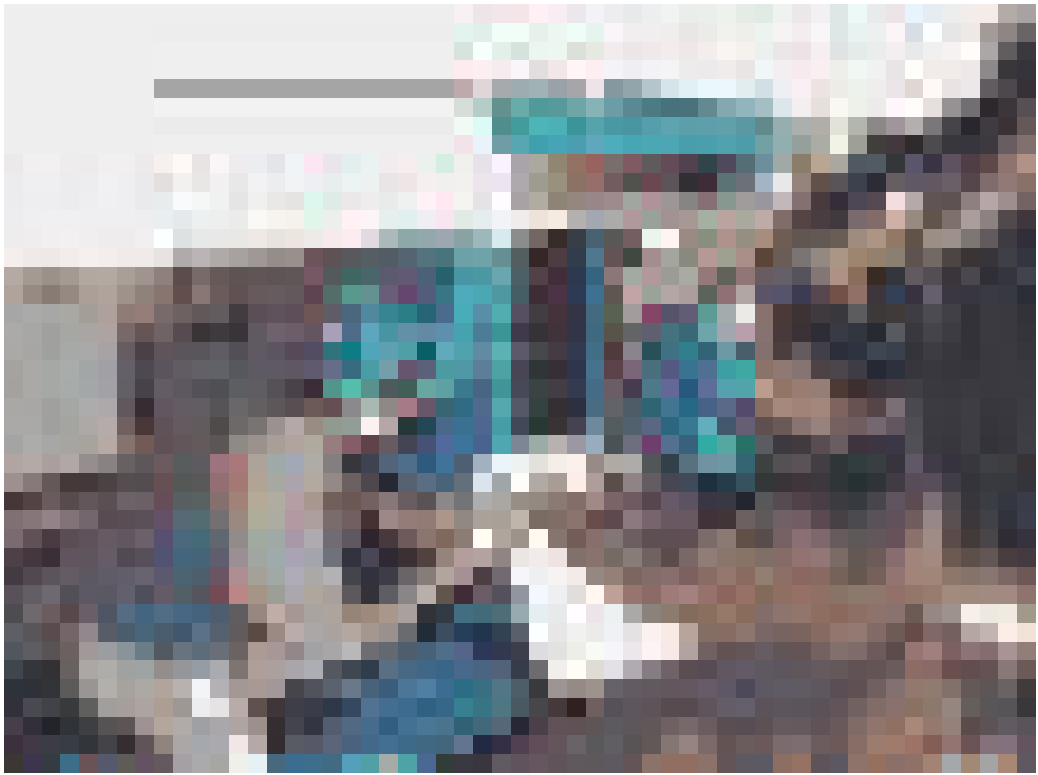}
\caption{The fiber-fed spectrograph is being installed to the SDSS telescope.
 }\label{fig:spectrograph_being_set}
\end{figure}

 By selecting targets from the photometric catalog produced by the imaging survey, 
 the SDSS observes spectra of $10^6$ bright
 galaxies to the depth of  $r^* \sim 17.77$ mag with median redshift of
 $z\sim$0.1 ($r^*\sim 19.5$ mag for luminous red galaxies,  reaching the redshift of $z\sim$0.45) 
 and $10^5$ brightest quasars to  $i^* \sim 20.0$ mag.
 The spectra are observed, 640 at a time (with a total integration time of
 45 minutes) using a pair of double fiber-fed spectrographs shown in
 Figures \ref{fig:spectrograph} and \ref{fig:spectrograph_being_set}. 
 The wavelength 
 coverage of the spectrographs is continuous from
 about 3800 \AA{} to 9200 \AA{},
 and  the wavelength resolution, 
 $\lambda/\delta\lambda$, is 1800 .
 The fiber diameter is 0.2 mm ($3''$ on the sky), 
 and adjacent fibers cannot be located
 more closely than $55''$ on the sky.
 The throughput of the spectrograph will be better than 25\% over  4000
 \AA{} to 8000 \AA{} excluding the loss due to the telescope and
 atmosphere.
 
\begin{figure}[h]
\includegraphics[scale=0.7,angle=0]{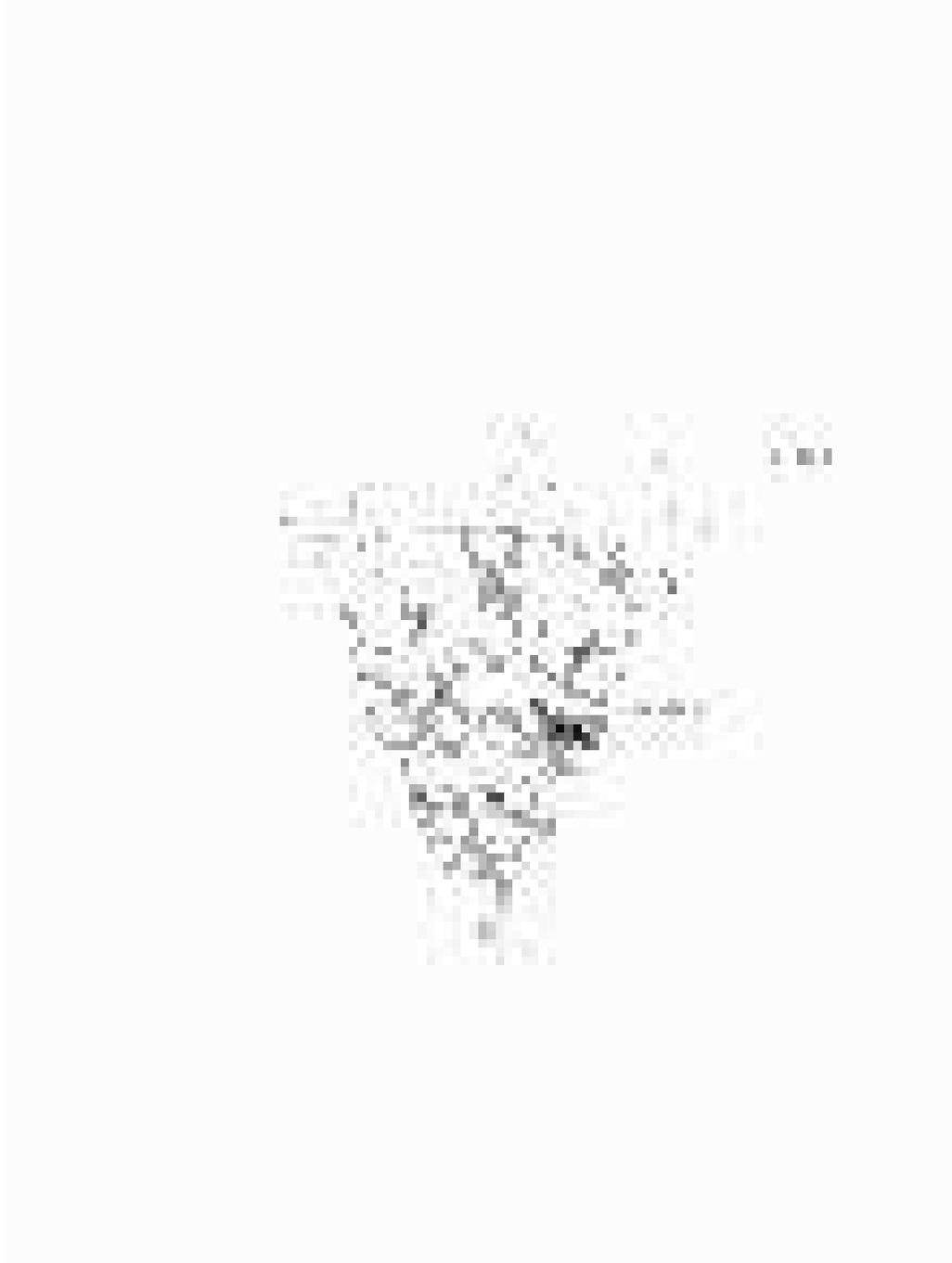}
\caption{A cone diagram; redshift distribution of galaxies observed with
 the SDSS spectrograph. The equatorial data in the range of  145.1
 $<$R.A.$<$236.1 are plotted.
 }\label{fig:cone}
\end{figure}

 Two samples of galaxies are selected from the objects classified as
 ``extended''. The main galaxy sample consists of $\sim$ 900,000
 galaxies with $r^*<$17.77.  This magnitude limit was chosen to satisfy
 the desired target density of 90 objects per square degree. In
 selecting target galaxies, the SDSS uses Petrosian magnitude (Petrosian
 1976), which is based on the aperture defined by the ratio of local
 surface brightness
 within an  annulus to the average surface brightness inside that radius, providing
 redshift independent estimate of total magnitude. The SDSS also applies
 a surface-brightness limit at $\mu_{r*}<$24.5 mag arcsec$^{-2}$, in
 order not to waste fibers on galaxies too faint to observe. This
 surface brightness cut eliminates just 0.1\% of galaxies that would
 otherwise be targeted for observation. Galaxies in the main galaxy
 sample have a median redshift of $z\sim$0.104. See Strauss et
 al. (2002) for more details of the main galaxy target selection.
 We show an example redshift distribution of the equatorial data of the
 SDSS commissioning phase in Figure \ref{fig:cone}.

 The SDSS observes additional $\sim$100,000 luminous red galaxies (LRG). For
 luminous red galaxies, redshift can be well measured with the SDSS
 spectrograph to $r^*\sim$19.5 mag due to their intrinsic brightness and
 their strong absorption lines due to high metallicity. Galaxies located
 at the dynamical centers of clusters often have these
 properties. Reasonably accurate ($\Delta z\sim$0.03) photometric
 redshifts can be determined for these galaxies, allowing the selection
 by magnitude and $g,r,i$ color of an essentially distance-limited
 sample to a redshift of $z\sim$0.45. See Eisenstein et al. (2001) for
 more details of the LRG target selection.

 150,000 quasar candidates are selected from cuts in multi-color space
 (Richards et al. 2002)
 and by identifying sources from the FIRST radio catalog (Becker et
 al. 1995). Due to their power-law continua and the strong Ly$\alpha$
 emission, quasars have $ugriz$ colors quite distinct from those of the
 vastly more numerous stars over most of their redshift range (Fan
 1999). Thus, among point sources,  quasar candidates are selected for spectroscopic
 observations as outliers  from the stellar locus in color-color
 space. The SDSS compile a sample of quasars brighter than $i^*\sim$19 at
 $z<$3; at redshift between 3 and 5.2, the limiting magnitude will be
 $i^*\sim$20. Point sources brighter than $i^*\sim$20 that are FIRST sources
 are also selected. Based on the commissioning data, it is estimated
 that $\sim$65\% of quasar candidates are genuine quasars. Comparison
 with samples of known quasars indicates that the completeness is $\sim$90\%.
 This sample will be orders of magnitude larger than any existing quasar
 catalog, and will be invaluable for quasar luminosity functions,
 evolution and clustering studies as well as providing sources for
 followup absorption-line observations.

 In addition to the above three classes of spectroscopic targets, which
 are designed to provide statistically complete samples, the SDSS
 also obtains spectra of many thousands of stars and for various
 serendipitous objects when remaining fibers are available.

 In this work, we mainly use a subsample of the main galaxy sample.

\clearpage

\chapter{The Cut \& Enhance Galaxy Cluster Catalog}
\label{chap:CE}

\section{Introduction}\label{Jun 28 09:03:18 2001}

 Clusters of galaxies are 
 the most massive virialized systems known and
 provide powerful tools in the study of cosmology and extragalactic
   astronomy.
 For
example, clusters are efficient tracers of the large--scale structure in the
Universe as well as  a useful tool to determine the amount of dark matter on Mpc scales
 (Bahcall 1998; Carlberg et al. 1996; Borgani \& Guzzo 2000 and Nichol 2001 and references
therein).
 Furthermore, clusters provide a laboratory to study a
large number of galaxies at the same redshift and thus assess the effects of
dense environments on galaxy evolution, for example,  the morphology--density
relation (Dressler et al.\ 1980, 1984, 1997), the Butcher--Oemler effect (Butcher
\& Oemler 1978, 1984) and the density dependence of the luminosity function of
galaxies (Garilli et al. 1999).
 In recent years, surveys of clusters of galaxies have been used extensively in
constraining cosmological parameters such as $\Omega_m$, the mass density
parameter of the universe, and $\sigma_8$, the amplitude of mass fluctuations
at a scale of 8 $h^{-1}$ Mpc (see
 Oukbir \& Blanchard 1992; Viana \& Liddle
1996, 1999; Eke et al. 1996; Bahcall, Fan \& Cen 1997; Henry 1997, 2000;
Reichart et al. 1999 as examples of an extensive
literature on this subject).  Such constraints are achieved through the
comparison of the evolution of the mass function of galaxy clusters, as
predicted by the Press-Schechter formalism (see Jenkins et al. 2001 for the
latest analytical predictions) or simulations (e.g., Evrard et
 al. 2001 and Bode et al. 2001),
with the observed abundance of clusters as a function of redshift. Therefore, to obtain
robust constraints on $\Omega_m$ and $\sigma_8$, we need large samples of
clusters that span a large range in redshift and mass as well as possessing a
well--determined selection function (see Nichol 2001).

 Despite their importance, existing catalogs of clusters are limited in both
their size and quality. For example, the Abell catalog of rich clusters (Abell
1958), and its southern extension (Abell, Corwin and Olowin 1989), are still
some of the most commonly used catalogs in astronomical research even though
they were constructed by visual inspection of photographic plates.  Another
large cluster catalog by Zwicky et al.\ (1961-1968) was similarly constructed
by visual inspection.  Although the human eye can be efficient in detecting
galaxy clusters, it suffers from subjectivity and incompleteness. 
 For cosmological studies, the major disadvantage of visually
constructed catalogs is the difficulty to quantify selection bias
and thus, the selection function. 
 Furthermore, the response of photographic
 plates is not uniform. 
 Plate-to-plate sensitivity variations can
 disturb the uniformity of the
 catalog.
 To overcome these problems,
 several cluster catalogs have been constructed using
 automated
 detection methods on CCD imaging data. 
 They
 have been, however,  restricted to small areas
 due to the lack of large--format CCDs.
 For example, the PDCS catalog (Postman et al.\ 1996) only covers 5.1 deg$^2$
 with 79 galaxy clusters. 
 The need for a uniform, large cluster catalog is strong.
 The Sloan Digital Sky Survey (SDSS; York et al.\ 2000)
 is the largest CCD imaging  survey currently underway
  scanning 10,000 deg$^2$ centered approximately on the North Galactic
 Pole, and thus,
 offers the opportunity 
 to produce the largest and most uniform galaxy cluster catalog in
 existence.
 
 The quantity and quality of the SDSS data demands the use of sophisticated
cluster finding algorithms to help maximize the number of true cluster
detections while suppressing the number of false positives.
 The history of the automated cluster finding methods goes back to Shectman's
count-in-cell method (1985). He counted the number of galaxies in cells
 on the sky to estimate the galaxy density. Although this provided important
progress over the visual inspection, the results depend on the size
and position of the cell.
 Currently the commonly used automated cluster finding method is the Matched
 Filter technique (MF; Postman et al.\ 1996; Kawasaki et 
 al.\ 1998;  Kepner et al.\ 1999; Schuecker \& Bohringer 1998; Bramel et
 al.\ 2000; Lobo et al.\ 2000; da Costa et al.\ 2000 and Willick et al.\ 2000).
 The method assumes a filter for the 
 radial profile of galaxy clusters and for the luminosity function of
 their members.
 It then selects clusters from imaging data by maximizing the likelihood
 of matching the data to the cluster model.
Although the method has been successful, galaxy clusters that do
not fit the model assumption (density profile and LF) may be missed.
  We present here a
new cluster finding method called the {\it Cut-and-Enhance } (CE)
method.
 This new algorithm is semi--parametric and is designed to be as
simple as possible using the minimum number of assumptions on
cluster properties. In this way, it can be sensitive to all types of galaxy
overdensities even to those that may have recently under--gone a merger and
therefore, are highly non--spherical.
 One major difference between the CE and previous cluster finders is
 that the CE
 makes full use of colors of galaxies, which become available as a
 result of  the
 advent of the accurate CCD photometry of the SDSS data.  
 We apply this detection method 
 on 350 deg$^2$ of the SDSS commissioning
 data and  construct the large cluster catalog.
 The catalog ranges from rich clusters to more numerous poor clusters
 of galaxies over this area.  We also determine the selection function of the CE method.

In Section\ \ref{sec:data}, we describe the SDSS commissioning data. 
In Section\ \ref{sec:method}, we describe the detection strategy of the CE method. 
In Section\ \ref{sec:monte}, we present the performance test of the CE method and selection
function using Monte Carlo simulations.
In Section\ \ref{sec:visual}, we visually check the success rate of the CE method.
In Section\ \ref{sec:compare}, we compare the CE method with the
 other detection methods applied to the SDSS
data. 
In Section\ \ref{sec:summary}, we summarize the results.


\section{The SDSS Commissioning Data }\label{sec:data}

 The data we use to construct the SDSS Cut \& Enhance galaxy cluster catalog 
are equatorial scan data taken in September 1998 and
March 1999 during the early part of the SDSS commissioning phase.
A contiguous area of 250 deg$^2$ (145.1$<$RA$<$236.0, $-$1.25$<$DEC$<$+1.25) and 150 deg$^2$ (350.5$<$RA$<$56.51, $-$1.25$<$DEC$<$+1.25)  were obtained during four nights,
where seeing varied from 1.1'' to 2.5''. 
 Since we intend to use the CE method at the faint end of imaging data,
 we include galaxies to
$r^*$=21.5 (Petrosian magnitude), which is the star/galaxy separation
limit (Scranton et al. 2002).  
 The details about SDSS commissioning data are
 described in Stoughton et al. (2002).

\section{The Cut \& Enhance Cluster Detection Method}\label{sec:method}

\subsection{Color Cut}\label{color-cut}

 The aim of the CE 
 method is to construct a
 cluster catalog that has as little bias as possible by minimizing the
 assumptions on cluster properties.
 If a method assumes
 a luminosity function or radial profile, for example, the resulting clusters will
be biased to the detection model used.
 We thus exclude all such assumptions except for a generous color cut.
 The assumption on colors of cluster galaxies appears to be robust,
  as all the
galaxy clusters appear to have the same general color-magnitude relation
(Gladders et al.\ 2000). 
 Even a claimed ``dark cluster''
(Hattori et al.\ 1997) was found to have a normal color magnitude relation.
(Benitez et al.\ 1999; Clowe et al.\ 2000; Soucail et al.\ 2000).
 A tight color-magnitude relation of cluster galaxies is known as follows;
 among the various galaxy populations within a cluster,
 (i.e., spiral, elliptical, dwarf, irregular),
 bright red elliptical galaxies have similar color and 
 they populate a red ridge-line 
 in the color-magnitude diagram.
 Bower, Lucey, \& Ellis (1992) obtained high precision $U$ and $V$ photometry
of spheroidal galaxies in two local clusters, Virgo and Coma, observing
 a very small scatter, $\delta(U-V)<0.035$ rms, in the color-magnitude relation. 
 Ellis et al.\ (1997)
 studied the $U-V$ color-magnitude relation at high redshift ($z\sim$0.54) and found a
scatter of $<$ 0.1 mag rms. Similarly,  Stanford et al.\ (1998) studied optical-infrared
colors ($R-K$) of early-type (E+S0) galaxies in 19 galaxy clusters out to $z=$0.9
and found a very small dispersion of $\sim$0.1 mag rms in the optical-infrared 
colors.

\begin{figure}
\includegraphics[scale=0.7]{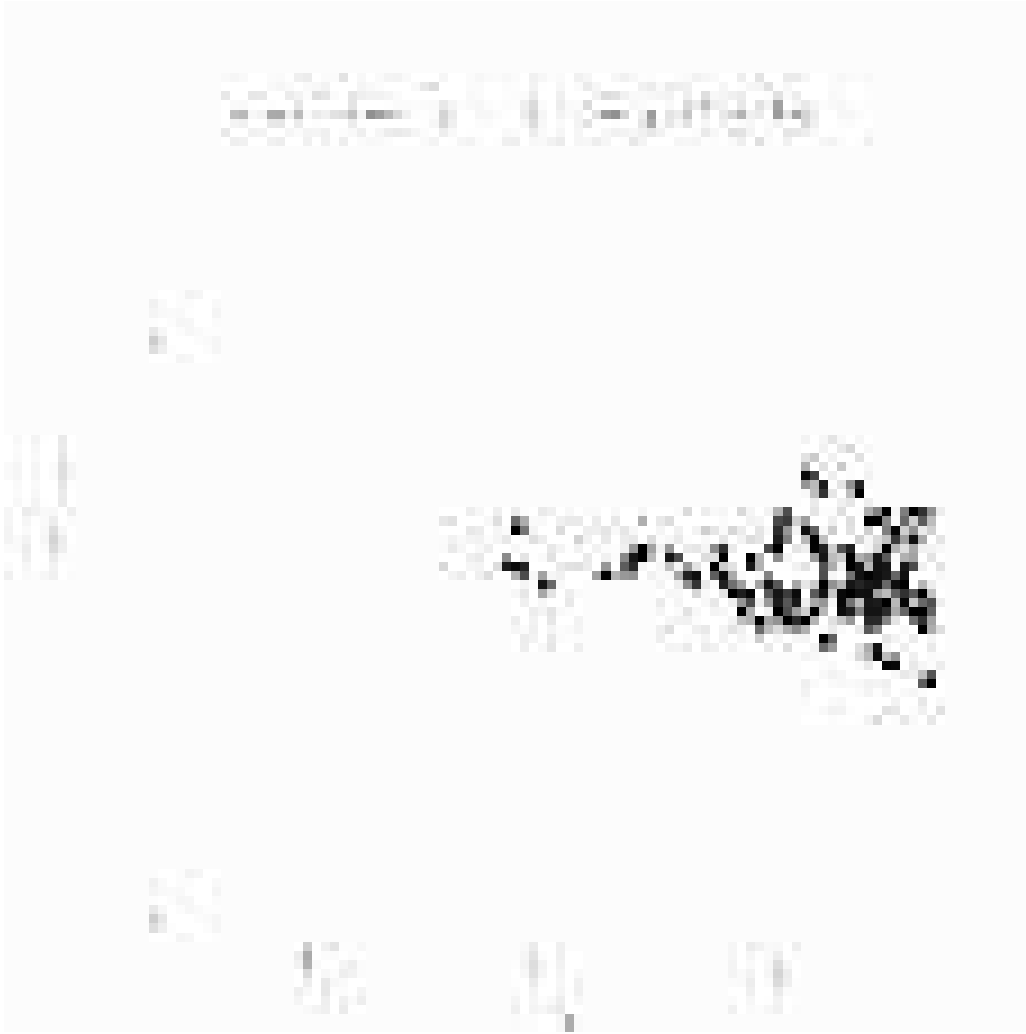}
\caption{
$r-i$ color-magnitude diagram of A168.
$r-i$ color is plotted against $r$ magnitude for confirmed member
 galaxies of A168.
Colors and magnitude are taken from the SDSS commissioning data by matching
 up the positions with  the spectroscopic observation of Katgert et al.\ (1998).
The standard errors of colors estimated by the reduction
 software are shown as error bars.
$r-i$ color cut bins are superimposed on the color-magnitude relation of Abell 168.
The horizontal dotted lines are the borders of the
 color cuts.
}\label{fig:grcut168}
\end{figure}

\begin{figure}
\includegraphics[scale=0.7]{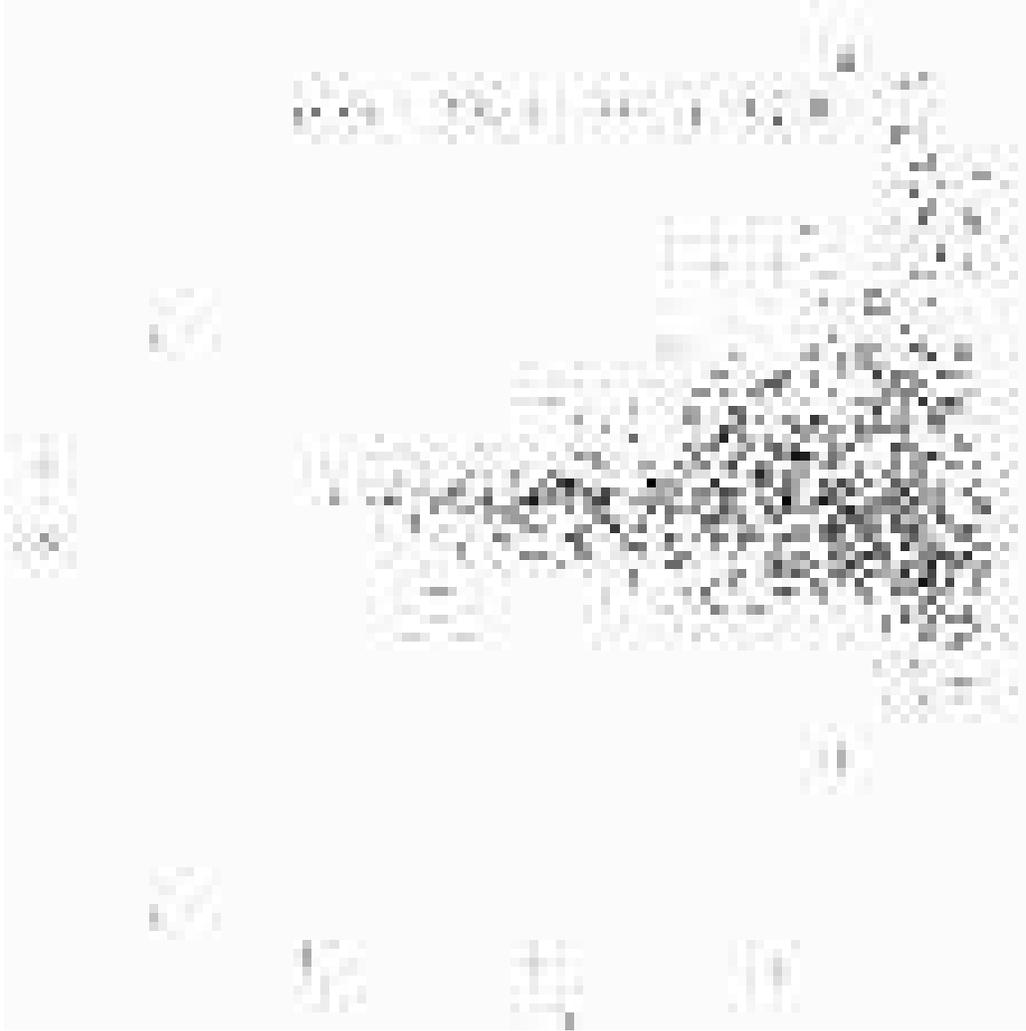}
\caption{
 $g-r$ color-magnitude diagram.
 $g-r$ color cut bins are
 superimposed on the color-magnitude relation of Abell 1577.
 The abscissa is the $r$ apparent magnitude. The ordinate is $g-r$ color. 
 Galaxies in the SDSS fields covering A1577 ($\sim8.3\times10^{-2}$
 deg$^2$) are plotted with the dots. 
 The horizontal dashed lines are the borders of the
 color cuts.
}\label{fig:grcut}
\end{figure}

\begin{figure}
\includegraphics[scale=0.7]{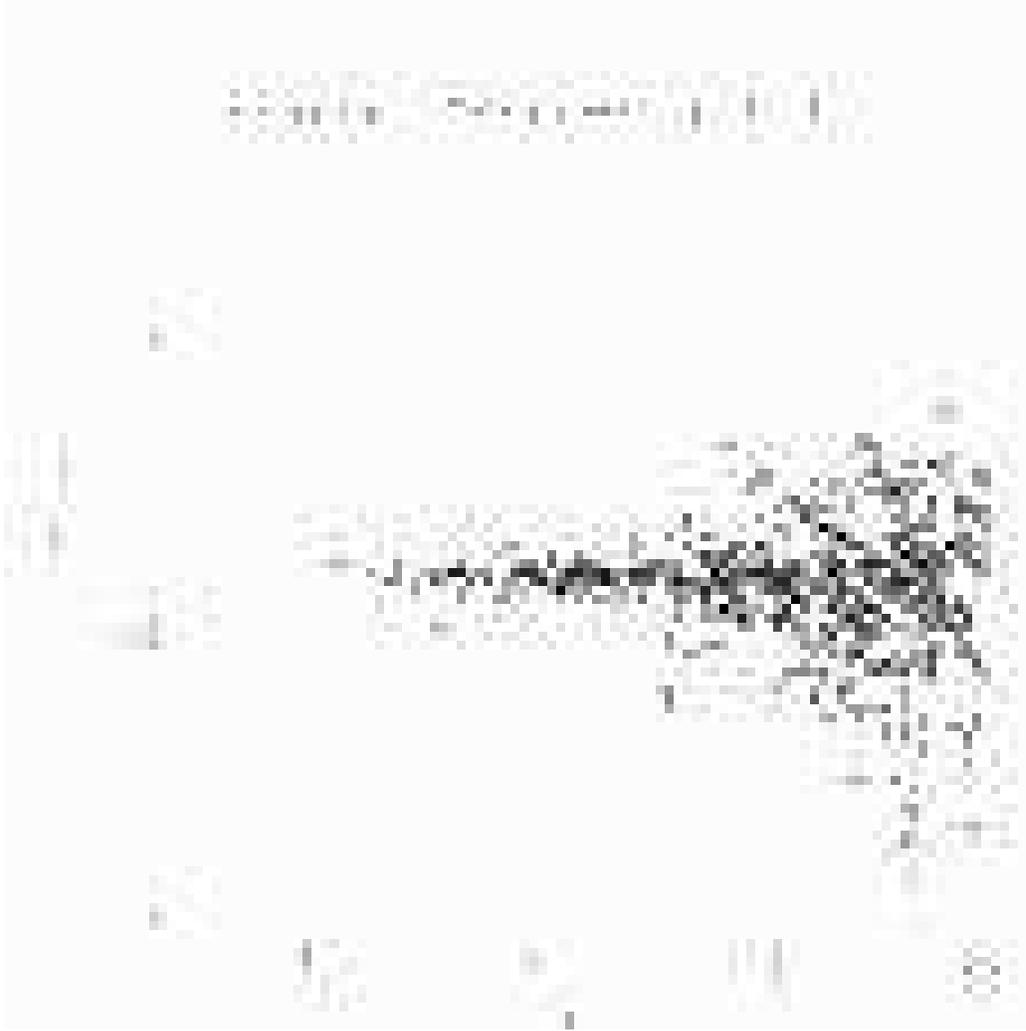}
\caption{
$r-i$ color-magnitude diagram.
$r-i$ color cut bins are
superimposed on the color-magnitude relation of Abell 1577.
 The abscissa is the $r$ apparent magnitude. The ordinate is $r-i$ color. 
Galaxies in the SDSS fields  covering A1577 ($\sim8.3\times10^{-2}$
 deg$^2$) are plotted with the dots. 
Colors and magnitudes are taken from the SDSS commissioning data.
The horizontal dashed lines are the borders of the
 color cuts.
}\label{fig:ricut}
\end{figure}

\begin{figure}
\includegraphics[scale=0.7]{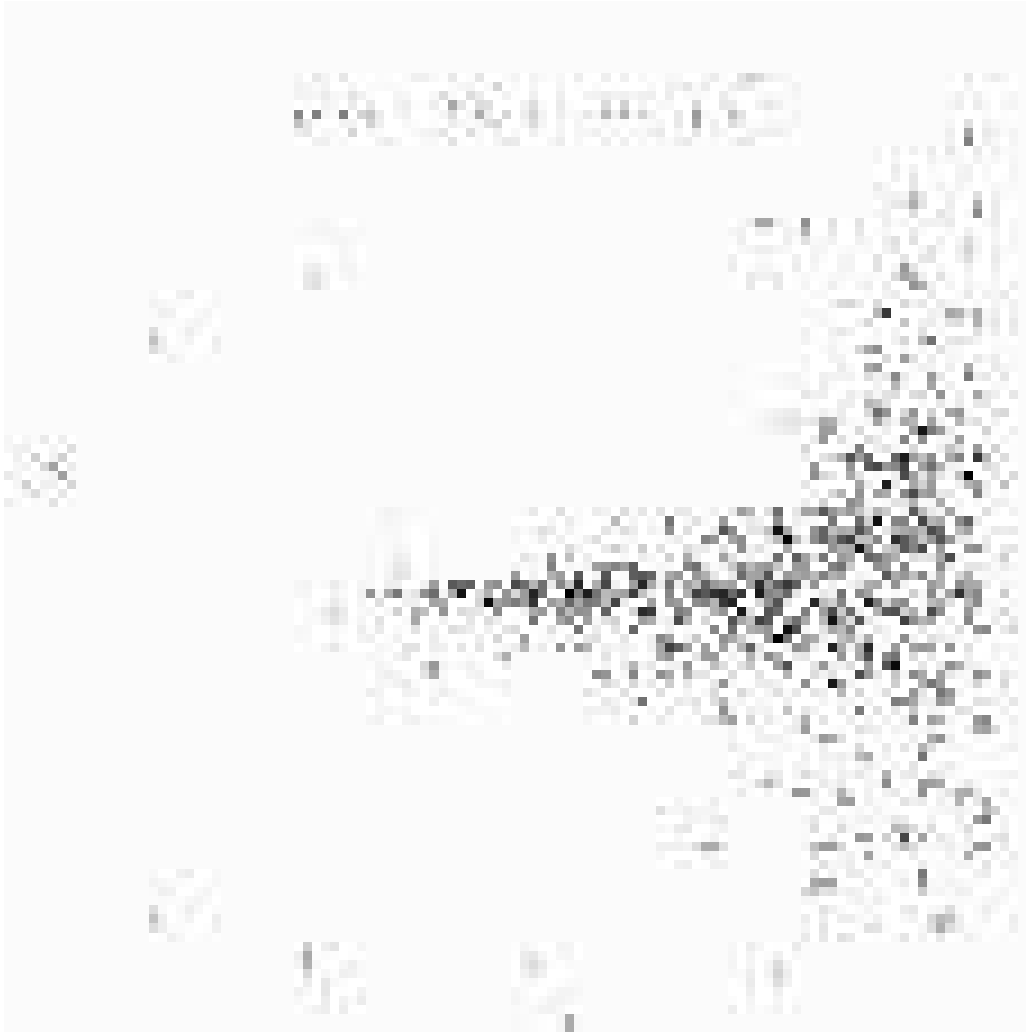}
\caption{
Same as Figure \ref{fig:grcut}, but for $i-z$.
}\label{fig:izcut}
\end{figure}

 Figure\ \ref{fig:grcut168} shows a color-magnitude diagram in ($r-i$
 vs. $r$) using SDSS data for galaxy members in the cluster A168 ($z=$0.044).
The cluster member galaxies are identified by matching 
 the positions of galaxies in the SDSS commissioning data with the
 spectroscopic observation of 
Katgert et al.\ (1998). 
The error bars show the standard errors of $r-i$ color estimated by
the SDSS reduction
 software (Lupton et al.\ 2001). 
The red ridge-line of the color-magnitude relation
is seen at $r^*-i^*\sim$0.4 from $r^*=17.5$ to $r^*=20$.
The scatter is 0.08 mag from the brightest to $r^*=18$.
Figure\ \ref{fig:grcut} shows the color-magnitude diagram (in $g-r$ vs
$r$) for all galaxies in the SDSS fields ($\sim8.3\times10^{-2}$ deg$^2$) 
that contain Abell 1577. A1577 has a redshift of $z\sim0.14$ and Abell richness class $\sim1$.
Figure\ \ref{fig:ricut} and Figure\ \ref{fig:izcut} show color-magnitude
 diagrams for the same field in $r-i$ and $i-z$ colors, respectively.
 All the galaxies in the region of the imaging data  are included.
 Even without spectroscopic information,
 the red ridge-line of the color-magnitude relation is clearly visible
 as the  horizontal distribution in the figure.
The scatter in the color-magnitude relation is the largest in $g-r$
  since the difference of the galaxy spectral energy distribution due to  the age or
  metallicity difference is prominent around
  $3500\sim5000$\AA.
The color distribution is much wider at faint
 magnitudes, partly because fainter galaxies have larger color errors,
and partly because of the increase in the number of both background
 galaxies and cluster galaxies.

 In the color-magnitude relation, 
 fainter galaxies are  known to have slightly bluer color than bright
 galaxies (Kodama et
 al.\ 1998).
 However, this tilt is relatively small in the SDSS color bands.
 The tilt and its scatter in the case of A1577 (Figure\ \ref{fig:grcut})
 is summarized in Table \ref{tab:cm-tilt}.
 The tilt is small in $g-r$ and $r-i$ ($\sim$0.08), and even smaller 
 in $i-z$ (0.0018). These values are much smaller than the color cuts
 of the CE method. The scatters are also small: 0.081, 0.040 and 0.033 in $g-r$, $r-i$ and
 $i-z$, well smaller than the bin size of color cuts of the CE method as
 described in Section \ref{sec:ce_color-color-cut}. 
 The small scatter of $<$0.1 mag is also often reported in  previous 
 work (Bower et al.\ 1992; Ellis et al.\ 1997; Stanford et al.\ 1998).
  Since the tilt of the color-magnitude relation is smaller than the
 scatter in the SDSS color bands, it is sufficient for this work to  treat a color cut as
 essentially independent of  magnitude of galaxies as described below. 

\begin{figure}
\begin{center}
\includegraphics[scale=0.7]{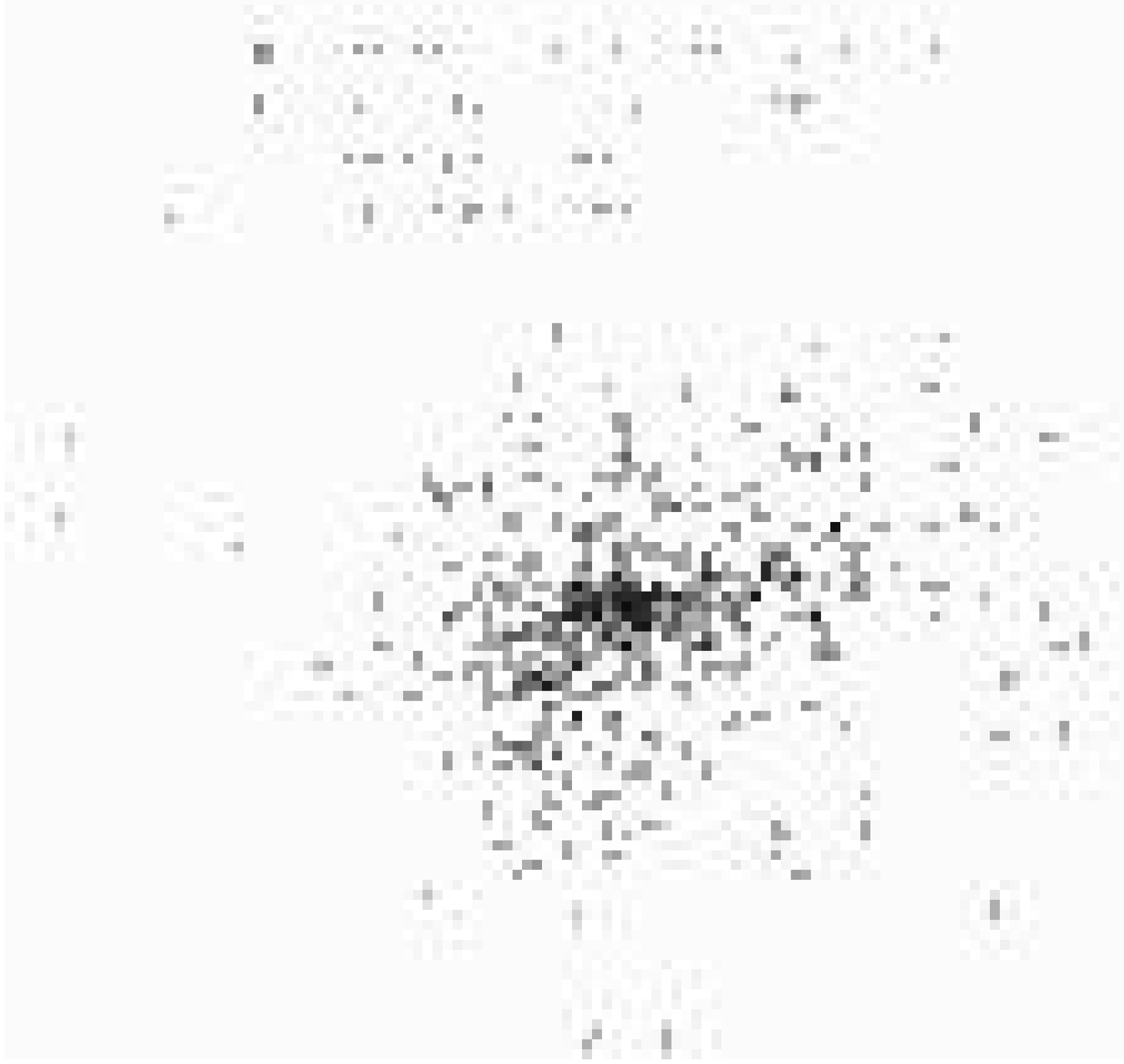}
 \caption{
 $g-r-i$ color-color boxes to find galaxy clusters.
 The abscissa is the $g-r$ color. 
 The ordinate is $r-i$ color.
The low-$z$ $g-r-i$ box is  drawn with the dashed lines.
The high-$z$ $g-r-i$ box is  drawn with the dotted lines. 
  Galaxies brighter than $r^*=22$ in the SDSS fields
 ($\sim8.3\times10^{-2}$deg$^2$)  which covers A1577 are plotted with
 small dots. 
 The triangles show the 
 color prediction of elliptical galaxies (Fukugita et al.\ 1995).
 } \label{fig:griboxes1}
\end{center}
\end{figure}

  Colors of the ridge-line of the color-magnitude relation become redder
 and redder with an increasing redshift.
 Figure\ \ref{fig:griboxes1} presents the color-color diagram,
 $g-r$ vs $r-i$, for all 
 galaxies brighter than $r^*$=22 in the SDSS fields
 that covers A1577,
 as well as the color predictions of elliptical galaxies at different
 redshifts (triangles; Fukugita et al.\ 1995).   
 The $g-r$ color becomes monotonously redder from $z=$0 to $z=$0.4. 
 Similarly   $r-i$ reddens monotonously. 
 At $z\sim0.4$, the 4000 $\AA$  
 break of an elliptical galaxy
 crosses the border between $g$ and $r$ bands, and 
 appears as a sharp turn in the color at this redshift (Figure\
 \ref{fig:griboxes1}). By using this color change, we can reject
 foreground and background galaxies, and thus can select galaxies likely to be
 in a certain redshift range as described below. This is a significant
 advantage in having multi-color data since optical cluster finders have
 suffered chance projections of galaxies on the sky.

 To select galaxies with similar colors (and thus likely to belong to the
 same cluster),
 we divide the $g-r$ vs. $r$ color-magnitude diagram into 11 bins. 
 These bins are shown in Figure\ \ref{fig:grcut} as the horizontal dashed lines.
 The bins are not tilted because the tilt is smaller than the scatter in
 the SDSS color bands (see above), and 
 because we wish to minimize the assumptions used for cluster selection.
  Since colors of the ridge-line of the color-magnitude relation become redder
 and redder with an increasing redshift, we use bluer color cuts to
 target low redshift clusters and redder color cuts to target high
 redshift clusters.
 We use two bins as one color cut in order to produce overlap in the
 color cuts;
 the cut is shifted by one bin each time we step to a higher redshift
(redder cut). 
 Similarly, we use ten color cuts in both $r-i$ (shown in Figure\ \ref{fig:ricut}
 as the dashed lines) and ten color
 cuts in $i-z$ (shown in Figure\ \ref{fig:izcut} as the dashed lines).
 The width of the bins in $g-r$, $r-i$ and $i-z$ color are 0.2 mag,
 0.1 mag and 0.1 mag, respectively.
 The width of the $r-i$ and $i-z$ bins is smaller than the $g-r$
 width since the colors of elliptical galaxies have less scatter in $r-i$ and
 $i-z$ than in $g-r$. 
 When detecting cluster candidates, the above color cuts in the three
 colors are applied independently. 
 To prevent faint galaxies with large color errors from weakening a signal of
 a cluster,  we exclude galaxies with color errors larger than the size of
 the color bin. 
 The standard color error estimated by the SDSS reduction software
 at $r^*=21.5$  (the limiting magnitude used in the CE method)
 is 0.20$\pm$0.09, 0.16$\pm$0.06 and 0.26$\pm$0.1 in $g-r$, $r-i$
 and $i-z$, respectively. In $g-r$ and $r-i$, the color error
  is smaller than the size of the color cut box. In $i-z$, the color error at
 $r^*=21.5$ is slightly larger than the size of the color cut boxes (0.2
 mag). At slightly brighter magnitude of $r^*$=20.5, however, the
 errors of $i-z$ is 0.11$\pm$0.05.  

\begin{figure}
\begin{center}
\includegraphics[scale=0.7]{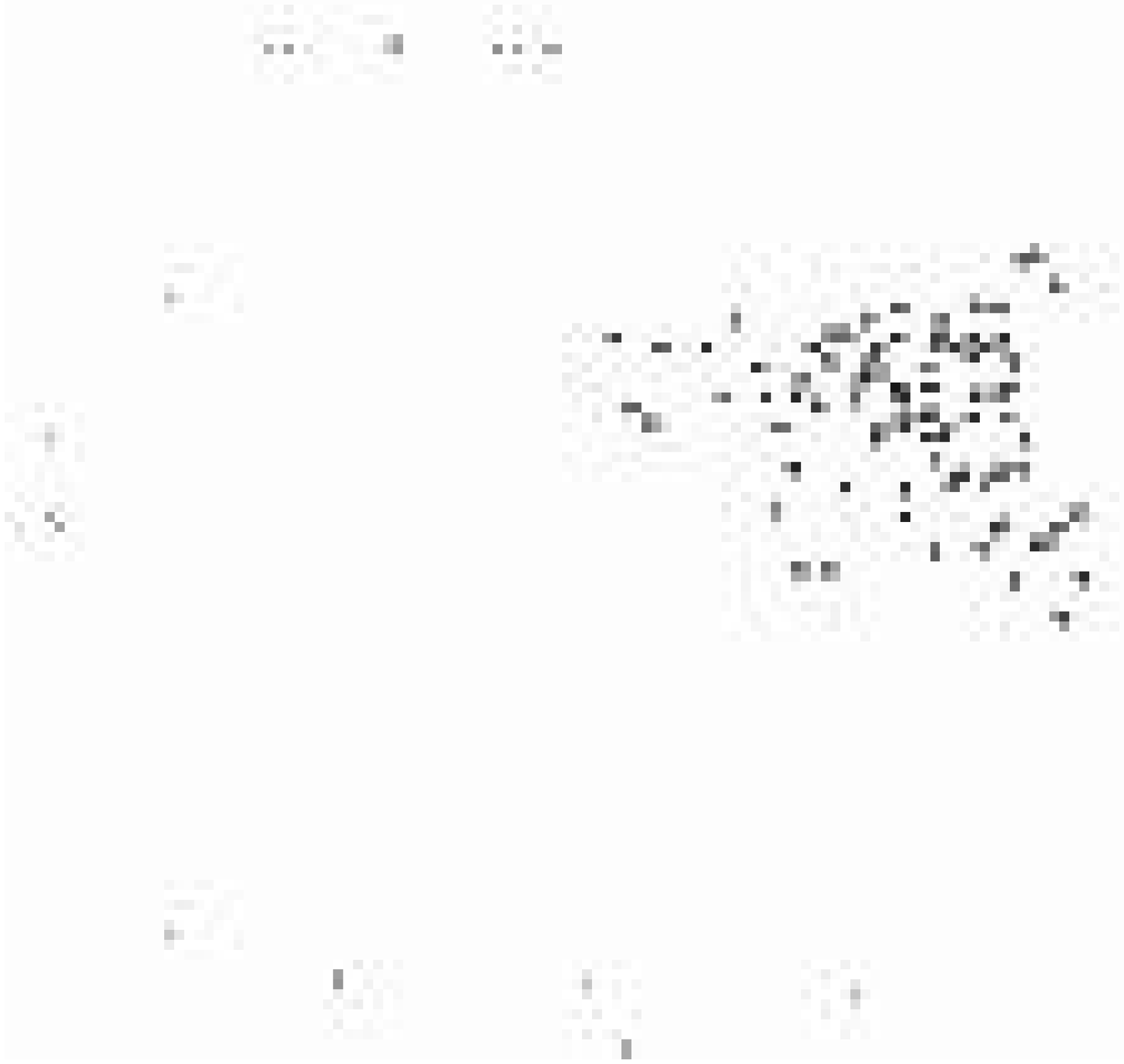}
 \caption{
 An example of color-cut capturing color-magnitude relation.
 Galaxies within 1.5 $h^{-1}$Mpc aperture around RXJ0256.5+0006
 ($z=$0.36) are plotted as black dots.
 Distribution of all the galaxies in the SDSS commissioning data is
 drawn as contours.
 The $g-r$ color-cut successfully captures the red-sequence of RXJ0256.5+0006
 and remove the foreground galaxies bluer than the sequence.
 } \label{fig:RXJ0256_gr}
\end{center}
\end{figure}

\begin{figure}
\begin{center}
\includegraphics[scale=0.7]{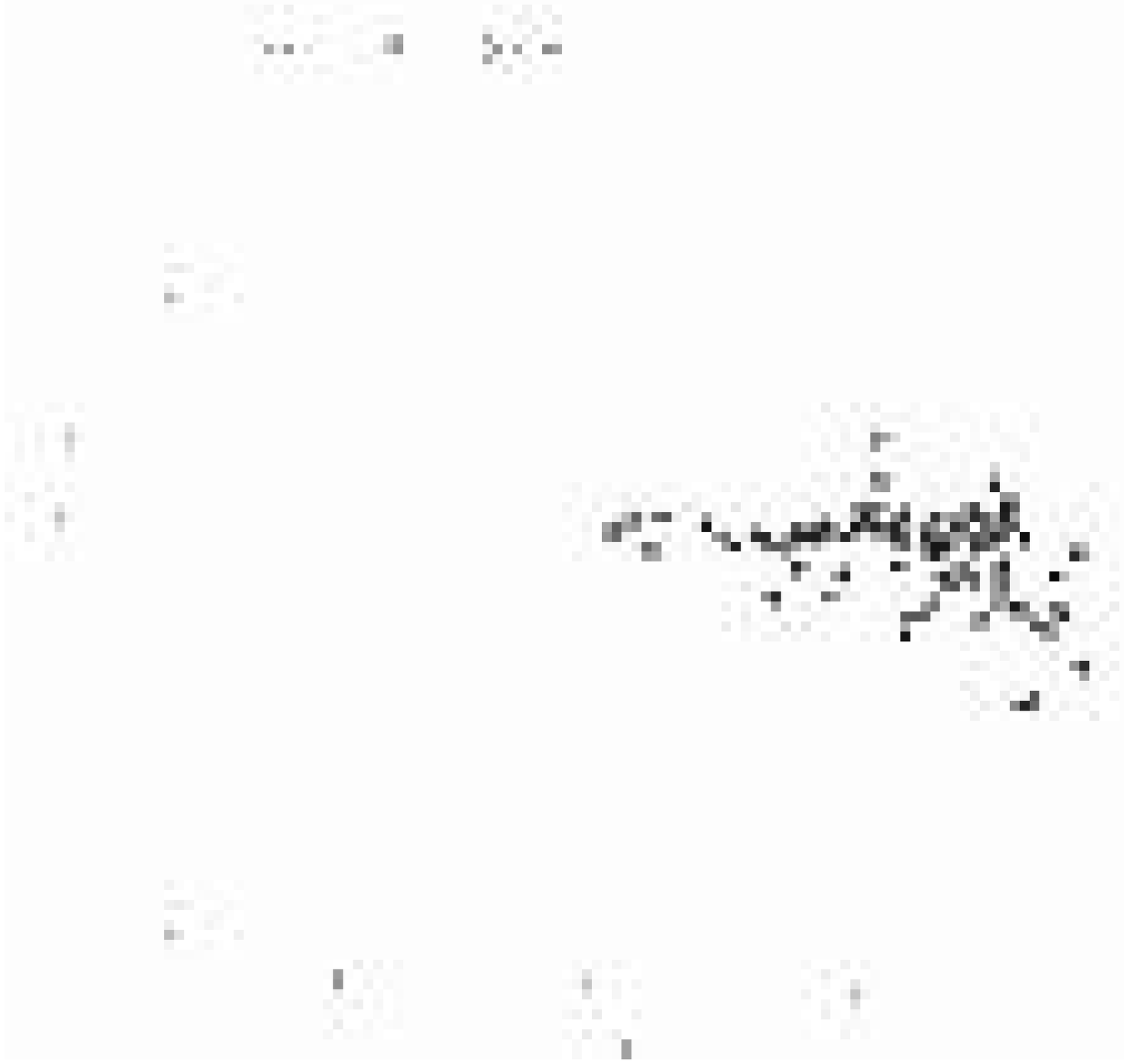}
 \caption{
 An example of color-cut capturing color-magnitude relation.
 Galaxies within 1.5 $h^{-1}$Mpc aperture around RXJ0256.5+0006
 ($z=$0.36) are plotted as black dots.
 Distribution of all the galaxies in the SDSS commissioning data is
 drawn as contours.
 The $r-i$ color-cut successfully captures the red-sequence of RXJ0256.5+0006
 and removes the foreground galaxies bluer than the sequence.
 } \label{fig:RXJ0256_ri}
\end{center}
\end{figure}

 In Figure \ref{fig:RXJ0256_gr} and Figure \ref{fig:RXJ0256_ri}, we
 demonstrate the effect of the color cut. Black dots are the galaxies
 within 2.7' (1.5$h^{-1}$Mpc at $z=$0.37) from the center of
 RXJ0256.5+0006 (Romer et al. 2001) . No background and foreground correction are applied. 
 Contours represent the distribution of all the galaxies of the SDSS
 imaging data. The corresponding color cuts to the redshift of the cluster are drawn in each figure. 
 In each case, the color cuts capture the red-sequence of RXJ0256.5+0006 successfully
 and reject foreground galaxies as designed. In fact, we show
in Table
 \ref{tab:RXJ0256_color_cut}, 
the  fraction of galaxies inside of the color
 cut for both in cluster region and outside of cluster region.
 As shown in Figure \ref{fig:RXJ0256_gr},
 Figure \ref{fig:RXJ0256_ri} and Table \ref{tab:RXJ0256_color_cut}, indeed
 the fraction in the color cut 
 increases dramatically from 13.5\% to 36.9\% in $g-r$ cut and from 42.4\% to
 62.1\% in $r-i$ cut. The efficiency of color cut increases as we see
 higher redshifts since the colors of cluster galaxies are further away
 from the color distribution of foreground galaxies.
 The upper left panel in Figure \ref{fig:beforecut_RXJ} shows the galaxy
 distribution of the SDSS commissioning data around RXJ0256.5+0006 
 before applying any color cut while
 the upper right panel shows it after applying the $g-r$ color cut at
 the cluster redshift. These two panels illustrate the power of color
 cuts in enhancing a  cluster signal.

\subsection{Color-color Cut}\label{sec:ce_color-color-cut}

When more than two colors are available,
 it is more
effective to select galaxies in color-color space. 
 We thus added four additional color-color-cut boxes to enhance the contrast
 of galaxy clusters.
The cuts are low-$z$ and high-$z$  boxes in
$g-r-i$ space 
and in $r-i-z$ space, as shown in Figure\ \ref{fig:griboxes_a168} and Figure\ \ref{fig:colors_of_Elliptical_a168}.
 These color boxes are based on the fact that cluster
 galaxies concentrate 
 in specific regions in
 color-color space (Dressler \& Gunn 1992).
 In Figure\ \ref{fig:griboxes_a168}, we show the $g-r$ vs. $r-i$ 
  color-color diagram of A168 for
 spectroscopically confirmed member galaxies (Katgert et al.\ 1998) brighter than $r^*$=21. 
 The low-$z$ $g-r-i$ color-color-cut box is shown with the dashed lines and
 the high-$z$ $g-r-i$ color-color-cut box is shown by the dotted
 lines.
 The triangles present the color prediction as a function of redshift
 for elliptical galaxies with a redshift step of $\Delta z=$0.1 (Fukugita et al.\ 1995). 
 The relatively wide distribution of the dots in the plots represents
 the fact that cluster galaxies are not all elliptical galaxies, but in
 reality, the mixture of different type of galaxies. 
 Similar results are shown in Figure\ \ref{fig:colors_of_Elliptical_a168}
 for the $r-i-z$ color-color diagram of A168.
 Member galaxies of A168 ($z=$0.044; Struble \& Rood 1999) are well
 included in the low-$z$ $g-r-i$ and $r-i-z$ boxes.

Figure\ \ref{fig:griboxes1} is the $g-r-i$ color-color diagram of galaxies
 (brighter than $r^*$=22) in the 
 SDSS fields covering A1577 ($z=$0.14).
The low-$z$  and high-$z$ color-color-cut boxes are also
 shown.
The triangle points show the color prediction for 
 elliptical galaxies. 
Figure\ \ref{fig:colors_of_Elliptical} represents similar results in the 
 $r-i-z$ color-color space for the same field.
 Even though both cluster members and field galaxies are included in the plot,
 the concentration of cluster galaxies inside the low-$z$ boxes is
 clearly seen.

 The boundaries of color-color cuts are chosen based on the spectroscopic observation
 of Dressler \& Gunn (1992) and the color prediction of elliptical galaxies
 (Fukugita et al.\ 1995). We reject galaxies that have standard color
 errors larger than the size of the color-color boxes.
The standard color error at $r^*$=21.5 (the limiting magnitude of CE method.)
 is 0.20$\pm$0.09, 0.16$\pm$0.06 and 0.26$\pm$0.1 in $g-r$, $r-i$
 and $i-z$, respectively.
The smallest size of the color-color boxes is the $r-i$ side of the low-$z$
 $g-r-i$ box, which is 0.34 in $r-i$.
 The standard color error is smaller than the color cut boxes even
 at $r^*$=21.5.
 In Figure \ref{fig:beforecut_comparison} the upper left panel shows  the galaxy
 distribution of the SDSS commissioning data in 23.75 deg$^2$ before
 applying any cut. The upper right panel shows the galaxy distribution
 after applying the $g-r-i$ color-color cut.
 Abell clusters in the region are shown their position in numbers.
 It illustrates the color cut enhancement of nearby clusters. 
 We used RXJ0256.5+0006 ($z=$0.36) to numerate the fraction of galaxies inside of the color
 cut for both in cluster region 
 and outside of cluster region in Table
 \ref{tab:RXJ0256_color_cut}.  Indeed, 
 the fraction of galaxies in the color cut
 increases from 48.8\% to 58.3\% in $g-r-i$ cut and from 65.7\% to
 76.7\% in $r-i-z$ cut. Since the color cuts has overlaps at
 $z\sim$0.4, $g-r-i$ high-$z$ cut also increases the fraction of
 galaxies. This example shows that the signal of galaxy clusters,
 indeed,  increases by applying the color-color cuts.




 We thus use 30 color cuts and four color-color cuts independently to
 search for cluster signals.
 We then merge 34 cluster candidate lists into a final cluster catalog.
 Because of star/galaxy separation limit of the SDSS data, we do not use galaxies fainter
 than $r^*$=21.5 . 
 The only main assumption made in the CE detection method is
 these generous color cuts.


 In Figure \ref{fig:color-cut-test}, we plot the color prediction of galaxies
 with evolving model with star formation (filled pentagons)  and the same model 
 without star formation (filled square) from $z=$0 to $z=$0.6 (PEGASE
 model, Fioc, M., \& Rocca-Volmerange 1997).  The model galaxies with
 star formation are the extreme star forming galaxies.  
 Filled triangles show the color prediction of elliptical galaxies with
 a redshift step of $\Delta z=$0.1 (Fukugita et al. 1995).
  Black dots are the galaxies around
 Abell 1577, for reference.
  Although the evolving model steps outside of the high-$z$ color cut 
 box at $z\sim$0.6, the CE is designed to detect galaxy clusters if
 enough red galaxies (shown as small triangles) are in the color cut by weighting the galaxies with
 similar color. 
 In fact, randomly chosen 100 spectroscopic galaxies with
 0.4$<z\leq$0.5 shown in small green triangles
 are well within the high-$z$ box.
 As seen in the real catalog in Section \ref{sec:real-catalog} , due to the
 magnitude limit of SDSS, it is difficult to find many clusters beyond
 $z\sim$0.4 . 
 On the other hand, if we move the color cut bluer, we increase the contamination 
 from $z\sim$0.3 galaxies (which are well within the magnitude limit of
 SDSS).
 By compromising these two effects, we optimize the color cut criteria.

\subsection{Enhancement Method}\label{sec:enhance}

 After applying the color cuts, we use a special enhancement method to
 galaxies within the cut in order to enhance the signal to noise ratio of clusters further.
 Among galaxies within a certain color cut (or color-color cut), we find
 all pairs of galaxies within five arcmin; this scale 
 corresponds to the size of a galaxy cluster at $z\sim$0.3 .
 Selecting larger separations blurs high $z$ clusters,
 while smaller separations  weaken the signal of low $z$ clusters.
 We empirically investigated several separations 
 and found 5' to be the best.
 Using the angular distance and color difference of each pair of galaxies,
 we distribute a Gaussian cloud representing the density of galaxies
 on the center position of each pair. 
 The width of the Gaussian cloud is set to be an angular separation of the pair.
 The volume of a Gaussian cloud is given by its weight ($W$), 
 which is calculated as follows:

\begin{equation}\label{equation}
 W =  \frac{1}{\Delta r + 1''} \times \frac{1}{\Delta (g-r)^2 +  2.5\times10^{-3}},
\end{equation}
 where $\Delta r$ is an angular separation between the two galaxies and 
$\Delta$($g-r$) is their color difference. Small softening parameters
 are empirically determined and  added in the denominator of each
term to avoid the weight becoming infinite.  
 This enhancement method provides stronger weights to pairs which
are closer both in angular space and in color space,
 thus are more likely to appear in real galaxy clusters. 
 Gaussian clouds are distributed to each  30''$\times$30'' cell on the sky.
 The 30'' cells are much smaller than sizes of galaxy
 clusters (several arcmins even at $z\sim$0.5).

\begin{figure}
\includegraphics[scale=0.3]{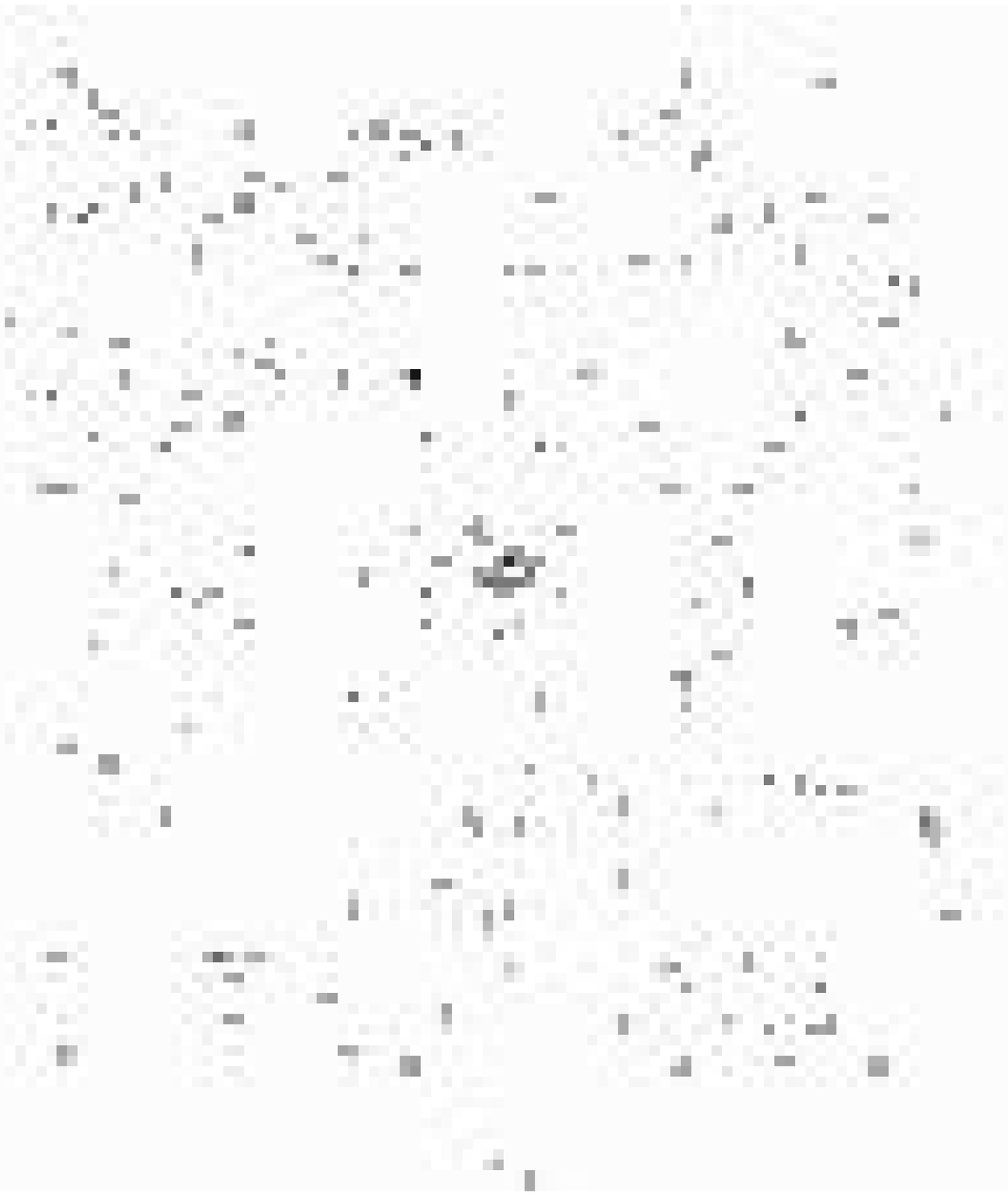}
\includegraphics[scale=0.3]{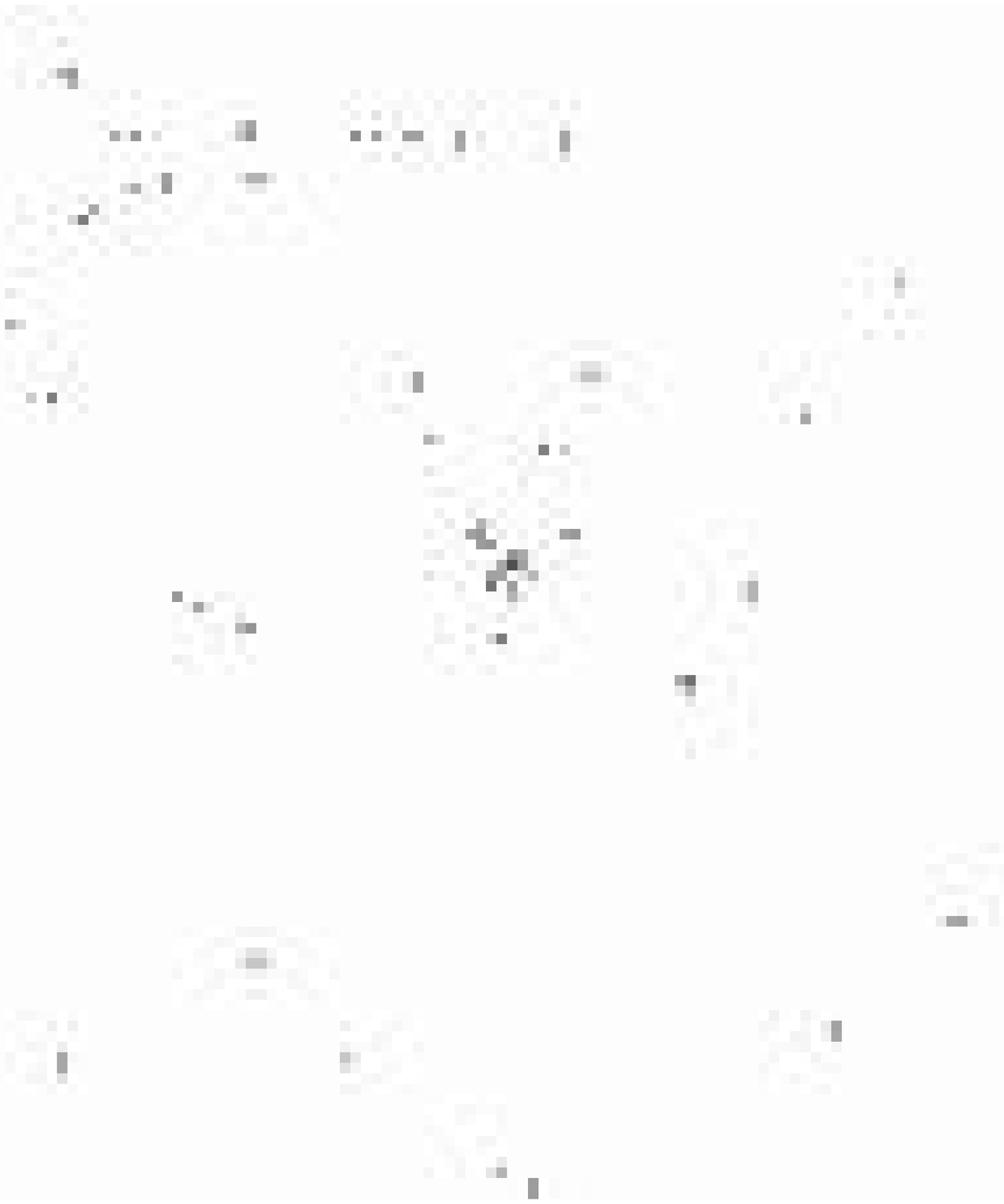}
\centering{
\includegraphics[scale=0.35]{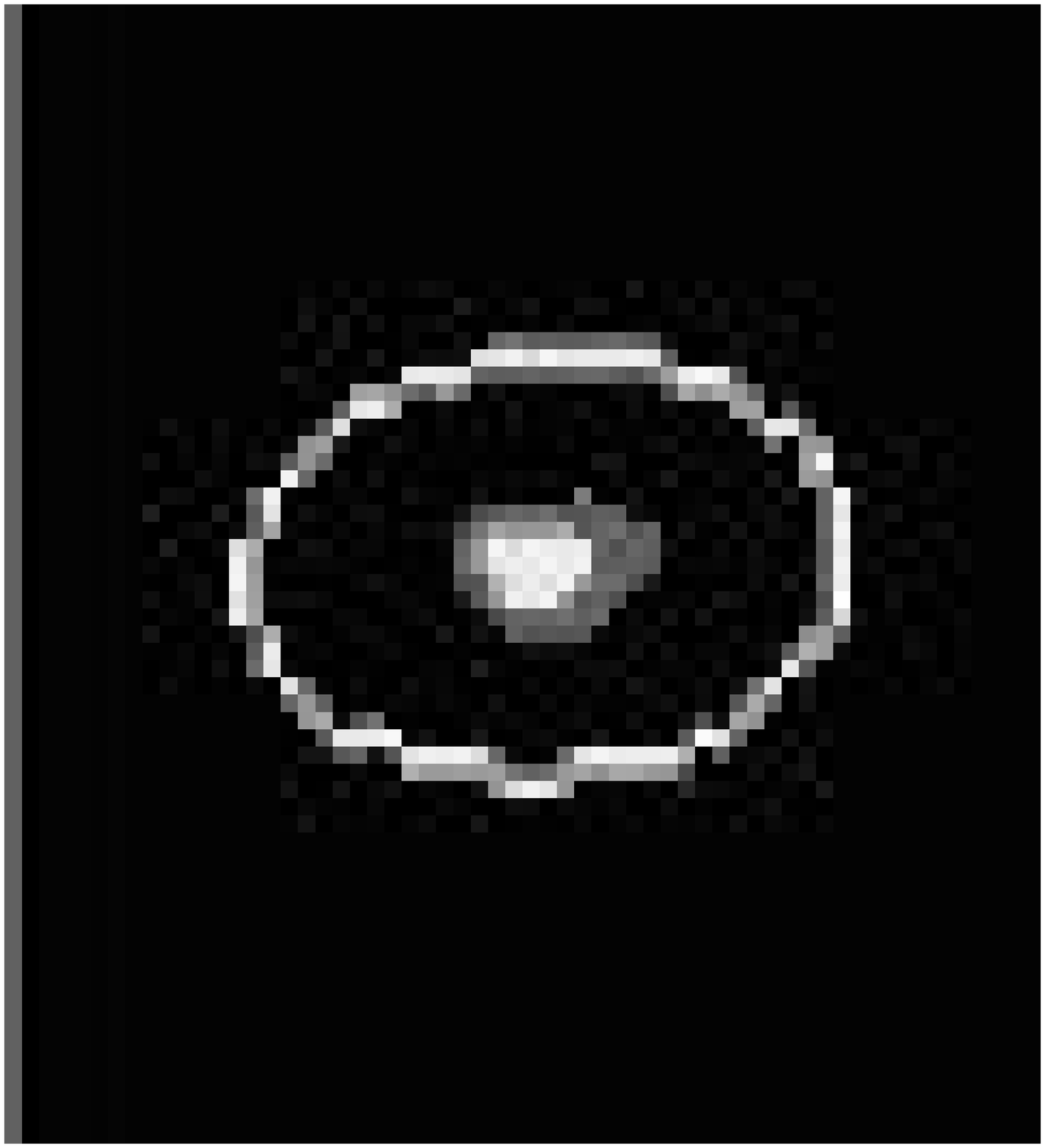}
}

\caption{
 The distribution of galaxies brighter than $r^*$=20.0 around
 RXJ0256.5+0006. The upper left panel shows the distribution before
 applying any color cut. The upper right panel shows the distribution after applying
 $g-r$ color cut. The lower panel shows the enhanced density map.
 The color cut removes foreground and background galaxies as designed.
 RXJ0256.5+0006 is successfully detected as circled with the white line.
}\label{fig:beforecut_RXJ}
\end{figure}

\begin{figure}
\begin{center}
\includegraphics[scale=0.7]{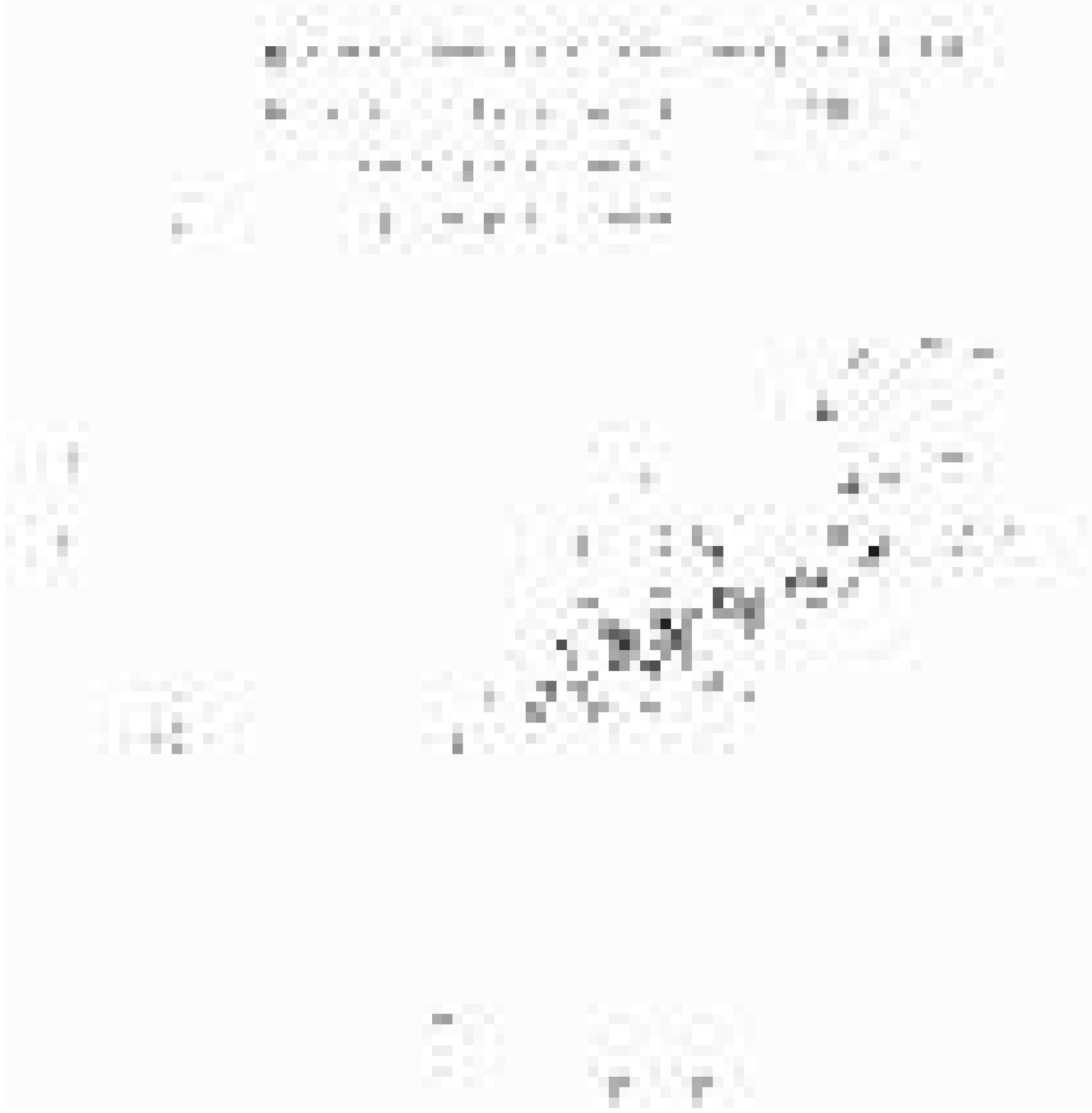}
 \caption{
Color-color diagram of spectroscopically confirmed member galaxies of
 A168.
 The abscissa is $g-r$ color. 
 The ordinate is $r-i$ color.
The low-$z$ $g-r-i$ box is  drawn with the dashed lines.
The high-$z$ $g-r-i$ box is  drawn with the dotted lines. 
 Galaxies brighter than $r^*=$21 which matched up the spectroscopically
 confirmed galaxies (Katgert et al.\ 1998) are plotted with the dots. 
 The triangles show the 
 color prediction of elliptical galaxies (Fukugita et al.\ 1995).
} \label{fig:griboxes_a168}
\end{center}
\end{figure}

\begin{figure}
\begin{center}
\includegraphics[scale=0.7]{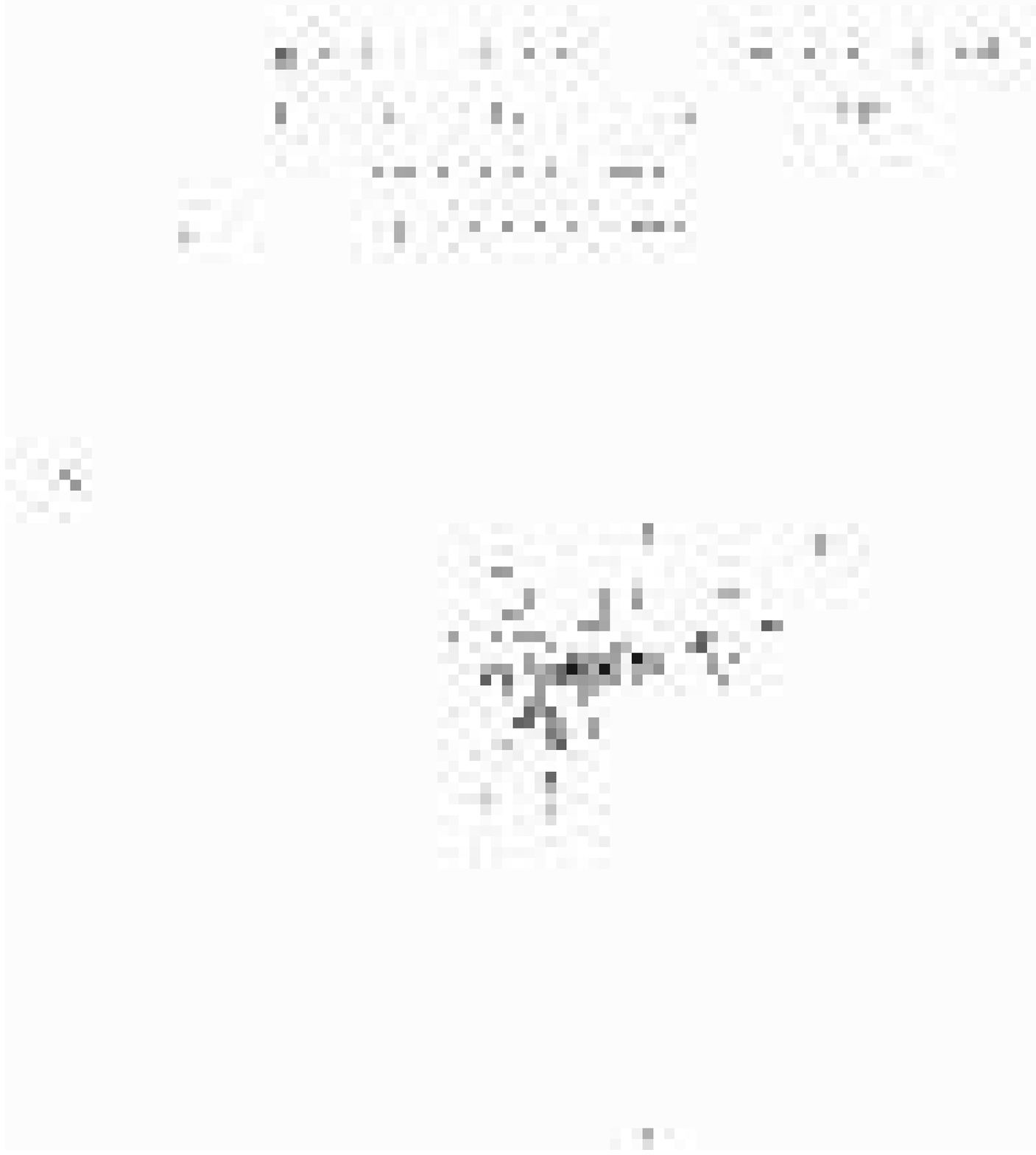}
\caption{
 Color-color diagram of spectroscopically confirmed member galaxies of
 A168.
 The abscissa is $r-i$ color. 
 The ordinate is $i-z$ color.
 The low-$z$ $r-i-z$ box is  drawn with the dashed lines.
 The high-$z$ $r-i-z$ box is  drawn with the dotted lines. 
 Galaxies brighter than $r^*=$21 which matched up the spectroscopically
 confirmed galaxies (Katgert et al.\ 1998) are plotted with the dots. 
 The triangles show the 
 color prediction of elliptical galaxies (Fukugita et al.\ 1995).
}\label{fig:colors_of_Elliptical_a168}
\end{center}
\end{figure}

\begin{figure}
\begin{center}
\includegraphics[scale=0.7]{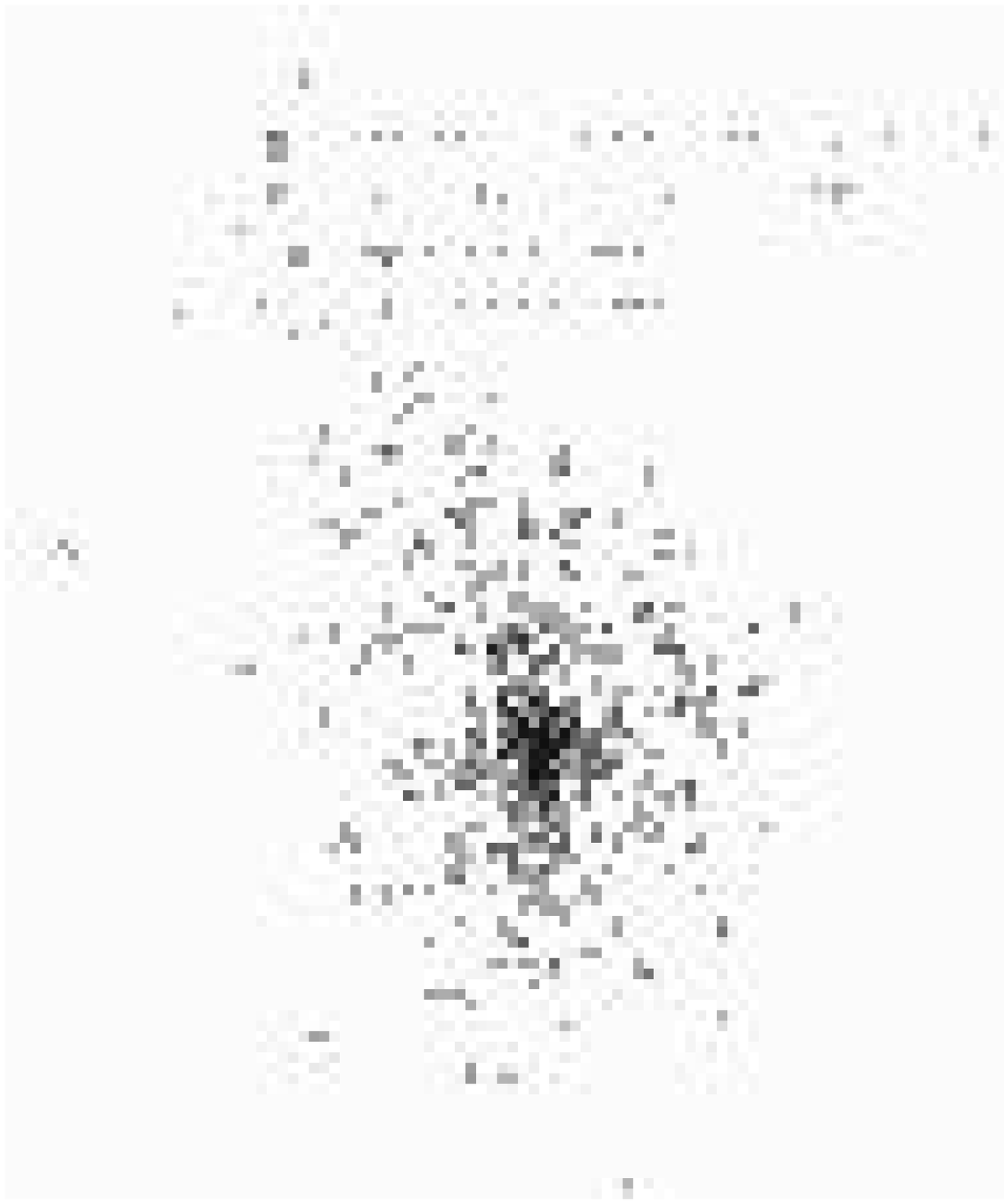}	
       \caption{
 $r-i-z$ color-color boxes to find galaxy clusters.
 The abscissa is $r-i$ color. 
 The ordinate is $i-z$ color.
The low-$z$ $r-i-z$ box is  drawn with the dashed lines.
The high-$z$ $r-i-z$ box is  drawn with the dotted lines. 
 Galaxies brighter than $r^*$=22 in the SDSS fields
 ($\sim8.3\times10^{-2} deg^2$)  which covers A1577 are plotted with the
 small dots.
 The triangles show the 
 color prediction of elliptical galaxies (Fukugita et al.\ 1995).
}	      \label{fig:colors_of_Elliptical}
\end{center}
\end{figure}

\begin{figure}
\includegraphics[scale=0.38]{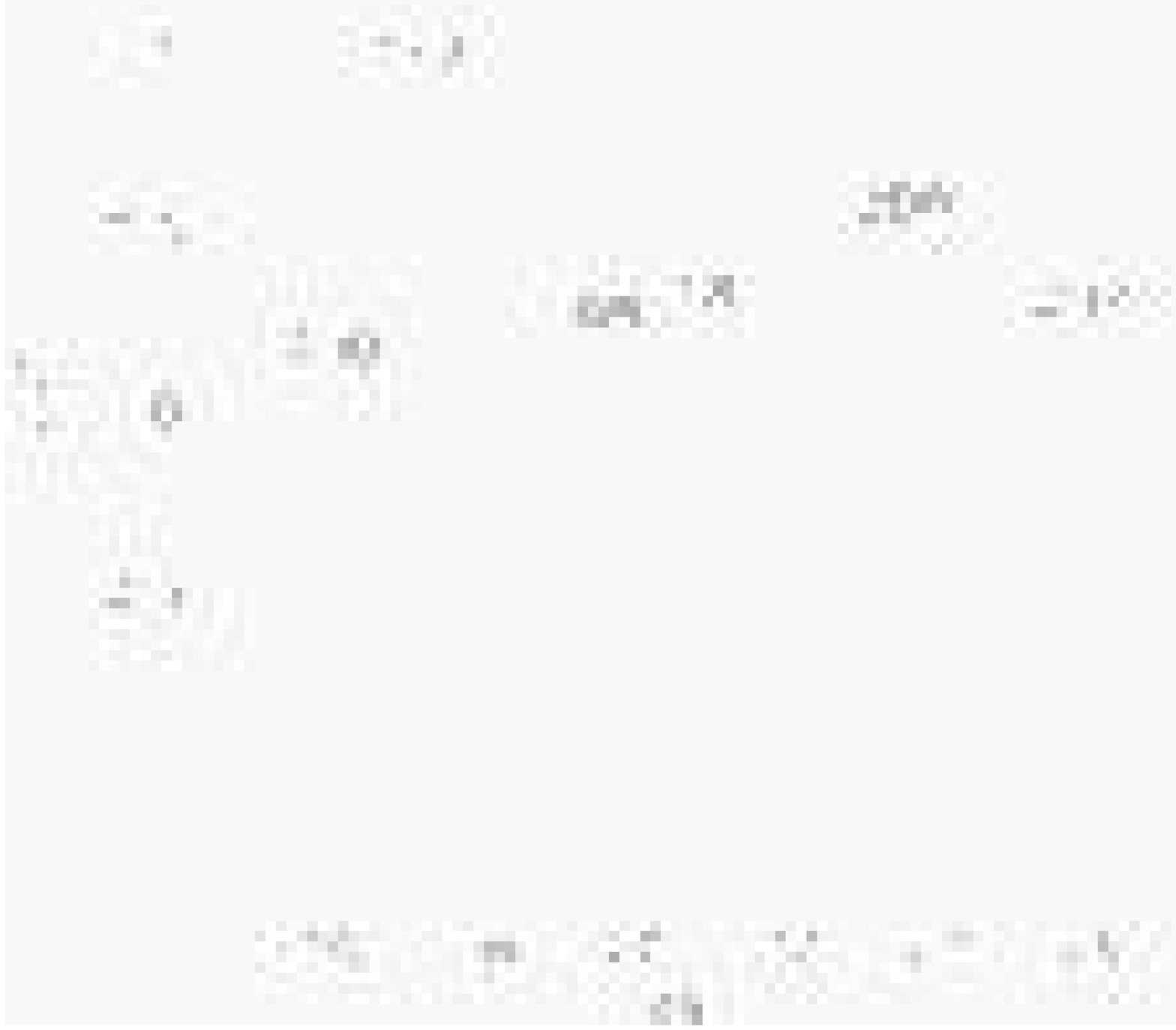}
\includegraphics[scale=0.38]{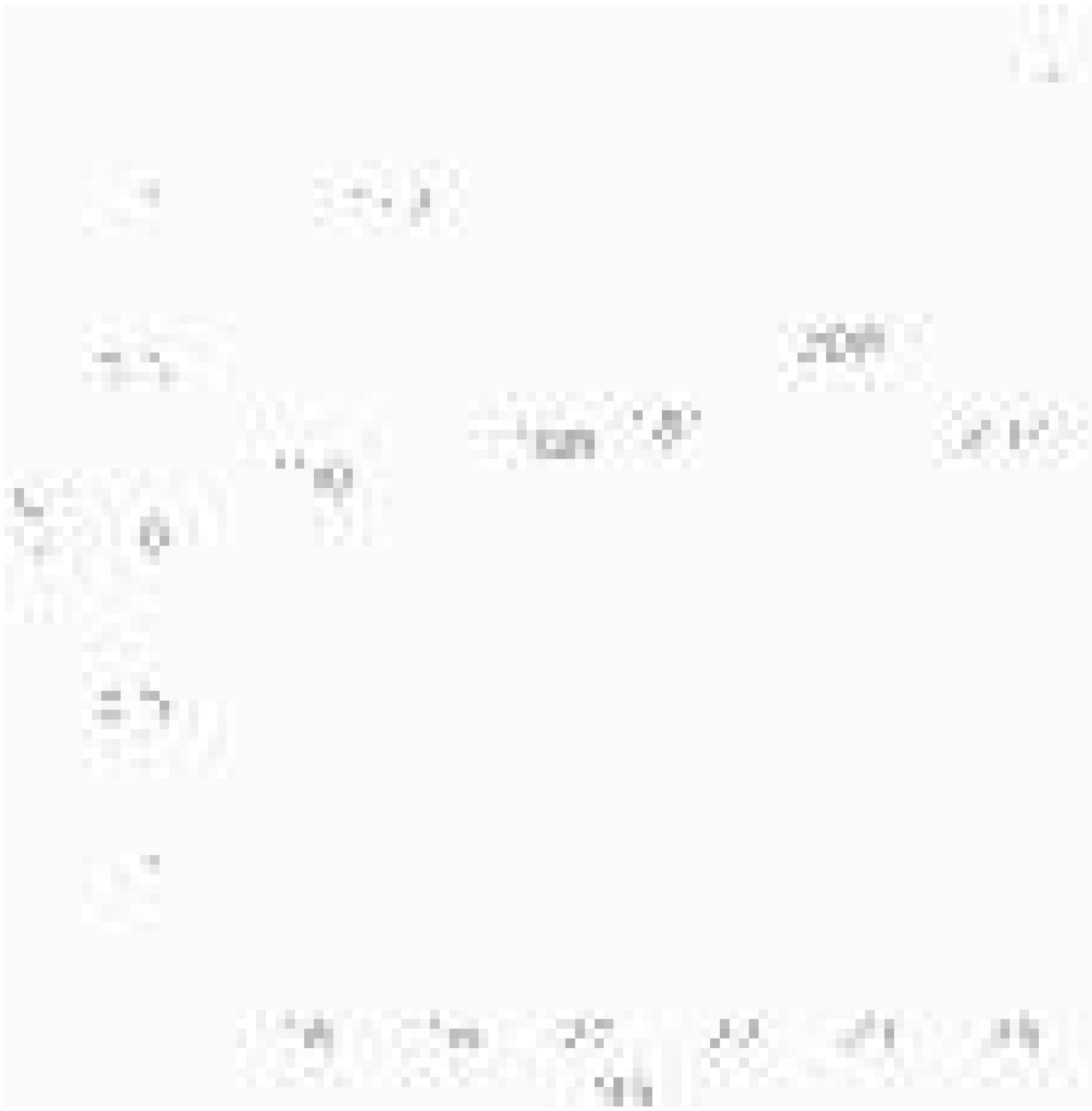}
\centering{\includegraphics[scale=0.45]{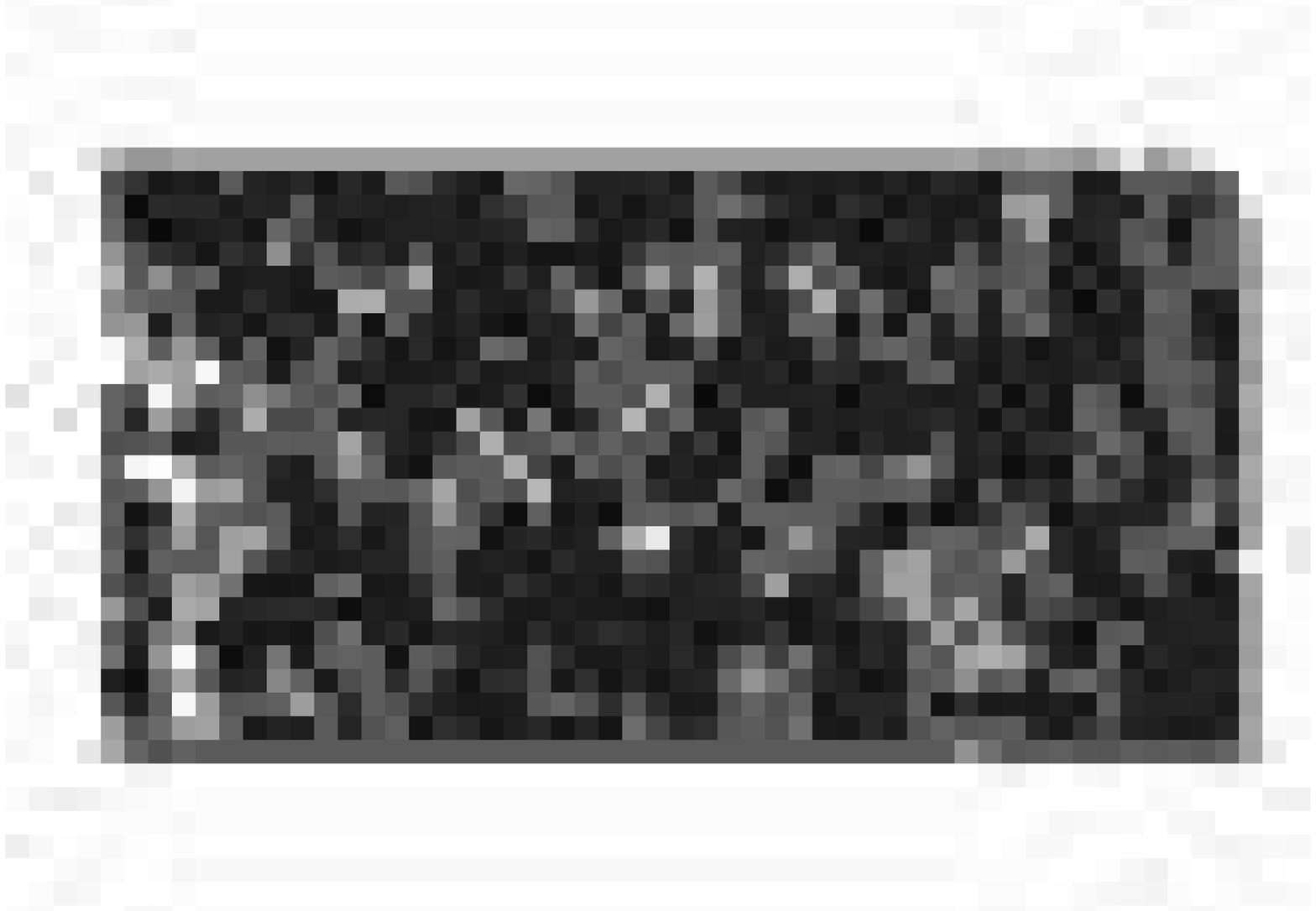}}
\caption{
The distribution of galaxies brighter than $r^*$=21.5.
The upper left panel shows the distribution before applying any cut.
The upper right panel shows the distribution after applying
 $g-r-i$ color-color cut. The numbers show the positions of Abell clusters.
 The lower panel shows the enhanced density map in $g-r-i$ color-color
 cut. Detected clusters are circled with the white lines.
}\label{fig:beforecut_comparison}
\end{figure}

 An enhanced density map is obtained by summing
 up the Gaussian clouds for every pair of galaxies within 5'.
 The lower panels in Figure \ref{fig:beforecut_RXJ} and
 \ref{fig:beforecut_comparison} present such enhanced density maps of
 the region in their upper panels.
 RXJ0256.5+0006 is successfully enhanced in
 Figure \ref{fig:beforecut_RXJ}.  Figure  \ref{fig:beforecut_comparison}
 illustrates how the CE method finds galaxy clusters in a larger region. 
 The advantage of this enhancement method in addition to the color
 cuts is that it makes full use of color similarity of
 cluster galaxies. 
 The color cuts are used to reduce foreground and background 
galaxies and to enhance the signal of clusters. 
Since the color-magnitude relation of cluster galaxies is frequently tighter than the width
of our color cuts, the use of the second term in equation (\ref{equation})  
- the inverse square of the color difference - further enhances the signal of cluster,
 in spite of the larger width of the color cuts.
 Another notable feature is that the enhancement method is adaptive,
 that is, large separation pairs have a large Gaussian and small separation
 pairs have a sharp, small Gaussian. In this way, the enhancement method
 naturally fit to any region with any number density of 
galaxies in the sky. Therefore, it is also easy to apply it to data from another 
telescope with different depth and different galaxy density.
 Another benefit of the enhancement method is that it includes 
a smoothing scheme. And thus, 
 conventional detection methods  commonly used in astronomical community
 can be used to detect clusters in the enhanced density map.
 The enhancement method uses an angular separation in the computation of $W$-values. 
 This might bias our catalog against nearby clusters ($z<$0.1),
 which have a large angular extent (and thus are given less $W$).
 However, these nearby clusters are few in the SDSS commissioning data due to the small
 amount of the volume probed in the nearby universe. 
 In addition, these nearby clusters will also be well sampled in the SDSS
 spectroscopic survey with fiber redshifts, 
 and will thus be detected in the SDSS 3D cluster selection.
 (the CE cluster detection method is
 intended to detect clusters using imaging data).
 These nearby clusters do not have a significant effect on angular or
 redshift-space correlations because the number of such clusters is a
 small fraction of any large volume-limited sample.

\subsection{Detection}\label{sec:detection}


 We use SourceExtractor (Bertin et al.\ 1996) to detect clusters
 from the enhanced density map discussed in Section\ref{sec:enhance}.  
 SourceExtractor identifies high density peaks  above a given threshold
 measuring the background and its fluctuation locally.
 The threshold selection determines the number of clusters obtained.
 A high threshold selects only rich clusters.
 We tried several thresholds, examining the
 colored image, color-magnitude and color-color diagrams of  the
 resulting cluster catalog. The effect of changing threshold is 
 summarized in Table \ref{tab:sigma_test}. The numbers of clusters
 detected are not very sensitive to the threshold\footnote{The numbers
 of detection can increase or decrease with increasing sigma because the
 following two effects cancel 
 out each other. (1) Lower threshold detects faint sources and thus
 increases the number of detections. (2) Higher threshold deblends the peaks and
 increases the number of the detections.}.  
 Based on the above, we have selected the threshold to be 
 six times the background
 fluctuation, it is  the threshold which yields a large number of clusters
while the spurious detection rate is still low.  

 Monte Carlo simulations are sometimes used  to decide the optimal threshold,
 where most true clusters are recovered while
 the spurious detection rate is still low.
   However, the simulations reflect an ideal situation, and they are 
inevitably different from true data; for example, a uniform background
 cannot represent the true galaxy distribution with its large scale
  structure. There are always clusters which do not match the
   radial profile or luminosity function assumed in Monte Carlo
   simulations 
 and this may affect the optimization of the threshold.
 The optimal threshold in Monte Carlo simulation differs from
 the optimal threshold in the real data.
 Therefore, we select the threshold empirically using the actual data and later
 derive the selection function using Monte Carlo simulation.

At high redshifts ($z>$0.4), the number of galaxies within the color
cuts is small; therefore the 
 rms of the enhanced density map is generally too low and
the clusters detected at high redshift have
 unusually high signal. 
 To avoid such spurious detections,
we applied another threshold at maximum absolute 
flux of 1000 in the enhanced density map\footnote{FLUX\_MAX+BACKGROUND=1000, where FLUX\_MAX and
 BACKGROUND are the parameters of Source Extractor.
  FLUX\_MAX+BACKGROUND is the highest
 value in the pixels within the cluster.  It is an absolute value, and 
not affected by rms value.}.
Spurious detections with high signal would generally have low values of
this parameter
because they are not true density peaks.
 The threshold of maximum absolute flux of 1000 can thus reject spurious detections.
The value is determined by investigating the image, color-magnitude and
 color-color diagrams 
 of the detected clusters and iterating the detection with different values of
 the maximum absolute flux threshold. The effect of changing the
 absolute flux threshold is summarized in Table  \ref{tab:fluxmax}.

 To secure the detection further, at all redshifts, we demand at least
 two detections in the 34 cuts. 
 Since the cluster galaxies have similar colors 
 in all $g-r$, $r-i$ and $i-z$ colors; 
 real clusters should thus be
 detected in at least two color cuts. 
 This requirement significantly reduces spurious detections.  


\subsection{Merging}
 We apply the procedure of cut (Section \ref{color-cut},
 \ref{sec:ce_color-color-cut}), enhance (Section \ref{sec:enhance}) and
 detection (Section \ref{sec:detection}) to all of the 34 color
cuts (30 color cuts + four color-color cuts) independently. 
After creating the 34 cluster lists, we 
merge them into one cluster catalog. 
We regard the detections within 1.2 arcmins as one cluster.
 To avoid two clusters with different redshifts being merged into one cluster
due to the chance alignment,
 we do not merge clusters  
 unless the successive two color cuts in a certain color both detect it.

 An alternative way to merge clusters would be 
 to merge only those clusters which are detected 
in the consistent color cut in all $g-r$, $r-i$ and $i-z$ colors,
 using 
 the model of the elliptical galaxy colors.
 However, the catalog will be biased against clusters which have
 different colors than the model ellipticals.
 In order to minimize the assumptions on cluster properties 
 we treat the three color space, $g-r$, $r-i$ and $i-z$ , independently.

\subsection{Redshift and Richness Estimation}\label{sec:real-catalog}

  One of the very important parameters for various scientific researches using
 a cluster catalog is redshift and richness.
 We estimate redshift and richness of each cluster as
 follows.
 Instead of the richness estimator introduced by  Abell (1958),
 we count the number  of galaxies inside the detected cluster radius 
which lie in the two magnitude range (in $r$) from $m_3$ (the third
 brightest galaxy) to $m_3$+2 (CE richness).
 The difference from Abell's richness is that he used a fixed 1.5 $h^{-1}$Mpc as
a radius.  
 Here we use the detection radius of the cluster detection algorithm
 which can be larger or smaller than
Abell radius, typically slightly smaller than  1.5 $h^{-1}$Mpc.
 Since there are a significant variety in a size of real clusters,   a
 varying radius can measure more representative cluster richness than a
 fixed radius. 
 The background galaxy count is subtracted using the average galaxy counts
in the SDSS commissioning data.

\begin{figure}
\begin{center}
\includegraphics[scale=0.7]{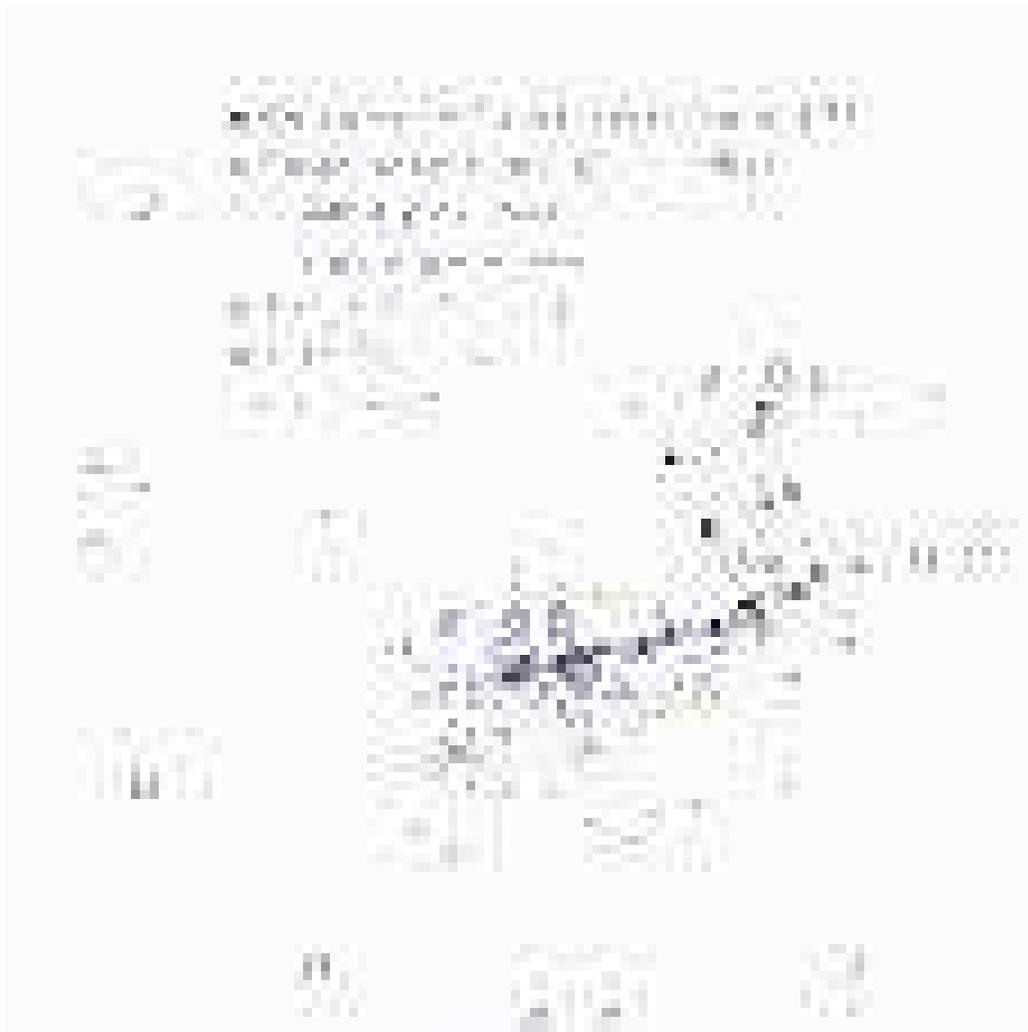}
\caption{
Evaluation of high-$z$ color cut. 
The filled triangles show the color prediction for elliptical galaxies (Fukugita et al. 1995).
The open triangles show the color prediction of non-star-forming galaxies of
 the PEGASE model (Fioc \& Rocca-Volmerange 1997).
The open squares show the color prediction of star-forming galaxies of the PEGASE model.
The black dots are the galaxies around A1577.
high-$z$ color cut and low-$z$ color cut are drawn with the dashed and
 the long-dashed lines.
}\label{fig:color-cut-test} 
\end{center}
\end{figure}

 For the redshift estimates, we use 
the strategy of the redshift estimation of the maxBCG technique
(Annis et al. in prep.). 
 We count the number of galaxies within the detected radius 
 that are brighter than $M^*_{r^*}=-$20.25 at each redshift 
 and are within a color range of $\pm$0.1 mag in $g-r$ around the color
 prediction for elliptical galaxies  (Fukugita et al.\ 1995).
 This procedure is iterated for each redshift step of $\Delta z$= 0.01.
 After subtracting average background number counts from each bin,
the redshift of the bin that has the largest number of galaxies is
taken as an  estimated cluster redshift. 
The estimated redshifts are calibrated using the spectroscopic redshifts
from the SDSS spectroscopic survey. Our redshift estimation depends on
the model of Fukugita et al. (1995), but the difference from other
models are not so significant, as  seen in 
the difference between filled squares (PEGASE model) and
filled triangles (Fukugita et al. 1995) of Figure \ref{fig:color-cut-test}.
If a cluster has enough elliptical galaxies, the redshift of the cluster
is expected to be well measured.
 If a cluster is, however, dominated only by spiral galaxies,
 the redshift of the cluster might be underestimated,  as can be judged  in the
 difference between the pentagons and the triangles in Figure \ref{fig:color-cut-test}.
 Figure\ \ref{fig:zaccuracy.eps} shows redshift accuracy of the method.
 The estimated redshifts are plotted against observed redshifts from the
 spectroscopic observation.
 The redshift of the SDSS spectroscopic galaxy within the detected
radius and with the nearest spectroscopic redshift to the estimated redshift is
adopted as a real redshift. In the fall equatorial region,
 699 clusters have spectroscopic redshifts.
 The correlation between true and estimated redshifts is excellent:
 the rms scatter is $\delta z$=$\pm$0.0147 for $z<$0.3
clusters, and $\delta z$=$\pm$0.0209 for $z>$0.3 clusters.
 The triangles in Figure \ref{fig:zaccuracy.eps} show 15 Abell clusters whose spectroscopic redshift is
 available in the literature.
 There are three outliers at low
 spectroscopic redshifts. CE counterparts for these three  clusters all have
 very small radii of several arcmin. Since these Abell
 clusters are at $z<$0.1, 
 the discrepancy is probably not in the redshift estimation 
 but rather in the too small detection radius.

 By applying this CE method including redshift and richness estimation
 to the SDSS commissioning data, 
  we construct the SDSS CE galaxy cluster catalog. The SDSS
 CE cluster catalog contains 4638 galaxy clusters. 
 The catalog is available at the
 following website: http://astrophysics.phys.cmu.edu/$\sim$tomo
 \footnote{Mirror sites are available at
 http://sdss2.icrr.u-tokyo.ac.jp/$\sim$yohnis/kokki/public\_html/ce/index.html
 , and 
 http://indus.astron.s.u-tokyo.ac.jp/$\sim$yoh/ce/index.html
 }

\section{Monte Carlo Simulation}\label{sec:monte}

 For statistical study using a cluster catalog, it is of extreme
 importance to know completeness and contamination rate of the catalog.
 In this section, we examine the performance of the CE
 method and determine the selection function using extensive Monte Carlo
 simulations. We also perform false positive tests.

\subsection{Method}

 We perform Monte Carlo simulations both with a real background using the SDSS
 commissioning data and with a shuffled background. 
 For the real background, we randomly choose an 1 deg$^2$ region of the SDSS
 data with seeing better than 1.7''.\footnote{
 Although the SDSS survey criteria for seeing
is better than 1.5'', some parts of the SDSS commissioning data have seeing worse than 2.0''.
It is expected that the seeing is better than
1.5'' for all the data after the survey begins.}
 For the shuffled background,
 we re-distribute all the
 galaxies in the above 1 deg$^2$ of SDSS data randomly in position, but keep
their colors and magnitudes unchanged.

 Then, we place artificial galaxy clusters on these backgrounds.
 We distribute cluster galaxies randomly using a King profile (King
 1966; Ichikawa
 1986) for the  radial galaxy density, 
 with concentration index of 1.5 and cut
 off radius of $2.1h^{-1}$Mpc, which is the size of Abell 1577 (Struble \& Rood  1987). 
 For colors of the artificial cluster galaxies, we use the color and magnitude
distribution of Abell 1577 (at $z\sim0.14$, Richness$\sim1$) as a model.
 We choose the SDSS fields
which cover the entire Abell 1577 area and count the number of galaxies in
each color bin. The size of the bins is 0.2 magnitude in both colors and
magnitude. 
The color and magnitude distribution spans in four dimension space, $r$,
$g-r$, $r-i$ and $i-z$.
 We count the number of field galaxies using a field of the same size
 near (but different from) the Abell 1577 field and subtracted the distribution of field
galaxies from the distribution of galaxies in the Abell 1577 fields.  
 The resulting  
color distribution is used 
as a model for the artificial galaxy clusters. 
 Artificial galaxy colors are assigned randomly so that they reproduce the overall color
 distribution of Abell 1577. 
 The distribution is linearly interpolated when allocating colors and
 magnitudes to the galaxies.

 For the high redshift artificial clusters, we apply $k$-correction and
 the color prediction of elliptical galaxies from Fukugita et al.\
 (1995).
  For the color prediction, only  the color difference, not the absolute
 value, is used due to the uncertainty in zero points.
 Galaxies which become fainter than $r^*$=21.5 by applying cosmological
 dimming and a $k$-correction
 are not used in the simulations.

\subsection{Monte Carlo Results}

 First, we run a Monte Carlo simulation with only the background,
  without any
 artificial clusters, in order to measure the detection rate of the
simulation itself.
The bias detection rate is defined as the percentage 
in which any detection is found within 1.2 arcmins from the
position where we later place an artificial galaxy cluster.
The main reason for the false detection is 
that a real cluster sometimes comes into the detection 
position, where an artificial cluster is later placed.
 This is not the false detection of the CE method but rather
  the noise
 in the simulation itself.
The bias detection rate with the real SDSS background is
4.3\%.
 This is small relative to the actual cluster detection rate discussed below.
 The bias detection rate using the shuffled background is lower (2.4\%),  as
 expected.

 We are now ready to perform simulations with an artificial cluster.
 We run Monte Carlo
 simulations with a set of artificial clusters with redshifts ranging 
 from $z=$0.2 to $z=$0.6, and with richnesses of $Ngal$= 40, 60, 80
 and 100, at each redshift ($Ngal$ is the number of galaxies
 inputted into an artificial cluster. If a galaxy becomes fainter than $r^*=21.5$, it is
not counted in the CE detection method
 even if it is included in $Ngal$).
 For each set of parameters, a simulation is iterated 1000 times. 
 In Figure\ \ref{fig:ngal-ab1ellrich}, we compare $Ngal$ 
 with cluster richness
 where richness is defined  as the number of galaxies 
 within the two magnitude range fainter than the third brightest galaxy, located within the
 cluster radius that the CE method returns (Section \ref{sec:method}). 
 The error bars are 
 $1\sigma$ standard  error. $Ngal$=50 corresponds to  Abell richness
 class $\sim$1.

 Figure\ \ref{fig:monte-recovery-simu} shows the recovery rate in the Monte
 Carlo simulations 
 on the real background.
 The  recovery rate is shown in percentage as a function of redshift.
 Each line represents input clusters of different richness, $Ngal$=100,
 80, 60 and 40 (top to bottom).
 Since the false detection rate in the simulation with real
 background is 4.3\%, all the lines converges to 4.3\% at high redshift.
 The detection rate drops suddenly at $z=$0.4 because  at this point, a
 large fraction of the cluster member galaxies are lost as a result of  
 the magnitude limit at $r^*=21.5$. Roughly speaking, this apparent
 magnitude limit determines the depth of the SDSS cluster catalog.
 $Ngal$=80  clusters are recovered 
 $\sim$80\% of the time to $z<$0.3, dropping to $\sim$40\%
 beyond $z\sim$0.4.
 Clusters of the lowest richnesses, 
 $Ngal$=40 clusters are more difficult to detect, as expected.
 The recovery rates of $Ngal$=40 clusters are  
 less than 40\% even at $z=$0.3.
 Strangely, the recovery rate for $Ngal$=100 at $z=$0.2 is not 100\%.
 However, if we widen the detection radius from 1'.2 to 5'.4, the recovery rate
 reassuringly increases to 100\%.  
 Since the radius of 5'.4 is fairly smaller than
 the size of A1577 (11' at $z=$0.2; Struble \& Rood 1987),
 the reason may be that 
 a real cluster (in the real background) 
 happens to be located close to an artificial cluster.
 In such a case, the detected
 position may  be shifted away by more than 
 the detection radius (1'.2) from the cluster center, 
 resulting in the recovery rate of less than 100\% within 1'.2 .

 Figure\ \ref{fig:z-posi-simu} shows the positional accuracy of the detected
 clusters in the Monte Carlo simulation with the real SDSS
 background, as a function redshift and richness.
  The 1$\sigma$ positional errors of the detected clusters is shown.
 Note that since the CE does not detect many clusters beyond
 $z=$0.4 in the SDSS data, there is not much importance in discussing the position accuracy
 of beyond  $z=$0.4 with this data.
 The positional accuracy is
 better than 1' until $z=$0.4 in all the richness ranges used.
 The accuracy is nearly independent of the redshift because the high redshift
 clusters are more compact than the low redshift ones. 
 This partially cancels the effect of losing  more galaxies at high redshift
 due to the flux limit of the sample.
 The positional accuracy roughly corresponds to the mesh size of the enhancement
 method, 30''.
 As expected, the positional accuracy is worse for high redshift poor
 clusters ($z=$0.4 and $Ngal\leq$60).
 The statistics for these objects are also worse; 
 the detection rate of $Ngal$=60 and 40 clusters are less than 20\% at
 $z=$0.4 (Figure \ref{fig:monte-recovery-simu}).

 Figure\ \ref{fig:monte-recovery-uni} presents the recovery rate of artificial
 clusters in Monte Carlo simulations with the shuffled background. The
 recovery rates are slightly better than those with the real background since
 no real clusters are in the shuffled background. Again, the recovery
 rates drop sharply at $z=$0.4. 
 The $Ngal$=100  clusters are recovered with $\sim$90\% probability to
 $z\sim$0.3 and $\sim$40\%  at $z\sim$0.4.
 At $z\leq$0.3,  $Ngal>$40 clusters are recovered at $>$40\%.
 Figure\ \ref{fig:z-posi-uni} shows the positional accuracy of the detected
 clusters in the simulations (with shuffled background).
 The results are similar to those with the real background.
 The positional accuracy is
 better than 40'' until $z=$0.3 for all the richnesses.

\subsection{False Positive Tests}\label{sec:false-positive}
 
 In order to test false positive rate, we prepared 
 four sets of data: 
 1) Real SDSS data of 25 deg$^2$.
 2) Position of galaxies in the same 25 deg$^2$ are 
 randomized (galaxy colors untouched.) 
 3) Colors of galaxies are shuffled. (galaxy position untouched.) 
 4) Color is shuffled and position is smeared (5'). Galaxy colors are randomized 
 and positions are randomly distributed in the way that galaxies still lie within 5'
 from their  original positions. This case is intended to include large scale
 structure without galaxy clusters.
 The results are shown in Figure \ref{fig:tim_test}. 
  The solid line represents the results with the real
 data. The dotted line represents the results with the position shuffled
 data. The long-dashed line is for the color shuffled data. For color shuffled 
 data, we subtracted the detections in the real data, because they still
 contains real clusters in them.  
 It is consistent with the generous color cuts of the CE method that 
 many clusters can still be detected in the color shuffled data.
 The short-dashed line is for the color shuffled smeared data.
 In Figure \ref{fig:tim_test_ratio}, the fraction to the real data is plotted 
 against CE richness. 
 The promising fact is that not so many sources are detected from 
 the position shuffled data. The fraction to the real data is less than 20\% at
 CE richness $>$20. 
 More cluster candidates are detected from the color shuffled data 
 and the smeared data but this does not mean the false positive rate of
 the CE
 is as high as those values since
 smeared data still contains a
 structure larger than 5', and they can be real clusters. 
 To conclude, our simulations show that for clusters with richness$>$10, over 70\% of 
 CE clusters are likely to be real systems (as shown by the color \& position
 shuffled simulations.)

\section{Visual Inspection}\label{sec:visual}

 To investigate  whether the detected clusters are true clusters or spurious
detection, spectroscopic observations are necessary.
 Although  large spectrometers which can observe the spectra of many
galaxies at one time are becoming available (e.g., SDSS; 2dF),
 spectroscopy of large number of galaxies is still time consuming.
Since the SDSS CE cluster  catalog 
will have more than 100,000 galaxy
clusters when the survey is complete, it is in fact impossible to 
spectroscopically confirm all the clusters in 
the catalog.
As a preliminary check of our method, 
we visually inspect all the CE clusters within a given area
(right ascension between 16 deg and 25.5 deg and declination between
$-$1.25 deg and +1.25 deg, totaling  23.75 deg$^2$. The region in Figure  \ref{fig:beforecut_comparison}).
 A total of 278 CE
 galaxy clusters are located within this area (after removing clusters
 touching the region's borders). 
 Out of the 278 CE 
 galaxy clusters, 
 we estimate that 
10 
are false detections.
 Since the strategy of the CE method is to detect 
 every clustering of galaxies, 
 we call every angular clustering of galaxies with the same color  
a successful detection here. (As we show in Section \ref{sec:false-positive},
 30\% of clusters could be false detections, such as chance projections.)

 Among the 10 false detections, three are  bright big galaxies deblended
 into several pieces. 
 In the other cases,
 a few galaxies are seen but not an apparent cluster or group.
 (In one case a rich cluster exists
 about six arcmin from the false detection).
 The 10 false detections are summarized in Table \ref{tab:ce-miss-new}.
 $\sigma$ (column[1]) is the significance of the detection; 
CE richness (column[2])  is its richness;
$z$ (column[3]) is the color estimated redshift;
 and comments are given in column[4].

 As the successful examples, 
 we show two typical examples of clusters detected only with the CE method
but not with the other methods (discussed below).
 One is a clustering of blue galaxies. 
Since the CE method does not reject blue
spiral galaxies, it can detect clustering of several blue spiral
galaxies. 
 Indeed, some of the detected clusters that we visually inspected 
 are clustering of blue galaxies.
The other is a clustering of numerous faint elliptical galaxies;
 in these regions faint elliptical galaxies 
 spread out over a large area ($\sim$ 0.01 deg$^2$) but with no bright cluster galaxies. 
CE method detects these
regions successfully with a large radius. Figure\ \ref{fig:dwarf-region}
shows the true color image of one of these clusters with numerous faint elliptical
galaxies.
 Figure\ \ref{fig:success-cluster} shows a typical galaxy cluster
 successfully detected with the CE method.

\section{Comparison with Other Methods}\label{sec:compare}

 At the time of writing, the SDSS collaboration has implemented several
independent cluster finding methods and have run these algorithms on the SDSS
commissioning data. These methods include the Matched Filter (MF; Kim
et al.\ 2001), the Voronoi Tessellation (VTT; Kim et al.\ 2001), and the
maxBCG
technique (Annis et al. in prep.). Therefore, we have the unique opportunity to
compare the different catalogs these algorithms produce to further understand
each algorithm and possible differences between them. (also see Bahcall
et al. 2003 for comparisons of SDSS cluster catalogs.)

Here we provide a comparison between the CE method and the MF, VTT and maxBCG
techniques using a small sub--region of the SDSS data, i.e., 23.75 deg$^2$
of commissioning data with RA between 16 and 25.5 degrees and Declination
between $-$1.25 and +1.25 degrees (the region in Figure \ref{fig:beforecut_comparison}). We first matched the CE catalog with each of
the other three catalogs using a simple positional match criterion of less
than six arcmins.  The number of matches between the CE and other catalogs
varies significantly because each cluster--finding algorithm has a different
selection function. At present, the selection functions for all these
algorithms are not fully established so we have not corrected for them in this
comparison.
 Although each algorithm measures cluster richness and redshift in its
 own way, the scatter between the measurement is large and it makes the
 comparison difficult.
 Therefore, we re-measured richness and redshift of the MF, VTT and maxBCG
 clusters using the CE method
to see the richness and redshift dependence of the comparison.

In Table \ref{tab:number2}, we list the number of clusters each method finds
in our test region (Column 2 called ``N detection''). We also list in column 3
 the number of clusters found in common between the CE method
 and each of other methods discussed above. 
 Columns 4 and 5 of Table \ref{tab:number2} give the detection rates of
 other methods with respect to the CE method and those of the CE method
 with respect to other methods, respectively.
For comparison, in Table
\ref{tab:other}, we also compare the number and percentage of matches found
between the VTT, MF and maxBCG technique.
 These two tables illustrate that the
overlap between all four algorithms is between 20 to 60\% which is simply a
product of their different selection functions. Furthermore, we note we have
used a simplistic matching criteria which does not account for the cluster
redshift or the errors on the cluster centroids. Future SDSS papers will deal
with these improvements (Bahcall et al. 2003). Tables \ref{tab:number2} \& \ref{tab:other} show that
the CE and the maxBCG methods detect overall more clusters than the other methods,
 i.e., 363 and 438 clusters respectively, compared with 152 and 130
clusters for the MF and the VTT respectively.  This difference in the number of
clusters found is mainly due to differences in the thresholds used for each of
these algorithms. As illustrated in Figure\ \ref{fig:rich}, a majority of the
extra clusters in the maxBCG and CE catalogs are low richness systems.
 As seen in Figure \ref{fig:redshift}, these low richness
systems appear to be distributed evenly over the entire redshift range of the
CE catalog, i.e., out to $z\simeq 0.4$.

\subsection{Comparison of the Matched Filter and the CE Methods}\label{sec:mf-ce}

We focus here on the comparison between the CE and the MF (see Kim et al. 2001).
In Figure\ \ref{fig: 0012030-tomomf-rate-z-rich.eps}, we show the
fraction of the
MF clusters found in the CE catalog. We also split the
sample as a function of CE richness.  In Figure\ \ref{fig:
0012030-mftomo-rate-z-rich.eps}, we show the reverse relationship, i.e.,
the fraction of the CE clusters found by the MF as a function of estimated redshift
and CE richness. These figures show that there is almost complete overlap
between the two catalogs for the highest richnesses systems at the
highest redshift bin (there are, however, only 5 systems with $z>0.3$ in the MF catalog).
At low redshifts ($z<0.3$), the overlap decreases, e.g., only 60\% of
the MF
clusters are found in the CE catalog. To understand this comparison further, we
visually inspected all the clusters found by the CE method that were missing
 in the MF catalog.  As expected, most of these systems were compact ($\sim$
1 arcmin) groups of galaxies.

Finally, in Figure\ \ref{fig:elong}, we plot the distribution of elongations
(major to miner axis ratio) for both the whole CE clusters
as well as just the CE clusters in the MF catalog. This plot shows that a
majority of clusters in both samples have nearly spherical morphologies with
the two distributions in good agreement up to the elongation of 3 (axes ratio of 3 to 1).
However, there is a tail of 11 CE clusters which extends to higher
elongations 
that are not seen in the CE plus the MF sub--sample. 
However, this is only $\sim$3\% of the CE clusters.

\subsection{Comparison of the maxBCG and the CE Methods}

In Figure\ \ref{fig:0012030-tomojim-rate-z-rich.eps}, we show the fraction of
 the maxBCG clusters which are found in the CE catalog, while in Figure\ \ref{fig:
0012030-jimtomo-rate-z-rich.eps}, we show the reverse relationship, i.e.,
the fraction of the CE clusters found in the maxBCG catalog. In both figures, we
divide the sample by estimated redshift and observed CE richness.  First, we
note that the matching rate of the maxBCG clusters to the CE is $\sim70$\% or better
for clusters with a richness of $>20$ at all redshifts.  For the low
richness systems, the matching rate decreases at all redshifts.  To further
understand the comparison between these two samples of clusters, we first
visually inspected all clusters detected by the CE method but were missing
from the maxBCG sample and found them to be blue, nearby poor clusters.  This
is a reflection of the wider color cuts employed by the CE method which allows
the CE algorithm to include bluer, star--forming galaxies into its color
criterion. The maxBCG, however, is tuned specifically to detect the E/S0
ridge--line of elliptical galaxies in clusters. We also visually inspected all
maxBCG clusters that were not found by the CE method and found these systems
to be mostly faint high redshift clusters whose members mostly have fallen below the
magnitude limit used for the CE method ($r^*$=21.5).

\subsection{Comparison of the VTT and the CE Methods}

In Figure\ \ref{fig: 0012030-vtttomo-rate-z-rich.eps}, we show the fraction of
 the VTT clusters which were found by the CE as a function of estimated redshift
 and CE richness. Figure\ \ref{fig: 0012030-tomovtt-rate-z-rich.eps} shows the
 fraction of the CE clusters found in the VTT catalog as a function of estimated redshift and
 CE richness.  Because the CE method detects twice as many clusters as
 does the VTT,
 the matching rate is higher in Figure\ \ref{fig:
 0012030-vtttomo-rate-z-rich.eps} than in Figure\ \ref{fig:
 0012030-tomovtt-rate-z-rich.eps}, showing that the CE catalog contains a high
 fraction of the VTT clusters.  In Figure.\ \ref{fig:
 0012030-tomovtt-rate-z-rich.eps}, the matching rate of low richness clusters
 improves at higher redshift because the poor clusters, which the VTT does not
 detect become fainter and therefore both methods can not detect these
 clusters at high redshift.

\section{Summary} \label{sec:summary}

 We have developed a new cluster finding method, the CE method.
 It uses 30 color cuts in the color-magnitude diagrams and four
 color-color cuts in the color-color diagrams
 to enhance the contrast of galaxy clusters over the background galaxies.
 After applying the color and color-color cuts,
 the method uses the color and angular separation weight of galaxy pairs
 as an
 enhancement method to increase the signal to noise ratio of  galaxy
 clusters.
 We use the Source Extractor to  detect galaxy clusters from the enhanced density maps.
 The enhancement and detection are performed for
 every color cut and every color-color cut, producing 34 cluster lists, which are then merged into
 a single cluster catalog.

 Using the Monte Carlo simulations with a real SDSS background
 as well as a shuffled background,
 the CE method is shown to
 have  the ability to
 detect rich clusters ($Ngal$=100) to $z\sim0.3$ with $\sim$80\%
 completeness.
 The completeness drops sharply at $z$=0.4 due to the flux limit of the
 SDSS imaging data.
 The positional accuracy is better than 40'' for clusters of all richnesses examined
 at $z\leq$0.3. 
 The false positive test shows that over 70\% of clusters are likely to be real systems for CE richness $>$10. 
 We apply the CE method to the SDSS commissioning data and
 produce the SDSS CE cluster catalog containing 4638 galaxy
 clusters in $\sim$350 deg$^2$. 
  We compare the CE clusters  with other cluster detection methods:
 the MF, the maxBCG and the VTT. 
 The SDSS CE cluster catalog developed in this work is a useful tool to study both cosmology and property of clusters and cluster galaxies.

\clearpage



\clearpage

\clearpage

\clearpage

\begin{figure}
\begin{center}
\includegraphics[scale=0.7]{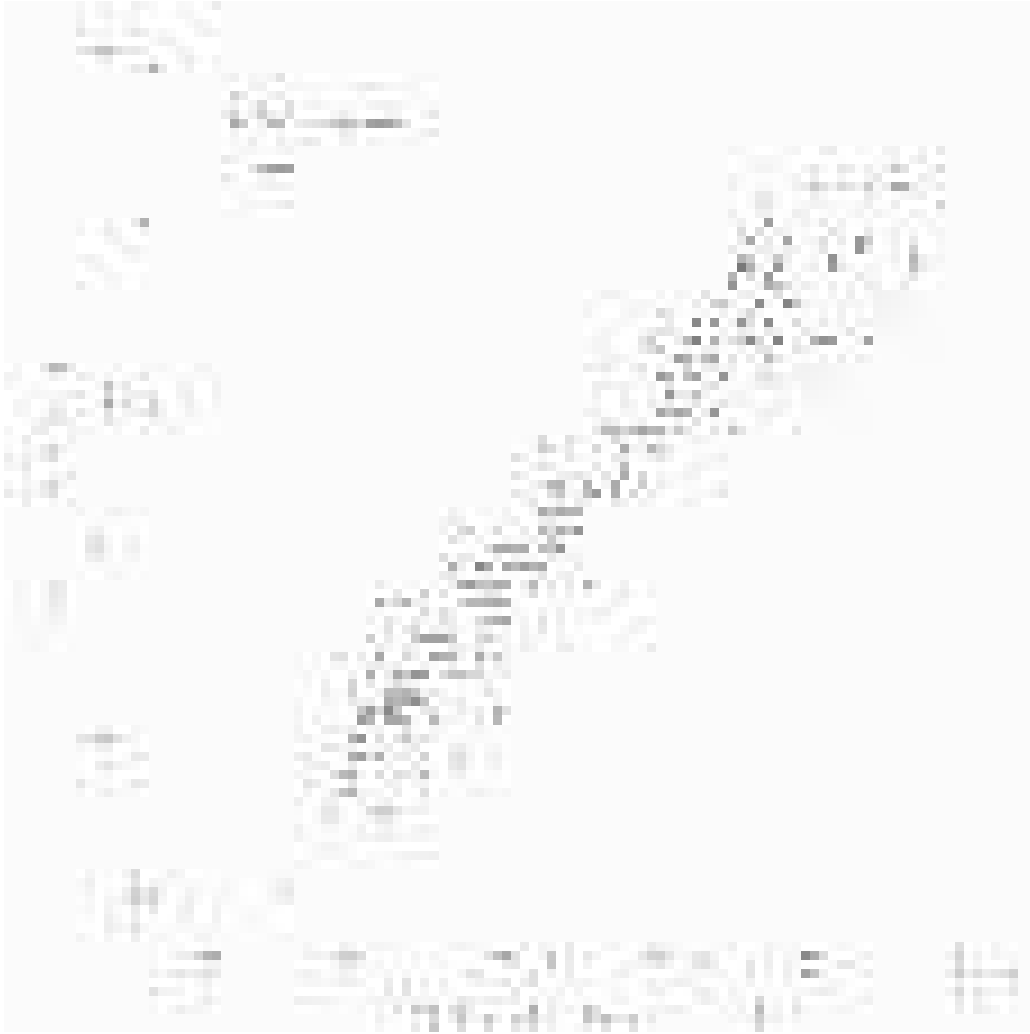}
 \caption{
 The redshift estimation accuracy. 
 The estimated redshifts are plotted against spectroscopic redshifts.
 Abell clusters are plotted with the triangles.
 Dots are the redshifts from SDSS spectroscopic galaxies.
 Extensive outliers $\delta z>$0.1 are removed.
 The dispersion is 0.0147 for $z<$0.3 and 0.0209 for $z>$0.3 .
} \label{fig:zaccuracy.eps}
\end{center}
\end{figure}

\begin{figure}
\begin{center}
\includegraphics[scale=0.7]{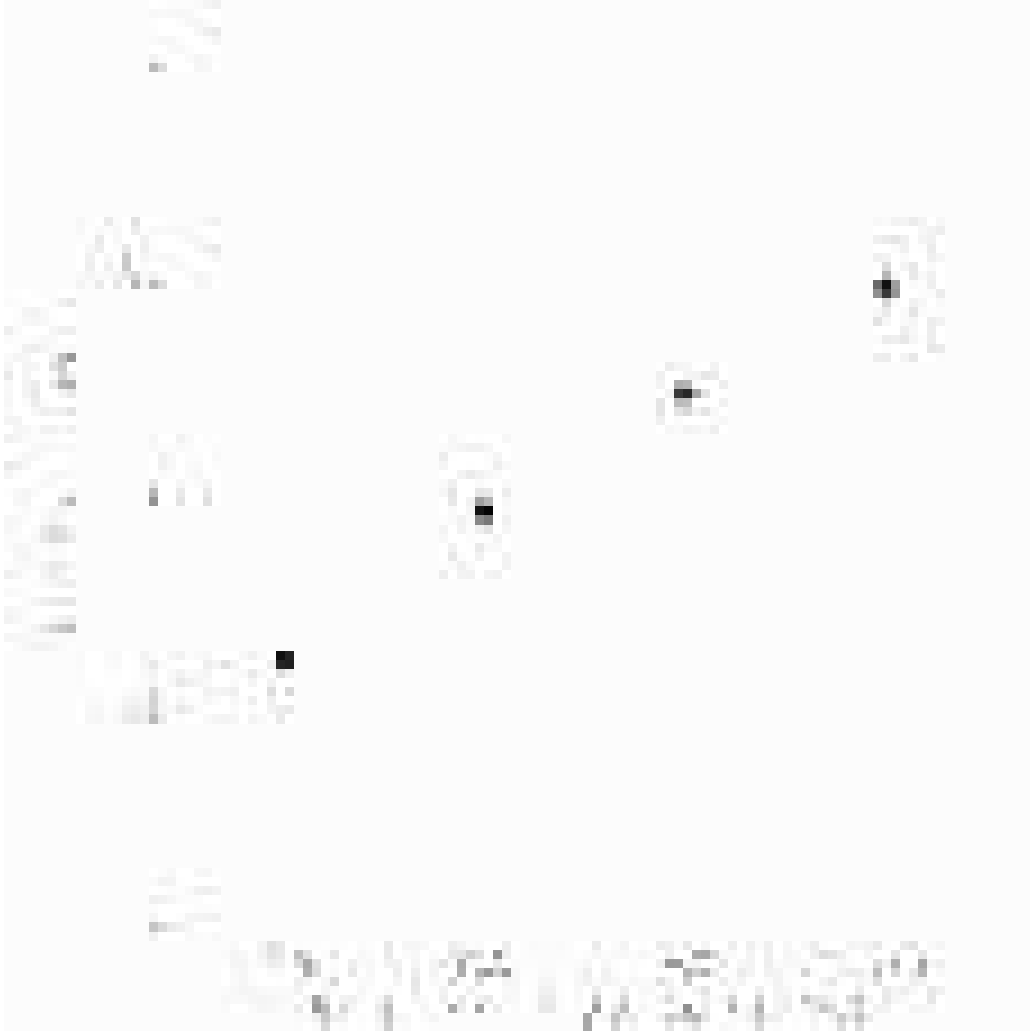}
\caption{
 The number of galaxies fed ($Ngal$) v.s. CE richness. The number of galaxies inputted into
 an artificial cluster is compared with CE richness (the number
 of galaxies within the detected radius whose magnitude is between the magnitude of the third
 brightest galaxy and 
the magnitude fainter than that by two mag). The error bars show 1$\sigma$ standard
 error.}
\label{fig:ngal-ab1ellrich}
\end{center}
\end{figure}

\begin{figure}
\begin{center}
\includegraphics[scale=0.7]{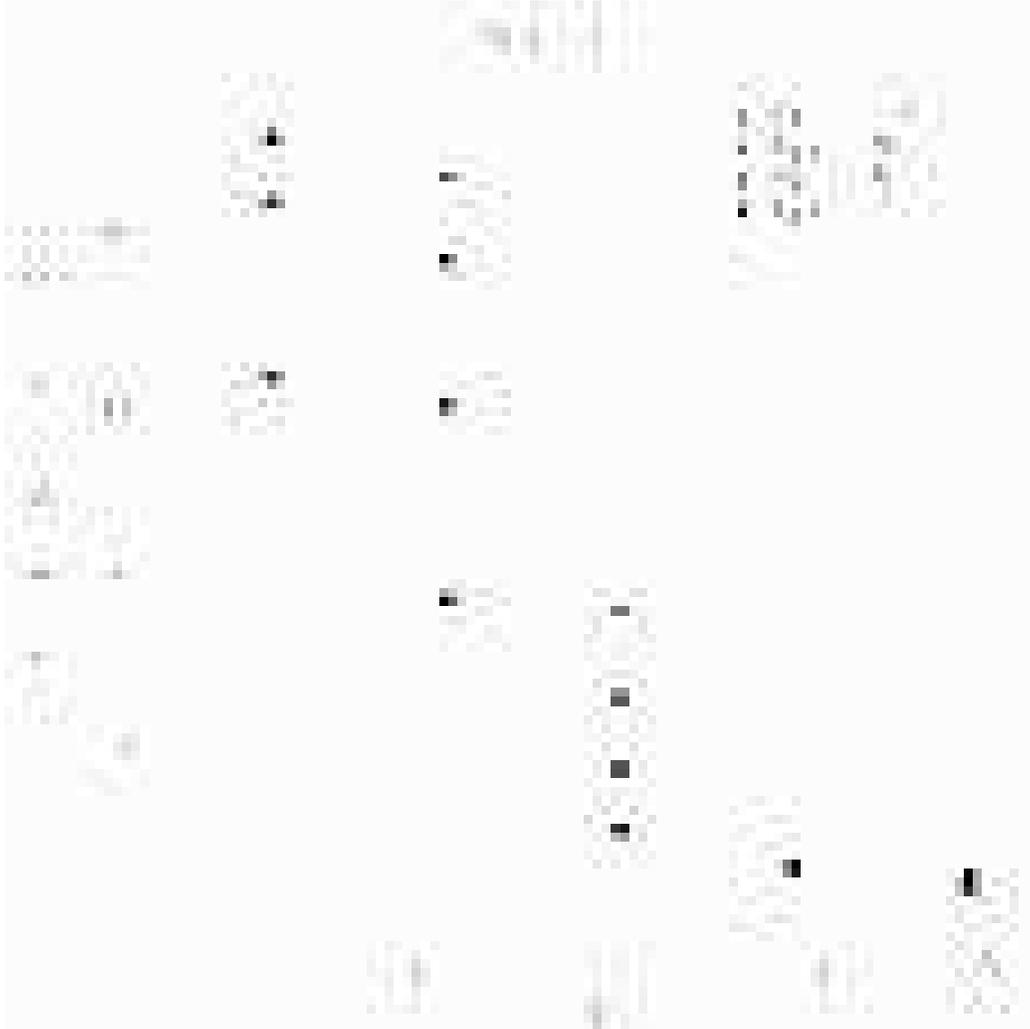}
\caption{
 Recovery rate in Monte Carlo simulation with the real SDSS background.
 Recovery rate is plotted against redshift.
 The artificial clusters are added on the real SDSS background randomly
 chosen from the SDSS commissioning data. The detection is iterated 1000
 times for each data point. Even at $z=$0.3, $Ngal$=80 cluster is detected with more
 than 75\% probability.
}\label{fig:monte-recovery-simu}
\end{center}
\end{figure}

\begin{figure}
\begin{center}
\includegraphics[scale=0.7]{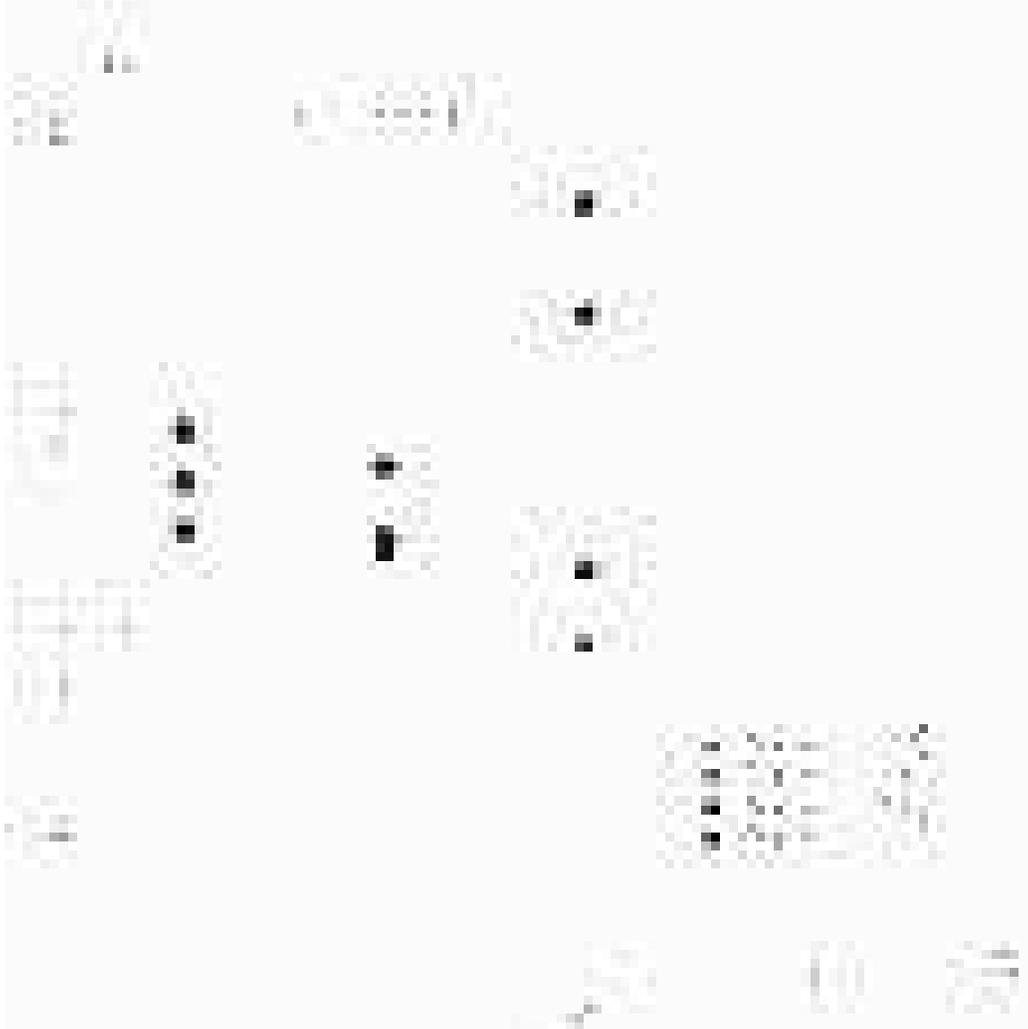}
\caption{
Positional accuracy with the real SDSS background. The positional accuracy is
 almost constant against redshift because the more distant cluster is more compact in
 angular space. Positional accuracy of $\sim$30'' is reasonable considering
 that the mesh size of the enhancement method is 30''.
}\label{fig:z-posi-simu}
\end{center}
\end{figure}

\begin{figure}
\begin{center}
\includegraphics[scale=0.7]{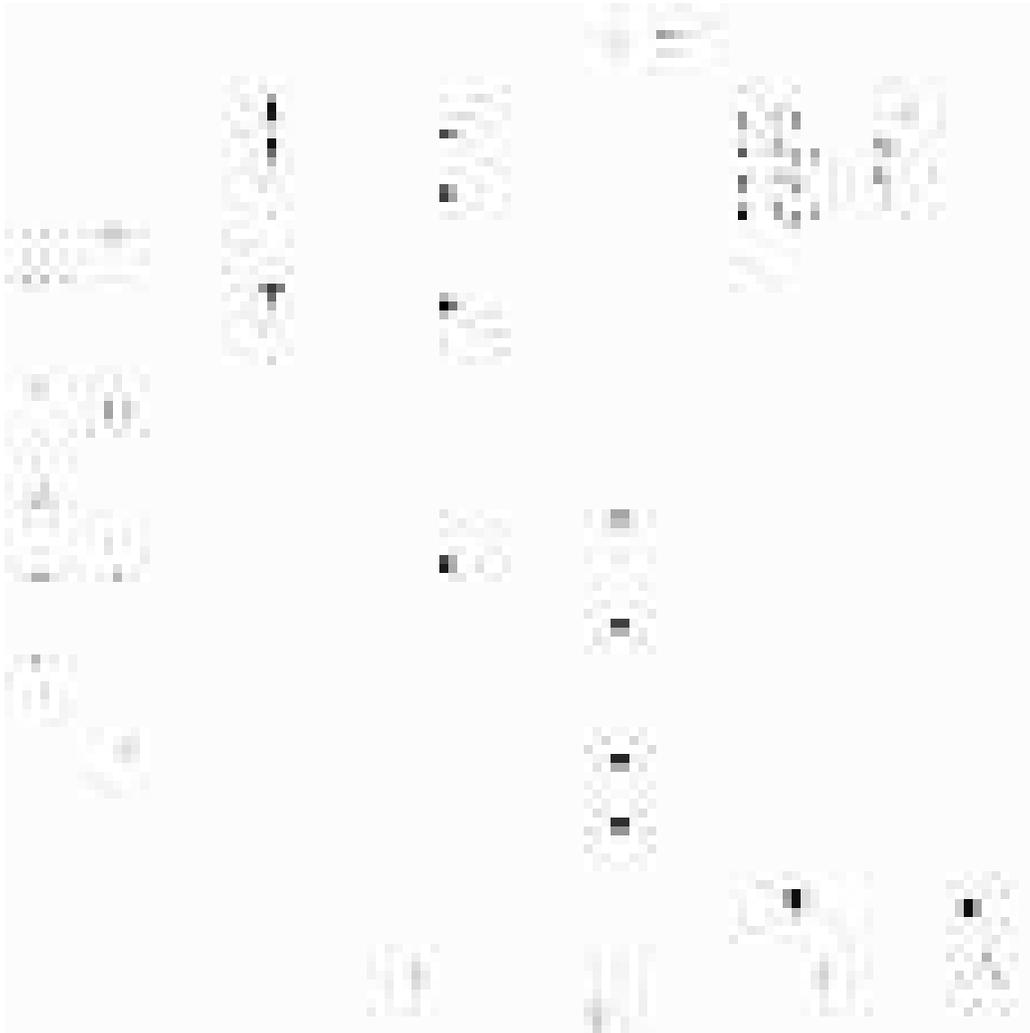}
\caption{
Recovery rate in Monte Carlo simulation with the shuffled background.
The artificial clusters are added on the shuffled background randomly
 chosen from the SDSS commissioning data. The detections are iterated 1000
 times. 
}\label{fig:monte-recovery-uni}
\end{center}
\end{figure}

\begin{figure}
\begin{center}
\includegraphics[scale=0.7]{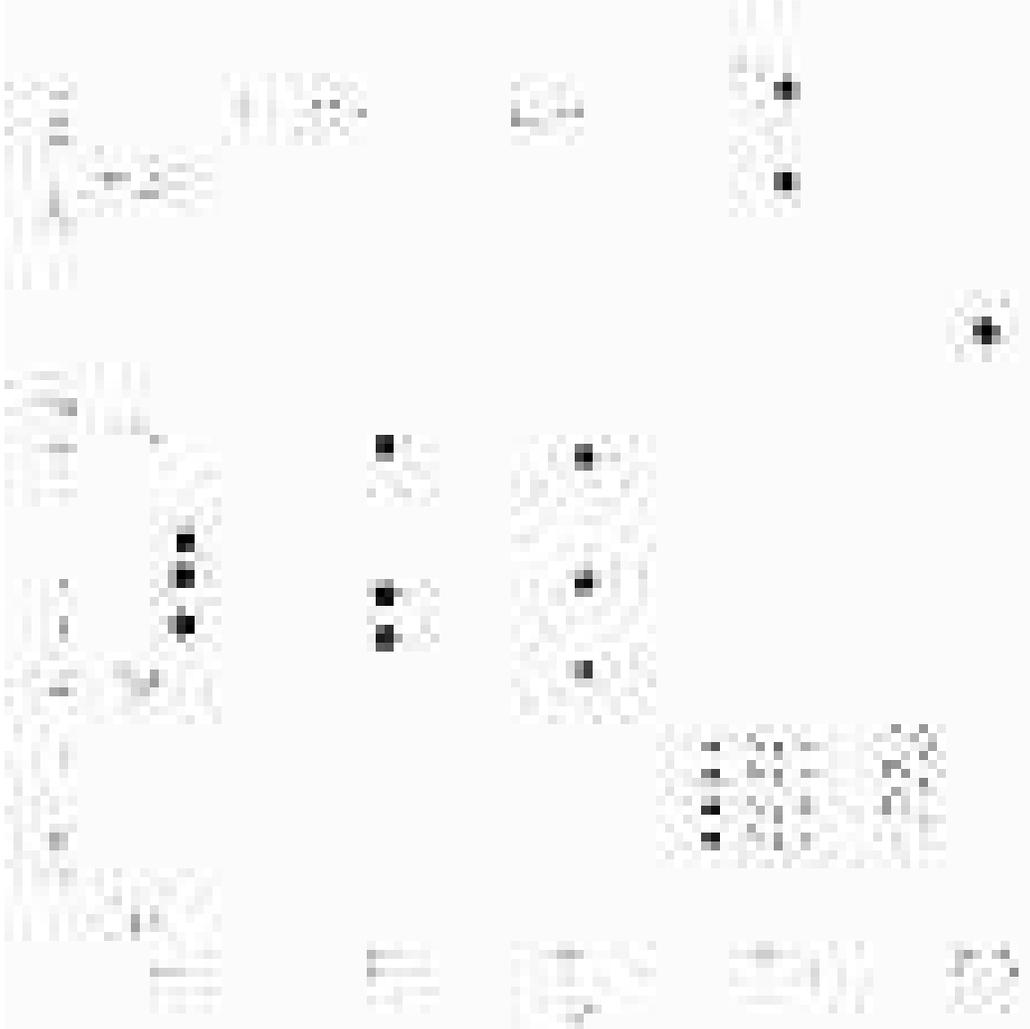}
\caption{
Positional accuracy with the shuffled background. The positional accuracy is
 almost independent of redshift because the more distant cluster is more compact in
 angular space. Positional accuracy of $\sim$ 30'' is good considering
 that the mesh size of the enhancement method is 30''.
}\label{fig:z-posi-uni}
\end{center}
\end{figure}

\begin{figure}
\begin{center}
\includegraphics[scale=0.7]{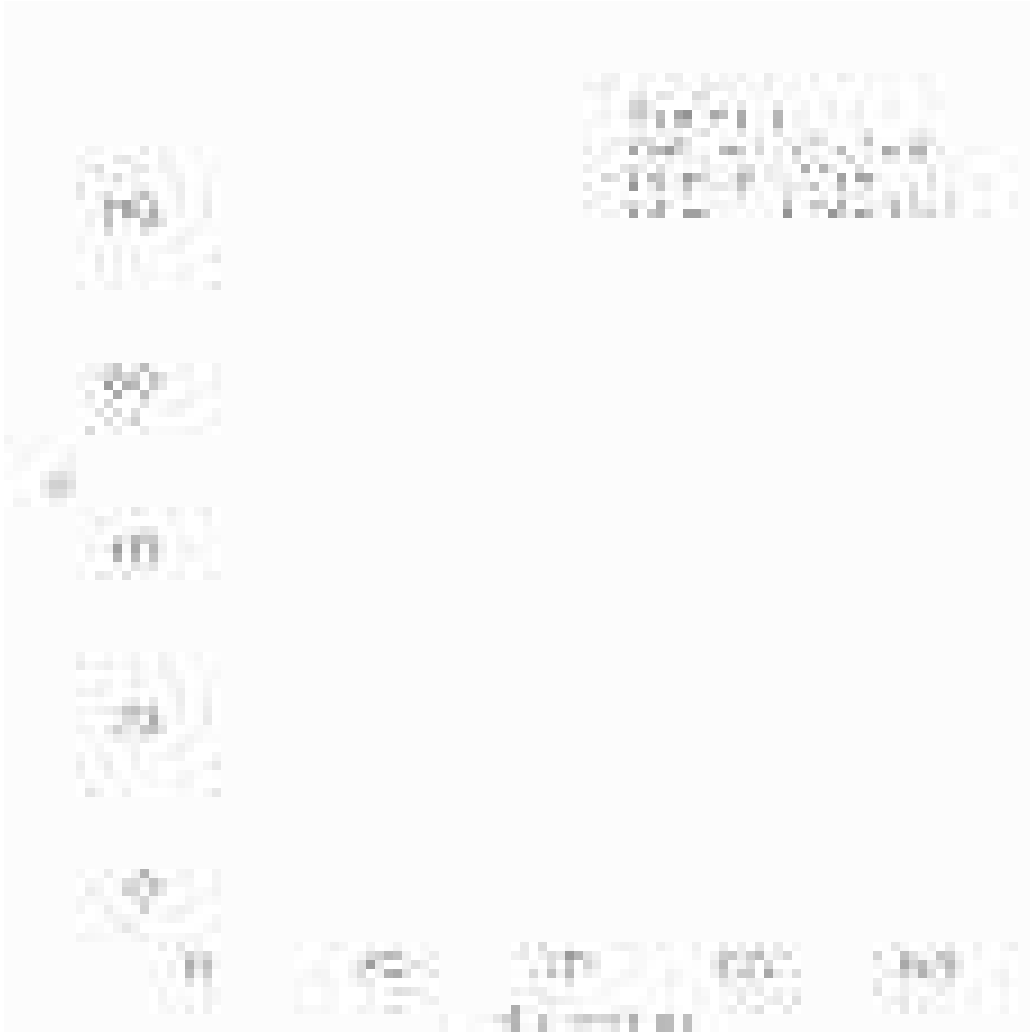}
\caption{
 False positive tests. Detection tests are performed using 23.75 deg$^2$ of
 the SDSS commissioning data. Number of detected clusters is plotted against
 the CE richness. 
 The solid line represents the results with real
 data. The dotted line represents the results with position shuffled
 data. The long-dashed line is for color shuffled data subtracting the
 detection from the real data. The short-dashed line is for color shuffled
 smearing data.}\label{fig:tim_test}
\end{center}
\end{figure}

\begin{figure}
\begin{center}
\includegraphics[scale=0.7]{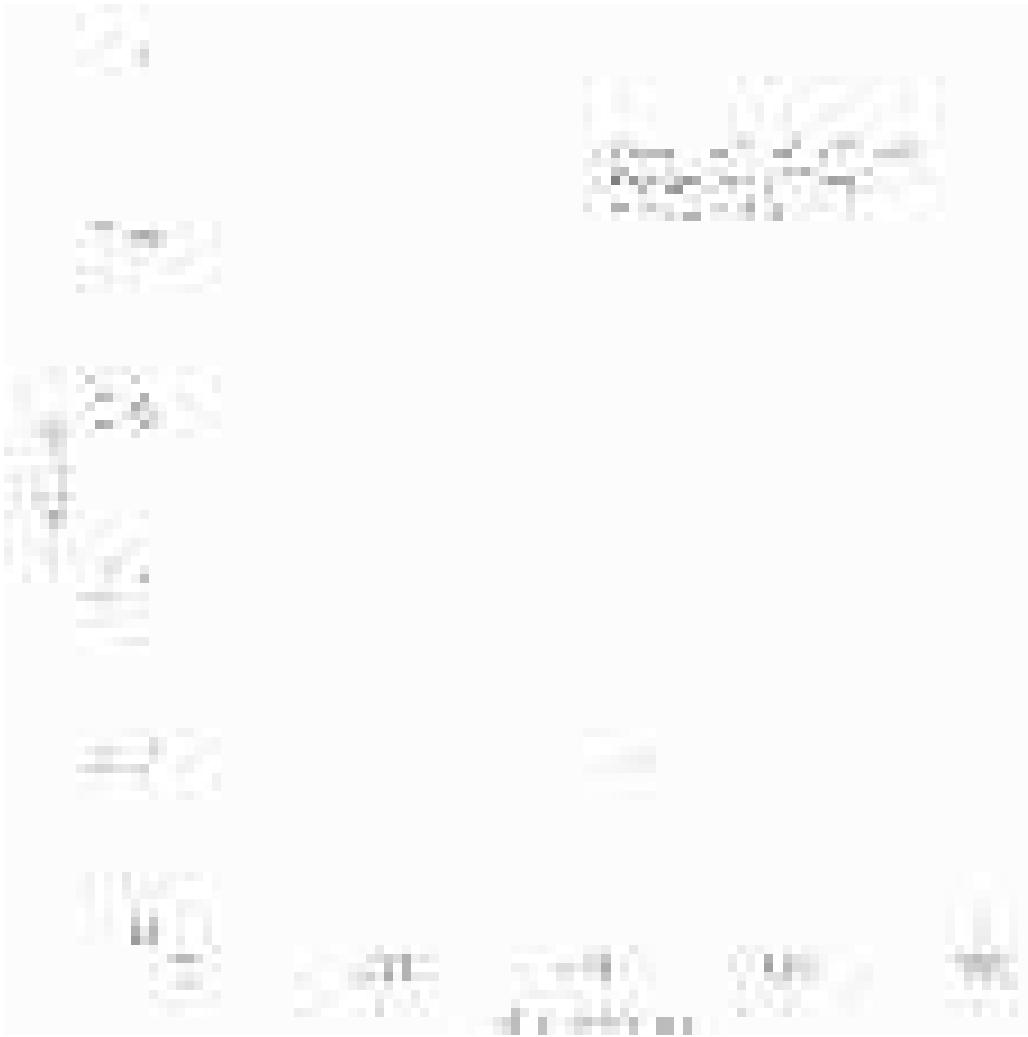}
\caption{
 False positive tests. Detection tests are performed using 23.75 deg$^2$ of
 the SDSS commissioning data. 
 The fraction of false detections to the real data is plotted against the
 CE richness.
 Each line represents the ratio to the real
 data at the richness bin.
 The dotted line represents the results with position shuffled
 data. The long-dashed line is for color shuffled data subtracting the
 detection from the real data. The short-dashed line is for color shuffled
 smearing data.}\label{fig:tim_test_ratio}
\end{center}
\end{figure}

\begin{figure}
\begin{center}
\includegraphics[scale=0.7]{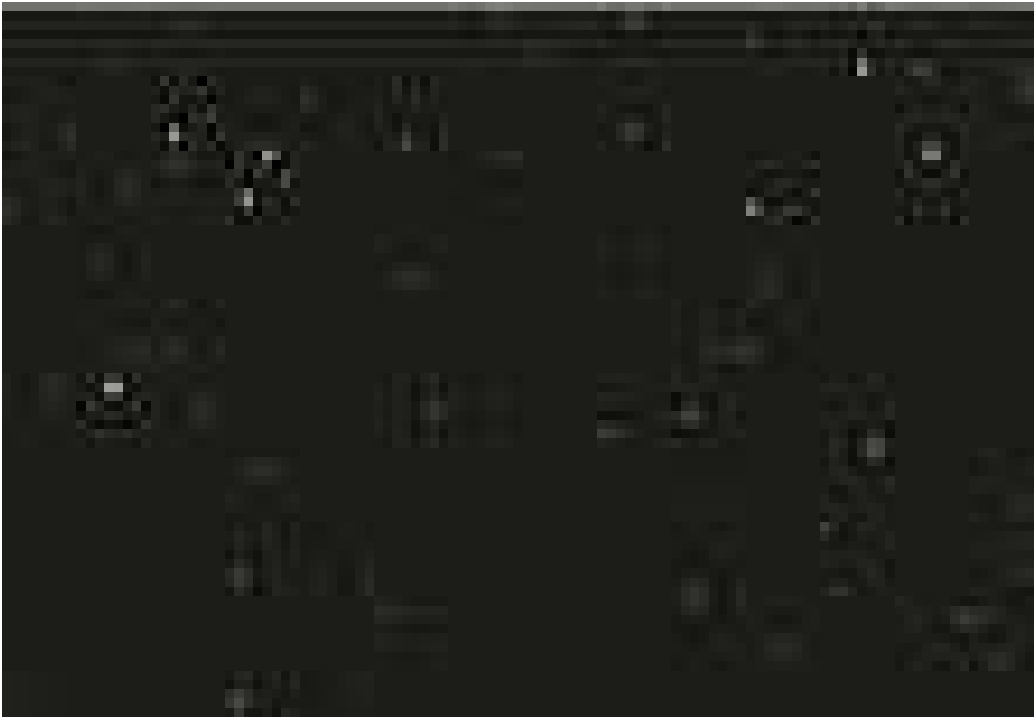}
 \caption{
 A  successful detection with the CE method.
 The image is 6'$\times$13' true color image of the SDSS commissioning
 data.
There are many faint galaxies in the region.
CE method has the ability to detect
the region in the sky where many faint galaxies are clustering.
This cluster was found only with the CE method.
} \label{fig:dwarf-region}
\end{center}
\end{figure}

\begin{figure}
\begin{center}
\includegraphics[scale=0.7]{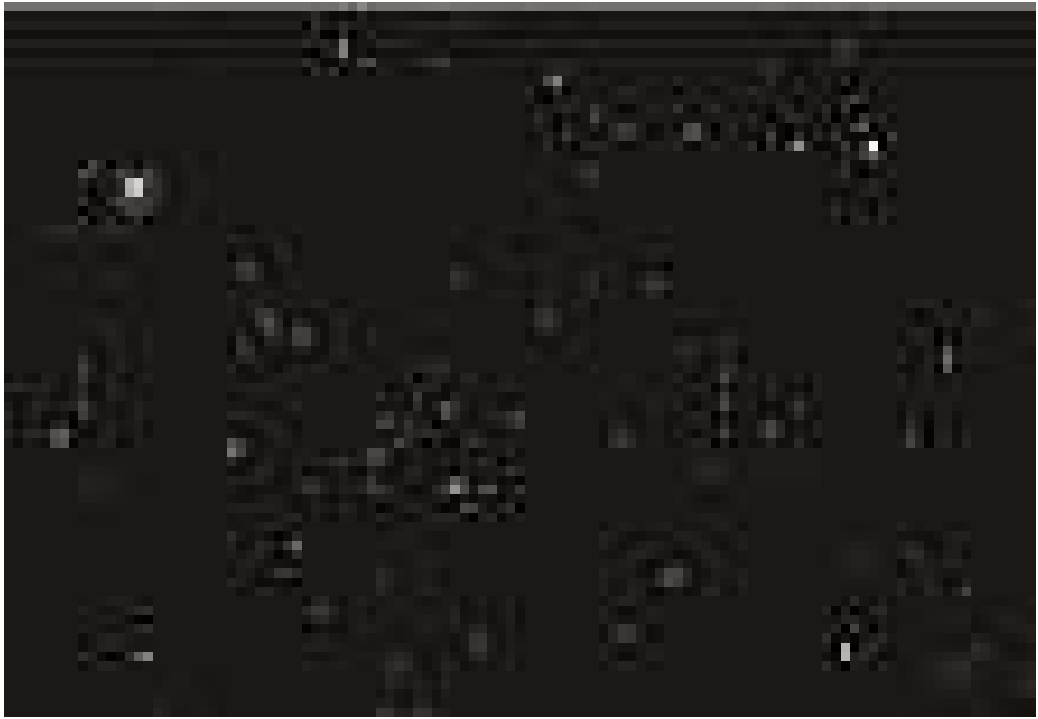}
 \caption{
 A successful detection with the CE method.
The image is 6'$\times$13' true color image of the SDSS commissioning data.
The cluster position and radius is shown with the yellow circle.
} \label{fig:success-cluster}
\end{center}
\end{figure}

\begin{figure}
\includegraphics[scale=0.7]{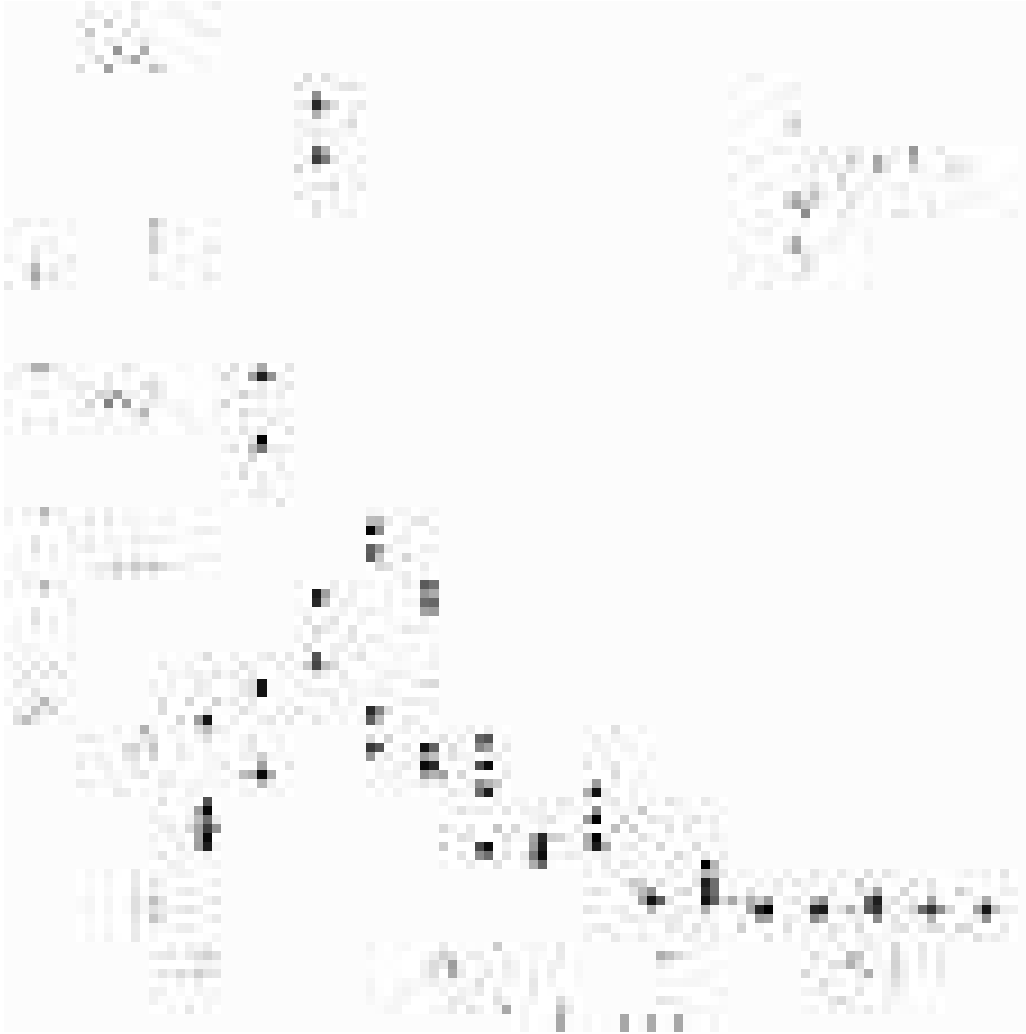}
\caption{
Comparison of four catalogs by richness. 
The abscissa is the richness of the cluster.
The ordinate is the number of the detected clusters.
The CE clusters are drawn with the solid lines.
The maxBCG clusters are drawn with the dotted lines.
Matched Filter clusters are drawn with the short-dashed lines.
Voronoi tessellation clusters are drawn with the long-dashed lines.
 The CE and the maxBCG detect poor clusters (richness $<$20) more than the MF or
 the VTT. 
}\label{fig:rich}
\end{figure}

\begin{figure}
\includegraphics[scale=0.7]{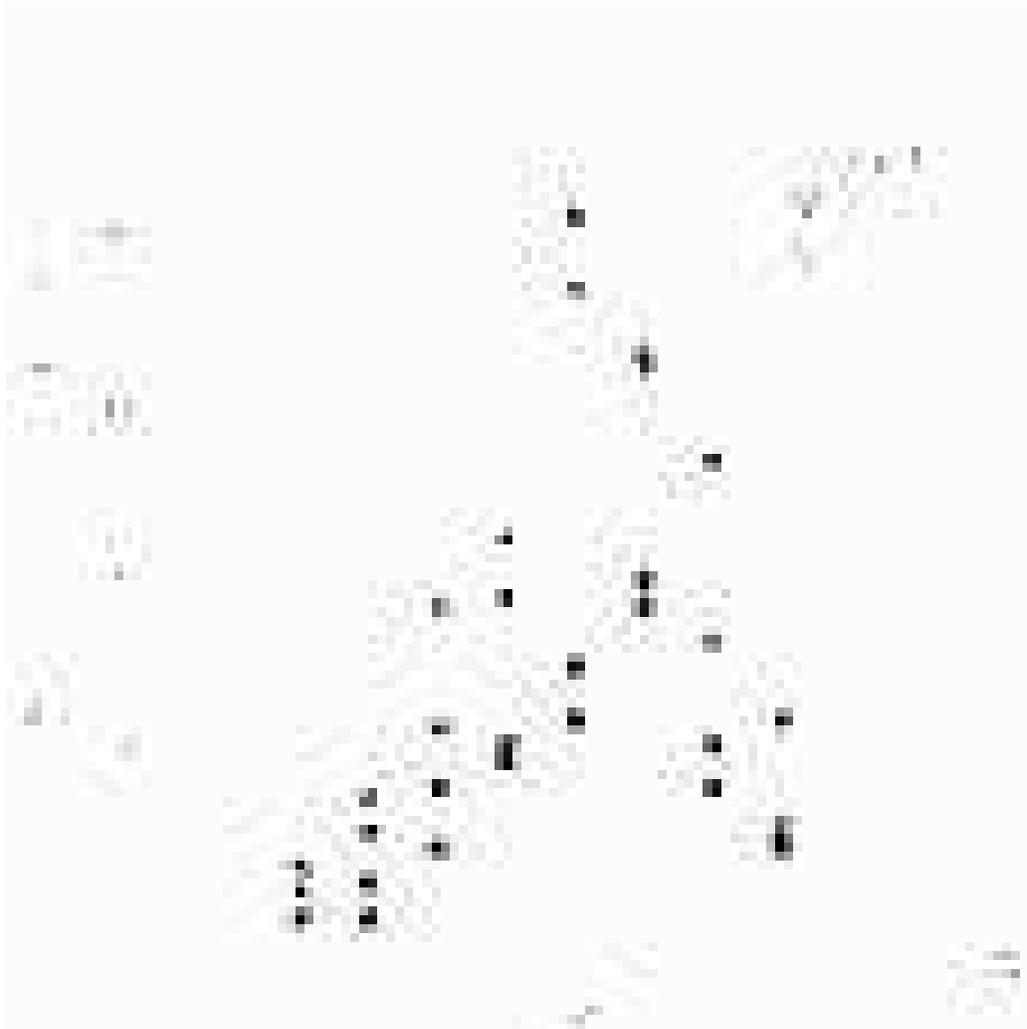}
\caption{
Comparison of four catalogs by redshift.
The abscissa is the redshift of the clusters.
The ordinate is the number of the clusters.
CE clusters are drawn with the solid lines.
The maxBCG clusters are drawn with the dotted lines.
The Matched Filter clusters are drawn with the short-dashed lines.
The Voronoi tessellation clusters are drawn with the long-dashed lines.
The redshift is estimated photometrically (described in Section \ref{sec:method}).
}\label{fig:redshift}
\end{figure}

\clearpage

\begin{figure}\begin{center}	       
\includegraphics[scale=0.7]{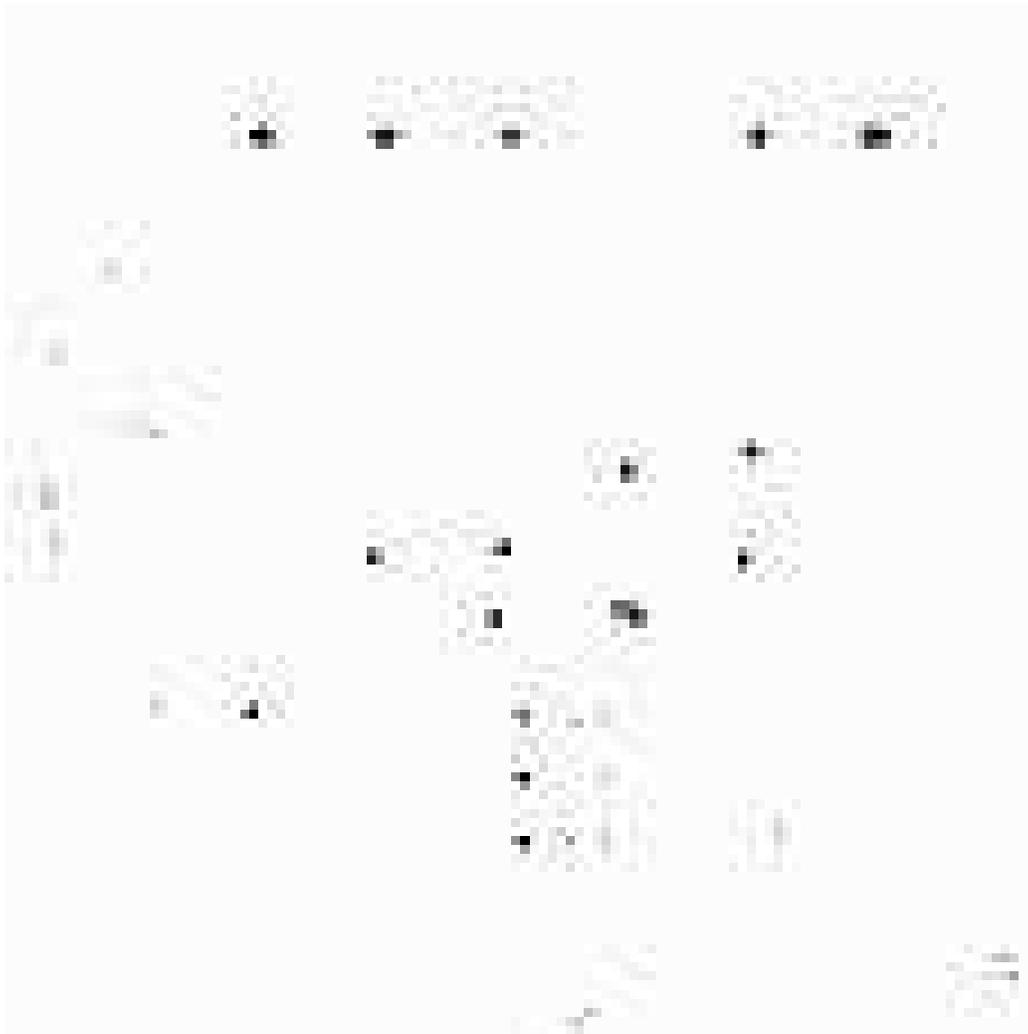}
	       \caption{
	       Comparison of the MF with the CE.  The abscissa is the estimated
redshift. The ordinate is the rate of the MF clusters which are found
in the CE catalog to the number of the CE clusters.
 CE richness 0$\sim$20 is plotted with the solid lines. CE richness 20$\sim$40
is plotted with the dotted lines. CE richness 40$\sim$60 is plotted with the
	       dashed lines.
The error bars for richness 40$\sim$60 clusters are large and 
omitted for clarity (The error is 80\% at $z=$0.3).  
The data for  richness 20$\sim$40 and 40$\sim$60 are shifted in redshift
	       direction by 0.01 for clarity.
}	       \label{fig: 0012030-tomomf-rate-z-rich.eps}
	      \end{center}
\end{figure}

\clearpage

\begin{figure}\begin{center}	       
\includegraphics[scale=0.7]{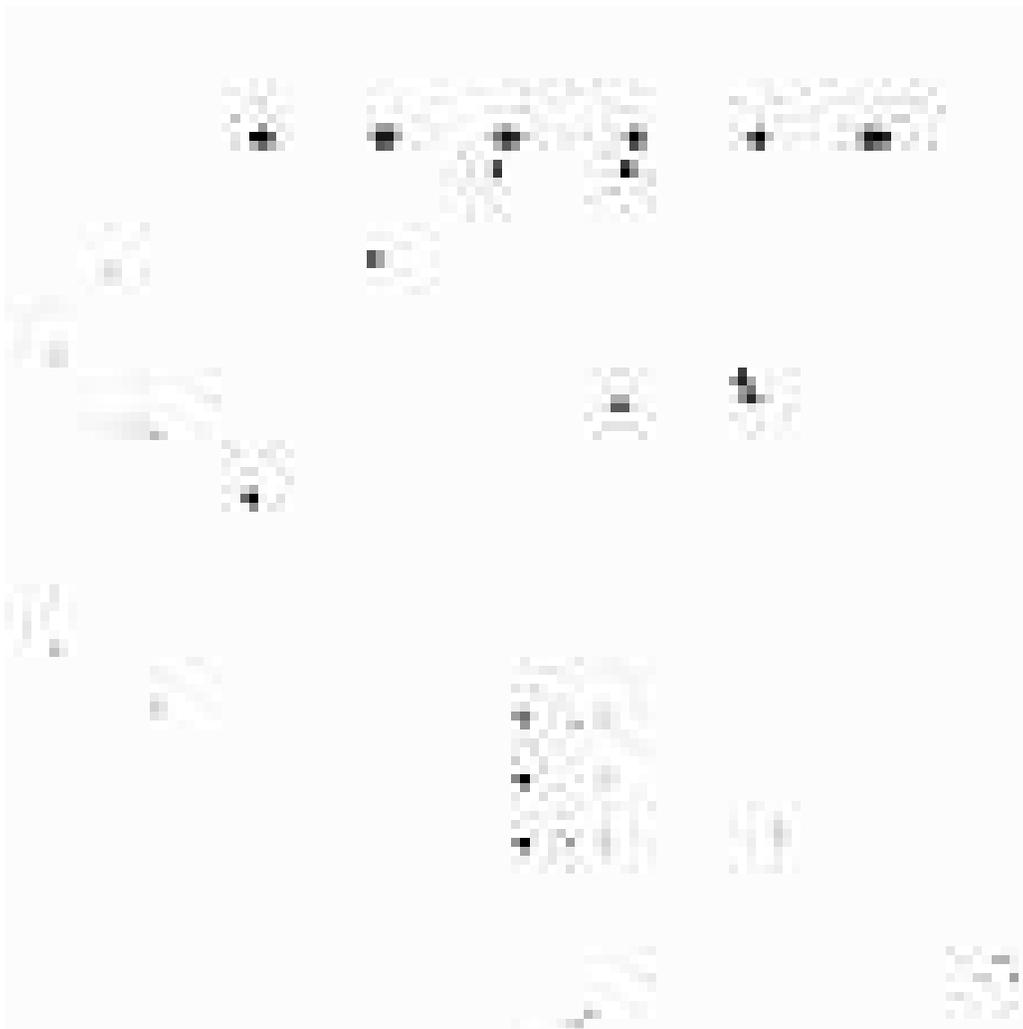}
	       \caption{
	       Comparison of the CE with the MF. The abscissa is the estimated
redshift. The ordinate is the rate of the CE clusters which are found
in the MF catalog to the number of the CE clusters. 
 Matching rate is low for poor clusters indicating
 	     the  CE detects poor clusters more.
The error bars for richness 40$\sim$60 clusters are large and 
omitted for clarity (The error is 80\% at $z=$0.3).
The data for  richness 20$\sim$40 and 40$\sim$60 are shifted in redshift
	       direction by 0.01 for clarity.
	      }	       \label{fig: 0012030-mftomo-rate-z-rich.eps}
\end{center}
\end{figure}

\begin{figure}
\begin{center}
\includegraphics[scale=0.7]{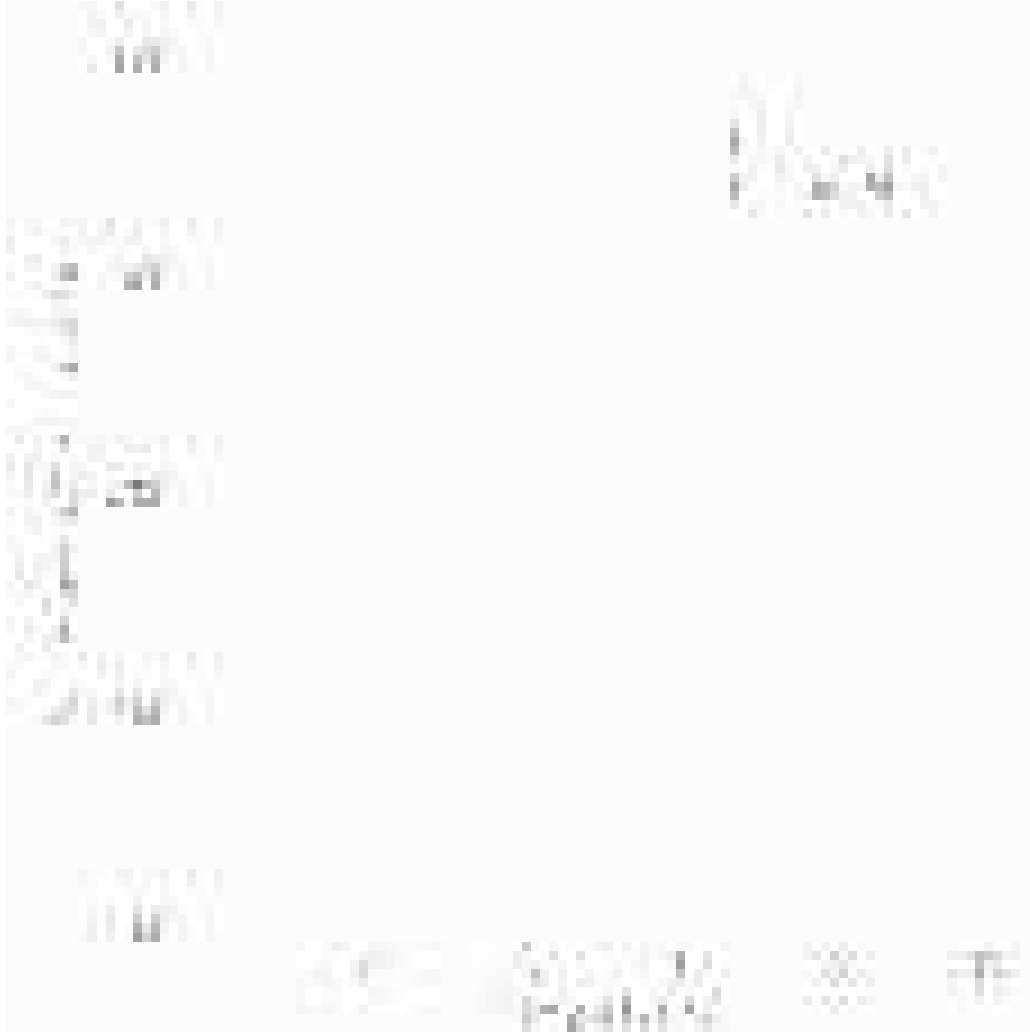}
\caption{
 Elongation distribution of the detected clusters.
 The number of the clusters is plotted against
 elongations (ratio of the major axis to the minor axis).
The solid line is for the clusters detected with the CE method.
The dotted line is for the  the clusters detected with both the MF and the CE method, which is shifted by 0.01 for clarity.
}\label{fig:elong}
\end{center}
\end{figure}

\clearpage

\begin{figure}[h]
\includegraphics[scale=0.7]{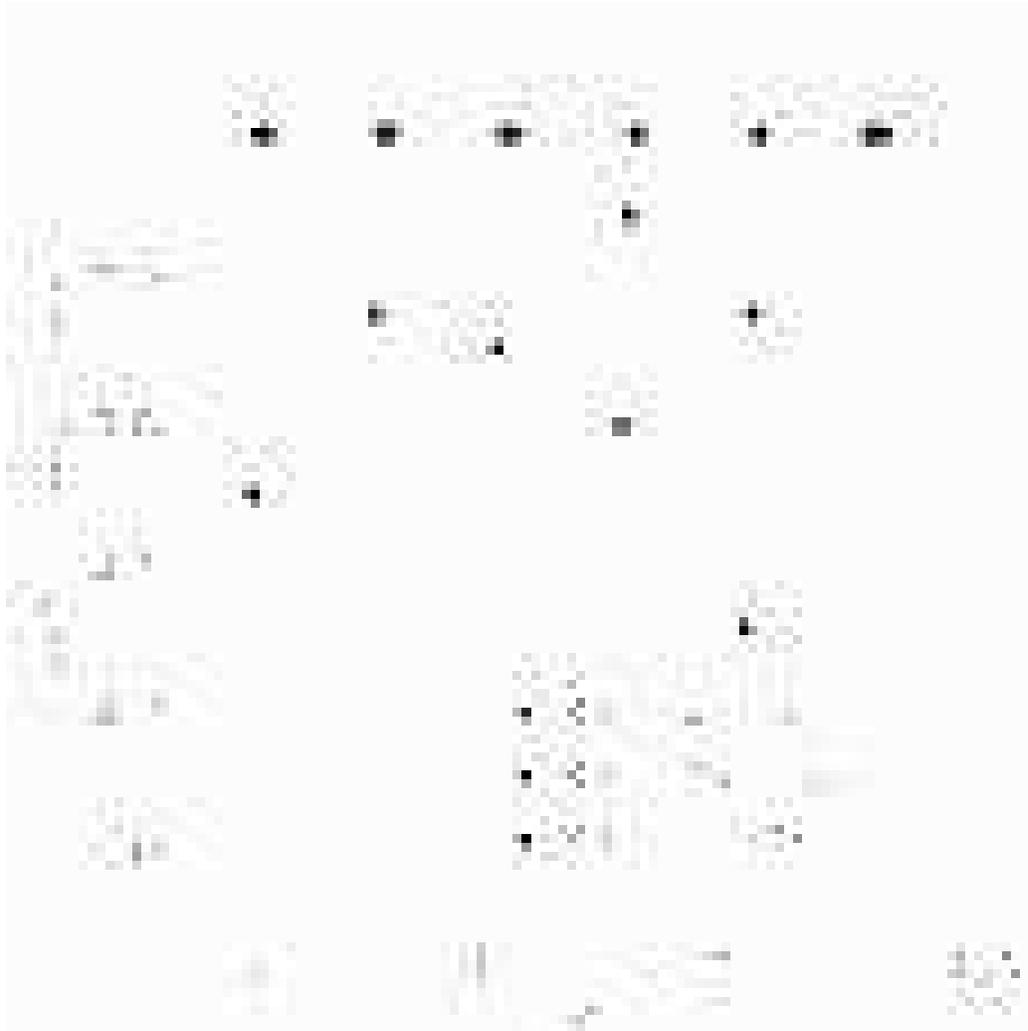}
\caption{
Comparison of the maxBCG clusters with the CE catalog.
 The abscissa is the color estimated
redshift. The ordinate is the ratio of the maxBCG clusters which are found
in the CE catalog to the number of the maxBCG clusters. The
 error bars for richness 40$\sim$60 clusters are large and omitted for
 clarity (The error is 80\% at $z=$0.3).
 The data for  richness 20$\sim$40 and 40$\sim$60 are shifted in redshift
	       direction by 0.01 for clarity.}\label{fig:0012030-tomojim-rate-z-rich.eps}
\end{figure}

\begin{figure}[h]
\includegraphics[scale=0.7]{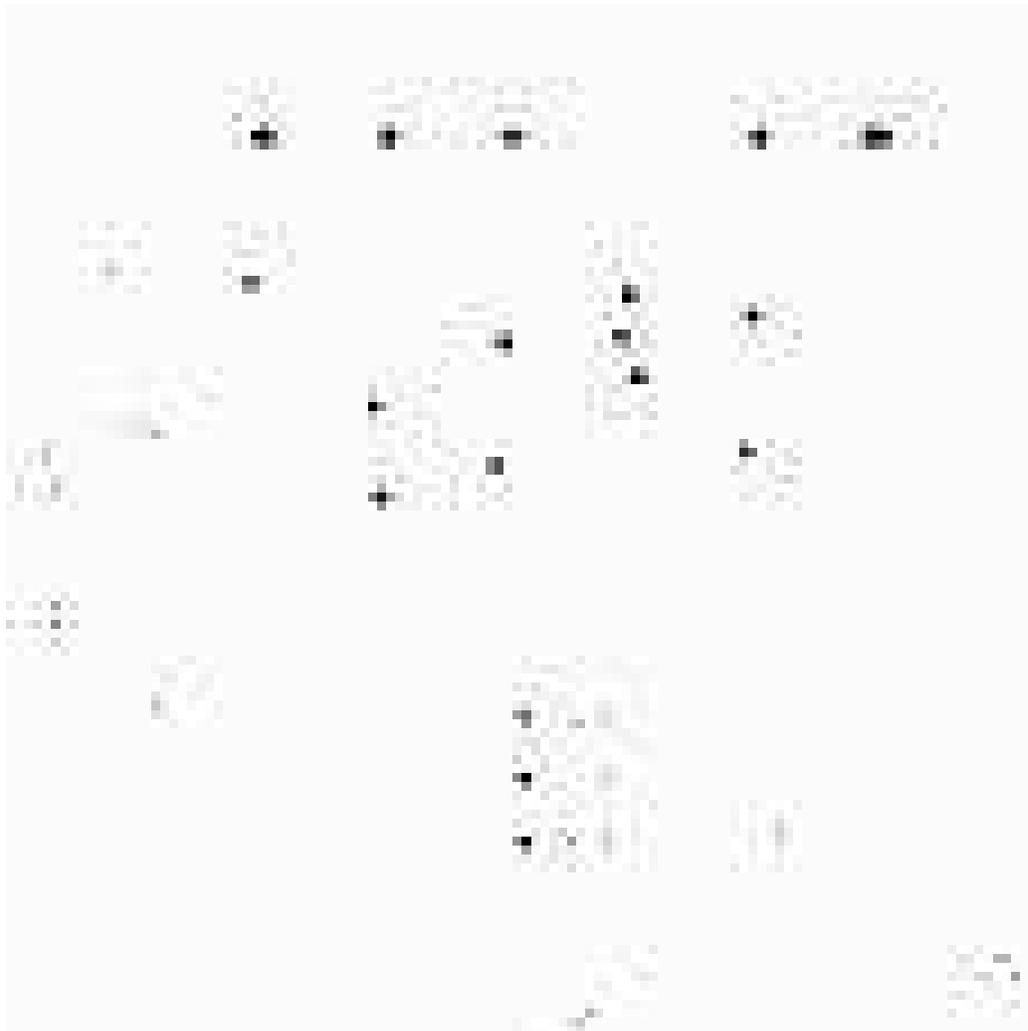}
\caption{
Comparison of the CE method with the maxBCG method. The abscissa is the estimated
redshift. The ordinate is the rate of the CE clusters which are found
in the maxBCG catalog to the number of the CE clusters. 
The error bars for richness 40$\sim$60 clusters are large and omitted for clarity.
The data for  richness 20$\sim$40 and 40$\sim$60 are shifted in redshift
	       direction by 0.01 for clarity.}\label{fig: 0012030-jimtomo-rate-z-rich.eps}
\end{figure}

\clearpage

\begin{figure}[h]
\includegraphics[scale=0.7]{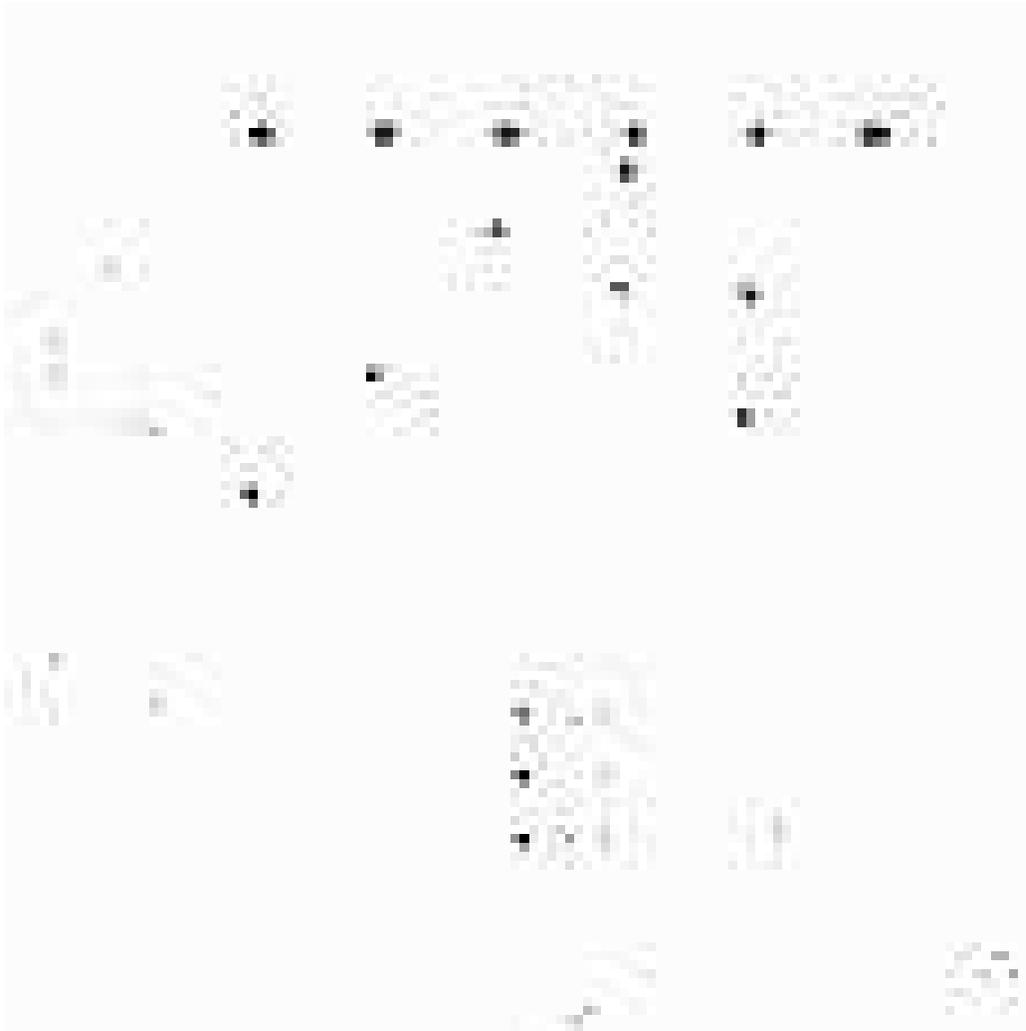}
\caption{
Comparison of the CE method with the VTT method. 
  The abscissa is the estimated
redshift. The ordinate is the rate of the CE clusters which are found
in the VTT catalog to the number of all CE clusters. 
 The CE detects twice as many clusters as the VTT does. The error bars
 for richness 40$\sim$60 clusters are 
large and omitted for clarity.
 The data for  richness 20$\sim$40 and 40$\sim$60 are shifted in redshift
	       direction by 0.01 for clarity. }\label{fig: 0012030-vtttomo-rate-z-rich.eps}
\end{figure}

\begin{figure}
\includegraphics[scale=0.7]{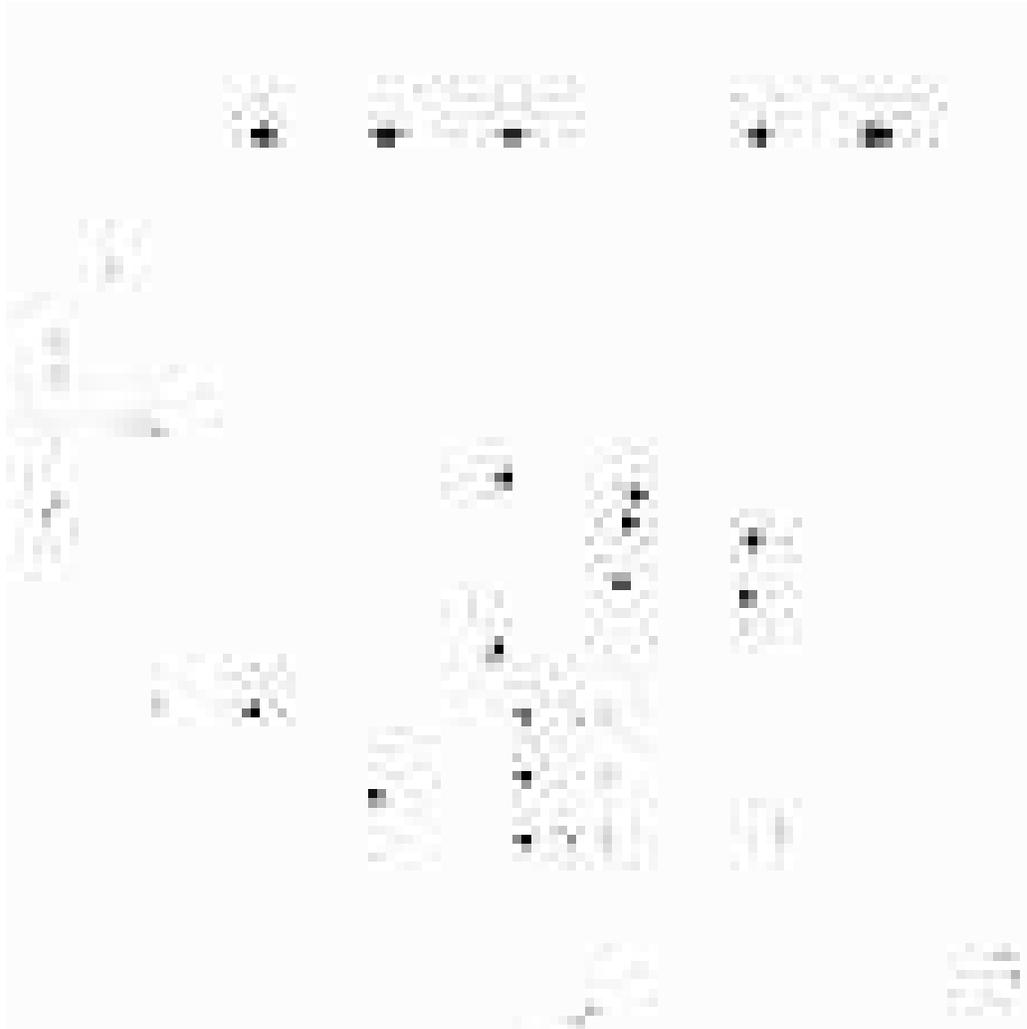}
\caption{
Comparison of the VTT method with the CE method. 
 The abscissa is the estimated
redshift. The ordinate is the ratio of the VTT clusters which are found
in the CE catalog to the number of the VTT clusters.
 Note that the CE method detects twice as many as the VTT does. The
 error bars for richness 40$\sim$60 clusters are large and omitted for
 clarity. 
 The data for  richness 20$\sim$40 and 40$\sim$60 are shifted in redshift
	       direction by 0.01 for clarity.}\label{fig: 0012030-tomovtt-rate-z-rich.eps}
\end{figure}

\clearpage

\begin{table}[h]
\caption{\label{tab:cm-tilt}
 Tilt and scatter of color-magnitude relation of A1577.
}
\begin{center}
\begin{tabular}{lllll}
\hline
Color & Tilt (color/mag) & (magnitude range)  & Scatter (mag) &  (magnitude range) \\
\hline
\hline
$g^*-r^*$ &  0.0737 & $r^*<$19  & 0.081 & $r^*<$17   \\
$r^*-i^*$ &  0.0898 & $r^*<$19  & 0.040 & 18$<r^*<$19   \\
$i^*-z^*$ &  0.0018 & $r^*<$21  & 0.033 & 18$<r^*<$19   \\
\hline
\end{tabular}
\end{center}
\end{table}

\bigskip

\begin{table}[h]
\caption{\label{tab:RXJ0256_color_cut}
 The fraction of galaxies in the color-cut both for galaxies inside of the
 RXJ0256.5+0006 ($z=$0.36) and outside of RXJ0256.5+0006.
}
\begin{center}
\begin{tabular}{llll}
\hline
Color cut & In cluster region(\%)  & Outside of cluster (\%) &   \\
\hline
\hline
$g-r$ &  36.9$^{+7.0}_{-6.0}$ & 13.57$\pm$0.03    \\
$r-i$ &  62.1$^{+8.8}_{-7.7}$ & 42.35$\pm$0.06    \\
$i-z$ &  59.2$^{+8.6}_{-7.6}$ & 44.55$\pm$0.06    \\
$g-r-i$ &  58.3$^{+8.5}_{-7.6}$ & 48.77$\pm$0.06   \\
$r-i-z$ &  76.7$^{+10.7}_{-7.7}$ & 65.68$\pm$0.07     \\
$g-r-i$ high-$z$&  29.1$^{+6.3}_{-5.3}$ & 10.86$\pm$0.03    \\
$r-i-z$ high-$z$&  6.8$^{+3.7}_{-2.5}$  & 9.94$\pm$0.02    \\
\hline
\end{tabular}
\end{center}
\end{table}

\clearpage

\begin{table}[b]
\caption{
Sigma cut test. Sigma is a threshold when detecting clusters in the
 enhanced density map. The test is performed using a 23.75 deg$^2$ region in the data.
}\label{tab:sigma_test}
\begin{flushleft}
\begin{tabular}{l|lllll}
\hline
Sigma    & 2 &  4 & 6 & 8 & 10 \\
\hline
N detection &  402 &   437   &  453 &  434 & 415   \\
\hline
\end{tabular}
\end{flushleft}
\end{table}

\begin{table}[b]
\caption{
Test of fluxmax cut. The test is performed using a 23.75 deg$^2$
 region in the data. A number of detections is shown as a function of
 fluxmax threshold.
}\label{tab:fluxmax}
\begin{flushleft}
\begin{tabular}{l|lllll}
\hline
fluxmax    & 500 &  750 & 1000 & 1500 & 2000 \\
\hline
N detection &  890 &   655   &  464 & 260 &10   \\
\hline
\end{tabular}
\end{flushleft}
\end{table}

\begin{table}[h]
\caption{\label{tab:ce-miss-new}
10 false detections of the CE method.
The region examined is RA between 16 and 25.5 deg, DEC between $-$1.25 and $+$1.25 deg
 (23.75 deg$^2$).
 $\sigma$ (column [1]) is the significance of the detection. 
CE richness (column [2])  is the richness of the detection.
$z$ (column [3]) is the color estimated redshift of the detection.
 Comments on the detection are in column [4].
}
\begin{center}
\begin{tabular}{llll}
\hline
$\sigma$ & richness  & $z$ & comment \\
\hline
\hline
12.39 &  31 & 0.22 & looks like a blank field.\\
7.85 & 21 & 0.18 &  looks like  a blank field.\\
4.80 & 11 & 0.10 & looks like  a blank field.\\
16.74 & 16 & 0.18 & looks like  a blank field.\\
56.85 & 8 & 0.04 & a big galaxy.\\
9.35 & 7& 0.00 &   looks like  a blank field.\\
7.06 & 1& 0.04 &  eight blue galaxies.\\
11.0 & 14 & 0.04 &  looks like  a blank field.\\
 4848.39 & 17 & 0.12 & a big galaxy.\\
25.20 & 7 & 0.00 & a big galaxy.\\
\hline
\end{tabular}
\end{center}

\end{table}

\begin{table}[b]
\caption{\label{tab:number2}
Ratio of number of clusters detected with the MF, the maxBCG and the VTT
 to the CE clusters.
The region used is RA between 16 and 25.5 deg, DEC between $-$1.25 and
 $+$1.25 deg (23.75 deg$^2$). 
Column 1 lists the name of each method. 
N detection (column [2]) is the numbers of clusters detected with each
 method.
Common detection (column  [3]) is the number of clusters detected with both
 the method and the CE method.
Rate to the CE (column  [4]) is the percentage of the number of detections with each
 method divided by those  with the CE method (CE in
 the table).
Rate to the method (column [5]) is the percentage of the number of detections with
 the CE method divided by those with  each method.
}
\begin{flushleft}
\begin{tabular}{llllll}
\hline
\hline
 & N detection  & Common detection &
 Rate to CE (\%)  & Rate to the method (\%) \\
\hline
MF & 152 & 116  &   32.0 & 76.3  \\
\hline
maxBCG & 438 & 183 & 50.4 & 41.8\\
\hline
VTT & 130 & 96 & 26.4 &73.8\\
\hline
\hline CE & 363 & - & - &- \\
\hline
\end{tabular}
\end{flushleft}
\end{table}

\begin{table}[b]
\caption{\label{tab:other}
The comparison of the detected clusters with the methods other than the Cut \&
 Enhance method.
Column 1 and row 1 denote the name of each method.
 The parentheses in row 1 are the total numbers of the clusters detected
 with each method in the region RA 
 between 16 and 25.5 deg, DEC between $-$1.25 and  $+$1.25 deg (23.75
 deg$^2$).
Rows  2$\sim$4 list the numbers of the clusters detected with both of
 the two methods
 (column 1 and row 1) and their percentage to the methods in column 1. 
}
\begin{flushleft}
\begin{tabular}{l|lll}

\hline
 & MF &  VTT & maxBCG \\
 & (152) &  (130) & (438) \\
\hline
MF &  - &   39.4\% (60)   & 59.2\% (90)   \\
\hline
VTT &  45.5\% (60) & - & 65.4\% (85) \\
\hline
maxBCG &  20.5\% (90) & 19.4\% (85)&-\\
\hline
\end{tabular}
\end{flushleft}
\end{table}


\chapter{Composite Luminosity Functions}
\label{chap:LF}

\section{Introduction}

 Luminosity function (LF) of galaxies within clusters of
galaxies is a key tool for understanding the role of the environment on galaxy
formation and evolution.  The shape of the cluster LF as a function of
the galaxy colors and morphologies, as well as a function of the cluster radius or local density, can provide strong observational constraints
on the theories of galaxy formation. For example, Springel et al. (2001)
recently showed that semi--analytical models of hierarchical structure formation could
now explain both the shape of the composite cluster LF ($B$--band LF of Trentham 1998) and the morphology--radius relationship of Whitmore et al. (1993)
using just a simple prescription for the properties of galaxies in clusters
based on their merger and cooling rates (see also Okamoto, Nagashima
2001; Diaferio et al. 2001). Empirically, there is also growing evidence
for a correlation between 
the shape of the cluster LF and the underlying cluster properties.  Phillipps
et al. (1998) and Driver et al. (1998) show that more
evolved clusters, based on either their density profile or the presence of a
cD galaxy, have flatter faint-end slopes, which they attribute to the
disruption of faint galaxies in the cores of such evolved systems [see the
earlier theoretical work on galaxy cannibalism by Hausman \& Ostriker
(1978)]. In summary, the LF of galaxies in clusters as a function of both the
galaxy and cluster properties is a powerful observational test for theories
of galaxy formation and evolution. The reader is referred to the seminal
review by Binggeli, Sandage, and Tammann (1988), which is still relevant today.

In this Chapter, we present an analysis of the composite cluster LF
based on the commissioning data of the Sloan Digital Sky Survey (SDSS; see
Gunn et al. 1998; York et al. 2000; Stoughton et al. 2002). This analysis has several key
advantages over previous studies of the composite cluster LF, including
accurate multi-color CCD photometry for all galaxies (in optical passbands
$u,g,r,i$ and $z$; Fukugita et al. 1996), large aerial
 coverage, thus enabling us to make a local correction for the projected field
LF, and finally, the availability of several objectively-measured galaxy
properties like morphology.  Furthermore, we selected our clusters from
the SDSS Cut and Enhance (CE) cluster catalog of Goto et al. (2002a;
Chapter \ref{chap:CE}), which has
two major benefits over previous cluster samples used for LF studies. First,
the CE catalog was objectively constructed using the latest cluster-finding
algorithms, and therefore has a well-determined selection function (see Goto
et al. 2002a; Chapter \ref{chap:CE}). Secondly, CE has obtained an
accurate photometric redshift for 
each cluster based on the observed color of the E/S0 ridge-line using
 the maxBCG method (J. Annis et al. in preparation). The error on
this cluster photometric redshift is only $\delta z=0.015$ for $z<0.3$
clusters (see Figure. 14 of Goto et al. 2002a; Figure \ref{fig:zaccuracy.eps}
 in Chapter \ref{chap:CE}) and, as we show herein, is
accurate enough to allow us to determine the composite LF for a large sample
of CE clusters without the need for spectroscopic redshifts. Thus, our analysis
of the composite cluster LF is based on one of the largest samples of
 clusters  to date.

We present this work now to provide a low-redshift benchmark for
on-going multi-color photometric studies of high redshift clusters of
galaxies.  With the advent of large-area CCD imagers on large telescopes, the
number of distant clusters with such data will increase rapidly over the next
few years; e.g., Kodama et al. (2001) recently presented large-area
multi-color CCD photometry for the distant cluster A 851 ($z=0.41$) using
Suprime-Cam on the Subaru Telescope. Gladders \& Yee (2000) searched
distant clusters over 100 deg$^2$ of CCD data. 

 This Chapter is organized as follows: In
section \ref{lf_method} we describe the methods used to construct the composite LF of CE
clusters and show our results as a function of passband and morphology. In
section \ref{lf_discussion} we test the robustness of the analysis, and
in section \ref{lf_conclusion} we summarize our work. Throughout this
 Chapter, we use $h_0$ = 0.7, $\Omega_{\rm{M}}$ = 0.3 and $\Omega_{\Lambda}$ = 0.7.

\section{SDSS Data}
\label{May 29 19:19:42 2003}

In this section, we outline the data used in this Chapter. The photometric data
used herein were taken from the SDSS commissioning data, as discussed by York et
al. (2000). Our analysis focuses on the 150 deg$^2$ contiguous area made
 up
from the overlap of SDSS photometric runs 752 \& 756,
 i.e., 145.1
$<$R.A.$<$
236.1 and 
-1.25$<$Dec.$<$ 
 +1.25. This is a subset
of the SDSS Early Data Release, as discussed in Stoughton et al. (2002), and
similar to the data used by Scranton et al. (2002) for studying the angular clustering of SDSS galaxies. This photometric data
 reach 5$\sigma$ detection limits for point sources of 22.3, 23.3, 23.1,
 22.3 and 20.8 mag in the $u,g,r,i$ and
 $z$ passbands, respectively (for an airmass of 1.4 and $1''$
seeing)\footnote{The photometry obtained at
this early stage of SDSS is denoted $u^*,g^*,r^*,i^*$, and $z^*$ to stress the
preliminary nature of the calibration.}.
 The photometric uniformity of the data across the whole area is less
than 3\% [see Hogg et al. (2001) and Smith et al. (2002) for photometric calibration],
 while the star-galaxy separation is robust to $r^*\simeq$21.0 (Scranton
 et al. 2002). The SDSS is
 significantly better than previous photographic surveys, which suffer from
 larger plate-to-plate photometric fluctuations and a lower dynamic range
 [see Lumsden et al. 1997 for the problems associated with photographic studies
of the cluster composite LF].  For each galaxy, we used the model
magnitude computed by the PHOTO data analysis pipeline, which has been shown
by Lupton et al. (2001) and Stoughton et al. (2002) to be the optimal
magnitude for faint SDSS galaxies. It is also close to the total magnitude for
the fainter SDSS galaxies.  For a full discussion of the photometric data, and
the galaxy parameters derived from that data, we refer the reader to Lupton et
al. (2001) and Stoughton et al. (2002).

The clusters used here were drawn from the large sample of CE clusters
presented in Goto et al. (2002a;
Chapter \ref{chap:CE}), which were selected over the same photometric
runs of 752 \& 756. We only selected the richer systems which were determined
 by the number of galaxies brighter than $-$18th magnitude, ($N_{-18}$).
 The CE clusters used here satisfy the following conditions:

1) Number of galaxies brighter than $-$18th magnitude in the $r$ band ($N_{-18}$) $>$ 20,

2) 0.02$<z<$0.25.

  Condition 1 was used to select richer systems. $N_{-18}$ is defined
as the number of galaxies brighter than $-$18th magnitude in the $r$ band
after subtracting the background using the method described in section
\ref{lf_method}. Galaxies within 0.75 Mpc from
a cluster center were used.
 Condition 1 is useful to avoid letting small groups with only a few very bright
 galaxies dominate the composite LFs in the weighting scheme (The
 weighting scheme is explained in detail in section \ref{lf_method}). Even
 though we used $N_{-18}>$20 as a criteria to select our clusters, we
 show in section \ref{lf_discussion} that our composite LFs were not
 affected by this richness criteria. Since the high redshift clusters ($z\sim$0.3)
 are not imaged to the fainter galaxies, we restricted our clusters to be
 in the range  0.02$<z<$0.25. In total, 204 clusters satisfy these criteria.
 
\section{Analysis and Results}
\label{lf_method}

\subsection{Construction of the Composite Cluster LF}

 We discuss here the construction of the composite luminosity function of
 galaxies within the subsample of the CE clusters discussed above. The first
 critical step in such an analysis is the subtraction of  the background and
 foreground contamination.  Ideally, one would wish to do this via
 spectroscopic observations, but since the CE cluster catalog contains
 $\sim$2000 clusters in the region used, it is not feasible to observe
 all clusters spectroscopically.
 Therefore, we must make a statistical correction based on the
 expected contamination from projected field galaxies. One of the main
 advantages of the SDSS data is that such a correction can be estimated
 locally (i.e., free from any galaxy number count variances due to the large-scale
 structure)
 for each cluster since we possess all the photometric data, to the same depth
 and in the same filter set, well outside of the cluster. 
 Indeed, such local background subtraction was thought to be ideal in
 previous work, but was not possible due to the small coverage of the sky.

 For the composite cluster LF, we only used galaxies within 0.75 Mpc of the
 cluster centroid. This radius was 
 determined empirically so as not to lose
 statistics by using a too small radius,
 and not to lose the contrast of clusters
 against the background by using a too large radius.
 Foreground and background contamination were corrected for
 using an annulus around each cluster with an inner radius of 1.5 Mpc and an
 outer radius of 1.68 Mpc. These radii represent a compromise between having as
 large an aperture as possible to avoid removing legitimate cluster galaxies,
 while still providing an accurate estimate of the local projected field
 population.  Since the background/foreground galaxies are themselves highly
 clustered, it is important to obtain as local an estimate as possible.  The
 photometric redshift of each cluster was used to convert these metric
 apertures into angular apertures.  The center of each cluster was taken from
 the CE catalog, and estimated from the position of the peak in the
 enhanced density map of Goto et al. (2002a;
Chapter \ref{chap:CE}). The cluster centroids were expected
 to be determined with an accuracy better than $\sim$40 arcsec through Monte-Carlo simulation.  When an annulus touches the boundary
of the SDSS data, we corrected for contamination using the number-magnitude
relationship of the whole data set instead (this only affected a few of the
clusters used here).
 
 Since each sample cluster has a different redshift, each cluster
 reaches the SDSS apparent magnitude limit at different absolute magnitudes. Also, because
 they have various richnesses, the number of galaxies in each cluster
 is different.  To take these different degrees of completeness into account,
 we followed the methodology of Colless (1989) to construct the composite cluster LF.
 The individual cluster LFs are weighted according to the cluster
 richness and the number of clusters which contribute to a given bin. This is
 written as

 \begin{equation} 
    N_{cj}=\frac{m_j}{N_{c0}}\sum_{i}\frac{N_{ij}}{N_{i0}},   
 \end{equation}  

\noindent where $N_{cj}$ is the number of galaxies in the $j$th bin of the
 composite LF, $N_{ij}$ is the number in the $j$th bin of the $i$th cluster LF,
 $N_{i0}$ is the normalization of the $i$th cluster LF, and was measured to be
 the field-corrected number of galaxies brighter than $M_{r^*}=-18$,
 $m_j$ is the number of clusters contributing the $j$th bin and, finally,
 $N_{c0}=\sum_{i}N_{i0}$.  The formal errors on the composite LF were computed
 using

 \begin{equation} 
    \delta N_{cj}=\frac{N_{c0}}{m_j}   \left[\sum_{i} \left(\frac{\delta N_{ij}}{N_{i0}}\right)^2 \right]^{1/2}   ,
 \end{equation} 

 \noindent where $\delta N_{cj}$ and $\delta N_{ij}$ are the errors on the $j$th bin for
 the composite and $i$th cluster, respectively.  In this way we can take
 into account the different degrees of completeness.

 Like other authors, we discarded the brightest cluster galaxy (BCG)
 within 0.75 Mpc of the cluster centroid when constructing the composite
 LF, since such BCGs tend not to follow the cluster LF.  We only used SDSS galaxies
 brighter than $r^*$=21.0, since this is the limit of the SDSS star--galaxy
 separation (Scranton et al. 2002; Lupton et al. 2001). This
 magnitude limit and weighting scheme combined with our cosmology enabled us to dig LF down to $M_{r^*}$=$-$17.5 .  When converting apparent to absolute magnitudes, we assumed a $k$-correction for the early-type galaxy given by Fukugita et al. (1995).

 In Figure \ref{fig:all.eps}, we show the composite
 LF of the 204 CE clusters discussed above. We present one composite
 LF for each of the five SDSS passbands.  We also present in Table
 \ref{tab:5color} the best-fit parameters from a fit of a Schechter
 function to these data. For a comparison, we also show the field values as
 derived by Blanton et al. (2001; corrected for $h_0$ = 0.7). 
 In Figure \ref{fig:all.eps}, field LFs normalized to cluster LFs are shown by dotted lines.
 As expected, the $M^*$ for our cluster LFs
 is significantly brighter (by 1 -- 1.5 magnitudes depending on the
 bands) than those seen for the field LFs in all five bands.
 Furthermore, the faint end slopes ($\alpha$) of the cluster composite
 LFs are much flatter that those seen for the field LFs. This is
 especially noticeable for the redder passbands ($i$ and $z$)
 while the slope of the cluster LF systematically flattens from the $u$
 passband to the $z$ passband. 

 These results are consistent with the hypothesis that the cluster LF has two
 distinct underlying populations, i.e., the bright end of the LF is
 dominated by bright early types that follow a Gaussian-like luminosity
 distribution, while the faint-end of the cluster LF is a steep
 power-law-like function dominated by star-forming (bluer) galaxies.
 Binggeli et al. (1988) originally suggested this hypothesis, 
 and the recent work of Adami et al. (2000), Rakos et al. (2000) and
 Dressler et al. (1999) supported this idea. 
 Particularly, Boyce et al. (2001) showed LF of Abell 868 is made up
 of three different populations of galaxies; luminous red and two
 fainter blue populations. 
  The idea is illustrated by the fact that the cluster LFs in the redder passbands, which
 are presumably dominated by the old stellar populations of the early types,
 have much brighter $M^*$'s and significantly shallower slopes than those
 measured in the bluer passbands. Those results can also be interpreted
 as showing that bright elliptical galaxies are more populated 
 in dense regions, like inside of clusters. The results are consistent with
 the morphology-density relation advocated by Dressler et al. (1980, 1997).

\subsection{The Composite Cluster LF as a Function of Morphology}\label{sec:morph}

 One of the key aspects of the SDSS photometric data is the opportunity to
 statistically study the distribution of galaxies as a function of their
 morphology. In this subsection, we discuss the composite cluster LF as a function
 of morphology using three complementary methods for determining the
 morphological type of each galaxy. These include:  i) the best-fit de
 Vaucouleur or exponential model profile; ii) the inverse of concentration
 index and iii) the $u-r$ color of the galaxies.
  We present all three methods, since at present it is unclear which
 method is the most successful in separating the different morphological galaxy
 types. Also, each method suffers from different levels of contamination, and
 the differences in the methods can be used to gauge the possible systematic
 uncertainties in the morphological classifications. We discuss the three
 methods used in detail below.

 The first method we consider here uses the de Vaucouleur and
 exponential model fits of the galaxy light profiles measured by the SDSS photometric
 pipeline ($PHOTO$ R.H. Lupton et al., in preparation) to broadly
 separate galaxies into the late and early-type. If
 the likelihood of a de Vaucouleur model fit to the data is higher than the
 that of an exponential model fit, the galaxy is called a late-type, and vice versa. Galaxies that have the same likelihoods for both
 model fits are discarded. In Figure \ref{fig:exp_dev.ps}, we present the
 composite cluster LF of late-type and early-type galaxies (as defined using
 the model fits above) for all five SDSS passbands. In Table \ref{tab:dev_exp},
 we present the best-fit Schechter function parameters to these data, and show
 the fits in Figure \ref{fig:exp_dev.ps}.

  The second method uses the inverse of the concentration index, which is defined
 as $C$ = $r_{50}/r_{90}$, where $r_{50}$ is the radius that contains
 50\% of the Petrosian flux and $r_{90}$ is the radius that contains
 90\% of the Petrosian flux (see R.H. Lupton et al., in preparation;
 Stoughton et al. 2002). Both
 of these parameters are measured by the SDSS PHOTO analysis pipeline for each
 galaxy. The concentration parameter used here ($C$) is just the inverse of the
 commonly used concentration parameter, and thus early-type galaxies have a
 lower $C$ parameter than late-type galaxies.  The correlation of
 $C$ with visually-classified morphologies has been studied in detail by
 Shimasaku et al. (2001) and Strateva et al. (2001). They found that galaxies
 with $C<$0.4 are regarded as early-type galaxies, while galaxies with
 $C\geq$0.4 are regarded as late-type galaxies.  Therefore, in Figure
 \ref{fig:cin.ps}, we show the composite cluster LF of late-type and
 early-type galaxies as defined using this second method for all five SDSS
 passbands. In Table \ref{tab:5color_cin}, we present the best-fit Schechter
 function parameters to these data.

 The third method used here for morphological classification was to use the
 observed $u-r$ color of the galaxy which has been proposed by Strateva et
 al. (2001). Using the fact that $k$-correction for $u-r$ is almost
 constant until $z$ = 0.4, they showed that galaxies shows a clear bimodal distribution in
 their $u-r$ color and $u-r$ = 2.2 serves as a good classifier of morphology
 until $z\sim$ 0.4 by correlating $u-r$ classification with visual
 classifications. Therefore, we have classified galaxies with $u-r<$ 2.2 as
 early-type and galaxies with $u-r\geq$ 2.2 as late-type.  Figure
 \ref{fig:ur.ps} shows the composite cluster LF for both types of galaxies
 along with their best-fit Schechter functions (in all five passbands). The
 best-fit Schechter parameters are summarized in Table \ref{tab:5color_ur}.

 As expected, there are  noticeable differences between early-type and
 late-type galaxies in these three morphological
 classifications, as portrayed by the differences in their composite LFs (see
 Figures. \ref{fig:exp_dev.ps}, \ref{fig:cin.ps}, and \ref{fig:ur.ps}).  However,
 it is worth stressing here the similarities between the methods. For example,
 the faint end slope of the LF is always shallower for early-type galaxies
 than late-type, regardless of the passband and methodology. Also, the faint end
 slope for early-type galaxies decreases steadily toward the redder
 passbands, while the faint-end slope for the late-type galaxies is nearly
 always above $-1$ and consistent with (or steeper than) the field LF in most
 passbands. These observations are again qualitatively in agreement with the
 hypothesis that the bright end of the cluster LF is dominated by bright, old
 early-types, while the faint-end of the
 cluster LF represents late-type galaxies maybe in greater numbers
 than the average field. This model is in agreement with hierarchical models of
 structure formation and the model for the tidal disruption of dwarf galaxies
 by the dominant early types.

\section{Discussion}
\label{lf_discussion}

 In this section, we discuss various tests which we have performed on our
 measurement and results.

\subsection{Monte-Carlo Simulations}\label{sec:lf_monte}

 To test the robustness of our methods, we performed Monte-Carlo
 simulations which involved adding artificial clusters to the SDSS data and
 computing their composite LF using the same algorithms and software as used on
 the real data.  Our model for the artificial clusters was constructed using
 the SDSS data on Abell 1577 (at $z\sim0.14$, richness$\sim1$). We used
 the method described in Goto et al. (2002a;
Chapter \ref{chap:CE}) to make artificial clusters. 
 The radial profile for the artificial clusters was taken to be a King profile (Ichikawa
1986) with a concentration index of 1.5 and a cut-off radius of 1.4 Mpc,
 which is the size of Abell 1577 (Struble \& Rood 1987).  The
 color-magnitude distributions for the artificial clusters were set to be the
 observed, field-corrected, color-magnitude distributions of Abell 1577
 binned into 0.2 magnitude bins in both color and magnitude.
 From this model, we then constructed artificial clusters as a function of
 the redshift and richness. For redshift, we created clusters at
 $z$ = 0.2, 0.3, 0.4 and 0.5, ensuring that we properly accounted for the cosmological
 effects; i.e., the clusters became smaller, redder and dimmer with
 redshift.  We used the $k$-corrections for an early type spectrum.  For
 richness, we change the number of galaxies within each cluster
 randomly between 10 and 50. For each redshift, we
 created 100 clusters (400 clusters in total). The galaxies within these
 artificial clusters were distributed randomly in accordance with the radial and
 color-magnitude distributions discussed above. We made no attempt to
 simulate the density-morphology relation nor the luminosity segregation in
 clusters.

 The artificial clusters were randomly distributed within the real SDSS imaging
 data and we constructed a composite LF for these clusters using exactly the same
 software as for the real clusters. Since the artificial clusters were all made
 from the same luminosity distribution, the composite LF should therefore look
 very similar in shape as the original input LF. Figure
 \ref{fig:tfake_1577_ab.ps} shows the  result of our Monte-Carlo
 simulations. The histogram shows the original absolute magnitude
 distribution of Abell 1577 after a field correction, while the symbols show
 the composite luminosity functions which we constructed as a function
 of the input redshift. The composite LFs, indeed, lie on top of 
 the original input LF, suggesting that our code can successfully
 recover the input LF from the data through fore/background subtraction
 at any redshift.

\subsection{Check of Photometric Redshifts}
\label{sec:lf_photoz_specz}

 One of the most innovative parts of this analysis is the use of photometric
 redshifts to determine the composite luminosity function of clusters. As
 demonstrated in Goto et al. (2002a;
Chapter \ref{chap:CE}), the accuracy of photometric redshift is
 excellent ($\delta z$ = $\pm$0.015 for $z<$0.3) and this method will certainly
 be used in the near future as the number of clusters with photometric
 redshifts will increase rapidly, far quicker than the number of clusters with
 spectroscopically confirmed redshifts. 

 To justify our use of photometric redshifts, since all previous composite
 cluster LF's used spectroscopic redshifts, we constructed a
 composite LF using only the clusters with spectroscopically confirmed
 redshifts. We derived our spectroscopic redshift for CE clusters by matching
 the SDSS spectroscopic galaxy data with our CE clusters. This was achieved by
 searching the SDSS spectroscopic galaxy sample for any galaxies within the CE
 cluster radius and within $\delta z$ = $\pm$0.01 of the photometric redshift of the
 cluster. 
 The radius used here was from Goto et al. (2002a;
Chapter \ref{chap:CE}). If multiple
 galaxies satisfied this criteria, the closest spectroscopic redshift to the photometric
 redshift was adopted. The number of clusters with spectroscopic redshifts
 was 75 out of 204 at the date of this writing.

 The results of this test are shown in Figure \ref{fig:all.eps}
 (in the bottom right-hand panel).  Also the parameters for the best-fit
 Schechter functions are given in Table \ref{tab:5color}.
 and referred to as
 $r^*$(spec). We only performed this test for the $r$
 passband. The slope and characteristic magnitude of the best-fit Schechter
 function for the spectroscopically determined LF is in good agreement with
 that derived using photometric redshifts. As can be seen in Table \ref{tab:5color},
 both $M_r^*$ and the slope agree within the error. This test shows that
 we can truly construct composite LFs using photometric redshift of clusters.

\subsection{Test of Cluster Centroids}
 
 One key aspect of measuring the composite cluster LF is the choice of
 the cluster centroid.
   To test the effect of different cluster centroids on the composite LF, we constructed a
 composite cluster LF using the position of brightest cluster galaxies (BCGs) as a
 centroid instead of the peak in the enhanced density map, as discussed in Goto
 et al. (2002a;
Chapter \ref{chap:CE}). The BCGs have been determined to be the brightest galaxy
 among galaxies whose absolute magnitudes computed with the cluster
 redshift are fainter than $-$24th magnitude within 0.75 Mpc from the
 cluster center. 
 Galaxies brighter than $-$24th magnitude are regarded as being foreground galaxies.
 The mean offset between the BCG position and the centroid previously used is 1.02 arcmin. 
  Table \ref{tab:bcg_center} lists the parameters of the best-fit Schechter functions
 to the five SDSS passbands, which should be compared to the values obtained
 using the optical centroid given in Table \ref{tab:5color}.
 In all five bands, the characteristic magnitudes and slopes agree very
 well within the error. 
 This test shows that our composite LFs are not dependent on the method
 of center determination.

\subsection{Test of Background Subtraction}

 Since we constructed composite LFs from 2-dimensional projected sky image, 
 subtraction of fore/background galaxies played an important role in this work.
 We test here the effect of making a global background subtraction for all
 clusters instead of the local background subtraction discussed above. 
 We use the number-magnitude relation of all the galaxies in the entire
 150 deg$^2$ region as the global background.  Table \ref{tab:global} gives
 the best-fit Schechter parameters of composite LFs constructed using
 global background subtraction.
 Compared with Table \ref{tab:5color},
 again, every Schechter parameter agrees very well within 1 $\sigma$.
 Although we use annuli around clusters to subtract the background to
 avoid the large-scale structure disturbing the measurement of composite
 LFs, this test shows that our composite LFs are not dependent on
 background subtraction. Valotto et al. (2001) showed that
 a statistical background subtraction can not re-produce composite LFs
 using a mock galaxy catalog constructed from a large $N$-body
 simulation. Our result, however, combined with the fact that we derived the
 same LF as input through a Monte-Carlos simulation (in subsection
 \ref{sec:lf_monte}), supports that our composite LFs are not subject to
 background subtraction.

\subsection{Test of Cluster Richness}

 Another aspect we are concerned about is our choice of cluster richness criteria.
 To test this, we construct composite LFs of different subsample with
 $N_{-18}>$20 and    $N_{-18}>$40, given in Table \ref{tab:richness}. 
 $N_{-18}$ here is defined as the number of galaxies brighter than $-$18th
 magnitude after subtracting the background in the way we construct
 composite LFs.  $N_{-18}>$20 is used to
 construct composite LFs, as mentioned in section \ref{lf_method}.  In Table
 \ref{tab:richness}, even though $M^*$ is slightly brighter and
 the slope is slightly steeper for the richer sample, they agree within 1 $\sigma$. 
 The steepening of the slopes can be interpreted as a bias in selecting
 richer systems using $N_{-18}$,  i.e., clusters with steeper
 tails tend to have a larger value of $N_{-18}$.
 This, however, confirms  that our composite LFs are not dependent on
 the  richness criteria we chose.

\subsection{Comparison with Other LFs}

 As the final test of our composite cluster LF, we compare our composite
 LFs with previous work. First, we must be careful to match the different
 cosmologies used by various authors as well as the different photometric
 passbands. To facilitate such a comparison, therefore, we present in Table \ref{tab:previous} the best-fit Schechter function parameters for our
 composite LF, but calculated for each author's cosmology and passband using the
 color corrections of Fukugita et al. (1995) and Lumsden et al. (1992).

 In the case of the three $b_j$ photographic surveys of Colless (1989),
  Valotto et al. (1997),  and Lumsden et al. (1997), we find a significantly brighter $M^*$ than these
 studies as well as a much shallower slope.
 We also tried to fit a Schechter function using their $\alpha$ value
  for the slope, but $M^*$'s become even brighter. The fits are not 
 good when fixed $\alpha$'s are used. 

 Lugger (1989) found   $M_R$=$-$22.81$\pm$0.13  and
 $\alpha$=$-$1.21$\pm$0.09 by re-analyzing nine clusters presented in
 Lugger (1986). The slope is steeper  and $M^*$ is
 slightly brighter than our results. When we fix the slope with her
 value at $\alpha$=$-$1.21, the two LFs agree well.

 Garilli et al. (1999) studied 65 Abell and X-ray selected samples of galaxies
 in the magnitude range of -23.0$<M_r<$-17.5  and found that
 $M_r^*$=$-$22.16$\pm$0.15 and $\alpha$=$-$0.95$\pm$0.07 (in isophotal
 magnitudes).
  This slope is steeper
 than ours. A possible difference with ours is that they used the color condition to select
 cluster galaxies. $M^*$ is in agreement with our results within the
 error. We also tried to fit a Schechter function with a fixed value of
 $\alpha$ = 0.84. $M^*$ became brighter by 0.18 mag although the fit was poor.  
 
 Paolillo et al. (2001) studied composite LF of 39 Abell clusters using the
 digitized POSS-II plates. They obtained $M^*$=$-$22.17$\pm$0.16 in $r$.
 The slope is $\alpha$=$-$1.11$_{-0.09}^{+0.07}$.
  Although the slope differs significantly,
 $M^*$ agrees well compared with our composite LF.

 Yagi et al. (2002a,b) observed 10 nearby clusters with their Mosaic CCD
 camera to derive composite LF. Their best-fit Schechter parameters are
 $M^*$=$-$21.1$\pm$0.2 and  $\alpha$=$-$1.49$\pm$0.05 in $R$. 
 They also studied type-specific LF using exponential and $r^{1/4}$
 profile fits to classify the galaxy types. They derived $M^*$=$-$21.1
 and  $\alpha$=$-$1.49 for exponential galaxies and $M^*$=$-$21.2
 and  $\alpha$=$-$1.08 for $r^{1/4}$ galaxies. Considering that they
 derived composite LFs using the data taken with different instruments
 analyzed in a different way, it is reassuring that they
 reached the same conclusion as our results discussed in subsection
 \ref{sec:morph}, i.e., exponential galaxies have the steeper faint end
 tail than $r^{1/4}$ galaxies, while their $M^*$ are almost the same.

 Concerning the disagreement of our LFs with previous studies, various differences
 in measuring composite LFs may be the reason.
  The possible sources of differences are different ways of weighting, different ways of
 background subtraction, and different depths of the luminosity function. 
 The sample clusters, themselves, should have, to some extent, different
 richness distributions.
 For $M^*$, although we tried to transform our magnitude into their
 magnitude, the color conversion between SDSS bands and others might not be
 accurate enough.  Thus, the difference with the previous studies is not
 necessarily a mistake in the analysis, but rather it represents a different way of
 analysis. Throughout our analysis discussed in section \ref{lf_method} we carefully
 used exactly the same way to construct the various composite LFs. We thus keep our
 composite LFs internally consistent.

\section{Summary}\label{lf_conclusion}

 We studied the composite LF of the 204 SDSS CE galaxy clusters.  The over-all
 composite LF is compared with other composite LFs.  Comparing it to the field
 luminosity function, a tendency of a brighter $M^*$ and a flatter slope is
 seen. This is consistent with our understanding that cluster regions are
dominated by brighter galaxies than field galaxies.
 We divided the composite LF by galaxy morphology in three ways. In all three
cases, we found that early-type galaxies have flatter slopes than
late-type galaxies. 
 These observations are in agreement with the
 hypothesis that the bright end of the cluster LF is dominated by bright, old
 early-types, while the faint-end of the
 cluster LF represents late-type galaxies.
 This is also consistent with the morphology--density relation
originally advocated by Dressler (1980). 
 We also studied these composite
LFs in five SDSS color bands. The slopes become flatter and flatter
toward the redder color bands. This again suggests that cluster regions are
dominated by elliptical galaxies with old stellar population.
These composite LFs provide a good low redshift benchmark
to study higher-redshift clusters in the future.
 Since the data used in this work came from 2\% of the SDSS data, further studies with
large SDSS data will increase the statistical significance on these topics as
the SDSS proceeds.

\newpage

\begin{figure}[h]
\begin{center}
\includegraphics[scale=0.2]{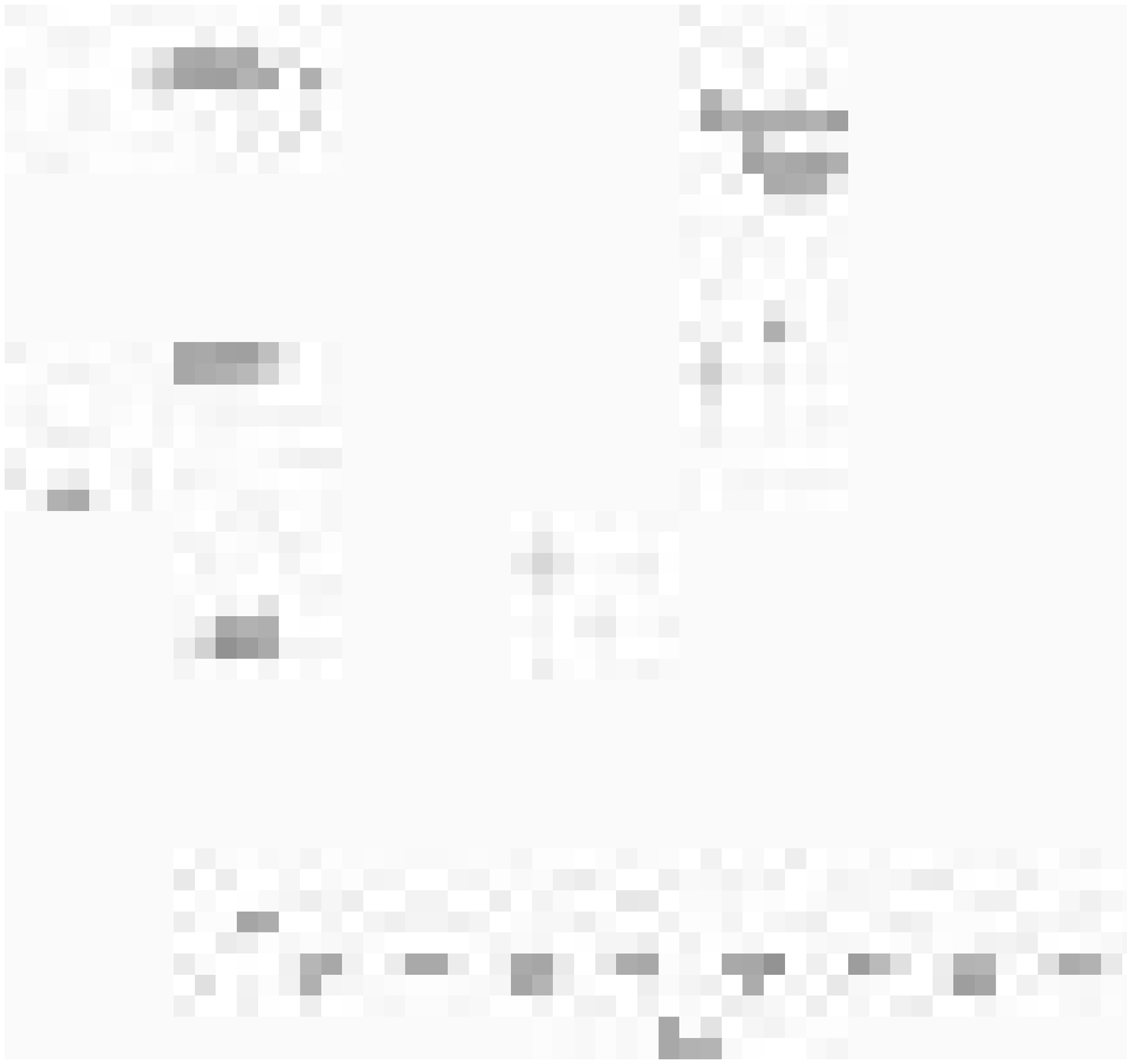}
\includegraphics[scale=0.2]{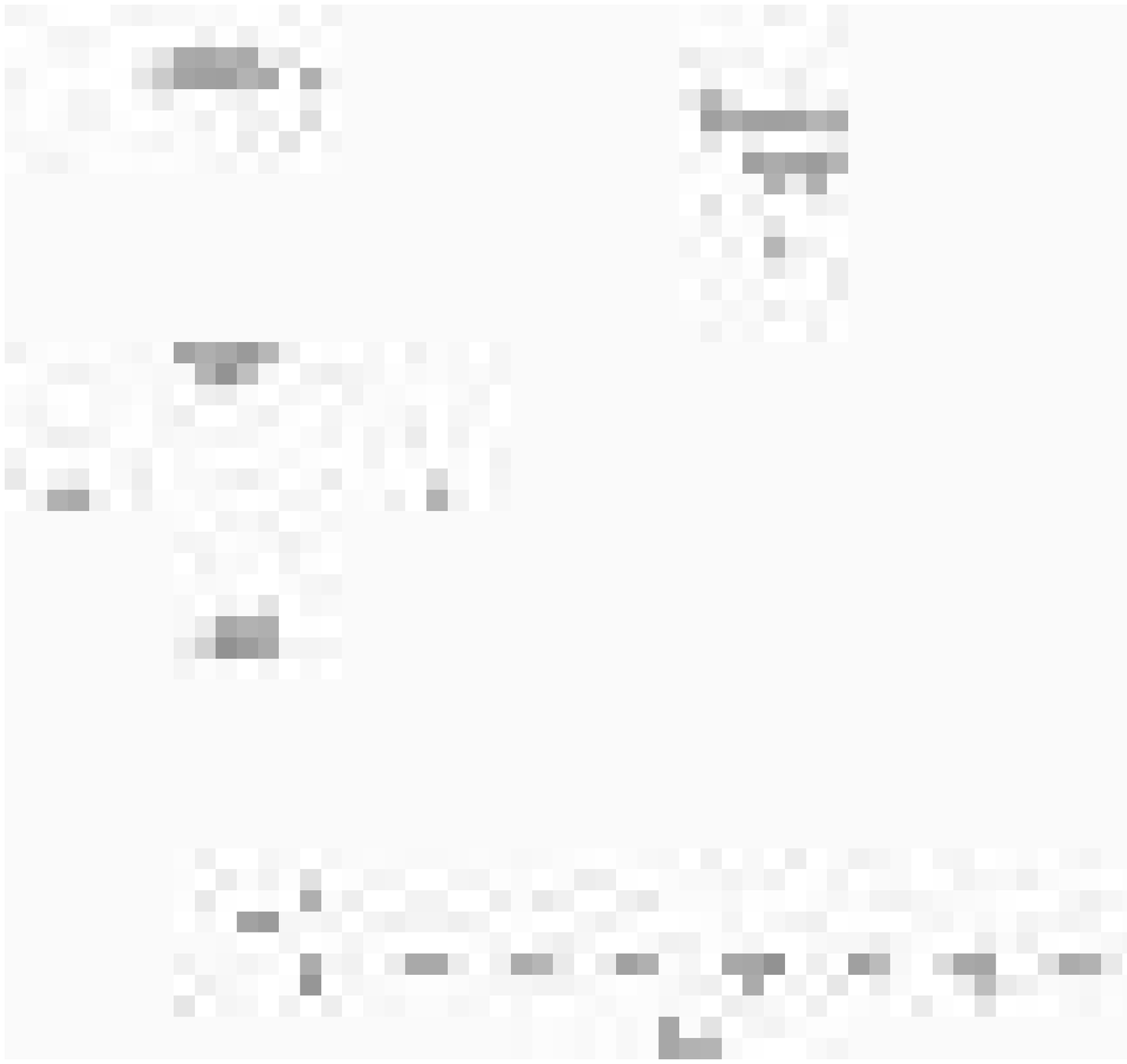}
\includegraphics[scale=0.2]{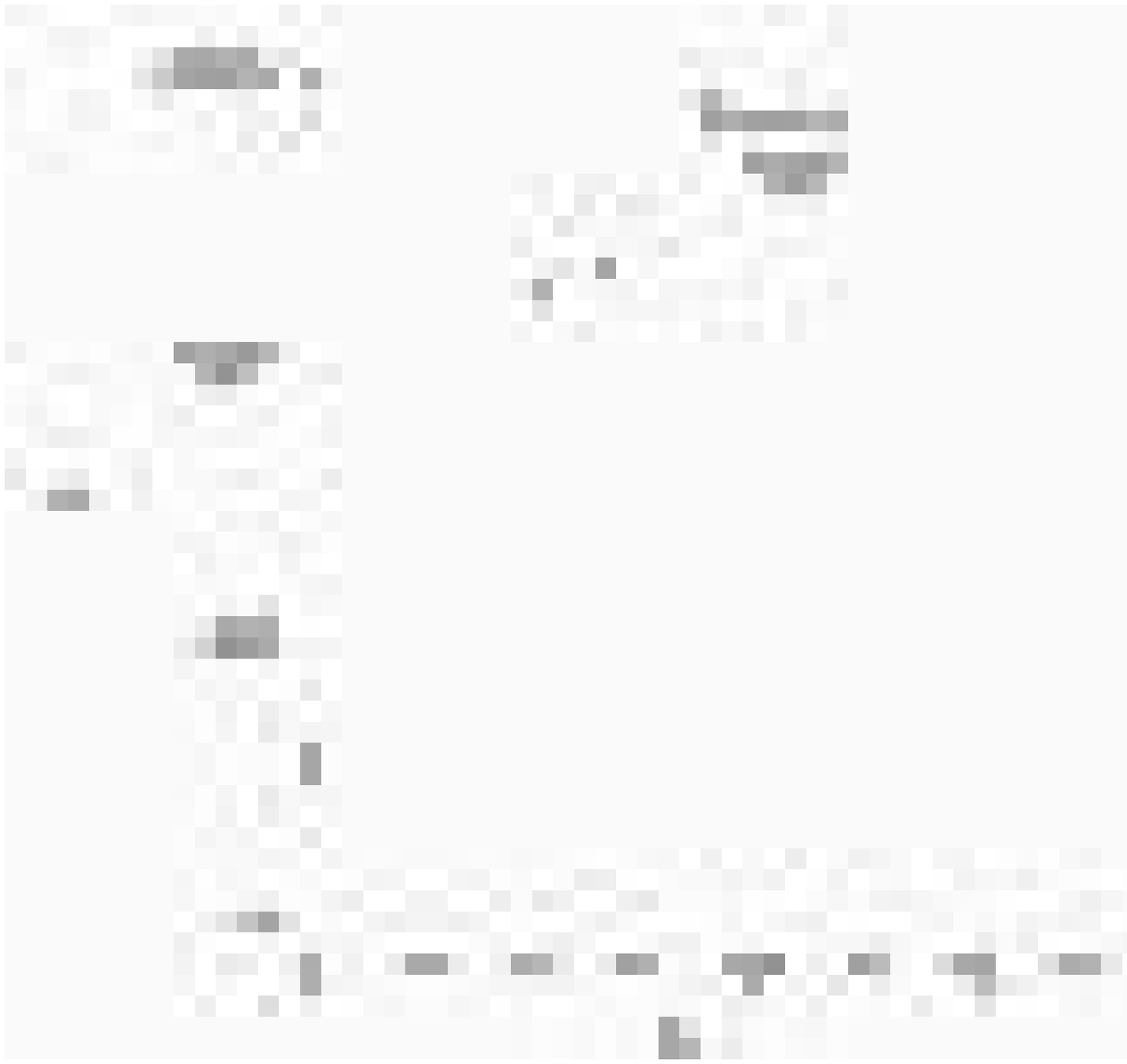}
\end{center}
\begin{center}
\includegraphics[scale=0.2]{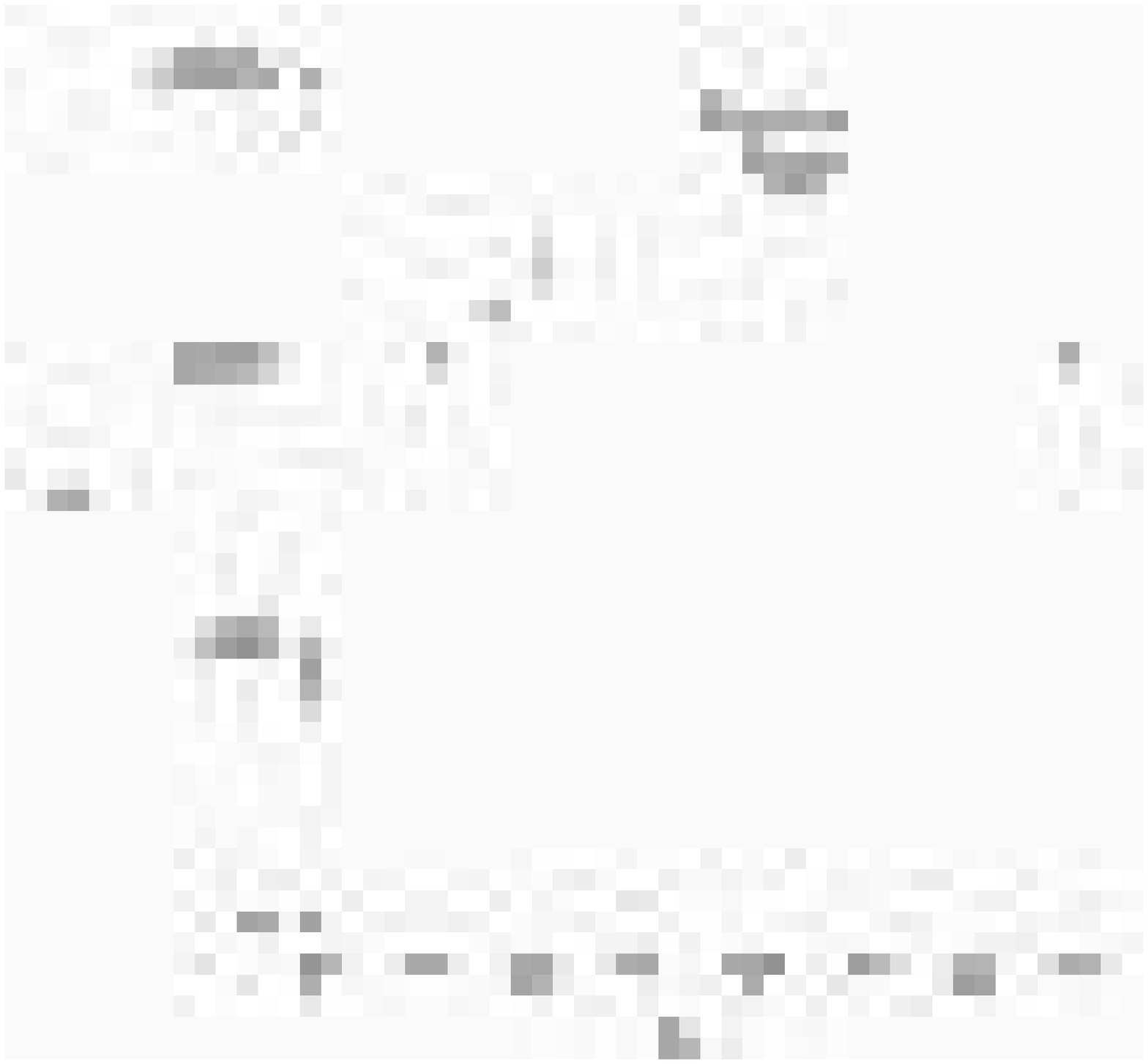}
\includegraphics[scale=0.2]{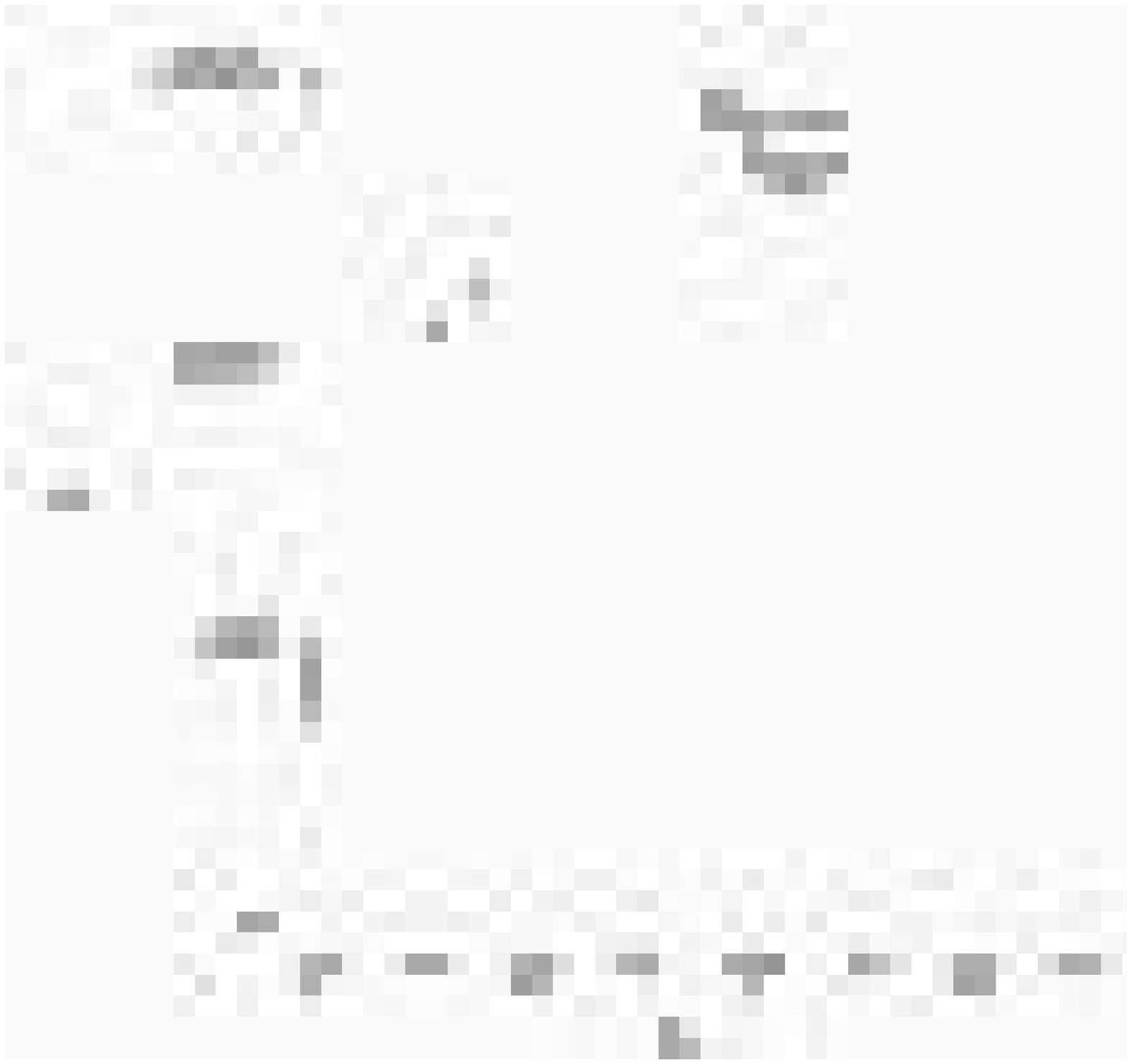}
\includegraphics[scale=0.2]{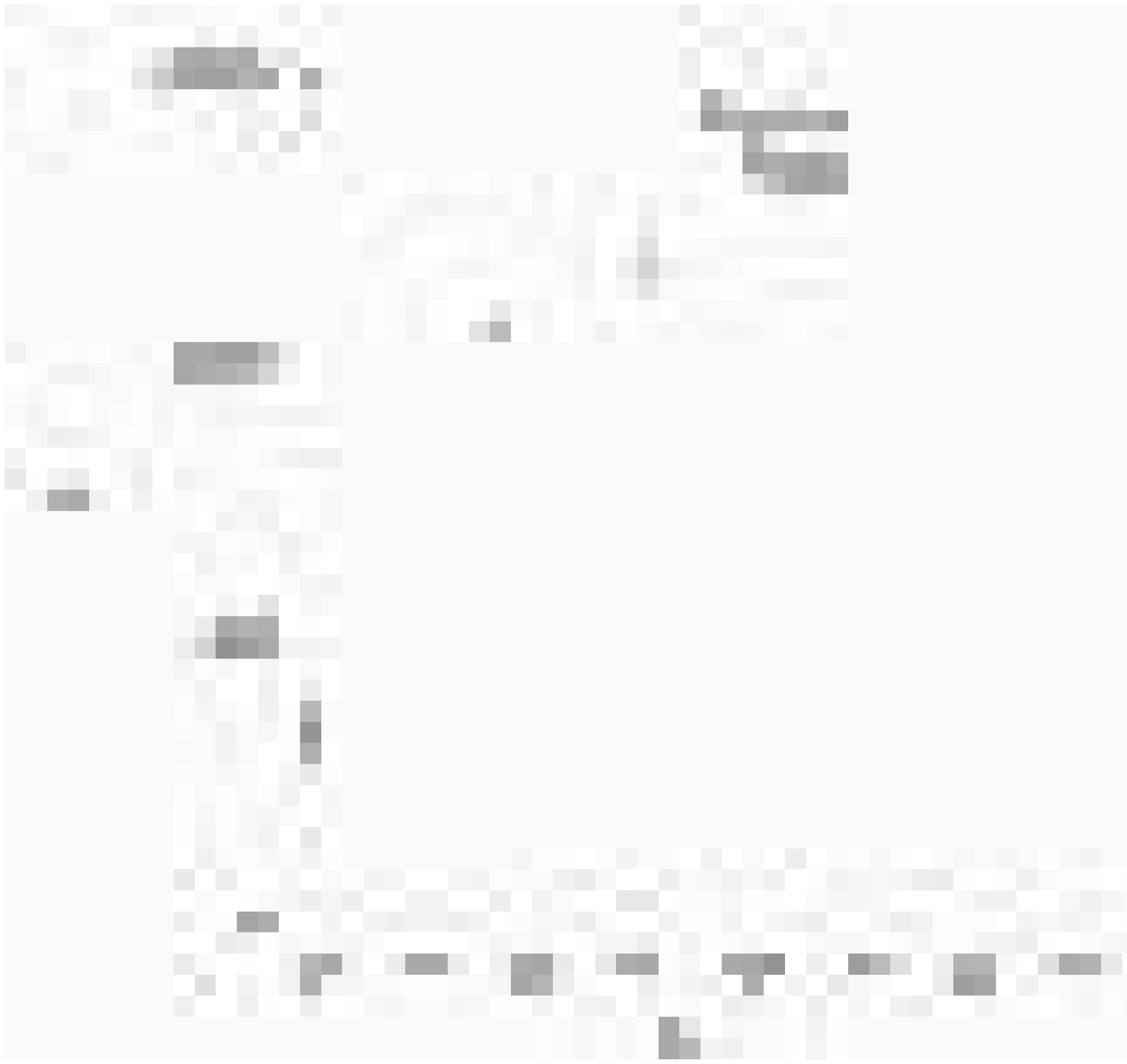}
\end{center}
\caption{
\label{fig:all.eps}
 Composite LF of 204 galaxy clusters from the SDSS CE galaxy cluster catalog
 in five SDSS bands. 
The solid line is the best-fit Schechter functions. The y-axis is arbitrary.
 The dotted line is the field LFs from Blanton et al. (2001) re-scaled to
 our cosmology. The normalization of field LFs was adjusted to match the
 cluster best-fit
 Schechter functions.
 The lower-right panel is for 75 clusters with spectroscopic redshifts
 (see Section \ref{sec:lf_photoz_specz}). The best-fit Schechter parameters are summarized in Table \ref{tab:5color}.
}
\end{figure}

\clearpage
\begin{figure}[h]
\begin{center}
\includegraphics[scale=0.2]{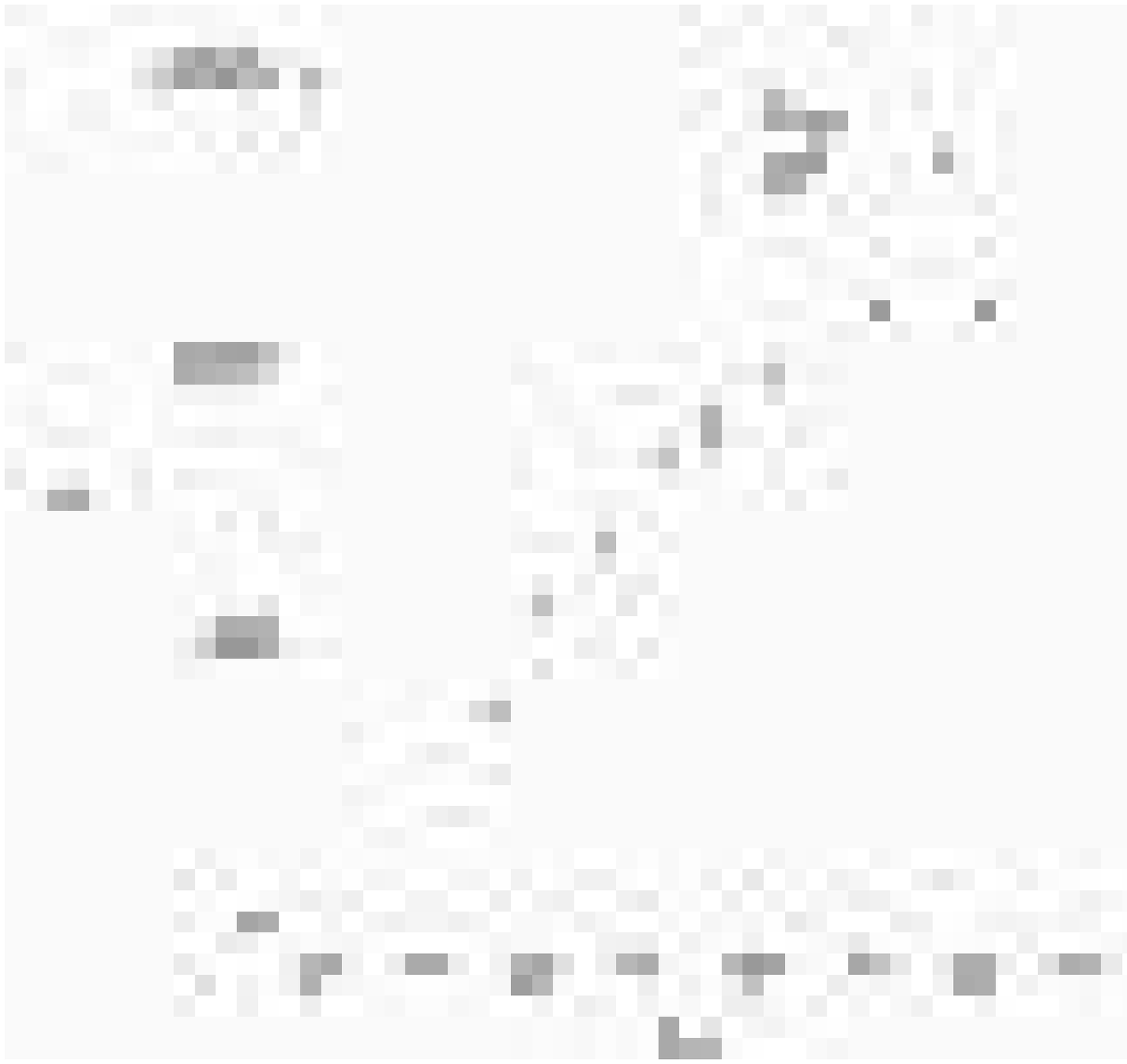}
\includegraphics[scale=0.2]{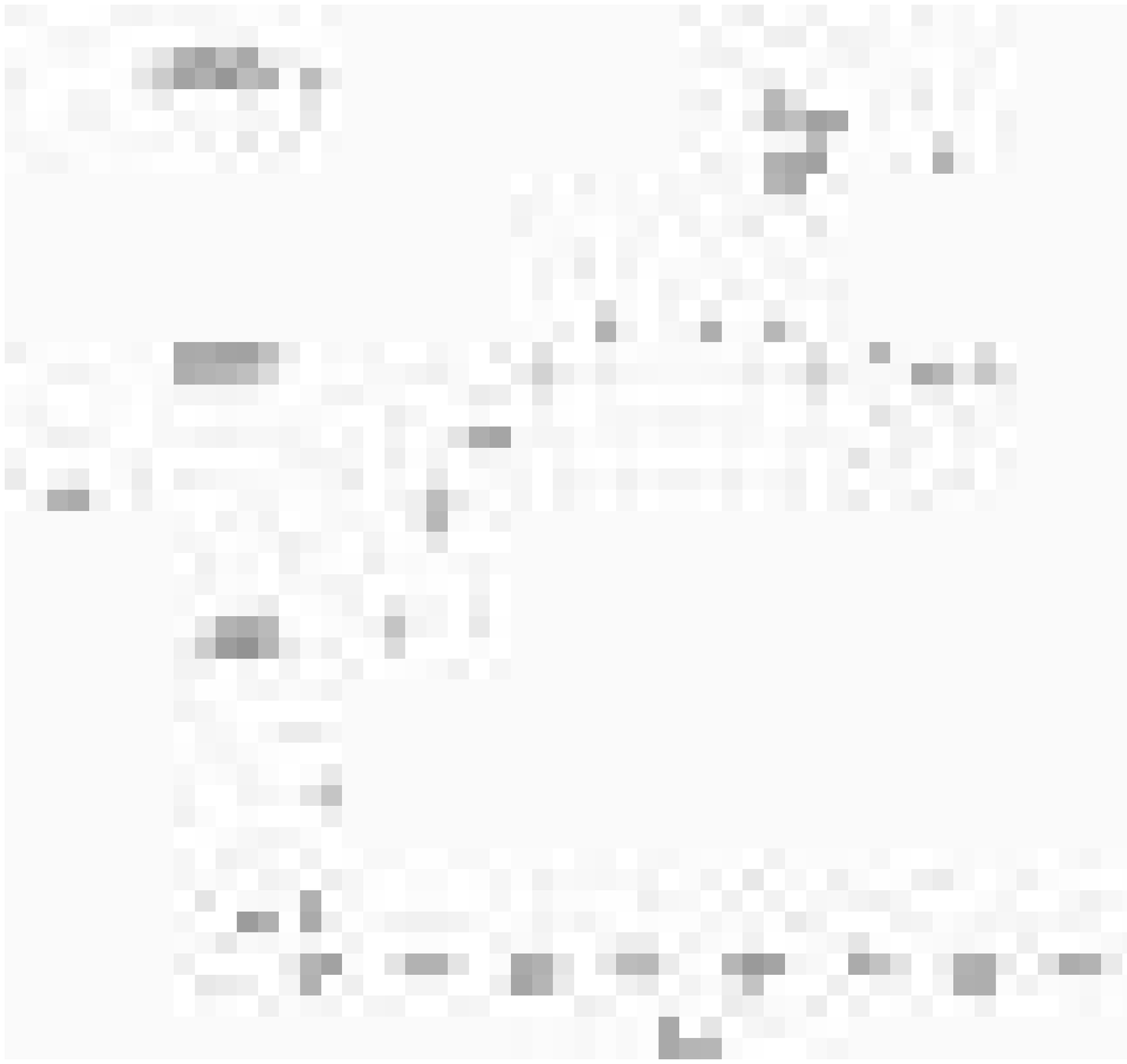}
\includegraphics[scale=0.2]{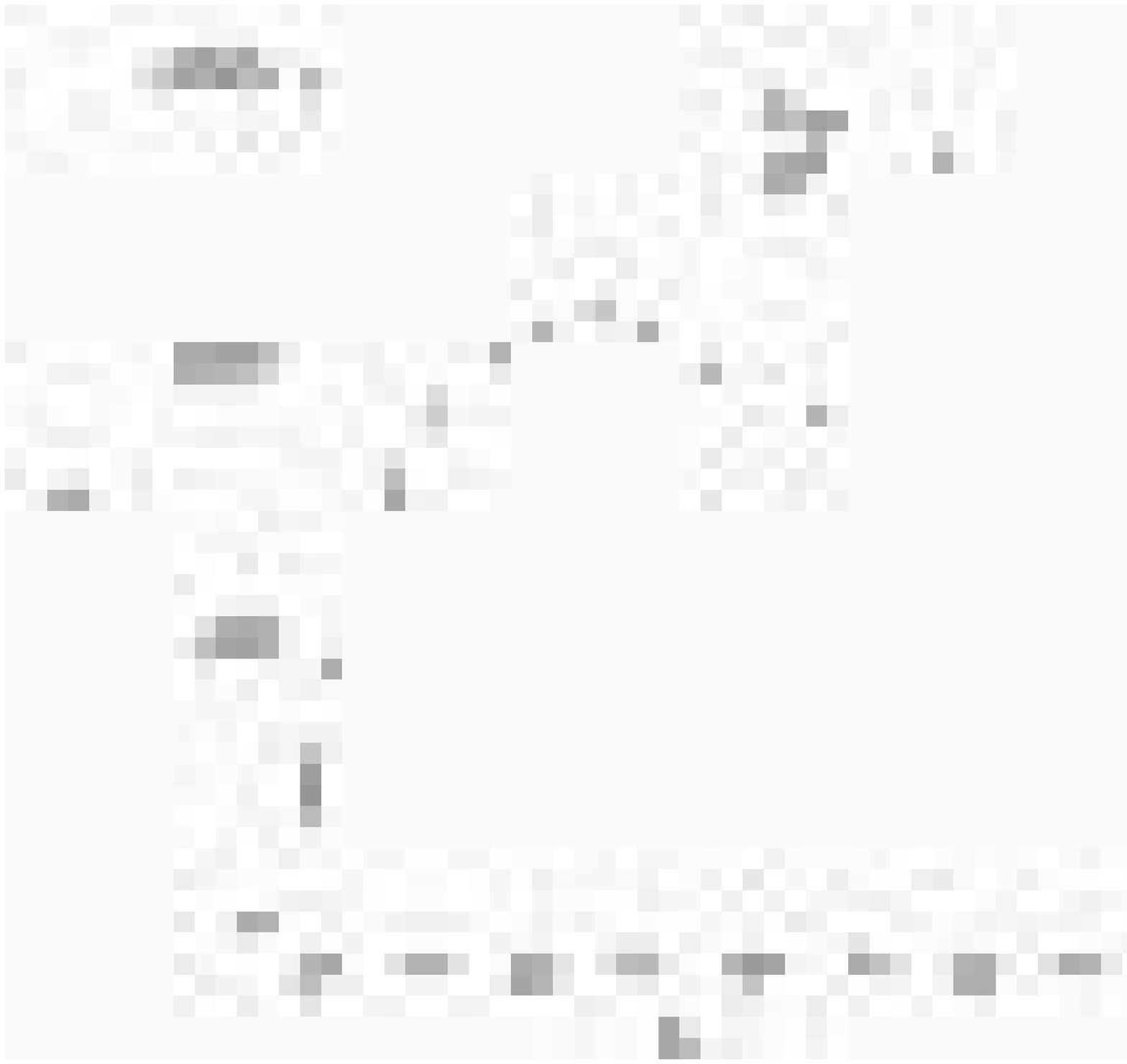}
\end{center}
\begin{center}
\includegraphics[scale=0.2]{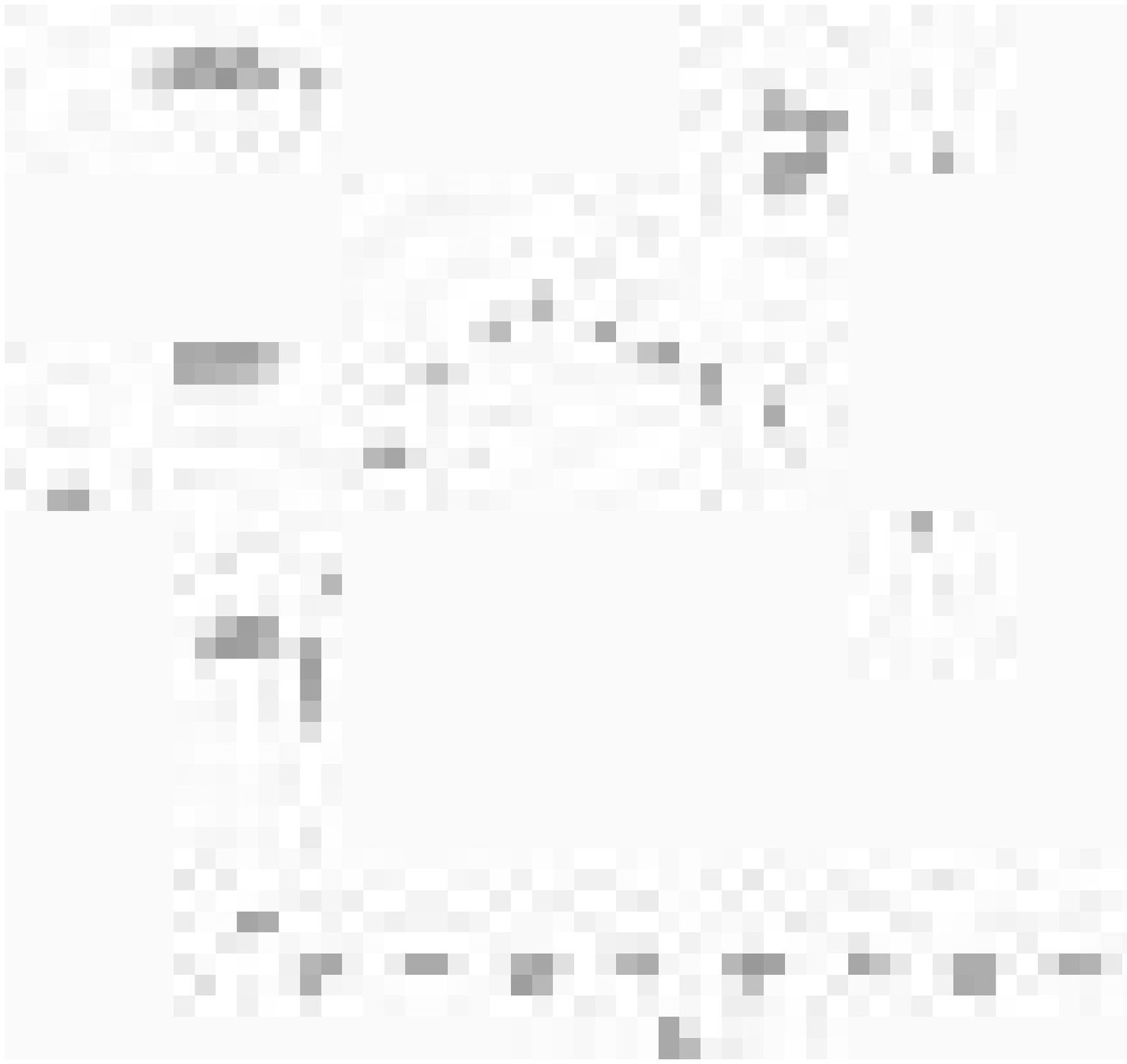}
\includegraphics[scale=0.2]{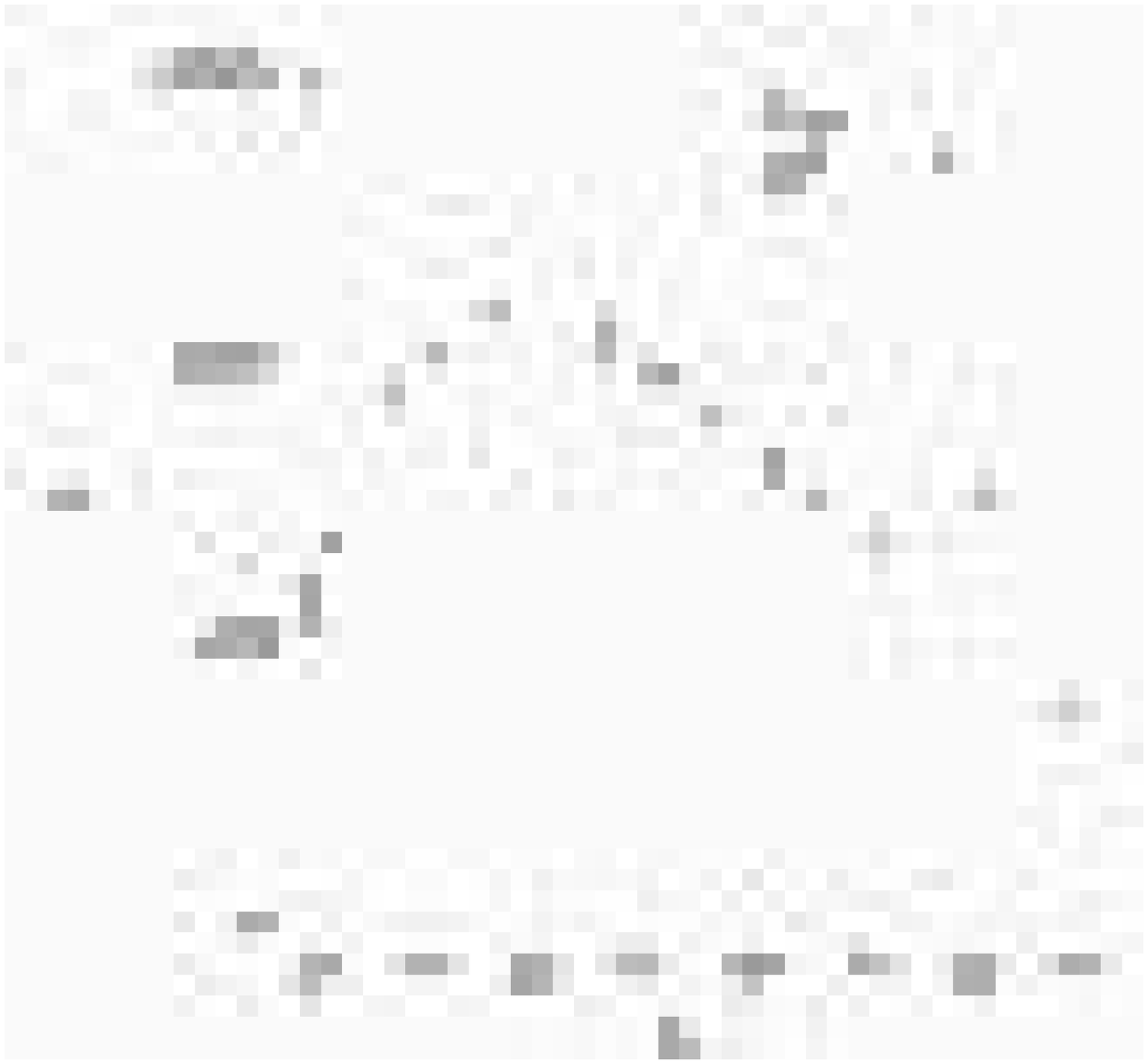}
\end{center}
\caption{
\label{fig:exp_dev.ps}
 Composite luminosity functions of de Vaucouleur galaxies and exponential
 galaxies. The galaxies are divided into two subsamples using profile
 fitting. The lines show the best-fit Schechter functions (solid for de
 Vaucouleur galaxies, dotted for exponential galaxies). The y-axis is arbitrary.
 de Vaucouleur galaxies always have a brighter $M^*$ and a flatter faint end
 tail. The best-fit Schechter parameters are summarized in
 Table \ref{tab:dev_exp}.
}
\end{figure}

\clearpage
\begin{figure}[h]
\begin{center}
\includegraphics[scale=0.2]{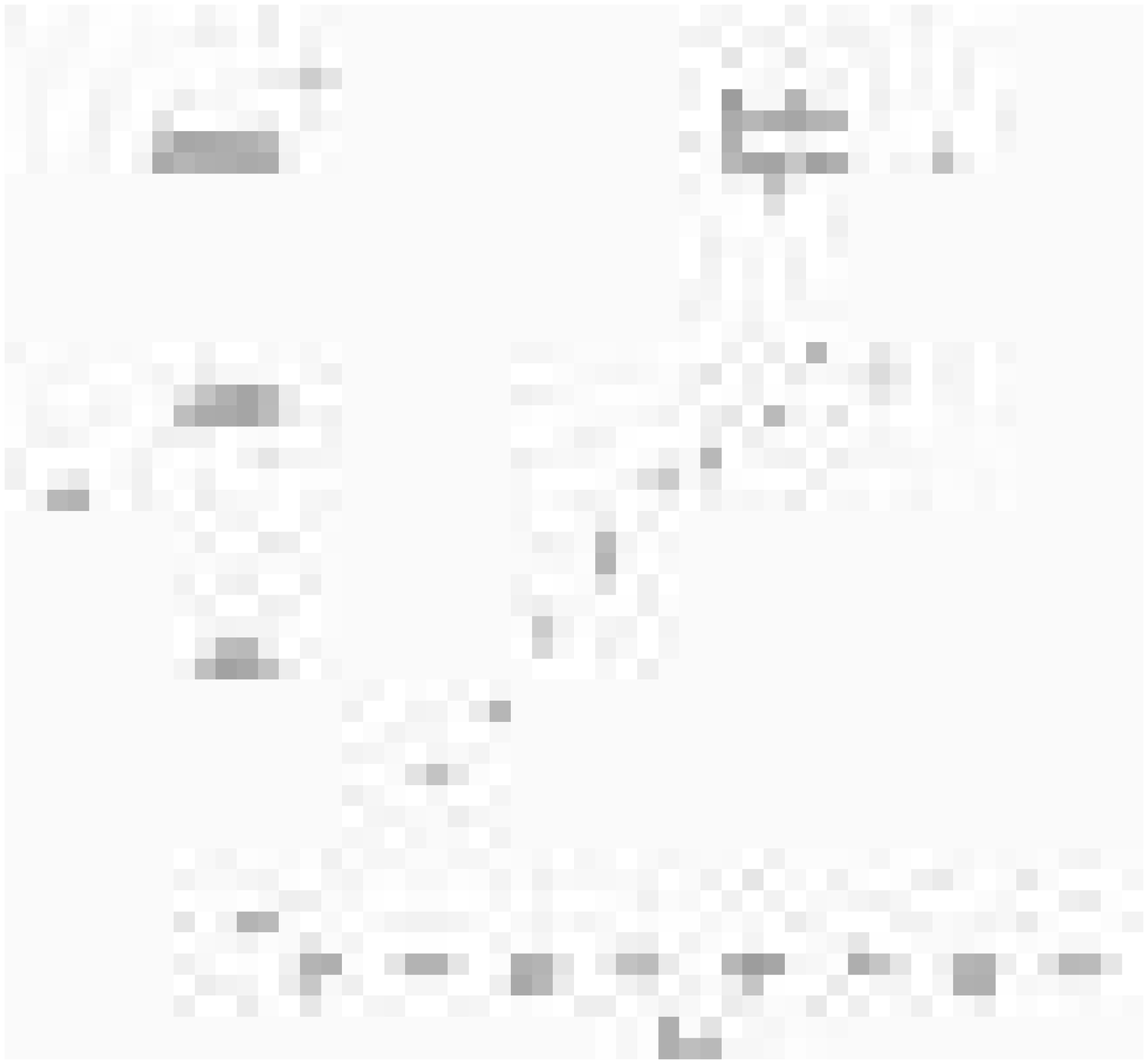}
\includegraphics[scale=0.2]{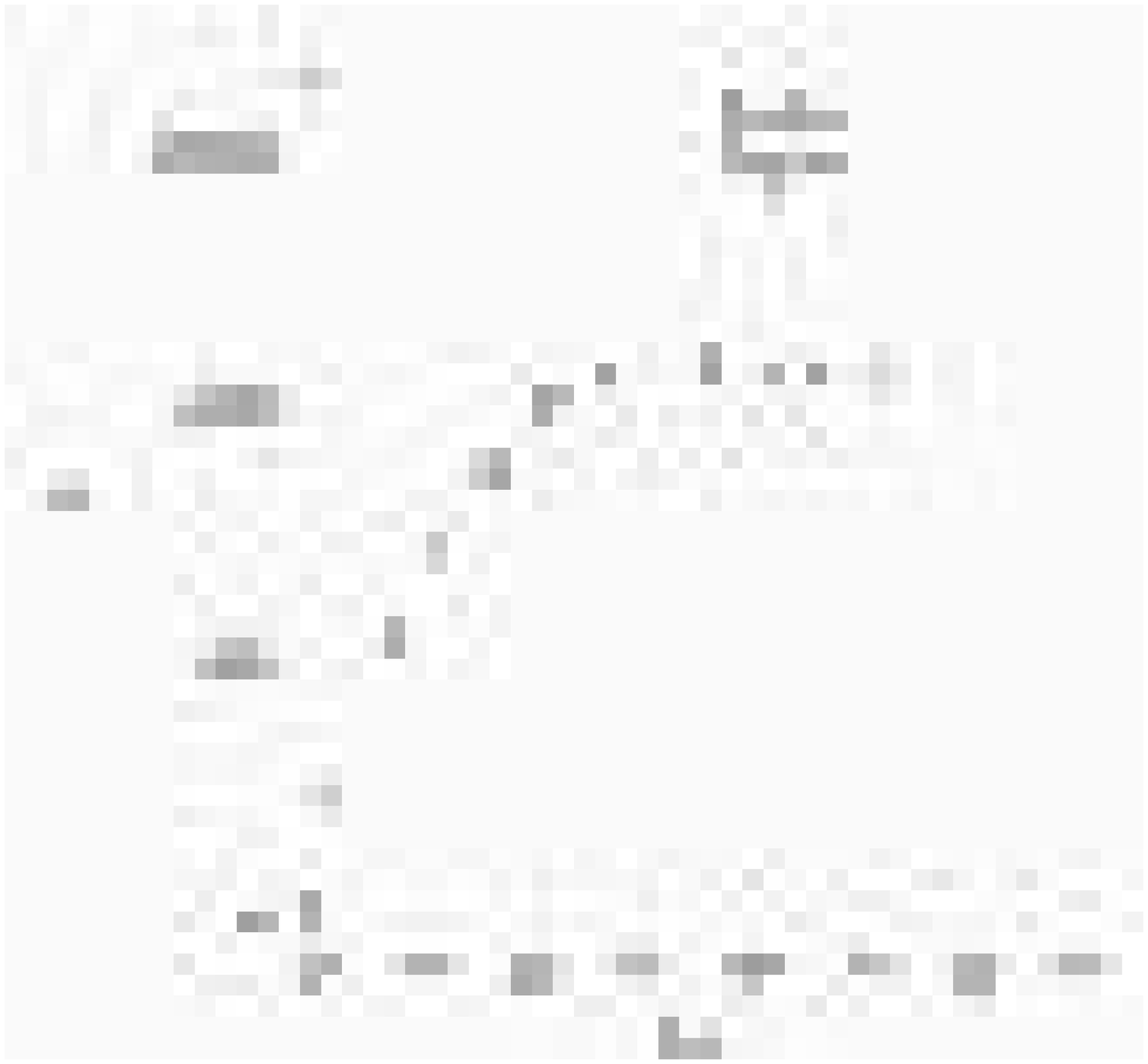}
\includegraphics[scale=0.2]{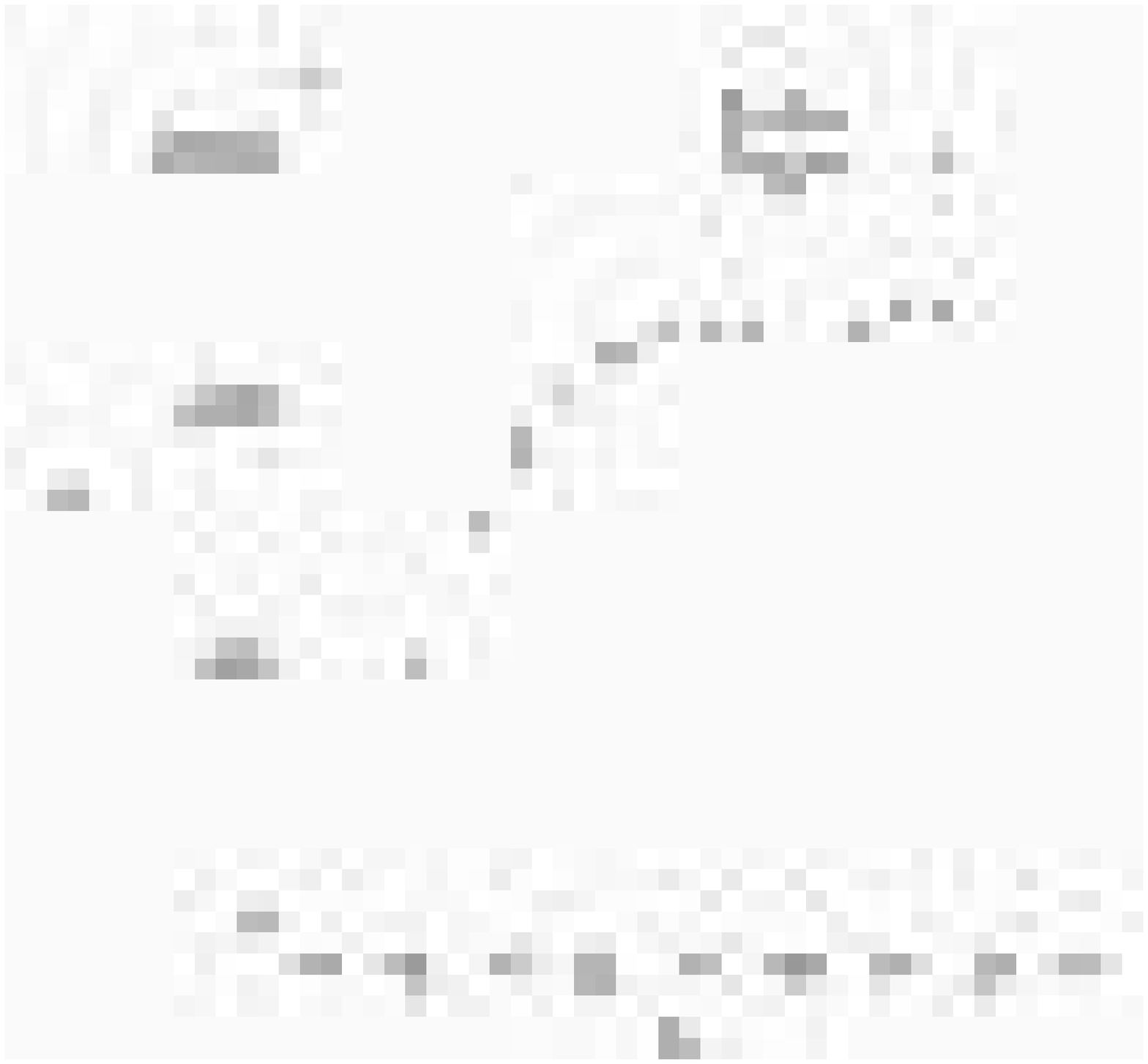}
\end{center}
\begin{center}
\includegraphics[scale=0.2]{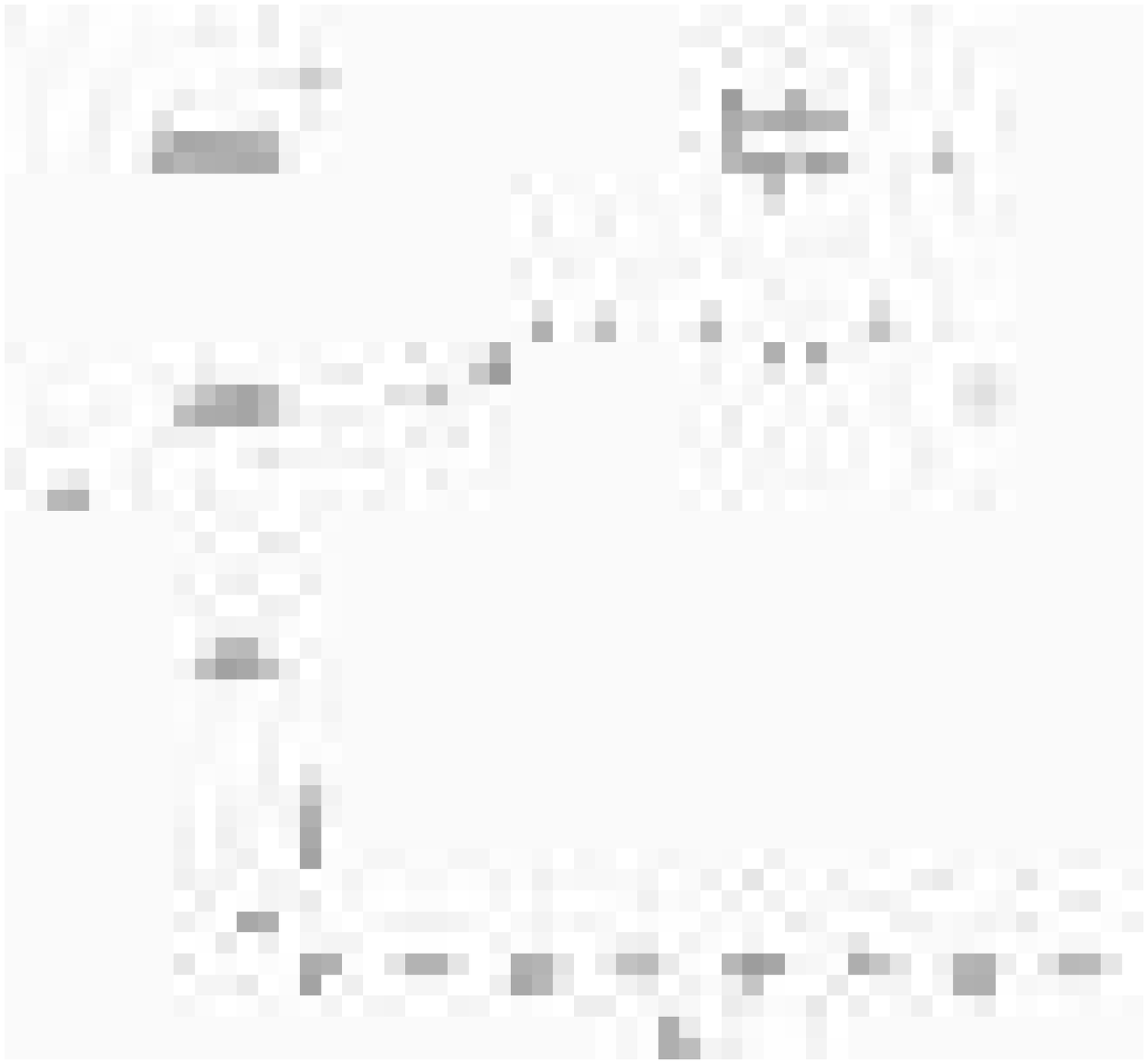}
\includegraphics[scale=0.2]{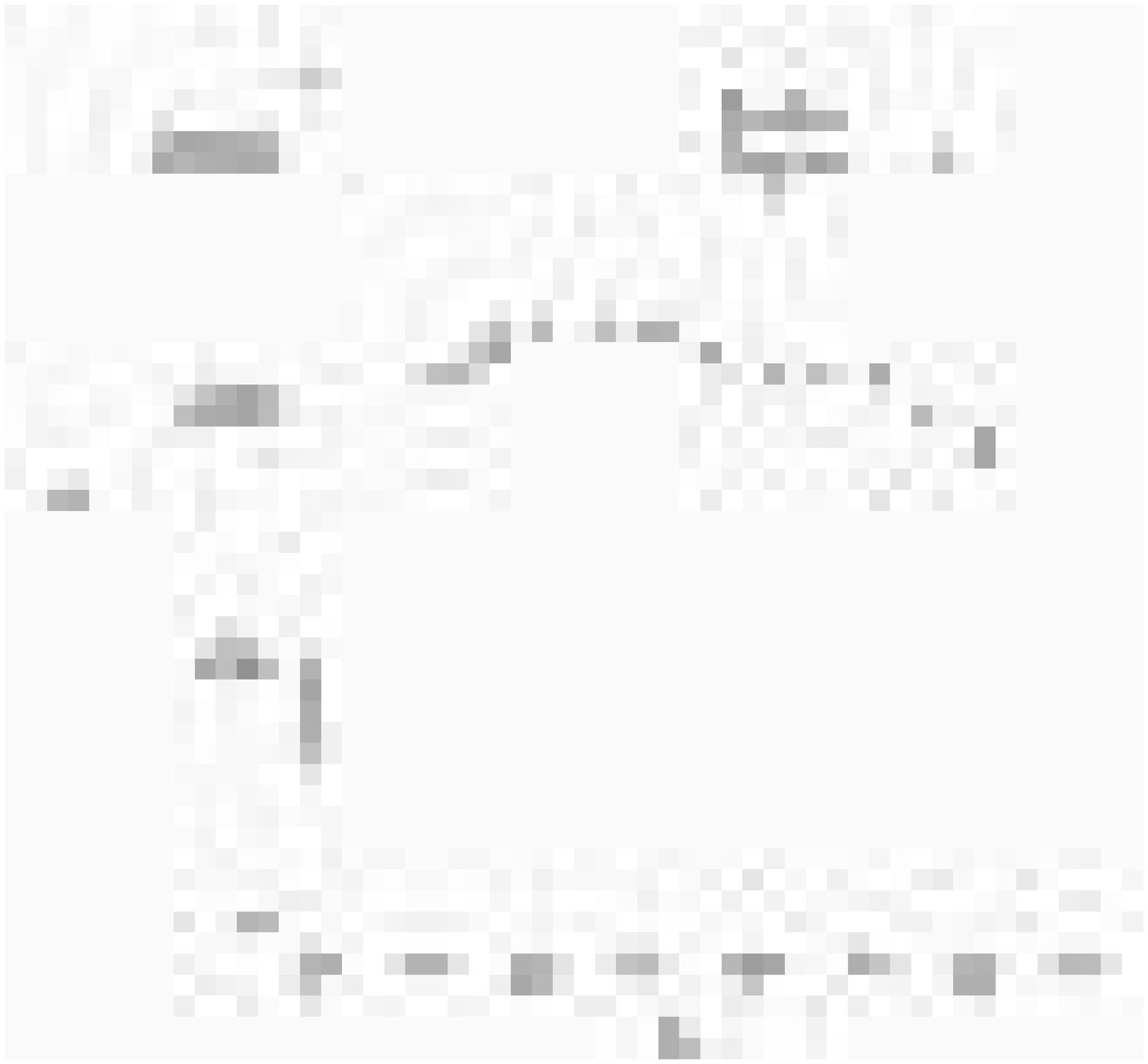}
\end{center}
\caption{
\label{fig:cin.ps}
Composite luminosity functions of high-concentration and low-concentration
 galaxies. The concentration index ($C$) used here is  the ratio of the 50\%
 Petrosian flux radius to the 90\% Petrosian flux radius. 
 In this figure, early-type galaxies have $C<$0.4, and
 late-type galaxies have $C\geq$0.4. Early-type galaxies
 have flatter faint end tails in all five bands. 
 Lines are the best-fit Schechter functions (solid for $C<$0.4, dotted
 for $C\ge$0.4). The y-axis is
 arbitrary. 
 The best-fit Schechter parameters are summarized in
 Table \ref{tab:5color_cin}.
}
\end{figure}

\clearpage

\begin{figure}[h]
\begin{center}
\includegraphics[scale=0.2]{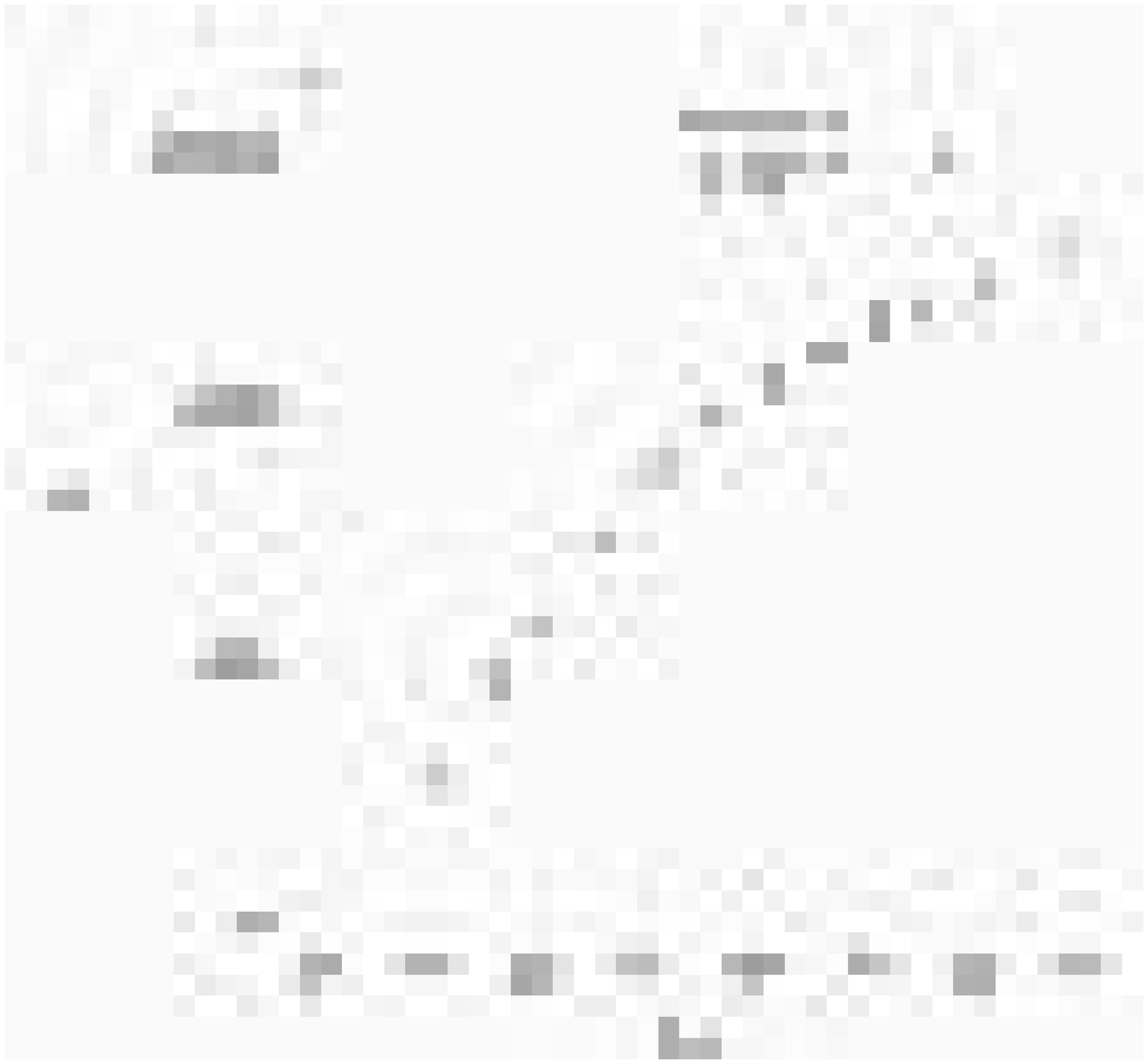}
\includegraphics[scale=0.2]{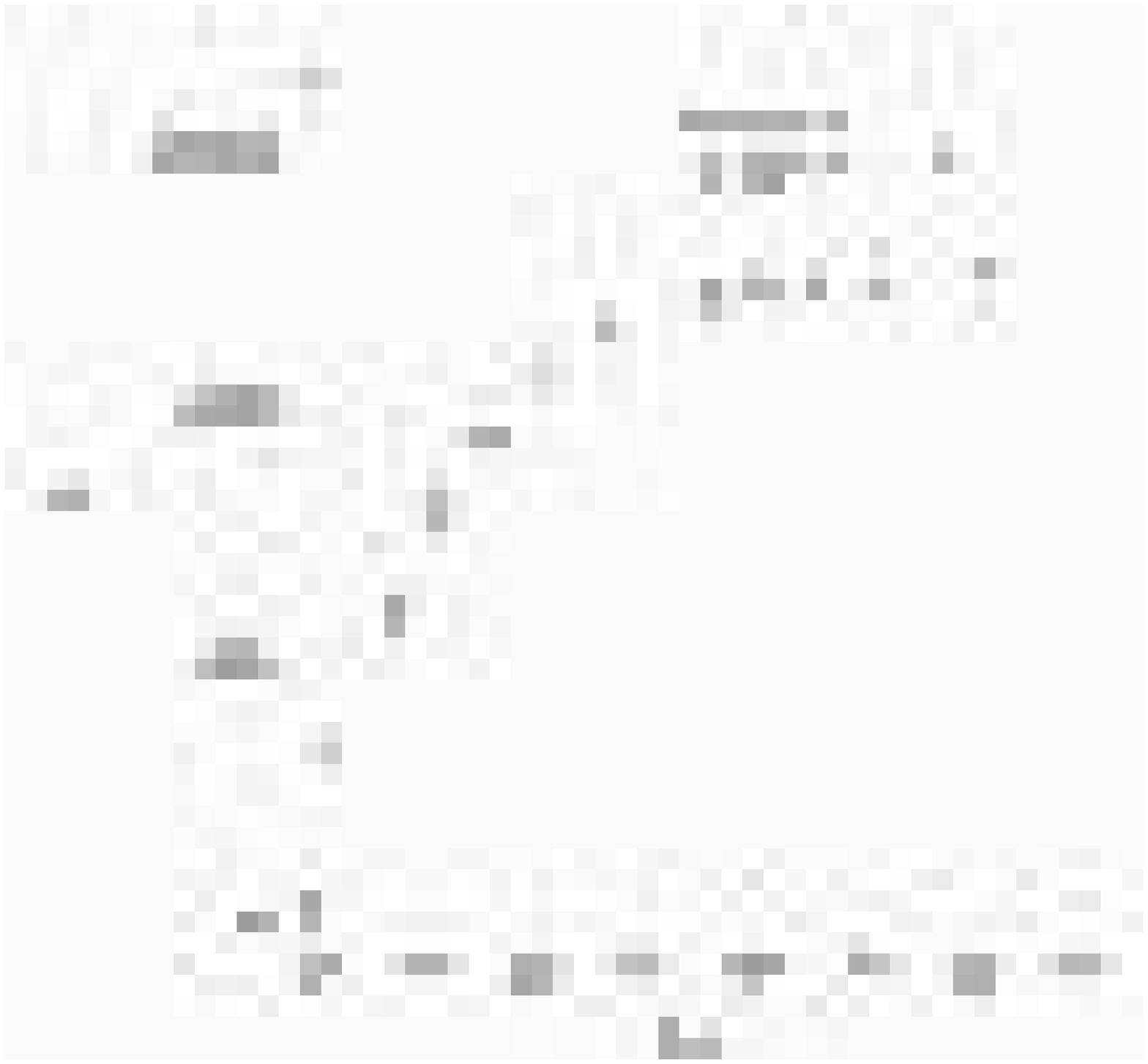}
\includegraphics[scale=0.2]{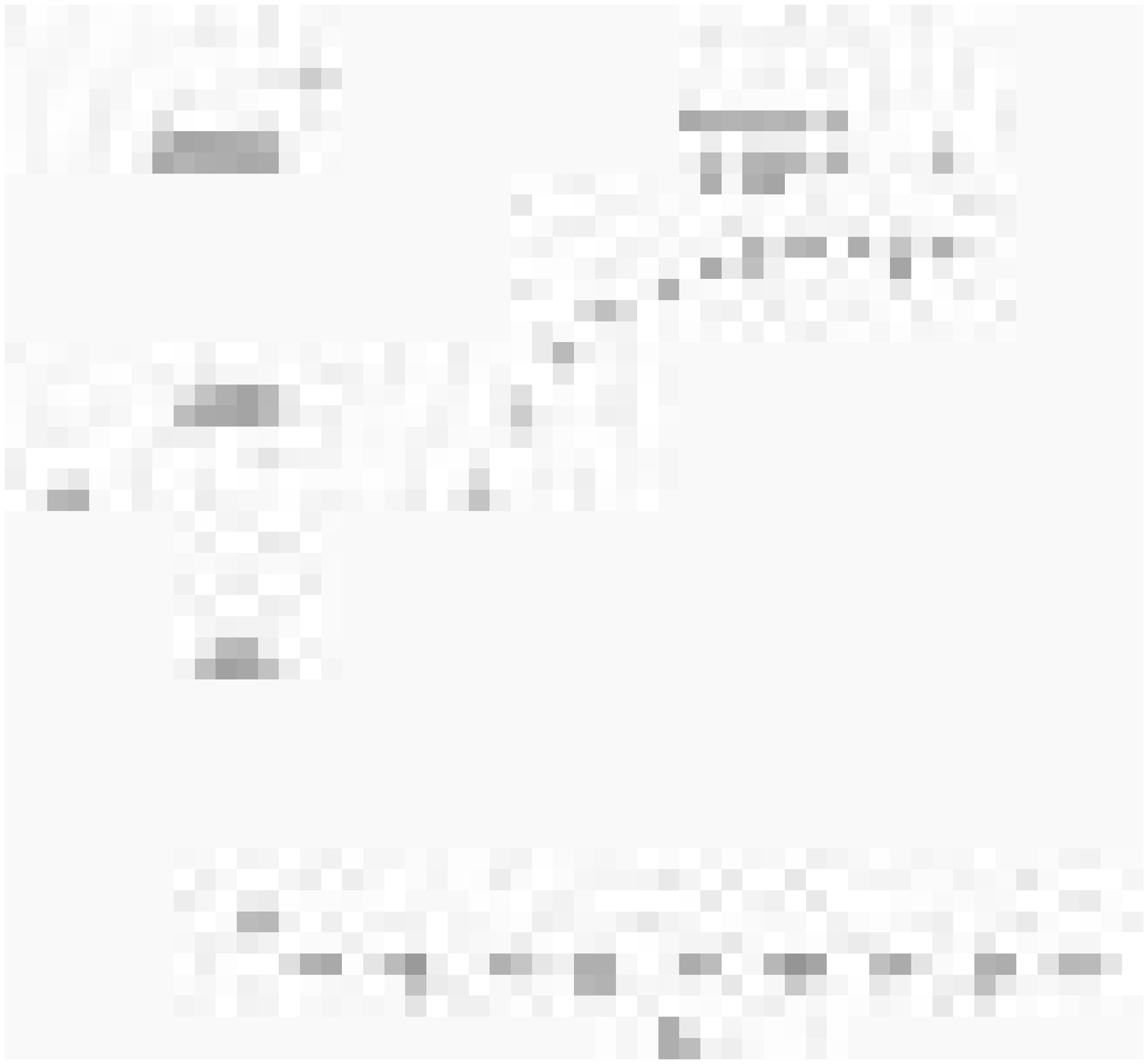}
\end{center}
\begin{center}
\includegraphics[scale=0.2]{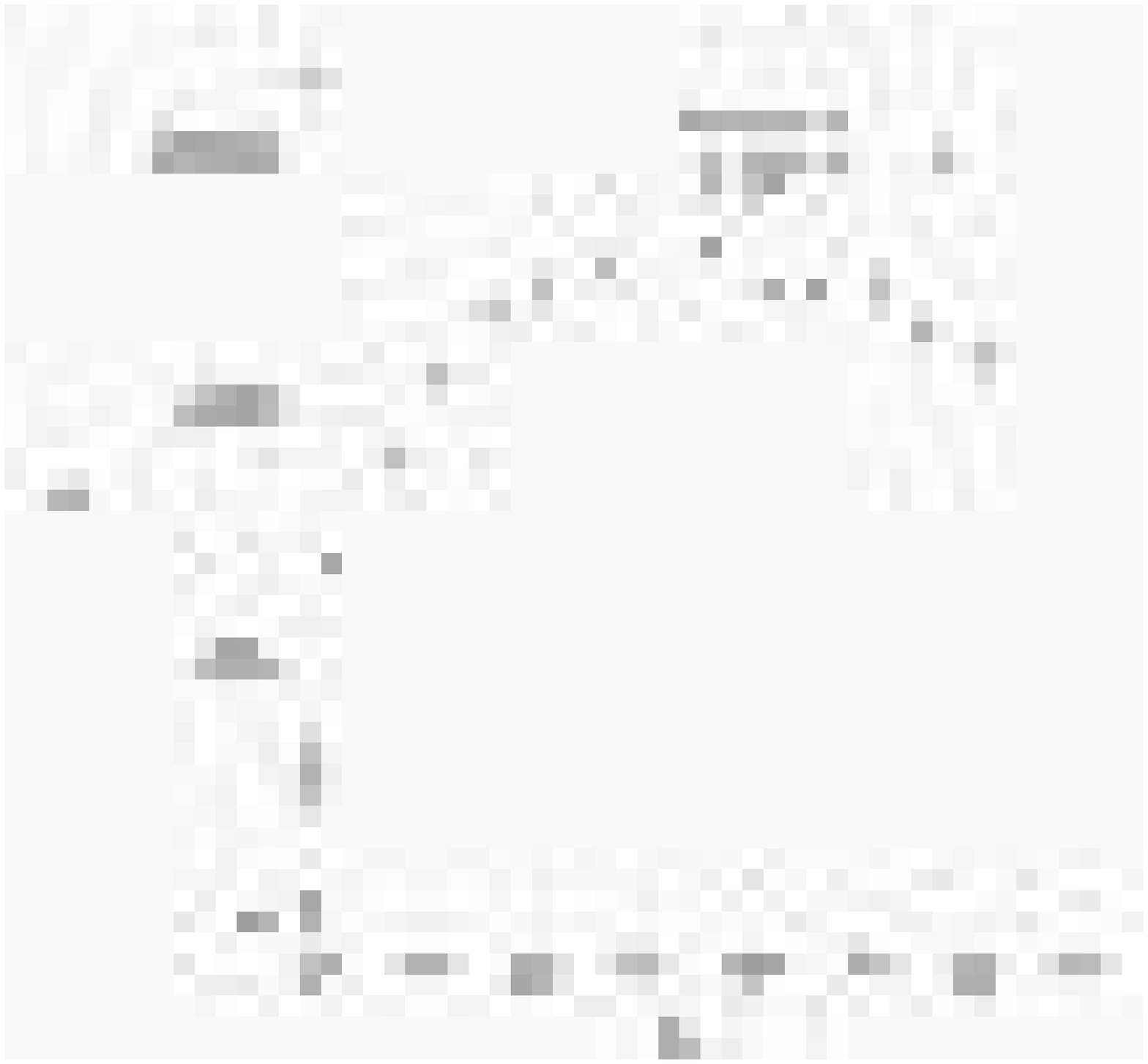}
\includegraphics[scale=0.2]{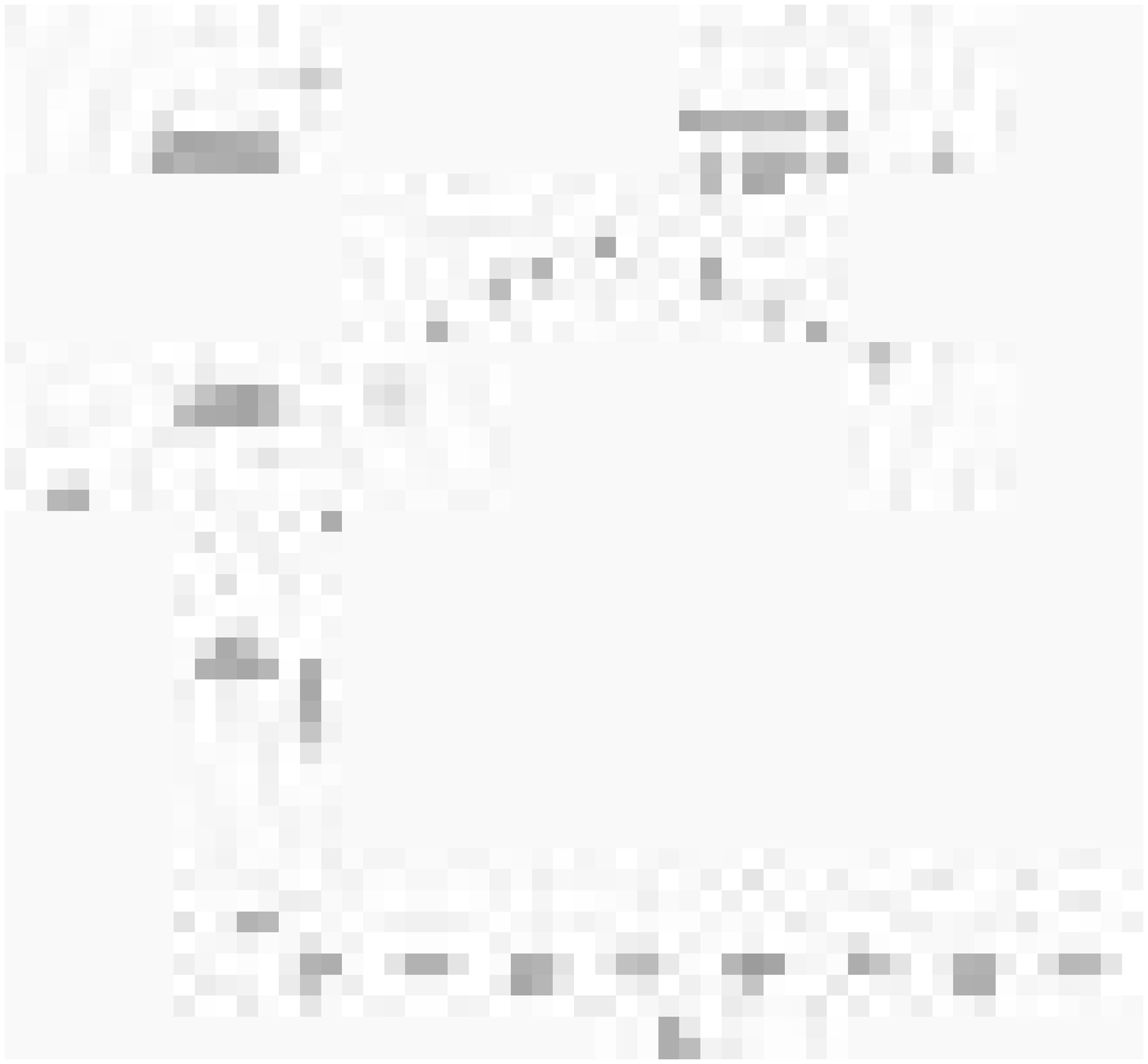}
\end{center}
\caption{
\label{fig:ur.ps}
Composite luminosity functions of $u-r<$ 2.2 (late-type) and
 $u-r\geq$ 2.2 (early-type)
 galaxies. Early-type galaxies
 have flatter faint end tails in all five bands. 
 The lines are the best-fit Schechter functions (solid for $u-r<$ 2.2,
 dotted for $u-r\geq$ 2.2). The y-axis is
 arbitrary.  The best-fit Schechter parameters are summarized in Table \ref{tab:5color_ur}.
}
\end{figure}

\begin{figure}[h]
\begin{center}
\includegraphics[scale=0.7]{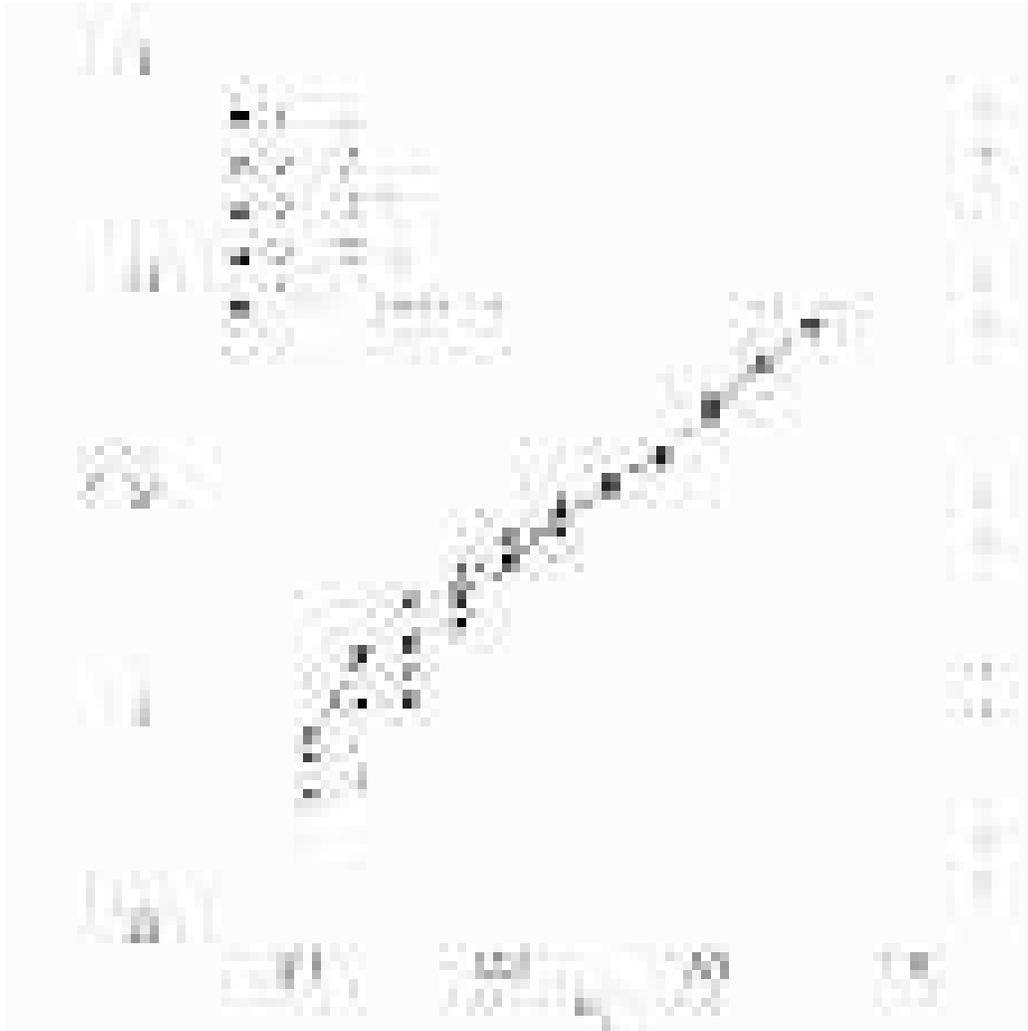}
\end{center}
\caption{
\label{fig:tfake_1577_ab.ps}
 Results of a Monte-Carlo simulation to test the robustness of the
 weighting scheme. 
 The histogram shows the luminosity function of model cluster A 1577.
 The circles, triangles, squares and pentagons represent the
 composite luminosity function at each redshift ($z$ = 0.2, 0.3. 0.4 and
 0.5, respectively) constructed with 100 fake clusters at each
 redshift.  The hexagonals show the composite luminosity function from all 400 fake
 clusters distributed on the real SDSS data. 
}
\end{figure}

\clearpage

\begin{table}[h]
\caption{
 Best-fit Schechter parameters of the composite luminosity function of five
 SDSS bands. The field values are from Blanton et al. (2001), whose
 parameters were shifted to match our cosmology.
 Galaxies within 0.75 Mpc from the cluster center are used. 
}\label{tab:5color}
\begin{center}
\begin{tabular}{lllll}
\hline
Band & $M^*$ & $\alpha$& Field  $M^*$ &  Field $\alpha$\\
\hline
\hline
$u$ & $-$21.61$\pm$0.26 & $-$1.40$\pm$0.11 & $-$19.11$\pm$0.08 & $-$1.35$\pm$0.09    \\
$g$ & $-$22.01$\pm$0.11 & $-$1.00$\pm$0.06 & $-$20.81$\pm$0.04 & $-$1.26$\pm$0.05    \\
$r$ & $-$22.21$\pm$0.05 & $-$0.85$\pm$0.03 & $-$21.60$\pm$0.03 & $-$1.20$\pm$0.03    \\
$i$ & $-$22.31$\pm$0.08 & $-$0.70$\pm$0.05 & $-$22.03$\pm$0.04 & $-$1.25$\pm$0.04    \\
$z$ & $-$22.36$\pm$0.06 & $-$0.58$\pm$0.04 & $-$22.32$\pm$0.05 & $-$1.24$\pm$0.05    \\
$r$(spec) & $-$22.31$\pm$0.13 & $-$0.88$\pm$0.07 & $\cdots$ & $\cdots$   \\
\hline
\end{tabular}
\end{center}
\end{table}

\begin{table}[h]
\caption{
\label{tab:dev_exp}
Best-fit Schechter parameters for de Vaucouleur and exponential galaxies in
 five SDSS bands. The galaxies are divided into two subsamples using profile
 fitting. Galaxies within 0.75 Mpc from the cluster center are used. 
}
\begin{center}
\begin{tabular}{lllll}
\hline
Band & $M^*$ (deV) & $\alpha$ (deV)&  $M^*$(exp) &  $\alpha$ (exp)\\
\hline
\hline
$u$ & $-$21.64$\pm$0.30 &  $-$1.41$\pm$0.12 &  $-$21.45$\pm$0.13 & $-$1.27$\pm$0.07    \\
$g$ & $-$21.92$\pm$0.11 & $-$0.73$\pm$0.07 & $-$21.89$\pm$0.13 & $-$1.20$\pm$0.06   \\
$r$ & $-$22.01$\pm$0.07 & $-$0.37$\pm$0.06 & $-$21.73$\pm$0.12 & $-$1.04$\pm$0.06     \\
$i$ & $-$22.13$\pm$0.07 & $-$0.25$\pm$0.06 & $-$21.69$\pm$0.13 & $-$0.80$\pm$0.08    \\
$z$ & $-$22.24$\pm$0.06 & +0.12$\pm$0.06 & $-$21.76$\pm$0.11 & $-$0.65$\pm$0.07    \\
\hline
\end{tabular}
\end{center}
\end{table}

\begin{table}[h]
\caption{
\label{tab:5color_cin}
Best-fit Schechter parameters for low concentration (early-type) and high
 concentration (late-type) galaxies in
 five SDSS bands. The concentration index here is  the ratio of 50\%
 Petrosian flux radius to 90\% Petrosian flux radius. 
 Early-type galaxies have a concentration of $<$0.4, and
 late-type galaxies have a concentration of $\geq$0.4.  Galaxies within
 0.75 Mpc from the cluster center are used. 
}
\begin{center}
\begin{tabular}{lllll}
\hline
Band & $M^*$ (Early) & $\alpha$ (Early)&  $M^*$(Late) &  $\alpha$ (Late)\\
\hline
\hline
$u$ &  $-$21.42$\pm$0.24 & $-$1.28$\pm$0.12 & $-$21.82$\pm$0.11 & $-$1.42$\pm$0.06    \\
$g$ &  $-$22.05$\pm$0.11 & $-$0.89$\pm$0.07 & $-$22.26$\pm$0.11 & $-$1.36$\pm$0.05    \\
$r$ & $-$22.31$\pm$0.06 & $-$0.92$\pm$0.04 & $-$22.24$\pm$0.12 & $-$1.32$\pm$0.06     \\
$i$ & $-$21.97$\pm$0.09 & $-$0.59$\pm$0.10 &    $-$22.02$\pm$0.13 & $-$1.04$\pm$0.08    \\
$z$ & $-$22.08$\pm$0.09 & $-$0.47$\pm$0.09 &    $-$22.09$\pm$0.12 & $-$0.87$\pm$0.07    \\
\hline
\end{tabular}
\end{center}
\end{table}

\begin{table}[h]
\caption{
\label{tab:5color_ur}
Best-fit Schechter parameters for $u-r>$ 2.2 (early type) and $u-r\leq$
 2.2 (late type) galaxies in
 five SDSS bands. Galaxies within 0.75 Mpc are used. 
}
\begin{center}
\begin{tabular}{lllll}
\hline
Band & $M^*$ (Early) & $\alpha$ (Early)&  $M^*$(Late) &  $\alpha$ (Late)\\
\hline
\hline
$u$ &  $-$21.65$\pm$0.26 & $-$1.47$\pm$0.11 &  $-$21.78$\pm$0.13 &  $-$1.37$\pm$0.07    \\
$g$ & $-$22.04$\pm$0.10 & $-$1.03$\pm$0.06 & $-$22.30$\pm$0.09 & $-$1.38$\pm$0.05    \\
$r$ & $-$22.29$\pm$0.04 & $-$0.97$\pm$0.02 & $-$22.22$\pm$0.12 & $-$1.41$\pm$0.06     \\
$i$ & $-$21.91$\pm$0.08 & $-$0.58$\pm$0.07 & $-$22.17$\pm$0.16 &  $-$1.23$\pm$0.08    \\
$z$ & $-$21.93$\pm$0.07 & $-$0.36$\pm$0.08 & $-$22.14$\pm$0.19 & $-$1.08$\pm$0.09    \\
\hline
\end{tabular}
\end{center}
\end{table}

%

\begin{table}[h]
\caption{
\label{tab:bcg_center}
Best-fit Schechter parameters for galaxies using positions of brightest cluster
 galaxies as a  center in five SDSS bands. The mean deviation from the CE
 center  is 1.02 arcmin.
}
\begin{center}
\begin{tabular}{lll}
\hline
Band & $M^*$  & $\alpha$ \\
\hline
\hline
$u$ & $-$21.84$\pm$0.16 & $-$1.43$\pm$0.07\\
$g$ & $-$22.16$\pm$0.15 & $-$1.05$\pm$0.07 \\
$r$ & $-$22.29$\pm$0.05 & $-$0.91$\pm$0.03 \\
$i$ & $-$22.31$\pm$0.06 & $-$0.73$\pm$0.03 \\
$z$ & $-$22.18$\pm$0.07 & $-$0.55$\pm$0.07 \\
\hline
\end{tabular}
\end{center}
\end{table}

\begin{table}[h]
\caption{
\label{tab:global}
 Best-fit Schechter parameters for galaxies using global background
 subtraction  in five SDSS bands. Instead of the annuli around the
 cluster, the global background was used to subtract the background
 galaxies to see the dependence on the background subtraction. 
}
\begin{center}
\begin{tabular}{lll}
\hline
Band & $M^*$  & $\alpha$ \\
\hline
\hline
$u$ &  $-$21.77$\pm$0.17 & $-$1.47$\pm$0.07 \\
$g$ &  $-$22.01$\pm$0.12 & $-$1.06$\pm$0.07 \\
$r$ &  $-$22.20$\pm$0.05 & $-$0.90$\pm$0.03 \\
$i$ &  $-$22.24$\pm$0.07 & $-$0.72$\pm$0.04 \\
$z$ &  $-$22.10$\pm$0.06 & $-$0.50$\pm$0.06 \\
\hline
\end{tabular}
\end{center}
\end{table}

\begin{table}[h]
\caption{
\label{tab:richness}
 Best-fit Schechter parameters in
 the $r$ band for galaxies using richer systems.
 The best-fit Schechter parameters for $N_{-18}>$20 and  
 $N_{-18}>$40  subsamples are shown. $N_{-18}$ is defined as
 the number of galaxies brighter than $-$18th magnitude after subtracting the background.
}
\begin{center}
\begin{tabular}{llll}
\hline
Band & $M^*$  & $\alpha$ & $N$(cluster) \\
\hline
\hline
$N_{-18}$ $>$20 &  $-$22.21$\pm$0.05 & $-$0.85$\pm$0.03 & 204  \\
$N_{-18}$ $>$40 &  $-$22.29$\pm$0.06 & $-$0.90$\pm$0.04 & 120  \\
\hline
\end{tabular}
\end{center}
\end{table}

\begin{table}[h]
\caption{\label{tab:previous}
 Comparison with previous studies on the composite luminosity function. The CE
 composite LFs (this work) was re-calculated using each
 author's cosmology. The magnitude was transformed using data from Fukugita et
 al. (1995) and Lumsden et al. (1992).
}
\begin{center}
\begin{tabular}{llllll}
\hline
Paper & $M^*$ & $\alpha$ & Band & Ncluster & Cosmology  \\
\hline
\hline 
CE   & 	$-$22.21$\pm$0.05  &   $-$0.85$\pm$0.03 & $r$ & 204 &
 $\Omega_{\rm{M}}$=0.3 $\Omega_{\Lambda}$=0.7 $H_0$=70 \\
\hline
Colless 89 &	$-$20.04 &	$-$1.21  &	$bj$ & 14
 rich & 	$H_0$=100 $q_0$=1	\\
(CE)  &	$-$21.58$\pm$0.12	& $-$0.93$\pm$0.06 & $bj$ & 204 &
 $H_0$=100 $q_0$=1	\\
 (CE) &	  $-$22.20$\pm$0.12	  &	$-$1.21 fixed  &	$bj$
 & 204 & $H_0$=100 $q_0$=1 	\\
\hline
Lugger 89 &	$-$22.81$\pm$0.13 	& $-$1.21$\pm$0.09  &  $R$ (PDS)   &  9 &	$H_0$=50	\\
(CE)  &	$-$22.49$\pm$0.06	& $-$0.69$\pm$ 0.05 & $R$ (PDS)  &  204 &
 $H_0$=50 $q_0$=0.5	\\
(CE)  &	$-$22.77$\pm$0.17 & $-$1.21 fixed & $R$ (PDS)  &  204 &	$H_0$=50 $q_0$=0.5	\\
\hline
Valotto 97 &	$-$20.0$\pm$0.1 	& $-$1.4$\pm$0.1  &  $bj$   &  55 Abell APM &	$H_0$=100 \\
(CE)  &	$-$21.58$\pm$0.12	& $-$0.93$\pm$0.06 & $bj$   &  204 & $H_0$=100 $q_0$=1	\\
(CE)  &	$-$22.69$\pm$0.23 & $-$1.4 fixed & $bj$  &  204 &	$H_0$=100 $q_0$=1	\\
\hline
Lumsden 97 &	$-$20.16$\pm$0.02  &	$-$1.22$\pm$0.04  &	$bj$
 & 22 rich & $H_0$=100 $q_0$=1 	\\
 (CE) &	  $-$21.58$\pm$0.12	  &	$-$0.93$\pm$0.06  &	$bj$	 & 204 &
 $H_0$=100 $q_0$=1 \\
 (CE) &	  $-$22.22$\pm$0.10	  &	$-$1.22 fixed  &	$bj$
 & 204 & $H_0$=100  $q_0$=1	\\
\hline
Garilli 99 & $-$22.16$\pm$0.15 & $-$0.95$\pm$0.07 & $r$ (CCD)  & 65 Abell
 X-ray & $H_0$=50 $q_0$=0.5 \\
 (CE) &	 $-$22.15$\pm$0.06 & $-$0.69$\pm$ 0.05 &	$r$ (CCD)   & 204 & $H_0$=50 $q_0$=0.5 \\
 (CE) &	 $-$22.28$\pm$0.05 & $-$0.84 fixed &	$r$ (CCD)   & 204 & $H_0$=50 $q_0$=0.5 \\
\hline
Paolillo 00 &	$-$22.26$\pm$0.16 	& $-$1.11  &  $r$ (POSSII)  &  39 Abell &	$H_0$=50 $q_0$=0.5	\\
(CE)  &	$-$22.15$\pm$0.06	& $-$0.69$\pm$ 0.05 & $r$ (POSSII)  &  204 &
 $H_0$=50 $q_0$=0.5	\\
(CE)  &	$-$22.55$\pm$0.12 & $-$1.11 fixed & $r$ (POSSII)  &  204 &	$H_0$=50 $q_0$=0.5	\\
\hline
Yagi 02 &	$-$21.3$\pm$0.2 	& $-$1.31$\pm$0.05  &  $R_C$  &  10 Abell &	$H_0$=100 $q_0$=0.5	\\
(CE)  &	$-$21.89$\pm$0.10 & $-$1.03$\pm$ 0.05 & $R_C$   &  204 &
 $H_0$=100 $q_0$=0.5	\\
(CE)  &	$-$22.55$\pm$0.14   & $-$1.31 fixed & $R_C$  &  204 &  $H_0$=100 $q_0$=0.5	\\
\hline
\end{tabular}
\end{center}
\end{table}

\clearpage


\chapter{The Morphological Butcher-Oemler Effect}
\label{chap:BO}

\section{Introduction}\label{bo_intro}
  
 The Butcher-Oemler effect was first reported by Butcher \& Oemler
 (1978, 1984) as an increase in the fraction of blue galaxies ($f_b$) toward
 higher redshift in 33 galaxy clusters  over the redshift range
 0$<z<$0.54.
 Butcher and Oemler's work made a strong impact since it showed direct evidence for the
 evolution of cluster galaxies. Much work regarding the nature of
 these blue galaxies followed. Rakos \& Schombert (1995) found that the
 fraction of blue galaxies increases from 20\% at $z=$0.4 to 80\% at
 $z=$0.9, suggesting that the evolution in clusters is even stronger than
 previously thought. Margoniner \& De Carvalho (2000) studied 48
 clusters in the redshift range of 0.03$<z<$0.38 and detected a strong
 Butcher-Oemler effect consistent with  that of Rakos \& Schombert (1995). 
 Despite the trend with redshift,  almost all previous work has reported 
 a wide range of blue fraction values at a fixed redshift. 
 In particular, in a large sample of 295 Abell clusters, Margoniner et al. (2001)
 not only confirmed the existence of the Butcher-Oemler effect,
 but also found the blue fraction depends on cluster richness.

 Although the detection of the Butcher-Oemler effect has been claimed in
 various studies, 
 there have been some suggestions of strong selection biases in the cluster samples.
 Newberry, Kirshner \& Boroson (1988) measured velocity
 dispersions and surface densities of galaxies in clusters and found a marked difference between local clusters and intermediate
 redshift clusters. More recently, Andreon \& Ettori (1999)
 measured X-ray surface brightness profiles, sizes, and luminosities of the
 Butcher-Oemler sample of clusters and concluded that the sample is not
 uniform. The selection bias, thus, could mimic evolutionary
 effects. Smail et al. (1998) used 10 X-ray bright clusters in the
 redshift range of 0.22$\leq z\leq$0.28 and found
 that the clusters have only a small fraction of blue galaxies.  The
 Butcher-Oemler effect was not observed with their sample. 
 Similarly, galaxies in radio selected groups are not significantly
 bluer at higher redshifts (Allington-Smith et al. 1993). 
 Garilli et al. (1996) observed 67 Abell and X-ray selected clusters and
 found no detectable Butcher-Oemler effect at $z<$0.2. Fairley et
 al. (2002) studied eight X-ray selected galaxy clusters and found no
 correlation of blue fraction with redshift.   Rakos \& Schombert (1995)'s
 sample was selected from the catalog compiled by Gunn, Hoessel \& Oke
 (1986) using photographic plates taken only in two color bands.
 The sample
 thus have a possible bias against red, high redshift clusters.
 In addition to the possible sample selection biases, with the exception of Margoniner et
 al. (2001), the number of clusters in the previous works was small, consisting of a few to dozens of clusters. Therefore the statistical
 uncertainty was large. Many authors also noted
 that cluster-to-cluster variation of the fraction of blue galaxies is considerable.
 The need for a larger, more uniform sample of clusters has been evident. 

 There have been various attempts to find another
 physical mechanism causing the large scatter which has been seen in
 almost all previous work.
 Wang \& Ulmer (1997) claimed the existence of a correlation between the blue fraction
 and the ellipticity of the cluster X-ray emissions in their sample of
 clusters at 0.15$\leq z\leq$0.6. Metevier, Romer \& Ulmer (2000) showed that two clusters with a bimodal X-ray surface brightness profile have an unusually high blue fraction
 value and thus do not follow the typical Butcher-Oemler relation. They
 claimed that the Butcher-Oemler effect is an environmental phenomenon
 as well as an evolutionary phenomenon.  Margoniner et al. (2001) found
 a richness dependence in the sense that richer
 clusters have smaller blue fractions.  They claimed that this
 richness dependence causes a large scatter in the blue fraction--redshift
 diagram. Therefore, it is of extreme interest to explore an origin
 of the scatter in the blue fraction despite the redshift trend.

 At the same time,  various studies using morphological information have
 reported a similar evolutionary effect in cluster galaxies. 
 Dressler et al. (1997) studied 10 clusters at
 0.37$<z<$0.56 and found a steep increase in the S0 fraction toward $lower$
 redshift, compared to nearby clusters studied earlier (Dressler 1980). 
 Couch et al. (1994,1998) studied three clusters at $z=$0.31 and found
 their S0 fraction to be consistent with the 
 trend observed by Dressler et
 al. (1997). Fasano et al. (2000) observed nine clusters at intermediate
 redshifts (0.09$<z<$0.26) and also found an increase in the S0 fraction
 toward lower redshift. 
 It has been proposed that the increase in the S0 fraction is caused by the
 transformation of spiral galaxies into S0 galaxies through a process
 yet unknown. These studies, however, need to be pursued further, considering that most
 of the previous work was based on morphological galaxy
 classification by eye. Although it is an excellent tool to
 classify galaxies, manual classification could potentially have unknown
 biases. (A detailed comparison of human classifiers can be found in
 Lahav et al. 1995). 
 A machine based, automated classification 
 would better control biases and would allow a reliable determination of
 the completeness and false positive rate. A further reason to
 investigate the evolution of cluster galaxies is the sample size. 
 The morphological fraction studies of clusters reported so
 far are based on only dozens of clusters. Furthermore, the clusters
 themselves have intrinsic variety in the fraction of blue/spiral
 galaxies as reported by various Butcher-Oemler and morphological analyses
 listed above.  Since several authors have  suggested that the fraction of
 blue galaxies depends on cluster richness, it is important
 to use a uniform cluster sample, preferably selected by an automated
 method with a well known selection function.

 Various theoretical models have been proposed to explain the Butcher-Oemler
 effect and the increase of the S0 fraction. These models include ram pressure stripping
 of gas (Spitzer \& Baade 1951; Gunn \& Gott 1972; Farouki \& Shapiro
 1980; Kent 1981; Abadi, Moore \& Bower 1999; Fujita \& Nagashima 1999;
 Quilis, Moore \& Bower
 2000), galaxy infall (Bothun \& Dressler 1986; Abraham et al. 1996a; Ellingson et al. 2001),
 galaxy harassment (Moore et al. 1996, 1999), cluster tidal forces (Byrd
 \& Valtonen 1990; Valluri 1993),  enhanced star formation (Dressler \&
 Gunn 1992),
 and 
  removal and consumption of the gas (Larson,
 Tinsley \& Caldwell 1980; Balogh et al. 2001; Bekki et al. 2002). It is, however, yet
 unknown exactly what processes play major roles in changing the color and
 morphology of cluster galaxies. 
 To derive a clear picture explaining 
 the evolution of cluster galaxies, it is important to clarify both the
 Butcher-Oemler effect and the S0 increase, using a large and uniform
 cluster sample in conjunction with a machine based morphological classification. 

 With the advent of the Sloan Digital Sky Survey (SDSS; York et al. 2000),
 which is an imaging and spectroscopic survey of 10,000 deg$^2$ of the
 sky, we now have the opportunity to overcome these limitations. The SDSS Cut
 \& Enhance galaxy cluster catalog (Goto et al. 2002a; Chapter \ref{chap:CE}) provides a large
 uniform cluster catalog with a well defined selection function.
 The CCD-based, accurate photometry of the SDSS (Fukugita
 et al. 1996; Hogg et al. 2002; Smith et al. 2002) and the wide coverage of the SDSS
 on the sky allow accurate estimation of blue fraction with robust
 local background subtraction. Although the SDSS is a ground based
 observation, the state-of-the-art reduction software and the accuracy
 of CCD data make it possible to derive morphological classification in an
 automated way  (Lupton et al. 2001, 2002). By using the SDSS data
 set, we are able to study one of the largest samples to date --- 514 clusters --- to
 the depth of $M_{r^*}=-$19.44 ($h$=0.75).  

 The purpose of this chapter is as follows. We aim to confirm or disprove the existence
  of the Butcher-Oemler effect using one of the largest, most 
 uniform cluster samples. At the same time, we
 hope to shed light on the morphological properties of the Butcher-Oemler
 galaxies using morphological parameters derived from the SDSS data. 
 Finally we investigate the origin of the scatter in the galaxy type
 fraction versus redshift
 relation, in hope of gaining some understanding
 about the physical processes responsible for the scatter. 

 Since it is known that field galaxies also evolve both
 morphologically (Schade et al. 1996;  Brinchmann et al. 1998; Lilly et
 al. 1998; Kajisawa et al. 2001; Abraham et al. 2001) and
 spectroscopically 
 (Madau et al. 1996; Lilly et al.1996; Hammer et al. 1997; Treyer et al. 1998;  Cowie et al. 1999; Sullivan et al. 2000; Wilson et al. 2002),   it is of extreme
 importance to compare the evolution of cluster galaxies with that of field galaxies to
 specify a responsible physical mechanism. 
 It is  possible that the Butcher-Oemler effect and morphological
 transition of cluster galaxies are more commonly happening including
 the field region of the universe, thus a cluster specific mechanism is
 not responsible for the evolution of galaxies. However, since the SDSS spectroscopic data
 are not deep enough to probe cosmologically interesting time scale, we
 leave it to future work. 
  
  The Chapter is organized as follows: In
 Section \ref{bo_data}, we describe the SDSS data and the Cut \& Enhance cluster
 catalog. In Section \ref{bo_method}, we analyze the late type fraction,
 both spectrally and morphologically. In Section \ref{bo_discussion}, we discuss
 the possible caveats and underlying physical processes in the evolution of galaxies.  In
 Section \ref{bo_conclusion}, we summarize our work and findings.
 The cosmological parameters adopted throughout this chapter are $H_0$=75 km
 s$^{-1}$ Mpc$^{-1}$, and ($\Omega_m$,$\Omega_{\Lambda}$,$\Omega_k$)=(0.3,0.7,0.0).

\section{Data}\label{bo_data}

  The galaxy catalog used here is taken from the Sloan Digital Sky Survey (SDSS)
 Early Data Release (see Fukugita et al. 1996; Gunn et al. 1998;  Lupton
 et al. 1999, 2001, 2002; York et al. 2000; Hogg et al. 2001; Pier et al. 2002; Stoughton et al. 2002 and Smith et al. 2002 for more detail of the SDSS data). 
 We use equatorial scan data, a contiguous area of 250 deg$^2$
 (145.1$<$RA$<$236.0, $-$1.25$<$DEC$<$+1.25) and 150 deg$^2$
 (350.5$<$RA$<$56.51, $-$1.25$<$DEC$<$+1.25).
 The SDSS imaging survey observed the region to depths of 22.3, 23.3, 23.1, 22.3
 and 20.8 in the $u,g,r,i$ and $z$ filters, respectively. (See
 Fukugita et al. 1996 for the SDSS filter system; Hogg et al. 2002
 and Smith et al. 2002 for its calibration). 
 Since the SDSS photometric system is not yet finalized, we refer to the
 SDSS photometry presented here as $u^*,g^*,r^*,i^*$ and $z^*$. We
 correct the data for galactic extinction determined from the maps given
 by Schlegel, Finkbeiner \& Davis (1998).  
 We include galaxies to $r^*$=21.5 (Petrosian magnitude), which is the
 star/galaxy separation limit (studied in detail by Scranton et
 al. 2002) in the SDSS data. 

 The galaxy cluster catalog used here is a subset of the SDSS Cut \& Enhance
 galaxy cluster catalog (Goto et al. 2002a; Chapter \ref{chap:CE}). There are 4638 clusters in
 the equatorial region (See Kim et al. 2003;  Bahcall
 et al. 2003; Annis et al. in prep. and Miller et al. in prep. for other works on the SDSS
 galaxy clusters). Besides the uniformity of the catalog with its well
 defined selection function, the catalog has very good photometric redshifts,
 $\delta z$=0.015 at $z<$0.3, which enables us to use a large sample
 of clusters (Goto et al. 2002a; Chapter \ref{chap:CE}; see also Gal et al. 2000 and Annis et
 al. 2003 for photometric redshift methods for clusters). We use
 clusters in the redshift range of 0.02$\leq z\leq$0.3 
 and galaxies brighter than $M_{r^*}=-19.44$, which corresponds to $r^*$=21.5
 at $z=$0.3. 
 Since several authors in previous work claimed that
  biases in sample cluster selection can mimic the evolutionary effect, it
 is important to control cluster richness well. We use clusters with
 more than 25 member galaxies between $M_{r^*}=-24.0$ and  $-19.44$
 within 0.7 Mpc from the cluster center 
 after fore/background subtraction, as explained in the next
 Section. The large areal coverage of the SDSS data 
 enables us to subtract the fore/background counts reliably. We can thus
  control the richness of the sample clusters well. The criteria
 leave 514 clusters in the region.  

 We stress the importance of the uniformity of the cluster
 catalog. Although the Abell cluster catalog (Abell 1958; Abell, Corwin
 and Olowin 1989) has been used in many studies, it was constructed by eye,
 and is sensitive to projection effects.
 When comparing clusters at different redshifts, it is particularly important
 to ensure that the data quality and cluster selection techniques are
 uniform, to avoid the introduction of potential selection biases.
%
 The SDSS Cut \& Enhance cluster catalog used here is constructed using
 only the 
 data taken with the SDSS telescope. Also, clusters
 are detected using a single algorithm (CE) throughout the
 entire redshift range (0.02$\leq z\leq$0.3). Combined with the well controlled
 richness criteria, our cluster sample is not only one of the biggest
 but also one of the most statistically uniform cluster catalogs. 
  To study colors of galaxies in clusters, it is also important to use a
 cluster catalog created without targeting the red sequence of color
 magnitude relation of  cluster galaxies. For example, Gladders \&
 Yee(2000) and Annis et 
 al. (2003) use a color filter targeting the red sequence of clusters
 and find galaxy clusters efficiently without suffering from 
 projection effects. These techniques, however, can potentially have
 biases with regards to the colors of detected galaxy clusters, since
 clusters with a strong red sequence 
 is more easily detected.  They may not, therefore, be ideally
 suited to a Butcher-Oemler type of analysis.   In contrast, the SDSS Cut \& Enhance
 cluster catalog does not pick red galaxies selectively, and is therefore more
 suitable for this study. 
  (Note that the the Cut \& Enhance method does use generous color cuts.  
   Therefore it is not completely free from color originated biases
   although the color cut is designed to be wide 
   enough to include blue galaxies in clusters.)
 In previous work, 
 clusters have often been detected using data from only one or two
 color bands.  This can introduce a bias, since
 higher redshift clusters are redder and fainter than lower
 redshift clusters. 
 The SDSS Cut \& Enhance cluster catalog detects clusters using
 four bands of the SDSS data ($g,r,i$ and $z$), which minimizes the bias
 against redshift. 


\section{Analysis and Results}\label{bo_method}

 We compute the fraction of late-type galaxies in four different ways
 using conventional blue fraction, $u-r$ color, profile fitting, and a
 concentration parameter. 

\subsection{Fore/Background subtraction}\label{bg}

 Before proceeding to the computation of the late-type fractions,
 we describe the statistical fore/background
 subtraction method we use in common to all the four ways.  
 In counting galaxies, all galaxies are assumed to be at the cluster
 redshift to calculate absolute magnitudes. Then galaxies whose
 absolute magnitudes lie  between
 $M_{r^*}=-24.0$ and $-19.44$ are used in the analysis. 
 We count the number of late-type/total galaxies within 0.7 Mpc from the
 center of each cluster. 
 Valotto et al. (2001) claimed that the global background
  correction can not correct background contamination appropriately.
   Following the claim, we use a local background correction.  The
  number of late-type/total galaxies in fore/background is estimated in
 the same absolute magnitude range
  using an annular area around each cluster with an inner 
 radius of 2.1 Mpc and an outer radius of 2.21 Mpc.    
  The annulus-based fore/background subtraction enables us to
 estimate the fore/background locally, minimizing variations in galaxy
 number counts due to the large scale structure. 
  When an outer annulus touches the boundary of the region, a
 fore/background count is globally subtracted using galaxy number counts
 in the entire 400 deg$^2$ region by adjusting it to the angular area each
 cluster subtends. 
 This fore/background subtraction is used in the analyses described in
   subsections \ref{f_b}-\ref{bo_scatter}. 
 The fraction of blue/late-type galaxies, $f_{late}$, and its error,
  $\delta f_{late}$, are computed 
 according to the following equations.
  \begin{equation} 
    f_{late}=\frac{N_{c+f}^{late}-N_{f}^{late}}{N_{c+f}^{all}-N_{f}^{all}},\label{frac}
 \end{equation}
  \begin{equation} 
    \delta f_{late}=   f_{late}\times \sqrt{\frac{ N_{c+f}^{late}+N_{f}^{late}}{(  N_{c+f}^{late}-N_{f}^{late}  )^2}  +    \frac{N_{c+f}^{all}+N_{f}^{all}}{(N_{c+f}^{all}-N_{f}^{all})^2}   - \frac{2(\sqrt{N_f^{late}\times N_f^{all}}+ \sqrt{N_{c+f}^{late} \times N_{c+f}^{all}})}{(N_{c+f}^{late}-N_f^{late})(N_{c+f}^{all}-N_f^{all})} }
\label{frac_err}
 \end{equation}
 \noindent where $N_{c+f}^{late}$ and $N_{f}^{late}$ represent numbers
 of blue/late-type galaxies in a cluster region and a field region,
 respectively. $N_{c+f}^{all}$ and $N_{f}^{all}$ represent numbers
 of all galaxies in a cluster region and a field region, respectively.
 The equation (\ref{frac_err}) assumes that $N^{late}$ and $N^{all}$ are not
 independent. We explain the derivation of  equation (\ref{frac_err}) in
 Section \ref{bo_error}.

 \subsection{Errors on Blue/Late Type Fractions}\label{bo_error}

 In this section we summarize how we derived eq. (\ref{frac_err}) to
 calculate errors on blue/late type fractions. 
  To begin with, we assume the following.

 \begin{itemize}
  \item Number of galaxies in a cluster region ($N_{c+f}^{all}$) follows
	Poisson statistics.
  \item Number of galaxies in a certain area of field region ($N_{f}^{all}$) follows
	Poisson statistics. 
  \item $N_{c+f}^{*}$ and $N_{f}^{*}$ are independent of each other.
  \item Number of blue/late type galaxies in a cluster region ($N_{c+f}^{late}$) is
	strongly correlated with number of all galaxies in that region
	($N_{c+f}^{all}$).
  \item  Number of blue/late type galaxies in a certain field region ($N_{f}^{late}$) is
	strongly correlated with number of all galaxies in that region
	($N_{f}^{all}$).
 \end{itemize}

 And we clarify the definition of our notation.
 In this section, $\delta A $ means a deviation of a sampled value
 from an expectation value, $E(A)$. \\
 $\delta A$=$A-E(A)$, where A is each data value.\\
 $E(A)$ and $\delta A$ satisfy the following relations.\\
 $E(\delta A)$=0, $E(\delta A)^2$=$\sigma^2$,
 and if A and B are independent, $E(\delta A \delta B)$=0.
 Note that the equation \ref{frac_err} is not a deviation of
 a single sample but the expectation value estimated from the sample,
 and should be written as $E(\delta f_{late}^2)$ if we write rigidly.

 Under these assumptions, the error of late type fraction, 
 $\delta f_{late}$, become
 \begin{equation} \label{xy}
 \delta f_{late}^2 =   (X/Y)^2 \times ((\delta X^2/X^2) + (\delta Y^2/Y^2) -2(\delta X \delta Y/XY)),
 \end{equation}
 where $X$=$N_{c+f}^{late}-N_{f}^{late}$,
 $Y$=$N_{c+f}^{all}-N_{f}^{all}$, and $f_{late}$ = $X/Y$. 

Since $N_{c+f}^{all}$ and $N_{f}^{all}$ follow Poisson statistics,
         and they are independent. 
         \begin{equation}
	       E(\delta Y^2) = N_{c+f}^{all}+N_{f}^{all}\end{equation} 
         Similarly, when  $N_{c+f}^{all}$ follows Poisson statistics,
         $N_{c+f}^{late}$ also follows  Poisson statistics
         since  $N_{c+f}^{late} \sim N_{c+f}^{all}\times f_{late}$.
	 Therefore,
 	 \begin{equation}
	 E(\delta X^2) = N_{c+f}^{late}+N_{f}^{late}.\end{equation}
 
           Deriving the cross term at the end of the equation is not
             so straightforward. 
	     The cross term is expanded as 
\begin{equation}
               \delta X \delta Y 
=  \delta(N_{c+f}^{late}-N_{f}^{late})\delta(N_{c+f}^{all}-N_{f}^{all})\end{equation}\begin{equation}
	     =(\delta(N_{c+f}^{late})-\delta(N_{f}^{late}))(\delta(N_{c+f}^{all})-\delta(N_{f}^{all}))\\\end{equation}\begin{equation}
	     =\delta(N_{c+f}^{late})\delta(N_{c+f}^{all})
	     -\delta(N_{f}^{late})\delta(N_{c+f}^{all})
	     -\delta(N_{c+f}^{late})\delta(N_{f}^{all})
	     +\delta(N_{f}^{late})\delta(N_{f}^{all})\end{equation}

	     Since we assume $N_{f}^{late}$ and $N_{c+f}^{all}$,
	     $N_{c+f}^{all}$ and $N_{f}^{late}$ are both independent,
	 \begin{equation}
	 E(-\delta(N_{f}^{late})\delta(N_{c+f}^{all}))=0 \end{equation}

	     and
\begin{equation}
E(-\delta(N_{c+f}^{late})\delta(N_{f}^{all}))=0. \end{equation}

	     Therefore,  
\begin{equation} 	      
  E(\delta X \delta Y) =
  E(\delta(N_{c+f}^{late})\delta(N_{c+f}^{all})+\delta(N_{f}^{late})\delta(N_{f}^{all}))\\
\end{equation}

  Since we assume that $N_{c+f}^{late}$ and $N_{c+f}^{all}$, or
  $N_{f}^{late}$ and $N_{f}^{all}$ strongly correlate,
	 we can approximate that\\
\begin{equation}
 	  E(\delta(N_{c+f}^{late})\delta(N_{c+f}^{all}))=
	  \sigma(N_{c+f}^{late})\sigma(N_{c+f}^{all}) =
\sqrt{N_{c+f}^{late}}\sqrt{N_{c+f}^{all}} \end{equation}

	  and,\\
\begin{equation}\label{field_approximation}
 	  	  E(\delta(N_{f}^{late})\delta(N_{f}^{all}))=
	  \sigma(N_{f}^{late})\sigma(N_{f}^{all}) =
\sqrt{N_{f}^{late}}\sqrt{N_{f}^{all}}\\ \end{equation}

	 Therefore we obtain,
	 \begin{equation}\label{deltaxy}
 	   E(\delta X \delta Y) =  \sqrt{N_{c+f}^{late}}\sqrt{N_{c+f}^{all}} +\sqrt{N_{f}^{late}}\sqrt{N_{f}^{all}}
	  \end{equation}

            By substituting eq. (\ref{deltaxy}) for $\delta X \delta Y$
            in eq. (\ref{xy}) , we derive
            eq. (\ref{frac_err}).

	However, this is not the only way to estimate the error.  Actually, the
  correlation between  $N_{f}^{late}$ and $N_{f}^{all}$ is not so
  obvious since late type fraction in the field and that in the cluster
  region might be different. Although we regard the difference is so small
  that we can assume the eq.(\ref{field_approximation}),
  if we assume that $N_{f}^{late}$, $N_{f}^{early} (= N_{f}^{all} - N_{f}^{late})$,
  $N_{c+f}^{late}$, and $N_{c+f}^{early} (= N_{c+f}^{all} - N_{c+f}^{late})$ 
  are independent, we derive,
\begin{equation}
 	  E(\delta(N_{c+f}^{late})\delta(N_{c+f}^{all}))=
          E(\delta(N_{c+f}^{late})(\delta(N_{c+f}^{late})+\delta(N_{c+f}^{early})))=
	  \sigma(N_{c+f}^{late})^2 = N_{c+f}^{late} \end{equation}
	  and,\\
\begin{equation}
 	  	  \delta(N_{f}^{late})\delta(N_{f}^{all})=
	  \sigma(N_{f}^{late})^2 = N_{f}^{late}. \end{equation}
Then, the expectation value of $\delta X \delta Y$  becomes,
 
\begin{equation} E(\delta X \delta Y) = N_{c+f}^{late}+N_{f}^{late}\end{equation}
  In this case, eq. (\ref{frac_err}) becomes,
\begin{equation} 
    E(\delta f_{late}) =  f_{late}\times \sqrt{\frac{ N_{c+f}^{late}+N_{f}^{late}}{(  N_{c+f}^{late}-N_{f}^{late}  )^2}  +    \frac{N_{c+f}^{all}+N_{f}^{all}}{(N_{c+f}^{all}-N_{f}^{all})^2}   - \frac{2(N_{c+f}^{late}+N_{f}^{late})}{(N_{c+f}^{late}-N_f^{late})(N_{c+f}^{all}-N_f^{all})} }
\label{frac_err_referee}
 \end{equation}

\subsection{Blue Fraction}\label{f_b}
  
 The blue fraction of galaxy clusters ($f_b$) is measured as the fraction
 of galaxies bluer in $g-r$ rest frame color than the color of the ridge
 line of the cluster by 
 0.2 mag. This color criterion is equivalent to
 Butcher \& Oemler's (1984) 0.2 mag in $B-V$ and Margoniner et al.'s (2000,
 2001) 0.2 mag in $g-r$.  
 The color of the ridge line is measured from the color-magnitude
 diagram using the same color-magnitude box used in measuring
 photometric redshift (Goto et al. 2002a; Chapter \ref{chap:CE}). 
 The colors of the ridge lines are confirmed
 to agree with empirical color of elliptical galaxies  observed in the
 SDSS at the same redshift with less than 0.05 difference in $g-r$ color
 (Eisenstein, private communication). 
 We also use Fukugita et
 al.'s (1995) model of an elliptical galaxy and a galaxy bluer than it by
 0.2 mag in  $g-r$. By redshifting these two galaxies, we measure
 $\delta (g-r)$ in the observed frame, which corresponds to the
 restframe  $\delta (g-r)$=0.2 . In calculating $f_b$, we count 
 galaxies within 0.7 Mpc from the center of each cluster, which is the
 same radius as Margoniner et al. (2000, 
 2001), and corresponds to the average radius of Butcher \& Oemler
 (1984). (We explore possible caveats in using fixed radius in Section \ref{bo_varying}.)
 Galaxies between $M_{r^*}=-24.0$ and $M_{r^*}=-19.44$ are counted. The
 latter value corresponds to $r^*$=21.5 at $z$=0.3 for an average $k$-correction of all types of
 galaxies (Fukugita et al. 1995). Compared to the field luminosity function
 of Blanton et al. (2001), this includes galaxies as faint as $M^*_r$+1.36.  
 Fore/background galaxies are statistically subtracted in the way described in Section \ref{bg}.
 
 The lower left panel of Figure \ref{fig:bo} shows $f_b$ as a function of  redshift. 
 The error in $f_b$ is estimated using equation (\ref{frac_err}) and
 the median values of the errors in $f_b$ and $z$ are shown in the upper
  left corner of the plot.  
 Dashed line shows the weighted least-squares fit to the data.
 Solid lines and stars show the median values of the data. 
 The scatter is considerable, but both of the lines show a clear increase of $f_b$
 toward higher redshift. The  Spearman's correlation coefficient is
 0.238 with significance of more than 99.99\% as shown in Table
  \ref{tab:correlation}. The correlation is weak, but
 of high significance.  The lower left panel of Figure \ref{fig:kstest}
  further clarifies the evolution effect. A dashed and a solid line show 
  normalized distributions of $f_b$ for clusters with $z\leq$0.15 and
  0.15$<z\leq$0.3, respectively. The two distributions are significantly
  different at the 98\% level, as determined by a Kolomogorov-Smirnov test.
 The slope shown with the dashed line in Figure \ref{fig:bo}
 rises up to $f_b\sim$0.2 at $z$=0.3 (look back time of $\sim$3.5 Gyr),
 which is consistent with previous
 work such as Butcher \& Oemler (1978,1984), Rakos \& Schombert (1995), and
 Margoniner et al. (2000, 2001), within the scatter. 
 We conclude that the Butcher-Oemler effect is seen in 
 the SDSS Cut \& Enhance galaxy cluster catalog.

   We caution readers on the systematic uncertainties in measuring $f_b$.
  Marzke et al.(1994,1997,1998), Lin et al. (1999) and Blanton et
  al. (2001)  showed that luminosity functions of galaxy clusters depend
  significantly on galaxy type, in such a way that
  the bright end of the  cluster luminosity function is dominated by
  redder galaxies and the  faint end is dominated by bluer galaxies. 
  Boyce et al.(2001) and Goto et al. (2002b; Chapter \ref{chap:LF}) showed that a similar
  tendency exists for cluster luminosity functions. 
  This difference in luminosity functions leads to a different blue fraction depending
  on  the absolute magnitude range used. Furthermore, if the radial
  distributions of blue and red galaxies are different (e.g., Kodama et
  al. 2001), the $f_b$ measurement
  depends heavily on the radius.  When comparing with previous work, therefore,
  it is important to take account of the exact method used to calculate $f_b$.
  We discuss the
  uncertainty in measuring blue fractions further in Section \ref{bo_discussion}.

\subsection{Late Type Fraction Using $u-r<$2.2}\label{u-r}

 Recently Shimasaku et al. (2001) and Strateva et al. (2001) showed that
 the SDSS $u-r$ color correlates well with galaxy morphologies.
 In this section we use $u-r$
 color to separate early($u-r\geq$2.2) and late ($u-r<$2.2) type galaxies
 as proposed by Strateva et al. (2001). Note that although $u-r$ color
 is claimed to correlate well with galaxy types, it is still a color
 classifier and thus different from the morphological parameters
 we investigate in the following two sections.
 The methodology used to measure late type fraction is similar to the one we use
 to measure $f_b$. We regard every galaxy with $u-r<$2.2 as a late type
 galaxy. We define $f_{u-r}$ as the ratio of the number of late type
 galaxies to the total number of galaxies within 0.7 Mpc from the cluster center.
 Fore/background subtraction is performed in a way described in Section \ref{bg}.   
 
 The upper left panel of Figure \ref{fig:bo} shows $f_{u-r}$
 as a function of redshift. The error in $f_{u-r}$ is calculated using equation
 (\ref{frac_err}) and the median  values of the errors in $f_{u-r}$ and $z$
 are shown in the upper left corner
 of the
 panel. The dashed line shows the least square fit to the data.
   The solid lines and stars show the median values of the data. 
 As in the case of $f_b$, the scatter is considerable, but the weak
 increase of the late type galaxies is seen. The  Spearman's correlation coefficient
 is 0.234 and is inconsistent with zero at greater than 99.99\% confidence
 level (Table \ref{tab:correlation}). Again, weak but significant
 correlation is found. The upper left panel of Figure \ref{fig:kstest}
 shows distributions of $f_{u-r}$ for $z\leq$0.15 clusters and
 0.15$<z\leq$0.3 clusters with a dashed and solid line, respectively. A Kolomogorov-Smirnov
 test shows that the distributions are different with more than a 99\%
 significance. 
 In addition to the increase in $f_b$ shown in the last section,
 the increase in $f_{u-r}$ provides further evidence of color
 evolution of cluster galaxies.  Furthermore, since $u-r$ color of galaxies is
 sensitive to a galaxy's morphology as shown in Figure 6 of Strateva et
 al (2001), it suggests possible evolution of morphological types of galaxies
 as well.
 We investigate the morphological evolution of galaxies in clusters in the next subsection.

\subsection{Late Type Fraction Using Profile fitting} \label{exp}

  One of the purposes of this chapter is to determine if there is a
  morphological change of galaxies in clusters as a function of
  redshift. 
  The SDSS photometric pipeline (PHOTO; Lupton et al. 2002) fits a de Vaucouleur
 profile and an exponential profile to every object detected in the SDSS imaging
 data and returns the likelihood of the fit. By comparing the likelihoods
 of having an exponential profile against that of a de Vaucouleur profile, we can classify
 galaxies into late and early types. In this section, we
 regard every galaxy that has an exponential likelihood higher than a de Vaucouleur
 likelihood in $r$ band as a late type galaxy. A galaxy with higher
 de Vaucouleur likelihood in $r$ band is regarded as an early type
  galaxy. 
 We define $f_{exp}$ in the same way as in previous subsections, i.e.,
 $f_{exp}$ is the ratio of the number of late type galaxies to
 the total number of galaxies within 0.7 Mpc. Fore/background counts are
  corrected using the method described in Section \ref{bg}. 

 The resulting  $f_{exp}$ is plotted in the lower right panel of Figure
 \ref{fig:bo}. 
  The error in $f_{exp}$ is estimated using equation (\ref{frac_err}) and
 the median values of the errors in  $f_{exp}$ and $z$ are shown in the upper
  left corner of the plot.  
 The dashed line shows the weighted least-squares fit. 
 The solid lines and stars show the median values of the data. 
 The scatter is considerable, but we see the increase of $f_{exp}$ toward the higher redshift. The
  Spearman's correlation coefficient is 0.194, which is inconsistent with zero at
 more than a 99.99\% confidence level (Table \ref{tab:correlation}).
 The upper right panel of Figure \ref{fig:kstest} shows the
 distributions of clusters with $z\leq$0.15 and
 with 0.15$<z\leq$0.3 with a dashed and solid line, respectively. 
 The two distributions show a difference of more than 99\% significance
 in a Kolomogorov-Smirnov test. 
 We emphasize that the galaxy
 classification used here is purely morphological --- independent
 of colors of galaxies. The fact that we still see the increase of the
 late type galaxies toward higher redshift suggests that these
 Butcher-Oemler type blue galaxies also change their morphological
 appearance as well as their colors. We also point out that the slope of the
 change is similar to that in the lower left panel of Figure
 \ref{fig:bo}, which is  $\sim$30\% between $z=$0.02 and $z=$0.3. 
 We note that there is a potential bias associated with the use of $r$
 band profile fitting throughout 
 the redshift range, since the $r$ band wavelength range at $z$=0.3 is almost that of $g$
 band at restframe. 
 We investigate this effect in Section \ref{bo_discussion}, and conclude that it is small. 
 Like the blue fraction, the morphological late-type fraction is also sensitive to the
 magnitude range considered.
  Binggeli et al. (1988), Loveday et al. (1992), Yagi et al. (2002a,b) and Goto et al. (2002b; Chapter \ref{chap:LF}) reported
 luminosity functions of elliptical galaxies have brighter
 characteristic magnitudes and flatter faint end tails compared to those
 of spiral galaxies in both field and cluster regions. Careful attention should be paid to 
 the magnitude range used in an analysis when fractions of spiral galaxies are compared. 
  We discuss the uncertainty further in Section \ref{bo_discussion}.

\subsection{Late Type Fraction Using Concentration Parameter} \label{Cin}

   As another morphological galaxy classification method, we use the inverse of
   the concentration parameter ($C_{in}$) advocated by Shimasaku et
   al. (2001) and Strateva et al. (2001). We define $C_{in}$ as the
   ratio of Petrosian 50\% radius to Petrosian 90\% radius in $r$
   band. They are the radii which
   contain 50\% and 90\% of Petrosian flux, respectively. (See Stoughton
   et al. 2002 for more details of Petrosian parameters). Since $C_{in}$
   is the inverse of a conventional concentration parameter, spiral
   galaxies have a higher value of $C_{in}$. Following Strateva et al. (2001), we
   use $C_{in}$=0.4 to divide galaxies into early and late type
    galaxies. Readers are referred to Morgan (1958,1959), Doi, Fukugita \& Okamura (1993) and
   Abraham et al. (1994, 1996) for previous usage of concentration of light
   as a classification parameter.    
   $f_{Cin}$ is defined as the ratio of the number of galaxies
   with $C_{in}>$0.4 to the total number of galaxies within 0.7 Mpc from
   the cluster center as
   in the previous subsections. Note that our early type galaxies with
   $C_{in}<$0.4 include S0 galaxies in addition to elliptical galaxies
   since discerning elliptical and S0 galaxies is very difficult with
   the SDSS data, in which the seeing is typically 1.5'' (See Shimasaku et
   al. 2001 and Strateva et al. 2001 for the correlation of $C_{in}$
   with an eye classified morphology).
   Fore/background
   number counts are corrected as described in Section \ref{bg}.
   The absolute magnitude range used is  
   $-24<M_{r^*}<-19.44$. 
 
 The upper right panel of Figure \ref{fig:bo} shows $f_{Cin}$  as a function
   of redshift. Since the classification using $C_{in}$=0.4 leans
   toward late type galaxies, the overall fraction is higher than the
   other panels in the figure. The increase of late type fraction,
   however, is clearly seen. The dashed line shows the weighted least-squares 
   fit.  The solid lines and stars show the median values of the data. 
  The error in $f_{Cin}$ is estimated using equation (\ref{frac_err}) and
 the median values of the errors in  $f_{Cin}$ and $z$ are shown in the upper
  left corner of the plot.  
 The  Spearman's correlation coefficient is
 0.223 with significance of more than 99.99\% as shown in Table
   \ref{tab:correlation}. The upper right panel of Figure
   \ref{fig:kstest} further clarifies the evolution effect. The
   distribution of $z\leq$0.15 clusters in a dashed line and the
   distribution of 0.15$<z\leq$0.3 clusters in a solid line show a
   difference with more than a 99\% significance level.   
   We stress that the galaxy classification
   based on this concentration parameter is purely a morphological one.
  In this morphological classification, we still see the increase of the late type fraction just
   like the increase of $f_b$ --- as if observing the morphological equivalence of the
   Butcher-Oemler effect.
   The increase in $f_{Cin}$ combined with the increase in $f_{exp}$ 
   provides rather firm evidence of morphological change in the
   Butcher-Oemler type
   galaxies. Possible caveats in the usage of $C_{in}$ and comparisons
   with previous works are discussed in Section \ref{bo_discussion}.

\subsection{On the Origin of the Scatter} \label{bo_scatter}

 In the last four sections, we observed the increase of late type
 fractions toward higher redshift in all the four cases. At the same time, we see
 a significant amount of scatter around the late type fraction
 v.s. redshift relations. Although the errors on these measurements are
 also large, this scatter might suggest that there might be one or more physical
 properties  which determine the amount of late-type galaxies in
 clusters. Table \ref{tab:error_comparison} compares median error sizes
 of $f_b,f_{u-r},f_{exp}$ and $f_{Cin}$ with scatters around the
 best-fit lines (dotted lines in Figure \ref{fig:bo}). 
 In fact, in all cases, real scatters are larger than the statistical errors.
 In the literature, several 
 correlations are proposed  such as those with
 X-ray shapes of clusters (Wang et al. 2001; Metevier et al. 2000), and 
 with  cluster richness (Margoniner et al. 2001). 
 Our cluster richnesses are plotted against redshift in Figure \ref{fig:z_rich}.
 Richnesses are measured as numbers of galaxies between $M_{r^*}=-24.0$ and
 $-19.44$ 
 within 0.7 Mpc from the cluster center after fore/background subtraction, as explained in
 Section \ref{bg}. In Figure \ref{fig:z_rich}, this richness has no
 apparent bias with redshift. 

 In Figure \ref{fig:okamura_rich}, the difference of the late type
 fraction from the best-fit line is plotted against cluster richness.
 The circles and solid lines show median values. 
 In all the panels, there is a clear tendency of richer clusters 
 having a lower fraction of late type galaxies. This tendency is in
 agreement with Margoniner et al. (2001) who found richer clusters
 had lower blue fractions. We further discuss the richness
 dependence of the late type fraction in Section \ref{bo_discussion} and
 Section \ref{bo_varying}.
  As an alternative parameter to X-ray shape, we plot the difference from
 the best-fit line against cluster elongation in Figure
 \ref{fig:okamura_elong}. The elongation parameter is taken from Goto
 et al. (2002a; Chapter \ref{chap:CE}), which is the ratio of major and minor axes
 in  their enhanced map to find clusters. Circles and solid lines show
 median values. No obvious trend is seen here. Our result
 seems to agree with Smail et al.'s (1997) caution that a correlation between $f_b$
 and cluster ellipticity found by Wang et al. (1997) could be due to a small
 and diverse sample.
 However, since distribution of
 galaxy positions might not represent cluster ellipticities measured
 with X-ray shape well, we do not
 conclude  that there is no dependence on cluster ellipticities. The
 dependence should be pursued further in the future, ideally using X-ray
 profile shape with a large sample of clusters.

\section{Discussion}\label{bo_discussion}

\subsection{Morphological $k$-correction}

 In the upper right panel of Figure \ref{fig:bo}, we use $C_{in}$
 (inverse of concentration index) in the $r$ band to classify galaxies throughout our redshift
 range (0.02$\leq z\leq$0.3). This could potentially cause redshift dependent
 biases in our calculation of $C_{in}$. Since the universe is expanding,
 by analyzing the observed $r$ band data, we are analyzing bluer
 restframe wavelengths in the higher redshift galaxies.
 In fact, the $r$ band at $z$=0.3 is almost $g$ band in the
 restframe.  Various authors have pointed out that galaxy
 morphology significantly changes according to the wavelength used
 (e.g., Abraham et al. 2001). To estimate how large this bias is, we
 plot the normalized distributions of $C_{in}$ in $g$ and $r$ bands  in
 Figure \ref{fig:cin_g_r} using
 the galaxies with 0.02$\leq z\leq$0.03 in the SDSS spectroscopic data (1336
 galaxies in total; See Eisenstein et al. 2001; Strauss et al. 2002 and
 Blanton et al. 2003 for the SDSS spectroscopic data). In this small
 redshift range, the color shift due to the 
 expansion of the universe is small. We use this redshift range to study the
 dependence of the $C_{in}$ parameter on the restframe wavelength. At
 $z$=0.3, $r$ band corresponds to restframe $g$ band. The solid
 and dashed lines show the distribution for $g$ and $r$ bands,
 respectively. The two distributions are not exactly the same, but the
 difference between the two distributions is small. We summarize the
 statistics in Table \ref{tab:cin_g_r}. 
 There are 802/1336 galaxies with $C_{in}>$0.4 in $g$ band, and 787/1336
 galaxies have $C_{in}>0.4$ in $r$ band. The difference is 15/1336 galaxies, which is 1.1\% of the
 sample.  In section  \ref{Cin}, the change in $f_{Cin}$ is $\sim$30\%.
 The effect of the morphological $k$-correction is therefore much smaller.  
 We also point out that this analysis assumes the largest difference in
 redshift (0.02$\leq z\leq$0.3),
 therefore it gives the upper limit of the bias. 
 Since the majority of our clusters are at 
 $z\sim$0.2, the wavelength difference between the observed and
 restframe bands is typically much smaller.
 We conclude that the effect of the morphological $k$-correction
 is much smaller than the change in $f_{Cin}$ we observed in Section \ref{Cin}.

  In Section \ref{exp}, we use the $r$ band fit for all galaxies in our
 sample. The same redshift effect could potentially bring bias to our
 analysis. In Table \ref{tab:exp_g_r}, we limit our galaxies to
 0.02$\leq z\leq$0.03 and count the fraction of late type galaxies in
 the $g$ and $r$ bands corresponding to the observed $r$ band at $z=$0.0
 and $z=$0.3, respectively. 
 We list the number of galaxies with
 exponential likelihood higher than de Vaucouleur likelihood in column 1, the
 total number of galaxies in column 2, and the ratio of columns 1 to 2 in column
 3. As shown in the 3rd row, the difference in the fraction of late type
 galaxies between $g$ band data and $r$ band data is only 2.5\%, which
 is much smaller than the $f_{exp}$ change we see in the upper right panel of
 Figure \ref{fig:bo} ($\sim$30\%). 
 We conclude that the change of
 $f_{exp}$ and $f_{Cin}$ is not caused by the small redshift bias in using $r$ band
 data throughout the redshift range. 
%


\subsection{Seeing Dependence}\label{bo_seeing_depdendence}

  Another possible source of bias in measuring $f_{exp}$ and
 $f_{Cin}$ is the dependence on the seeing, relative to the size of the galaxies. At
 higher redshift, the size of a galaxy is smaller and a seeing convolution
 could be more problematic.  Especially for the concentration
 parameter ($C_{in}$), galaxy light becomes less concentrated when
 the seeing size is comparable to the galaxy size, and thus,  the effect could
 cause a bias towards higher $C_{in}$ values.  
 To check this, we plot $f_{Cin}$ against the
 point-spread function (PSF) size in the $r$ band for two redshift limited samples in Figure
 \ref{fig:seeing_cin}.  Open squares and solid lines show
 the distribution and medians of low $z$ clusters ($z\leq$0.15). Filled
 triangles and dashed lines show the distributions and medians of high $z$
 clusters (0.24$<z\leq$0.3). 
 For the median measurements, bins are chosen so that equal
 numbers of galaxies are included in each bin.  1 $\sigma$ errors are shown as vertical bars.
 As expected, lower
 redshift clusters show almost negligible dependence on seeing
 size. Higher redshift clusters show about a 5\% increase in $f_{Cin}$ between the best
 and worst seeing size. The evolution effect we see in the upper right
 panel of Figure \ref{fig:bo} is more than 20\%. Furthermore, 
 as is seen from the distribution of seeing shown in Figure \ref{fig:bo_seeing_hist},
 87\% of our sample galaxies have seeing better than 2.0''. 
 Therefore we conclude that varying seeing causes
 a small bias which is significantly weaker than
 the evolution we find in Section \ref{bo_method}.   
 The effect of varying seeing is less significant for the $f_{exp}$
 parameter. In Figure \ref{fig:seeing_exp}, we plot $f_{exp}$ against
 seeing size for two redshift samples with the same redshift ranges and symbols
 as in Figure \ref{fig:seeing_cin}.  1 $\sigma$ errors, shown as
 vertical bars, are dominant.  There is no significant correlation of
 $f_{exp}$ with seeing size.

\subsection{Radius, Fore/background Subtraction and Cluster Centroids}
 
  Throughout the analyses in Section \ref{bo_method}, we use 
 a 2.1-2.21 Mpc annular region for fore/background
 subtraction. In return for
 taking cosmic variance into account, annular (local) background subtraction  
 has larger statistical errors than global background subtraction due to its smaller
 angular area coverage. However, the difference is not so large. In
 case of blue galaxy counts ($f_b$) in the background, the
 median Poisson (1 $\sigma$) uncertainty for global background is 12.2\%, whereas 1
 $\sigma$ variation of local background is 12.6\%. This increases
 the errors, but only by 0.4 points. The actual effect to the late-type
 fraction is plotted in Figure \ref{fig:bg_test}.  Solid lines show
 distributions for our default choice of 0.7 Mpc radii 
 and 2.1-2.21 Mpc annuli. Dashed lines show distributions for global
 fore/background subtraction, where fore/background subtraction is
 performed using global number counts of galaxies for all the clusters
 in the sample. A Kolomogorov-Smirnov test between two samples does not show any
 significant difference.

  For cluster radius, we use 0.7 Mpc, since we do not have
 information about the virial radii of each system.
 It is, however, ideal to use virial radii since, for example, in a standard cold dark
 matter cosmology, virial radii
 at a fixed mass scales as $\propto$ (1+$z$)$^{-1}$. 
  Another possible cause of uncertainty is the accuracy in deciding cluster
 centers. 
 In this work, a center position of each cluster is taken from
 Goto et al. (2002a; Chapter \ref{chap:CE}), and is estimated from the position of the peak in their
 enhanced density map. Although, from Monte-Carlo simulations, cluster centroids are expected
 to be determined with an accuracy better than $\sim$40 arcsec,
 the offsets have a possibility to introduce a bias in our analyses.
 We test 
 different choices of these parameters in Figure \ref{fig:bg_test}. 
 Dotted lines show distributions where radii change as
 0.7 $\times$ (1+$z$)$^{-1}$ Mpc assuming a standard cold dark matter
 cosmology.  Long dashed lines show distributions when the position of
 brightest cluster galaxy (within 0.7 Mpc and  $Mr<-$24.0) is used as a cluster center. 
 Kolomogorov-Smirnov tests show no significant difference in any of the above cases.
 In all cases, the probability that the distributions are different is
 less than 26\%.  Our results in Section \ref{bo_method} are thus not
 particularly sensitive to our choice of annuli, radii  or cluster centers.
 We further pursue the effect of radius dependence of blue/late type fractions in
 Section \ref{bo_varying}, and show that it does not change our main results.

\subsection{Comparison with Late-type Fraction from Spectroscopy}

 To further test our late-type fraction measurement, we compare
 the late-type fraction obtained from the SDSS spectroscopic data with
 that obtained from the SDSS imaging data in Figure
 \ref{fig:comparison_with_spectroscopy}. Since the SDSS spectroscopic
 data are limited to $r^*<$17.77 (Strauss et al. 2002), the comparison
 can be done only for clusters with $z<$0.06. In the literature, three
 clusters are found to satisfy these criteria in the region used in this
 study. These clusters include ABELL 295, RXC
 J0114.9+0024, and  ABELL 957. For these clusters, late-type fractions
 are measured in the same way as in section \ref{bo_method}. Late-type
 fractions from spectroscopy are measured using all the SDSS
 spectroscopic galaxies within 0.7 Mpc
 from the cluster center 
 and $\delta z=\pm$0.005 from the redshift of each cluster. Note that there is no
 fore/background correction for spectroscopic late-type fraction. In Figure
 \ref{fig:comparison_with_spectroscopy}, all points agree with each
 other within the error. The good agreement suggests that our fore/background
 subtraction technique described in section \ref{bg} works
 properly. It would be ideal to perform the same test for high redshift
 clusters as well. However, the SDSS spectroscopic data are not deep
 enough to perform the test for higher redshift clusters.

\subsection{The Butcher-Oemler Effect: Comparison with Previous Work}

  The Butcher-Oemler effect--- an increase in the ratio of blue galaxies in
  clusters as a function of redshift--- is strong evidence of direct evolution of
  the stellar populations in galaxies; it has been studied by numerous
  authors in the past.
 In this section,
 we compare our results with previous work.
 Since different authors use  
 different cluster samples, color bands, cosmology, absolute
 magnitude ranges and methods of fore/background subtraction,
 which could affect the comparison,
 we emphasize the differences in analysis by each author. Note that one
 important difference is that some previous work used
 a sample of quite rich clusters, e.g., clusters with more than 100 members
 in magnitude and radius ranges comparable to those adopted in this
 study. Poorest systems in our 
 sample have only 25 member galaxies after fore/background subtraction. 
 Thus, difference in cluster samples could cause a difference in results.

  Butcher \& Oemler (1978, 1984) studied 33 clusters between $z=$0.003 and
 $z=$0.54. They used galaxies brighter than $M_V=-$20 ($h$=0.5 and $q_0$=0.1)  within the circular area
 containing the inner 30\% of the total cluster population. They found
 $f_b$ increases with redshift for $z\geq$ 0.1. Their $f_b$ at $z=$0.3 is
 $\sim$0.15, which is slightly lower than our value. 
 Considering the large scatter in both their and our samples, we do not claim that our results
 are inconsistent with their value. Note that Andreon \& Ettori (1999)
 found a trend of increasing X-ray luminosity with increasing redshift
 in the sample clusters of Butcher \& Oemler (1984).    
 
 Rakos \& Schombert (1995) studied 17 clusters using Stromgren $uvby$
 filters. Due to the usage of the narrow band filters redshifted to the
 cluster distance, their study did need to use model-dependent $k$-corrections.
 However, their high-redshift cluster sample is drawn from that
 of Gunn, Hoessel \& Oke (1986) which is based on IIIa-J and IIIa-F
 photographic plates.  At $z>0.5$, these plates measure the rest-frame
 ultraviolet to blue region of the spectrum.  Thus the cluster catalog will
 be biased toward clusters rich in blue galaxies at high redshift.
 Rakos \& Schombert found $f_b\sim$0.25 at $z=$0.3, which is
 slightly higher than the estimation of Butcher \& Oemler (1984) but
 in agreement with our results. 

 Margoniner et al. (2000) studied 44 Abell clusters between $z=$0.03 and $z=$0.38. They used galaxies
 between $M_r=-21.91$ and $-17.91$ ($h=0.75$) within 0.7 Mpc of the cluster center.
 The fore/background counts are subtracted using five control fields. 
 Their results are more consistent with the steeper relation estimated in
 1995 by Rakos and Schombert than with the original one by Butcher and
 Oemler in 1984. The results are also consistent with ours. 
 Margoniner et al. (2001) extended their work to 295 Abell clusters
 and found $f_b$=(1.34$\pm$0.11)$\times z-$0.03 with a {\it r.m.s.}
 of 0.07, which is in agreement with our fitted function shown in Figure
 \ref{fig:bo}. 

 Ellingson et al. (2001) studied 15 CNOC1 clusters (Yee, Ellingson, \&
 Carlberg 1996) between $z=$0.18 and $z=$0.55. Since they used spectroscopically observed
 galaxies, they do not suffer from the fore/background
 correction (but see Diaferio et al. 2001). They used galaxies brighter than $M_{r}=-19.0$ within
 $r_{200}$ from the cluster center (with an average of 1.17$h^{-1}$
 Mpc). Their best fit shows $f_b\sim$0.15 at $z=$0.3. The scatter  in their Figure 1 and our data are both substantial. 
 Thus, we can not conclude that this value is inconsistent with our results. 
 
 All of these  authors found considerable scatter
 in $f_b$ v.s. $z$ plot as is seen in our Figure \ref{fig:bo}. 
 It is promising that our results are consistent with the previous
 work within the scatter, despite the differences in the radial coverage
 and magnitude ranges used. 

\subsection{The Morphological Butcher-Oemler effect}

 In Sections \ref{u-r}, \ref{exp}, and  \ref{Cin}, we found an
 increase in the fraction of late type galaxies selected by morphological parameters
 with increasing redshift --- as if the Butcher-Oemler effect is happening morphologically.
 Perhaps revealing this morphological Butcher-Oemler effect is the most striking result of this study.
 It suggests that the Butcher-Oemler blue galaxies
 change their morphology from late to early type at the same time that they
 change their color from blue to red. Although accurately quantifying the fraction of
 galaxies which experience the morphological Butcher-Oemler effect is
 difficult due to the considerable scatter in the data, our best-fit lines suggest 
 that $\sim$30\% of galaxies in clusters undergo this transition
 between $z=$0.3 and $z=$0.02. 

  In previous work, Dressler et al. (1997)
 found a deficit of S0 galaxies in 10 intermediate  ($z\sim$0.5)
 clusters by classifying galaxy morphology in the HST image 
 by eye. They claimed that many S0s needed to be added to
 reach the fraction of S0s found in present clusters (Dressler 1980). 
 Couch et al. (1994, 1998) also found an indication of morphological
 transformation in the Butcher-Oemler  galaxies by studying three rich
 clusters at $z=$0.31.  Later, Fasano et al. (2000) showed that spiral
 galaxies are, in fact, turning into S0 galaxies 
 by observing nine clusters at intermediate redshifts and analyzing them
 together with higher redshift clusters in the literature. Their galaxy morphology was
 also based on eye classification. 
 Our SDSS data are taken using ground based telescopes with moderate
 seeing ($\sim$1.5''), and thus do not allow us to separate S0 galaxies
 from elliptical galaxies as the $HST$ does.
 The advantage of our classification is its automated nature,
 which allows accurate reproducibility and quantification of
 systematic biases.
  In particular, it is easy to compute the completeness and contamination rate for
 the automated classification, based on simulations; for the present sample,
 the completeness and contamination rate of the parameters 
 are given in Shimasaku et al. (2001) and Strateva et
 al. (2001).  Furthermore,   an automated galaxy classification is easier to reproduce in
  future observational work and in detailed computer simulations.
 Although we can not distinguish S0s from ellipticals, the increase of blue
 fraction and increase of late type galaxies toward higher redshift is
 qualitatively consistent with the process of S0 production over the
 interval in cosmic time suggested by previous investigations.
 
 Various physical mechanisms could be the cause of the morphological
  and spectral Butcher-Oemler effects.
  Possible causes include ram pressure stripping of gas (Gunn \& Gott 1972; Farouki
 \& Shapiro 1980; Kent 1981; Abadi, Moore \& Bower 1999; Quilis, Moore \& Bower
 2000), galaxy infall (Bothun \& Dressler 1986; Abraham et al. 1996a; Ellingson et al. 2001), galaxy harassment via high speed impulsive encounters (Moore et al. 1996, 1999), cluster
 tidal forces (Byrd \& Valtonen 1990; Valluri 1993) which distort
 galaxies as they come close to the center, interaction/merging of
 galaxies (Icke 1985; Lavery \& Henry 1988; Bekki 1998), and removal \& consumption of the gas due to the cluster environment (Larson, Tinsley \& Caldwell 1980; Balogh et
 al. 2001; Bekki et al. 2002). Mamon (1992) and Makino \& Hut (1997) showed that
 interactions/mergers can occur in a rich cluster environment despite the
 high relative velocities. Shioya et al. (2002) showed that the
  truncation of star formation can  explain the decrease of S0 with
  increasing redshift. 
   It has been known that preheating of intergalactic medium
 can effect morphologies of galaxies by strangling the gas accretion (Mo
 \& Mao 2002; Oh \& Benson 2002). In fact, Finoguenov et al. (2003)
 found the filamentary gas in Coma cluster and predicted quiescent star
  formation in galaxy disks around the filament. 
  Although our results provide some important clues, pinpointing what
 processes are responsible in the morphological and spectral
 Butcher-Oemler effect is a more  difficult challenge. 

  Our results suggest that the cause is a process that affects both
 color and morphological appearance of galaxies at the same time. 
  Couch et al. (1998), Dressler et al. (1999) and Poggianti et al. (1999) found ``passive
 spirals'', which are galaxies with spiral morphology but without star
 formation.  They probably belong to the same population as ``anemic
 spirals''  found by van den Bergh (1976). 
 The mechanism creating ``passive spirals'' or ``anemic
 spirals'', however, affects only the
 color of galaxies and, thus, 
 probably is not the main mechanism that accounts for the entire effect.
  The increase of morphologically late type galaxies toward higher
 redshifts at the same time as the 
 increase of blue galaxies is consistent with mechanisms which affect the gas supply
 (e.g., ram-pressure stripping, galaxy infall). However, if the infalling
 rate of field galaxies (mostly blue/late type) is higher in the past,
 almost any of the mechanisms mentioned above can explain our
 observational results.
 Furthermore, although we
 discussed about cluster specific phenomena, it is also known that
 galaxies in the field region evolve as a function of redshift as well.
 (e.g., Hammmer et al. 1996; Lilly et al. 1996; Balogh et al. 1997,
 2002). The evolution of field galaxies needs to be compared
 with that of cluster galaxies further in detail.
 Therefore, it is still an open question what mechanism causes spectral
 and morphological evolution of cluster galaxies.


  The finding of a 30\% change of the fraction during the look back time
  of $\sim$3.5  Gyr could also give us an additional hint in finding an
  underlying   physical process.
  If the gas in spiral galaxies is removed very efficiently by some
 physical processes (e.g., ram-pressure stripping) or consumed rapidly by
 star formation, the spiral arms will disappear after several disk
 rotation periods, $\sim$ 1 Gyr (Sellwood \& Carlberg 1984). 
 Interaction/merger processes are quicker than gas removal processes
  ($\sim$0.5 Gyr; Mihos 1995). 
 Moore et al.'s (1996) simulation showed that the galaxy harassment phase
 lasts for several Gyr. Kodama et al. (2001) used the phenomenological simulations to show that the timescale of the morphological transformation from spiral to S0 is 1$\sim$3 Gyr. 
 For spectral change,  Shioya et al. (2002) showed that a disk needs 2-3 Gyr after the removal  of gas (or truncation of star formation) to show a k spectrum. Poggianti
  et al. (1999) compared the spectral and morphological properties of
  cluster galaxies and suggested that the timescale of the
  morphological transition is longer than that of the spectral transition.
  This difference in timescale is interesting since if one process is
  significantly quicker than the other, we might be able to see the time
  difference in the decreases of the fraction of between late type galaxies and
  blue galaxies, which will provide a strong constraint in the evolution history of the Butcher-Oemler galaxies.
 In Figure \ref{fig:bo}, we see a $\sim$30\%
 of change in both the photometric and morphological Butcher-Oemler effect
 between $z=$0.02 and $z=$0.30 ($\sim$3.5 Gyr). The
 scatter in our measurement, however, is considerable and our choices of 
 criteria between late and early type galaxies do not necessarily coincide
 with each other. It is thus not straightforward to convert the
  information to the time scale of the responsible physical process.
 In addition, to understand change in fraction of morphological and
  spectral late-type galaxies, the change in infalling rate of field
  galaxies needs to be understood as well.
 Since computer simulations have recently made dramatic progress,  
  in the near future it will become possible for
 state-of-the-art simulations to simulate both dynamical and
 spectral evolutions of cluster galaxies, plus infalling rate of field
  galaxies in order to compare the results with the
 observed trend. For example, such a simulation can be done by combining
  dynamical simulations (e.g., 
 Evrard 1991; Kauffmann et al. 1995; Bekki, Shioya \& Couch 2001;
  Vollmer et al. 2001; Bekki et al. 2002) with cluster
 phenomenological simulations (e.g., Abraham et al. 1996; Fujita
  1998,2001;  Balogh et al. 1999; Stevens, Acreman, \& Ponman 1999; 
 Balogh, Navarro, \& Morris 2000; Kodama \& Bower 2001).
 Figure \ref{fig:bo} in this work provides the interesting observational data
 to tackle with using such a simulation of cluster galaxy formation.

%
%
%
%
%
%
%

%
%

\subsection{Richness Dependence}

  In Section \ref{bo_scatter}, we observe the tendency of richer clusters
 to have smaller fractions of late type galaxies, by measuring the
 residuals from the best-fit relations as a function of cluster
 richness. Our result is consistent with Margoniner et al. (2001), who
 used a similar optical richness to find that poorer clusters tend to
 have larger blue fractions than 
 richer clusters at the same redshift. Figure \ref{fig:okamura_rich},
 however, still shows a significant amount of scatter, which might be
 suggesting the existence of another factor in determining the blue
 fraction in addition to redshift and richness.   
 The dependence of the late type fraction on cluster richness, however, provides
 another hint on the underlying physical processes. Since ram pressure
 is stronger in clusters with higher temperature at the same gas density, Fujita \&
 Nagashima(1999) theoretically predicted that if ram pressure is the only
 mechanism responsible for the evolution of galaxies in clusters, the 
 fraction of blue galaxies will always be higher in lower X-ray luminosity
 clusters, which usually have low temperatures. Our data shown in Figure
 \ref{fig:okamura_rich} are consistent with the prediction from their ram
 pressure stripping model. Although
 our richness is from numbers of galaxies in optical imaging data, it is reasonable to assume it correlates well with X-ray luminosity (Bahcall 1977;
 Bower et al. 1994). Then, the optical
 richness can be related to the gas temperature through the well known $L_{X}-T$
 relation (Mitchell, Ives, and Culhane 1977; Henry \& Arnaud 1991; Edge
 \& Stewart 1991; David et al. 1993; White, Jones, and Forman 1997; Allen
 \& Fabian 1998; Markevitch 1998; Arnaud \& Evrard 1999; Jones \& Forman
 1999; Reichart, Castander, \& Nichol 1999; Wu, Xue, and Fang 1999; Xue
 \& Wu 2000; and see the references therein).
 In a simple estimation, ram pressure is proportional to $\rho
 v^2$. $L_X$ is proportional to $\rho^2$. From the virial theorem, $v^2\propto T$.
 The $L_{X}-T$ relation studied by Xue \& Wu (2000) is $L_X\propto T^{2.8}$.
 Therefore, ram pressure is proportional to $\sim L_X^{0.86}$. 
 Combined with an assumption that optical richness scales with X-ray
 luminosity (see, e.g., Bahcall 1974; Jones \& Forman 1978; Bower et
 al. 1994; and Miller et al. in preparation),  
 Figure \ref{fig:okamura_rich} provides another hint that ram
 pressure stripping induces the evolution of cluster galaxies.

 In the literature, however, the dependence of blue fractions on cluster richness has been controversial. Bahcall (1977) studied 14 X-ray clusters and found that
 the fraction of spiral galaxies decreases with increasing X-ray
 luminosity.  
 Lea \& Henry(1988) observed 14 clusters in X-ray and found that 
 the percentage of blue objects in the clusters seems to increase with the X-ray luminosity.
 On the other hand, Fairley et al. (2002) studied eight X-ray selected
 clusters and did not find any dependence of blue fractions on X-ray
 luminosities. Balogh et al. (2002) studied 10 clusters at $z=$0.25 with low X-ray
 luminosity and found similar morphological and spectral properties of
 galaxies compared with clusters with high X-ray luminosity (Balogh et
 al. 1997). 
 In all cases, the results were based on a small sample
 of clusters.  We also point out that although our results are consistent with
 a ram-pressure stripping model, there is a possibility that other
 mechanisms could explain the phenomena. For example, richer clusters
 might have higher rate of merger/interaction due to their higher galaxy
 density. The same argument holds true for galaxy harassment. Thus, more
 study is needed to conclude about the physical mechanism causing the phenomena.
 In the near future, confirming the richness
 dependence using X-ray luminosities or velocity dispersions with a larger
 sample of clusters would offer us more insight on the subject.

\subsection{Varying Radius}\label{bo_varying}

 In Section \ref{bo_method}, we used a fixed 0.7 Mpc radius to measure
 blue/spiral fractions among cluster galaxies since it was difficult to
 measure virial radius for relatively poor clusters in our sample from the
 SDSS imaging data. 
 However, it is known that
 virial radius changes according to cluster richness; i.e., richer
 clusters have larger virial radius than poorer clusters. Therefore
 using a fixed radius could bring some bias associated with cluster
 richness. In this section we try to rectify this problem using cluster
 richness to calculate virial radius under a simple assumption. 
 We assume that our cluster richness (number of galaxies between $M_{r^*}=-24.0$ and $-19.44$
 within 0.7 Mpc 
 after fore/background subtraction) is proportional to volume of a cluster, and
 therefore proportional to $radius^3$. Since richness is a relatively
 easy parameter to measure from the imaging data, we use the following
 equation to calculate radius for each cluster.
 
  \begin{equation} 
    radius = 0.7\times(Richness/32)^{1/3}
 \end{equation}

 where median richness of our sample cluster is 32. The coefficient of
 the equation is adjusted so that median clusters in our sample have
 radius of 0.7 Mpc, which corresponds to the mean radius used in Butcher et
 al. (1978, 1984) and Margoniner et al. (2000, 2001). The distribution of
 radius calculated in this way is presented in Figure \ref{fig:new_rad}. As expected
 it has a peak at 0.7 Mpc. 
 Using this varying radius, we re-calculated all figures in Section
 \ref{bo_method}. Sample clusters are still required to have more than 25
 galaxies after fore/background subtraction within the new
 radius. Therefore the number of sample clusters are somewhat reduced to
 413 clusters. 
 Results are presented in Figures
 \ref{fig:bo_new}-\ref{fig:okamura_elong_new}. 
 Reassuringly, all 
 figures have the same trend as presented in Section
 \ref{bo_method}. Therefore the discussion in Section \ref{bo_discussion}
 still holds.  Although it is ideal
 to use virial radius to measure blue/spiral fractions of clusters, we
 regard that our analysis using fixed 0.7 Mpc radius is not hampered to
 the extend where our main conclusions change.

\section{Summary}\label{bo_conclusion}

  In this chapter, we have investigated the fraction of late type galaxies in
  four different ways using one of the largest, most uniform samples of
  514 clusters between 0.02$\leq z\leq$0.3 from the SDSS Cut \& Enhance galaxy cluster catalog.
  All the clusters selected here have more than 25 member galaxies within
  0.7 Mpc from the cluster center and between $M_{r^*}=-24.0$ and $-19.44$ after statistical
  local fore/background 
  subtraction. The following four different ways to estimate the fractions of late
  type galaxies are adopted: restframe $g-r$ color (a classical Butcher-Oemler estimator), $u-r$ color, concentration  index and de Vaucouleur/exponential profile fit. 
 The last two parameters are indicators of  galaxy
 morphologies (Shimasaku et al. 2001; Strateva et al. 2001).
 In all four cases, we observe an
  increase of the fraction of late type galaxies toward higher redshift
  with a significance of more than 99.99\% (Table \ref{tab:correlation}).   
 We draw the following conclusion from this work.

\begin{figure}[h]
\begin{center}
\includegraphics[scale=0.7]{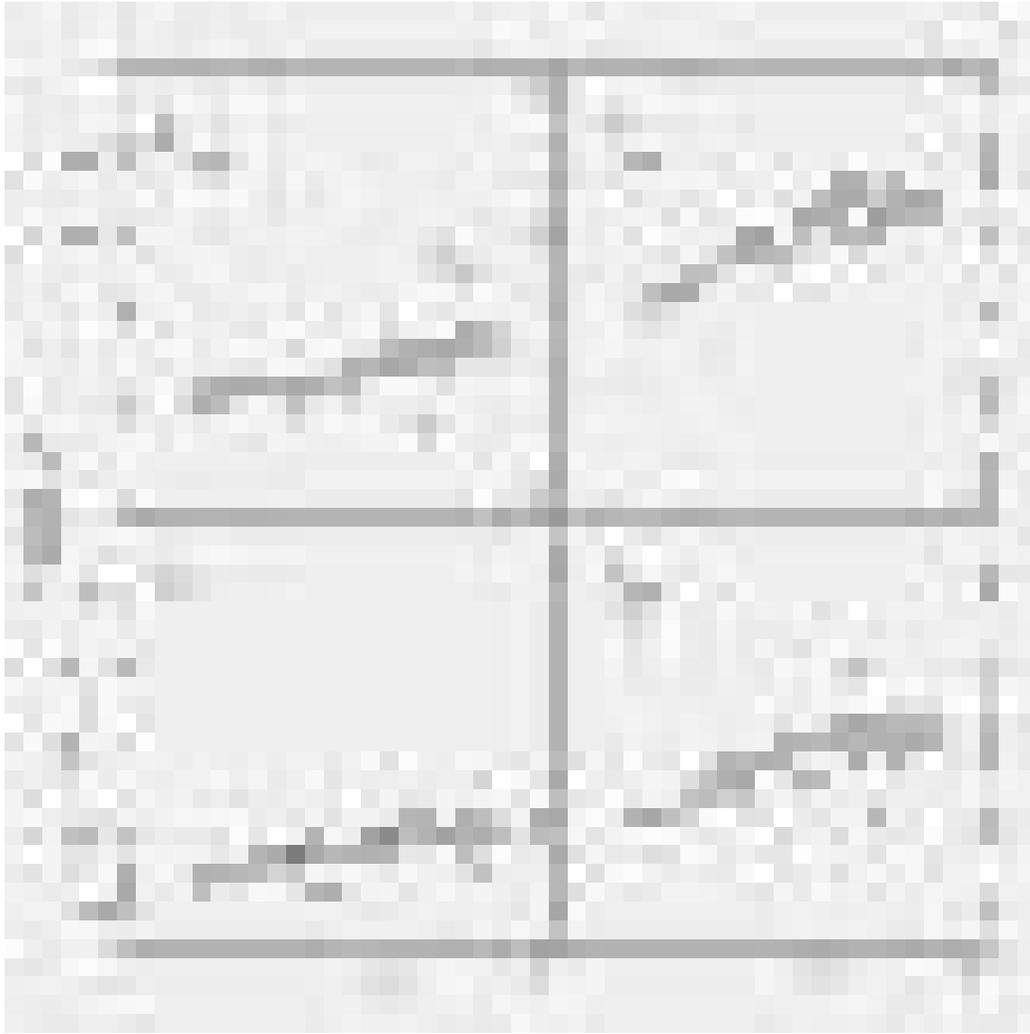}
\end{center}
\caption{\label{fig:bo}
 Photometric and morphological Butcher-Oemler effect from the 514 SDSS Cut \&
 Enhance clusters. 
$f_b$, $f_{Cin}$, $f_{exp}$ and $f_{u-r}$ are plotted against redshift. 
The dashed lines show the weighted least-squares fit to the data. The stars
 and solid lines show the median values. The median values of errors
 are shown in the upper left corners of each panel. The Spearman's correlation coefficients are shown in Table \ref{tab:correlation}.
}
\end{figure}

\begin{figure}[h]
\begin{center}
\includegraphics[scale=0.7]{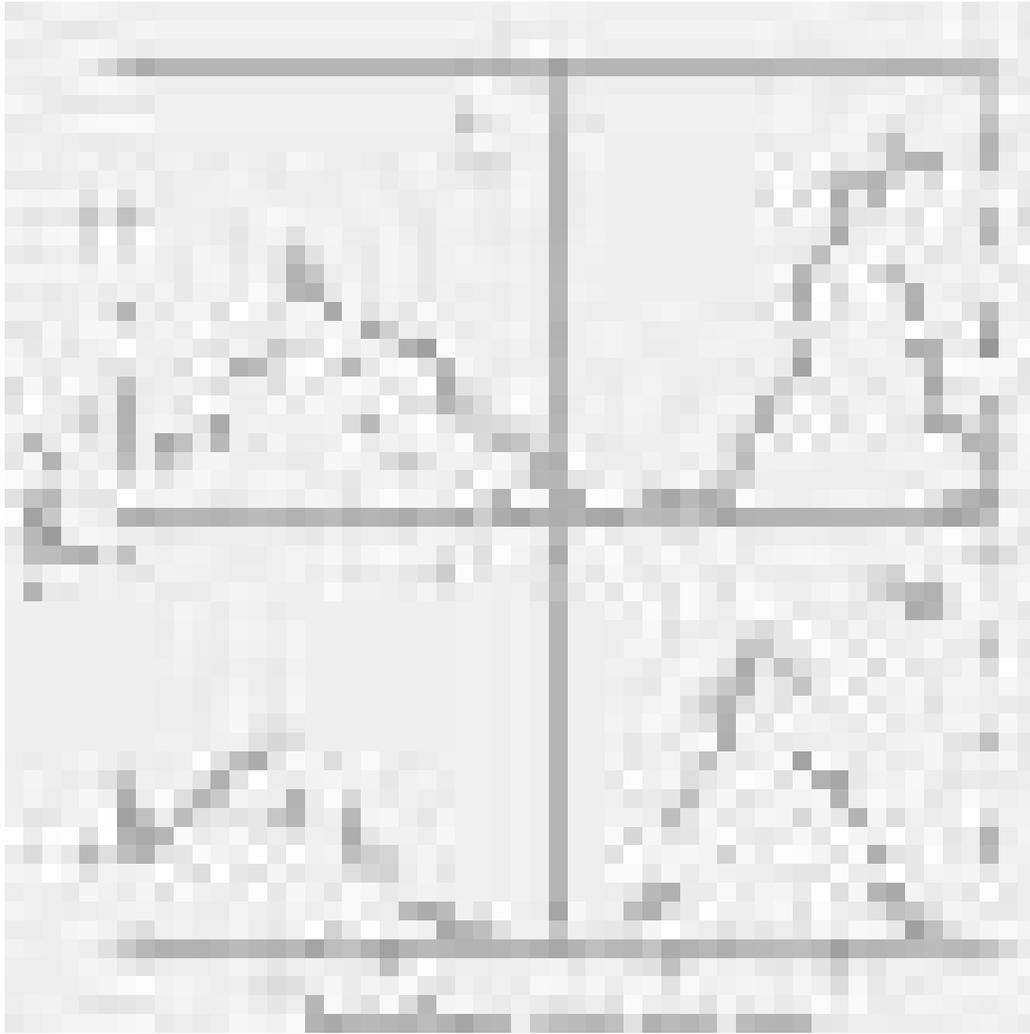}
\end{center}
\caption{\label{fig:kstest}
 Normalized distributions of late type fractions ($f_b$, $f_{Cin}$, $f_{exp}$ and $f_{u-r}$).
 The dashed lines show distributions of lower redshift clusters
 ($z\leq$0.15) and the solid lines show ones of higher redshift clusters
 (0.15$<z\leq$0.3).
 The results of Kolomogorov-Smirnov tests are shown in Table \ref{tab:kstest}. In all
 cases, Kolomogorov-Smirnov tests show the two distributions are significantly different. 
}
\end{figure}

\begin{figure}[h]
\begin{center}
\includegraphics[scale=0.7]{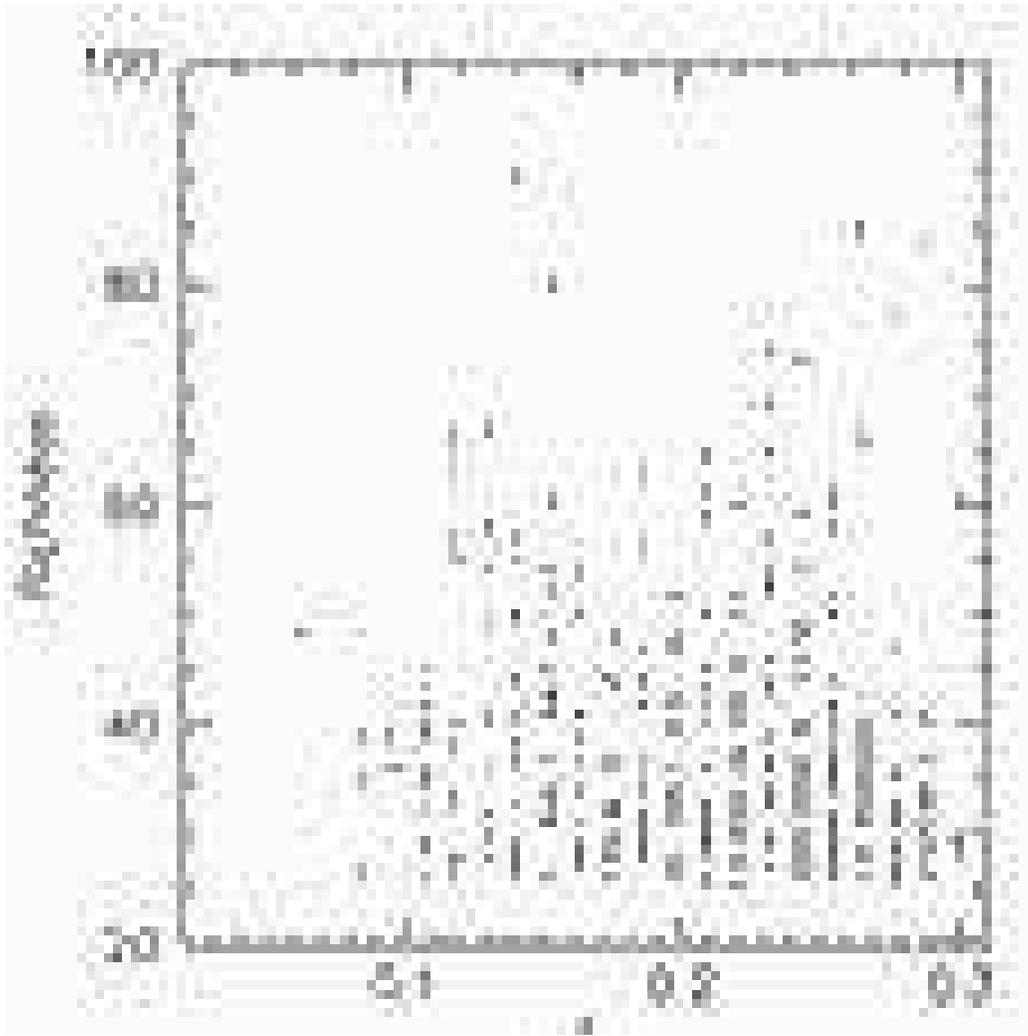}
\end{center}
\caption{
\label{fig:z_rich}
 Richness distribution as a function of redshift. Richnesses are measured
 as the number of galaxies brighter than $M_{r^*}=-19.44$ within 0.7 Mpc
 from the cluster center after
 fore/background subtraction. 
}
\end{figure}

 1. We confirm the presence of the Butcher-Oemler effect using $g-r$ color. The Butcher-Oemler effect is real and exists in our sample clusters as seen in the lower left panel of Figure \ref{fig:bo}. The slope of the increase is consistent with previous work although the scatter in the blue fraction is considerable. 
 Previous work also noted a large scatter in the fraction of blue galaxies. 
 The fraction of late type galaxies also shows a similar increase when
 we use a $u-r$ color cut. 


 2. We observe the morphological Butcher-Oemler effect as an
 increase of late type galaxies toward higher redshift, using pure morphological
 parameters such as a concentration parameter and de
 Vaucouleur/exponential profile fit. The rates of increase
 are consistent with previous work on the spiral to S0
 transition, albeit with  considerable scatter (Figure \ref{fig:bo}).
 The increase is also in agreement with the original Butcher-Oemler effect
 from $g-r$ color. Our results are consistent with the evolutionary
 scenario proposed by Dressler et al. (1997), Smail et al. (1997), Couch
 et al. (1998), and Kodama \& Smail (2001), 
 in which there is a progressive morphological conversion in
 clusters from spirals into E/S0's.  

\begin{figure}[h]
\begin{center}
\includegraphics[scale=0.7]{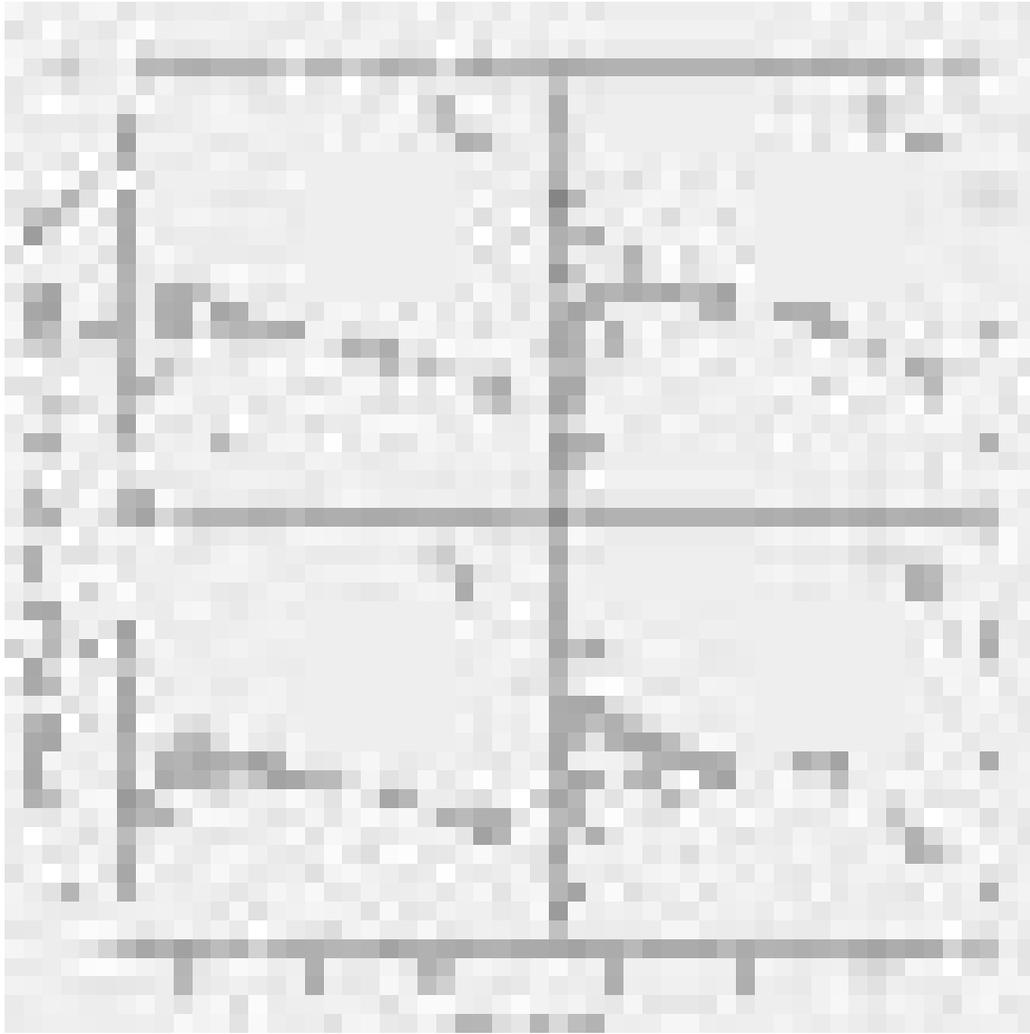}
\end{center}
\caption{
\label{fig:okamura_rich}
 The difference of the late type fractions from the best-fit lines as a
 function of redshift are plotted against cluster 
 richnesses. The solid lines and circles
 show the median values.
}
\end{figure}

\begin{figure}[h]
\begin{center}
\includegraphics[scale=0.7]{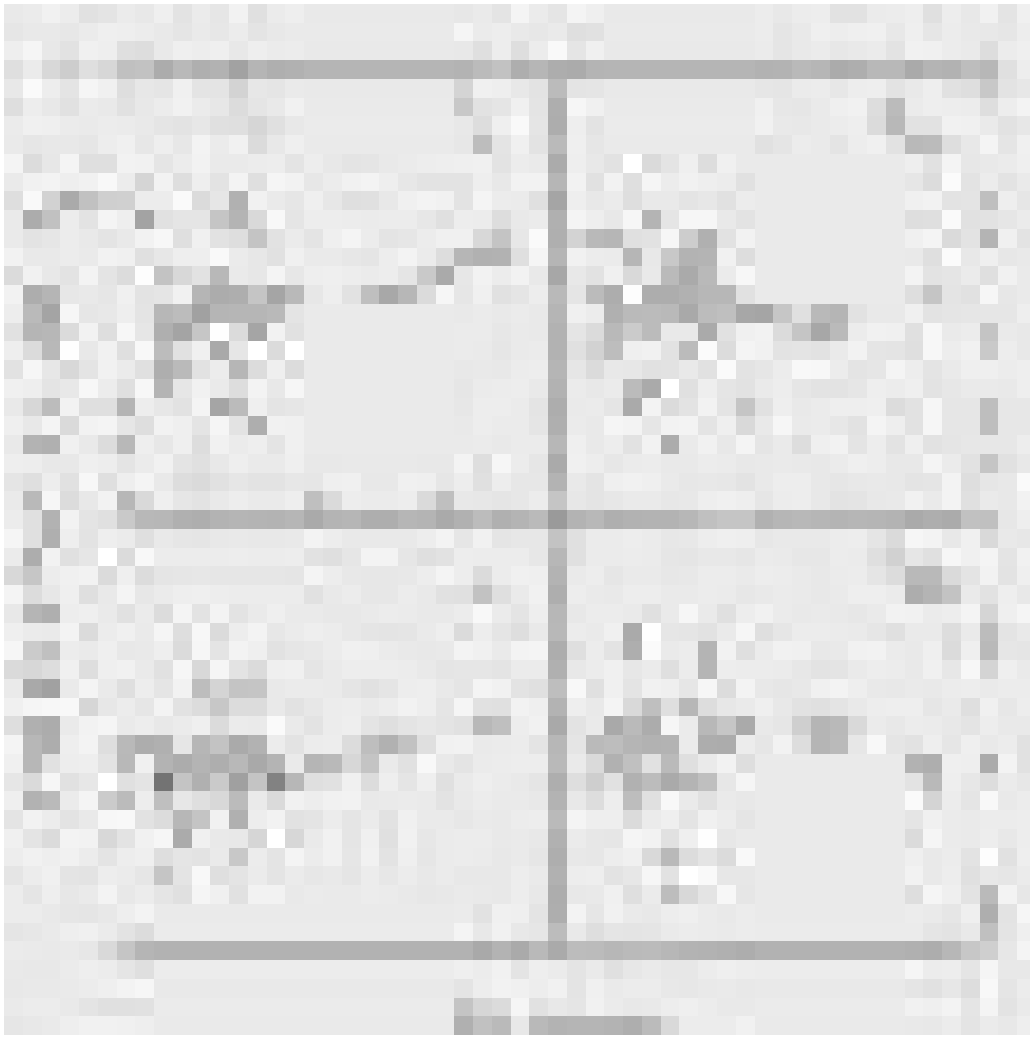}
\end{center}
\caption{
\label{fig:okamura_elong}
 The difference of late type fractions from the best-fit lines as a
 function of redshift are plotted against cluster elongation, which was
 measured as 
 a ratio of major axis to minor axis on an enhanced density map of Goto
 et al. (2002a; Chapter \ref{chap:CE}).  The solid lines and circles show median values.
}
\end{figure}

\begin{figure}[h]
\begin{center}
\includegraphics[scale=0.7]{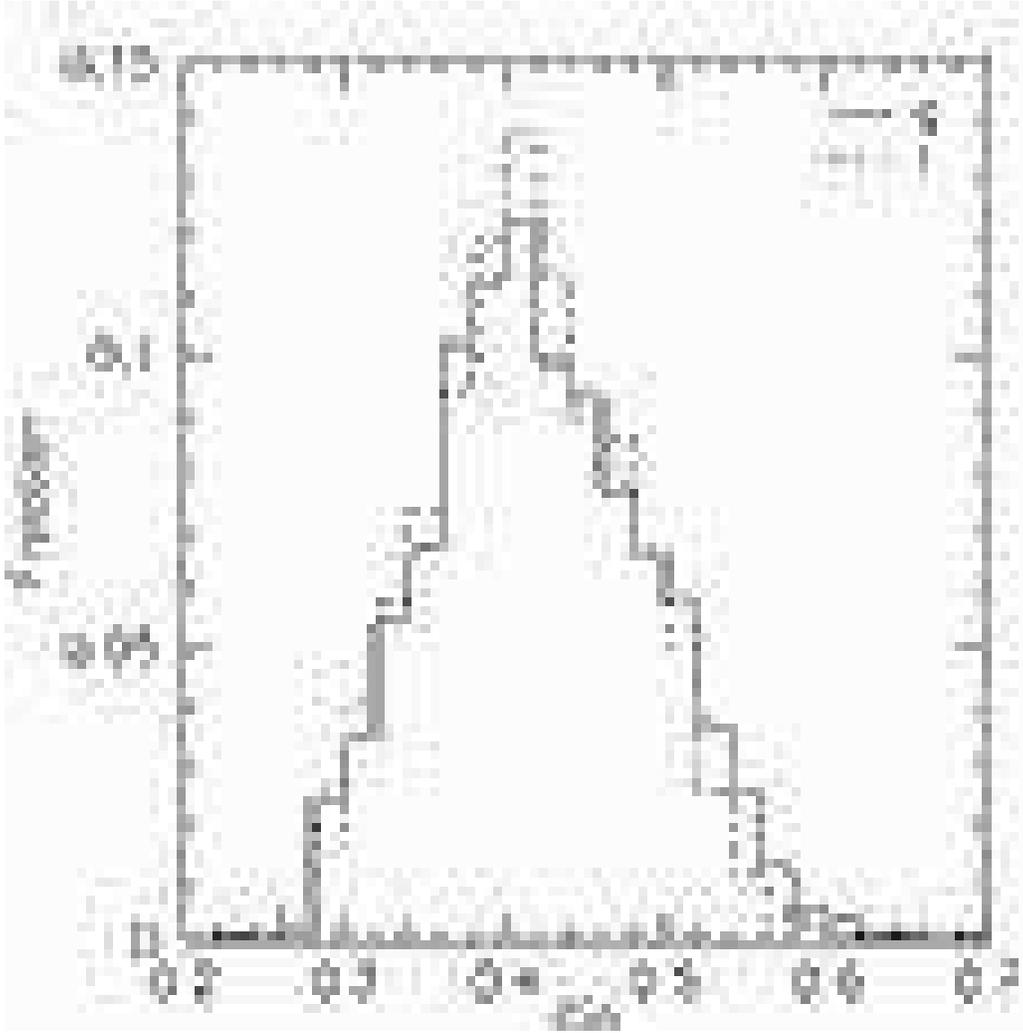}
\end{center}
\caption{
\label{fig:cin_g_r}
 The distribution of $Cin$, the inverse of the concentration index defined
 as the ratio of Petrosian 50\% flux radius to Petrosian 90\% flux
 radius, for 1336 local (0.02$\leq z\leq$0.03) SDSS galaxies. The solid line
 shows the distribution of $C_{in}$ measured in 
 the $g$ band image. The dashed line shows the distribution of $C_{in}$ measured in
 the $r$ band image. The difference between the $g$ band and $r$ band is
 marginal, assuring our usage of $r$ band $C_{in}$ in the upper right
 panel of Figure \ref{fig:bo} from $z=$0.02 to $z$=0.3. The
 statistics are summarized in Table \ref{tab:cin_g_r}.
}
\end{figure}

\begin{figure}[h]
\begin{center}
\includegraphics[scale=0.7]{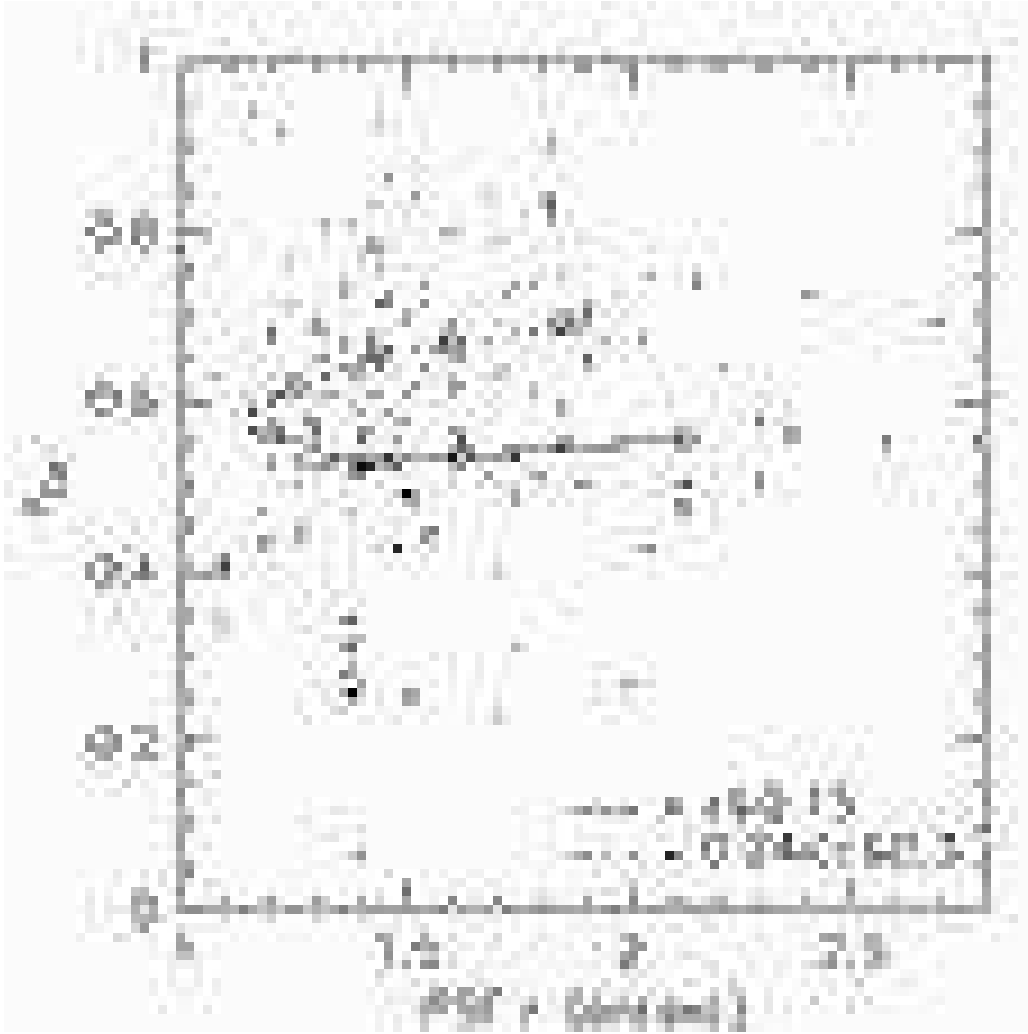}
\end{center}
\caption{
\label{fig:seeing_cin}
 The dependence of $f_{Cin}$ on seeing. The open squares and the solid lines show
 the distribution and medians of low $z$ clusters ($z\leq$0.15). The filled
 triangles and the dashed lines show the distribution and medians of high $z$
 clusters (0.24$<z\leq$0.3). The median bins are chosen so that equal
 numbers of galaxies are included in each bin.
}
\end{figure}

\begin{figure}[h]
\begin{center}
\includegraphics[scale=0.7]{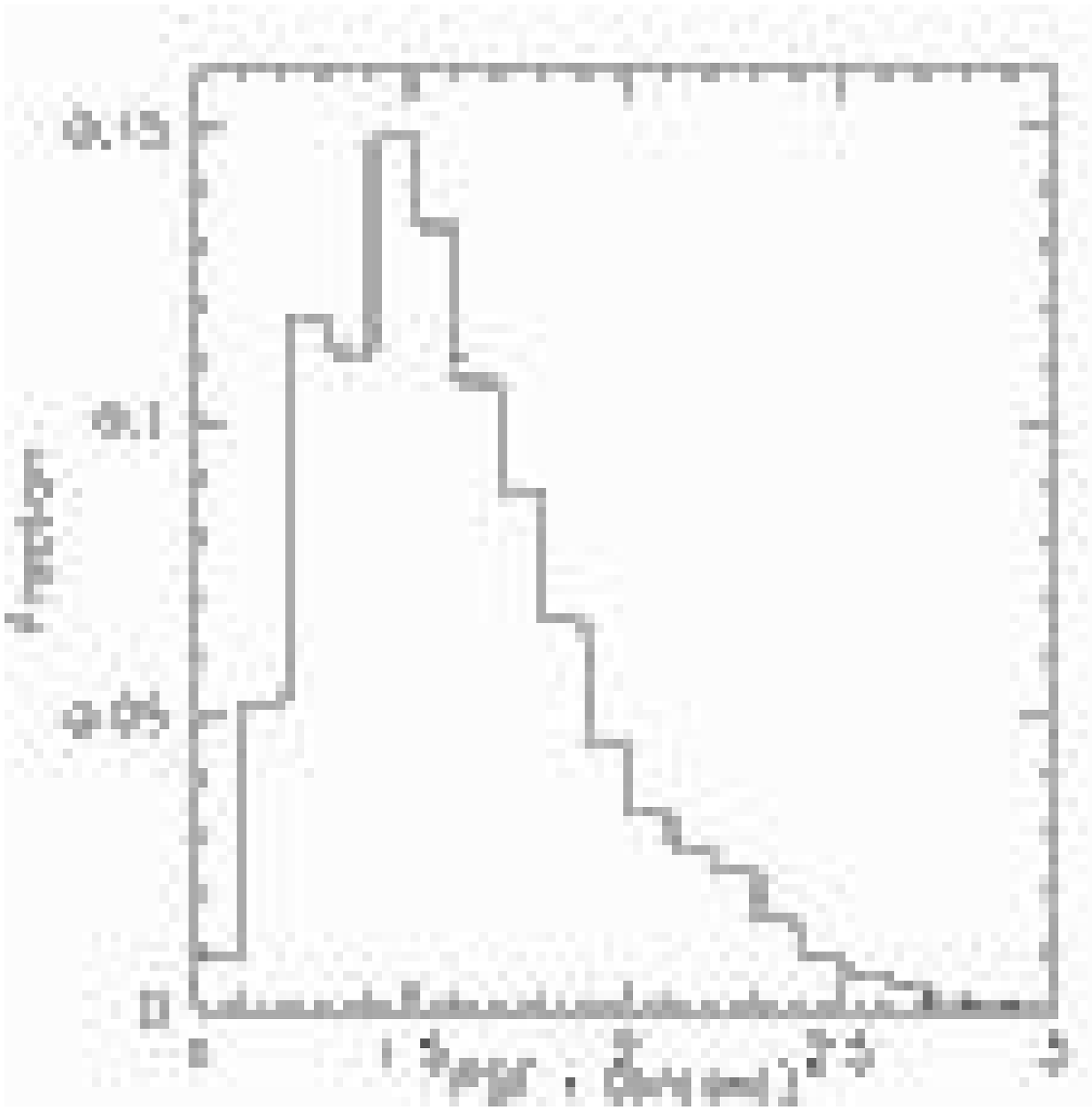}
\end{center}
\caption{
\label{fig:bo_seeing_hist}
The seeing distribution of all galaxies brighter than $r$=21.5. 87\% of all
 galaxies have seeing better than 2.0 arcsec. 
}
\end{figure}

\begin{figure}[h]
\begin{center}
\includegraphics[scale=0.7]{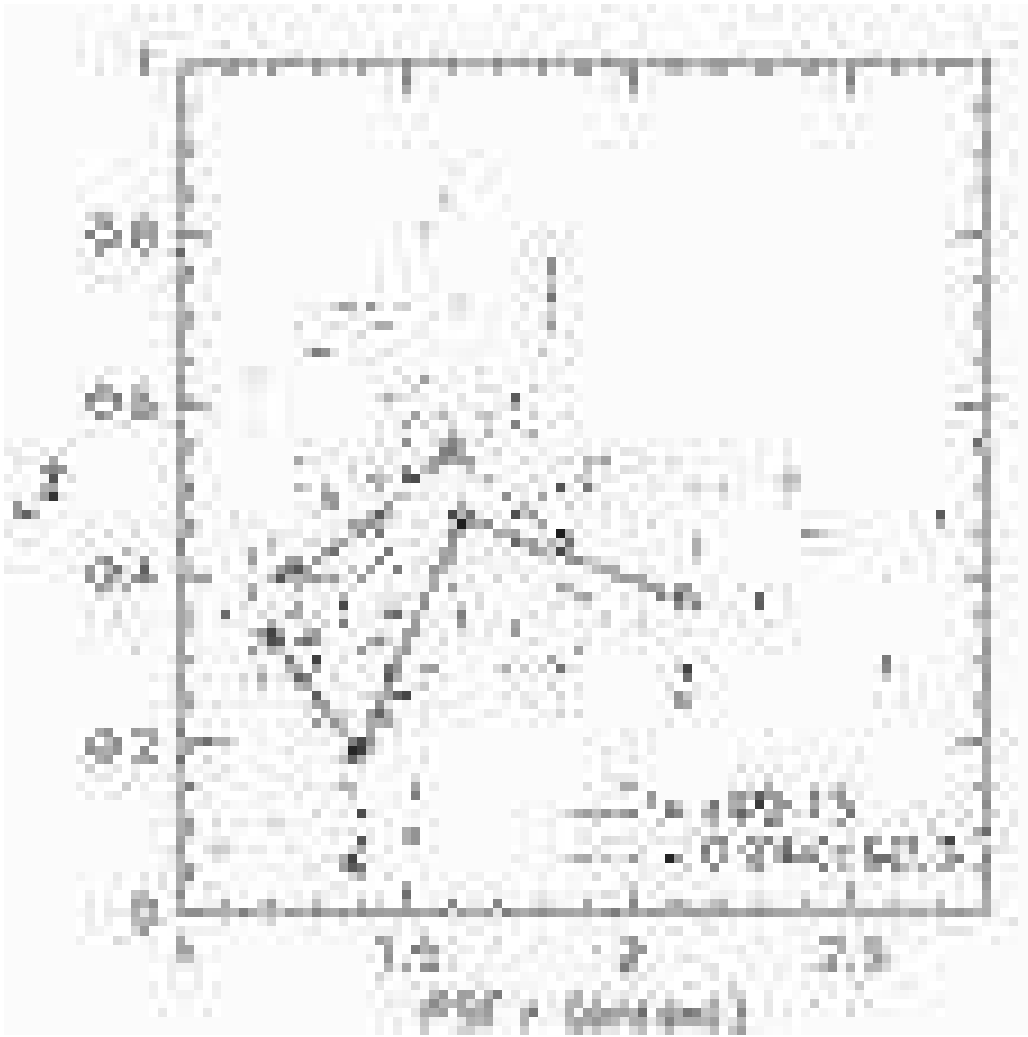}
\end{center}
\caption{
\label{fig:seeing_exp}
The dependence of $f_{exp}$ on seeing. The open squares and the solid lines show
 the distribution and medians of low $z$ clusters ($z\leq$0.15). The filled
 triangles and the dashed lines show the distribution and medians of high $z$
 clusters (0.24$<z\leq$0.3). Median bins are chosen so that equal
 numbers of galaxies are included in each bin.
 1 $\sigma$ errors shown as vertical bars are more dominant.
 There is no significant trend with seeing.
}
\end{figure}

\begin{figure}[h]
\begin{center}
\includegraphics[scale=0.7]{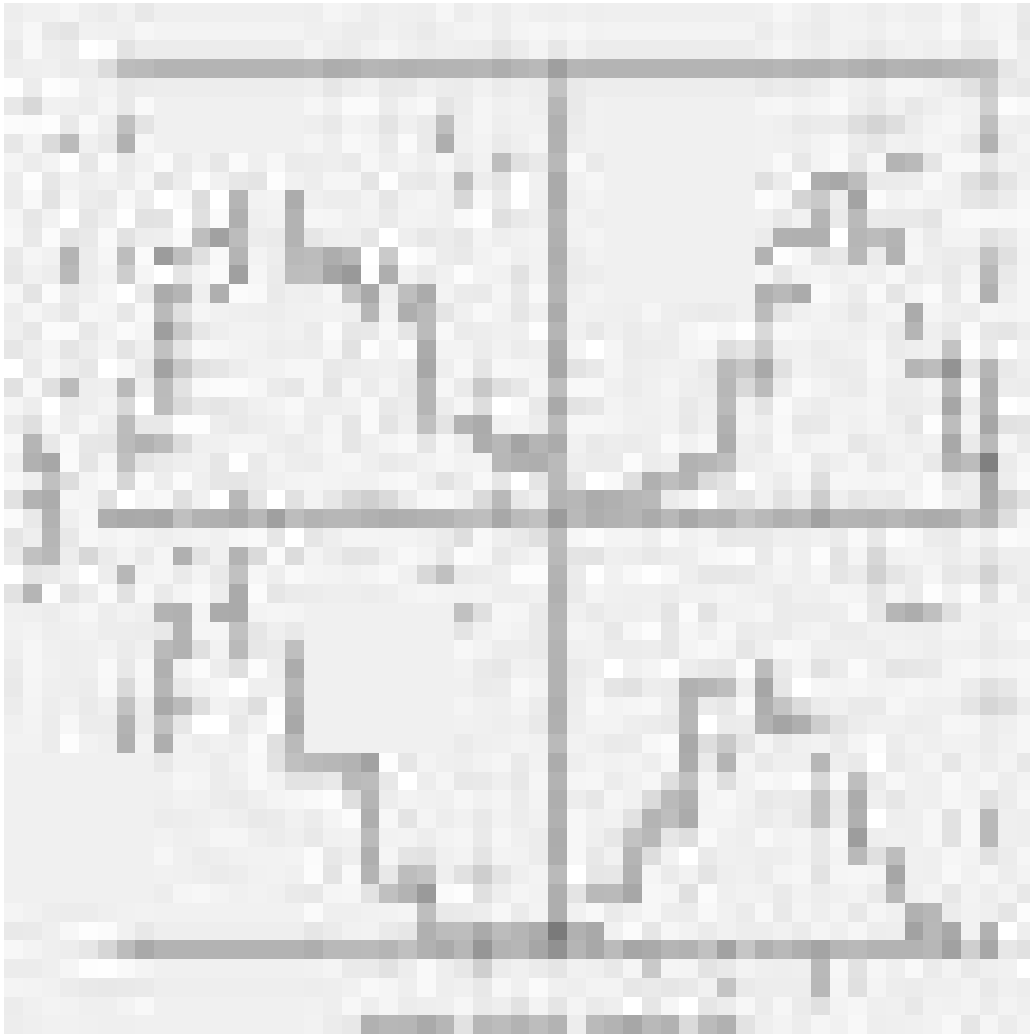}
\end{center}
\caption{
\label{fig:bg_test}
 Various systematic tests. The solid lines show distributions for
 2.1-2.21 Mpc annular fore/background subtraction. The dashed lines show
 distributions for a global background subtraction. The dotted lines show
 distributions using 0.7/(1+$z$) Mpc radius assuming a standard cold dark matter cosmology. 
 The long dashed lines show distributions using the brightest galaxy position as a cluster center. 
  In none of the cases does a Kolomogorov-Smirnov test show
 significant difference between the distributions (significance to be different is less than 26\% in all cases).
}
\end{figure}

\begin{figure}[h]
\begin{center}
\includegraphics[scale=0.7]{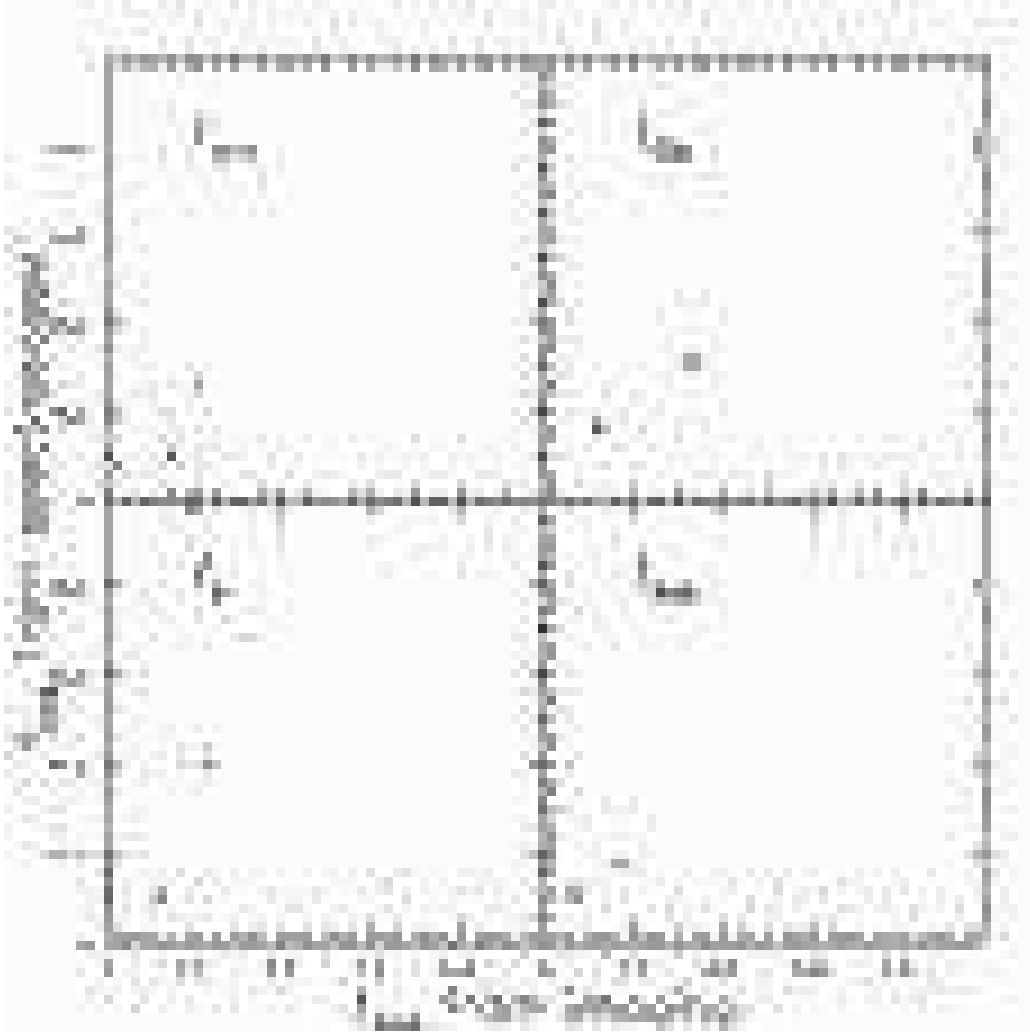}
\end{center}
\caption{
Comparison of the late-type fraction from imaging with that from spectroscopy. 
 Late-type fractions measured using spectroscopic data are plotted
 against that from imaging data for three clusters with $z<$0.06 (ABELL 295, RXC
 J0114.9+0024, and  ABELL 957). The dashed lines are drawn to guide
 eyes. All points agree with each other within the error. 
}\label{fig:comparison_with_spectroscopy}
\end{figure}

\begin{figure}[h]
\begin{center}
\includegraphics[scale=0.7]{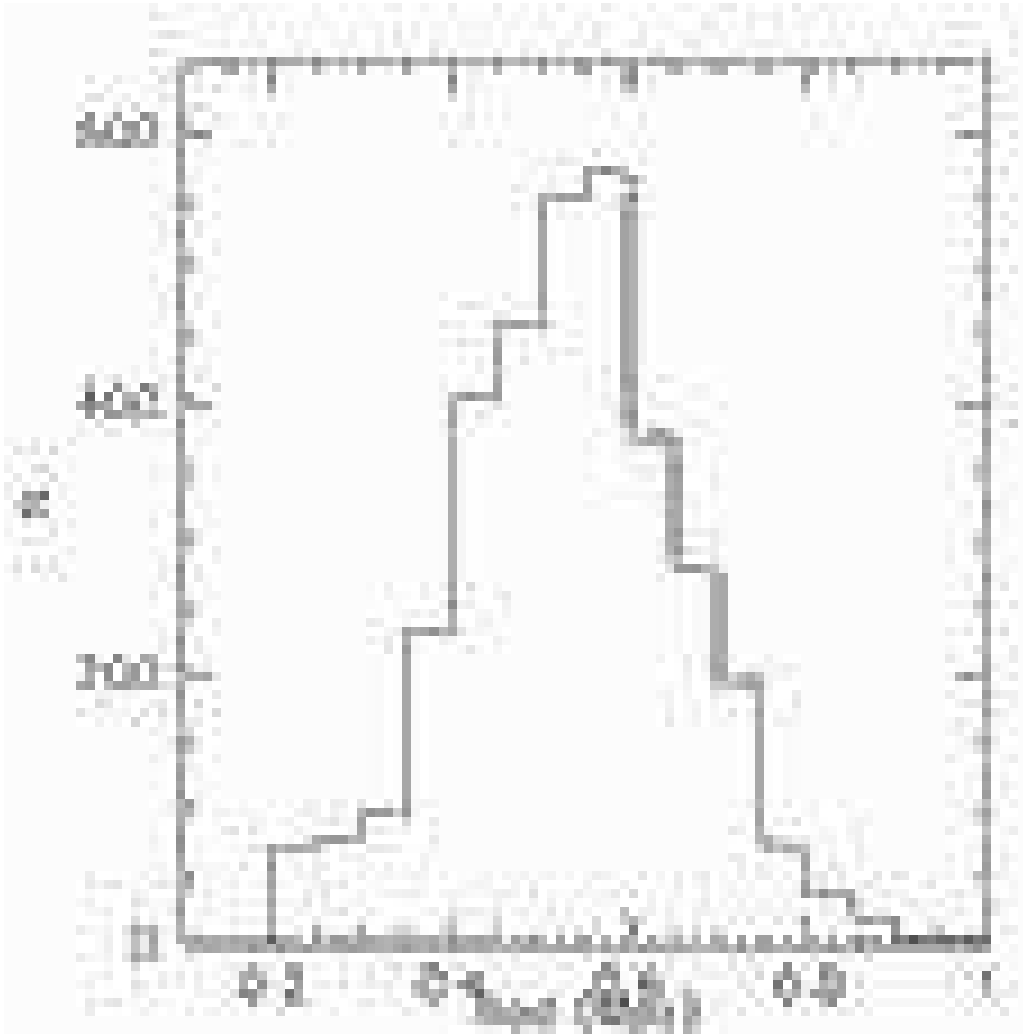}
\end{center}
\caption{
 Distribution of varying radius to measure blue/spiral fractions. It has
 a peak at 0.7 Mpc.
}\label{fig:new_rad}
\end{figure}

\begin{figure}[h]
\begin{center}
\includegraphics[scale=0.7]{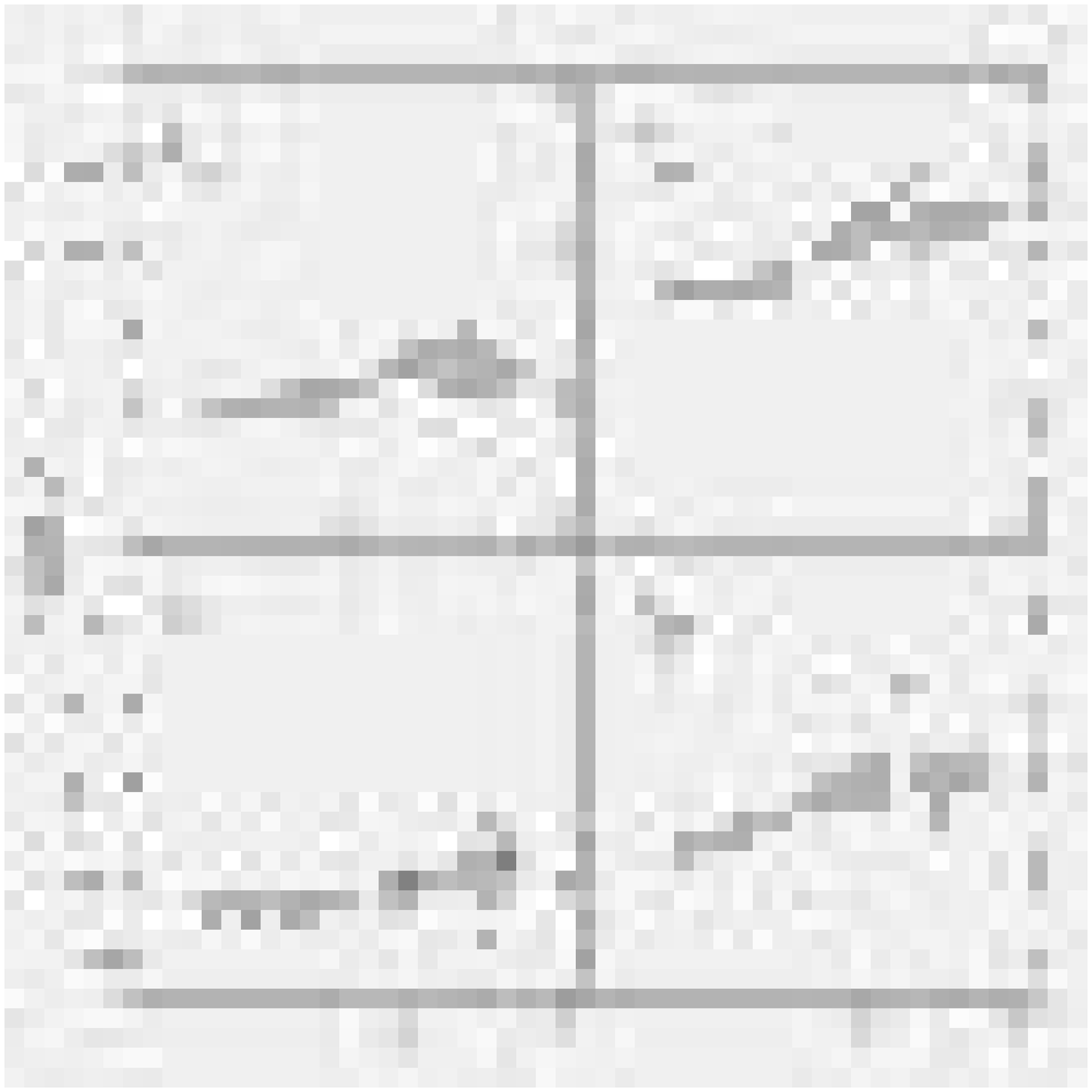}
\end{center}
\caption{
 The same as Figure \ref{fig:bo}, but measured with varying radius.
}\label{fig:bo_new}
\end{figure}

\begin{figure}[h]
\begin{center}
\includegraphics[scale=0.7]{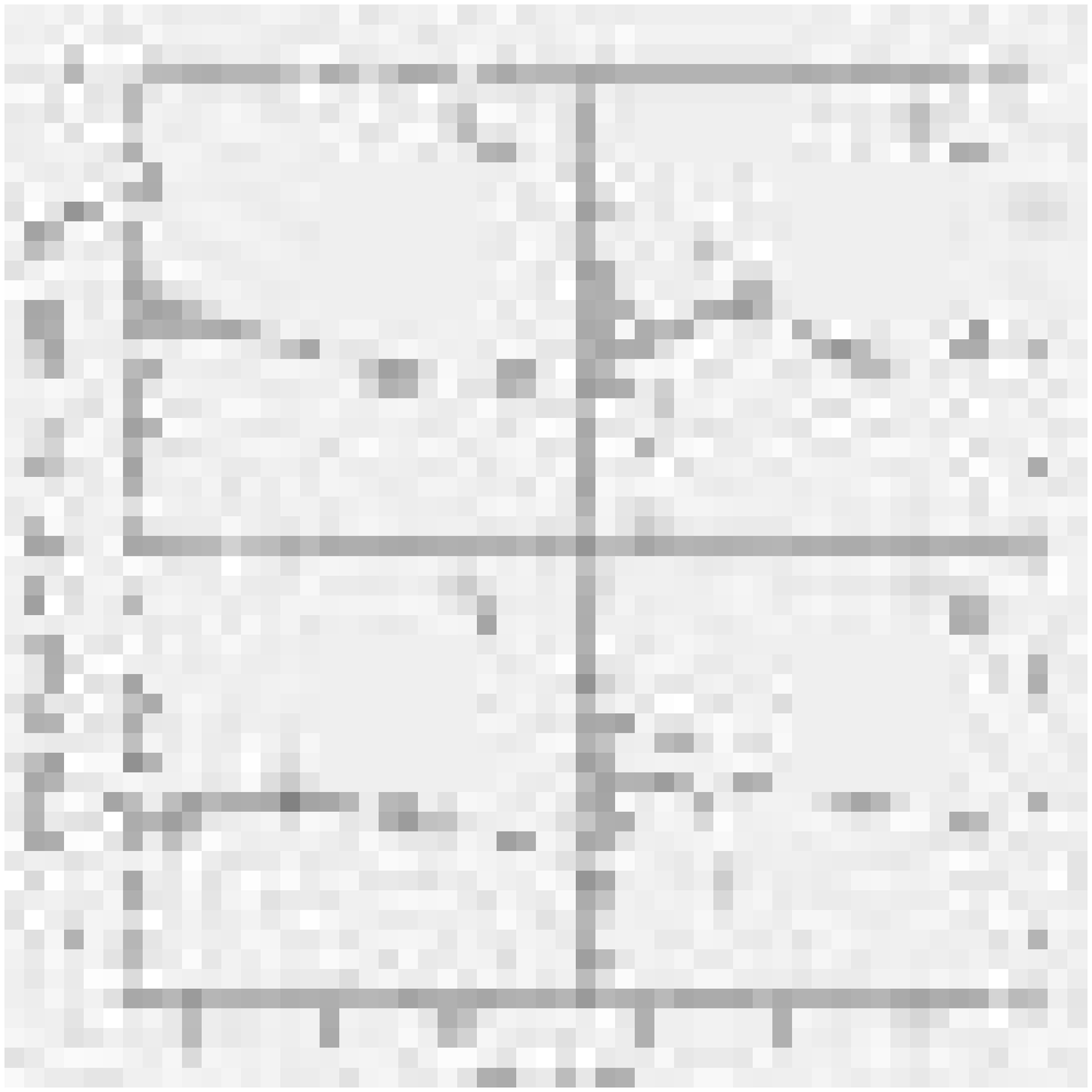}
\end{center}
\caption{
 The same as Figure  \ref{fig:okamura_rich}, but measured with varying radius.
}\label{fig:okamura_rich_new}
\end{figure}

\begin{figure}[h]
\begin{center}
\includegraphics[scale=0.7]{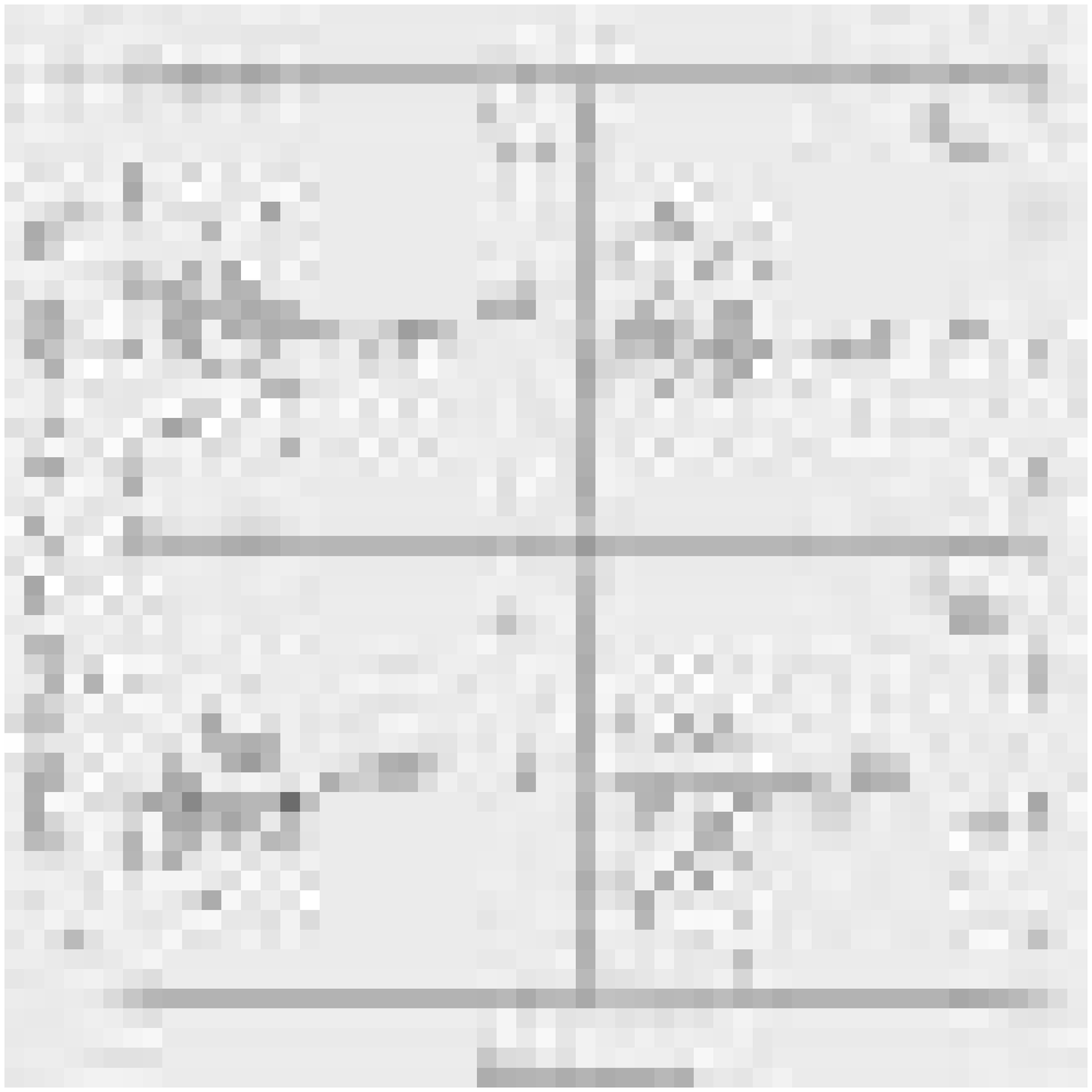}
\end{center}
\caption{
  The same as Figure \ref{fig:okamura_elong}, but measured with varying radius.
}\label{fig:okamura_elong_new}
\end{figure}

 3. We find a slight tendency for richer clusters to have lower values of
    the late type fraction (Figure \ref{fig:okamura_rich}).  This trend
    agrees with the ram pressure stripping model proposed by Bahcall (1977)
    and Fujita et al. (1999), in
    which galaxies in richer clusters are more affected by ram pressure due to
    their high temperature.

 Although our results 1,2, and 3 are all consistent with a ram-pressure
 stripping model, there still remains a possibility that other physical
 mechanisms are  responsible for the evolution of cluster galaxies.
 Thus, further study is needed both theoretically and observationally to
 reveal the underlying physical mechanism responsible for the evolution
 of cluster galaxies.
 Since this work is based on only 5\% of the whole SDSS data, an
 increase in the data will improve the statistical accuracy as the SDSS
 proceeds. Extending the work to higher redshifts using 4-8 m class
 telescopes will offer more insight on the origin and evolution of
 cluster galaxies.

\bigskip

\newpage

\newpage

\newpage

\newpage

\newpage

\newpage

\newpage

\newpage

\newpage
\newpage

\newpage

\newpage

%
%

\newpage

\newpage

\newpage

\newpage

\newpage
\clearpage

%
%
%

%

\begin{table}[h]
\caption{
\label{tab:correlation}
  Spearman's correlation coefficients between $z$ and fractions of late type
 galaxeis. 514 clusters with richness$>$25 are chosen as a sample.
}
\begin{center}
\begin{tabular}{llll}
\hline
  & Correlation coefficient  & Significance  & N clusters \\
\hline
\hline
 $f_b$      &        0.238 & 4.4$\times$10$^{-8}$ &         514\\ 
 $f_{u-r}$  &        0.234 & 7.6$\times$10$^{-8}$   &      514\\ 
 $f_{exp}$  &       0.194 & 9.6$\times$10$^{-6}$ &         514\\
 $f_{Cin}$  &       0.223 & 2.9$\times$10$^{-7}$  &         514\\ 
\hline
\end{tabular}
\end{center}
\end{table}

\begin{table}[h]
\caption{
\label{tab:kstest}
 Significances in Kolomogorov-Smirnov tests between distributions
 for $z\leq$0.15 and 0.15$<z<$0.3. In all cases, Kolomogorov-Smirnov tests show the distributions for
 the lower redshift sample and the higher redshift sample are significantly different. 
}
\begin{center}
\begin{tabular}{ll}
\hline
  & Significance  \\
\hline
\hline
 $f_b$      &  2.9$\times$10$^{-3}$ \\ 
 $f_{u-r}$  &  1.0$\times$10$^{-3}$ \\ 
 $f_{exp}$  &  3.4$\times$10$^{-4}$ \\
 $f_{Cin}$  &  2.9$\times$10$^{-3}$ \\ 
\hline
\end{tabular}
\end{center}
\end{table}

\begin{table}[h]
\caption{
Scatters in late-type fractions around the best-fit line are compared
 with median errors of late-type fraction calculated with equation (\ref{frac_err}). 
}\label{tab:error_comparison}
\begin{center}
\begin{tabular}{lllll}
\hline
     &  $f_b$  & $f_{u-r}$ & $f_{exp}$ & $f_{Cin}$\\
\hline
\hline
 Real scatter (1$\sigma$) &  0.169  & 0.183  & 0.171 & 0.163    \\
 Error estimate           &  0.078  & 0.069  & 0.050 & 0.089    \\
\hline
\end{tabular}
\end{center}
\end{table}

\begin{table}[h]
\caption{
\label{tab:cin_g_r}
Change in the fraction of galaxies with $C_{in}>$0.4 (late type) in two different filters($g,r$).
}
\begin{center}
\begin{tabular}{llll}
\hline
band & N($C_{in}>$0.4)  & N(total) & Percentage(\%) \\
\hline
\hline
 $g$ & 802 & 1336 &     60.0       \\
 $r$ & 787 & 1336 &     58.9       \\
\hline
 Difference & 15  & 1336   &  1.1\\ 
\hline
\end{tabular}
\end{center}
\end{table}

\begin{table}[h]
\caption{
\label{tab:exp_g_r}
Change in the fraction of galaxies with exponential fit likelihood
 greater than de Vaucouleur likelihood (late type) in two different
 filters($g,r$). Computations are performed using the local
 (0.02$\leq z\leq$0.03) SDSS galaxies. Since we discard the galaxies
 with the same likelihood in this analysis, the total number of galaxies
 in the sample are different in $g$ and $r$.
}
\begin{center}
\begin{tabular}{llll}
\hline
band & N(late)  & N(late + early) & Percentage(\%) \\
\hline
\hline
 $g$ & 503 & 804 &     62.6       \\
 $r$ & 476 & 792 &     60.1       \\
\hline
 Difference & -  & -   &  2.5\\ 
\hline
\end{tabular}
\end{center}
\end{table}


\chapter{The Morphology-Density Relation}
\label{chap:MD}
	
\section{Introduction}\label{md_intro}

 Morphological types of galaxies are one of the most basic properties
 and thus have been studied since the beginning of the extragalactic
 astronomy. However, it is still not very well understood where this
 diversity stems from. The existence of a correlation between galaxy
 morphology and local environment is a remarkable feature of galaxy population.
 Dressler (1980) studied 55 nearby galaxy clusters and found that
 fractions of elliptical 
 galaxies increase and that of spiral galaxies decrease with
 increasing local galaxy density in all clusters.
  The discovery left a great impact on
 astronomical community since it indicates that physical mechanisms that
 depend on environment of each galaxy mainly affect the final
 configuration of stellar component. Further observational constraints
 on the formation and evolution of galaxies were obtained by extending
 the analysis of the morphology-density relation to group of galaxies in
 the general field. Postman \& Geller (1984) extended morphology study
 to groups using the data from the CfA Redshift Survey (Huchra et
 al. 1983). 
  The relation was
 completely consistent with Dressler 
 (1980). At low densities, population fractions seemed to be independent
 of density below galaxy density $\sim$5 Mpc$^{-3}$. At high density,
 the elliptical fraction increased steeply above $\sim$3000 galaxies
 Mpc$^{-3}$. 
 Whitmore et al. (1993) re-analyzed the 55 nearby clusters
 (Dressler 1980) and argued that the morphology-density relation
 reflects a more fundamental morphology-radius relation; the
 correlation between morphology and cluster centric radius seems tighter
 than the morphology-density relation. This assertion is still
 controversial. 
 The opposite results on groups came from Whitmore et al. (1995). They analyzed
 the morphology-density relation in groups of galaxies by carefully
 removing cluster galaxies from their analysis and found that the
 relation is very weak or non-existent in groups. Helsdon \& Ponman
 (2002) also studied the morphology-density relation in groups using
 their X-ray bright group sample.
 
 Later the relation between morphology and density was traced back to
 higher redshift. 
 Dressler et al. (1997) studied 10 high redshift clusters at $z\sim$0.5
 and found that the morphology-density relation is strong for centrally
 concentrated clusters. However, the relation was nearly absent for less
 concentrated or irregular clusters. They also found that S0 fractions
 are much smaller than in nearby clusters, suggesting that S0 galaxies
 are created fairly recently ($z\leq$0.5). 
 Fasano et al. (2000) studied nine clusters at intermediate redshift 
 (0.1$\leq z\leq$0.25) and compared them with local (Dressler 1980) and
 high redshift clusters (Dressler et al. 1997). They found that the
 morphology-density relation exists in  high elliptical concentration clusters,
 but not in low elliptical concentration clusters.
 The finding is consistent with the results of Dressler et
 al. (1997). Considering that 
 low redshift clusters have the morphology-density relation regardless of
 the concentration of clusters, they suggested that spiral to S0 transition
 happened fairly recently (last 1-2 Gyr). They also plotted morphological
 fraction as a function of redshift and found that S0 fraction decreases
 with increasing redshift, whereas spiral fraction increases with
 redshift.  

 Hashimoto et al. (1999) used data from the Las Campanas Redshift Survey
 (LCRS; Shectman et al. 1996) to study the concentration-density
 relation. They found that the ratio of high to low concentrated
 galaxies decreases smoothly with decreasing density.    
 Dominguez et al. (2001) analyzed nearby clusters with X-ray and found
 that mechanisms of global nature (X-ray mass density) dominate in high
 density environments, 
 namely the virialized regions of clusters, while local galaxy density
 is the relevant parameter in the outskirts where the influence of
 cluster as a whole is relatively small compared to local effects.  
 Dominguez et al. (2002) studied groups in the 2dF Galaxy Group Catalog
 using PCA analysis of spectra as a galaxy classification and local galaxy
 density from redshift space as a measure of galaxy environment. 
 They found that both morphology-density relation and
 morphology-group-centric radius relation is clearly seen in high mass
 ($Mv\geq$10$^{13.5}M_{\odot}$) groups, but neither relation holds true for
 low mass ($Mv<$10$^{13.5}M_{\odot}$) groups. 
 These three studies made innovative step in terms of an analysis
 method, using automated morphological classification  and three dimensional density estimation.

   Various physical mechanisms have been proposed to explain the
  morphology-density relation.
  Possible causes include ram pressure stripping of gas (Gunn \& Gott 1972; Farouki
 \& Shapiro 1980; Kent 1981; Abadi, Moore \& Bower 1999; Quilis, Moore \& Bower
 2000), galaxy infall (Bothun \& Dressler 1986; Abraham et al. 1996a; Ellingson et al. 2001), galaxy harassment via high speed impulsive encounters (Moore et al. 1996,1999), cluster
 tidal forces (Byrd \& Valtonen 1990; Valluri 1993) which distort
 galaxies as they come close to the center, interaction/merging of
 galaxies (Icke 1985; Lavery \& Henry 1988; Bekki 1998), and removal \& consumption of the gas due to the cluster environment (Larson, Tinsley \& Caldwell 1980; Balogh et
 al. 2001; Bekki et al. 2002). Mamon (1992) and Makino \& Hut (1997) showed that
 interactions/mergers can occur in a rich cluster environment despite the
 high relative velocities. Shioya et al. (2002) showed that the
  truncation of star formation can  explain the decrease of S0 with increasing redshift. 
 Although these processes are all plausible, the effects provided by initial
  condition on galaxy formation could be also important. Since field
  galaxies have different population ratio compared with clusters (Goto
  et al. 2002b; Chapter \ref{chap:LF}), a change in
  infalling rate of field galaxies into clusters affects population
  ratio of various galaxy types (Kodama et al. 2001).
 Unfortunately, there exists little evidence demonstrating that any one
  of these processes is actually responsible for driving galaxy
  evolution. Most of these processes act over an extended period of
  time, while observations at a certain redshift cannot easily provide
  the detailed information that is needed to elucidate subtle and
  complicated processes.  

 To extract useful information from observational data, it is necessary
 to have detailed theoretical predictions.
 In recent years, due to the progress of computer technologies,
 it is becoming possible to simulate the morphology-density relations by
 combining semi-analytic modeling with N-body simulations of cluster
 formation.  
 Okamoto \& Nagashima (2001) simulated the morphology-density relation
 using a merger-driven bulge formation model. They found that elliptical
 fractions are well re-produced, but there remained a discrepancy on S0
 fractions.  Diaferio et al. (2001) also assumed that the
 morphologies of cluster galaxies are determined solely by their merging
 histories in the simulation. They used bulge-to-disc ratio to classify
 galaxy types and compared the cluster-centric radial distribution with
 those derived from the CNOC1 sample (Yee, Ellingson, \& Carlberg 1996). They
 found excellent agreement for bulge 
 dominated galaxies, but simulated clusters contained too few galaxies
 of intermediate bulge-to-disc ratio.  Springel et al. (2002) used a
 phenomenological simulation to predict the morphology-radius relation
 and compared it with Whitmore et al.(1993). Their morphological
 modeling is based on the merging history of galaxies. They found an
 excellent agreement with elliptical galaxy fractions, and some
 deficiency of S0 galaxies in the core of the cluster. Benson et
 al. (2002) combined  their N-body simulation with
 a semi-analytic model (Cole et al. 2000) to trace the
 time evolution of the morphology-density relation. Interestingly, they
 found that a strong morphology-density relation was well established by
 $z$=1. The relation was qualitatively similar to that at $z=$0. E/S0
 galaxies are treated as one population in their simulation.
 Three of above simulations
 suggest that (i) elliptical fractions are consistent with the merging
 origin; (ii) however, the deficit of S0 galaxies shows that processes other
 than major-merger might be important 
 for S0 creation. Therefore more than one mechanisms might be required to
 fully explain the morphology-density relation. These suggestions might be
 consistent with observational results from  Dominguez et al. (2001),
 who found two different key  parameters in cluster center and outskirts
 separately.  
  
  In the previous analysis of the morphology-density relation from
  observations, there have been two major difficulties; eye-based
  morphological classification and the density estimate from two
  dimensional imaging data. 
   Although it is an excellent tool to
 classify galaxies, manual classification could potentially have unknown
 biases (Lahav et al. 1995).  A machine based, automated classification 
 would better control biases and would allow a reliable determination of
 the completeness and false positive rate.
  Measuring local galaxy density from imaging data requires statistical
  background subtraction, which automatically introduces relatively large uncertainty
  associated with itself. 
  Furthermore deeper imaging data require
 larger corrections. Therefore three dimensional density
  determination from redshift data is preferred. 
    With the advent of the Sloan Digital Sky Survey (SDSS; York et al. 2000),
  which is an imaging and spectroscopic survey of 10,000 deg$^2$ of the
  sky, we now have the opportunity to overcome these limitations. The
  CCD imaging of the SDSS allows us to estimate morphologies of galaxies
  in an automated way (Yamauchi et al. 2003). Three dimensional density
  can be estimated from the redshift information. Due to the large area
  coverage of the SDSS, we are able to probe the morphology-density
  relation from cluster core regions to the field region without
  combining multiple data sets with inhomogeneous characteristics.
    The purpose of this chapter is as follows. We aim to confirm or
  disprove the morphology-density relation using the automated morphology
  and three dimensional density from the SDSS data. We also re-analyze
  the MORPHS data ($z\sim$0.5)
  using an automated morphology (Smail et al. 1997).
  By comparing them to the SDSS, we
  try to observe the evolution of the morphology-density relation.   
 Final goal of our investigation is to shed some light on the origin
  of the morphology-density relation.

 The Chapter is organized as follows: In
 Section \ref{sec:md_data}, we describe the SDSS data. 
 In Section \ref{sec:md_analysis}, we explain automated morphological classifications and
 density estimation.
 In Sections  \ref{sec:md_morphology_density} and \ref{Oct 30 18:22:01 2002} we present the results from the SDSS data. 
 In Section \ref{Oct 30 18:41:06 2002}, we present the results from the MORPHS data.
 In Section \ref{md_discussion}, we discuss the possible caveats and underlying physical
 processes which determines galaxy morphology.  In
 Section \ref{sec:md_conclusion}, we summarize our work and findings.
   The cosmological parameters adopted throughout this chapter are $H_0$=75 km
 s$^{-1}$ Mpc$^{-1}$, and ($\Omega_m$,$\Omega_{\Lambda}$,$\Omega_k$)=(0.3,0.7,0.0).

\section{The SDSS Data}\label{sec:md_data}
 
 The data we use to study the morphology-density relation are 
 from the Sloan Digital Sky Survey Early Date Release (SDSS EDR;
 Stoughton et al. 2002), which covers $\sim$400 deg$^2$ of the sky. 
 The imaging part of the SDSS observes the sky in five optical bands
 ($u,g,r,i,$ and $z$; Fukugita et al. 1996). 
 Since the SDSS photometric system is not yet finalized, we refer to the
 SDSS photometry presented here as $u^*,g^*,r^*,i^*$ and $z^*$.
 The technical aspects of the SDSS camera are described in Gunn et al.\ (1998).  
 The SDSS spectroscopic survey observes the spectra of essentially all
 galaxies brighter than  $r^*$=17.77.
 The target galaxies are selected from imaging part of the survey
 (Strauss et al. 2002). 
 The spectra are observed using a pair of double fiber-fed
 spectrographs obtaining 640 spectra per exposure of 45 minutes.
 The wavelength coverage of the spectrographs is continuous from about
 3800 \AA{} to 9200 \AA{}, and  the wavelength resolution,
 $\lambda/\delta\lambda$, is 1800. The fiber diameter is 0.2 mm ($3''$ on the sky). 
 Adjacent fibers cannot be located closer than $55''$ on the sky.
 The throughput of the spectrograph will be better than 25\% over  4000
 \AA{} to 8000 \AA{} excluding the loss due to the telescope and
 atmosphere. 
 (See Eisenstein et al. 2001; Strauss et al. 2002 and  Blanton et
 al. 2002 for more detail of the SDSS spectroscopic data).  

 We use galaxies in the redshift range 0.05$<z<$0.1 with a redshift
 confidence of $\geq0.7$ (See Stoughton et
 al. 2002 for more details of the SDSS parameters). 
 The galaxies are limited to 
 $Mr^*<-$20.5 and $r^*<$17.77, which gives us a volume limited sample with 7938
 galaxies. We correct magnitudes for Galactic extinction using reddening
 map of Schlegel, Finkbeiner \& Davis (1998). 
 We use $k$-correction given in Blanton et al. (2002)  to calculate
 absolute magnitudes.

\section{Analysis}\label{sec:md_analysis}

\subsection{Morphological classification}\label{sec:md_morphology}

\begin{figure}[h]
\includegraphics[scale=0.7]{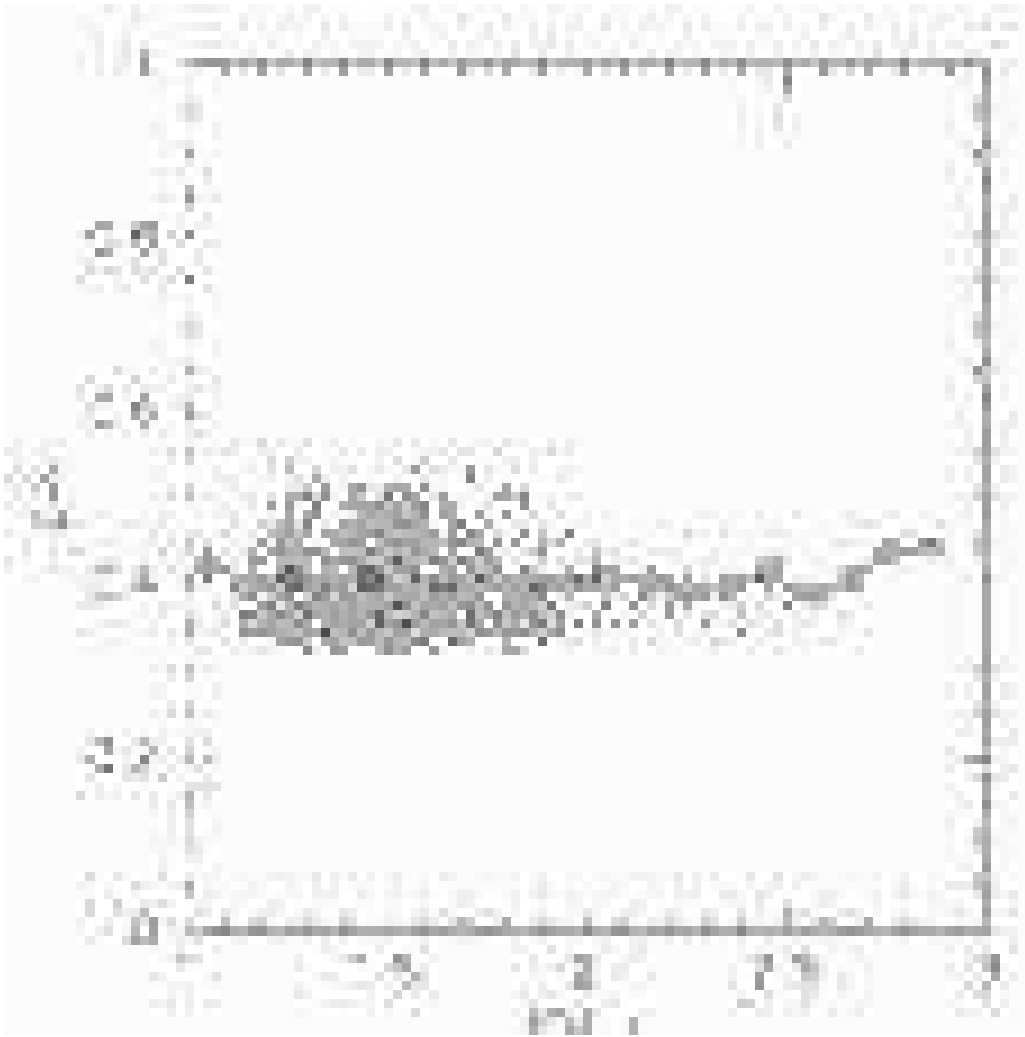}
\caption{
\label{fig:md_seeing_concent} 
 Seeing dependence of $Cin$ for 7938 galaxies used in the present
 study. The solid
 lines show medians. 87\% of our sample galaxies have seeing  
 between 1.2 and  2 arcsec, where seeing dependence of $Cin$ is negligible.  
}
\end{figure}

\begin{figure}[h]
\includegraphics[scale=0.7]{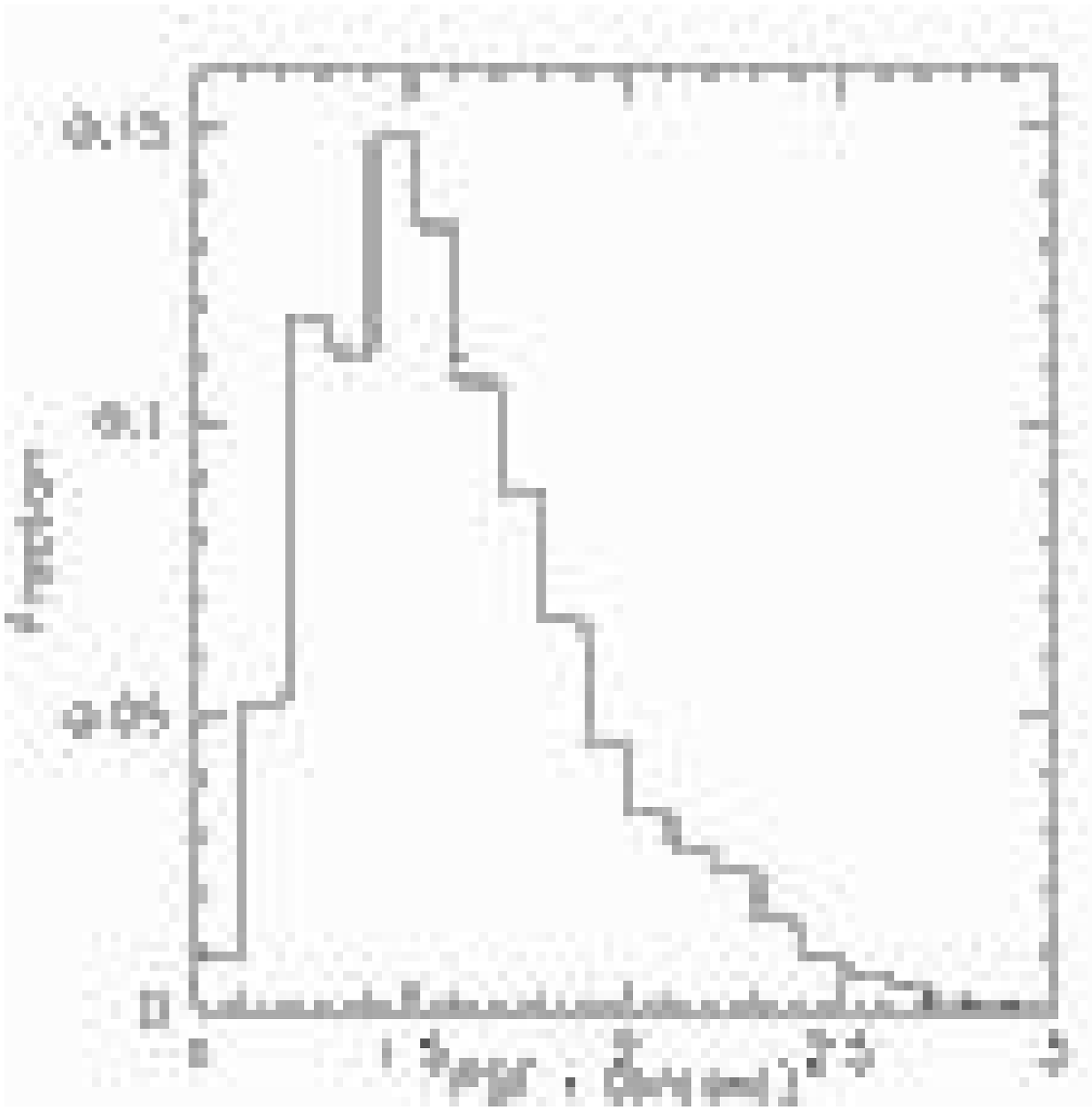}
\caption{
\label{fig:md_seeing_hist} 
 Distribution of seeing of the SDSS galaxies, measured in $r$ band. 
}
\end{figure}

 We use two different ways of classifying galaxy morphologies. The
 first one is concentration parameter $Cin$, which is defined as the ratio
 of Petrosian 50\% light radius to Petrosian 90\% light radius.
 Shimasaku et al. (2001) and Strateva et al. (2001) showed that this
 $Cin$ parameter correlates well with their eye-classified morphology
 (See Figure 10 of Shimasaku et al. 2001 and Figure 8 of Strateva et
 al. 2001).
 We regard galaxies with $Cin\geq$0.4 as late-type galaxies and ones with
 $Cin<$0.4 as early-type galaxies.  
  The criterion of $Cin$=0.4 is more conservative for late type
 galaxies. As shown by Shimasaku et al. (2001), $Cin$=0.4 provides
 late-type galaxy sample with little contamination and early-type
 galaxy sample with small contamination.
  The seeing dependence of $Cin$ is presented in Figure
 \ref{fig:md_seeing_concent} for our volume limited sample galaxies. As is
 shown in Figure \ref{fig:md_seeing_hist}, 
 87\% of our sample galaxies have seeing between 1.2 and 2 arcsec where
 the dependence of $Cin$ on seeing size is negligible. 

\begin{figure}[h]
\includegraphics[scale=0.7]{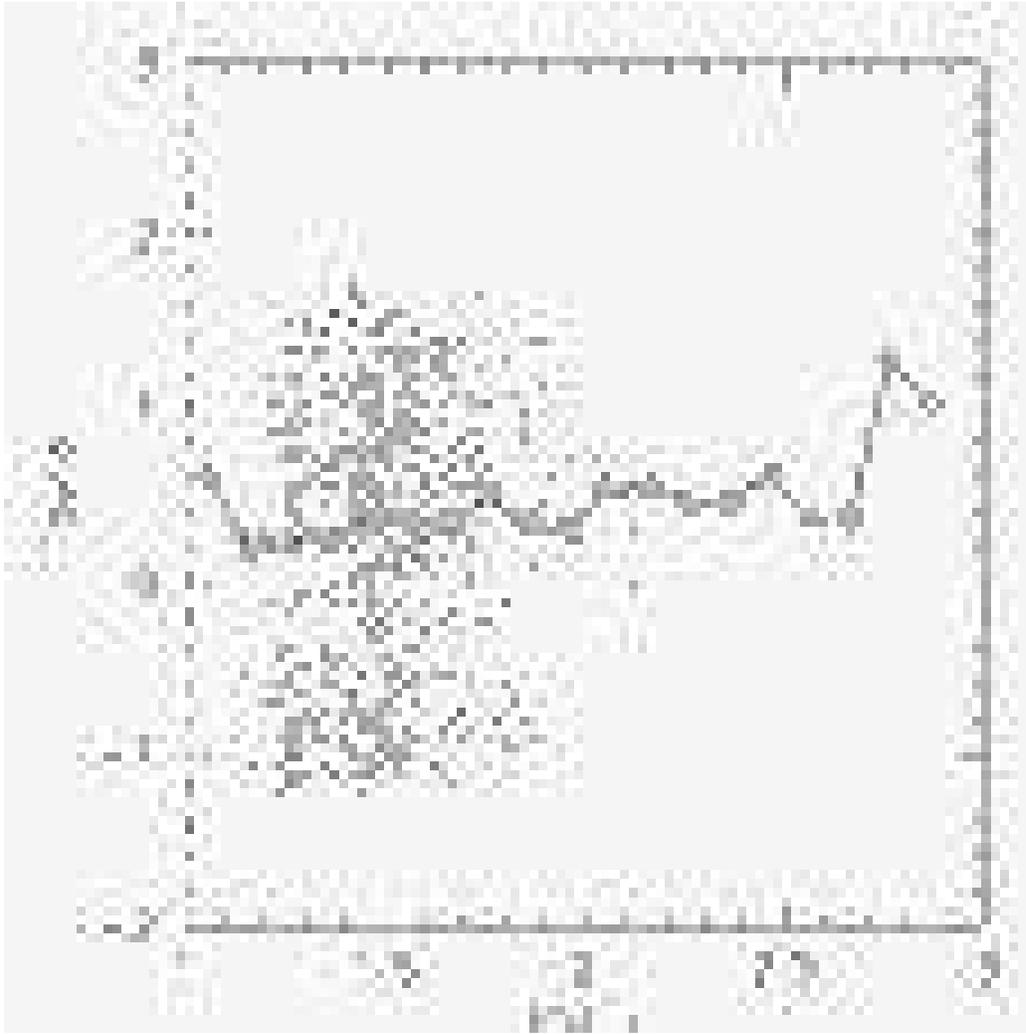}
\caption{
\label{fig:md_tauto_seeing} 
Seeing dependence of $Tauto$ for 7938 galaxies used in the present
 study. The solid lines show medians of the distribution. $Tauto$ is essentially independent of seeing size between 1.2 and 2 arcsec, where 87\% of our sample galaxies lie.
}
\end{figure}

\begin{figure}[h]
\includegraphics[scale=0.7]{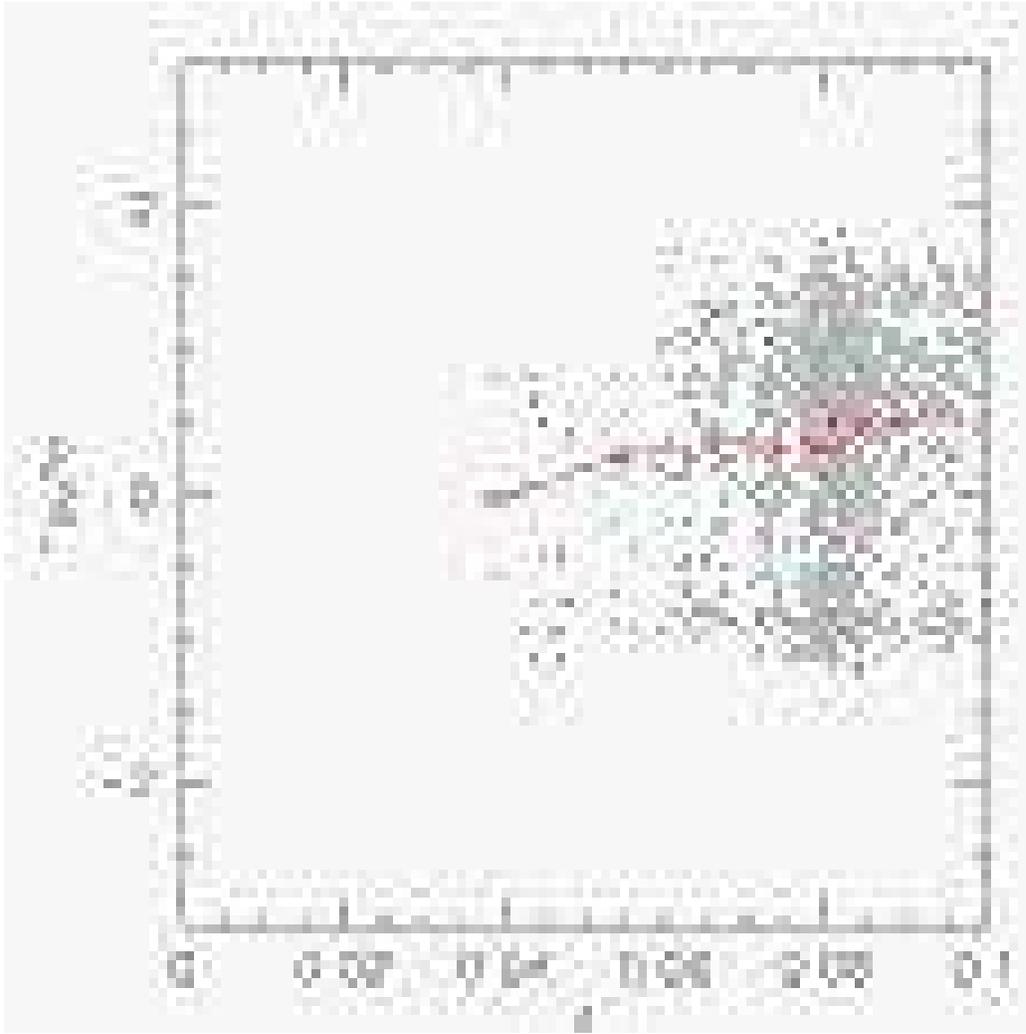}
\caption{
\label{fig:md_size_z} 
 Redshift dependence of $Tauto$ for 7938 galaxies used in the present
 study. The solid lines show medians of the 
 distribution, which are consistent with constant throughout the redshift
 range we use (0.05$<z<$0.1).
 $Tauto$ shows some deviation at lower redshift ($z<0.04$) since an apparent
 size of a galaxy on the sky radically increases at this low redshift.
}
\end{figure}

 The other classification uses morphological parameters measured by
 Yamauchi et al. (2003). We briefly summarize about their morphological
 classification. Details on the method and various systematic tests
 including completeness and contamination study are
 given in Yamauchi et al. (2003). The classification method consists of
 two parts. In the first part, concentration index $Ci$ is calculated as
 the ratio 
 of the Petrosian 50\% light radius and Petrosian 90\% radius
 as is for $Cin$ but the parameter is corrected for elongation of
 galaxies. The elongation correction prevent galaxies with low
 inclination (nearly  edge on galaxies) from being misclassified as
 early-type galaxies.   
 In the second part, coarseness of galaxies, $Cn$, is calculated as
 the ratio of {\it
   variants between a galaxy profile and the best fit to it}
  to {\it difference between the peak and bottom values of profile}.
 $Cn$ is sensitive to arm structures of spiral galaxies, and thus larger
 for spiral galaxies with a clear arm structure than  
 galaxies with a smooth radial profile such as ellipticals and S0s. This
 parameter, $Cn$, helps 
 classifying late-type galaxies further into two types of galaxies.
 Finally, $Ci$ and $Cn$ are combined to be a final morphological
 parameter, $Tauto$.  
 Both $Ci$ and $Cn$ are re-scaled so that their median values become 0.5, and then 
 divided by its standard deviation to be combined to the final parameter
 to classify morphologies as follows.

  \begin{equation} 
 Tauto = Ci(normalized) + Cn(normalized)
  \end{equation}
 $Tauto$ shows better correlation with eye classified morphology than
 $Cin$, as shown in Yamauchi et al. (2003). The correlation
 coefficient with eye-morphology is 0.89.
 Based on the $Tauto$ parameter, we divide galaxies into
 four sub-samples in this study.  We regard galaxies with $Tauto>$1.0 as late-type spiral
 galaxies, 0.1$<Tauto\leq$1.0 as early spirals,  -0.8$<Tauto\leq$0.1 as S0s
 and $Tauto<$-0.8 as elliptical
 galaxies. Among our sample galaxies, 549 galaxies have eye-classified
 morphologies (Shimasaku et al. 2001; Nakamura et al. 2003).
  In Table \ref{tab:md_completeness}, we quote
 completeness and contamination rate of these four types of galaxies
 classified by $Tauto$, using  eye-classified morphology.  Full
 discussion on contamination and completeness 
 will be given in detail in Yamauchi et al. (2003).    
 As shown in Figure \ref{fig:md_tauto_seeing}, the parameter is
 robust against seeing variance in our volume limited sample galaxies. 
 In Figure \ref{fig:md_size_z}, we plot $Tauto$ against redshift. Medians
 are shown in the solid line. In our redshift range (0.05$<z<$0.1),
 $Tauto$ is essentially independent of redshift.

 In Figure \ref{fig:md_iskra}, $Cin$ is plotted against $u-r$ color.
 Strateva et al. (2001) pointed out that $u-r$=2.2 serves as a good
 galaxy type classifier as well.  The
 distribution shows two peaks, one for elliptical galaxies at around
 ($u-r,Cin$)=(2.8,0.35), and one for spiral galaxies at around
 ($u-r,Cin$)=(2.0,0.45).  Our criterion at $Cin$=0.4 is located
 right between these peaks and separates these two populations
 well. 
 In Figure  \ref{fig:md_tauto_ur},
 $Tauto$ is plotted against $u-r$ color in four separate panels. 
 Due to the inclination
 correction of $Tauto$, two populations degenerated in $u-r$ color (around $u-r$=2.8)
 are now separated into elliptical and early spiral galaxies, which is
 one of the major improvements of $Tauto$ against $Cin$. Overplotted points are
 galaxies classified by eye. Upper left, upper right, lower left and
 lower right panels show elliptical,  S0, early-spiral and late-spiral
 galaxies classified by eye, respectively. Compared with $Tauto$
 criteria to separate the galaxies ($Tauto$=-0.8, 0.1 and 1.0), the
 figure suggests that our $Tauto$ criteria separate galaxies reasonably well.  
  The effect of inclination correction can be also seen
 in Figure \ref{fig:md_tauto_concent_each_type}, where $Tauto$ is plotted against
 $Cin$. In addition to the nice correlation between the two parameters,
 there are galaxies with high $Tauto$ and low $Cin$ values in the upper
 left part of the figure. These are galaxies correctly classified by $Tauto$
 due to its inclination correction. The correlation of $Tauto$ with eye
 morphology is studied in detail by Yamauchi et al. (2003). 
 In Figure \ref{fig:md_ha}, we plot H$\alpha$\ EW for four types of
 galaxies classified with $Tauto$. Later type galaxies show higher
 H$\alpha$ EWs, suggesting that our galaxy classification criteria work
 well.

\subsection{Density Measurements}
  
 We measure local galaxy density in the following way. 
 For each galaxy, we measure a projected distance to the 5th nearest
 galaxy within $\pm$1000 km/s in redshift space among the volume limited sample
 (0.05$<z<$0.1, $Mr^*<-$20.5). 
 The 
 criterion for redshift space ($\pm$1000 km/s) is set to be generous to
 avoid galaxies with a large peculiar velocity slipping out of the
 density measurement, in other words, not to underestimate the density
 of cluster cores. Then, the number of galaxies (5) within the distance
 is divided by 
 the circular surface area with the radius of the distance to the 5th
 nearest galaxy. 
 When the projected area touches
 the boundary of the data, we corrected the 
 density by correcting the area
 to divide. Since we have redshift information for all of the sample
 galaxies, our density measurement, $\rho$, is a
 pseudo-three dimensional density measurement and free from the
 uncertainty in background subtraction.  
 In Figure \ref{fig:md_density_distribution}, we present distributions of
 this local galaxy density for all 7938 galaxies, galaxies within 0.5 Mpc
 from a cluster and galaxies between 1 and 2 Mpc from a cluster. In
 measuring distance form a cluster, we use the C4 cluster catalog
 (Miller et al. in prep.). Part of the catalog is also presented in Gomez et
 al. (2003). For each galaxy, the distance from the nearest cluster
 center is measured on a 
 projected sky for galaxies within $\pm$1000 km/s from a cluster redshift.

\section{Results}\label{sec:md_results}
\subsection{The Morphology Density Relation for the SDSS data}\label{sec:md_morphology_density}
  
 In Figure \ref{fig:md_ann_cin}, we use $Cin$ to present the ratio of the
 number of early type galaxies to that of all galaxies as a
 function of the local galaxy density. The solid line shows the
 ratio of early type galaxies using $Cin$=0.4 as a separator,
 which separate elliptical galaxies and spiral galaxies well as shown in
 Figure \ref{fig:md_iskra}. 
 The fraction of
 early type galaxies clearly increases with increasing density. In the
 least dense region, only 55\% are early type, whereas in the densest
 region, almost 85\% are early type galaxies. Furthermore, it is
 interesting to see 
 that around galaxy density 3 Mpc$^{-2}$, the slope of the
 morphology-density relation abruptly becomes steeper.  3 Mpc$^{-2}$ is
 a characteristic density for cluster perimeter (1-2 Mpc; see Figure
 \ref{fig:md_density_distribution}) and also coincides with the density
 where star formation rate (SFR) of galaxies changes as studied by Lewis et
 al. (2002) and Gomez et al. (2003). 
  To see the dependence of the relation to the
 choice of our criterion ($Cin$=0.4), we use slightly different criteria for dashed
 and dotted lines, which use $Cin$=0.37 and $Cin$=0.43 as a criterion, respectively. 
 The former criterion is a little biased to spiral galaxies and the
 latter to elliptical galaxies. As is seen in the figure, both dotted
 and dashed lines show the morphology-density relation, but in somewhat
 flatter way than the solid line, indicating the effect of the
 contamination from spiral galaxies in case of $Cin$=0.43
 (incompleteness in case of $Cin$=0.37). In the density below 3
 Mpc$^{-2}$, the steepness of three slopes are almost identical. However, at the least
 dense region, only 65\% is elliptical galaxies in case of $Cin$=0.37,
 whereas almost 90\% is elliptical galaxies in case of  $Cin$=0.43.
 Thus, the absolute amount of ``elliptical'' galaxies is a strong
 function of the 
 $Cin$ criterion. Therefore careful attention to the $Cin$ criterion is
 needed when comparing to other work such as computer simulations and
 other observational data set.

 In Figure \ref{fig:md_md_ann_ytype}, the ratio of four morphological
 types of galaxies are plotted against galaxy density using
 $Tauto$. The short-dashed, solid, dotted and long-dashed lines represent elliptical,
 S0, early-spiral and late-spiral galaxies, respectively. The decline of
 late-spiral fraction toward high density is seen. Early-spiral
 fractions stays almost constant. 
 S0 fractions steadily increase toward higher density, but
 declines somewhat at the two highest density bins. Elliptical fractions
 show mild 
 increase with increasing density and radically increase at the two highest
 density bins.  
 In the figure, there exist two characteristic densities where the
 relation radically changes. Around galaxy density 1-2 Mpc $^{-2}$,
 corresponding to the cluster infalling region
 (Figure \ref{fig:md_density_distribution}), the slope for late-spiral
 suddenly goes down and  the slope for S0 goes up.
 At around galaxy density 6 Mpc $^{-2}$, corresponding to the
 cluster core region (Figure \ref{fig:md_density_distribution}), S0 fractions
 suddenly goes down and elliptical fractions show a sudden increase. 
 To clarify this second change in elliptical and S0 fractions, we plot
 S0 to elliptical number ratio against local galaxy density in Figure
 \ref{fig:md_md_es0}. As is seen in the previous figure, S0/E ratio
 remains almost constant from galaxy density 0.01 to 5 Mpc $^{-2}$, and it then suddenly
 declines after 6 Mpc $^{-2}$. We discuss the interpretation of this
 result in Section \ref{md_discussion}.

\subsection{Morphology-Radius Relation}\label{Oct 30 18:22:01 2002}
\begin{figure}
\centering{\includegraphics[scale=0.39]{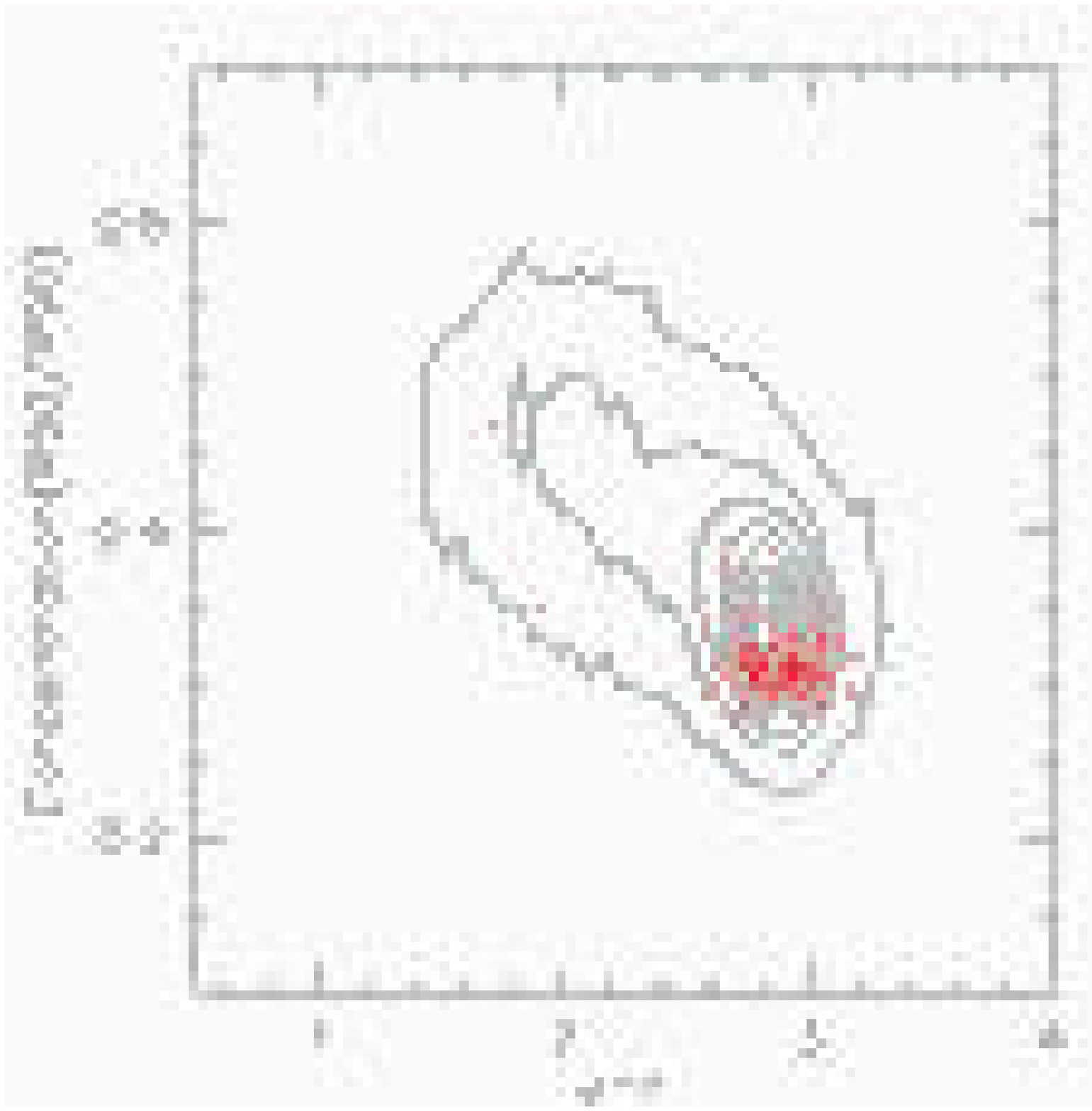}
\includegraphics[scale=0.39]{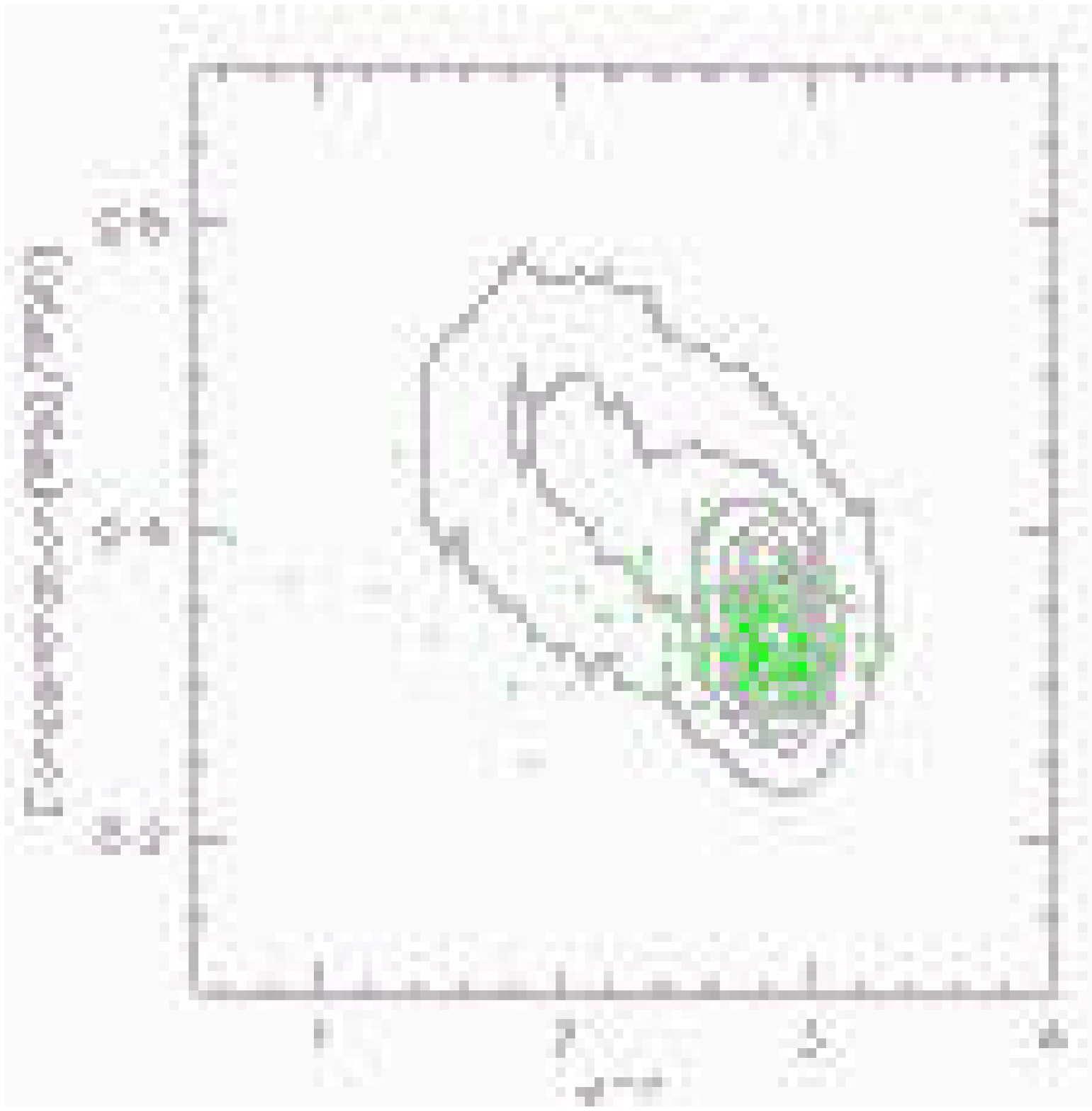}}
\centering{\includegraphics[scale=0.39]{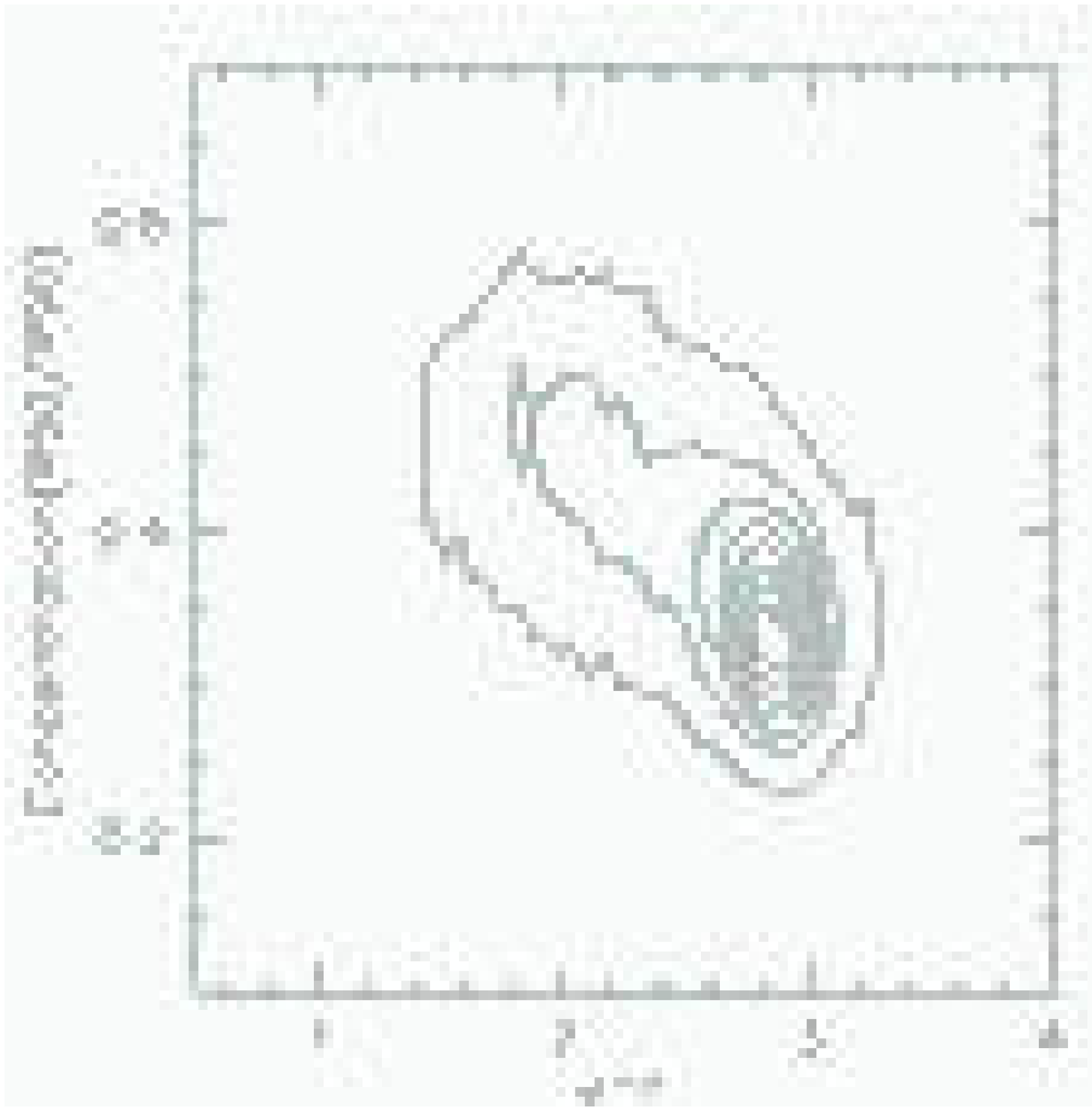}
\includegraphics[scale=0.39]{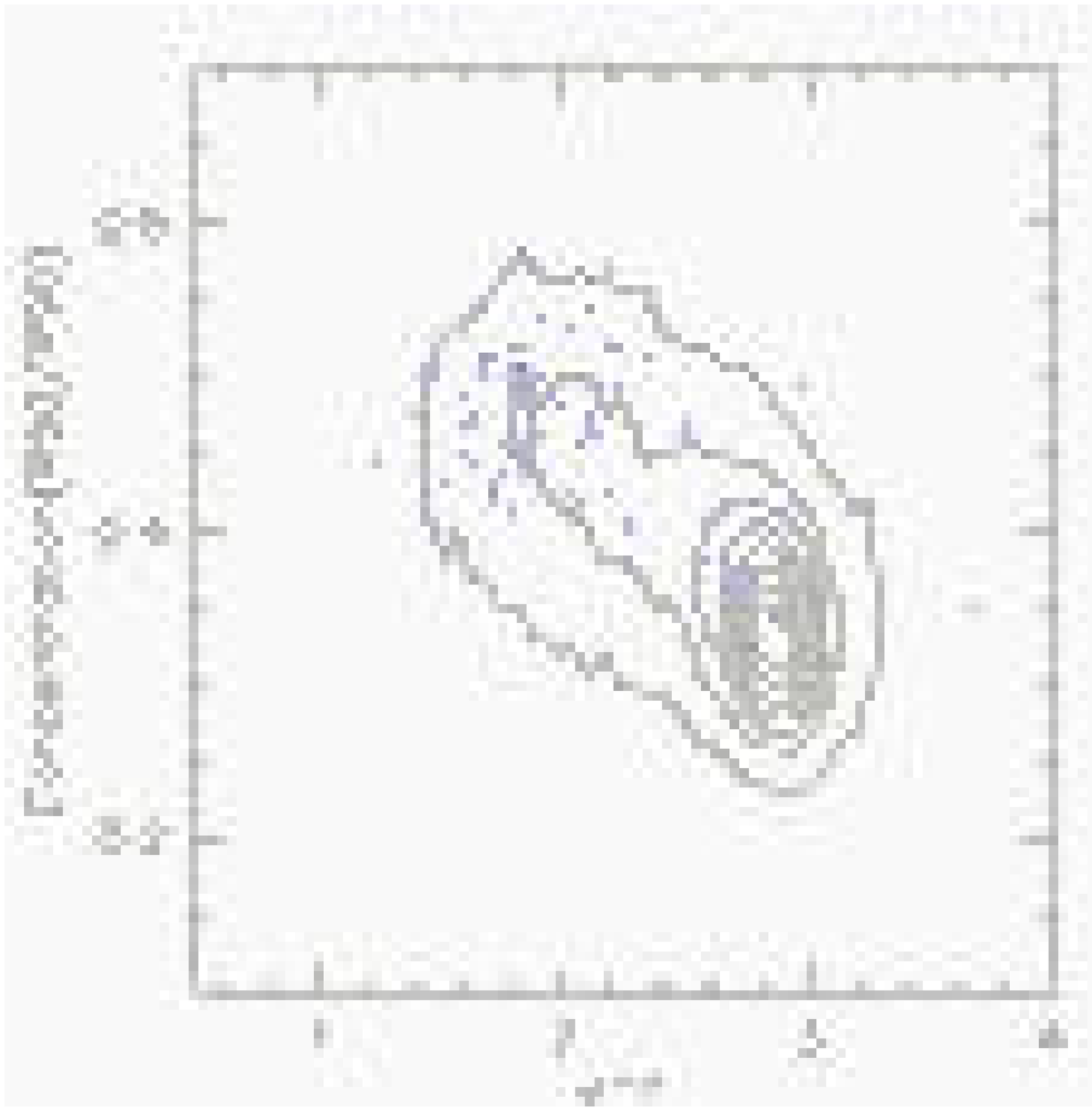}}
\caption{
\label{fig:md_iskra}
 $Cin$ is plotted against $u-r$. The contours show distribution of all 7938
 galaxies in the volume limited sample.  Points in each
 panel show the distribution of each morphological type of galaxies
 classified by eye (Shimasaku et al. 2001; Nakamura et
 al. 2003). Ellipticals are in the upper left panel. S0, Sa and Sc are
 in the upper right, lower left and lower right panels, respectively.   
}\end{figure}

\begin{figure}
\centering{\includegraphics[scale=0.39]{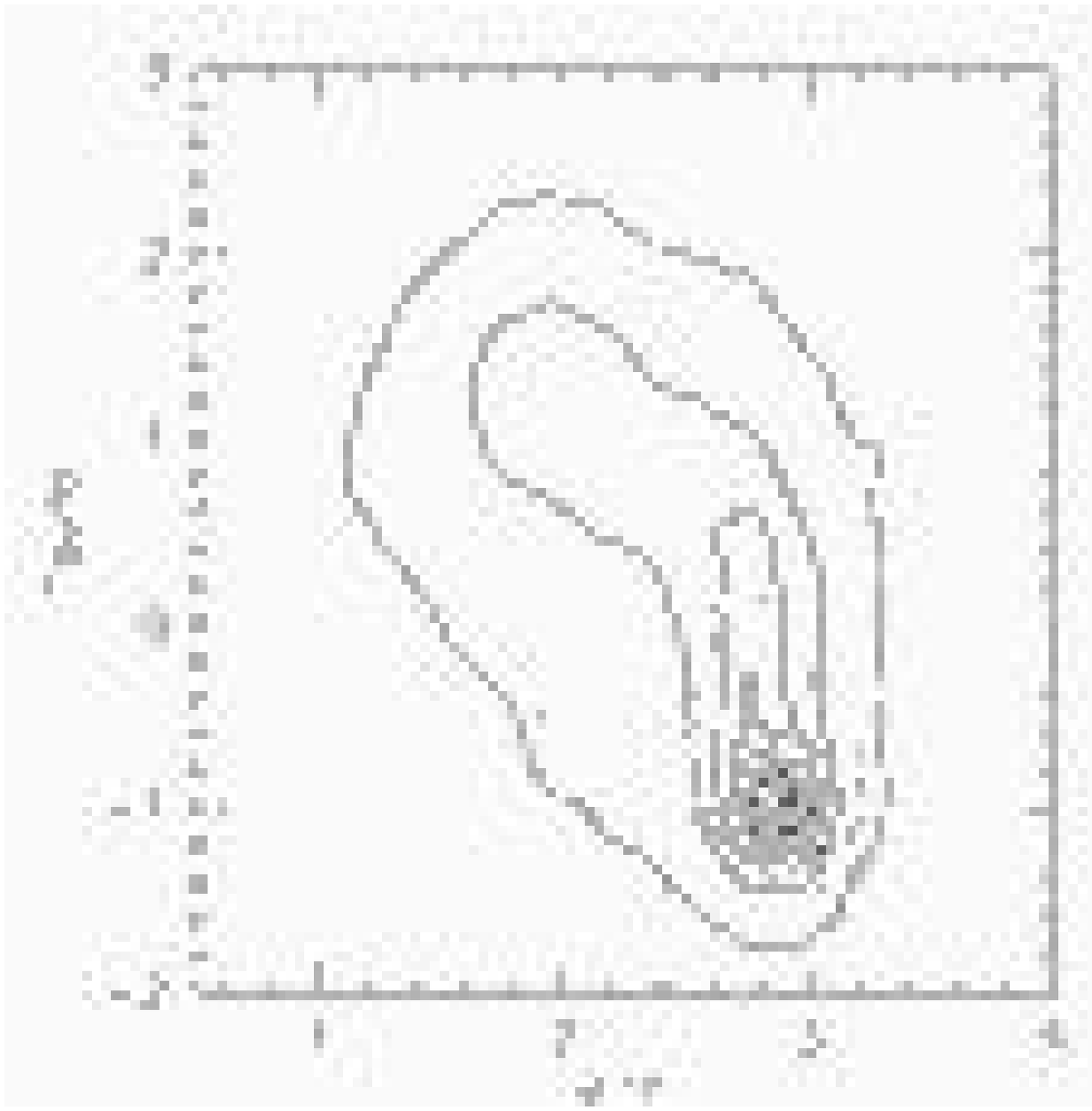}
\includegraphics[scale=0.39]{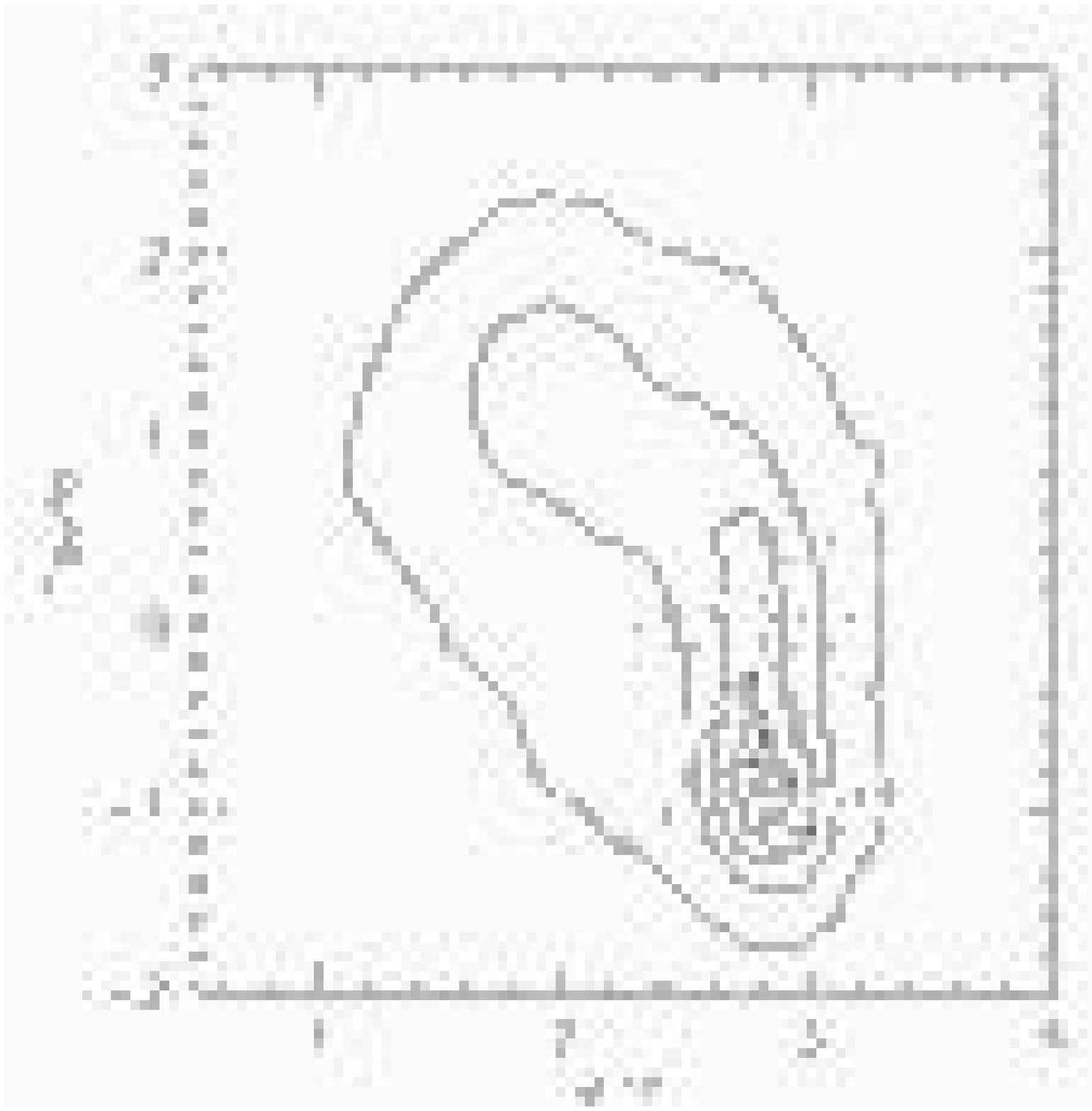}}
\centering{\includegraphics[scale=0.39]{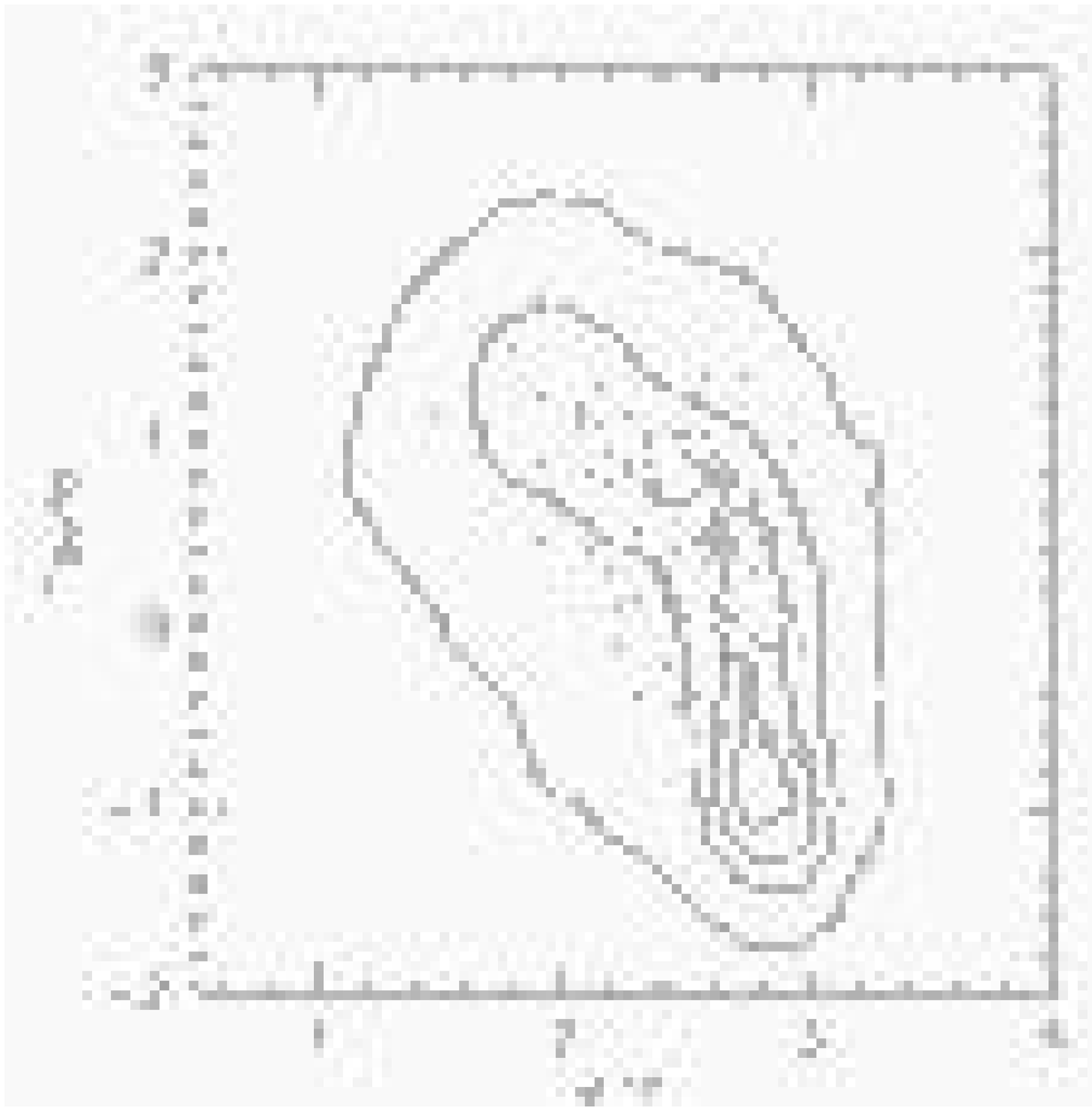}
\includegraphics[scale=0.39]{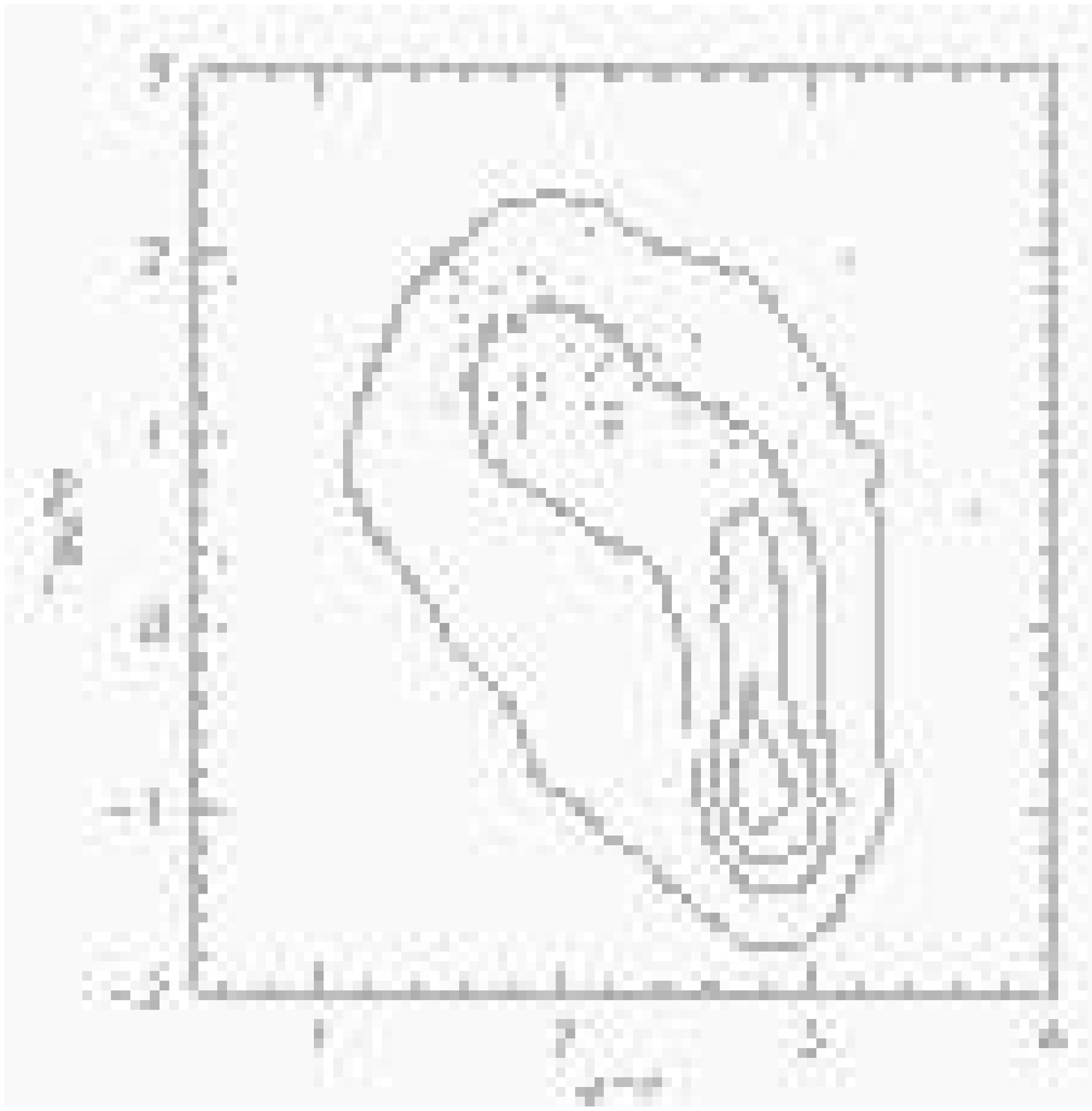}}
\caption{
\label{fig:md_tauto_ur}
 $Tauto$ is plotted against $u-r$. The extension of the distribution
 around $u-r$=2.8 is due to the inclination correction adopted in $Tauto$.
 The contours show distribution of all 7938
 galaxies in the volume limited sample.  Points in each
 panel show the distribution of each morphological type of galaxies
 classified by eye (Shimasaku et al. 2001; Nakamura et
 al. 2003). Ellipticals are in the upper left panel. S0, Sa and Sc are
 in the upper right, lower left and lower right panels, respectively.   
}\end{figure}

\begin{figure}
\centering{\includegraphics[scale=0.39]{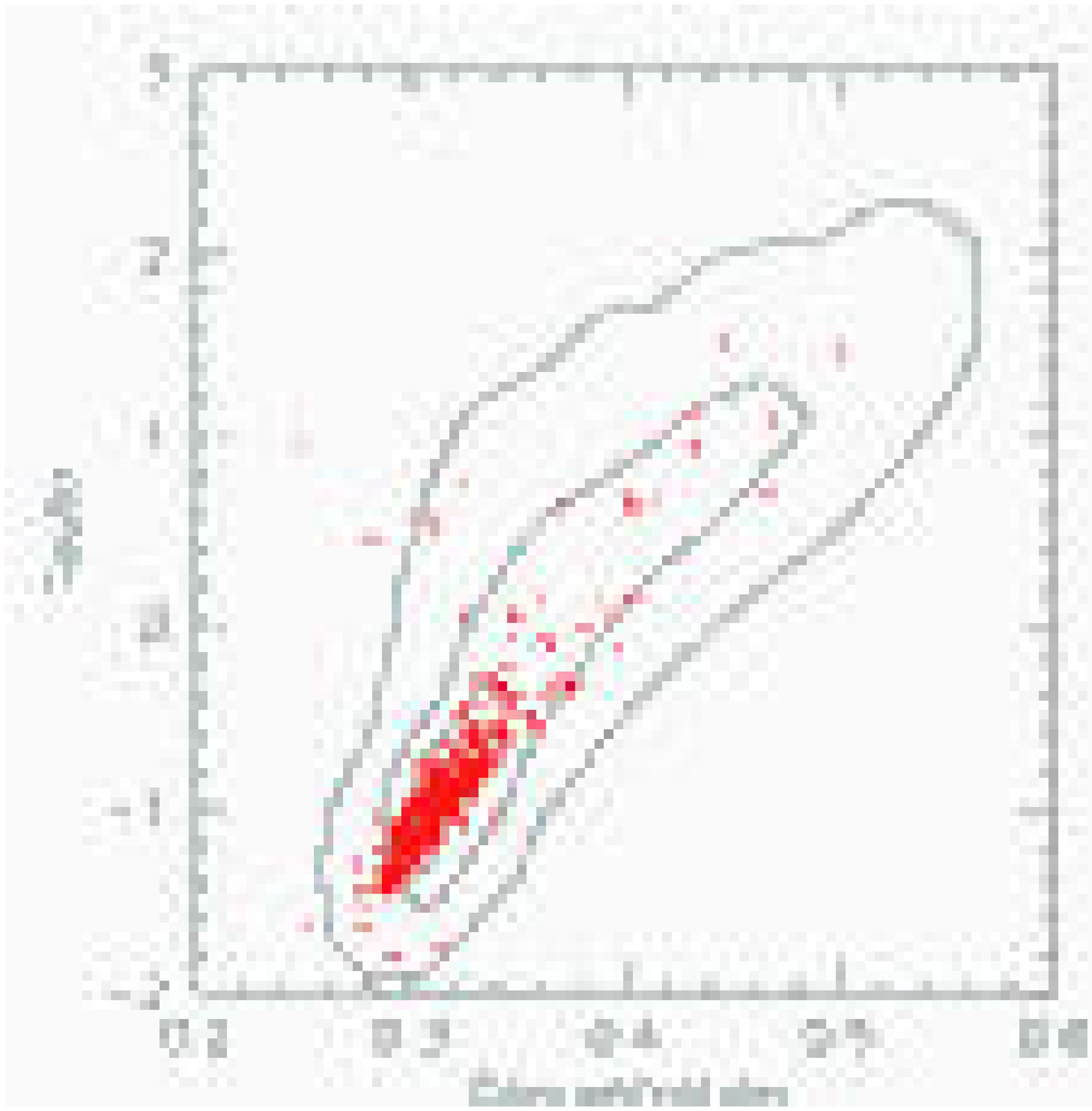}
\includegraphics[scale=0.39]{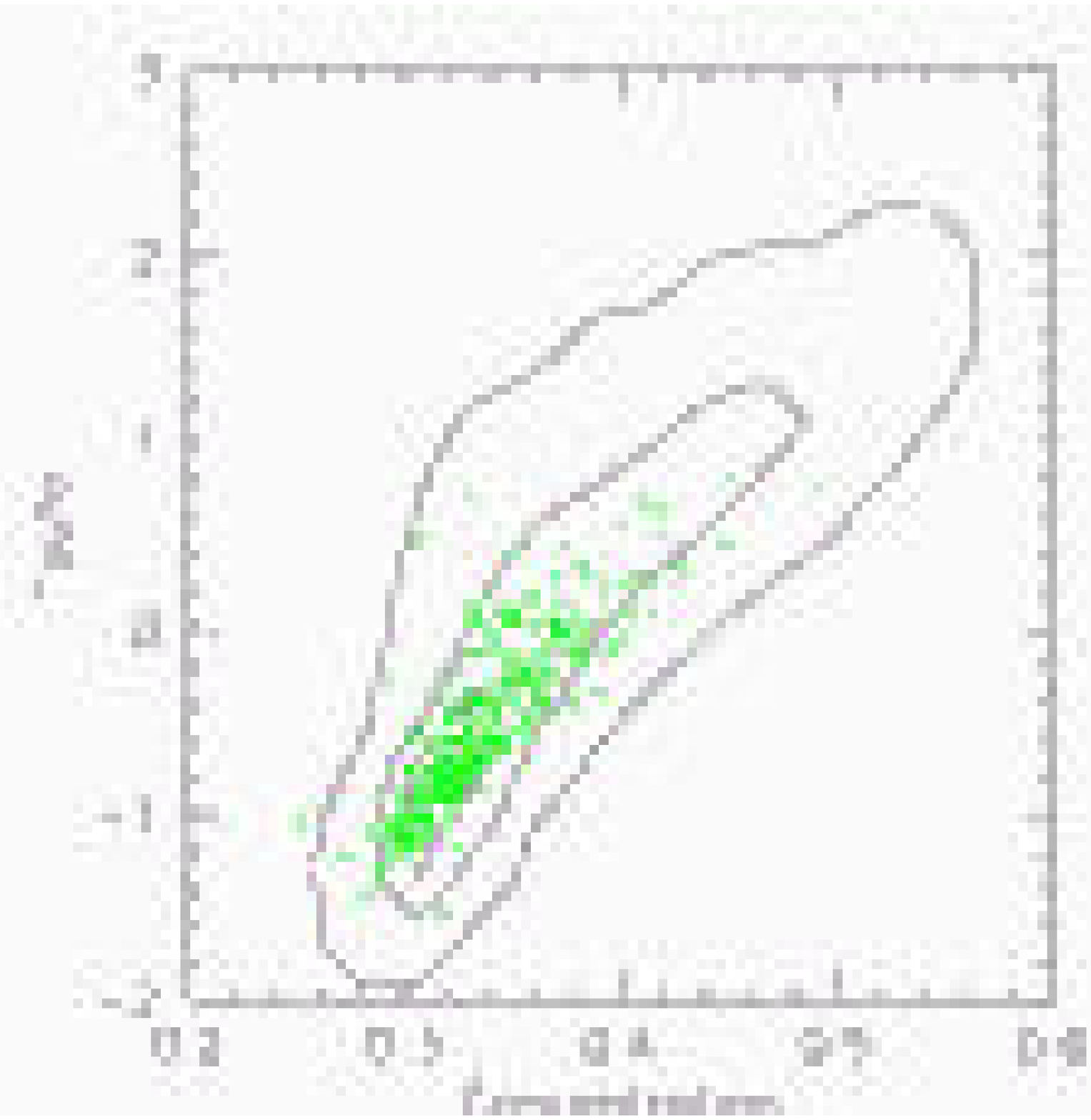}}
\centering{\includegraphics[scale=0.39]{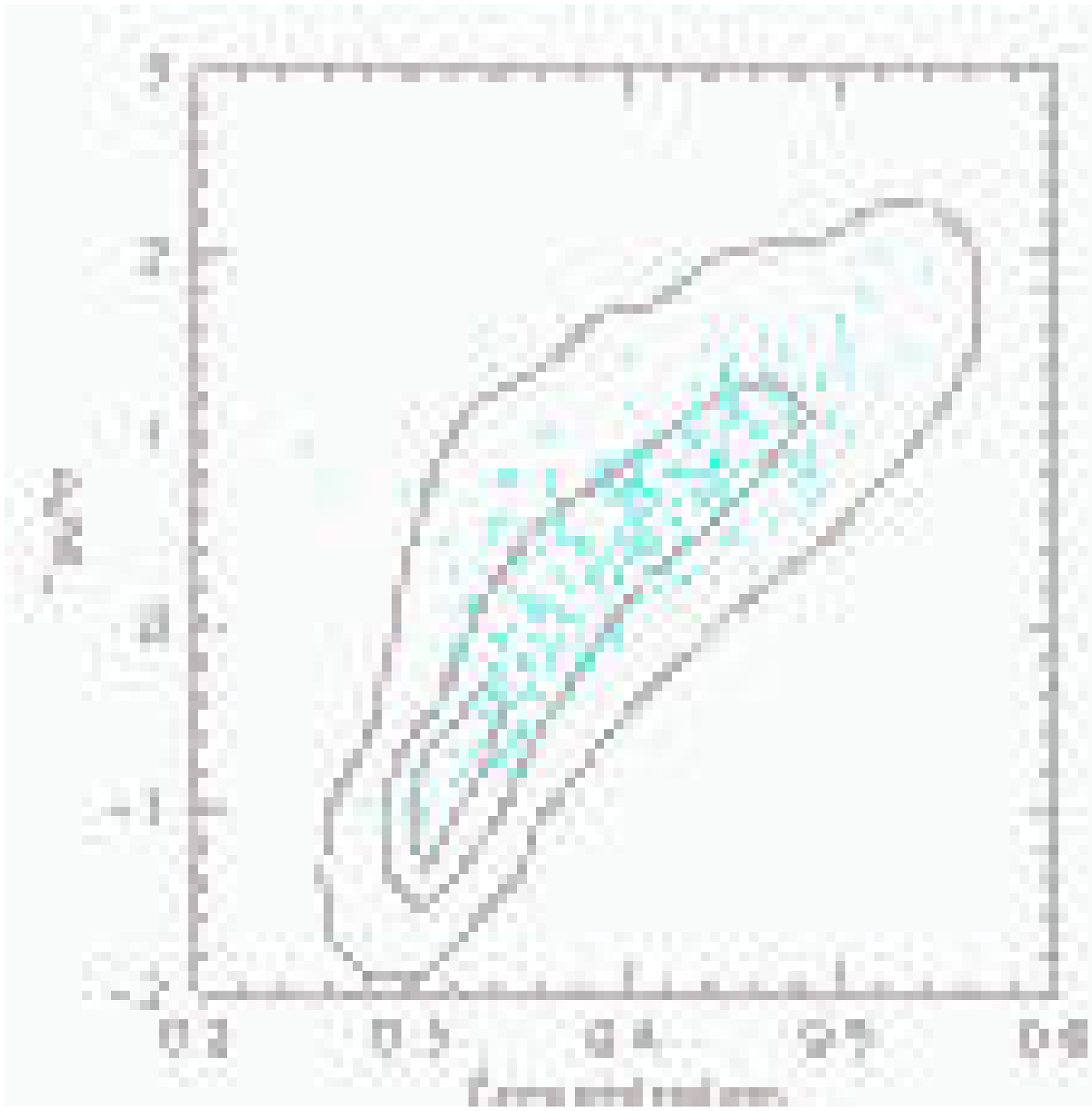}
\includegraphics[scale=0.39]{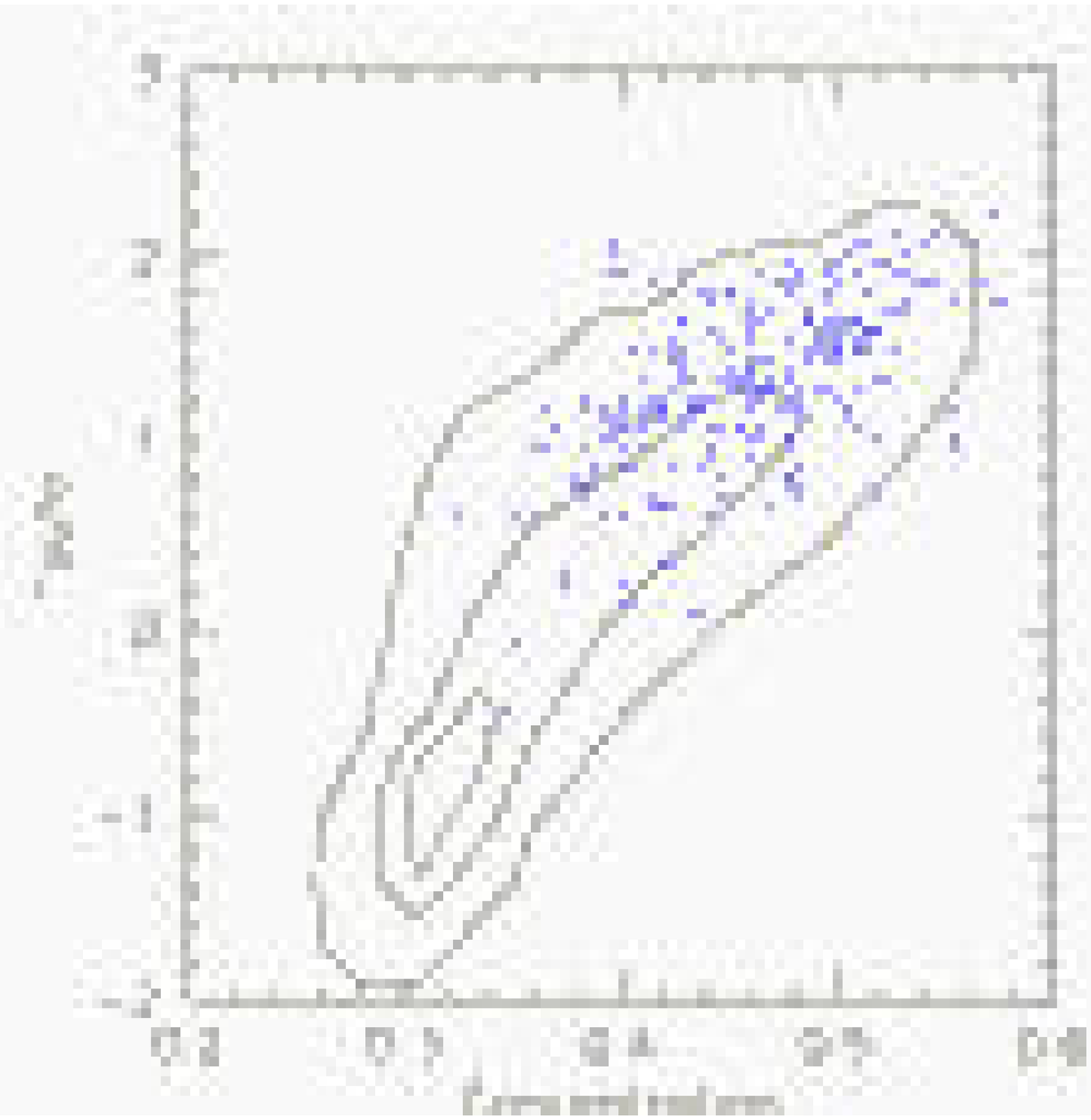}}
\caption{
\label{fig:md_tauto_concent_each_type}
 $Tauto$ is plotted against $Cin$. The contours show distribution of all 7938
 galaxies in the volume limited sample. A good correlation between
 two parameters is seen. The extension of the distribution toward the upper left
 corner is due to the inclination correction of $Tauto$. Points in each
 panel show the distribution of each morphological type of galaxies
 classified by eye (Shimasaku et al. 2001; Nakamura et
 al. 2003). Ellipticals are in the upper left panel. S0, Sa and Sc are
 in the upper right, lower left and lower right panels, respectively.   
}\end{figure}

\begin{figure}
\centering{\includegraphics[scale=0.39]{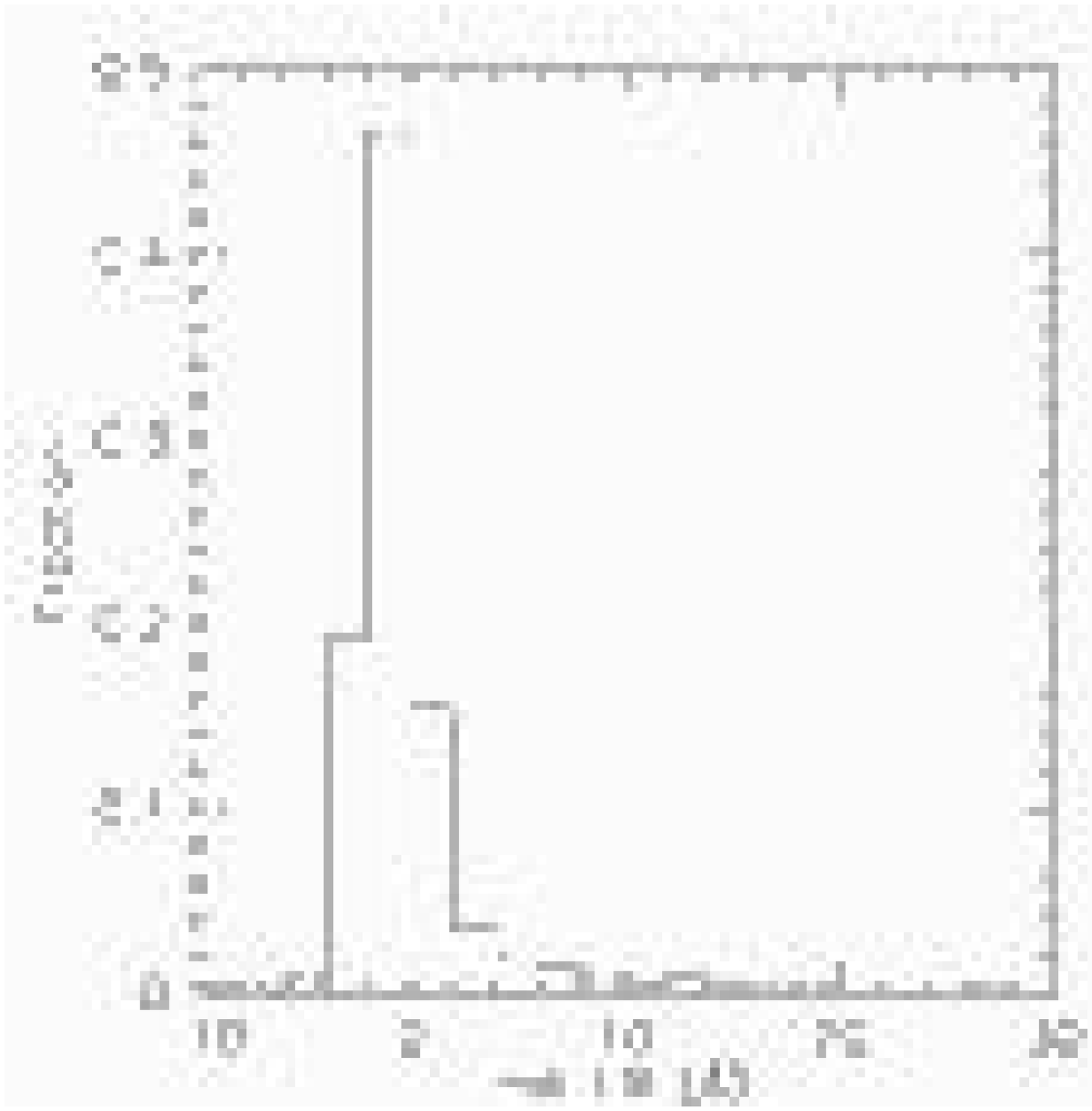}
\includegraphics[scale=0.39]{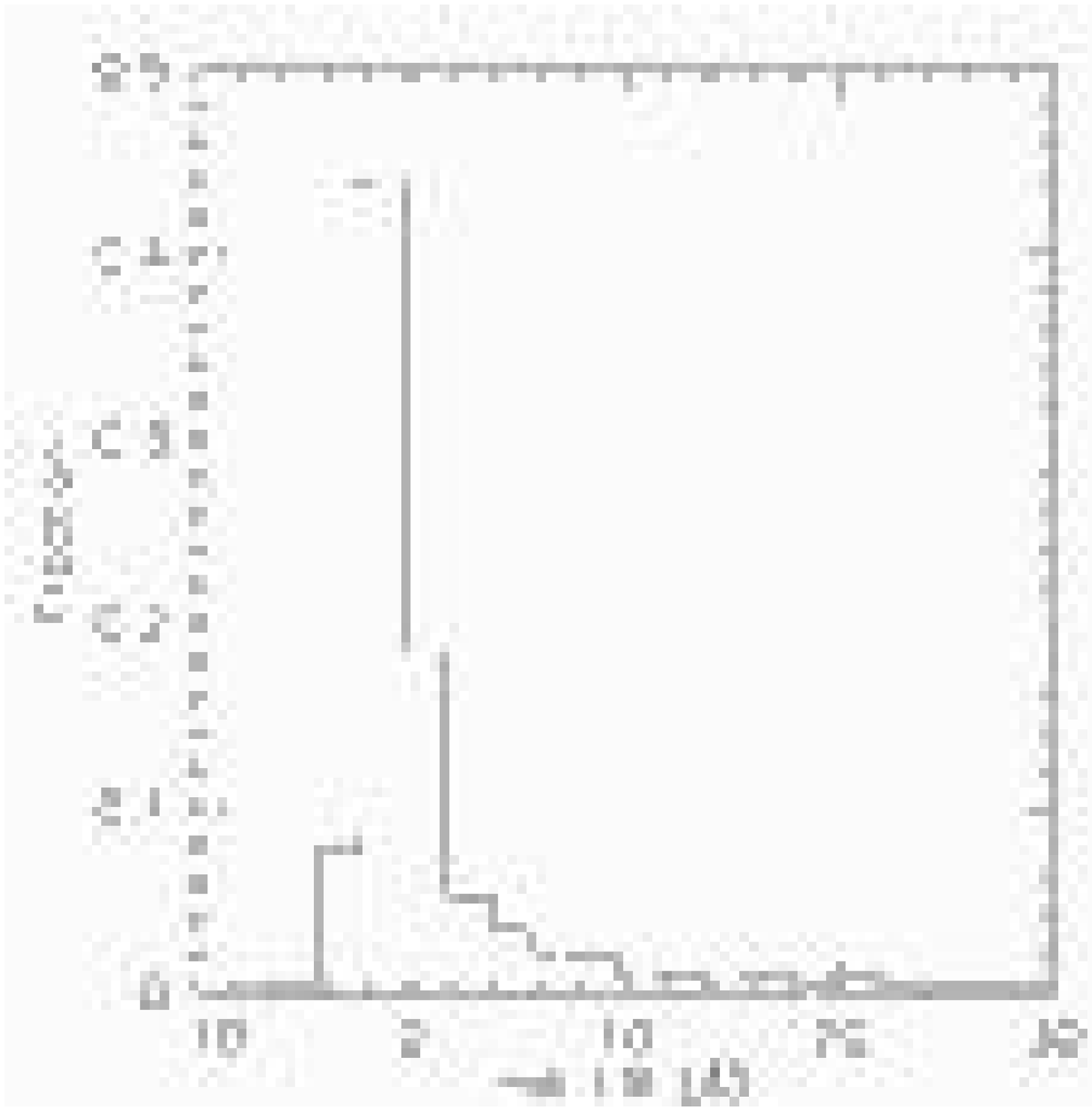}}
\centering{\includegraphics[scale=0.39]{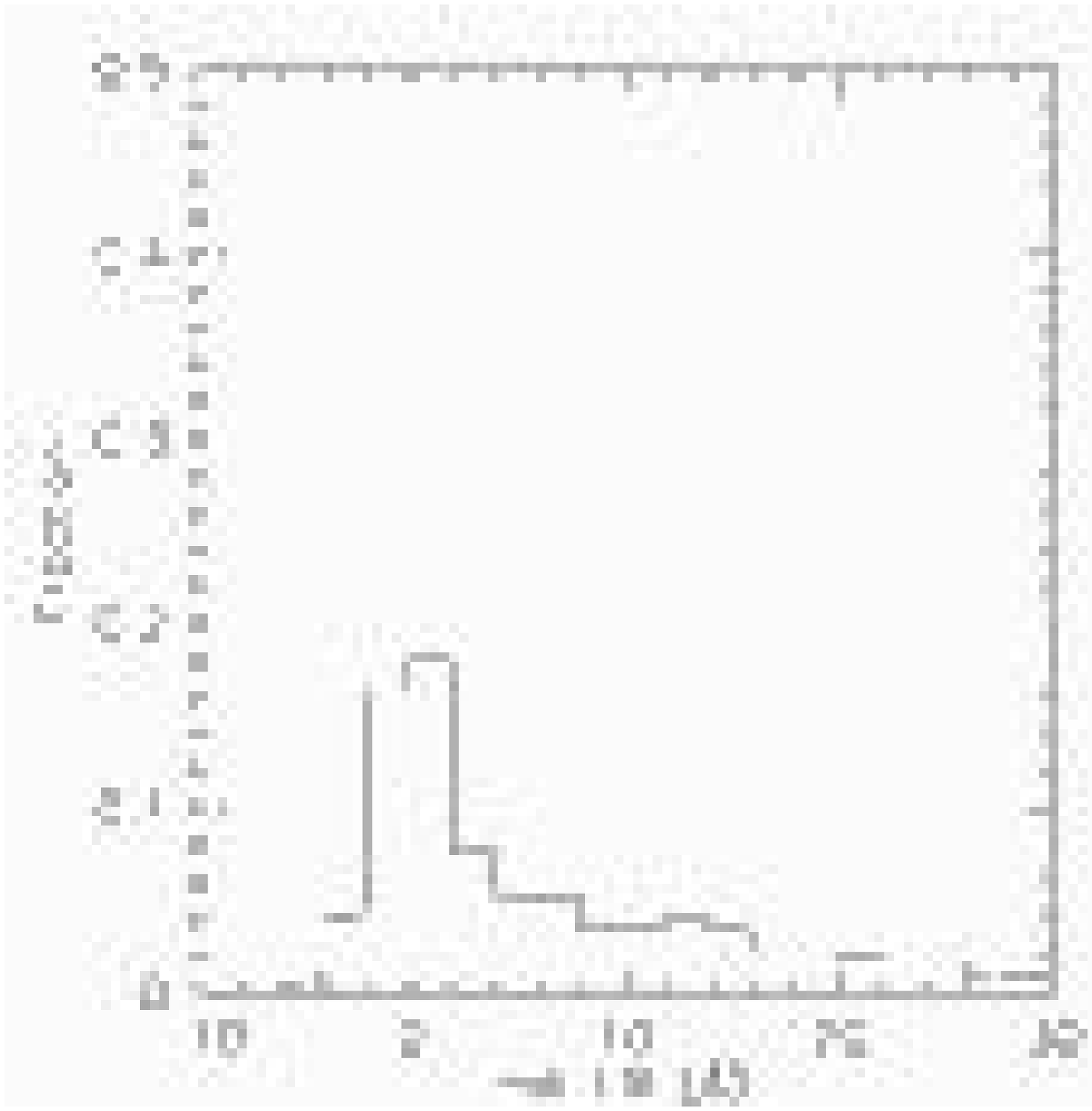}
\includegraphics[scale=0.39]{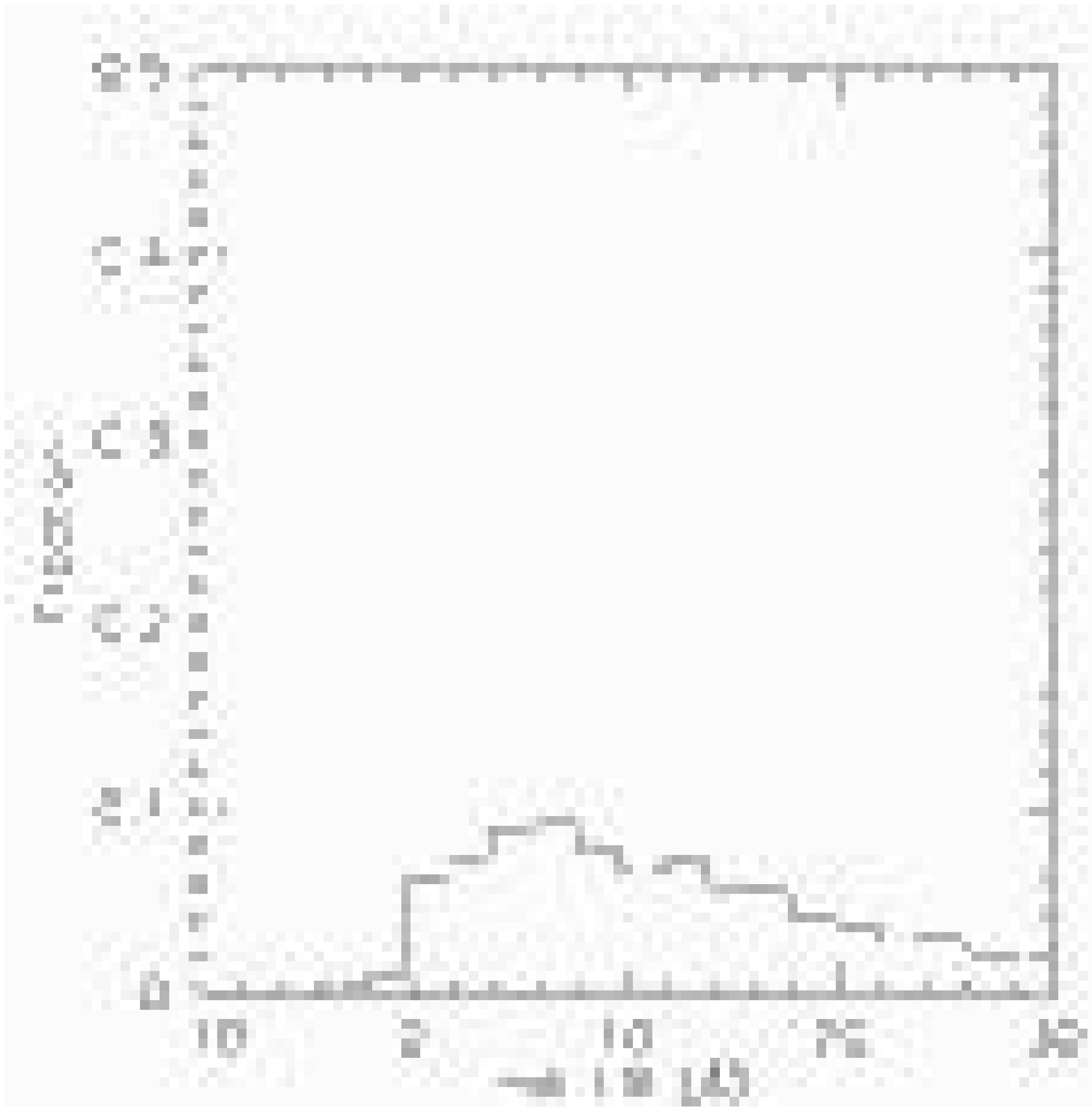}}
\caption{
\label{fig:md_ha}
 Distribution of H$\alpha$ EW for four types classified with
 $Tauto$. 
 The histogram in each
 panel shows the distribution of each morphological type of galaxies
 classified by $Tauto$. Ellipticals are in the upper left panel. S0, Sa and Sc are
 in the upper right, lower left and lower right panels, respectively.  
 An increase of H$\alpha$ EW
 toward later type galaxies suggests that our morphological
 classification works well. 
}\end{figure}

\begin{figure}[h]
\includegraphics[scale=0.7]{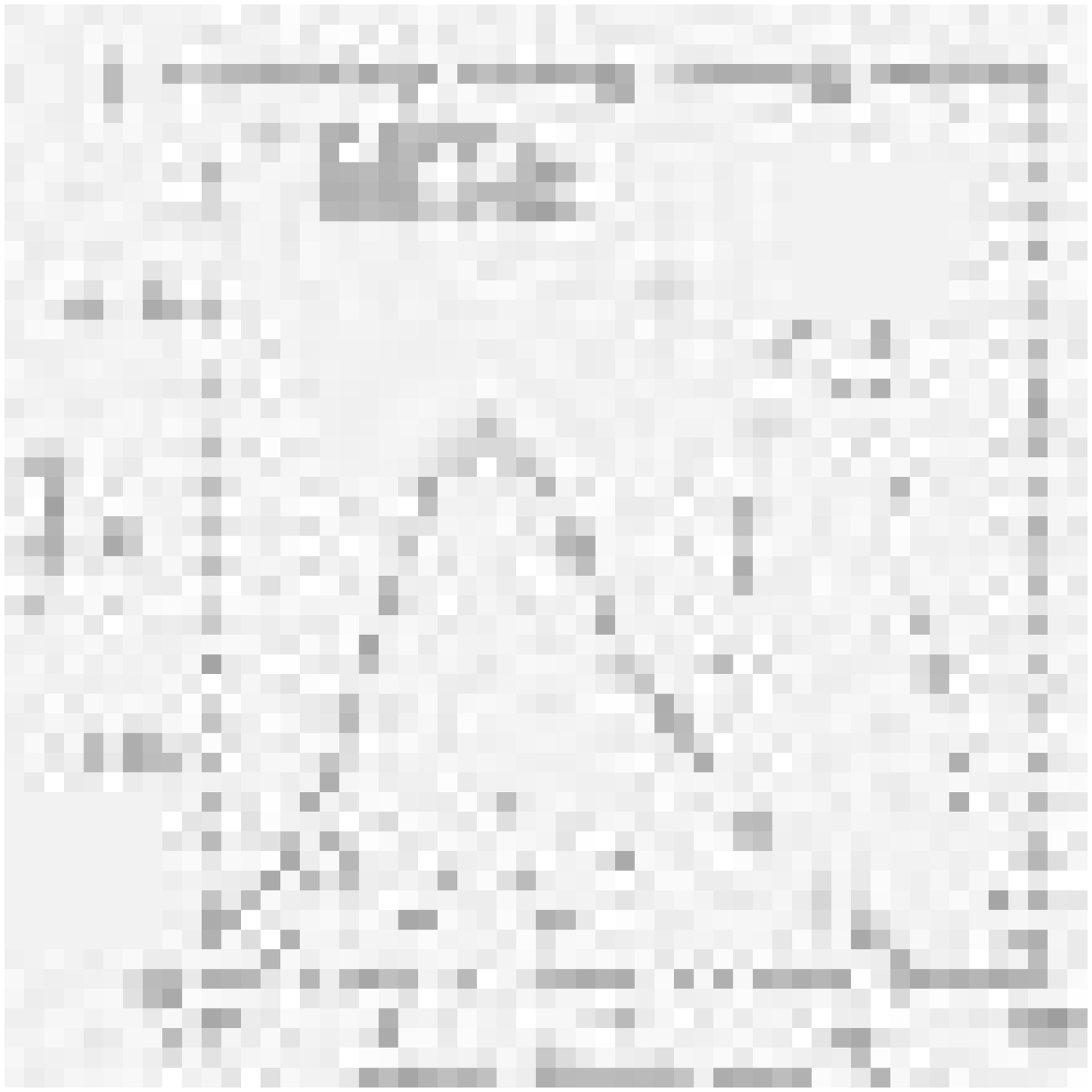}
\caption{
\label{fig:md_density_distribution} 
 Distribution of local galaxy density. The solid, dashed and dotted lines
 show distributions for all galaxies, galaxies within 0.5 Mpc from the nearest
 cluster and  galaxies between 1 and 2 Mpc from the nearest cluster, respectively.
}
\end{figure}

  In Figure \ref{fig:md_mr_cin}, we plot elliptical fractions classified with
  $Cin$, against a
  radius from the cluster center. We use the C4 galaxy cluster catalog
  (Miller et al. in prep.) when measuring the distance between a galaxy and
  the nearest cluster. 
  For each cluster, distances are measured by converting angular
  separation into physical distance for galaxies within
  $\pm$1000 km/s in redshift space. For each galaxy, the distance to the
  nearest cluster is adopted as a distance to a cluster. The distance is
  then expressed in units of virial radii
  using velocity dispersion
  given in Miller et al. in prep. and  the equation given in Girardi et
  al. (1998). 
  The  morphological fraction for each radius bin is measured in the same way
  as the last section; the solid, dashed and dotted lines represents
  different criteria, $Cin=$0.4, 0.37 and 0.43, respectively. As seen in
  Figure \ref{fig:md_iskra},   $Cin=$0.4 best separates elliptical and
  spiral galaxies. 
  In the figure, fraction of elliptical galaxies
  decreases toward larger distance from a cluster center. The relation
 becomes consistent with flat after 2 virial radius. 
  As in the case in Figure \ref{fig:md_ann_cin}, three criteria show
  similar slope. However, absolute amount of elliptical galaxies is a
  strong function of $Cin$ criteria. In case of $Cin$=0.4, elliptical
  fractions increase from 60\% to almost 90\% toward a cluster center.

  In Figure \ref{fig:md_mr}, we plot the morphology-cluster-centric-radius
  relation for four types of galaxies classified using
  $Tauto$. As is in Figure \ref{fig:md_md_ann_ytype}, the short-dashed, solid,
  dotted and long-dashed lines represent 
  elliptical,  S0, early-spiral and late-spiral galaxies, respectively.
  Fractions of late-spiral galaxies decrease toward smaller radius,
  whereas fractions of elliptical and S0 galaxies increase toward a
  cluster center. In the figure, three characteristic radii are
  found. Above 2 virial radii, four lines are consistent with flat,
  suggesting physical mechanisms responsible for the morphological change do not
  work beyond this radius. Between 0.3 and 1 virial radius, S0 fractions
  mainly increases toward a cluster center. Late and early spirals
  show corresponding decrease. Interestingly, S0 fractions increase more
  than elliptical fractions. Below 0.3 virial radius, elliptical fractions
  dramatically increase and S0 fractions decrease in turn. 
  To further clarify the change between S0 and elliptical fractions, we
  plot S0 to elliptical number ratio in Figure \ref{fig:md_mr_es0}. As is
  seen in the previous figure, the ratio slightly increase between 1 and 0.3
  virial radius toward a cluster center. At 0.3 virial radius, slope
  changes radically and the ratio decreases toward  a cluster center. We
  interpret these findings in Section \ref{md_discussion}. 

\subsection{Physical Sizes of Galaxies}
 It is important to understand relative galaxy sizes when we
 discuss transformation of galaxies.
 In Figure \ref{fig:md_size}, we plot physical galaxy sizes
 calculated using Petrosian 90\% flux radius in $r$ band, against
 $Tauto$. Above $Tauto$=0, galaxy sizes decrease with decreasing
 $Tauto$. However, below $Tauto$=0, galaxy sizes increase with decreasing
 $Tauto$. We discuss the result in conjunction with
 Figures. \ref{fig:md_md_ann_ytype} and \ref{fig:md_mr} in Section \ref{md_discussion}.

 %

\subsection{Comparison with the MORPHS Data}
\label{Oct 30 18:41:06 2002}
 In this section, we compare the morphology-density relation of the SDSS
 data ($z\sim$0.05) with that of the MORPHS data ($z\sim$0.5).
 The MORPHS data are used in Dressler et al. (1997) to study
 the morphology-density relation in high redshift clusters,
 and publicly available in Smail et al. (1997). The data consist of 10
 galaxy clusters at a redshift range $z=$0.37-0.55 as summarized in Table \ref{tab:md_morphs}. The sharp imaging
 ability of Hubble Space Telescope made it possible to measure galaxy
 morphology at this further away in the universe. We use concentration
 parameter given in Smail et al. (1997) as an automated morphology of
 the sample. As for the galaxy density, we count a number of galaxies
 brighter than $Mr^*=-$19.0 within 250 kpc and subtracted average galaxy
 number count of the area (Glazebrook et al. 1995).  Magnitude (either
 F702W or F814W) are
 $k$-corrected and transformed to the SDSS $r$ band using the relation
 given in Fukugita et al. (1995). 
 To make as fair comparison as
 possible, we re-measured the SDSS morphology-density relation using
 as similar criteria as possible. We re-measure galaxy density by
 counting galaxies within
 250 kpc and $\pm$1000 km/s and brighter than $Mr^*=$-19.0 in the SDSS
 data (0.01$<z<$0.054). The number of 
 galaxies are divided by the size of the area (250$^2$ $\pi$ kpc$^2$ if it does
 not go outside of the boundary. If it does, the area is corrected
 accordingly.) 
 We also match the criteria for both of concentration parameters. The
 concentration parameter of the SDSS is measured as the ratio of
 Petrosian 50\% light radius to 90\% light radius. The concentration
 parameter of the  
 MORPHS data is measured using the Source Extractor. Furthermore, the
 seeing size compared with typical galaxy size is not exactly the same
 between two samples. Therefore, we have to calibrate these two
 concentration parameters. 
 Fortunately, part of the SDSS galaxies are morphologically classified by
 eye in Dressler (1980). Since the MORPHS data are eye-classified by the same
 authors (Dressler et al. 1997), we regard these two eye-classified
 morphology essentially the same and use them to calibrate two
 concentration parameters. When we use the SDSS concentration criteria, $Cin<$0.4,
 it leaves 76\% of eye-classified elliptical galaxies (24\% contamination). By
 adjusting concentration parameter for the MORPHS data to these values
 (76\% and 24\%), we
 found that the MORPHS concentration parameter of 0.45 leaves 75\% of
 eye-classified elliptical galaxies. We regard this essentially the same
 criteria and adopt 0.45 as a criteria for the MORPHS concentration
 parameter which corresponds to that of the SDSS ($Cin$=0.4). In Figure
 \ref{fig:md_md_morphs}, we plot fractions of elliptical galaxies against
 galaxy density for both the SDSS and MORPHS data in the solid and dashed lines,
 respectively. Quite interestingly, two morphology-density relations lie
 on top of each other. Since the MORPHS data only exist for cluster
 region, we are not able to probe into as low density regions as in
 Figure \ref{fig:md_md_ann_ytype}. However, it shows that the
 morphology-density relation was already established at $z\sim$0.5.
 There is a sign of slight excess of elliptical galaxies in the SDSS
 data in two  highest density bins.

\section{Discussion}\label{md_discussion}

\subsection{Elliptical Fractions}
\begin{figure}[h]
\begin{center}
\includegraphics[scale=0.7]{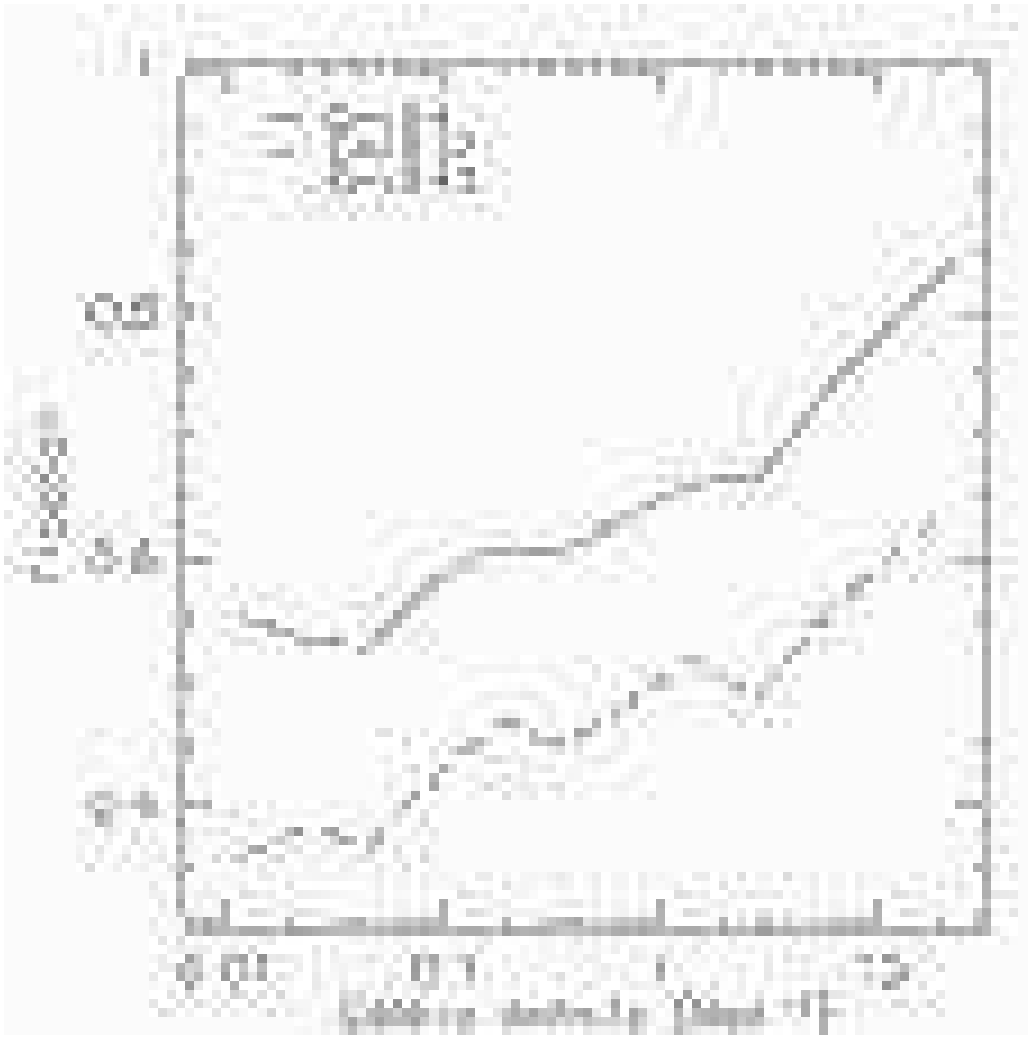}
\end{center}\caption{
\label{fig:md_ann_cin}
 The morphology-density relation for three criteria of $Cin$. Fractions of
 elliptical galaxies are plotted against local galaxy density. Three
 criteria are $Cin<$0.4, $Cin<$0.43 and $Cin<$0.37 in the solid, dashed and
 dotted lines, respectively.
}
\end{figure}

 In previous section, we have presented fractions of elliptical galaxies
 in several different ways. Since our morphological classification,
 density measurement are different from most of previous work, it is
 important to know how these elliptical fractions differ due to the
 choice of relevant parameters. We use the result of Whitmore et
 al. (1993) as a benchmark for our study since they applied various
 systematic corrections carefully 
 and the results are relatively widely used. Their result is elliptical:S0:spiral=18\%:23\%:59\%. 
   In Figure \ref{fig:md_ann_cin}, we have
 55\% of ellipticals in the least dense bin and 85\% of them in the
 densest bin.    In Figure \ref{fig:md_mr_cin}, we elliptical fractions vary from 60\% to
 90\%. Between these two figures, our values agree each other within the
 errors, suggesting
 our values are internally consistent. However, our elliptical fractions
 are slightly higher than the sum of ellipticals and S0s (41\%) in
 Whitmore et al. (1993). As noted in section  \ref{sec:md_analysis}, this comes
 from our choice of $Cin$=0.4 
 criteria slightly leaned toward spiral galaxies. Figure
 \ref{fig:md_ann_cin} showed that slight change in $Cin$ criteria can
 change absolute amount of elliptical galaxies dramatically, and thus,
 careful attention is needed when comparing our work with others. In our
 case, $Cin$=0.37 criteria shown in a dotted line in Figure
 \ref{fig:md_ann_cin} is closer to the classification of Whitmore et al. (1993).

   In Figure \ref{fig:md_md_ann_ytype}, we have 10\% of
 ellipticals in the least dense bin and 25\% in the densest bin. S0s are
 25\% in the least dense bin.
   In Figure \ref{fig:md_mr}, elliptical fractions  vary from 15\% to
 30\% and S0s vary from 25\% to 40\%. 
 Again our values
 are consistent within the errors internally. In addition, in these cases,
 values are only two sigma away from Whitmore et al. (1993). If we sum
 up 
 ellipticals and S0s, our values are 30\% and 35\%, whereas Whitmore
 reports 41\%. Therefore in case of $Tauto$ parameter, our choice of
 criteria is similar to that of Whitmore et al. (1993).

\subsection{The Morphology-Density Relation with $Cin$}
 
 In section \ref{sec:md_morphology_density}, two interesting results are
 found in the SDSS data using $Cin$=0.4 as a classification criterion;
   (i) Morphology-density relation exists in the SDSS data,
 however, it flatters out at low density. (ii) The characteristic galaxy
 density is at  around 3 Mpc$^{-2}$. In section \ref{Oct 30 18:22:01
 2002}, we analyzed 
 morphology-density relation in the view point of morphology-radius
 relation. The flattering at low density is seen as well and its turning
 point is at around 1 $R_{vir}$. 
  Since computer simulations sometimes use different magnitude range,
 different density measurement and different morphological
 classification (often bulge-to-disc ratio), it is difficult to do
 accurate direct comparison.
 However, both of morphology-density relation and morphology-radius relation are
 qualitatively in agreement with computer simulations such as Okamoto et
 al. (2001), Diaferio et al. (2002), Springel et al. (2000) and Benson
 et al. (2002).
  The flattering of morphology-density relation we saw in both Figures
 \ref{fig:md_ann_cin} and \ref{fig:md_mr_cin} is
 interesting since it suggests that whatever physical mechanism is
 responsible for morphological change of late-spiral galaxies in 
 dense regions, the mechanism starts working at galaxy density  $\sim$3
 Mpc$^{-2}$ or higher.

 Various mechanism can be responsible for the
 morphological change. 
  These include ram pressure stripping of gas (Spitzer \& Baade 1951;
 Gunn \& Gott 1972; Farouki \& Shapiro 
 1980; Kent 1981; Abadi, Moore \& Bower 1999; Fujita \& Nagashima 1999;
 Quilis, Moore \& Bower
 2000), galaxy infall (Bothun \& Dressler 1986; Abraham et al. 1996;
 Ellingson et al. 2001), 
 galaxy harassment (Moore et al. 1996,1999), cluster tidal forces (Byrd
 \& Valtonen 1990; Valluri 1993),  enhanced star formation (Dressler \&
 Gunn 1992), and   removal \& consumption of the gas (Larson,
 Tinsley \& Caldwell 1980; Balogh et al. 2001; Bekki et al. 2002).
  It is yet unknown exactly what processes play major roles in creating
 morphology-density relation. However, the mechanism must be the one that
 works at galaxy density 3 Mpc$^{-2}$ or higher. It is also interesting
 to note that this characteristic density coincides with the density
 where galaxy star formation rate abruptly drops (Lewis et al. 2002;
 Gomez et al. 2003). The coincidence suggests that the same mechanism
 might be
 responsible for both morphology-density relation and truncation of star
 formation rate.

\subsection{The Morphology-Density Relation with $Tauto$}

 In Figures \ref{fig:md_md_ann_ytype} and \ref{fig:md_mr}, we further studied
 the morphology-density and the morphology-radius relation using $Tauto$
 parameter, which allows us to divide galaxies into four categories
 (elliptical, S0, early-spiral and late-spiral). 
  In addition to the general trend found in the previous
 section, we found two characteristic changes in the relation at around
 galaxy density 2 and 6 Mpc$^{-2}$ or in terms of radius, 0.3 and 2
 virial radii. In the sparsest regions (below 2 Mpc$^{-2}$ or outside of
 2 virial radius), both 
 relations becomes almost flat, suggesting that the responsible physical
 mechanisms do not  work very well in these regions. In the intermediate
 regions (density between 2 and 6 Mpc$^{-2}$ or virial radius between
 0.3 and 2), S0 fractions dramatically increase toward denser or smaller
 radius regions, whereas  fractions of late-spiral galaxies
 decrease. In the densest regions (above 6 Mpc$^{-2}$ or inside of
 0.3 virial radius), interestingly S0 fractions decrease, and in turn,
 elliptical fractions radically increase suddenly. The change in the
 densest region are further confirmed in Figures \ref{fig:md_md_es0} and
 \ref{fig:md_mr_es0}, where we plotted S0 to elliptical number ratio as a
 function of density or cluster-centric-radius. In both Figures, S0 to
 elliptical ratio declines suddenly at the densest region. 

  The existence of two characteristic change in both the morphology-density
 and the morphology-radius relation suggests the existence of two
 different physical mechanisms responsible for each of the two morphological fraction
 changes. 
   In the intermediate region (density between 2 and 6 Mpc$^{-2}$
 or virial radius between 0.3 and 2), the mechanism creates S0 galaxies
 mostly, by reducing fractions of late-spiral galaxies. Although there is not much
 change in early-spiral fractions, perhaps it is natural to imagine that the
 mechanism turns late-spirals into early-spirals, and then early-spirals
 into S0s. As Figure \ref{fig:md_size} shows, median sizes of galaxies
 gradually declines from late-spirals to S0s, suggesting calm, gradual
 transformation of galaxies, maybe due to the truncation of star
 formation as observed at the same environment by Gomez et al. (2003)
 and Lewis et al. (2002). After the truncation of star formation, outer
 part of a galaxy disc fades 
 away as massive stars die. The plausible
 candidates of this mechanism includes ram-pressure stripping (Gunn \& Gott 1972; Farouki
 \& Shapiro 1980; Kent 1981; Abadi, Moore \& Bower 1999; Quilis, Moore \& Bower
 2000), unequal mass galaxy mergers (Bekki et al. 1998), galaxy
 harassment (Moore et al. 1999)   and  truncation of
 star formation due to the cluster environment (Larson, Tinsley \& Caldwell 1980; Balogh et al. 2001; Bekki et al. 2002).

  Very different consequences are found in the densest region
 (above 6 Mpc$^{-2}$ or inside of 
 0.3 virial radius), where the mechanism decrease S0 fractions and increase
 elliptical fractions. In Figure \ref{fig:md_size}, there is a significant
 increase in median galaxy sizes from S0s to ellipticals. Both of these
 observational results suggest that a very different mechanism from
 intermediate region is working in the densest region. Since galaxy size
 becomes larger from S0 to ellipticals, merging scenario is one of the
 candidate mechanisms. Computer simulations based on the galaxy merging
 scenario reported the deficit of S0 galaxies ( Okamoto et al. 2001;
 Diaferio et al. 2002; Springel et al. 2000; Benson 
 et al. 2002), which we might have seen
 observationally in the densest region of our data. 
 In previous work, Dominguez et
 al. (2002) also suggested that there are two mechanisms in the
 morphology-density relation; one with global nature and the other with
 local effects. Our findings of two characteristic changes in the
 morphology-density relation is perhaps an observational result of the same
 physical phenomena as Dominguez et al. (2001) noted from  a different point
 of view.

\subsection{Comparison with MORPHS data}  
  In section \ref{Oct 30 18:41:06 2002}, we compared the
  local morphology-density relation (SDSS; $z\sim$0.05) with that seen
  at higher
  redshift (MORPHS;0.37$<z<$0.5). Interestingly, two
  morphology-density relations agreed each other. The agreement suggests
  that morphology-density relation was already established at $z\sim$0.5
  as it is in the present universe, i.e., the origin of
  morphology-density relation stays in much higher redshift
  universe.  In the densest environments, there might be a sign of
  excess elliptical fractions in the SDSS than in the MORPHS. Although
  two data points agrees within the error, such an excess of elliptical
  galaxies might suggest additional formation of elliptical galaxies
  between $z=$0.5 and $z=$0.05.
  Little evolution of morphology-density relation is also
  interesting in terms of comparison with computer simulations. 
  Benson et al. (2002) predicted that the evolution of
  morphology-density relation will be seen as a shift 
  in morphological fractions
  without significant change in slope. However, a
  caveat is that absolute value of our Figure \ref{fig:md_md_morphs}
  depends solely on the calibration of concentration parameters of the
  both data, which by nature is difficult to calibrate accurately due to
  the large scatter in both concentration parameters. Therefore the
  results on evolution should not be over-interpreted.


%
%
%
%
%

\section{Summary}\label{sec:md_conclusion}

 We have studied the morphology-density relation and the
 morphology-cluster-centric-radius relation using a volume limited SDSS
 data (0.05$<z<$0.1, $Mr^*<-$20.5). Major improvements in this work are;
 (i) automated 
 galaxy morphology, (ii) three dimensional local galaxy density
 estimation, (iii) the extension of the morphology-density relation into
 the field region. Our findings are as follows.
   
 Both the morphology-density relation and the
 morphology-cluster-centric-radius relation are seen in the SDSS data for both of our
 automated morphological classifiers, $Cin$ and
 $Tauto$. 

  We found there are two characteristic changes in both the
 morphology-density and the morphology-radius relations, suggesting that
 two
 different mechanisms are responsible for the relations.
  In the sparsest regions (below 2 Mpc$^{-2}$ or outside of 2 virial
 radius), both relations are not evident, suggesting the
 responsible physical mechanisms require denser environment. The
 characteristic density (2 Mpc$^{-2}$) or radius (2 virial radius) 
 coincides with the sharp turn in the SFR-density relation
 (Gomez et al. 2003; Lewis et al. 2002), suggesting the same mechanism
 might be responsible for both the morphology-density relation and the SFR-density relation. 
   In the intermediate density regions, (density between 2 and 6
 Mpc$^{-2}$ or virial radius between 0.3 and 2), S0 fractions increase
 toward denser regions, whereas late-spiral fractions
 decrease. Considering the median size of S0 galaxies are smaller than
 that of late-spiral galaxies
 (Figure \ref{fig:md_size}) and star formation rate radically declines in
 these regions (Gomez et al. 2003; Lewis et al. 2002), the mechanism
 that gradually reduces star formation might be responsible for
 morphological changes in these intermediate density regions
 (e.g., ram-pressure stripping). The mechanism is likely to stop star formation in
 late-spiral galaxies, then late-spiral galaxies becomes early-spirals
 and eventually turns into smaller S0s after their outer discs and
 spiral arms become 
 invisible as young stars die. 
   In the densest regions (above 6 Mpc$^{-2}$ or inside of
 0.3 virial radius), S0 fractions decreases radically and elliptical
 fractions increase. This is a contrasting results to that in
 intermediate regions and it suggests that yet another mechanism is
 responsible for morphological change in these regions. Considering that
 the median sizes of elliptical galaxies are larger than that of S0
 galaxies, one of the candidate mechanisms is merging scenario, where
 merging of S0 galaxies 
 creates larger elliptical galaxies. The deficit of S0 galaxies at the
 densest regions are likely to be consistent with computer simulations
 based on merging scenario (Okamoto et al. 2001; Diaferio et
 al. 2002; Springel et al. 2000), which predicted 
 small fraction of S0 galaxies.  
  The existence of two different mechanisms itself is consistent with the
 discovery of Domiguez et al. (2001), where they found two different 
 parameters governing the morphology-density relation in cluster centers
 and outskirts  separately. 

%
%
%

 We also compared our morphology-density relation from the SDSS
 ($z\sim$0.05) with that of the MORPHS data ($z\sim$0.5). Two
 relations lie on top of each other, suggesting that the
 morphology-density relation was already established at $z\sim$0.5 as is
 in the present universe. In the densest bin, a slight sign of excess
 elliptical fraction was seen in the SDSS data, which might be
 indicating the formation of additional elliptical galaxies between $z=$0.5 and
 $z=$0.05.

\clearpage

\clearpage

\clearpage

\clearpage

\clearpage

\clearpage

\clearpage

\clearpage

\begin{figure}[h]
\begin{center}
\includegraphics[scale=0.7]{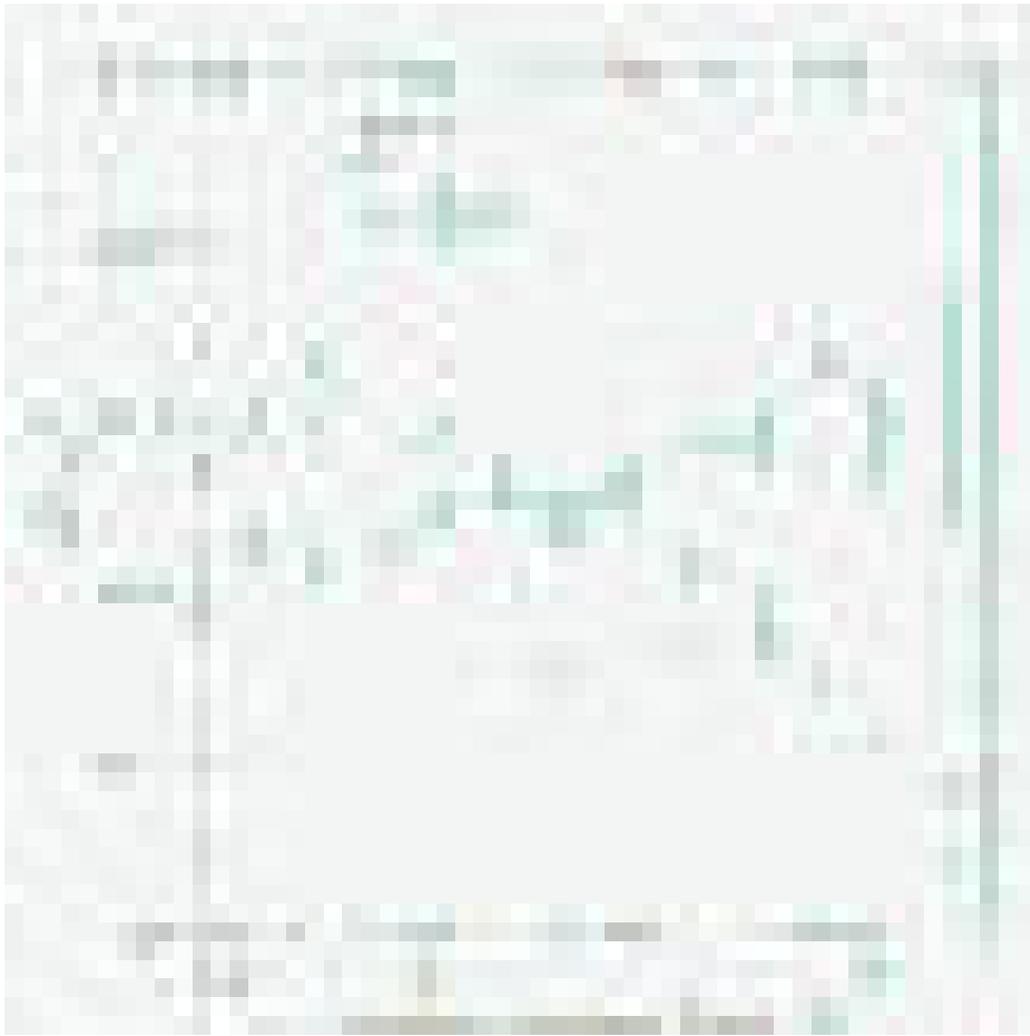}
\end{center}
\caption{
\label{fig:md_md_ann_ytype}
The morphology-density relation for four types of galaxies classified
 with $Tauto$.  
 The short-dashed, solid,
  dotted and long-dashed lines represent 
  elliptical,  S0, early-spiral and late-spiral galaxies, respectively.
}
\end{figure}

\clearpage

\begin{figure}[h]
\begin{center}
\includegraphics[scale=0.7]{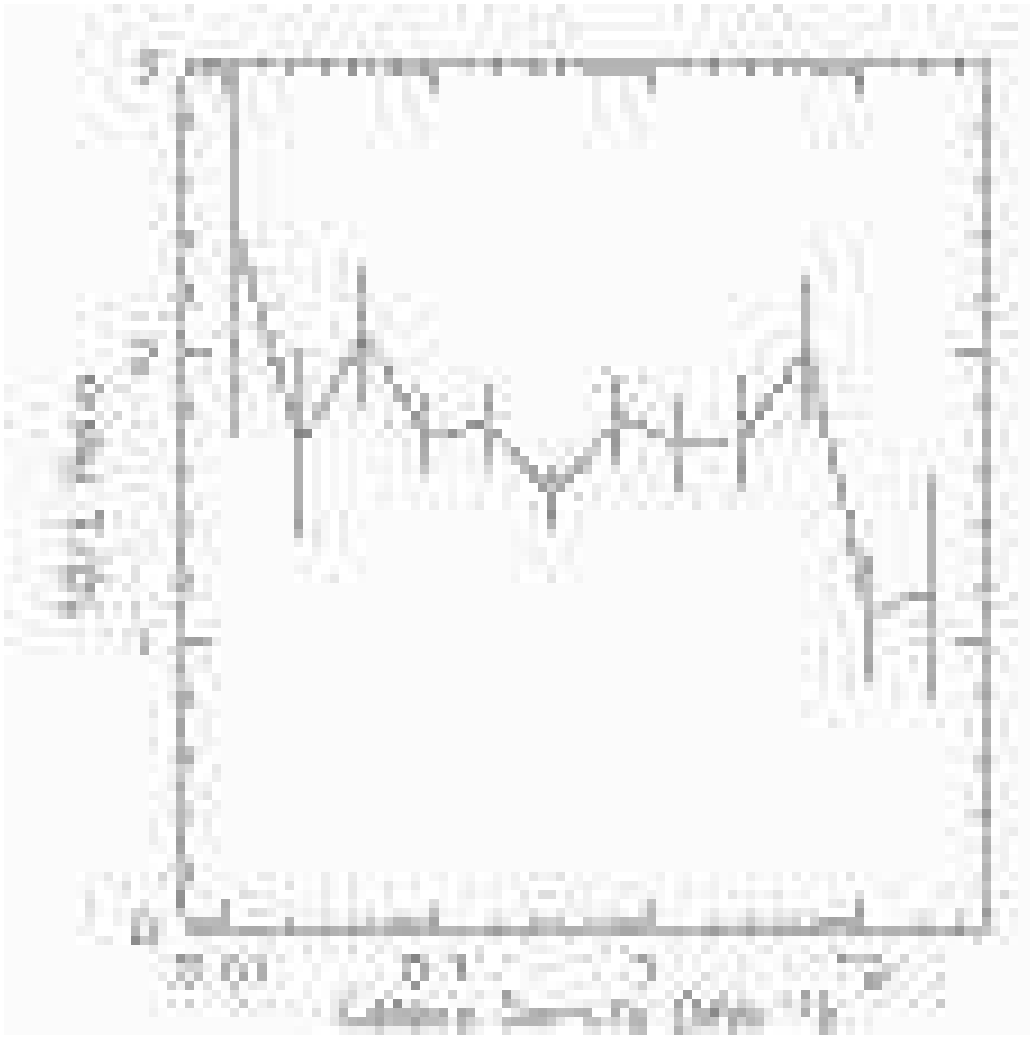}
\end{center}
\caption{
\label{fig:md_md_es0}
E/S0 number ratio as a function of local galaxy density.
}
\end{figure}

\clearpage

\begin{figure}[h]
\includegraphics[scale=0.7]{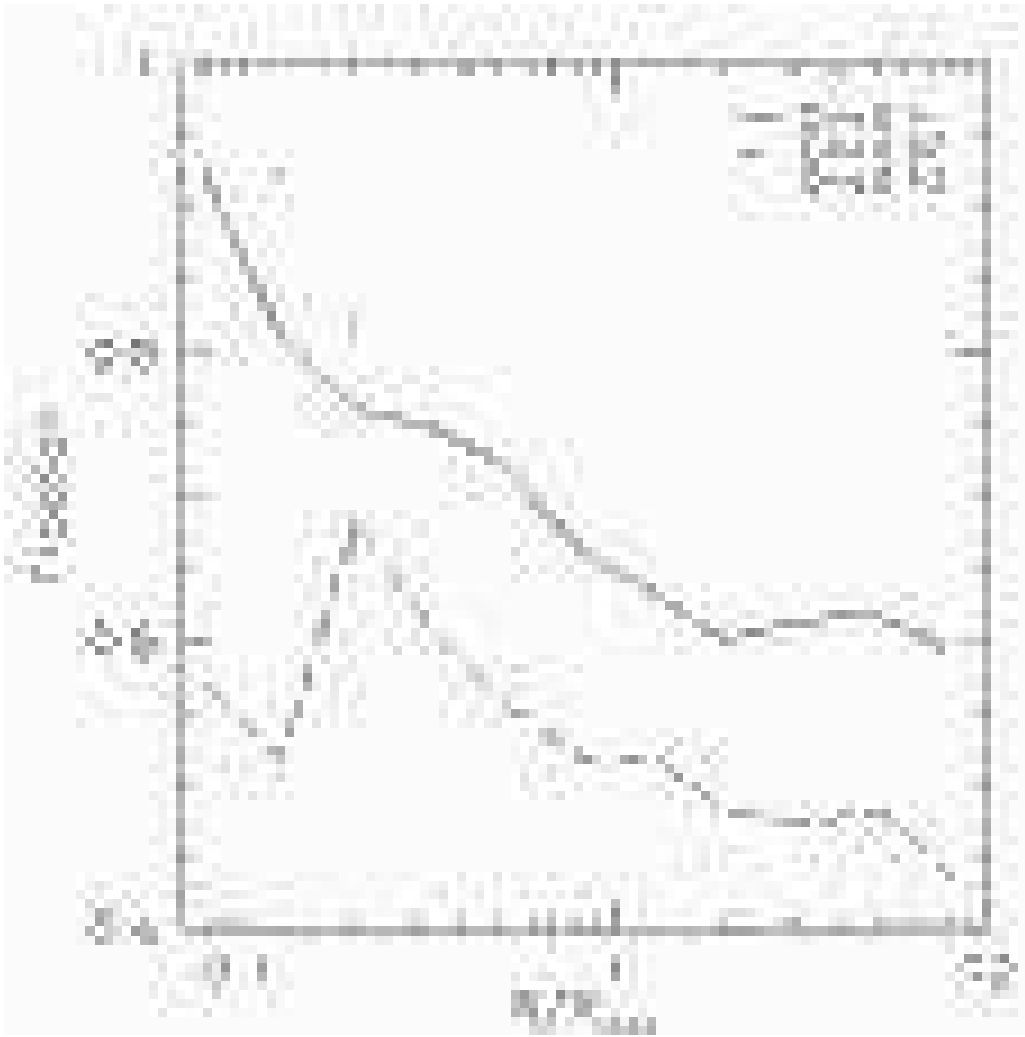}
\caption{
\label{fig:md_mr_cin} 
 The morphology-radius relation. Fractions of elliptical galaxies are plotted against cluster centric radius to the nearest cluster.  
 Criteria are $Cin<$0.4, $Cin<$0.43 and $Cin<$0.37 in the solid, dashed and
 dotted lines, respectively.
}
\end{figure}

\clearpage

\begin{figure}[h]
\includegraphics[scale=0.7]{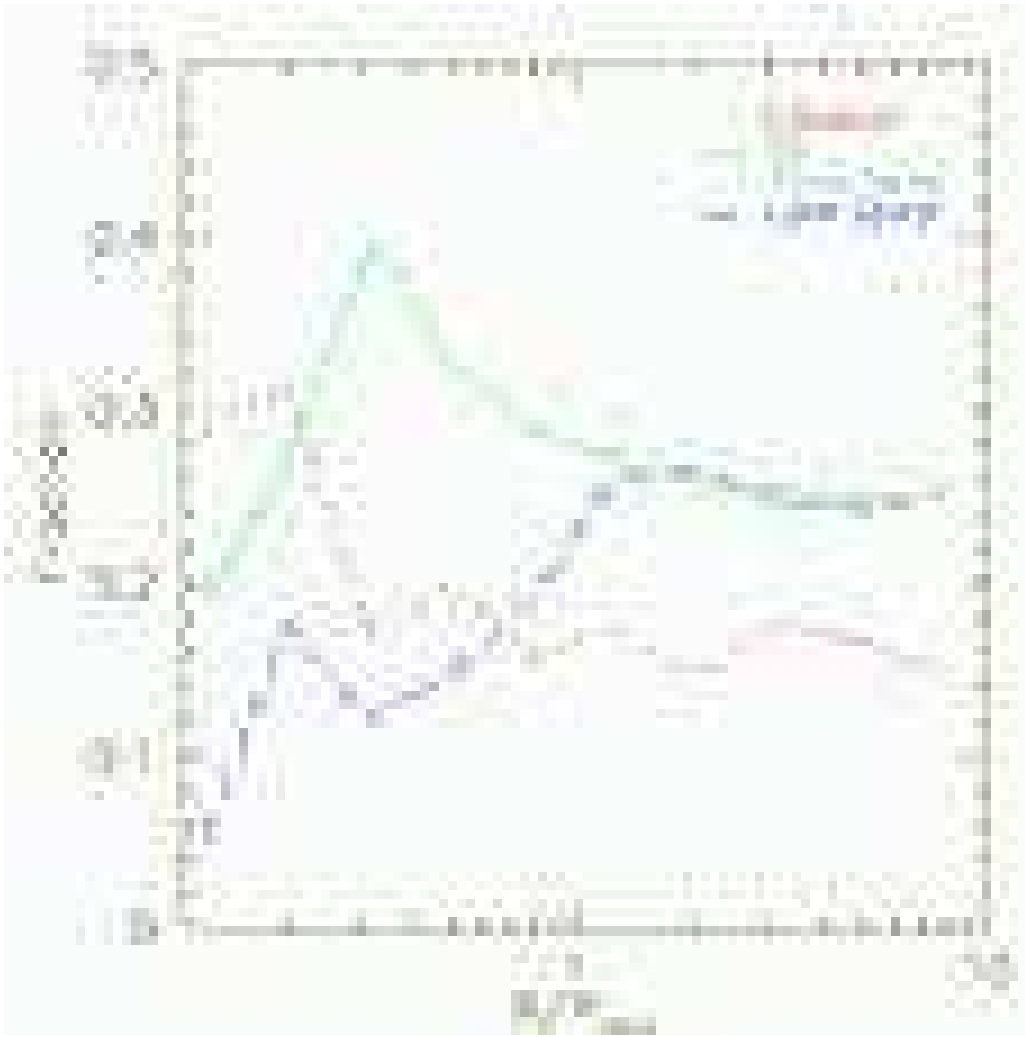}
\caption{
\label{fig:md_mr} 
 The morphology-radius relation. Fractions of each type of a galaxy is
 plotted against cluster centric radius to the nearest cluster. The short-dashed, solid,
  dotted and long-dashed lines represent 
  elliptical,  S0, early-spiral and late-spiral galaxies, respectively.
}
\end{figure}

\clearpage

\begin{figure}[h]
\includegraphics[scale=0.7]{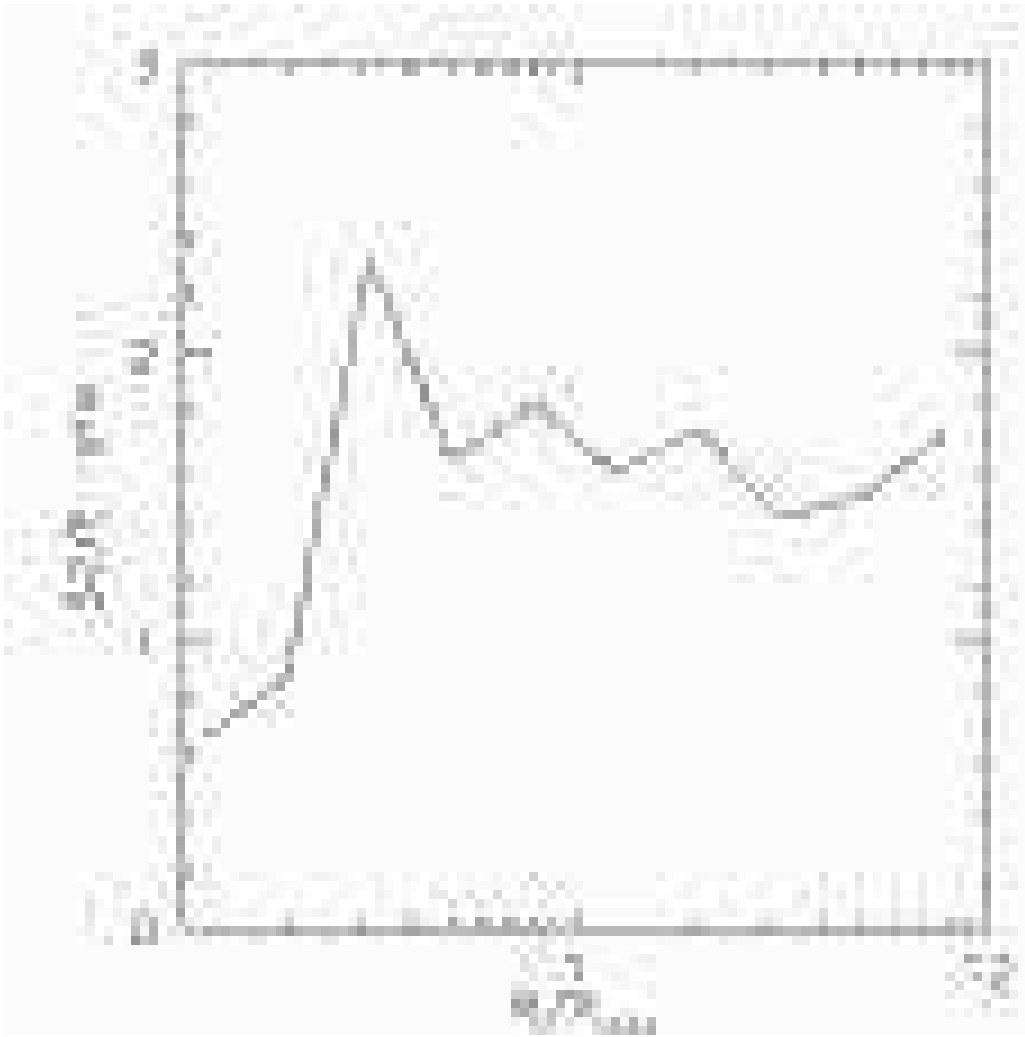}
\caption{
\label{fig:md_mr_es0} 
 S0 to elliptical number ratio is plotted against cluster centric
 radius. The ratio decreases at the cluster core region.
}
\end{figure}

\clearpage

%

%

%
%

\clearpage
\begin{figure}[h]
\includegraphics[scale=0.7]{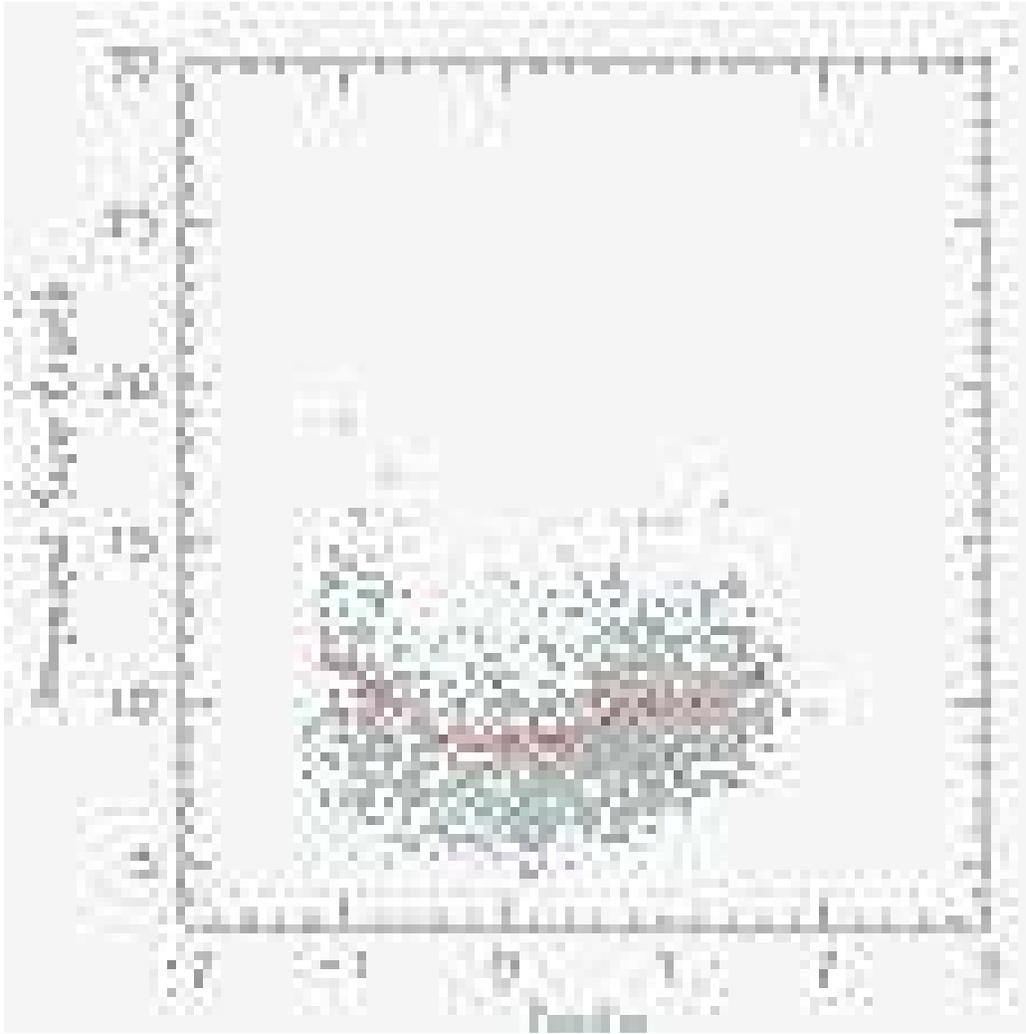}
\caption{
\label{fig:md_size} 
Physical sizes of all 7938 galaxies are plotted against $Tauto$. Petrosian 90\%
 flux radius in $r$ band is used to calculate physical sizes of
 galaxies. The solid line shows medians. It turns over around
 $Tauto\sim$0, corresponding to S0 population.
}
\end{figure}

\clearpage
\begin{figure}[h]
\includegraphics[scale=0.7]{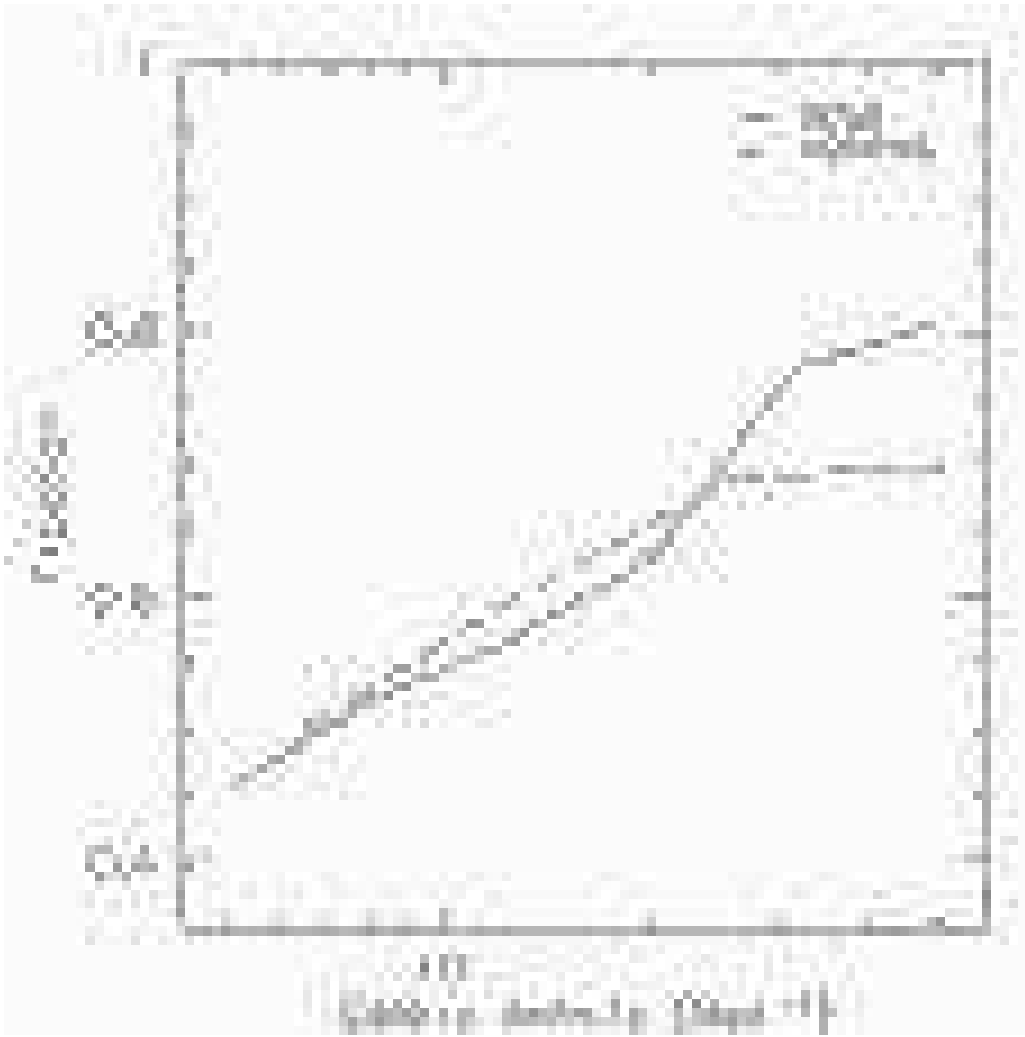}
\caption{
\label{fig:md_md_morphs} 
 Comparison of the morphology-density relations of the SDSS(low redshift)
 and the MORPHS(high redshift). Fraction of elliptical galaxies are
 plotted against local galaxy density within 250 kpc. The MORPHS data
 are plotted in the solid line, and the SDSS data are plotted in the dashed line.
}
\end{figure}

\clearpage

\begin{table}[h]
\begin{center}
\caption{
\label{tab:md_completeness}
 Completeness and contamination rate of our four sample of galaxies
 classified by $Tauto$ are calculated using eye-classified morphology.
}
\begin{tabular}{lll}
\hline
 Type  & Completeness (\%)  & Contamination (\%)\\
\hline
\hline
 Elliptical ($Tauto \leq$-0.8)         & 70.3 &  28.2   \\
 S0         (-0.8$\leq Tauto<$0.1)     & 56.4 &  56.5    \\ 
 Early Spiral  (0.1$\leq Tauto<$1.0)   & 53.1 &  24.1  \\ 
 Late Spiral    (1.0$\leq Tauto$)      & 75.0 & 45.9      \\ 
 \hline
\end{tabular}
\end{center}
\end{table}

\begin{table}
\begin{center}

\caption{
\label{tab:md_morphs}
 MORPHS cluster sample
}
{\small\scriptsize
\hspace{-1.truein}\begin{tabular}{lccrccccc}
\noalign{\medskip}
\hline\hline
\noalign{\smallskip}
{Cluster} & {R.A.} & {Dec.}  & {$z$} & filter  & {L$_X${\scriptsize (0.3--3.5)}} & $\sigma$ \hfil  \cr
\noalign{\smallskip}
\noalign{\hrule}
\noalign{\smallskip}
A370\#2 & 02~40~01.1 & $-$01~36~45 &  0.37 & F814W &  2.73&1350 [34]  \cr
Cl1447+23 & 14~49~28.2 & $+$26~07~57 &  0.37 & F702W &  ...\hfil & ... \cr
Cl0024+16 & 00~26~35.6 & $+$17~09~43 &  0.39 & F814W & 0.55 & 1339 [33]  \cr
Cl0939+47 & 09~43~02.6 & $+$46~58~57 &  0.41 & F814W &  1.05 & 1081 [31]  \cr
Cl0939+47\#2 & 09~43~02.5 & $+$46~56~07 &  0.41 & F814W & 1.05 & 1081[31]  \cr
Cl0303+17 & 03~06~15.9 & $+$17~19~17 & 0.42 & F702W & 1.05 & 1079 [21]  \cr
3C295 & 14~11~19.5 & $+$52~12~21 &  0.46 & F702W & 3.20 & 1670 [21]  \cr
Cl0412$-$65 & 04~12~51.7 & $-$65~50~17 &  0.51 & F814W & 0.08 & ...\hfil  \cr 
Cl1601+42 & 16~03~10.6 & $+$42~45~35 &  0.54 & F702W & 0.35 & 1166 [27]  \cr
Cl0016+16 & 00~18~33.6 & $+$16~25~46 &  0.55 & F814W & 5.88 & 1703 [30]  \cr
Cl0054$-$27 & 00~56~54.6 & $-$27~40~31 & 0.56 & F814W &  0.25 & ...\hfil   \cr
\hline\hline
\end{tabular}
}
\end{center}
\end{table}


\chapter{The Environment of Passive Spiral Galaxies}
\label{chap:PS}

\section{Introduction}

 Recent morphological studies of distant cluster galaxies revealed the
 presence of an unusual population of galaxies with a spiral morphology
 and lack of star-formation activity (Couch et al. 1998; Dressler et
 al. 1999; Poggianti et al. 1999). The origins of these ``passive
 spirals'' have remained a mystery since it
 has been difficult to understand the existence of such galaxies.  
  The phenomena suggest that star formation in these system
 has ended calmly, without disturbing their spiral arm structures. 
 Many people speculated that cluster related phenomena might be
 responsible for creation of passive spiral galaxies since they are found
 during the cluster studies.  However, it has not been well established if
 these phenomena are more relevant in clusters or they are common in the
 field regions as well, simply because it
 has been difficult to study this rare class of galaxies in the field
 region.    

    Also an existence of a
  similar type of galaxies has been reported. Galaxies with a low arm
  inter-arm contrast
  in their disks were classified as anemic by  van den Bergh
  (1976). He found the excess of anemic spiral galaxies in Virgo cluster.
  Various HI follow-up
  observations revealed lower gas density in anemic spiral galaxies,
  which, presumably, is the cause of  lower star formation rate and smoother spiral arms
 (Bothun \& Sullivan 1980; Wilkerson 1980; Phillipps 1988; Cayatte et al. 1994;
  Bravo-Alfaro et al. 2001).  
  Especially, Elmegreen et  al. (2002) found the gas surface density of anemic
  spirals is below the
  threshold for star formation (Kennicutt 1989), revealing low star
  formation in anemic spirals in fact comes from low gas density.
  Although the
  definition of anemic spiral galaxies is somewhat different from that
  of passive spirals, considering 
  similarities in properties (presence of spiral arms and lack of star
  formation), these two types of galaxies could be essentially the same
  population of galaxies, sharing the same nature and origin. 

  Various possible mechanism are proposed to explain these phenomena.
  Poggianti et al. (1999) found passive spiral galaxies in their
  sample of distant clusters and speculated that these findings show
  that the time scale of spectral change of cluster galaxies are shorter
  than the time scale of morphological change of galaxies. They proposed
  ram-pressure stripping (Gunn \& Gott 1976) as a possible physical
  mechanism responsible for these phenomena. Another possible cause is
  abrupt truncation of gas infall onto disks from the halo regions
  (Larson et al. 1980). Dynamical causes such as major galaxy merger 
  or harassment, which explain other properties of cluster
  galaxies very well (e.g., the Butcher-Oemler effect; Butcher \& Oemler 1978,
  1984), cannot explain these phenomena since such processes    
  disturb spiral arms and do not end up with passive spirals.
  A pioneering work to simulate passive spiral galaxies by combining
  numerical simulation and a phenomenological model was performed by Bekki
  et al. (2002). They demonstrated that halo gas stripping caused by
  dynamical interaction between halo gas and the hot ICM is a plausible
  mechanism.  
  Although these mechanisms are all plausible, the final conclusion has not yet drawn
  about what mechanisms are most responsible for these 
  phenomena.

  It is also interesting to investigate a possible link between passive
  spirals and statistical observational features of cluster galaxies. In
  cluster regions, it is known that fractions of  blue galaxies are
  larger at  higher redshifts. (the Butcher-Oemler effect; Butcher \& Oemler 1978,
  1984; Rakos \& Schombert 1995; Couch et al. 1994; 1998; Margoniner \& De
  Carvalho 2000; Margoniner et al. 2001; Ellingson  et al. 2001; Kodama
  \& Bower 2001; Goto et al. 2003a; Chapter \ref{chap:BO}). Cluster
  galaxies are also known to change their morphology, e.g., spiral to S0 transition
  during the cosmic time scale (Couch \&  Sharples 1987; Dressler
  et al. 1997;  Couch et al. 1998; Fasano et al. 2000;  Diaferio et
  al. 2001). 
  Goto et al. (2003a; Chapter \ref{chap:BO}) referred to this as the
  morphological Butcher-Oemler effect. If passive spiral
  galaxies are of cluster origin, they would fit well in both spectral
  and morphological evolution of cluster galaxies. They may be galaxies in transition
  between blue and red, or spiral and S0s.

 Since the Sloan Digital Sky Survey (SDSS; York et al. 2000) observes
 spectra of one million galaxies in one quarter of the sky. It provides
 us with the opportunity to study this interesting population of galaxies
 in all environments; from cluster core regions to general field
 regions. In addition, wide spectral coverage of 3800-9000 \AA\ allows us
 to study both [OII] and H$\alpha$ emission lines at the same time,
 which can reduce 
 possible biases from dust extinction and stellar absorption on the
 emission lines. 
 In this Chapter, we concentrate in revealing the environment of passive
 spiral galaxies. In Section \ref{sec:ps_data}, we explain the data used in the
 study. In Section \ref{sec:ps_selection}, we carefully define passive spiral galaxies. In
 Section \ref{sec:ps_density}, we present the environment of passive spiral galaxies. In
 Section \ref{May 29 10:07:03 2003}, we discuss the possible caveats and interpretation of the
 results. In Section \ref{sec:ps_conclusion}, we summarize our findings. The cosmological
 parameters adopted throughout this chapter are $H_0$=75 km
 s$^{-1}$ Mpc$^{-1}$, and ($\Omega_m$,$\Omega_{\Lambda}$,$\Omega_k$)=(0.3,0.7,0.0).

\section{Data}\label{sec:ps_data}

 In this section, we outline the data used in this chapter. 
  The galaxy catalog is taken from the Sloan Digital Sky Survey (SDSS;
  see Fukugita et al. 1996; Gunn et al. 1998;  Lupton 
 et al. 1999, 2001, 2002; York et al. 2000; Hogg et al. 2001; Pier et
  al. 2002; Stoughton et al. 2002; Strauss et al. 2002  and Smith et
  al. 2002 for more detail   of the SDSS data).
 The SDSS imaging survey observes one quarter of the sky to depths of 22.3, 23.3, 23.1, 22.3
 and 20.8 in the $u,g,r,i$ and $z$ filters, respectively (See
 Fukugita et al. 1996 for the SDSS filter system, Hogg et al. 2002
 and Smith et al. 2002 for its calibration). 
 Since the SDSS photometric system is not yet finalized, we refer to the
 SDSS photometry presented here as $u^*,g^*,r^*,i^*$ and $z^*$. We
  correct the data for galactic extinction determined from the maps
  given by Schlegel, Finkbeiner \& Davis (1998).  
 We include galaxies to $r^*$=17.7 (Petrosian magnitude), which is the
 target selection limit of the main galaxy sample of the SDSS spectroscopic survey.
 The spectra are obtained using two fiber-fed spectrographs (each with 320
 fibers) with each fiber subtending 3 arcseconds on the sky. 
 (We investigate aperture bias due to the limited size of the SDSS fiber
 spectrograph in Section \ref{sec:ps_aperture}). 
 The wavelength
 coverage of the spectrographs is 3800\AA\ to 9200\AA, with a spectral
 resolution of 1800. These spectra are then analyzed via the SDSS
 SPECTRO1D data processing 
 pipeline to obtain various quantities for each spectrum such as
 redshift, spectral classification and various line parameters. (see
  Stoughton et al. 2002; Frieman et al., in prep, for further 
 details). 
  The SDSS has taken 189,763 galaxy spectra as of the date of writing. Among
  them we restrict our sample to galaxies with S/N in $g$ band
  greater than 5 and with a redshift confidence of $\geq0.7$.
 Since we use concentration parameter in selecting passive
 spiral galaxies, we also remove galaxies with PSF size in $r$ band greater than
 2.0'' to avoid poor seeing mimicing less concentrated galaxies.  
  Then we make a
  volume limited sample by restricting our sample to 0.05$<z<$0.1 and
  $Mr^*<-$20.5. This magnitude limit corresponds to $M^*$+0.3 mag (Blanton
  et al. 2001). 
  The lower redshift cut is made to avoid strong aperture
  effects (see Gomez et al. 2003 for detailed investigation in aperture
  effects in the SDSS data). When calculating absolute magnitudes,  
  we use a  $k$-correction code provided by Blanton et al. (2002; v1\_11). 
  In this volume limited sample, there are 25,813 galaxies remained. 

\section{Selection of Passive Spiral Galaxies}\label{sec:ps_selection}

\subsection{Line Measurements}
 
 We measure [OII], H$\alpha$ and H$\delta$ equivalent widths (EWs) with the flux summing method
 as described in Goto et al. (2003b; Appendix \ref{EA1}). We briefly summarize the method
 here.  To estimate continuum, we fit a line using wavelength ranges
 around each line as listed in Table
 \ref{tab:ps_ps_wavelength}. The continuum values are weighted according to the
 inverse square of the errors during the fitting procedure. We then sum
 the flux in the wavelength range listed in the same table to obtain the
 equivalent width of the lines. For  H$\delta$ line, we used two
 different wavelength range; wider one for strong line and narrower one
 for weak line (details are described in Goto et al. 2003b; Appendix \ref{EA1} ). 
 Note that for H$\alpha$ line, we do not
 deblend adjacent [NII] lines. As a result, our H$\alpha$ equivalent
 width have contamination from [NII] lines. However, the contamination is
 less than 5\% from [NII](6648\AA) and less than 30\% from
 N[II](6583\AA).  These measurements show good
 agreement with measurements via Gaussian fitting (Goto et al. 2003b; Appendix \ref{EA1}). 

 We quantified errors of these measurement using spectra observed
 twice in the SDSS. The procedure is exactly the same as described in
 Goto et al. (2003b; Appendix \ref{EA1}). First, the difference of equivalent width are
 plotted against S/N of spectra . Then we fit 3rd polynomial to the 1
 $\sigma$ of the distribution. The polynomial is later used to assign
 errors to every spectra according to its S/N. The exact formula are
 given in Goto et al. (2003b; Appendix \ref{EA1}). Typical errors of high signal-to-noise
 spectra are  1.3, 1.0 and 0.4 \AA\
 for [OII], H$\alpha$ and H$\delta$ EWs, respectively (See Figures 9-11 of
 Goto et al. 2003b; Appendix \ref{EA1}).

\subsection{Selection Criteria}

\begin{figure}
\centering{\includegraphics[scale=0.39]{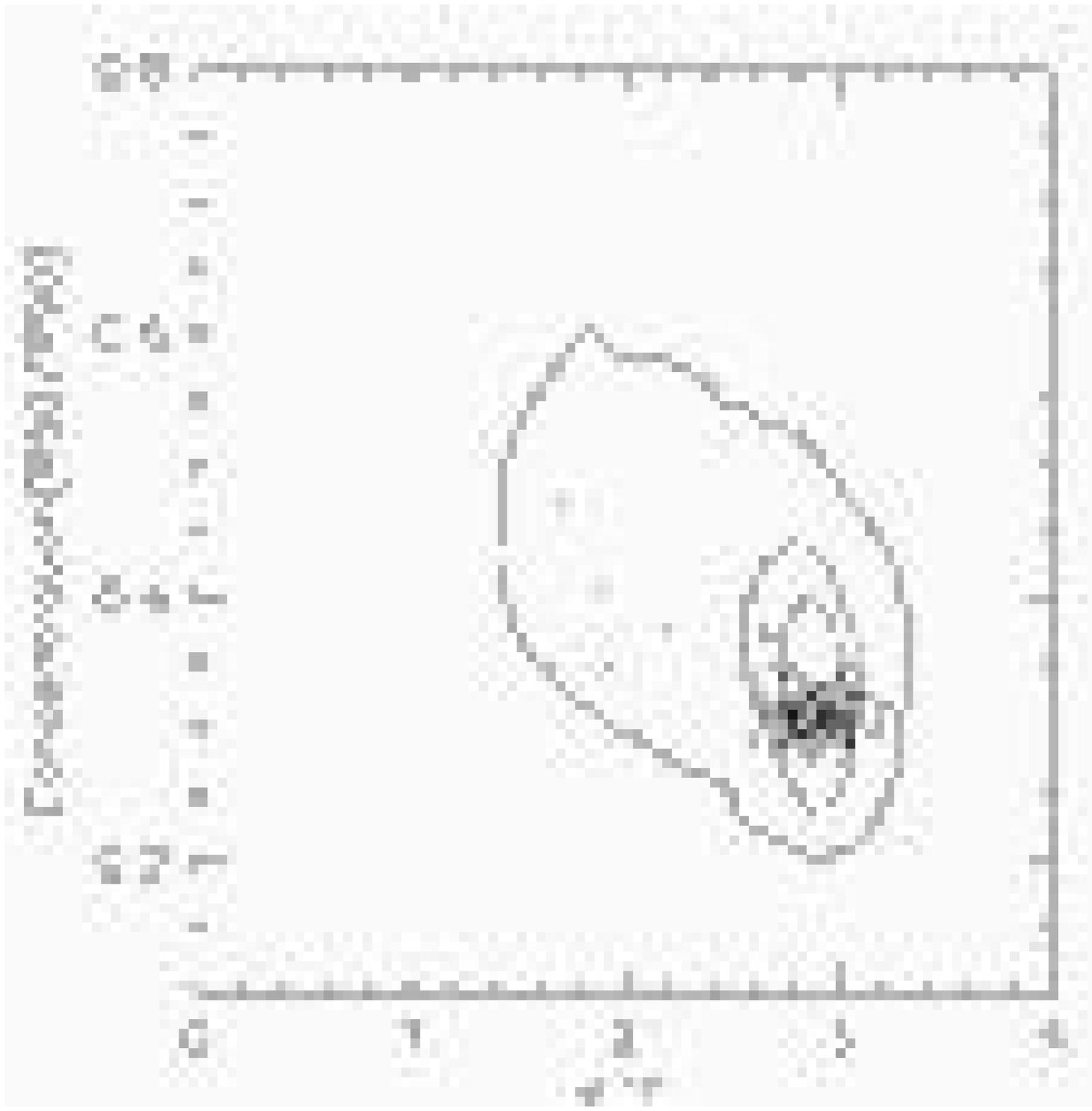}
\includegraphics[scale=0.39]{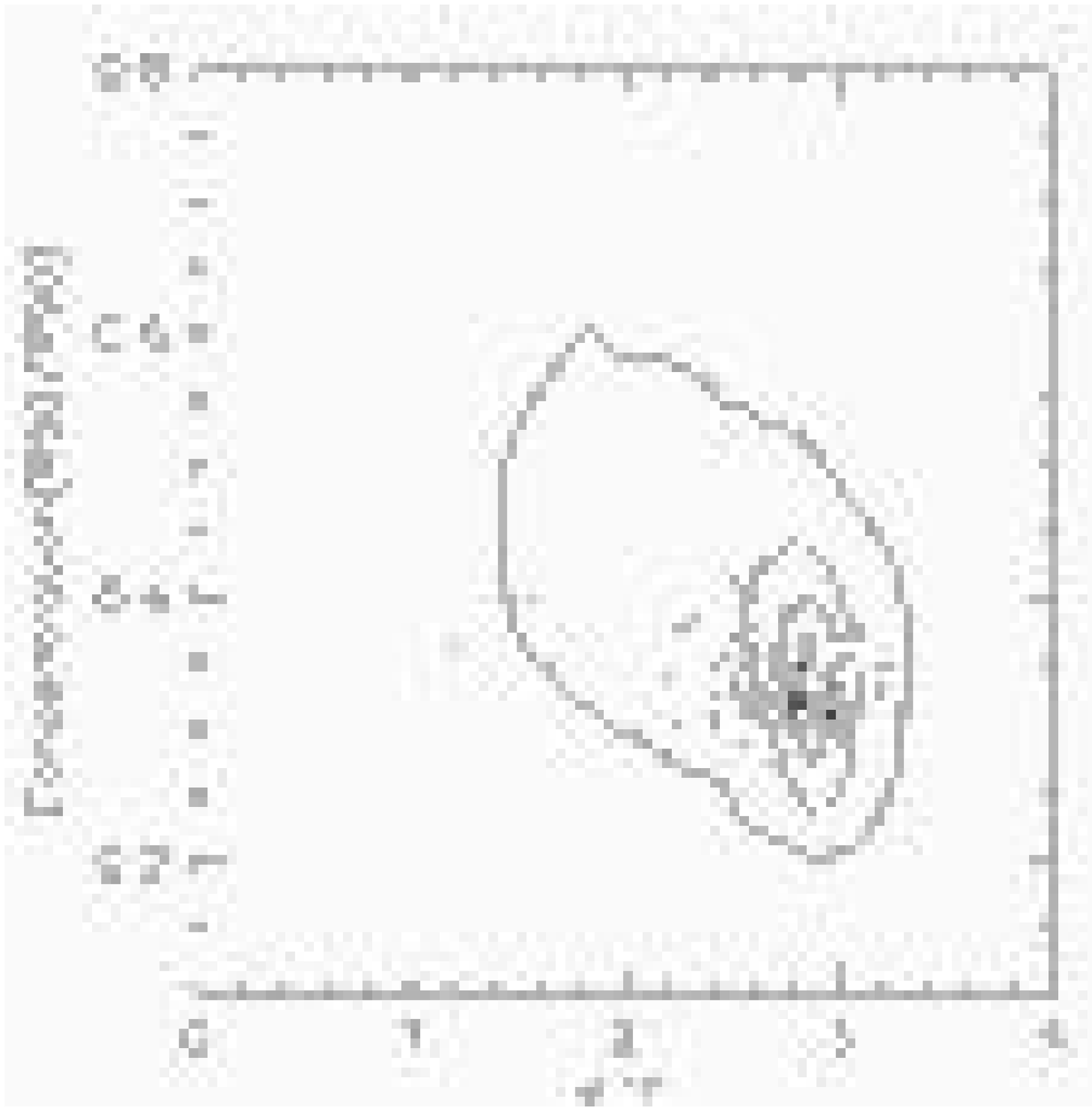}}
\centering{\includegraphics[scale=0.39]{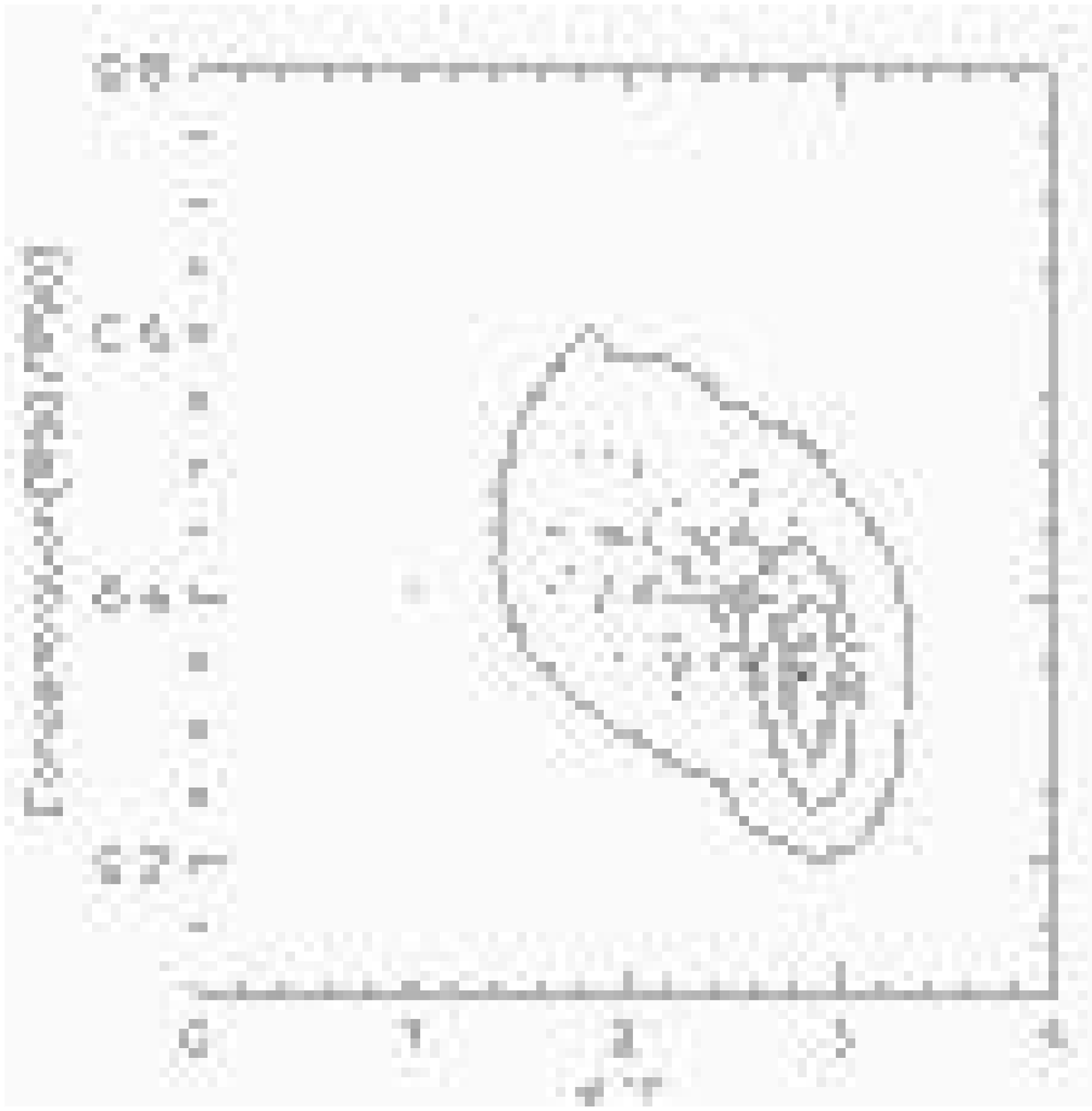}
\includegraphics[scale=0.39]{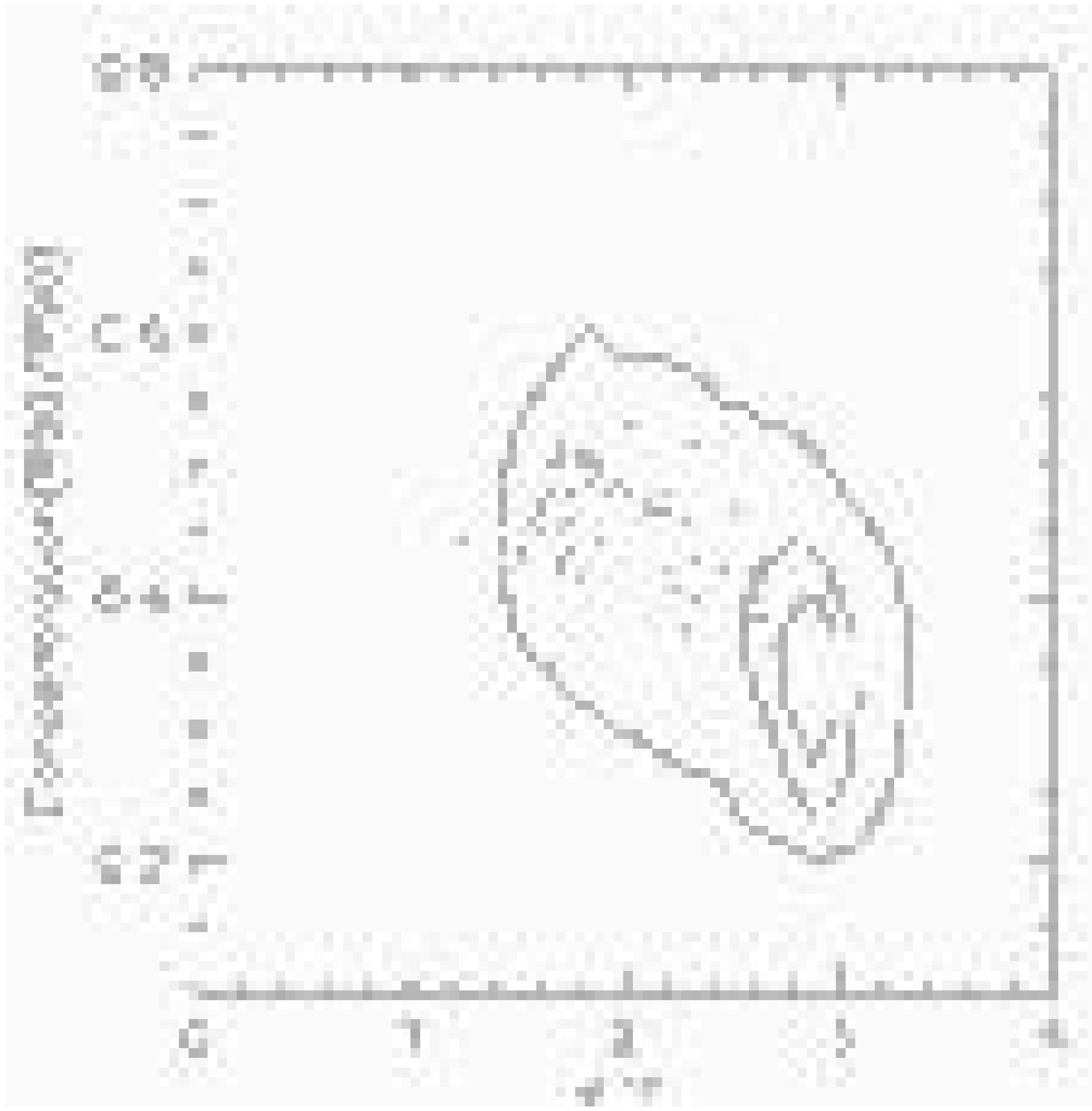}}
\caption{
\label{fig:ps_concent_each_type}
 $Cin$ is plotted against $u-r$. The contours show the distribution of all
 25813 galaxies in the volume limited sample (0.05$<z<$0.1 and
  $Mr^*<-$20.5). A good correlation between
 two parameters is seen. Points in each
 panel show the distribution of each morphological type of galaxies
 classified by eye (Shimasaku et al. 2001; Nakamura et
 al. 2003); Ellipticals are in the upper left panel. S0, Sa and Sc are
 in the upper right, lower left and lower right panels, respectively.   
}\end{figure}

 We select passive spiral galaxies using the following criteria. 
 Galaxies with the inverse of concentration parameter, $Cin>$0.5. 
 The inverse concentration parameter ($Cin$) is defined as the
   ratio of Petrosian 50\% light radius to Petrosian 90\% light radius
 in $r$ band 
 (radius which contains 50\% and 90\% of Petrosian flux, respectively).
 Shimasaku et  al. (2001) and Strateva et al. (2001) studied the
 completeness and 
 contamination of this parameter in detail. See Goto et al. (2002b; Chapter \ref{chap:LF}) and
 Gomez et al. (2003) for more usage of this parameter. The border line
 between spiral galaxies and elliptical galaxies are around
 $Cin$=0.33. Therefore $Cin>$0.5 picks up very less concentrated
 spiral galaxies. In different work, 549 galaxies in our volume limted sample
 were manually classified by Shimasaku et al. (2001) and Nakamura et al.
 (2003). In Figure \ref{fig:ps_concent_each_type}, we overplot
 eye-classified galaxies on a $Cin$ v.s. $u-r$ plane. Contours show the
 distribution of all galaxies in our volume limited sample. Note that
 our volume limited sample contains high fraction of concentrated
 galaxies as shown by contours
 due to its bright absolute magnitude limit (Yagi et al. 2002; Goto et
 al. 2002b; Chapter \ref{chap:LF}).
 In the top
 left panel, 
 eye classified ellipticals are overplotted. In the top right, bottom
 left, bottom right panels, eye-classified S0s, Sa-Sb, Sc or later are
 overplotted, respectively. As is shown in Strateva et al. (2001),
 $u-r$=2.2 also separates early and late-type galaxies well.
  As these panels show, few of elliptical or
 S0 galaxies have $Cin>0.5$. Therefore, we in fact are able to select
 spiral galaxies using $Cin$ parameter, without significant
 contamination from E/S0 population. We caution readers that the
 selection of less concentrated galaxies has a known bias against
 edge-on galaxies, in the sense that edge-on disc galaxies are excluded
 from our sample. The detailed investigation and correction of this
 bias will be presented in Yamauchi et al. (in prep.). However, we accept
 this bias in our sample selection since (i) the bias is independent of local
 galaxy environment; (ii) edge-on galaxies might be affected by larger
 amount of dust extinction, and thus could cast some doubts on truly
 passive nature of our sample galaxies.

\begin{figure}
\begin{center}
\includegraphics[scale=0.7]{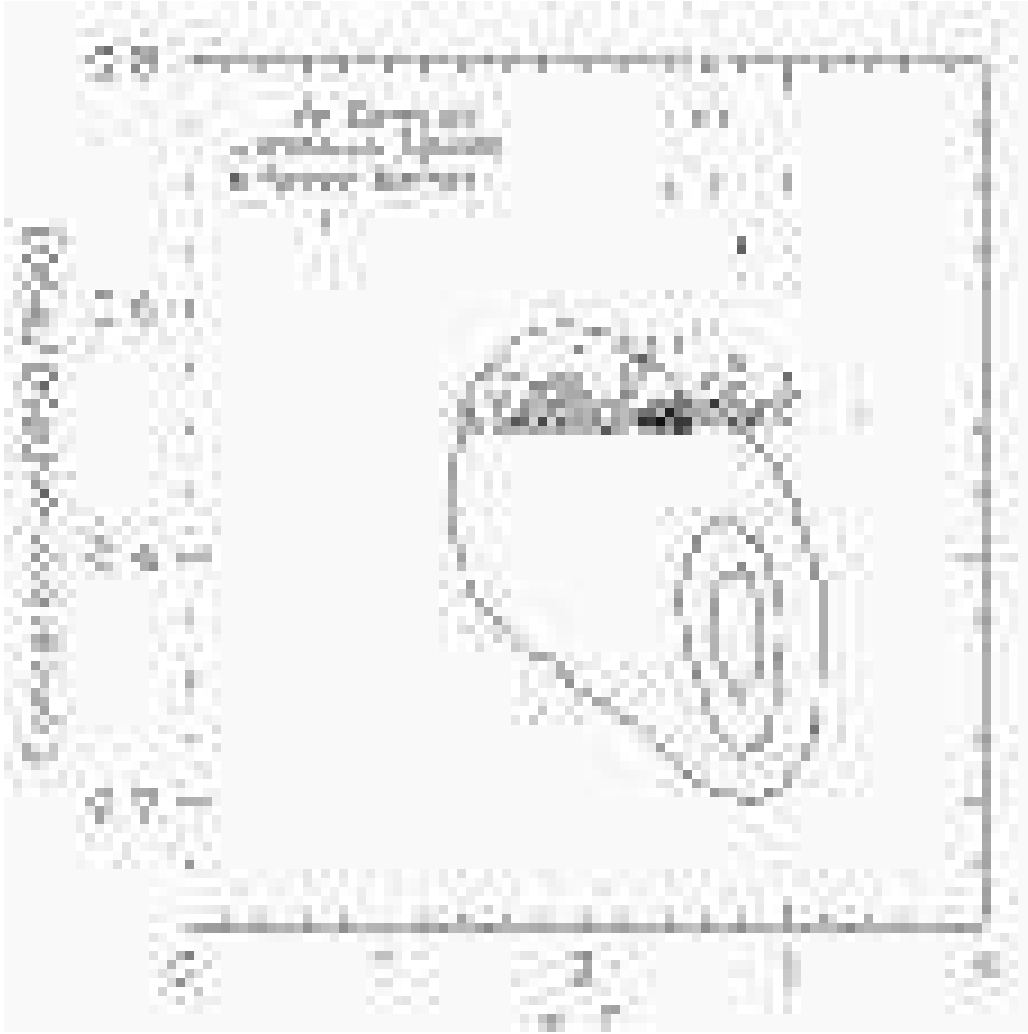}
\end{center}
\caption{
\label{fig:ps_ps_concent}
 The distribution of passive spiral galaxies in $Cin$ v.s. $u-r$
 plane. The contours show the distribution of all 25813 galaxies in our volume
 limited sample. The open circle and filled dots represent passive and
 active spiral galaxies, respectively.}
\end{figure}

   From spectral features, we select galaxies using the following
   criteria.\\
  \begin{equation} 
  [OII]\ EW - 1\sigma_{error}< 0 ,
 \end{equation}
  \begin{equation} 
  H\alpha\ EW - 1\sigma_{error}< 0 ,  
 \end{equation}
 where emission lines have positive signs. 
 In other words, we select galaxies with [OII] and H$\alpha$
 less than 1 $\sigma$ detection (in emission).
 A galaxy which satisfies both of the concentration and spectral
 criteria is regarded as a passive spiral galaxy in this work.
  Figure \ref{fig:ps_ps_concent} 
 shows the distribution of passive spiral galaxies in the $Cin$
 v.s. $u-r$ plane. 
 
  Figure \ref{fig:ps_image} shows example images (30''$\times$30'') of passive spiral
  galaxies. In Figure \ref{fig:ps_spectra}, corresponding spectra are
  shown. Unusual properties of these galaxies are already clear just by
  comparing these two figures, i.e., clear spiral arm structures are
  seen in the images, whereas there are no current star formation activity
  as shown by the lack of [OII] and H$\alpha$ emission lines in the spectra. It
  is interesting to study where these unique features originate from. 
  For a comparison purpose, we also select active (normal) spiral
  galaxies in our sample as galaxies with $Cin>$0.5 and 1 $\sigma$
  detections in both [OII] and H$\alpha$ in emission. We removed
  galaxies with AGN signature from the active spiral sample using a
  prescription given in Kewley et al. (2001) and Gomez et al. (2003).
  When a galaxy satisfies all three line ratio criteria to be an AGN (Figure 15, or
  eq. (5)(6)(7) of  Kewley et al. 2001), we removed it from our
  sample as an AGN.
 Images and spectra of active (normal) spiral galaxies are shown in 
   Figures \ref{fig:ps_as_image} and \ref{fig:ps_as_spectra}. Compared with these
  galaxies, passive spirals have smoother profile.
 
  Among 25813 galaxies in our volume limited sample (0.05$<z<$0.1 and
  $Mr^*<-$20.5), there are 73 (0.28$\pm$0.03\%) passive spiral galaxies in total. The
  number of active spirals is 1059 (4.10$\pm$0.12\%). Relatively small
  percentages stem from our stringent criteria for the inverse concentration
  parameter, $Cin$.

\section{Environment of Passive Spiral Galaxies}\label{sec:ps_density}

 \subsection{Local Galaxy Density}

 First, we aim to clarify the environment where passive spiral galaxies live. 
 For each galaxy in our volume limited sample (0.05$<z<$0.1 and $Mr^*<-$20.5), we measure the
 projected metric distance to the 5th nearest galaxy (within $\pm$3000 km/s of
 the galaxy in redshift space) within the same volume limited sample. If the projected area
 (in ${\rm Mpc^2}$), enclosed by the 5th nearest neighbor 
 distance, touched the boundaries of the SDSS data, we corrected the area
 appropriately for the amount of missing area outside the survey
 boundaries. Then, we divide 5 (galaxies) by the area subtended by the 5th nearest
 galaxy to obtain local galaxy density in Mpc$^{-2}$. 
   This methodology allows us to quote pseudo
 three-dimennsional local galaxy densities for all our galaxies in the volume
 limited sample and without requiring any corrections for background
 and foreground contamination.

 In Figure \ref{fig:ps_density}, we plot the local density distribution of passive
 spiral galaxies in the dotted line.  The solid line shows distribution of all
 galaxies in our volume limited sample. The long-dashed line shows distribution
 of cluster galaxies defined as galaxies within 0.5 Mpc from the nearest
 C4 galaxy cluster (Miller et al. 2003; Gomez et al. 2003) in angular
 direction and within
 $\pm3000{\rm km\, s^{-1}}$ from the redshift of the cluster. 
 Kolomogorov-Smirnov tests show that all the distributions are
 different from each other with more than 99.9\% significance level. 
 The distribution of passive spiral galaxies (the dotted line)
 are right in the middle of cluster galaxies and field
 galaxies. It is also shown  that passive spiral galaxies avoid the
 densest  cluster core regions, and at the 
 same time they do not show the same distribution as field galaxies.
 For a comparison, we plot a distribution of active (normal) spiral
 galaxies by the short-dashed line. Compared with that of all galaxies,
 the distribution
 slightly shifts to less dense environment, as expected from the
 morphology-density relation. The distribution of active spiral galaxies
 is different from that of passive spirals with more than 99.9\% significance. 

\subsection{Cluster Centric Radius}\label{sec:ps_radius}

 In Figure \ref{fig:ps_radius}, we plot the distribution of passive spiral
 galaxies as a function of cluster-centric radius. Here
 cluster-centric radius is measured as projected distance to the nearest
 cluster  within $\pm$3000 km/s from the cluster redshift.  Galaxies
 which do not have any cluster within $\pm$3000 km/s are not included in
 this analysis.  The cluster
 list is taken from Miller et al. (2003), which measures a cluster
 center as the position of the brightest cluster galaxy.
  The 
 physical distance is normalized to virial radius using the relation
 given in Girardi et al. (1998).
 We divide distributions by that of all galaxies and then normalize them to
 unity for clarity. Note that comparisons of fractions among different
 curves are meaningless due to this normalization.
  The dotted, hashed line shows the distribution of passive spiral galaxies. 
 The solid lines show that of active (normal) spiral galaxies.  The dashed line
 is for early-type galaxies selected using  $Cin$ parameter ($Cin<$0.33)
 with no constrains on emission lines. The
 fraction of early-type galaxies is higher at smaller
 cluster-centric-radii and that of spiral galaxies is higher at larger
 radii, which represents so-called ``the morphology-denstiy
 relation'' (Dressler et al. 1980,1997; Postman \& Geller 1984;  Fasano
 et al. 2000; Goto et al. 2003c; Chapter \ref{chap:MD}). Passive 
 spiral galaxies reside preferentially in 1-10 virial radius, which,
 along with the results showin in Figure \ref{fig:ps_density}, suggests
 that they live in cluster infalling regions.

\subsection{Photometric \& Spectroscopic Properties}

 In Figure \ref{fig:ps_gri}, we plot the distribution of passive spiral
 galaxies in restframe $g-r-i$ plane. Instead of Petrosian magnitude, we
 use model magnitude to compute colors of galaxies since signal-to-noise
 ratio is higher for model magnitude. In the SDSS, model magnitudes are
 measured using Petrosian radius measured in $r$ band. (Thus the same
 radius is used to measure model magnitudes in 5 filters. See Stoughton
 et al. 2002 for more details on the SDSS magnitudes.) 
 The observed color are $k$-corrected
 to the restframe using $k$-correction given in Blanton et
 al. (2002;v1\_11). Contours show the distribution of all 25813 galaxies in
 the volume limited sample for
 comparison. A peak of the contour around ($g-r$,$r-i$)=(0.75,0.4)
 consists of elliptical galaxies. The distribution of spiral galaxies
 extends to the bluer direction in both $g-r$ and $r-i$.
 Interestingly, passive spiral galaxies are almost as red as
 elliptical galaxies in $g-r$, reflecting truly passive nature of these
 galaxies. Note that colors are photometrically measured, thus they are free
 from the aperture bias. In $r-i$ color, some passive spirals
 are almost as blue as spiral galaxies.

 In Figure \ref{fig:ps_rk}, we present restframe $J-K$ v.s. $r-K$ colors of
 passive and active galaxies in the open and solid dots, respectively. 
 Infrared colors are obtained by matching our galaxies to the Two Micron
 All Sky Survey (2MASS; Jarrett et al. 2000) data. Infrared magnitudes
 are shifted to restframs using $k$-corrections given in Mannucci et al. (2001).
 Among our sample
 galaxies, 31/73 passive
 spirals (9317/25813 all galaxies) were measured with 2MASS.  
 As in the previous figure, the solid lines show distirbution of all galaxies in
 the volume limited sample.  Compared with the solid lines, active spirals
 show slightly bluer distribution in $r-K$. Passive spirals do not show
 significantly bluer distribution in $r-K$ than all galaxies.

 In Figure \ref{fig:ps_hd}, we plot distributions of H$\delta$ EW for
 passive spirals (dotted lines), active spirals (dashed lines) and all
 galaxies (solid lines) in the volume limited sample. H$\delta$ EWs are
 measured using the flux summing method discussed in Goto et
 al. (2003b; appendix \ref{EA1}). The flux summing method is robust for weak absorption
 lines and noisy spectra. In the figure, absorption lines have positive
 EWs. As shown in the figure, passive spirals have very weak H$\delta$
 absorption peaked around 0\AA.
 In contrast, active spirals have much stronger H$\delta$
 absorption, reflecting their strong star formation activity. 
 Since H$\delta$ absorption becomes strong only when A stars
 are dominant in galaxy spectra, it is a good indicator of
 post-starburst phase of a galaxy. For example, E+A galaxies (Zabludoff
 et al.1996; Balogh et al. 1999; Dressler et al. 1999; Poggianti et
 al. 1999; Chapters \ref{EA1} and \ref{EA2}) are thought
 to be a post-starburst galaxy since they do not have any current star
 formation (no [OII] nor H$\alpha$ in emission), but do have many A
 stars (strong H$\delta$ absorption). Such a phase can only appear when
 a starbursting galaxy truncates its starburst (a post-starburst
 phase). More details on  H$\delta$ strong galaxies can be found in Goto
 et al. (2003b; appendix \ref{EA1}). Therefore, small H$\delta$ EWs of passive spirals
 indicate that they are not in a post-starburst phase. The origin of
 passive spirals are likely to be different from that of E+A (or post-starburst)
 galaxies. Passive spirals seem to have stopped their star formation
 gradually rather than sudden truncation.

\section{Discussion}\label{May 29 10:07:03 2003}
\subsection{A Transient in Galaxy Evolution}\label{sec:ps_discussion}

 In Section \ref{sec:ps_selection}, we have selected unusual population of
 galaxies with spiral morphology and without emission lines such as
 H$\alpha$ and [OII]. The optical color-color diagram
 (Figure \ref{fig:ps_gri}) also revealed that these galaxies are as red as
 elliptical galaxies, reflecting  passive nature of these galaxies. 
 One possible explanation to these galaxies are heavy obscuration by
 dust. In such a case, passive spiral galaxies might have star formation
 activity just as normal galaxies, but the star formation
 activity might be hidden by dust.
  The scenario could be consistent with both of our observational results;
 lack of emission lines and red colors in optical. 
 However, in $r-K$ color (Figure \ref{fig:ps_rk}), passive spiral
 galaxies do not  appear to be much redder than normal galaxies.
  This is against dust enshrouded scenario which should results in very
 red $r-K$ color, and thus suggesting truly passive nature of these
 galaxies. In addition, it is not likely that very dusty galaxies
 preferentially live in cluster infalling regions.

 In Section \ref{sec:ps_density}, we revealed that passive spiral galaxies
 preferentially live in cluster infalling region, using both local
 galaxy density (Figure \ref{fig:ps_density}) and cluster-centric-radius
 (Figure \ref{fig:ps_radius}). This is direct evidence to connect
 the origins of these galaxies to cluster environment. The
 characteristic environments are 1$\sim$2 Mpc$^{-2}$ in local galaxy
 density and 1$\sim$10 virial radius in cluster-centric-radius. Quite
 interestingly, these environments coincide with characteristic density
 and radius where star formation rate declines toward cluster center or
 dense environment. Gomez et al. (2003) and Lewis et al. (2002) studied
 star formation rate in a galaxy as a function of cluster centric radius
 and local galaxy density and found that star formation rate declines
 around the same environment as we found in the present
 study. Furthermore, Goto et al. (2003c; Chapter \ref{chap:MD}) studied the morphology-density
 relation using the similar SDSS data (0.05$<z<$0.1 and $Mr^*<-$20.5)
 and an automated galaxy classification (Yamauchi et al. 2003).
 They found that the morphological 
 fraction of galaxies start to change approximately at the same
 environment as found in  our study; the fraction of S0 and elliptical galaxies
 start to increase (and Sc galaxies decrease) toward cluster center at $\sim$1 virial radius, or
 toward larger galaxy density at local galaxy density $\sim$1 Mpc$^{-2}$.
      These coincidences in the environment suggest that the same mechanism might be
 responsible for all the effects happening here; creation of passive
 spiral galaxies, decrease of galaxy star formation rate and
 morphological change in relative galaxy fraction. These coincidences
 might be explained naturally by the following interpretation;
  As galaxies
 approach this critical environment (1$\sim$2 Mpc$^{-2}$ or 1$\sim$10
 virial radius) , they stop their star formation as 
 seen in Gomez et al. (2003), by changing spiral galaxies into passive
 spiral galaxies as found in this study. If a spiral galaxy stops star
 formation calmly without its morphology disturbed, it is likely to
 develop to be a S0 galaxy (Bertin \& Romeo 1988; Bekki et al. 2002) as is seen in the
 morphology-density relation of Goto et
 al. (2003c; Chapter \ref{chap:MD}). According to this scenario, passive spirals are likely to
 be a population of galaxies in transition. 
      In addition, there have been many results from other observations
 and surveys, which supports 
 this scenario. Abraham et al.(1996) reported that cluster members
 become progressively bluer as a function of cluster-centric distance
 out to 5 Mpc in A2390 ($z$=0.23). 
  Terlevich, Caldwell \& Bower (2001) reported that $U-V$
 colors of early-type galaxies are systematically bluer at outside the
 core of Coma cluster.  
   Pimbblet et al.(2002) studied 11 X-ray luminous clusters
 (0.07$<z<$0.16) and found that median galaxy color shifts bluewards
 with decreasing local  galaxy density. 
   At higher redshift, Kodama et al. (2001) reported that colors of
 galaxies abruptly change at sub-clump regions surrounding a cluster
 at $z=$0.41. Although it is difficult to directly compare this
 environment with ours due to the different definitions of local galaxy
 density, it is highly possible that their color change happens at the
 same environment we found. 
  van Dokkum et al. (1998) found S0 galaxies in the outskirt of a
 cluster at $z=$0.33. These S0s show a much wider scatter in their colors
 and are bluer on average than those in cluster cores, providing
 possible evidence for recent infall of galaxies from the field. In
 addition, many studies reported that star formation in the cores of
 clusters is much lower than that in the surrounding field (e.g., Balogh
 et al. 1997,1998,1999; Poggianti et al. 1999; Martin, Lotz \& Ferguson 2000;
 Couch et al. 2001; Balogh et al. 2002).

 The existence of passive spiral
 galaxies also brings us a hint on the origin of these three phenomena.
 It supports a transformation of galaxies, which do not
 disturb arm structures of spiral galaxies. Possible preferred
 candidate mechanisms include ram-pressure stripping (Gunn \& Gott 1972; Abadi,
 Moore \& Bower 1999; Quilis, Moore \& Bower 2000) and simple removal of gas
 reservoir (Larson, Tinsley \& Cardwell 1980; Balogh et
 al. 1999). It has been known that preheating of intergalactic medium
 can effect morphologies of galaxies by strangling the gas accretion
 (strangulation; Mo
 \& Mao 2002; Oh \& Benson 2002). In fact, Finoguenov et al. (2003)
 found the filamentary gas in Coma cluster and predicted the existence
 of passive spirals around the filament. 
  Although the characteristic environment (1$\sim$2 Mpc$^{-2}$ or 1$\sim$10
 virial radius) might seem to be a little too low density environment
 for ram-pressure or strangulation to happen, it is possible for galaxy
 sub-clumps around a cluster to have local hot gas dense enough for
 stripping (Fujita et al. 2003). Indeed Kodama et al. (2001) found that galaxy colors change
 at such sub-clumps around a cluster.

 Perhaps major merger/interaction origins are less preferred since
 such dynamical processes disturb arm structures in spiral galaxies, and
 thus do not result in creating passive spirals. Weak H$\delta$
 absorption lines shown in Figure \ref{fig:ps_hd} also support quiescent
 transformation of galaxies. 
 However, we can not
 exclude a minor merger origin since such a process might be able to
 happen without disturbing spiral arms. 
 In their study of the morphology-density relation, Goto et al. (2003c; Chapter \ref{chap:MD})
 observed decrease of S0s and increase of 
 ellipticals at cluster cores (virial radius $<$ 0.3 or galaxy density
 $>$ 6 Mpc$^{-2}$), and they
 proposed that major merger/interaction might be
 dominant in cluster core regions.
 The proposal is consistent with
 our results, which showed  devoid of passive spiral galaxies within 0.6
 virial radius 
 or greater than $\sim$3 Mpc$^{-2}$ in local galaxy density. 
   On the other hand, some theoretical work predicts that 
 it is difficult to have frequent merger/interagtion in cluster
 cores since relative velocities of galaxies are so high in such regions
 (Ostriker 1980; Binney \& Tremaine 1987; Mamon 1992; Makino \& Hut 1997).   
 In such a case, S0s (or passive spirals) might simply fade away to be a small  elliptical galaxy.
  In summary, implication in cluster core regions is 
 either (i) passive spiral galaxies mergered into a large elliptical
 galaxies in cluster cores, or (ii) the disc of passive spiral galaxies completely fade
 away to become small elliptical galaxies.

 Also in terms of cluster galaxy evolution, passive spiral galaxies might fit
 well with the previous observational results. It has been known that
 fraction of blue galaxies are larger in higher redshift (the
 Butcher-Oemler effect; Butcher \& Oemler 1978, 1984; Couch \&
  Sharples 1987; Rakos \& Schombert
 1995; Couch et al. 1998; Margoniner \& De Carvalho 2000; Margoniner et al. 2001; Ellingson
 et al. 2001; Kodama \& Bower 2001; Goto et al. 2003a; Chapter \ref{chap:BO}) and 
 that the fraction of cluster spiral galaxies are also larger in the
 past (Dressler et al. 1997;  Couch et al. 1998; Fasano et al. 2000;
 Diaferio et al. 2001; Goto et al. 2003a; Chapter \ref{chap:BO}). Many people speculated the
 morphological transformation from spiral galaxies to S0 galaxies
 (e.g., Dressler et al. 1997; Smail et al. 1998; Poggianti et al. 1999; Fabricant, Franx, \&
 van Dokkum 2000; Kodama \& Smail 2001).
   Fraction of early-type galaxies in rich clusters are smaller in the
 past (Andreon, Davoust, \& Heim 1997; Dressler et al. 1997; Lubin et
 al. 1998; van Dokkum et al. 2000).  
   In morphological point of view, since passive spiral galaxies have
 already stopped their star 
 formation, in the near future, its disc structures including spiral
 arms will become fainter and 
 fainter, to be seen as a disc galaxy with smoother profile, i.e., possibly
 S0 galaxies. Spectrally, passive spirals are already almost as red as elliptical
 galaxies, but their spiral arms must have had star formation activity
 until recently, therefore a passive spiral galaxy itself must have been much
 bluer in the past, just like blue population of galaxies numerous in
 the higher redshift clusters. Therefore, although this is not direct
 evidence, it is very likely that passive spiral galaxies are a
 population of galaxies in transition, in the course of the Butcher-Oemler
 effect and morphological Butcher-Oemler effect. 

\subsection{Aperture Bias}\label{sec:ps_aperture}

 Since the SDSS spectroscopy is performed with a fiber spectrograph which
 captures light within 3 arcsecond aperture, aperture bias is a
 concern. Aperture bias could result in an increase of passive spiral
 galaxies with decreasing redshift since at lower redshift, 3 arcsecond
 fiber misses more light from a disc of a galaxy. Using the data from
 LCRS with a 3.5 arcsecond fiber spectrograph, Zaritsky et al. (1995)
 showed that at $z>$0.05, the spectral classifications of galaxies are
 statistically unaffected by aperture bias. Using the similar sample of
 the SDSS galaxies, Gomez et al. (2003) also
 limited their galaxies to $z>$0.05 and proved that aperture bias does
 not change their results. We followed these two authors and 
 limited our sample with $z>$0.05 to minimize this potential bias.
 In the main analysis of the chapter, there are several evidence
 suggesting that these passive spiral galaxies are not seriously biased
 by the aperture effect. In Figure \ref{fig:ps_gri}, passive spiral
 galaxies are much redder than normal galaxies. To calculate colors, we
 used a model magnitude 
 which uses Petrosian radius in $r$ for all colors (Stoughton et
 al. 2002), and thus is free from 3'' aperture bias.  
 Therefore the red colors of these galaxies suggest that they are truly
 passive systems, and not the artifact of the aperture effect. 
 Also in Figure \ref{fig:ps_density}, we compared density distribution of
 passive spirals with normal star forming spirals. The two distributions
 are statistically different. Again, if passive spirals are the artifact
 of aperture bias, the density distributions of star forming and passive
 spiral galaxies should be similar. Therefore this difference suggests
 that passive nature of these galaxies are truly unique to these
 galaxies.
 Figure \ref{fig:ps_color_gradient} shows difference in $g-r$ color between
  fiber magnitude (measured with 3'' aperture) and model magnitude
 (measured using Petrosian radius in $r$, usually larger than 3'',
 especially in low redshift) as a
 function of redshift. Contours show the distribution of all galaxies in
 our volume limited sample. The solid, dotted and dashed lines show medians
 of all galaxies, passive spirals and active spirals. Since both passive
 and active spirals are less concentrated, their medians have somewhat
 higher values than all galaxies. If aperture bias is severe,
 $\Delta(g-r)$ should be much larger in lower redshift since the
 difference between 3'' aperture and Petrosian radius of galaxies are
 larger. However, in Figure \ref{fig:ps_color_gradient}, $\Delta(g-r)$ of
 passive spirals is
 almost constant throughout the redshift range we used
 (0.05$<z<$0.1). The figure suggests that aperture effect is not a
 severe effect within the redshift range.
  In Figure \ref{fig:ps_aperture}, we present the fraction of passive spiral galaxies as a
 function of redshift. It clearly shows strong aperture effect at
 $z<$0.05. However, throughout this chapter, we limit our sample between $z=$0.05 and
 0.1, where fractions of passive spirals are consistent with constant
 within the error. It suggests that aperture bias is not strong within
 our sample. We end this section by quoting that Hopkins et al. (2003)
 compared star formation rate estimated from H$\alpha$ (SDSS data;
 subject to 3 arcsecond aperture
 bias) and that from radio flux (FIRST data; i.e., with no aperture
 bias), concluding that both star formation rate estimates agree with
 each other after correcting H$\alpha$ flux using the ratio of 3'' fiber
 magnitude to  Petrosian (total) magnitude in $r$ band.

\section{Summary}\label{sec:ps_conclusion}
 
 Using a volume limited sample of the SDSS data, we have studied the
 environment of passive spiral galaxies as a function of local galaxy
 density and cluster-centric-radius. Since passive spirals are only
 found in cluster regions in previous work, this is the first attempt to
 select passive spirals uniformly, in all the environment. It is found that passive
 spiral galaxies live in local galaxy density 1$\sim$2 Mpc$^{-2}$ and
 1$\sim$10 virial radius. Thus the origins of passive spiral galaxies
 are likely to be cluster related. These characteristic environments
 coincide with the environment where galaxy star formation rate suddenly
 declines (Lewis et al. 2002; Gomez et al. 2003) and the fractions of
 galaxy morphology start to deviate from the field value (Goto et
 al. 2003c; Chapter \ref{chap:MD}). Therefore it is likely that the same physical mechanism is
 responsible for all of these observational results; the
 morphology-density relation, the decline of star formation rate and the
 creation of passive spiral galaxies. The existence of passive spiral
 galaxies suggest that a physical mechanism that works calmly is
 preferred to dynamical origins such as major merger/interaction since such a
 mechanism can destroy spiral arm structures. 
  Passive spiral galaxies are likely to be a galaxy population in
 transition between red, elliptical/S0 galaxies in low redshift clusters
 and blue, spiral galaxies numerous in higher redshift clusters as seen
 in the Butcher-Oemler effect and the morphological Butcher-Oemler
 effect. 
 Computationally, simulating the evolution of passive spiral galaxies
 will bring more insight on the origin of cluster galaxy evolution. Such
 a simulation might be possible by combining a pioneering work by Bekki
 et al. (2002) with large cluster N-body simulation which can trace the
 evolution of cluster galaxies; (e.g., Diaferio et al. 2000; Benson et al. 2002).

\bigskip

\clearpage

\clearpage

\clearpage

\clearpage

\begin{figure}
\centering{
\includegraphics[scale=0.2]{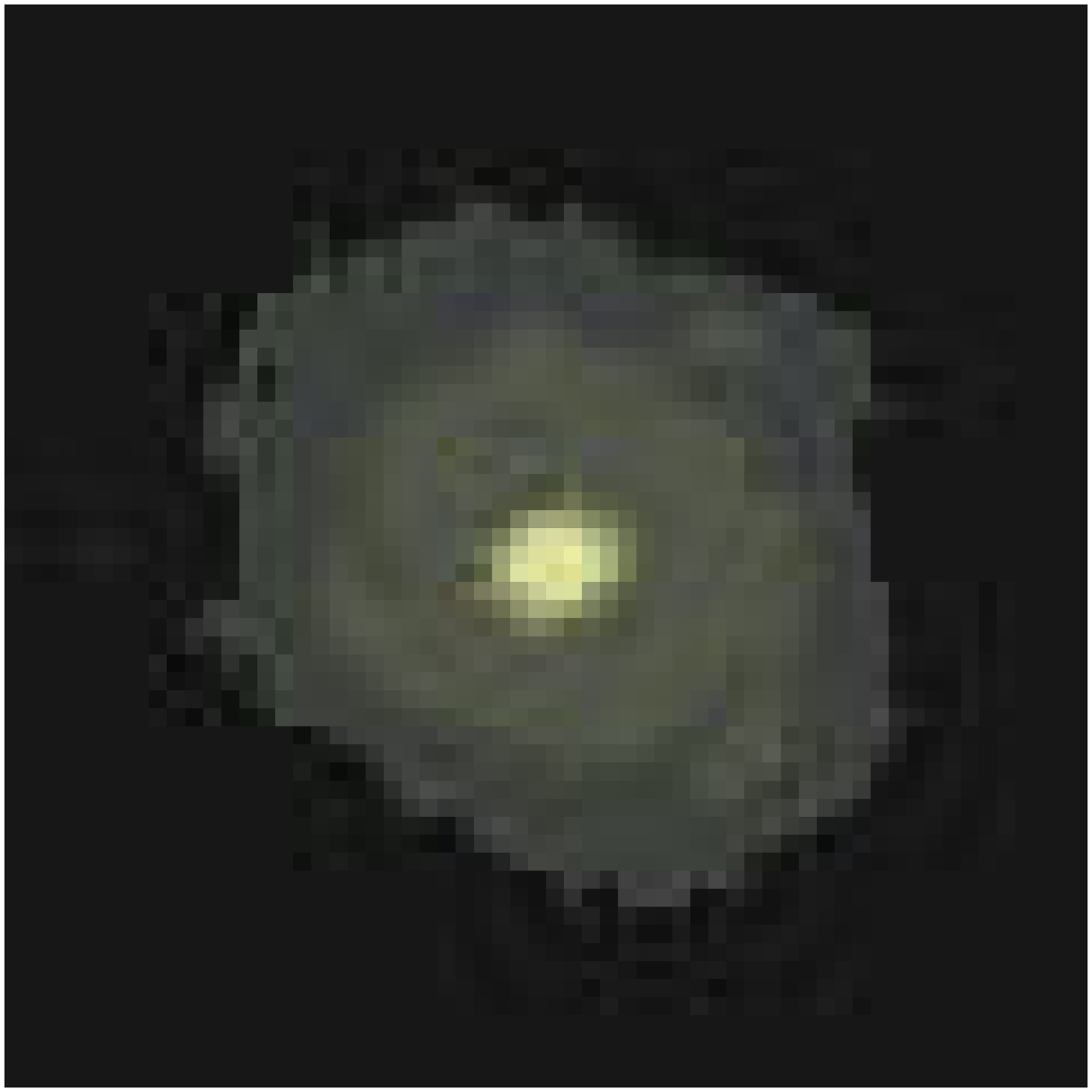} %
\includegraphics[scale=0.2]{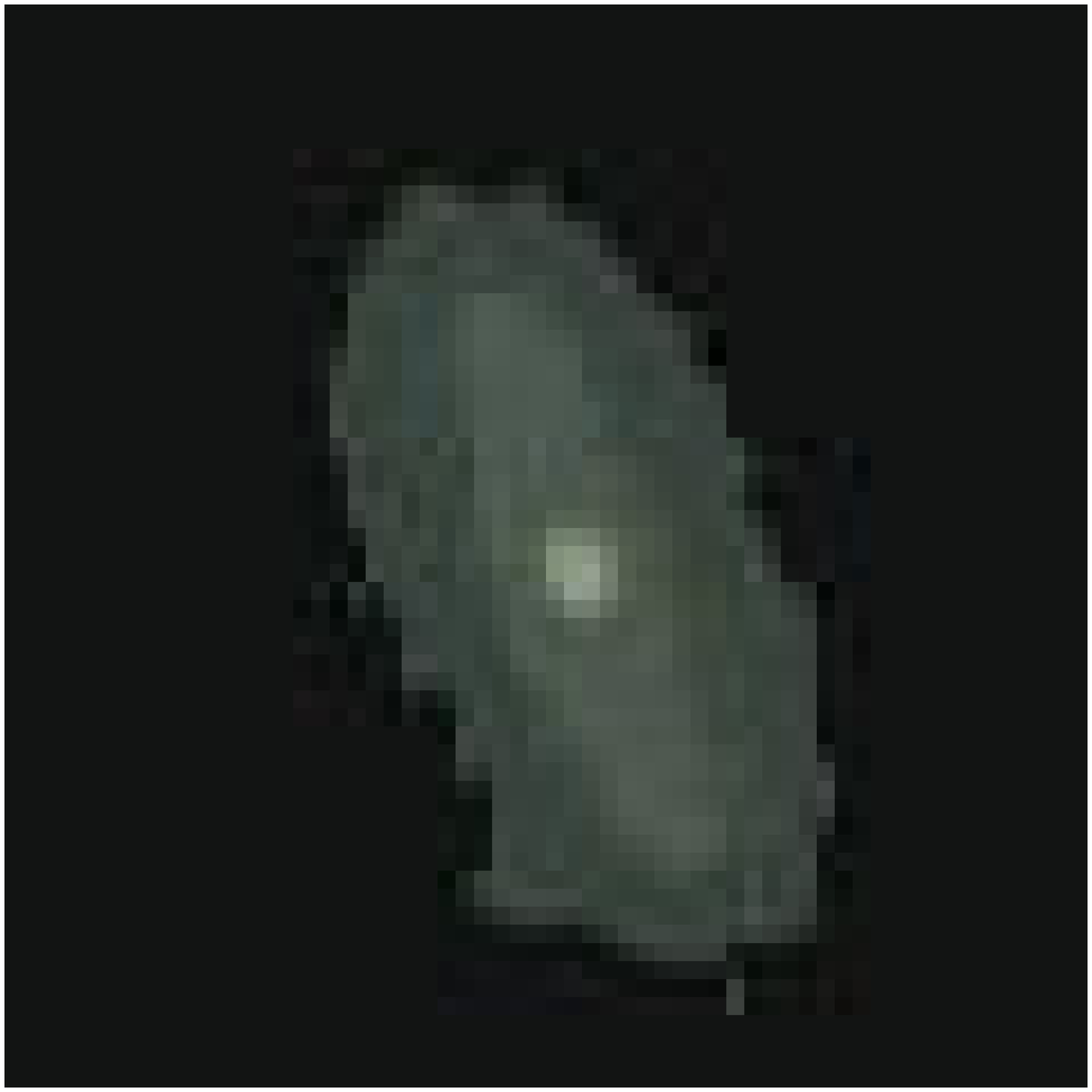}
\includegraphics[scale=0.2]{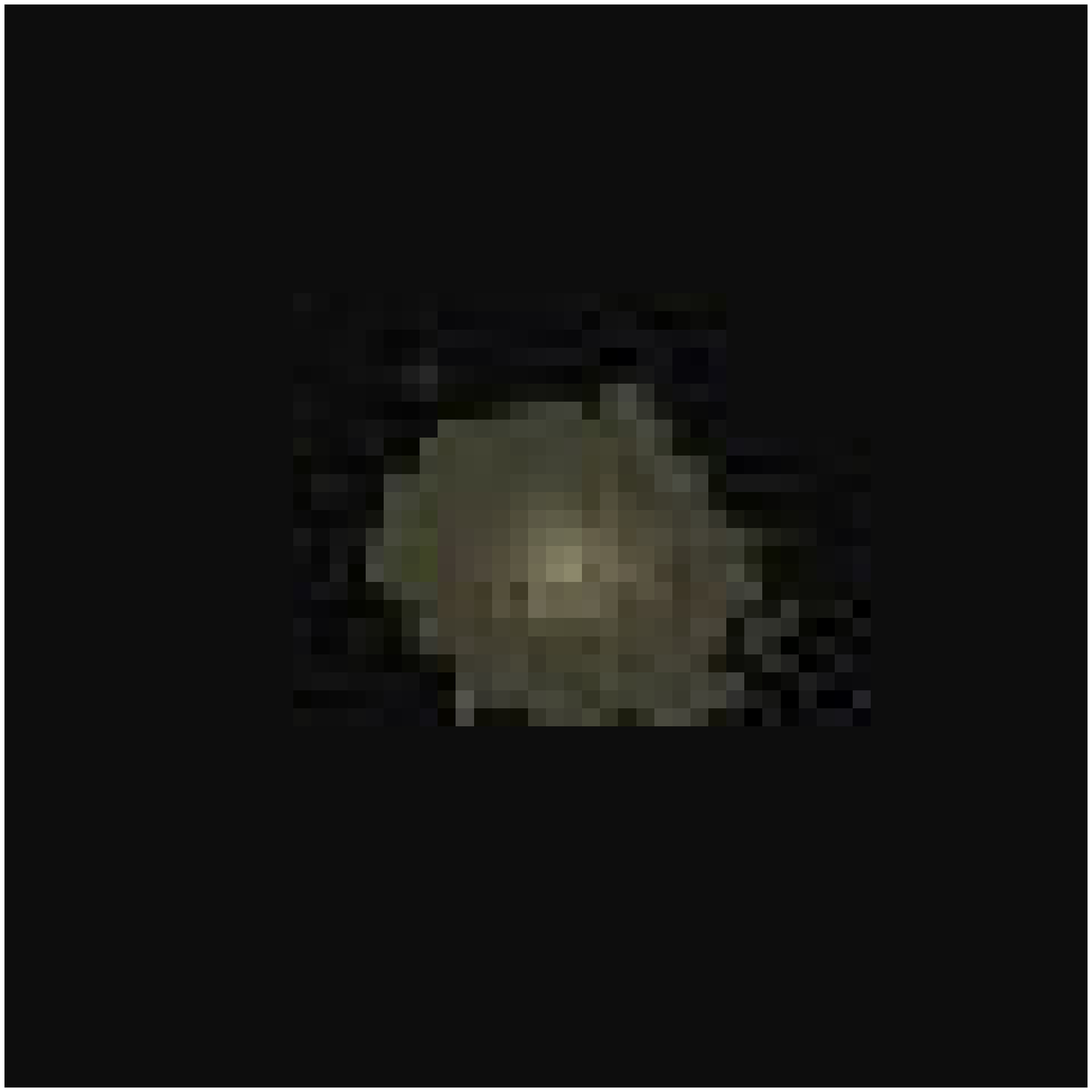}
}
\centering{
\includegraphics[scale=0.2]{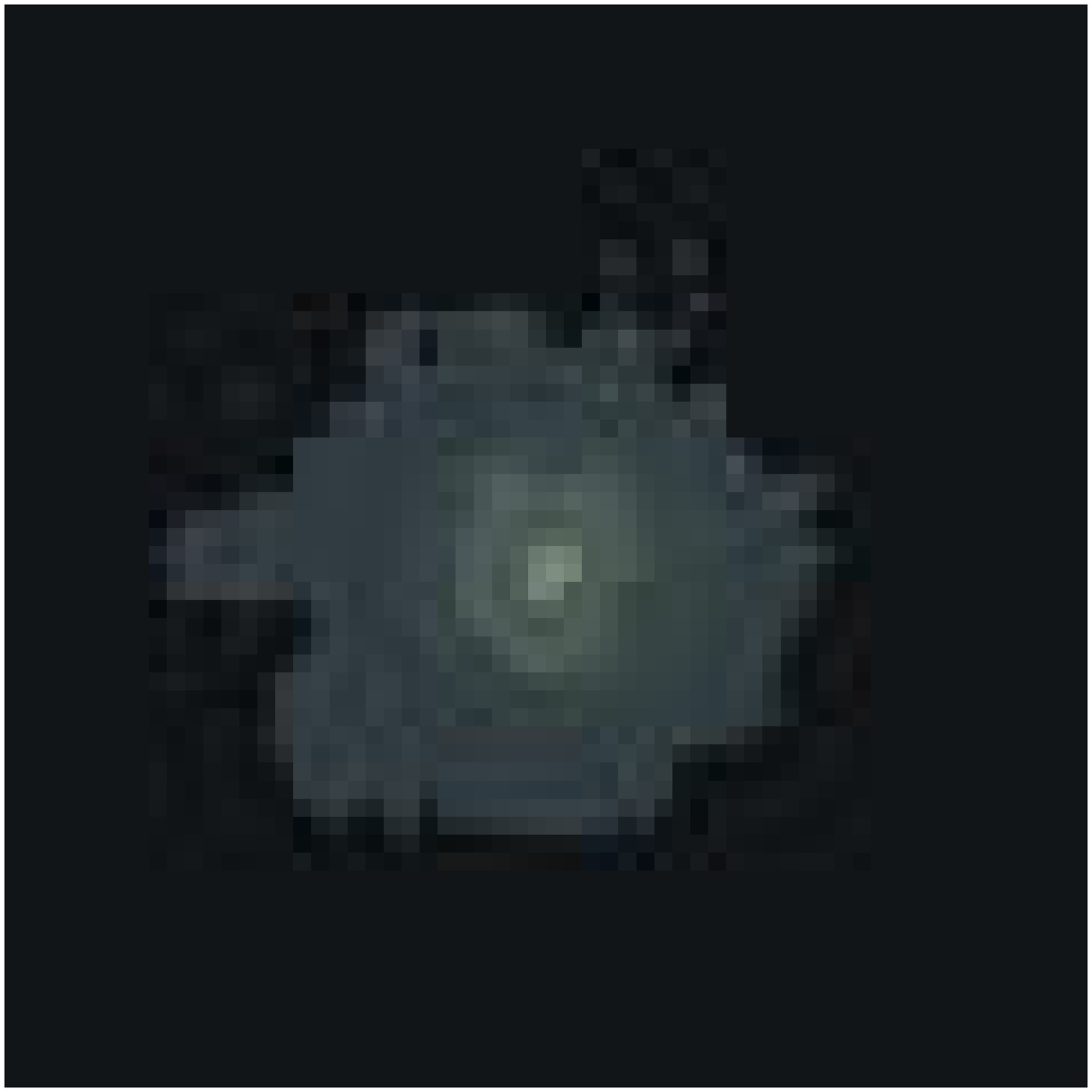}
\includegraphics[scale=0.2]{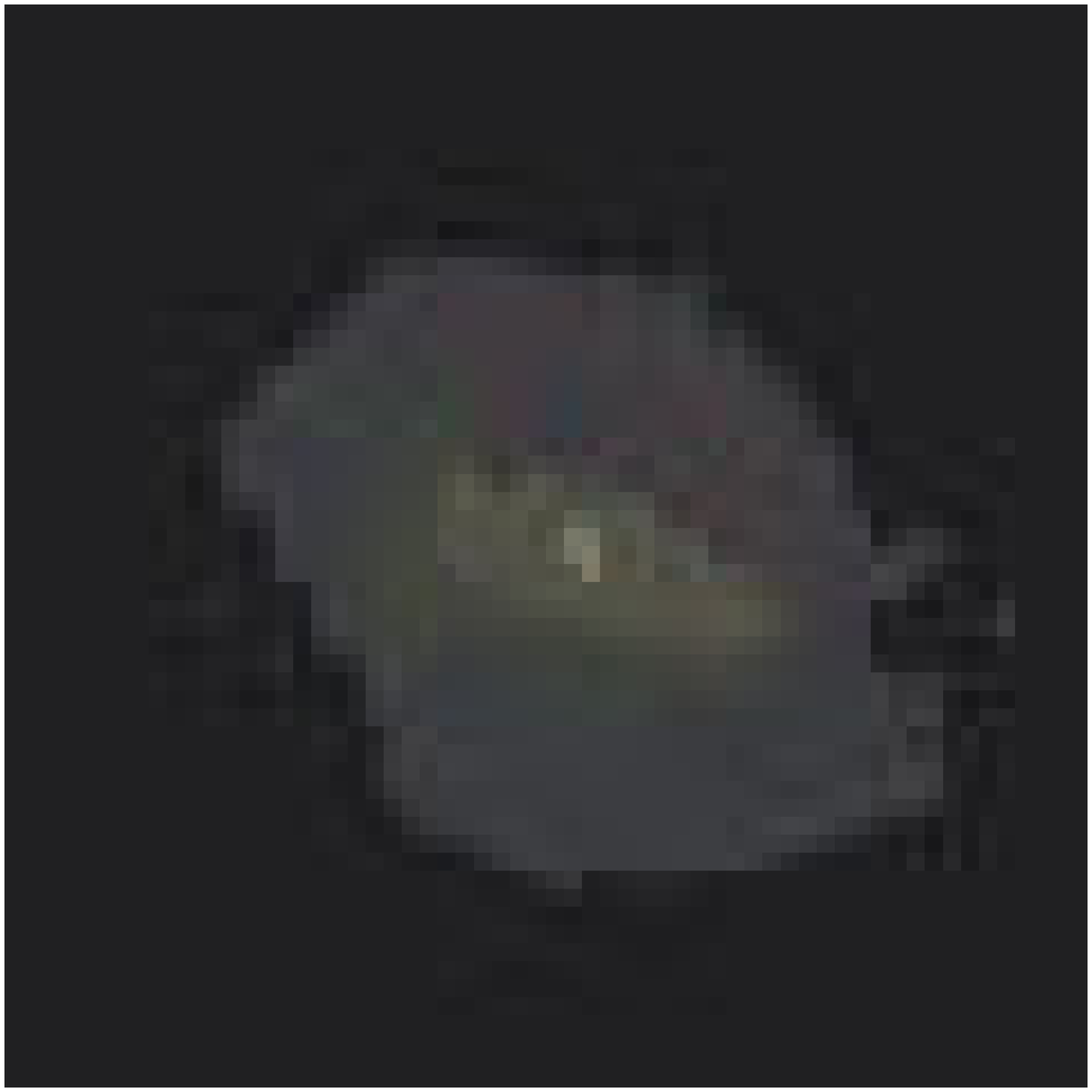}
\includegraphics[scale=0.2]{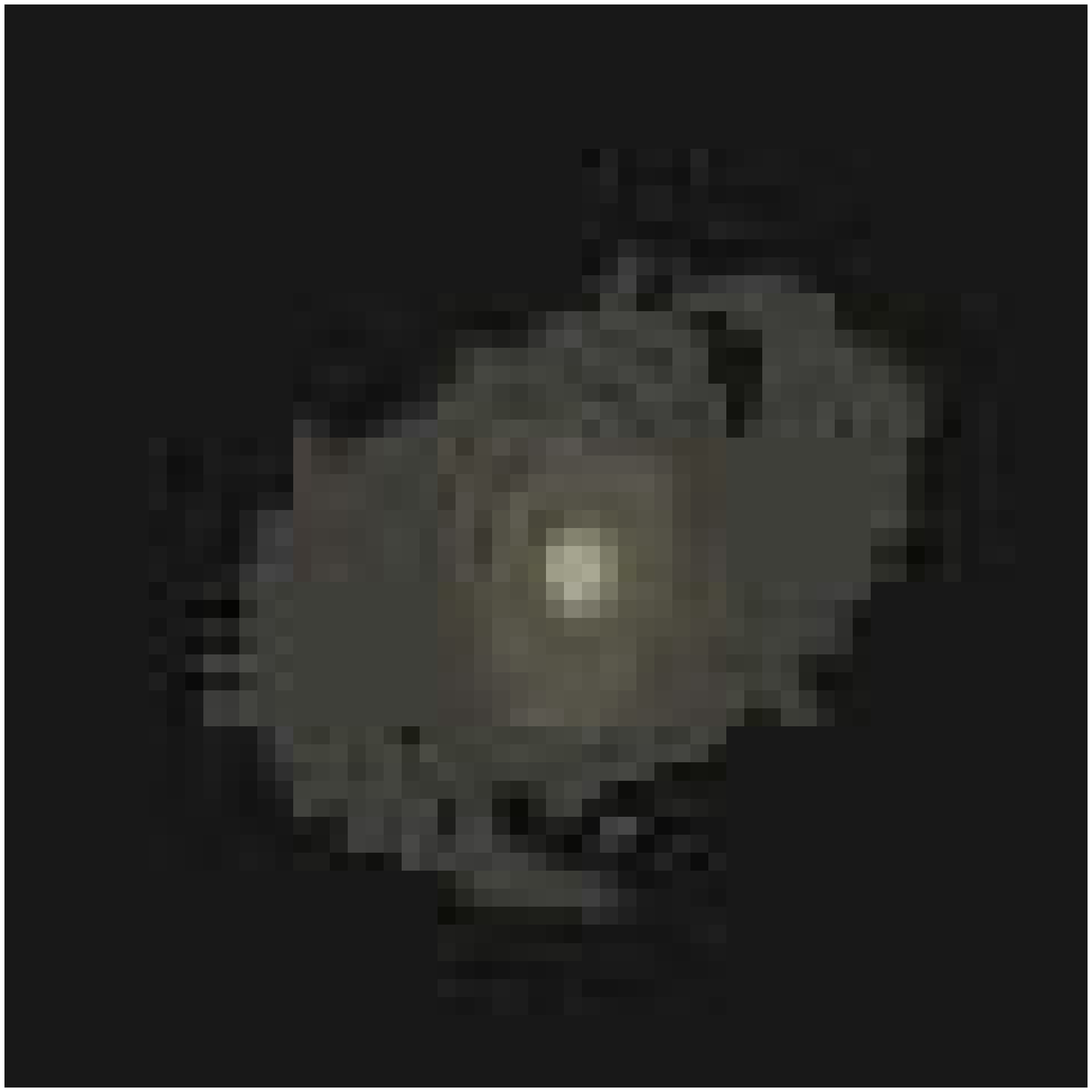}
}
\centering{
\includegraphics[scale=0.2]{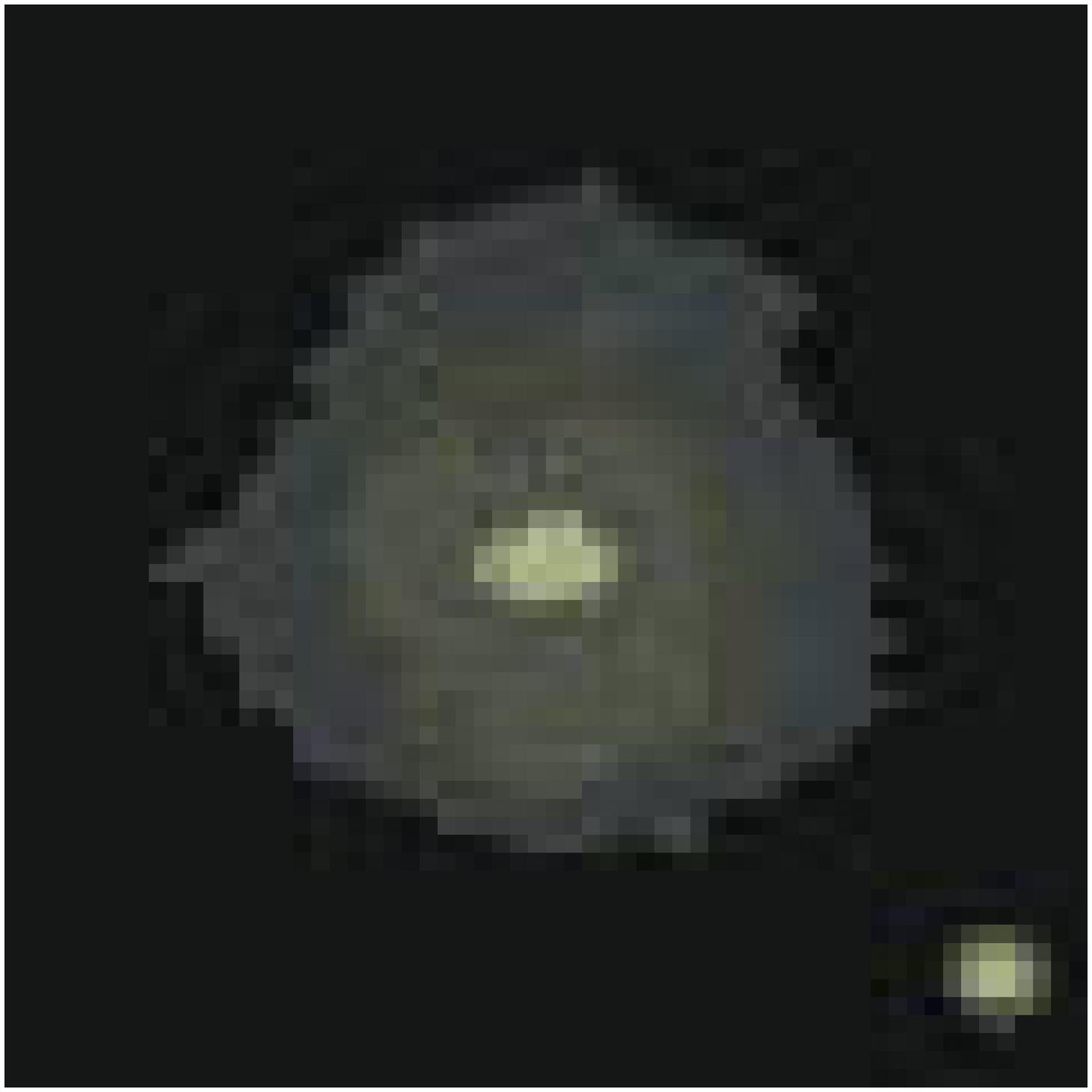}
\includegraphics[scale=0.2]{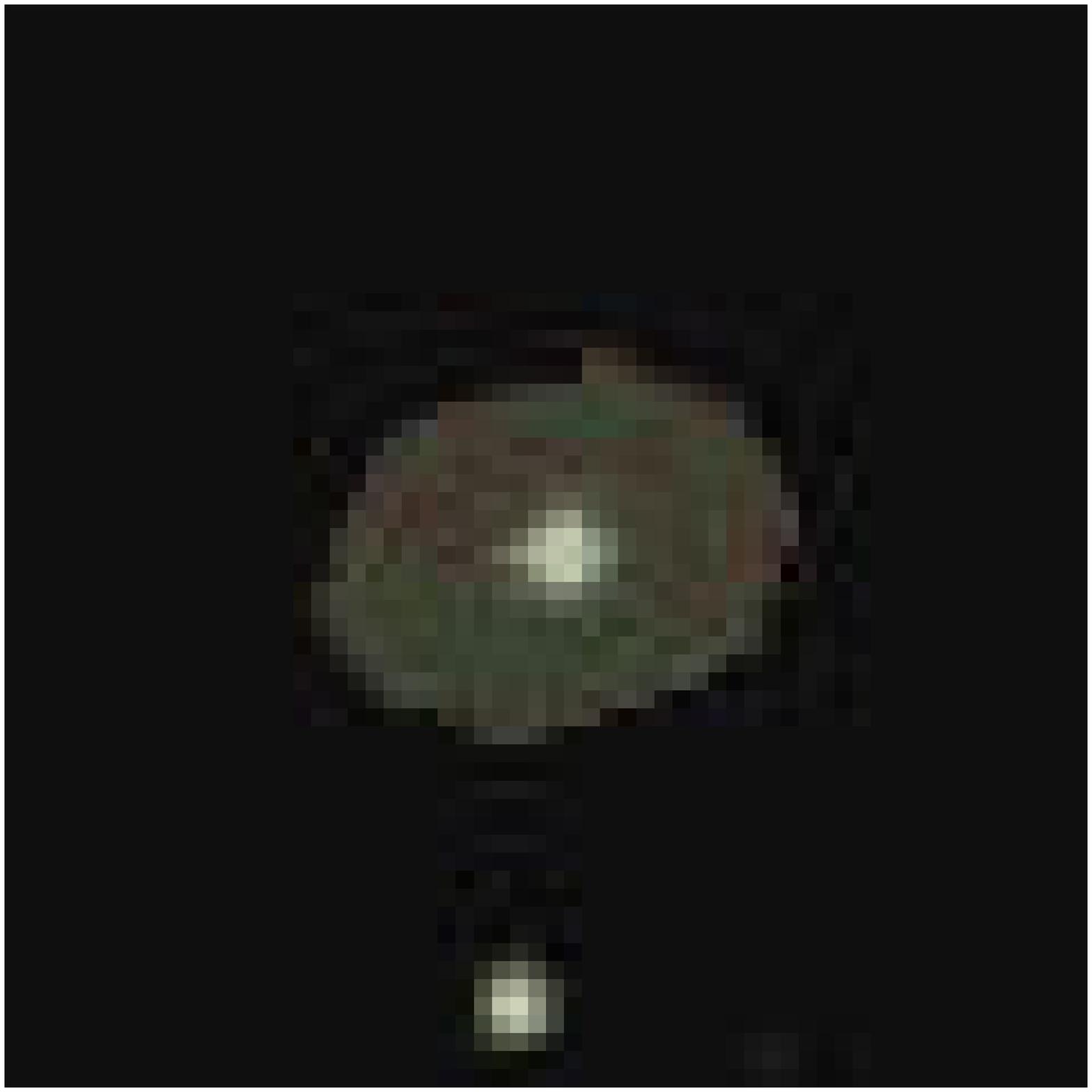}
\includegraphics[scale=0.2]{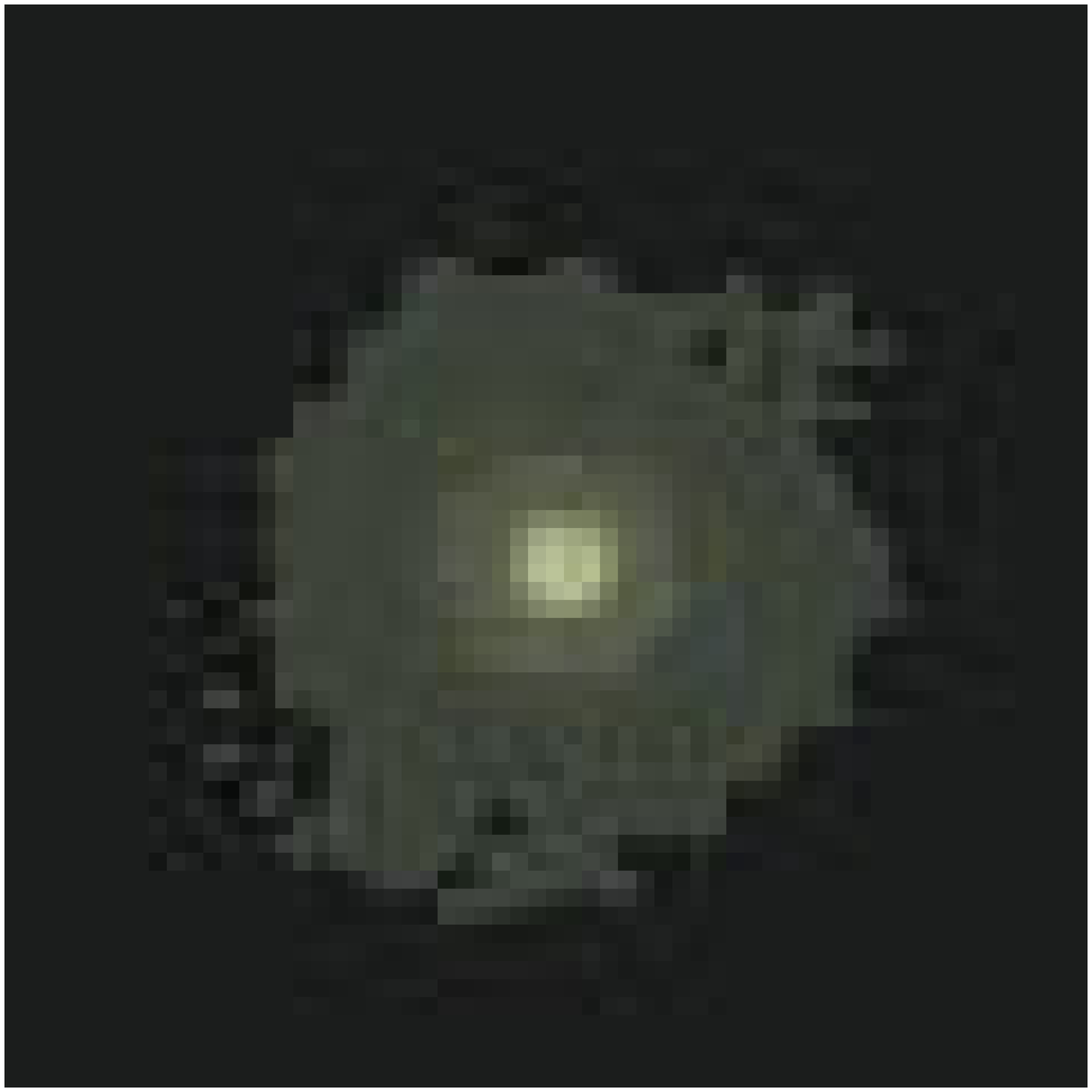}
}
\caption{
\label{fig:ps_image}
 Example images of passive spiral galaxies. Each image is a composite of
 the  SDSS $g,r$ and $i$ bands, showing
 30''$\times$30'' area of the sky with its north up.
 Discs and spiral arm structures are recognized.
}\end{figure}

\clearpage

\begin{figure}
\centering{
\includegraphics[scale=0.25]{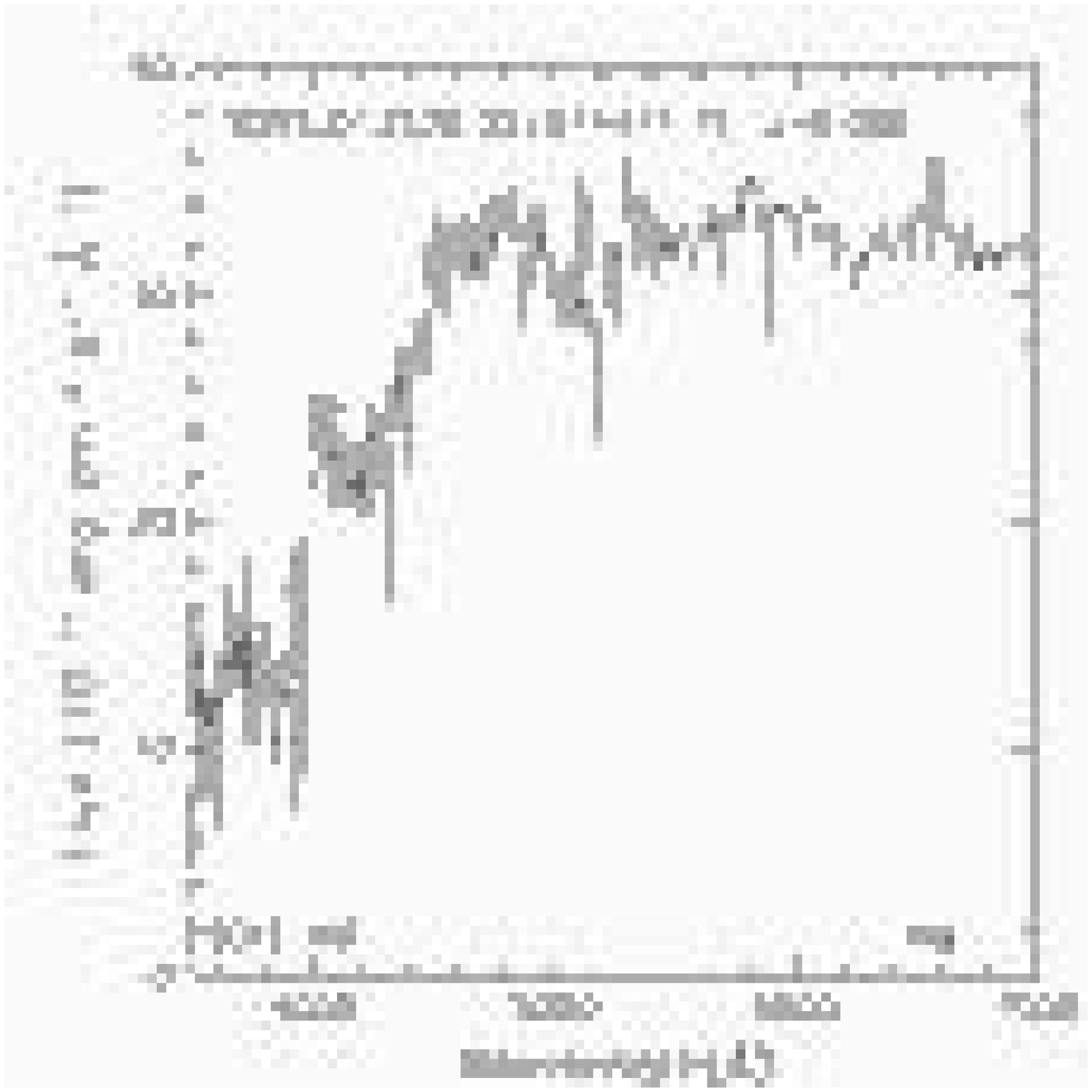} %
\includegraphics[scale=0.25]{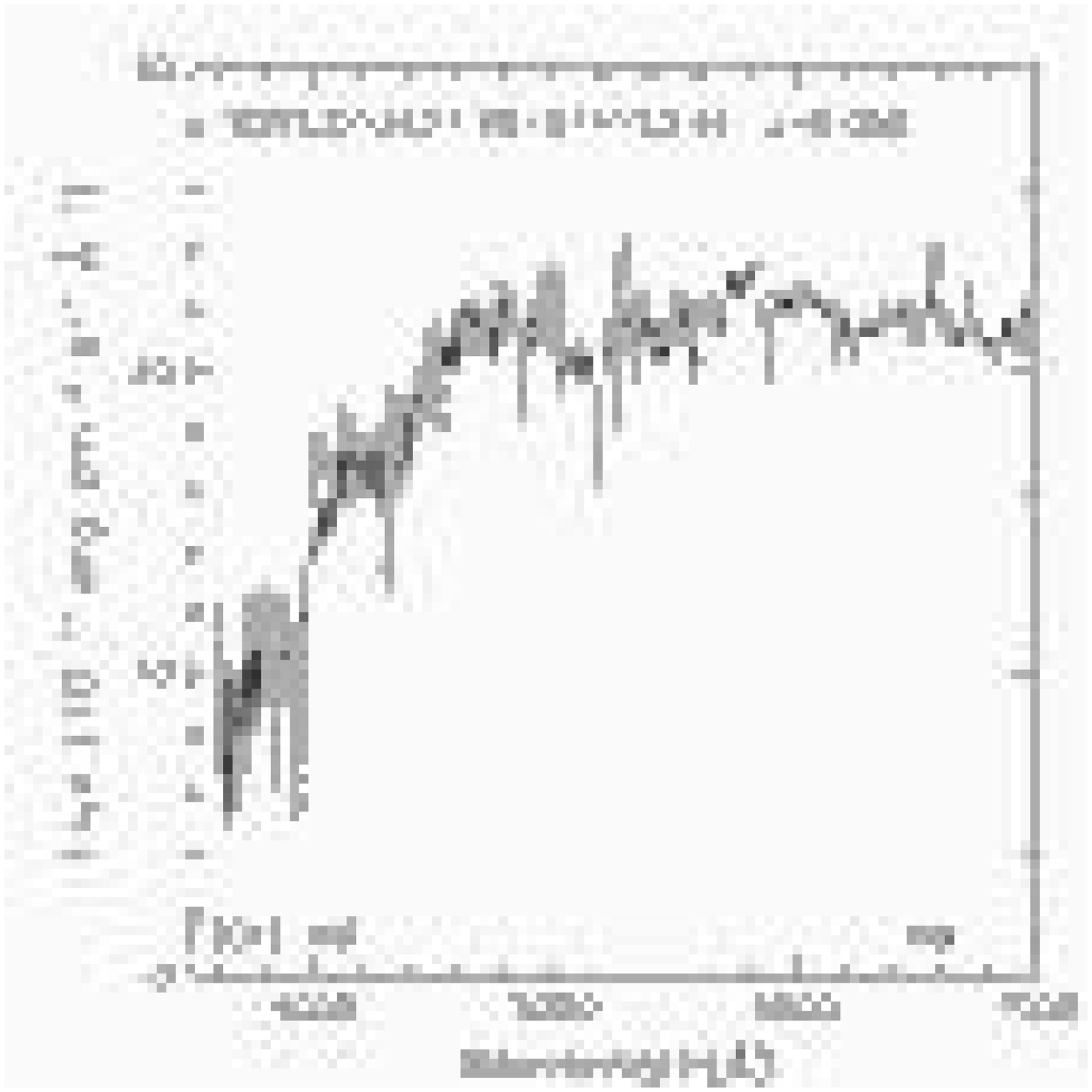}
\includegraphics[scale=0.25]{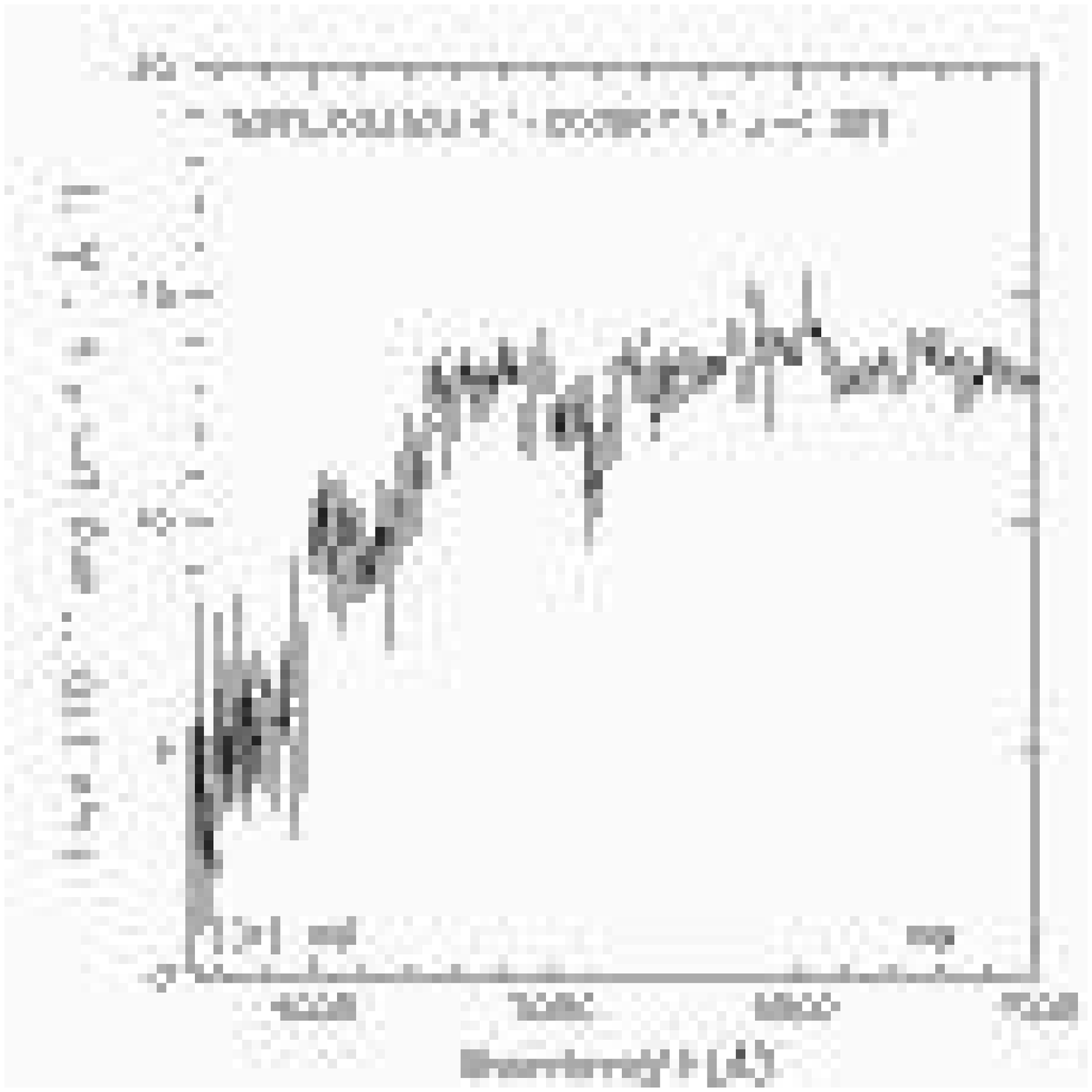}
}
\centering{
\includegraphics[scale=0.25]{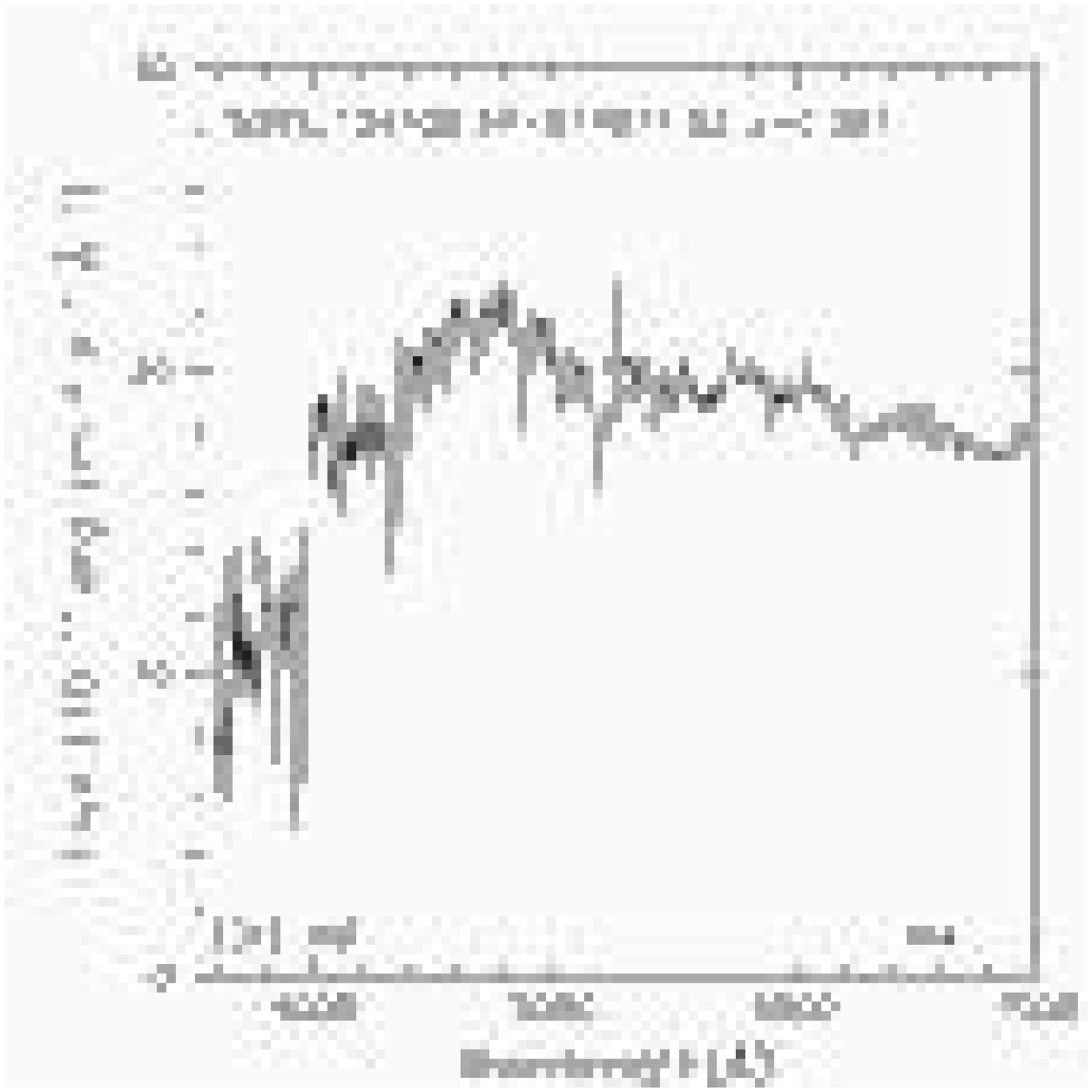}
\includegraphics[scale=0.25]{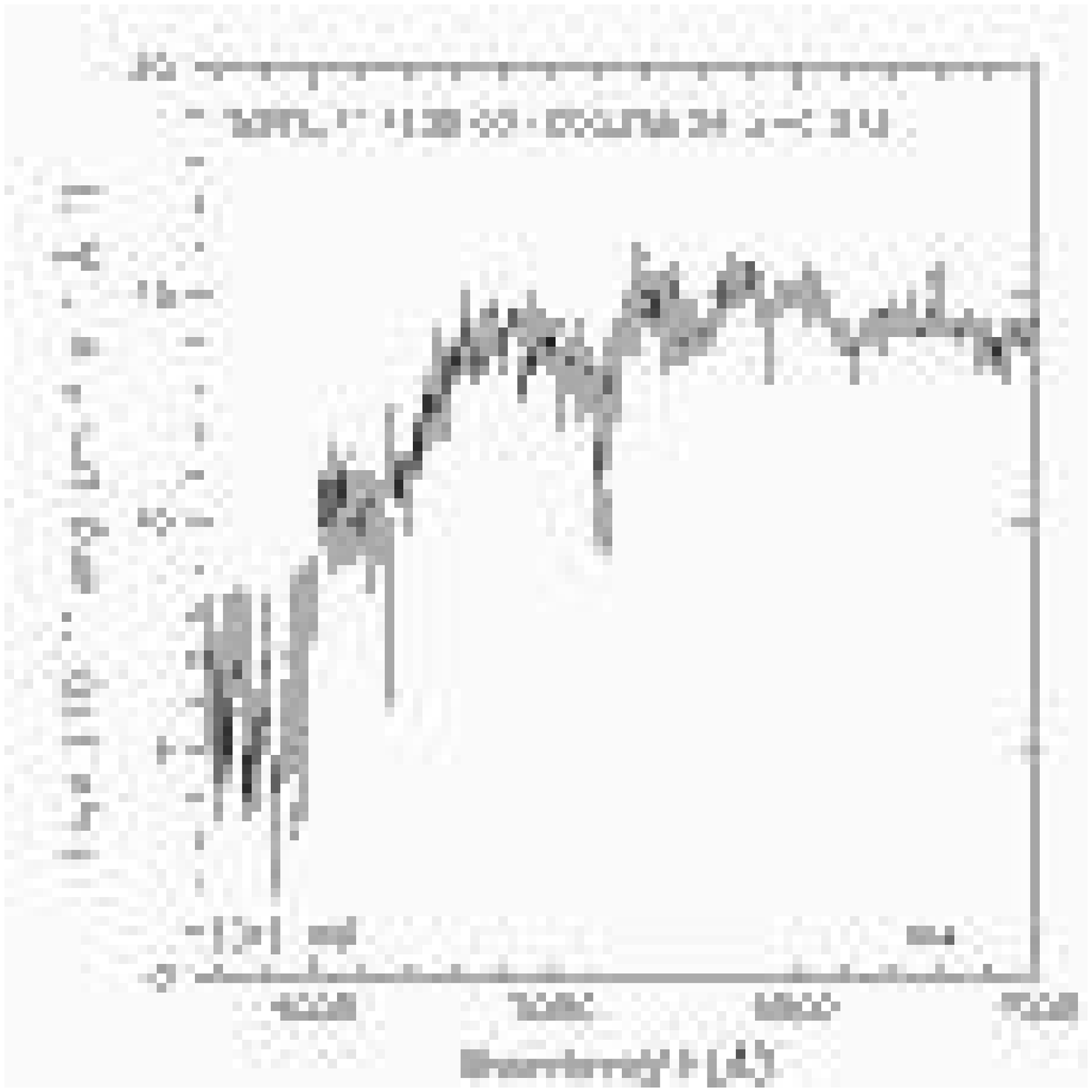}
\includegraphics[scale=0.25]{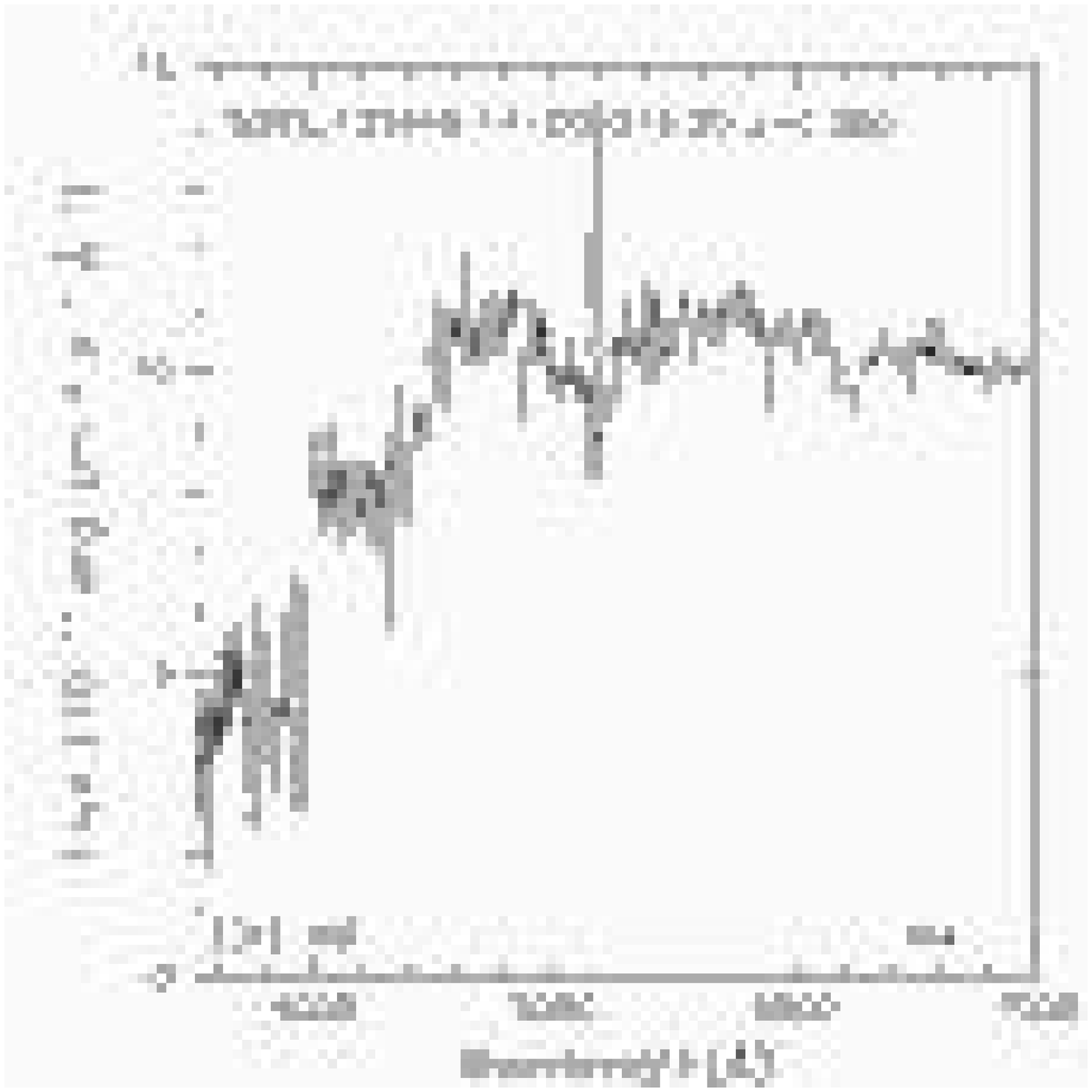}
}
\centering{
\includegraphics[scale=0.25]{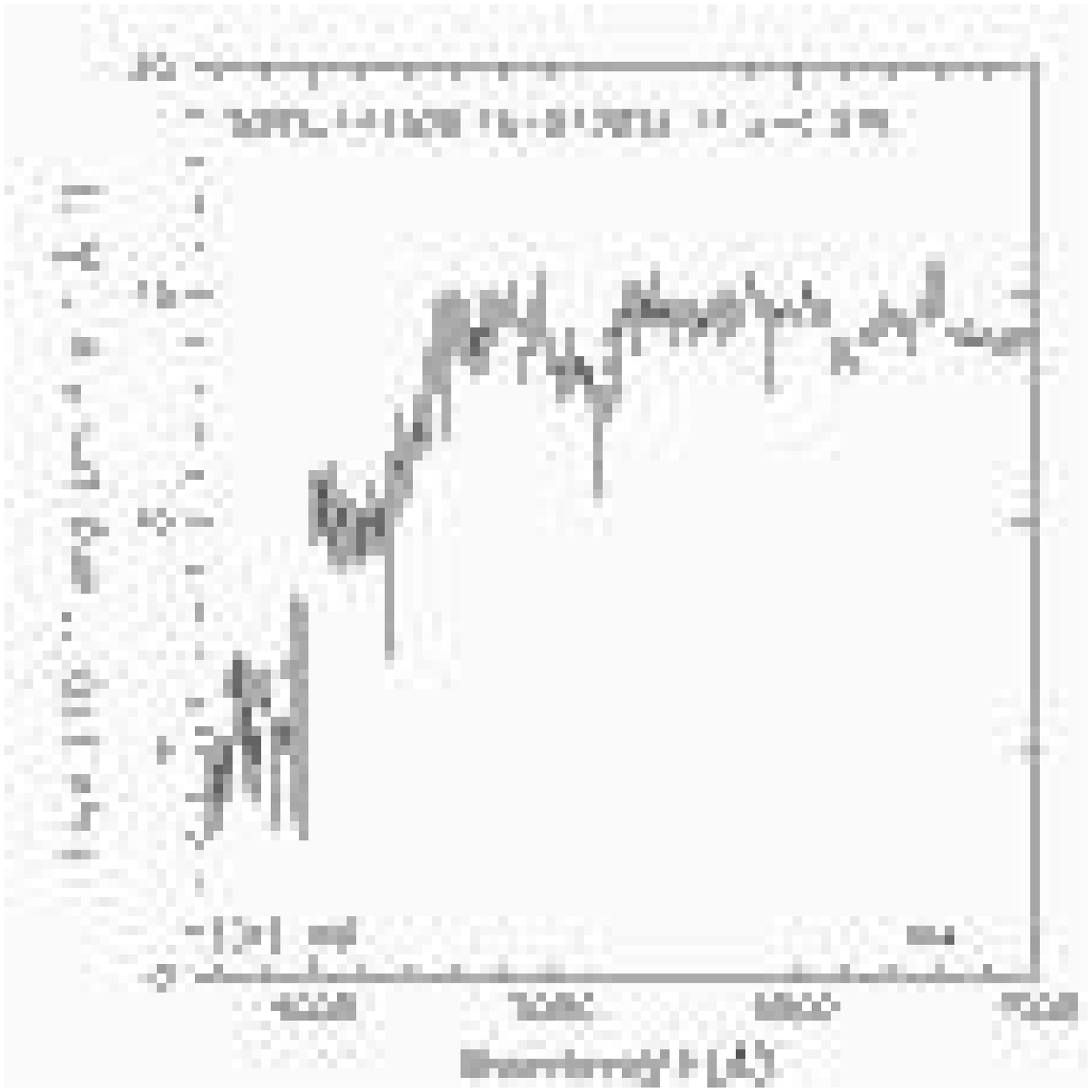}
\includegraphics[scale=0.25]{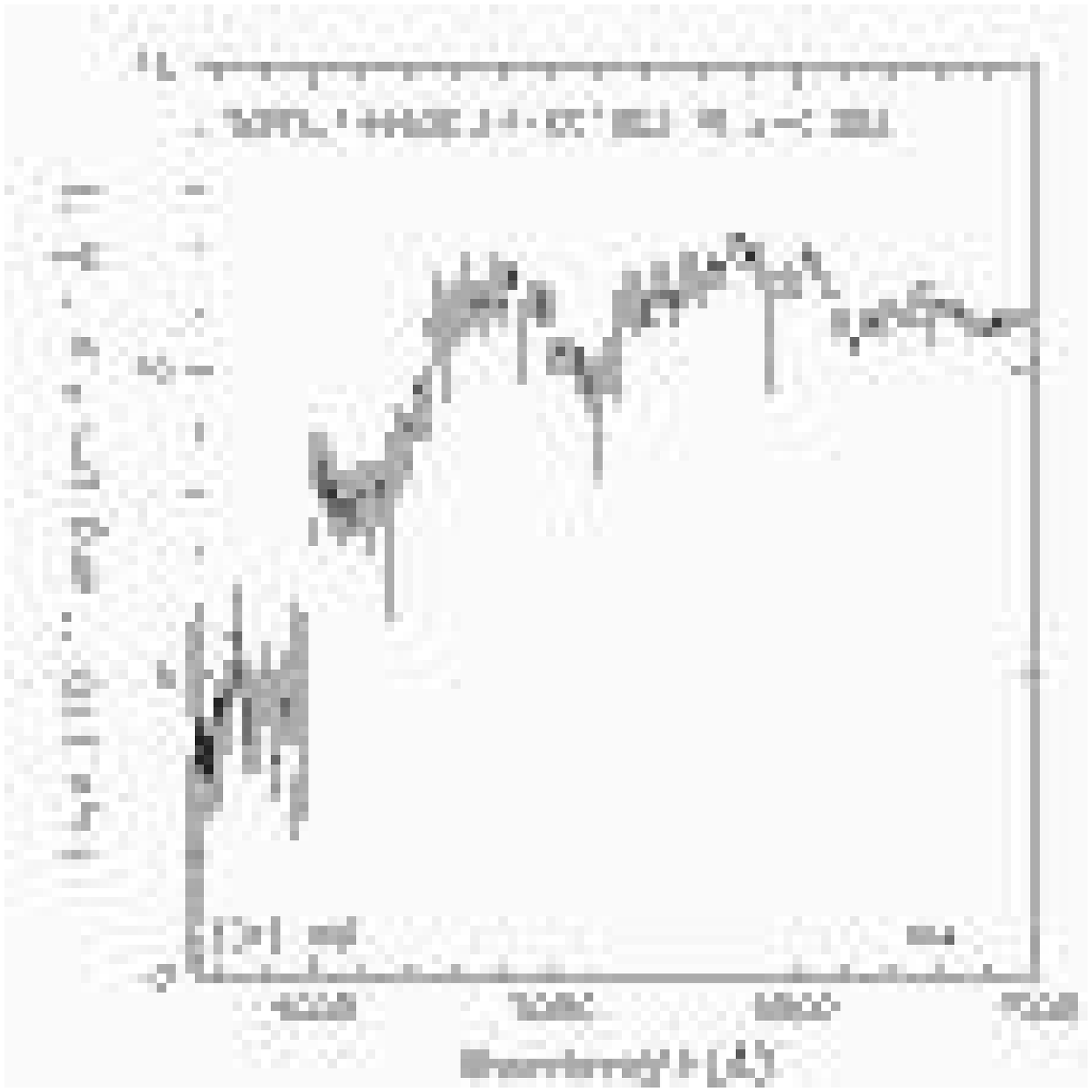}
\includegraphics[scale=0.25]{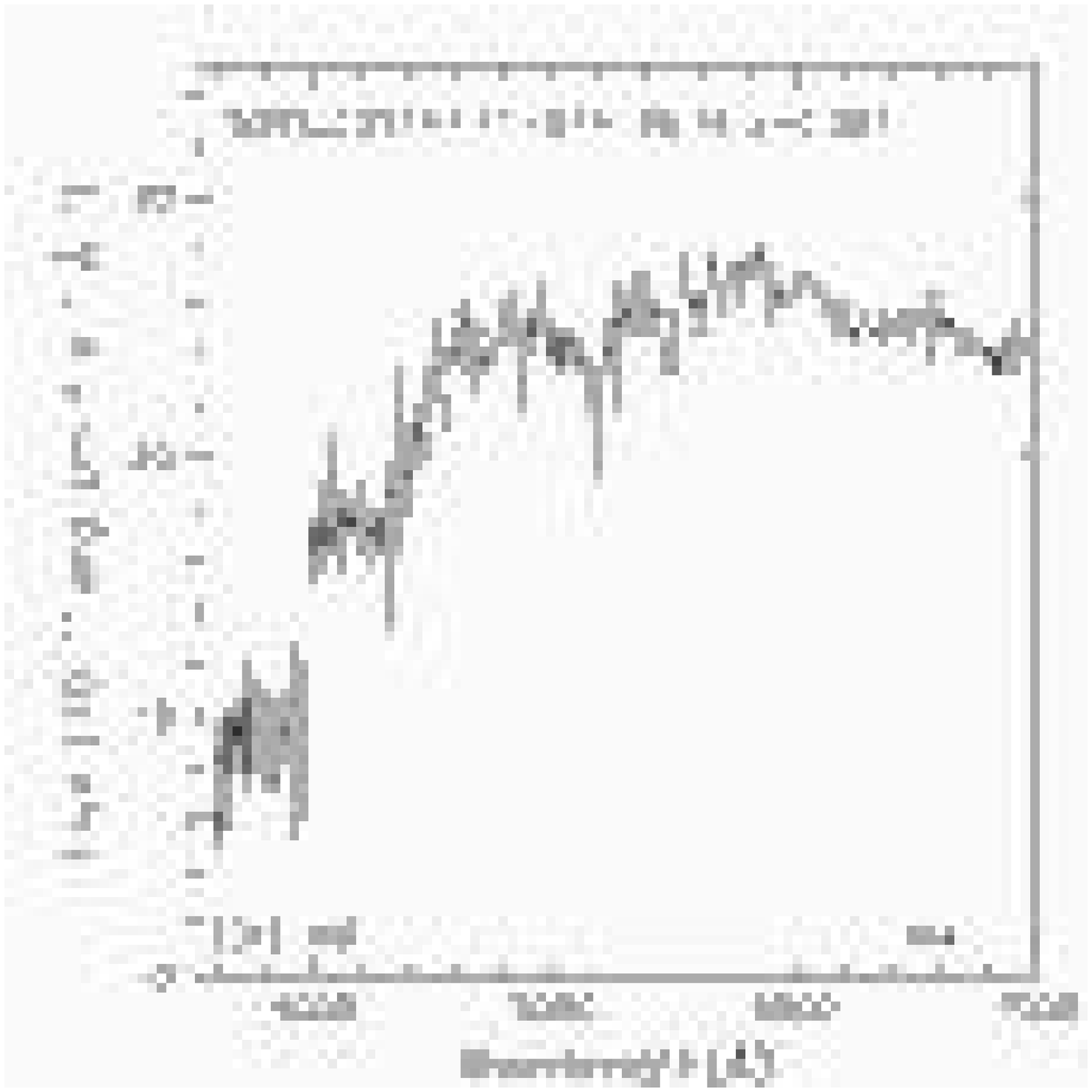}
}
\caption{
\label{fig:ps_spectra}
 Example restframe spectra of passive spiral galaxies. Spectra are
 shifted to restframe and smoothed using a 10\AA\ box.
 Each panel corresponds to
 that in Figure \ref{fig:ps_image}. 
}\end{figure}

\begin{figure}
\centering{
\includegraphics[scale=0.2]{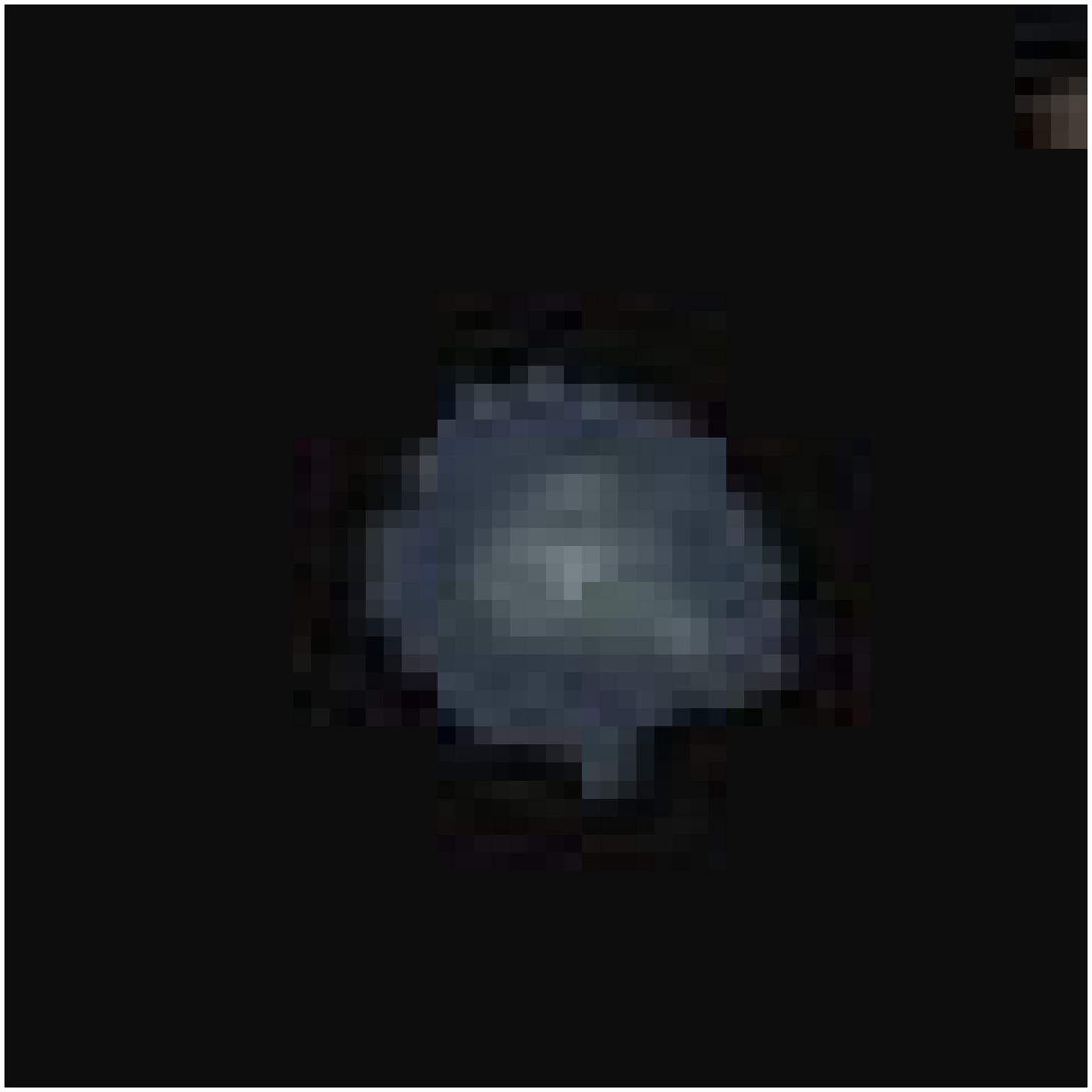} %
\includegraphics[scale=0.2]{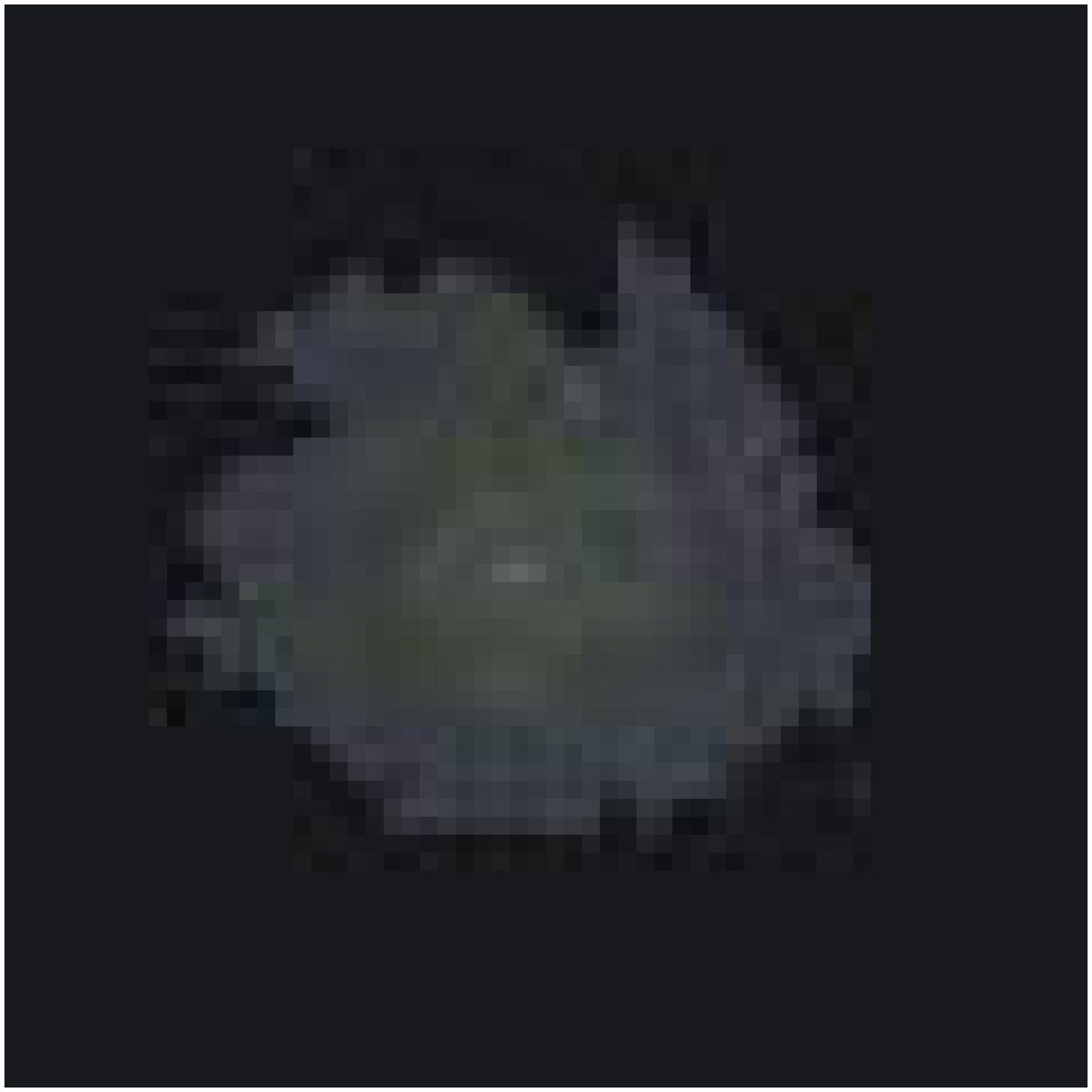}
\includegraphics[scale=0.2]{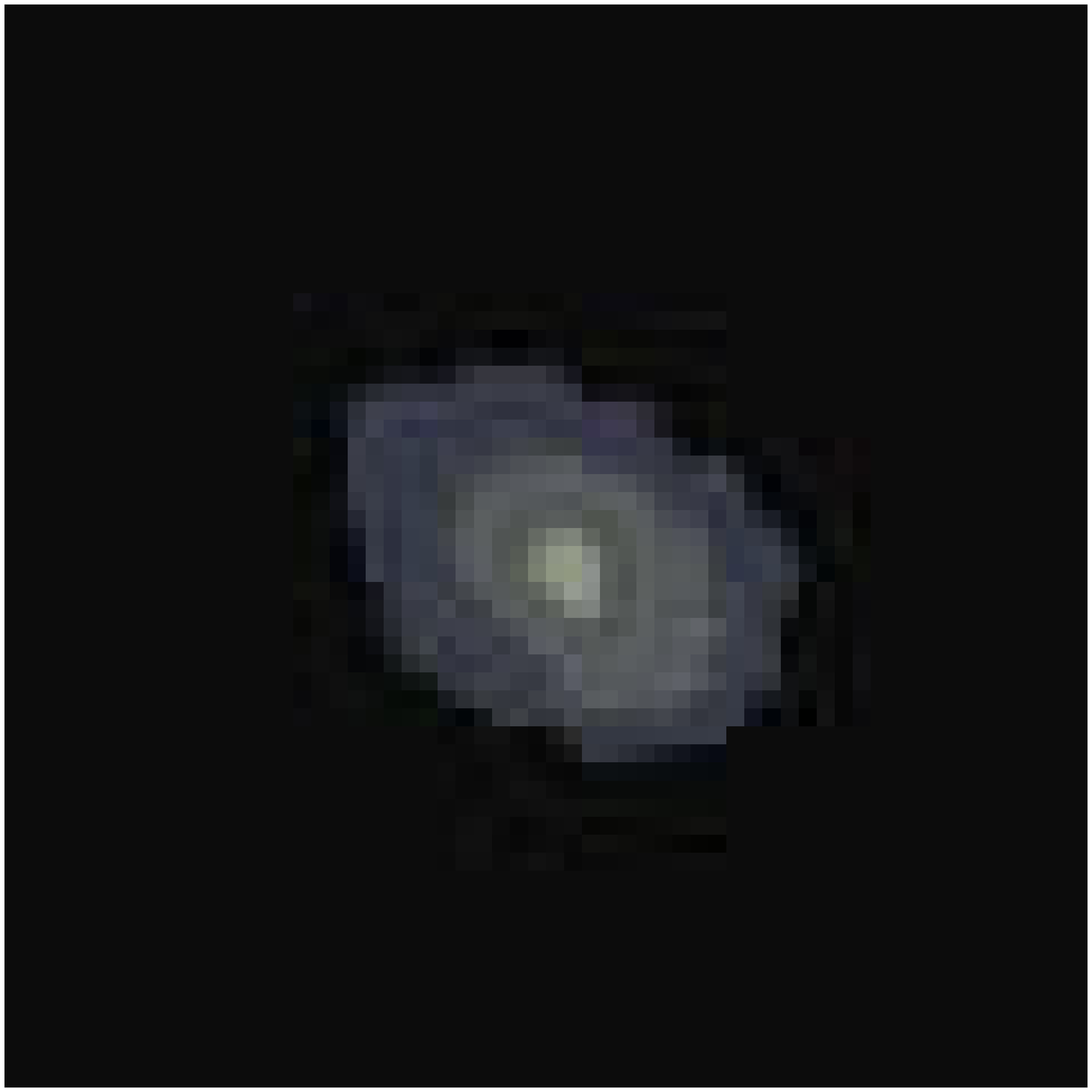}
}
\centering{
\includegraphics[scale=0.2]{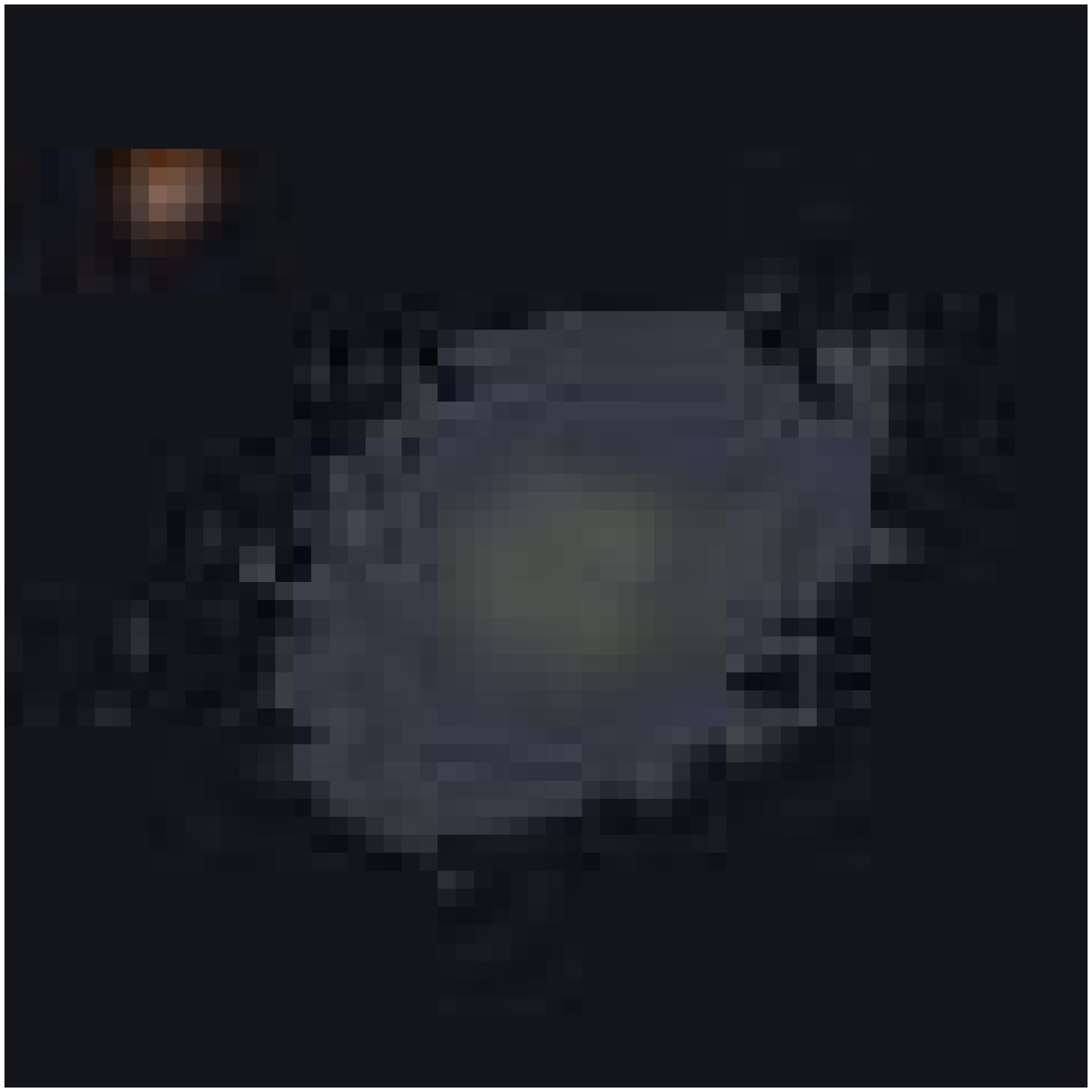}
\includegraphics[scale=0.2]{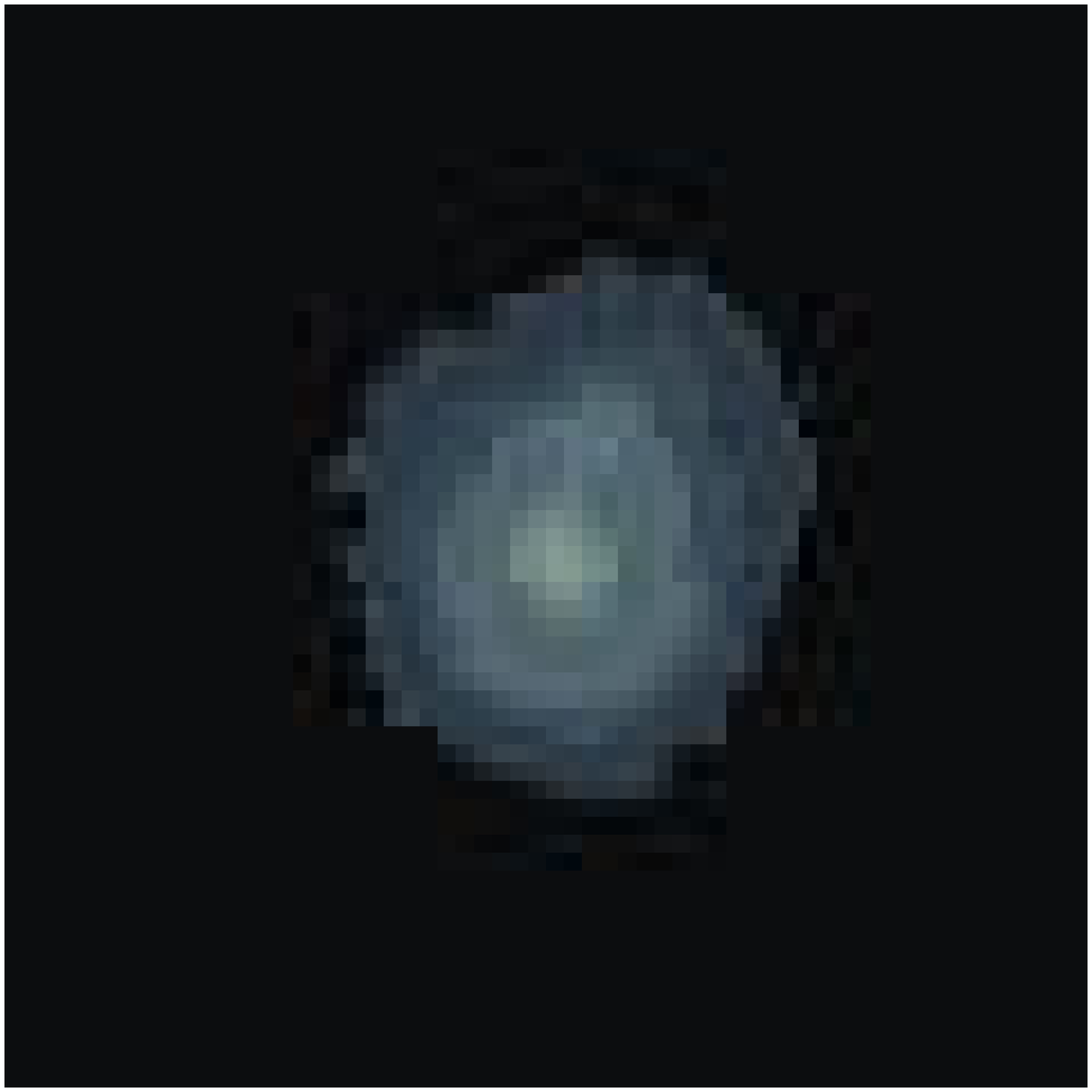}
\includegraphics[scale=0.2]{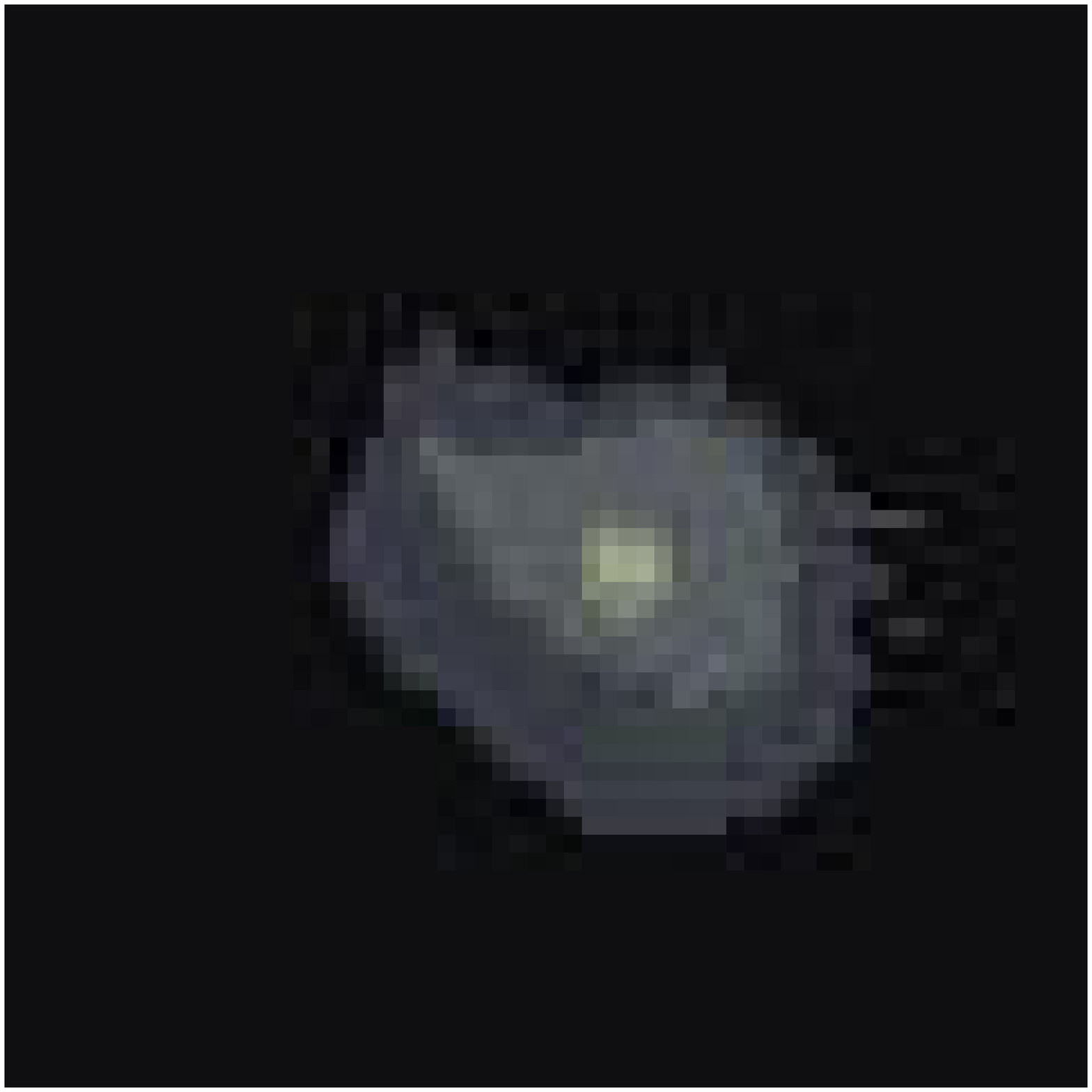}
}
\centering{
\includegraphics[scale=0.2]{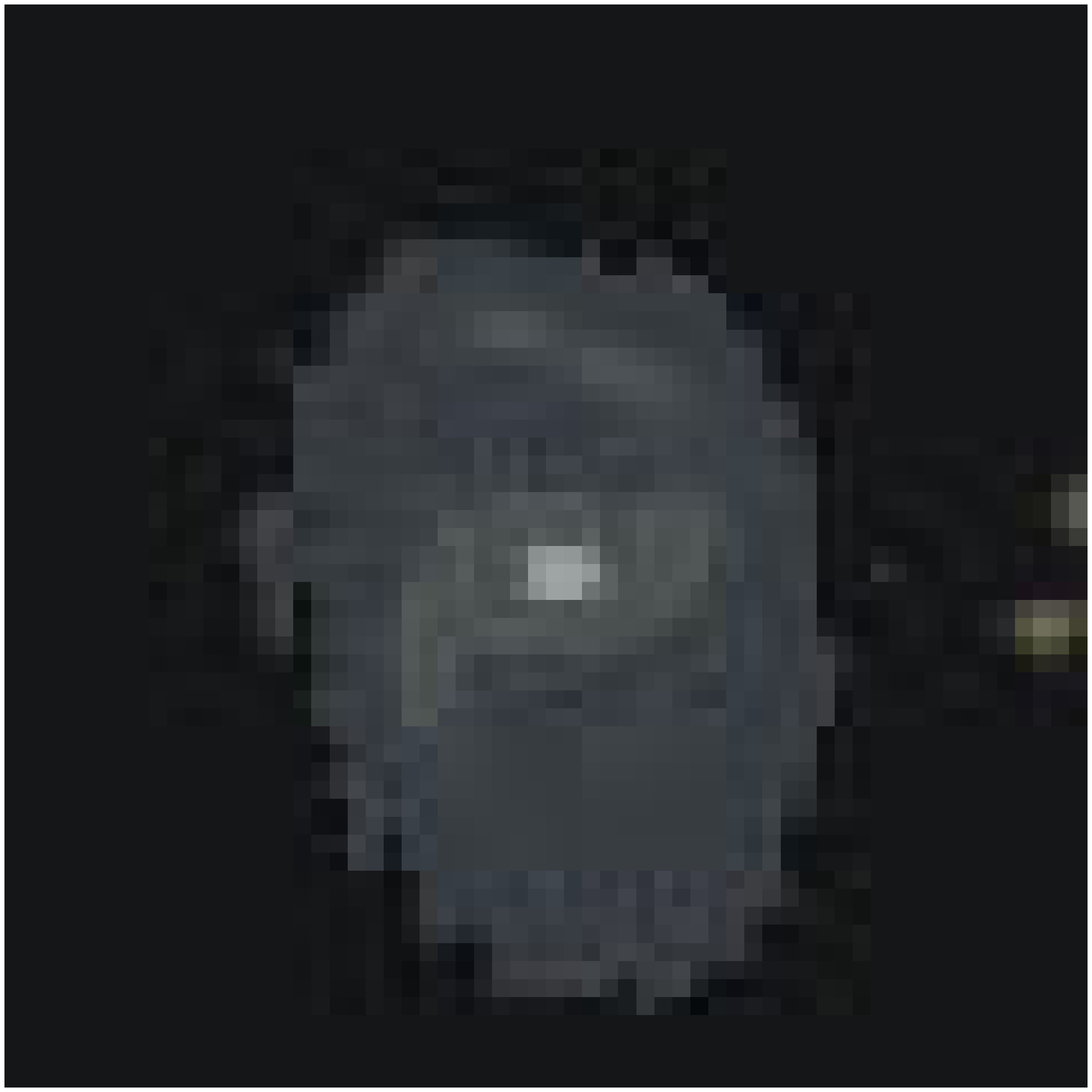}
\includegraphics[scale=0.2]{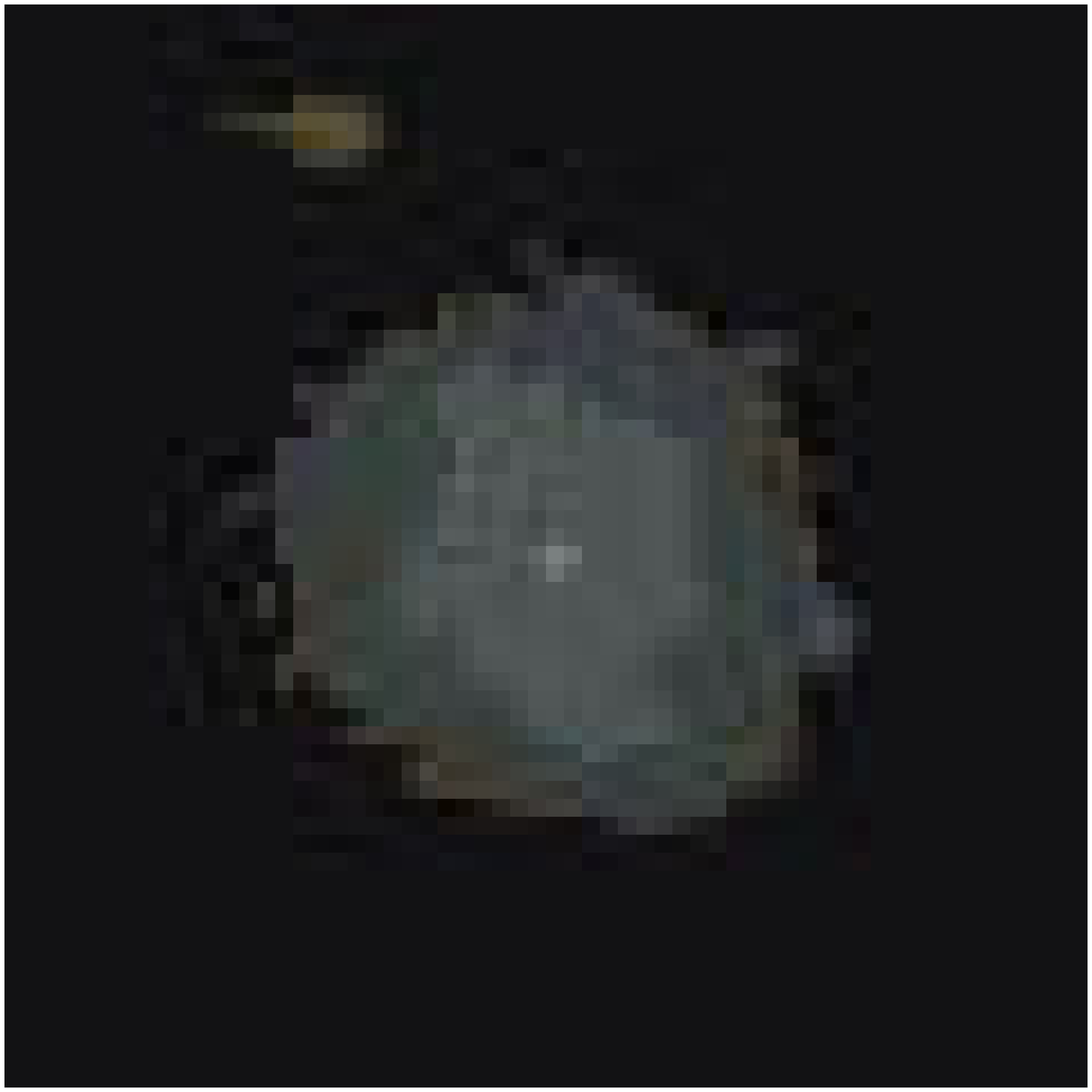}
\includegraphics[scale=0.2]{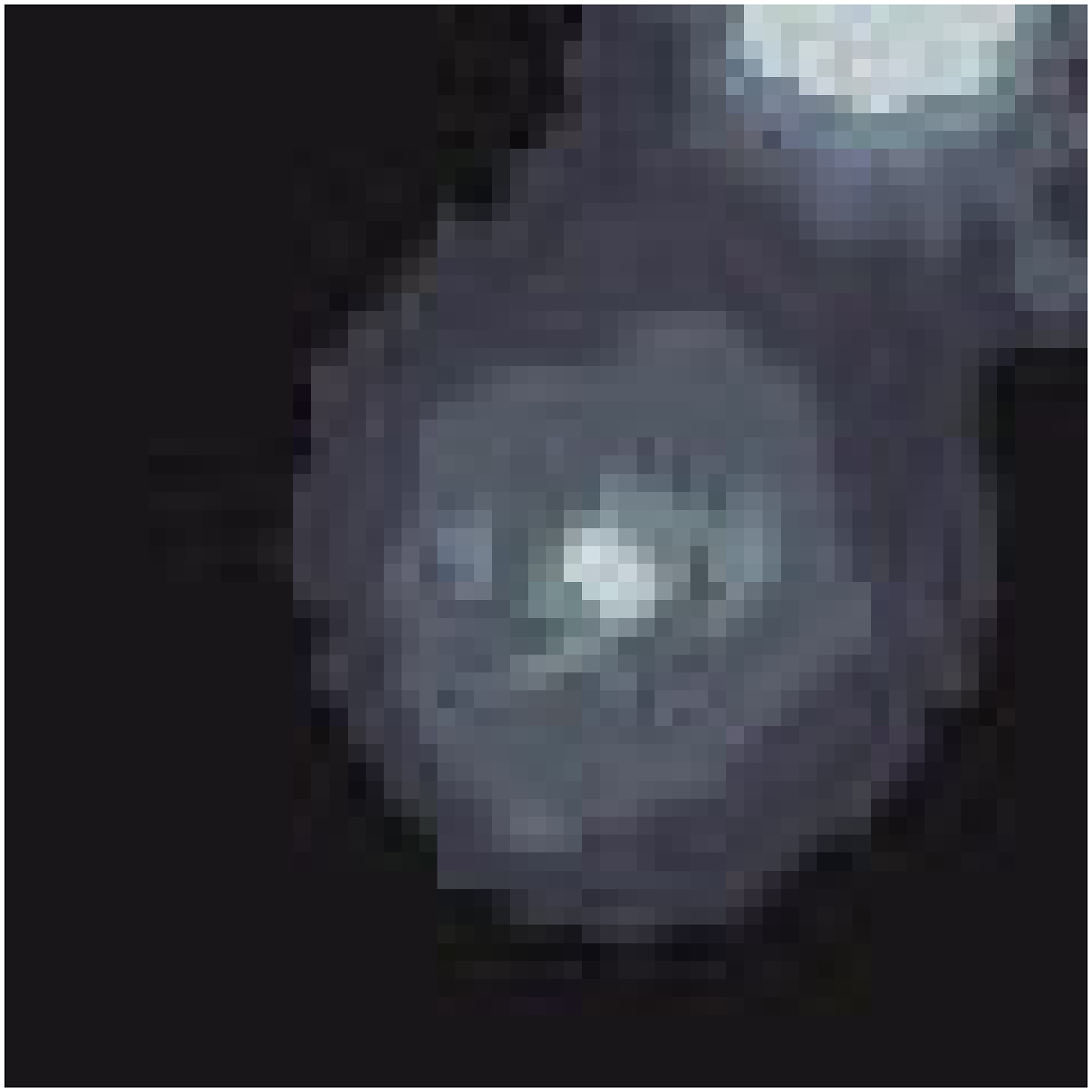}
}
\caption{
\label{fig:ps_as_image}
 Example images of active spiral galaxies. Each image is a composite of
 SDSS $g,r$ and $i$ bands, showing
 30''$\times$30'' area of the sky with its north up.
 Discs and spiral arm structures are recognized.
}\end{figure}

\begin{figure}
\centering{
\includegraphics[scale=0.25]{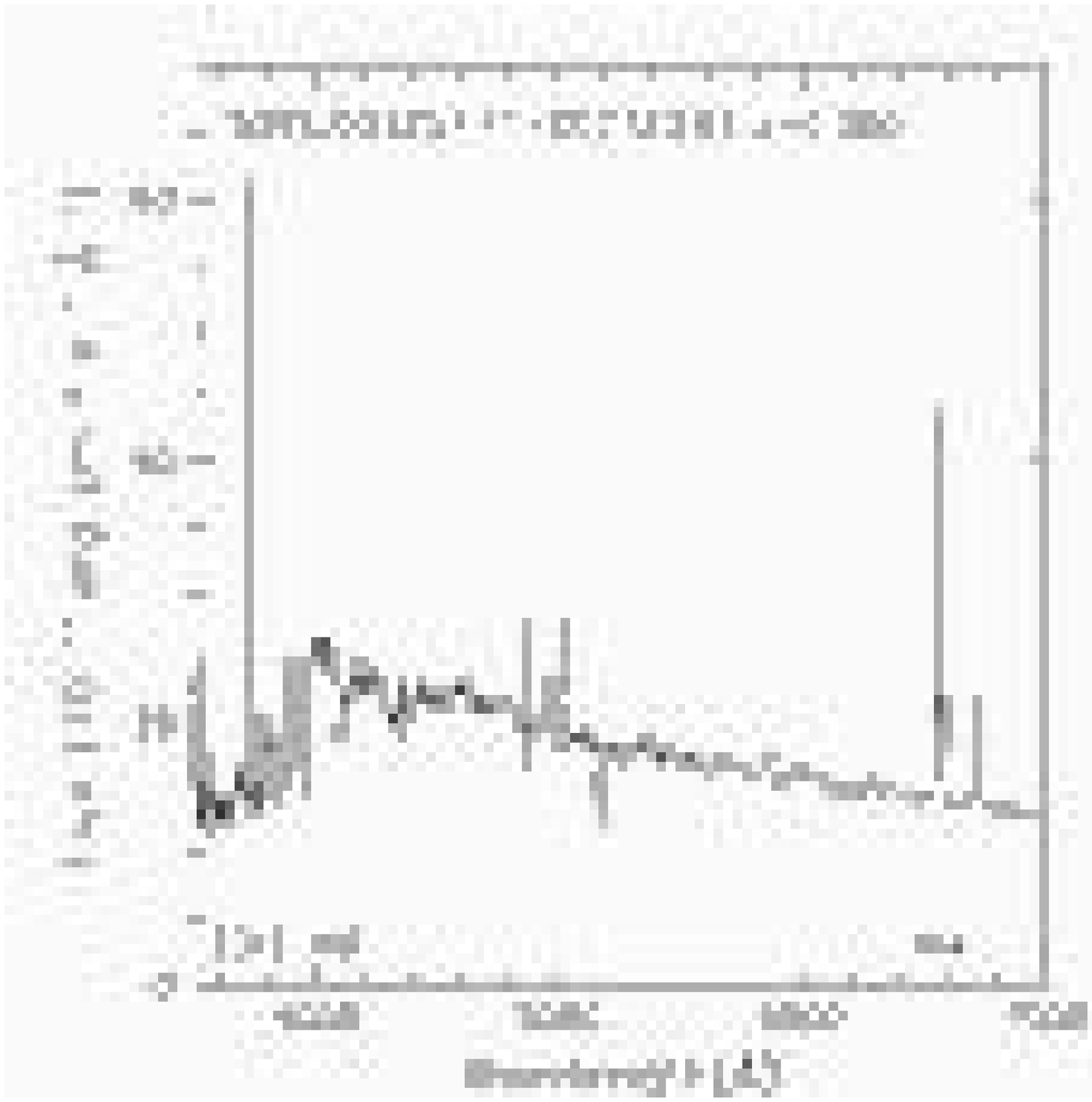} %
\includegraphics[scale=0.25]{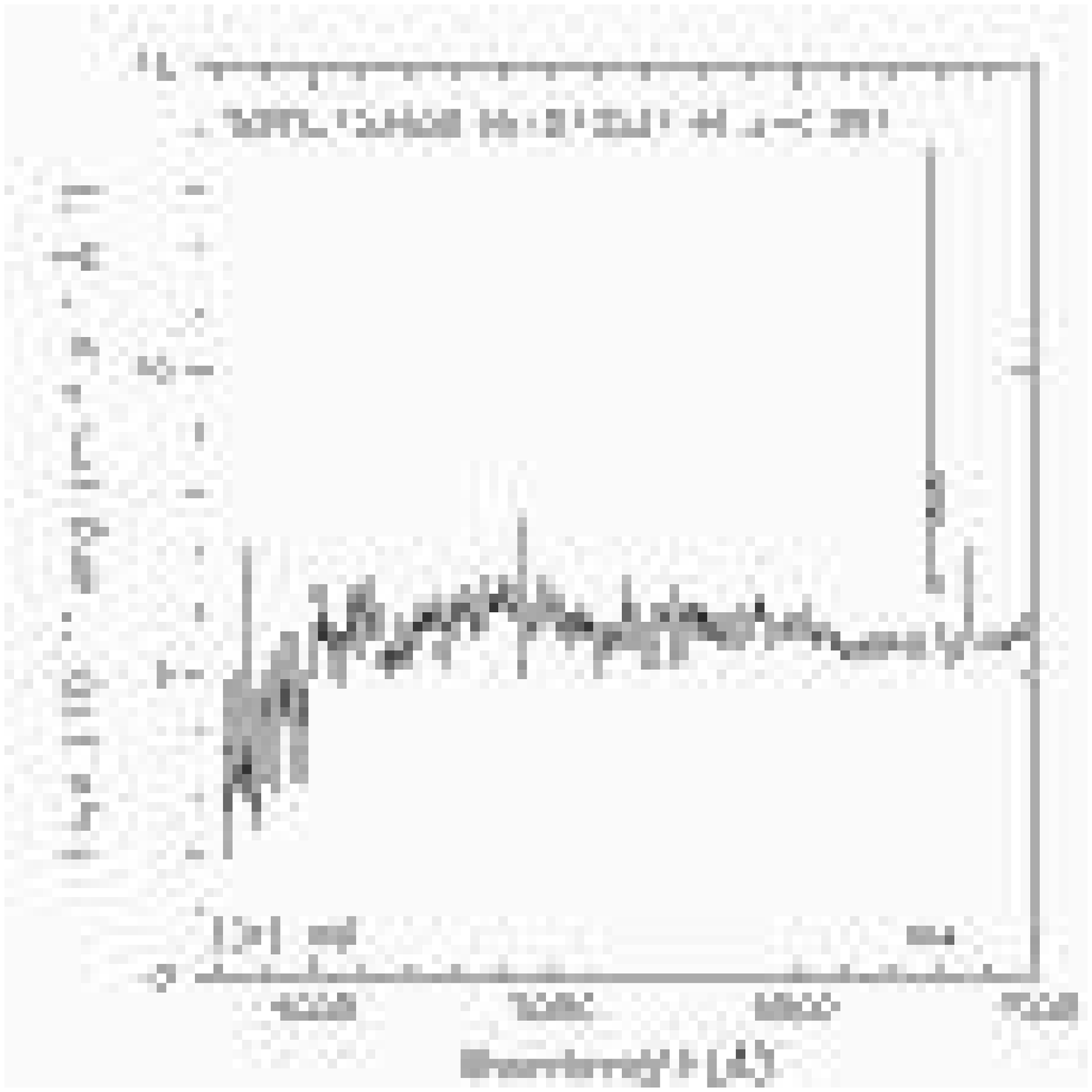}
\includegraphics[scale=0.25]{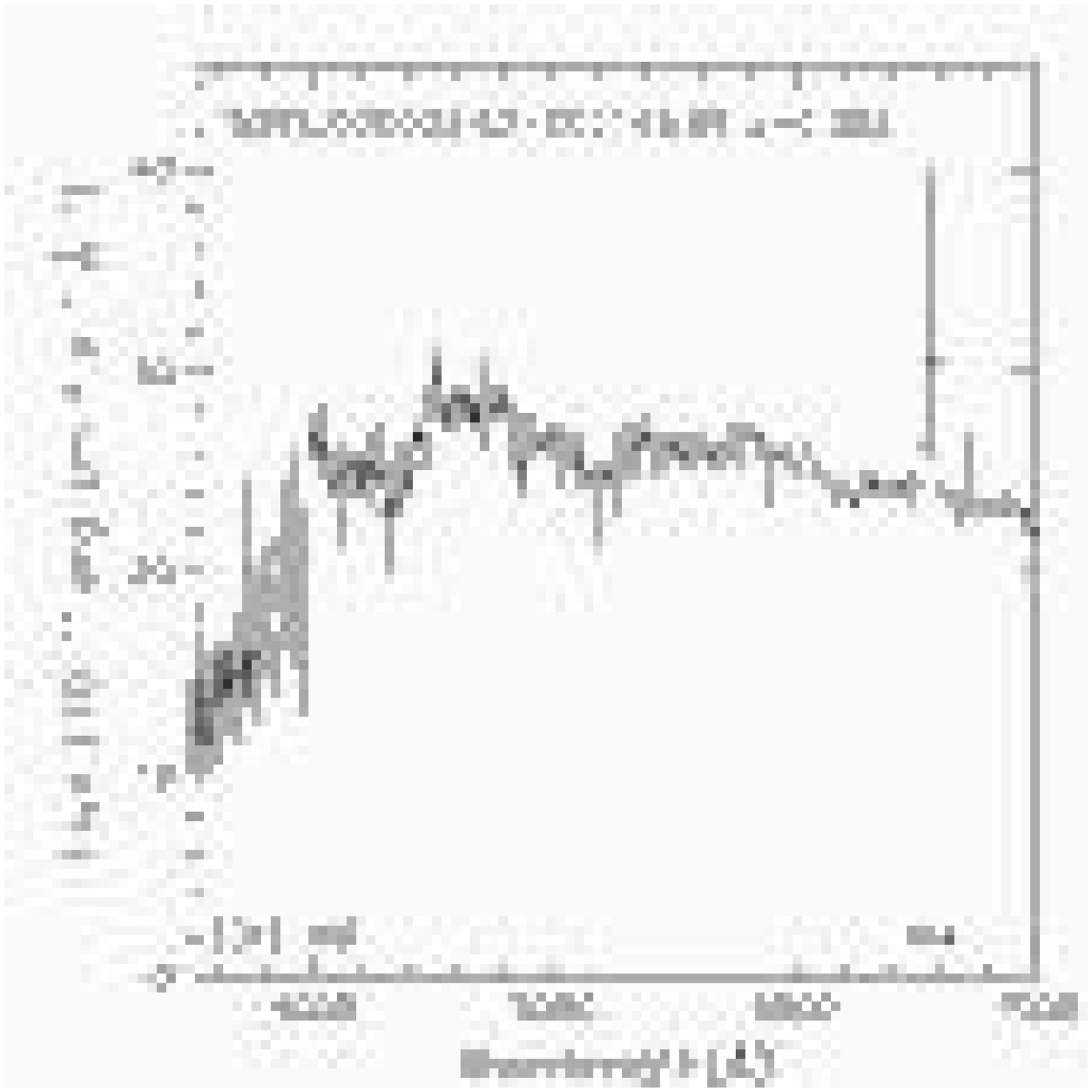}
}
\centering{
\includegraphics[scale=0.25]{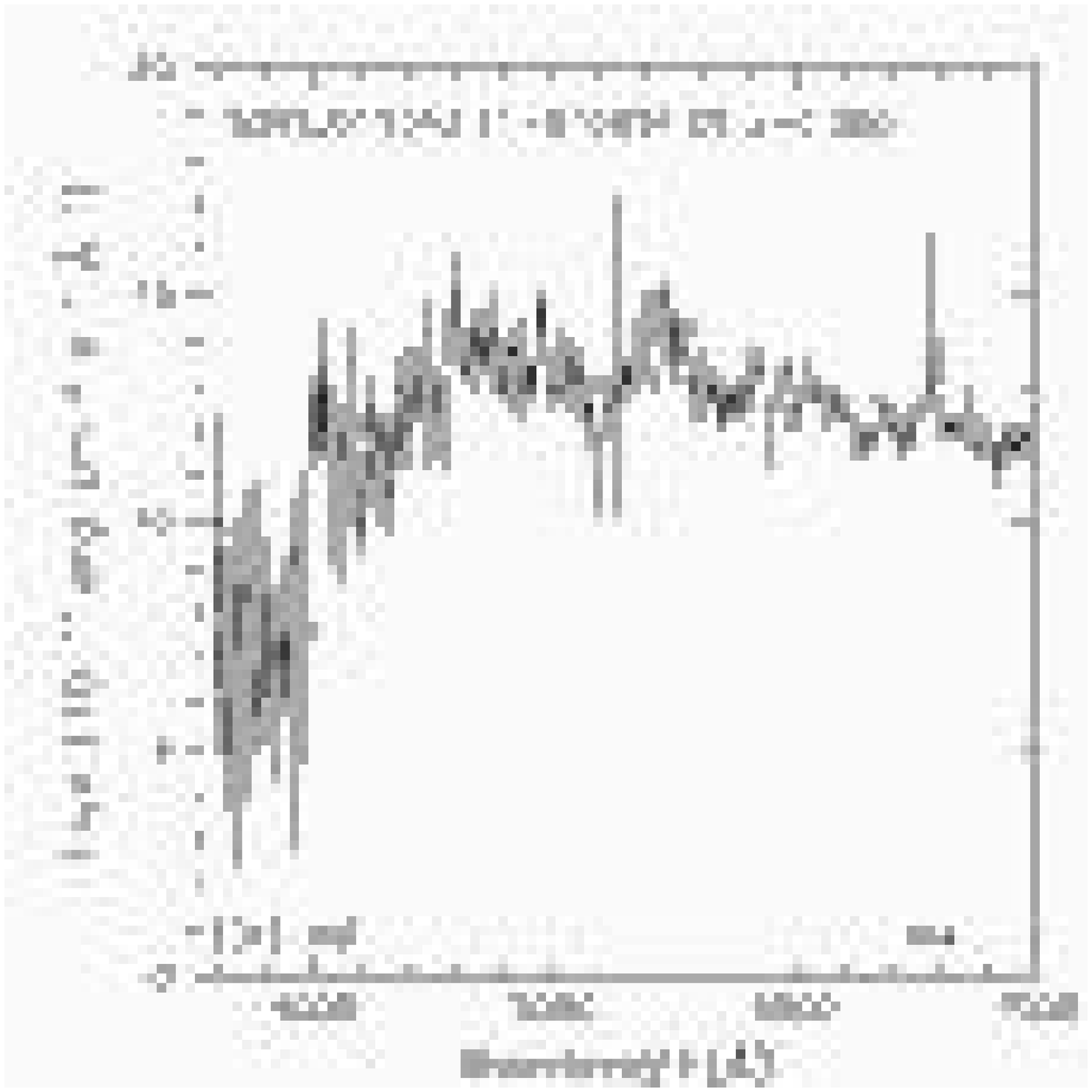}
\includegraphics[scale=0.25]{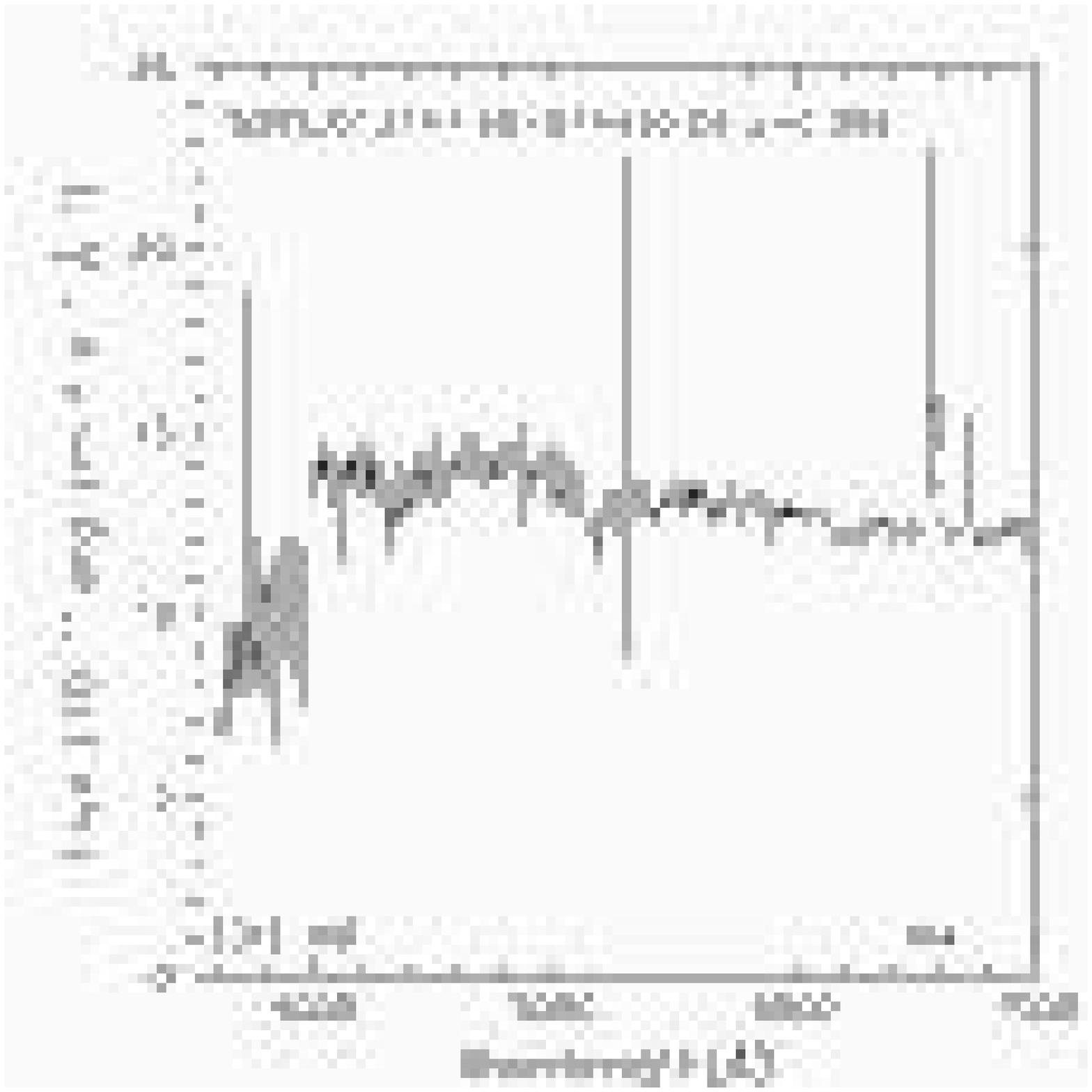}
\includegraphics[scale=0.25]{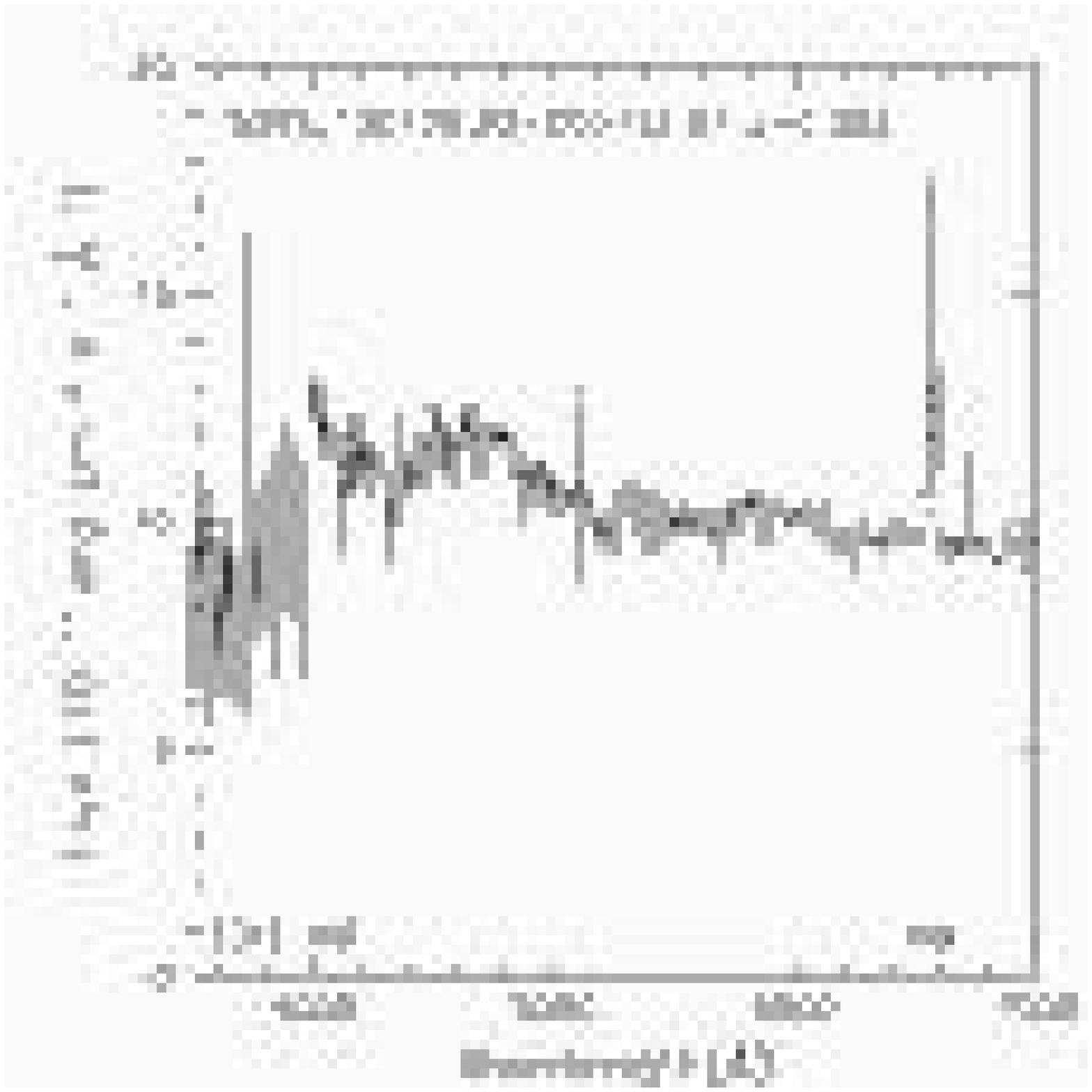}
}
\centering{
\includegraphics[scale=0.25]{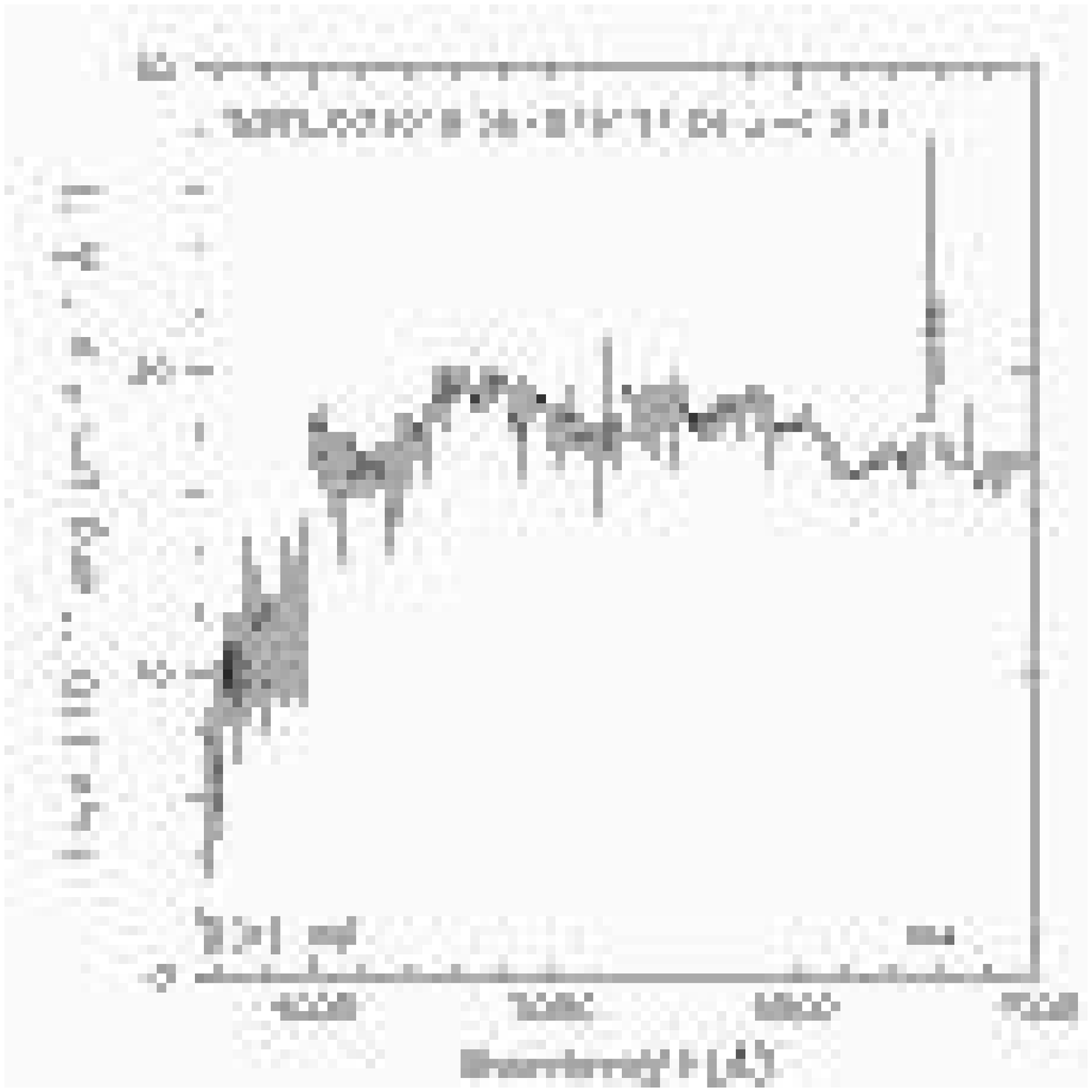}
\includegraphics[scale=0.25]{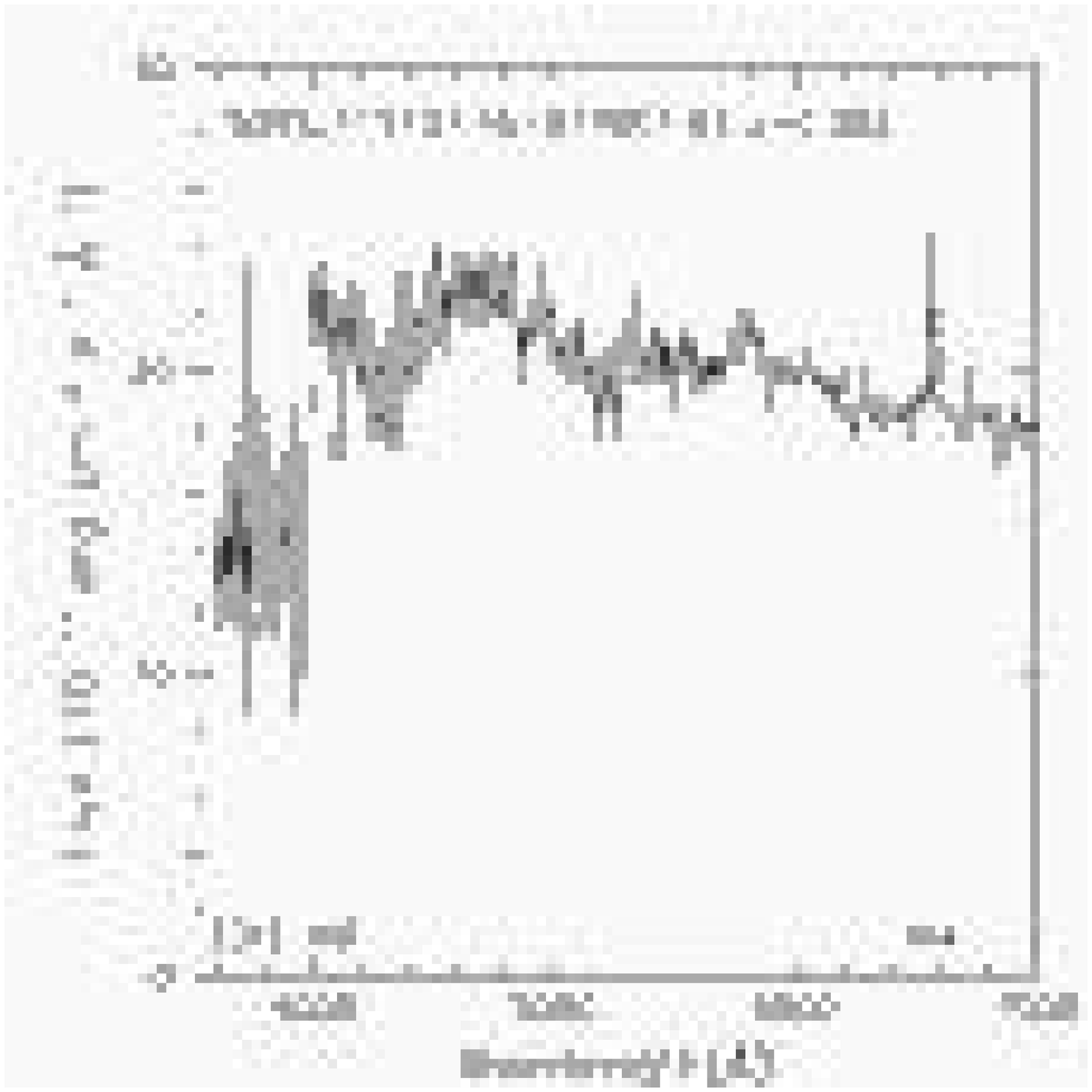}
\includegraphics[scale=0.25]{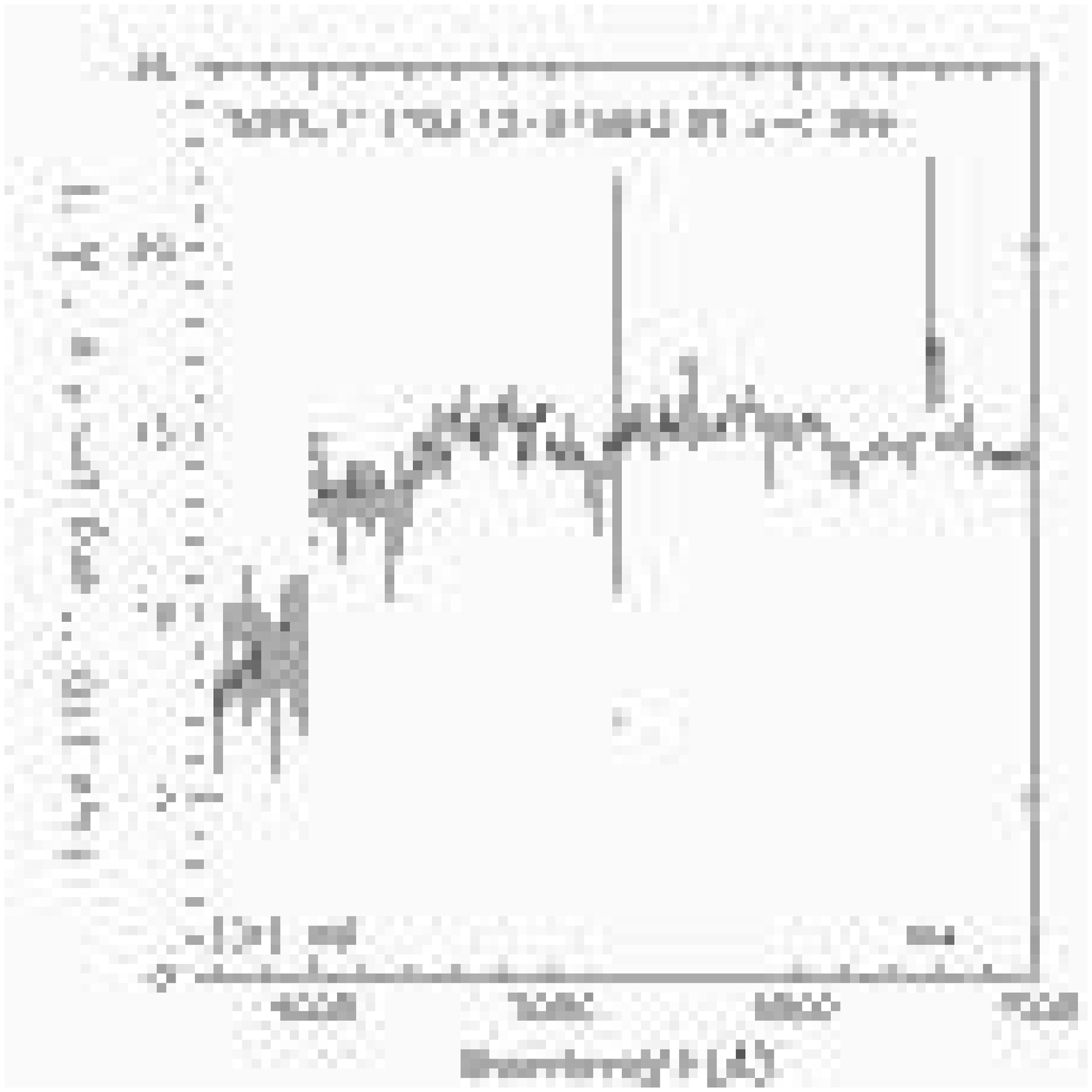}
}
\caption{
\label{fig:ps_as_spectra}
 Example restframe spectra of active spiral galaxies. Spectra are
 shifted to restframe and smoothed using a 10\AA\ box.
 Each panel corresponds to
 that in Figure \ref{fig:ps_as_image}. 
}\end{figure}

\clearpage

\begin{figure}
\begin{center}
\includegraphics[scale=0.7]{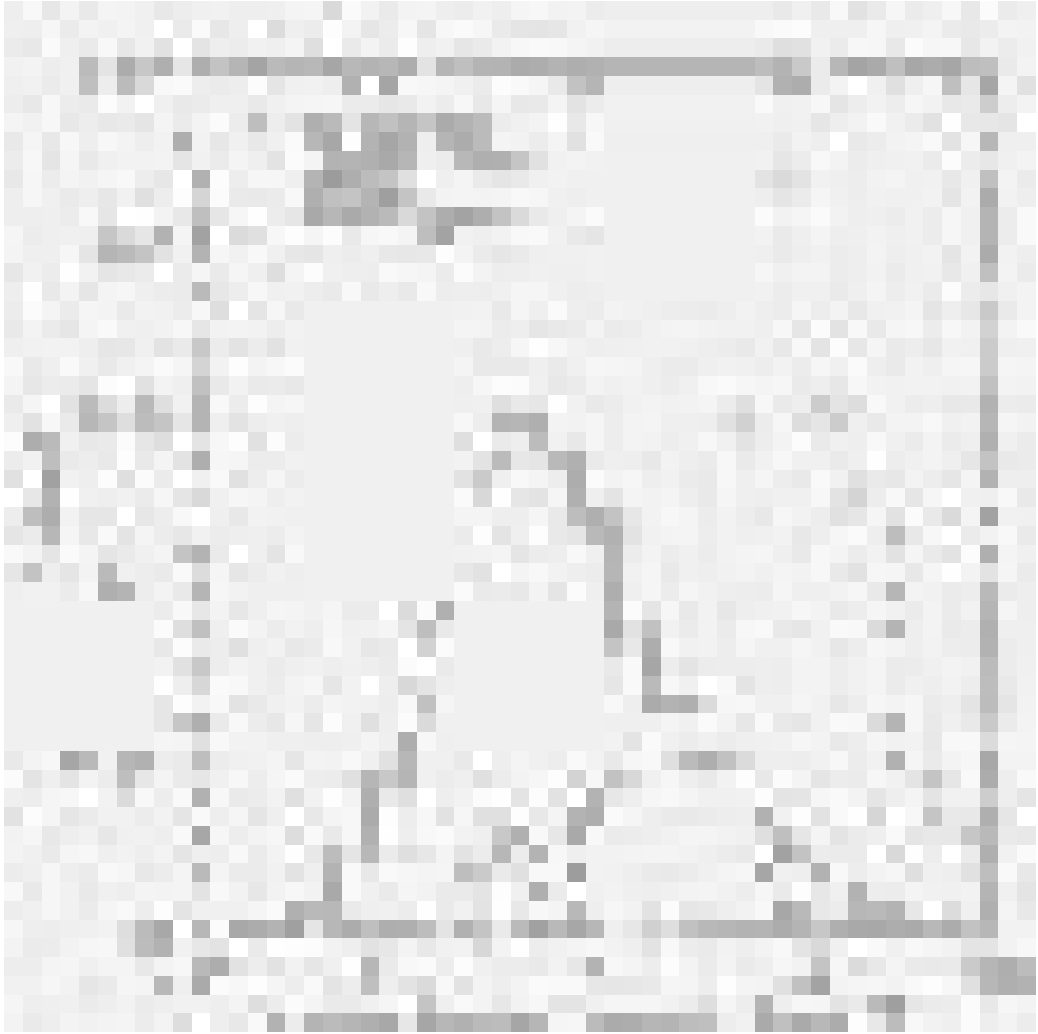}
\end{center}
\caption{
\label{fig:ps_density}
 The distribution of densities for passive spiral galaxies (hashed
 region) and all 25813 galaxies (solid line) in
 our volume limited sample.  A Kolomogorov-Smirnov test shows the distribution
 of passive spirals and that of 
 all galaxies are from different parent distributions. The long dashed line shows
 the distribution of cluster galaxies. The short dashed line shows that of
 active spiral galaxies. }
\end{figure}
\clearpage

\begin{figure}
\begin{center}
\includegraphics[scale=0.7]{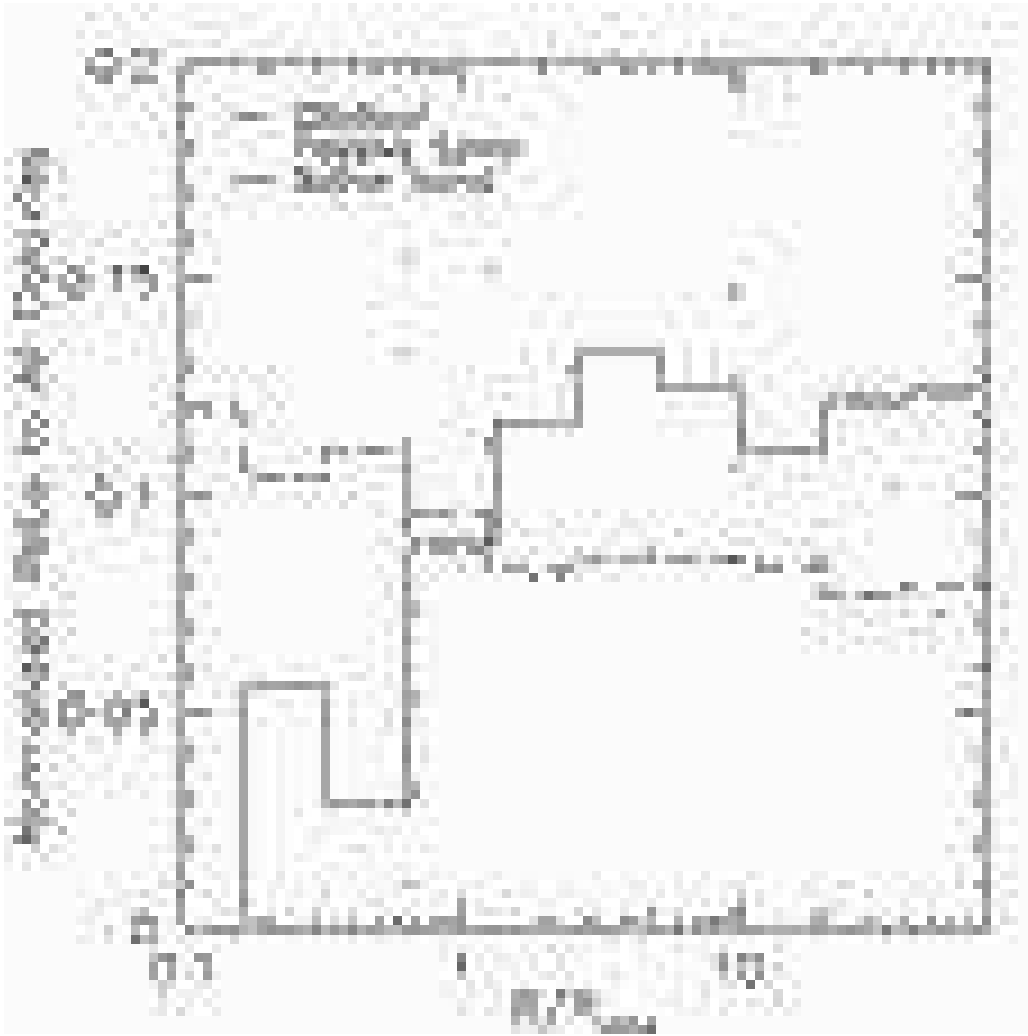}
\end{center}
\caption{
\label{fig:ps_radius}
 The distribution of passive spiral galaxies as a function of
 cluster-centric-radius. The dotted, dashed and solid lines show the
 distributions of passive spiral, elliptical and active spiral galaxies,
 respectively. The distributions are relative to that of all galaxies in
 the volume limited sample and normalized to be unity for clarity.
 The cluster-centric-radius is measured as the distance to a
 nearest C4 cluster (Miller et al. 2003) within $\pm$3000 km s$^{-1}$, and normalized by
 virial radius (Girardi et al. 1998). 
  }
\end{figure}
\clearpage

\begin{figure}
\begin{center}
\includegraphics[scale=0.7]{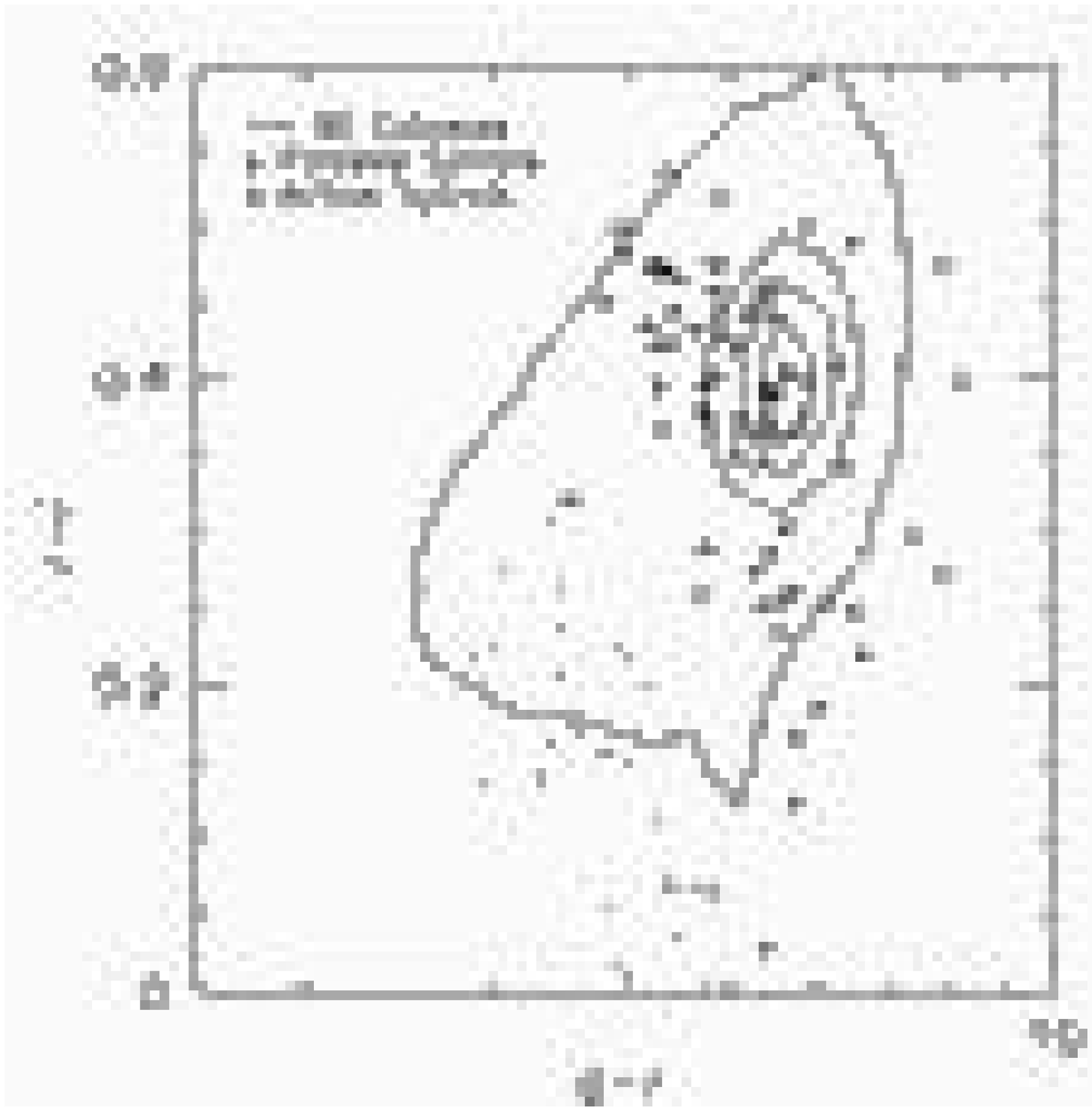}
\end{center}
\caption{
\label{fig:ps_gri}
 The distribution of passive spirals in restframe $g-r-i$
 plane. The contours show the distribution of all galaxies in our volume
 limited sample. 
 The open circle and filled dots represent passive and
 active spiral galaxies, respectively.
  }
\end{figure}
\clearpage

\begin{figure}
\begin{center}
\includegraphics[scale=0.7]{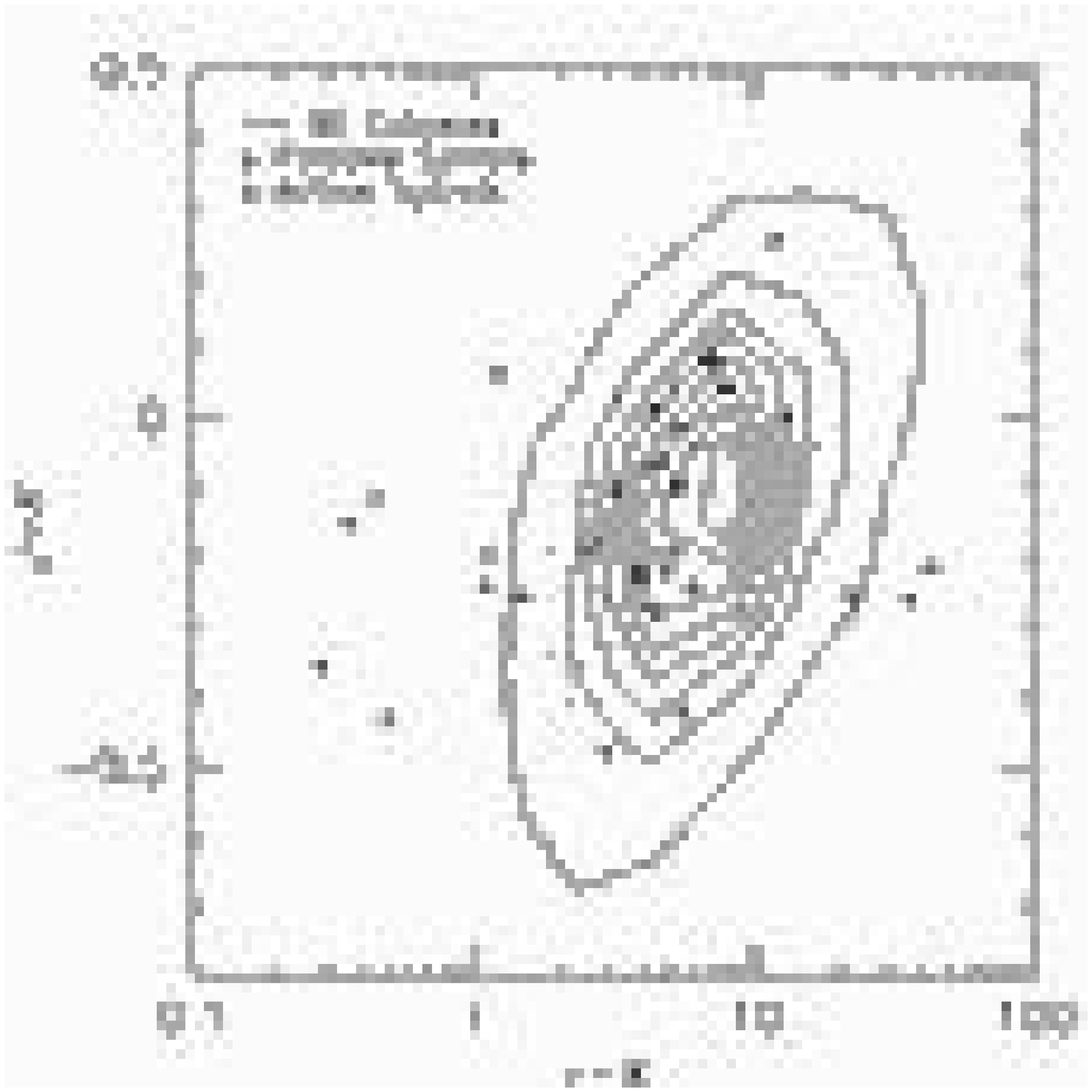}
\end{center}
\caption{
\label{fig:ps_rk}
 The distribution of passive spirals in restframe $J-K-r$
 plane. The contours show the distribution of all galaxies in our volume
 limited sample. 
 The open circle and filled dots represent passive and
 active (normal) spiral galaxies, respectively.
  }
\end{figure}
\clearpage

\begin{figure}
\begin{center}
\includegraphics[scale=0.7]{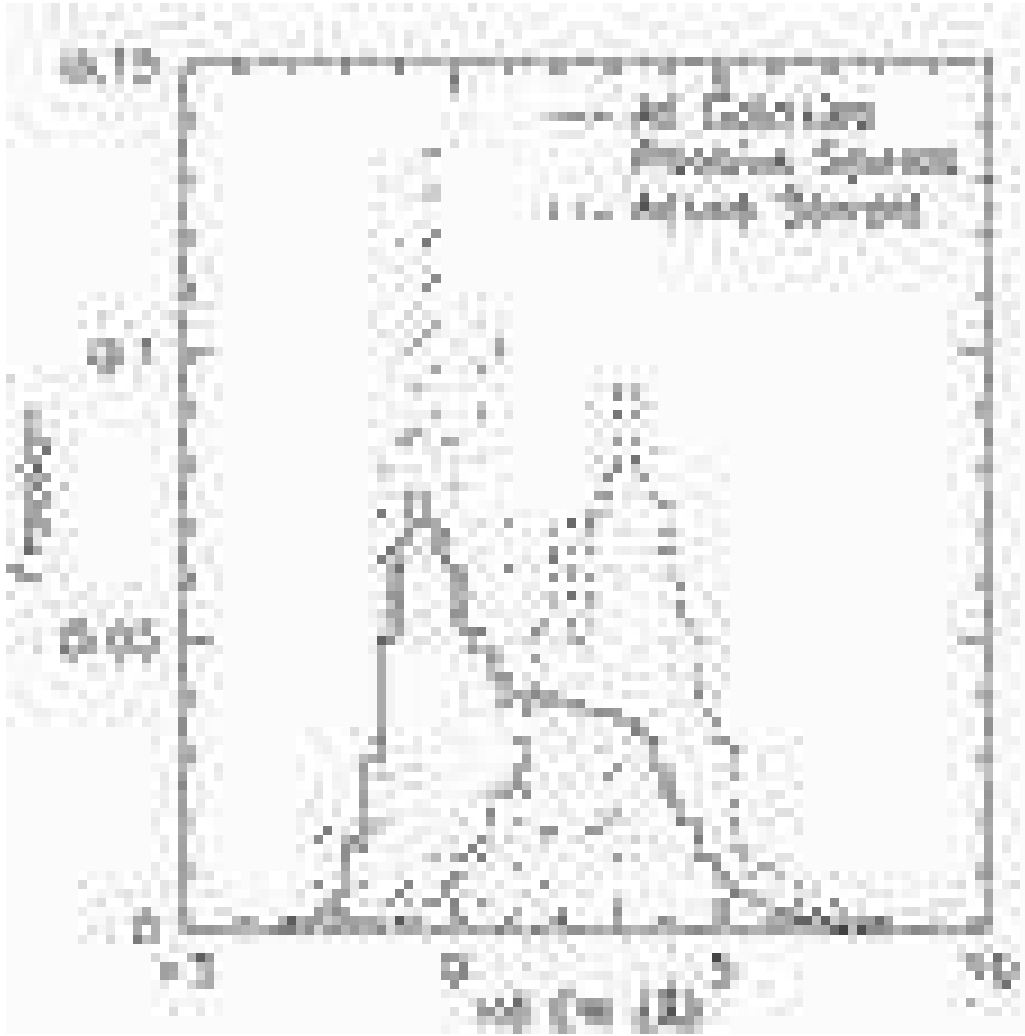}
\end{center}
\caption{
\label{fig:ps_hd}
 Distributions of H$\delta$ EWs of passive spiral galaxies, active
 spiral galaxies and all
 galaxies in the volume limited sample. The solid, dashed and dotted lines
 are for all galaxies, active spiral galaxies
 and passive spiral galaxies, respectively. Absorption lines are
 positive in this figure.  Passive spiral galaxies tend
 to have weak H$\delta$ absorption lines.
  }
\end{figure}
\clearpage

\begin{figure}
\begin{center}
\includegraphics[scale=0.7]{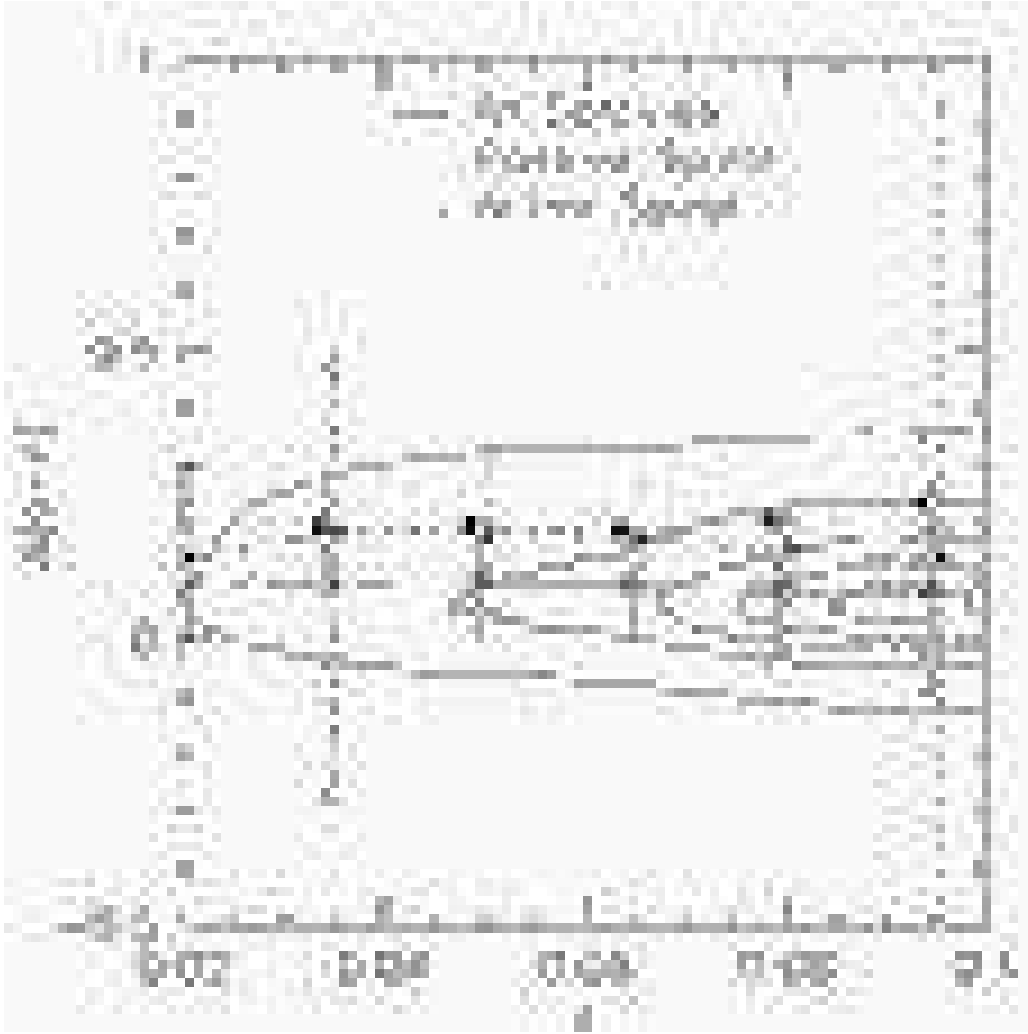}
\end{center}
\caption{
 Differences between fiber color (within 3'' aperture) and model color
 (using Petrosian radius measured in $r$) are plotted against
 redshift. The solid, dotted and dashed lines show medians of all galaxies,
 passive spirals and active spirals, respectively. 
 The difference $\Bigl(\Delta(g-r)\Bigr)$ should be smaller
 at higher redshift since 3'' fiber can collect larger amount of total galaxy
 light at higher redshift.
 Both passive and active spirals have larger
 $\Delta(g-r)$ than all galaxies since they are less concentrated. 
 Throughout the redshift range we used (0.05$<z<$0.1), $\Delta(g-r)$ of
 passive spirals is consistent with a constant within the error,
 suggesting that aperture effect is not a severe effect within the redshift
 range we used.
  }\label{fig:ps_color_gradient}
\end{figure}
\clearpage

\begin{figure}
\begin{center}
\includegraphics[scale=0.7]{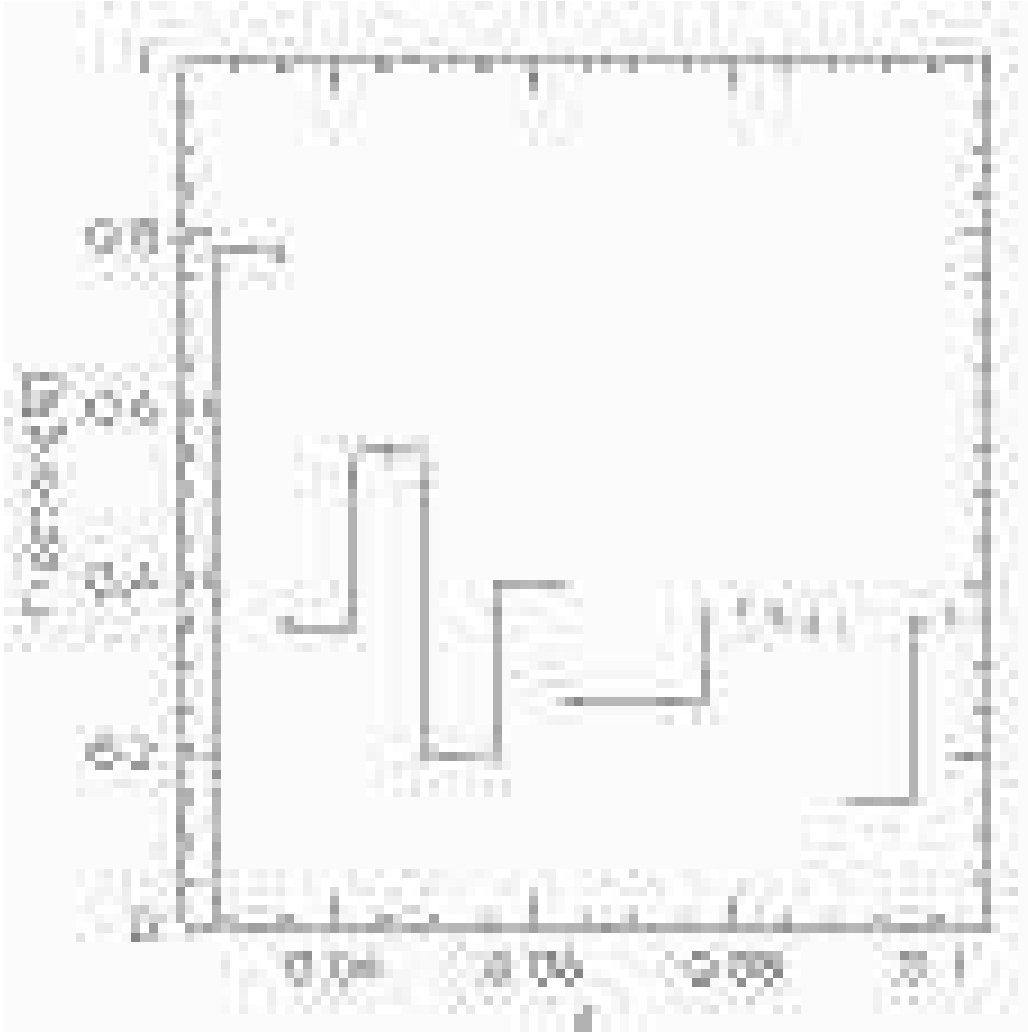}
\end{center}
\caption{
Fractions of passive spiral galaxies (in percentage) to all galaxies
 among the volume limited sample are
 shown as a function of redshift. Our sample includes passive spiral
 galaxies only between $z=$0.05 and $z=$0.1, where fractions are consistent
 with constant, suggesting aperture bias is not a strong effect in our
 sample.  
  }\label{fig:ps_aperture}
\end{figure}
\clearpage

\begin{table}[h]
\begin{center}
\caption{
 Wavelength ranges used to measure the equivalent widths of [OII],
 H$\alpha$  and  H$\delta$ lines. 
}\label{tab:ps_ps_wavelength}
\begin{tabular}{llll}
\hline
  & Blue continuum  & Line & Red continuum \\
\hline
\hline
$[OII]$            &  3653-3713\AA &    3713-3741\AA & 3741-3801\AA  \\ 
H$\alpha$          &  6490-6537\AA &    6555-6575\AA & 6594-6640\AA  \\ 
H$\delta$          &  4030-4082\AA &    4088-4116(4082-4122)\AA & 4122-4170\AA  \\ 
\hline
\end{tabular}
\end{center}
\end{table}


\chapter{Fate of Infalling Galaxies}
\label{chap:fate}
	 
 In this chapter, we consider all of our findings in the previous chapters and
 summarize a possible evolutionary history of infalling cluster
 galaxies. 

 \section{Summary of Our Findings}

 We have studied the environmental effects on galaxy evolution using the
 Sloan Digital Sky Survey (SDSS) data. Our main findings are as follows.

\begin{itemize}

 \item We have developed a new cluster finding method, the Cut \&
       Enhance (CE) method, 
 which uses color and angular separation to increase the signal to noise
       ratio of  galaxy 
 clusters.  
 We determine the selection function of the  Cut \& Enhance
  method via extensive Monte Carlo simulations, showing that
 the Cut \& Enhance method can
 detect rich clusters ($Ngal$=100) to $z\sim$0.3 with $\sim$80\%
 probability. 
 We apply Cut \& Enhance method to the SDSS commissioning data and
 produce an SDSS Cut \& Enhance cluster catalog containing 4638 galaxy
       cluster candidates in $\sim$350 deg$^2$. 
 The SDSS Cut \& Enhance cluster catalog developed in this work is a
       useful tool to study  cosmology as well as  properties of clusters and
       cluster galaxies.

%
%
%

 \item  We constructed composite luminosity functions (LFs) in the five
       SDSS bands, 
 $u,g,r,i$ and $z$, using the 204 SDSS CE clusters ranging from
 $z=$0.02 to $z=$0.25 .    The best-fit Schechter parameters are
 ($M^*$,$\alpha$)=($-$21.61$\pm$0.26, $-$1.40$\pm$0.11),($-$22.01$\pm$0.11,
 $-$1.00$\pm$0.06),($-$22.21$\pm$0.05, $-$0.85$\pm$0.03), ($-$22.31$\pm$0.08,
 $-$0.70$\pm$0.05) and ($-$22.36$\pm$0.06, $-$0.58$\pm$0.04) in  $u, g,r,i$ and
 $z$, respectively. 
  Compared with the field luminosity function, cluster LFs have  a
       brighter characteristic 
  magnitude ($M^*$) and a flatter slope  in the $g, r, i$ and $z$ band,
       suggesting that cluster galaxies have different evolutionary
       histories from field galaxies. 
 We also derived the type-specific composite LFs by dividing the cluster
       galaxies into early and late types using three different criteria 
 (concentration, galaxy profile fit and $u-r$ color). We found that
       early-type galaxies always have flatter slopes 
       than  late-type galaxies in all the three cases. 
           These observations are in agreement with the
 hypothesis that the bright end of the cluster LF is dominated by bright, old
 early-types, while the faint-end of the
 cluster LF represents late-type galaxies.

 \item  We investigate the evolution of the fractions of blue
 cluster galaxies as a function of redshift, using one of the largest,
 most uniform  cluster samples of 514 CE clusters in the range of 
 0.02$\leq z\leq$0.3. It is a significant improvement over previous
       work that both high and low redshift 
       clusters are selected from the same data using the same cluster
       finder, excluding richness and redshift related bias prevalent in
       previous work. 
 By selecting blue galaxies as those with restframe $g-r$ bluer by 0.2 than
       red-sequence or those with $u-r<$2.2,
 we found that blue fractions of cluster
       galaxies increase $\sim$20 points between $z=$0.02 and $z=$0.3 at the
       99.9\% significance level, confirming
 the presence of the Butcher-Oemler effect. 
 The results show that cluster galaxies do
 evolve by changing colors redder with decreasing redshifts.

 \item 
   We observed the morphological Butcher-Oemler effect as an
 increase of late-type galaxies toward higher redshifts, using pure morphological
 parameters such as a concentration parameter and de
 Vaucouleur/exponential profile fit. 
 The results confirm that cluster galaxies do
 evolve morphologically from late- to early-type, 
 in addition to the spectral evolution (the Butcher-Oemler effect). 
 Previously, such a study on morphological evolution of cluster galaxies has often
       been performed with a heterogeneous sample of low redshift and
       high redshift clusters  with eye-based morphology.
  This is the first result to show such morphological evolution with an
       automated morphology using a homogeneous sample of 514 clusters.

 \item  We found a slight tendency for richer clusters to have lower values of
    the late-type fraction (Figure \ref{fig:okamura_rich}).  This trend
       has significant implication for the underlying physical mechanism
       since it is expected in the ram-pressure stripping model proposed by Bahcall (1977)
    and Fujita et al. (1999) where
  galaxies in richer clusters are more affected by ram-pressure due to
    higher temperature of clusters.

 \item   We studied the morphology-density relation and the
 morphology-cluster-centric-radius relation using a volume limited SDSS
 data (0.05$<z<$0.1, $Mr^*<-$20.5).   
 Technical
 improvements compared with previous work are; (i) automated galaxy
 morphology classification capable to separate galaxies into four types, (ii) three
 dimensional local galaxy density 
 estimation free from the fore/background correction, (iii) the extension of the morphology-density study into the field region. 
 We found that there are two characteristic changes in both the
 morphology-density and the morphology-radius relations, suggesting two
 different mechanisms are responsible for the relations.
  In the sparsest regions (below 2 galaxy Mpc$^{-2}$ or outside of 2 $R_{vir}$), both of the relations  become flat, suggesting the
 responsible physical mechanisms require denser environment. 
    In the intermediate density regions, (density between 2 and 6
 galaxy Mpc$^{-2}$ or between 0.3 and 2 $R_{vir}$), S0 fractions increase
 toward denser regions, whereas late-spiral fractions
 decrease. 
Considering that the median size of S0 galaxies are smaller than
 that of late-spiral galaxies (Figure \ref{fig:md_size}), we propose that 
  the mechanism is likely to stop star formation in
 late-spiral galaxies, eventually turning them into S0 galaxies after their
 outer discs and spiral arms become 
 invisible as young stars die. For example, ram-pressure stripping is
 one of the candidate mechanisms. 
    In the densest regions (above 6 galaxy Mpc$^{-2}$ or inside of
 0.3 $R_{vir}$), S0 fractions decreases radically and elliptical
 fractions increase. The behavior of S0 fractions at the cluster cores
       is investigated for the first time in this work.
 The result in the core regions is in contrast to that in
 intermediate regions, and it suggests that yet another mechanism might be
 responsible for morphological change in these regions.
 The deficit of S0 galaxies at the
 densest regions are likely to be consistent with computer simulations
 based on merging scenario, which predicted 
 small fraction of S0 galaxies.

 \item 
 We  compared the morphology-density relation from the SDSS
 (0.01$<z<$0.054) with that of the MORPHS data ($z\sim$0.5). Two
 relations lie on top of each other, suggesting that the
 morphology-density relation was already established at $z\sim$0.5 as it
       is
 in the present universe. A slight sign of excess early-type fraction in
 the SDSS data in dense regions might be suggesting additional formation
 of elliptical 
 galaxies in the cluster core region between $z=$0.5 and $z=$0.05. 
  This work presented the morphology-density relation at $z\sim$0.5 with an
       automated morphology instead of eye-morphology for the first time.

 \item Using a volume limited sample of the SDSS data (0.05$<z<$0.1
       and $Mr^*<-$20.5), we studied the 
 environment of passive spiral galaxies as a function of local galaxy
 density and cluster-centric-radius. This is the first work that
       revealed the environment of passive spiral galaxies.
 It is found that passive
 spiral galaxies preferentially live in local galaxy density 1$\sim$2 galaxy Mpc$^{-2}$ and
 1$\sim$10 $R_{vir}$ (Figures \ref{fig:ps_density} and \ref{fig:ps_radius}). Thus
       the origin of passive spiral galaxies 
 is likely to be cluster related.  These characteristic environments
 coincide with the previously reported environment where galaxy star
 formation rate suddenly declines and the 
 morphology-density relation levels off (Figure \ref{fig:md_mr}).  
 In order to create passive spirals, 
 a physical mechanism that works calmly is
 preferred to dynamical origins such as major merger/interaction since such a
 mechanism can destroy spiral arm structures. 
  Compared with observed cluster galaxy evolution such as the
 Butcher-Oemler effect and the morphological Butcher-Oemler effect (Figure \ref{fig:bo}),
 passive spiral galaxies are likely to be a galaxy population in
 transition between red, elliptical/S0 galaxies in low redshift clusters
 and blue, spiral galaxies more numerous in higher redshift clusters.

 \item Based on our observational results, we propose a new picture
       of the fate of cluster infalling galaxies as follows. When a star forming
       galaxy falls into a cluster region, the galaxy is not affected by
       the cluster environment until it reaches  2 $R_{vir}$ or
       galaxy density $\sim$2 galaxy Mpc$^{-2}$
       (Figure \ref{fig:md_md_ann_ytype} and \ref{fig:md_mr}).  In the
       intermediate regions (1$\sim$2 galaxy
       Mpc$^{-2}$ or 1$\sim$10 $R_{vir}$;
       Figure \ref{fig:md_mr}),  the galaxy reduces its
       star formation rate through a gentle cluster-related process such as
       ram-pressure stripping, strangulation, evaporation or minor merger. Perhaps passive
       spiral galaxies (Figure \ref{fig:ps_image}) are in the intermediate
       stage of this transition, 
       which eventually become numerous S0 galaxies (Figure \ref{fig:md_mr})
       when outer discs become too faint to be visible. 
       In the cluster core regions (above 6 galaxy Mpc$^{-2}$ or inside of
 0.3 $R_{vir}$), bright elliptical galaxies become dominant
       (Figure \ref{fig:md_mr_es0}), perhaps originating from the
       merging/interaction at high redshift.

\end{itemize}

\section{Circumstantial Evidence of Cluster Galaxy Evolution}  
  In Chapter \ref{chap:LF} and Chapter \ref{chap:BO}, we found
  circumstantial evidence of cluster galaxy evolution.
  We presented composite LFs of
 cluster galaxies using one of the largest samples consisting of 204
  clusters in Chapter \ref{chap:LF}. 
 The results showed that the bright end of the cluster LFs
 is dominated by bright, old  early-type galaxies, while the faint-end of the
 cluster LF represents late-type galaxies. 
 Compared with the field LFs (Figure \ref{fig:all.eps}), the dominance of bright elliptical
  galaxies in clusters suggests that these cluster galaxies might have
 a different evolutionary history from field galaxies.
   In Chapter \ref{chap:BO}, we showed that the
 Butcher-Oemler effect exists in the SDSS clusters, in a sense that the
 fraction of blue galaxies are larger in higher redshifts. According to
 Figure \ref{fig:bo}, $\sim$20\% of galaxies became red since $z=$0.3. 
    Historically, there have been some doubts on the existence of the
 Butcher-Oemler effect (Newberry, Kirshner \& Boroson 1988;
 Allington-Smith et al. 1993; Garilli et al. 1996; Smail et
 al. 1998; Andreon \& Ettori 1999;   Fairley et al. 2002) although many
 people, on the other hand, claimed the existence of the Butcher-Oemler
 effect  (Butcher \& Oemler 1978, 1984; 
 Rakos \& Schombert 1995; Margoniner \& De Carvalho 2000; Margoniner et
 al. 2001). The previous work used only dozens of clusters, and
 thus had large statistical uncertainty, which surely was part of the reason
 why different people ended up with different answers. In contrast, our
 work used by far the largest sample of 514 clusters uniformly selected from
 the SDSS CE galaxy cluster catalog (Chapter \ref{chap:CE}). It is also
 of importance that low redshift clusters and high redshift clusters are
 selected from the same SDSS data using the same cluster finder (Chapter
 \ref{chap:CE}) in this work, rectifying the inhomogeneity of the
 previous samples.  
 Using the homogeneous sample, we firmly confirmed the existence of the
 Butcher-Oemler effect with negligible statistical uncertainty
 (Figure \ref{fig:bo}); the significance in correlation
 coefficient is greater than 99.99\%. 
 The result  is a direct evidence that cluster galaxies
 evolve by changing their colors from blue to red.
%

   In addition to the evolution in color, we showed that fractions of
  morphologically spiral galaxies are also 
  larger in higher redshift (the morphological Butcher-Oemler
 effect).
 This is the first detection of this morphological transition
 using an automated galaxy classification and a homogeneous sample
  (Figure \ref{fig:bo}).  Previous work used
 eye-classification and small number of clusters (a few to a few dozen) to probe morphological
  evolution, and thus had substantial uncertainty (Dressler et
  al. 1997; Fasano et al. 2000).  
 In addition to the spectral evolution of cluster galaxies (the
  Butcher-Oemler effect), our result
  implies that cluster galaxies also 
 change their morphology from 
 late to early type, possibly from spiral to S0 galaxies.  Figure
  \ref{fig:bo} shows more than 20\% of galaxies change 
  their morphology from less concentrated to concentrated between $z=$0.3
  and $z=$0. 

   Another important discovery in Chapter  \ref{chap:BO} is
 that, as a second parameter, blue/spiral fractions of cluster galaxies
 depend on the cluster richness, in the sense that richer clusters have
 smaller fractions of  blue/spiral galaxies
  (Figure \ref{fig:okamura_rich}). The result has significant
  implication for the underlying physical mechanism since 
 it is  expected in the ram-pressure stripping model by Fujita et
 al. (1999), where 
 ram-pressure stripping is stronger in clusters of higher X-ray temperatures.

\section{Morphological Evolution of Cluster Galaxies}
   To investigate these cluster galaxy evolution in more detail, we
 studied nearby galaxies with the SDSS spectroscopic information in
 Chapters \ref{chap:MD} and \ref{chap:PS}.
  We showed that there are two
 characteristic changes in the morphology-density relation using the
 volume limited sample (0.05$<z<$0.1 and  $Mr^*<-$20.5) consisting of
 the 7938 SDSS galaxies (Figure \ref{fig:md_mr}). 
              
         In the sparsest regions where the local galaxy density is less
 than 2 galaxy 
 Mpc$^{-2}$, or outside of 2 virial radius ($R_{vir}$), the
 morphology-density relation becomes less noticeable 
 (Figure \ref{fig:md_md_ann_ytype} and  Figure \ref{fig:md_mr}),
 suggesting that
 physical mechanisms responsible for the morphology-density relation
 do not work in the sparsest regions.  
   The characteristic density or radius where the morphology-density
 relation becomes flat (2 galaxy Mpc$^{-2}$ or 2 $R_{vir}$)
 coincides with the environment where the density-star formation rate (SFR)
 relation becomes flat. Lewis et al. (2002) and  Gomez et al. (2003)
 studied the correlation between local galaxy density and galaxy SFR to
 find galaxy SFR suddenly decreases at  
 around the same galaxy density we found.
 This coincidence suggests that  the same physical mechanism
 might be responsible for both the morphology-density relation and the SFR-density relation. 
              
      In the intermediate density regions (density between 2 and 6
 galaxy Mpc$^{-2}$ or  between 0.3 and 2 $R_{vir}$), we showed that S0
 fractions increase 
 toward denser regions, whereas late-spiral fractions
 decrease (Figures \ref{fig:md_md_ann_ytype} and \ref{fig:md_mr}). In this
 intermediate regions, it is found that the SFR of 
 galaxies starts to decrease toward denser environments  (Lewis et
 al. 2002; Gomez et al. 2003). Furthermore, in Chapter \ref{chap:PS}, we
 found that passive spiral  galaxies preferentially live in the same
 intermediate 
 regions (local galaxy density 1$\sim$2 galaxy Mpc$^{-2}$ or 1$\sim$10
 $R_{vir}$; Figures \ref{fig:ps_density} and \ref{fig:ps_radius}).
 Therefore, it is likely that the same physical mechanism is
 responsible for all of these observational phenomena (the
 morphology-density relation, the SFR-density relation and the creation
 of passive spirals) happening around
 this intermediate density regions, or in other words, cluster infalling
 regions.   

   In the densest regions (above 6 galaxy Mpc$^{-2}$ or inside of
 0.3 $R_{vir}$), S0 fractions decreases radically and elliptical
 fractions increase (Figure \ref{fig:md_mr_es0}). This is a contrasting
 result to that in 
 intermediate regions and it suggests that yet another mechanism might be
 responsible for morphological change in these regions. 

  Although the morphology-density relation has been studied by many authors
  in the past, it has been difficult to relate the correlation to underlying
  physical mechanisms. Benefitting from the large and high quality data
  from the SDSS,
 it has become possible,  for the first time,  to specify
  two characteristic environments where galaxy properties start to
  change, providing us with a great hint on the underlying physical mechanism.

\section{Possible Hypotheses}
 Considering all of these observational findings,  we would like to
 summarize a possible hypothesis which can 
 explain all of the above results. 
 From the results in
 Chapter \ref{chap:BO}, it is very likely that cluster
 galaxies change their morphology and colors during the course of their
 evolution. 

   In the infalling regions of clusters (below 2 galaxy Mpc$^{-2}$ or outside of 2 $R_{vir}$), fractions of late-type galaxies decrease and fractions of S0
 galaxies increase (Figure \ref{fig:md_md_ann_ytype} and
 Figure \ref{fig:md_mr}). Galaxies also decrease their SFR 
 abruptly in these regions (Lewis et al. 2002; Gomez et al. 2003). 
  It is also likely that the change in galaxy morphology and galaxy SFR in this intermediate
 density region is mainly responsible for the spectral and morphological
 evolution of cluster galaxies as shown in Figure \ref{fig:bo}. If a
 star-forming, spiral galaxy infalling to a cluster at high redshift is
 affected by 
 cluster environment at this intermediate regions as seen in Figures
 \ref{fig:md_md_ann_ytype}, \ref{fig:md_mr}, \ref{fig:ps_density} and
 \ref{fig:ps_radius},
 then, they are likely to become a red, elliptical-like galaxy at
 lower redshift. If 20$\sim$30 \% of cluster galaxies experience this
 process, the observed decrease of blue, spiral galaxies toward lower
 redshift (Figure \ref{fig:bo}) can naturally be explained.

\begin{figure}[h]
\includegraphics[scale=0.8,angle=0]{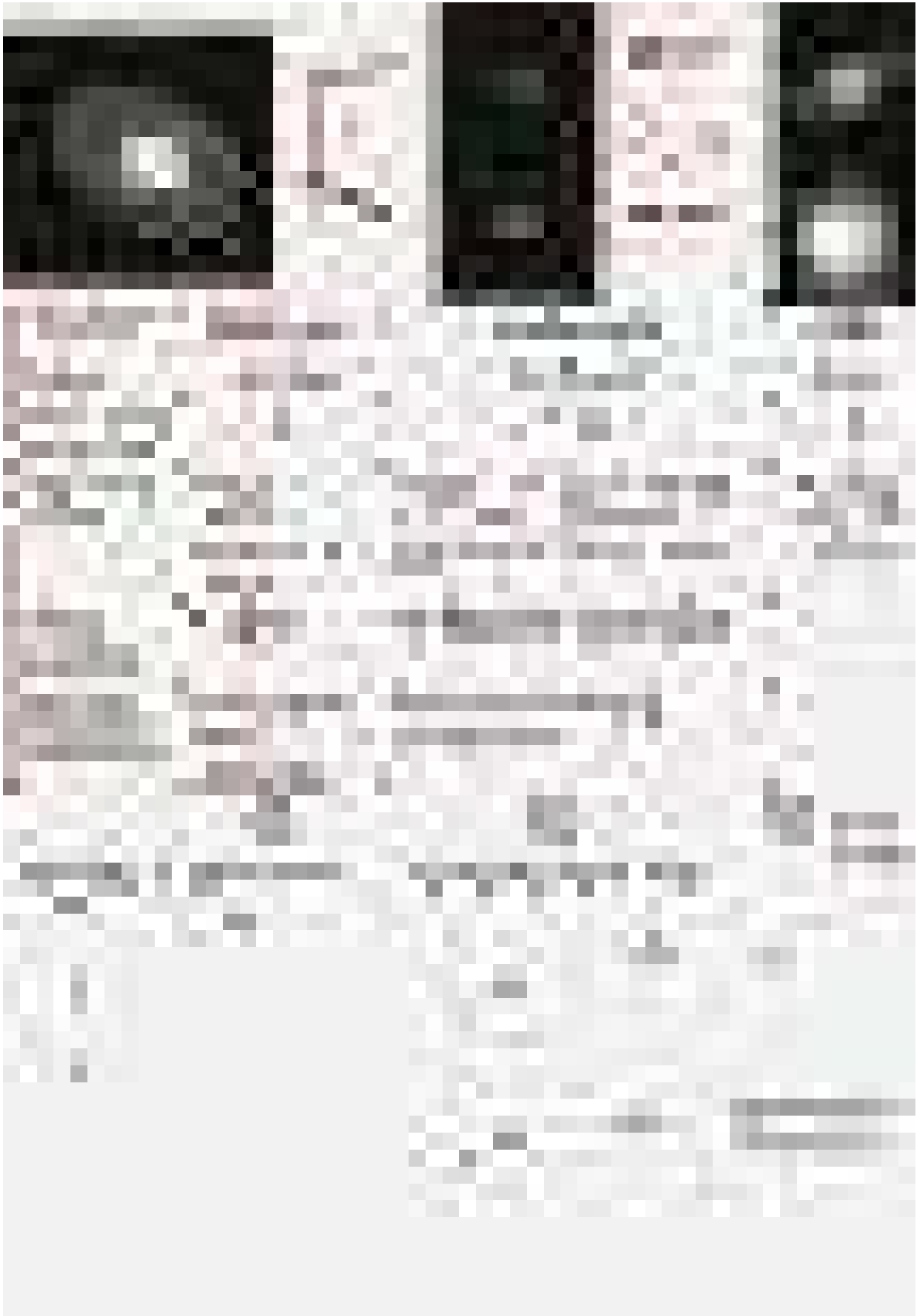}
\caption{
 A schematic illustration of the evolution of galaxies due to the
 cluster environment.
 }\label{fig:schematic}
\end{figure}

  Figure  \ref{fig:schematic} is a schematic illustration of the
 cluster-related environmental 
 effects on infalling galaxies.
 Possible physical
 mechanisms happening in the 
 regions include ram-pressure stripping, strangulation, galaxy mergers
 and evaporation. However, since this is still a relatively low density
 regions (local galaxy density $<$ 2 Mpc$^{-2}$), it is difficult for
 major mergers to take place frequently (Kodama et al. 2001; Treu et
 al. 2003). In addition, since major merger destroys spiral arms, it
 can not explain the existence of the passive spirals in this
 intermediate regions. 
  From the same reason, this
 cluster infalling regions are still too low gas density for ram-pressure
 stripping, strangulation or evaporation (interaction with intra-cluster
 medium) to be effective  (Balogh et
 al. 1997; Lewis et al. 2002). However, this 
 is the regions where overdensity of a few to a few dozen galaxies
 (cluster sub-clumps) can often be found. (e.g., Kodama et 
 al. 2001). These cluster sub-clumps might be a group of galaxies
 infalling into a cluster.
  Fujita et al. (2003) showed that in these cluster sub-clump regions,
 stripping can take place effectively, mainly due to higher mass density
 of the gas in
 sub-clumps even though their absolute mass is much smaller than clusters. 
 In fact, Kodama et al. (2001) found that $V-I$ colors of galaxies
 abruptly become redder 
 around these cluster sub-clump regions.
 If ram-pressure stripping
 (or strangulation), is responsible for galaxy evolution in these regions, it
 can also explain other observational results. Since it does not
 disturb spiral structures of disc galaxies, it can naturally create
 passive spiral galaxies (Figure \ref{fig:ps_image}; Chapter
 \ref{chap:PS}). Since 
 stripping is more effective in clusters 
 with higher X-ray temperature, it can explain richness dependence of
 blue/late type fractions (Figure \ref{fig:bo}; Chapter \ref{chap:BO}). 
 Stripped galaxies should be smaller than their progenitors,  just like 
 S0 galaxies were smaller than Sc galaxies in
 Figure \ref{fig:md_size}. Therefore, we propose that
 stripping (including strangulation) in sub-clump regions might be the physical mechanism happening in
 these regions although it is also difficult to exclude other possibilities.

  In the very cores of clusters (local galaxy density $\sim$6
 galaxy Mpc$^{-2}$ or within 0.3 $R_{vir}$), we observed that S0 fractions
 decrease, and in turn, 
 elliptical fractions radically increase  (Figure \ref{fig:md_mr}; Chapter
 \ref{chap:MD}). We also observed that passive spiral galaxies do not
 exist in cluster cores (Figure \ref{fig:ps_radius}; Chapter \ref{chap:PS}). Possible mechanism
 happening in these cluster cores must be the one that reduces passive
 spirals and increase elliptical fractions. Major galaxy mergers seem to
 be able to increase elliptical fractions by merging two or more S0
 galaxies into one giant elliptical galaxy. However, since the relative
 velocities between each galaxy is 
 high in cluster cores, it is difficult for gravitational interactions
 to happen frequently (Ostriker 1980; Binney  \& Tremaine 1987; Mamon
 1992; Makino \& Hut 1997). There have been several observational results
 reporting that cluster ellipticals have been in the core for a long time
 ($>$5 Gyr; van Dokkum et al. 1998). Therefore, cluster ellipticals might
 have been created in the cluster core regions through merging when a cluster
 itself was much younger ($z>$1). Since these giant ellipticals in
 cluster cores are numerous, the observational decrease of passive
 spirals and S0s at cluster cores  can be also explained 
 if S0s and passive
 spirals which stopped star formation in the intermediate regions
 become faint by the time they reach cluster cores, and drop out of the
 sample to increase the 
 dominance of giant elliptical galaxies.

\chapter{Conclusions}
\label{chap:summary}

 We have studied the environmental effects on galaxy evolution using the
 Sloan Digital Sky Survey (SDSS) data. Our main findings are as follows.

   We have developed a new cluster finding method, the Cut \&
       Enhance (CE) method,  which uses color and angular separation to
       create a uniform cluster catalog with high detection sensitivity.
     We apply Cut \& Enhance method to the SDSS commissioning data and
 produced an SDSS Cut \& Enhance cluster catalog containing 4638 galaxy
       cluster candidates in $\sim$350 deg$^2$. 

 Using this cluster catalog,  we constructed composite luminosity
 functions (LFs) of 204 SDSS CE clusters ranging from
 $z=$0.02 to $z=$0.25.  
  Compared with the field luminosity function, cluster LFs have a
 brighter characteristic magnitude ($M^*$) and a flatter slope  in the
 $g, r, i$ and $z$ band. We also found that early-type galaxies always
 have flatter slopes than  late-type galaxies in the clusters. 

 We investigate the evolution of the
 fractions of blue 
 cluster galaxies as a function of redshift, using 514 CE clusters in
 the range of  0.02$\leq z\leq$0.3.  
 This evolution has been investigated for
 the first time without any significant systematic bias  benefitting
 from  our
 large, uniform  cluster catalog. 
 By selecting blue galaxies as those with restframe $g-r$ bluer by 0.2 than
       red-sequence or those with $u-r<$2.2,
 we found that blue fractions of cluster
       galaxies increase $\sim$20 points between $z=$0.02 and $z=$0.3 at the
       99.9\% significance level, confirming
 the presence of the Butcher-Oemler effect.

   We observed the morphological Butcher-Oemler effect as an
 increase of late-type galaxies toward higher redshifts, using pure
 morphological parameters such as a concentration parameter and de
 Vaucouleur/exponential profile fit. In addition, we found a slight
 tendency for richer clusters to have lower values of  
    the late-type fraction (Figure \ref{fig:okamura_rich}).  
 
   We studied the morphology-density relation and the
 morphology-cluster-centric-radius relation using a volume limited SDSS
 data (0.05$<z<$0.1, $Mr^*<-$20.5).  Our results are based on the
 morphological classifier, $Tauto$, which correlates with
 eye-classified galaxy morphology.  If the galaxy classification with
 $Tauto$ corresponds to the physical galaxy classification, the
 following interpretation become possible.
 We found that there are two characteristic changes in both the
 morphology-density and the morphology-radius relations, suggesting two
 different mechanisms are responsible for the relations.
  In the sparsest regions (below 2 galaxy Mpc$^{-2}$ or outside of 2 $R_{vir}$), both of the relations  become flat, suggesting the
 responsible physical mechanisms require denser environment. 
    In the intermediate density regions, (density between 2 and 6
 galaxy Mpc$^{-2}$ or between 0.3 and 2 $R_{vir}$), intermediate
 fractions ($-0.8\leq Tauto<$0.1) increase 
 toward denser regions, whereas late-spiral fractions (1.0$\leq Tauto$)
 decrease. 
    In the densest regions (above 6 galaxy Mpc$^{-2}$ or inside of
 0.3 $R_{vir}$), intermediate fractions decreases radically and early-type
 fractions ($Tauto \leq -0.8$) increase.  

 We  compared the morphology-density relation from the SDSS
 (0.01$<z<$0.054) with that of the MORPHS data ($z\sim$0.5). Two
 relations lie on top of each other, suggesting that the
 morphology-density relation was already established at $z\sim$0.5 as it
       is in the present universe.

 We studied the  environment of passive spiral galaxies as a function of
 local galaxy 
 density and cluster-centric-radius. 
 It is found that passive
 spiral galaxies preferentially live in local galaxy density 1$\sim$2
 galaxy Mpc$^{-2}$ and 
 1$\sim$10 $R_{vir}$ (Figures \ref{fig:ps_density} and
 \ref{fig:ps_radius}). 

 Throughout the work, we have revealed the characteristic environments
 where cluster galaxies 
 evolve. These environments provide strong constraints in specifying the
 underlying 
 physical mechanisms that govern cluster galaxy evolution.


\appendix
\chapter{A Catalog of H$\delta$-strong Galaxies}\label{EA1}

\section{Introduction}\label{ea1_intro}

 Presence of a strong \hd\, absorption line (equivalent width of $>$
5\AA) in the spectrum of a galaxy is an indication that the spectral energy
distribution of that galaxy is dominated by A stars. Models of galaxy
evolution indicate that such a strong \hd\, line (in the spectrum of a galaxy)
can only be reproduced using models that include a recent burst of star
formation, followed by passive evolution, because any on--going star--formation in
the galaxy would hide the \hd\, absorption line due to emission--filling (of
the \hd\, line) and the dominance of hot O and B stars, which have
intrinsically weaker \hd\, absorption than A stars (see, for example, Balogh
et al. 1999; Poggianti et al. 1999). Therefore, the existence of a strong \hd\,
absorption line in the spectrum of a galaxy suggests that the galaxy has
undergone a recent transformation in its star--formation history. In the
literature, such galaxies are called ``post--starburst'', ``E+A'', or
H$\delta$--strong galaxies. Exact physical mechanism(s) responsible for
the abrupt change in the star formation history of such galaxies remains
unclear. These galaxies have received much attention as they provide an
opportunity to study galaxy evolution {\it ``in action''}.

\hd--strong galaxies were first discovered by Dressler \& Gunn (1983,
1992) in their spectroscopic study of galaxies in distant, rich clusters of
galaxies.  They discovered cluster galaxies that contained strong Balmer
absorption lines but with no detectable \oii\, emission lines. They named such
galaxies ``E+A'', as their spectra resembled the superposition of an
elliptical galaxy spectrum and A star spectrum. Therefore, E+A galaxies were
originally thought to be a cluster--specific phenomenon and several physical
mechanisms have been proposed to explain such galaxies.  For example,
ram--pressure stripping of the interstellar gas by a hot, intra--cluster
medium, which eventually leads to the termination of star formation once all
the gas in the galaxy has been removed, or used up for star formation (Gunn \& Gott 1972; Farouki
\& Shapiro 1980; Kent 1981; Abadi, Moore \& Bower 1999; Fujita \& Nagashima
1999; Quilis, Moore \& Bower 2000). Alternative mechanisms include high--speed
galaxy--galaxy interactions in clusters (Moore et al. 1996, 1999) and
interactions with the gravitational potential well of the cluster (Byrd \&
Valtonen 1990; Valluri 1993; Bekki, Shioya \& Couch 2001).

To test such hypotheses, Zabludoff et al. (1996) performed a search for
E+A galaxies in the Las Campanas Redshift Survey
(LCRS; Shectman et al. 1996) and found that only 21 of the 11113 LCRS galaxies
they studied satisfied their criteria for a E+A galaxy. This work
clearly demonstrates the rarity of such galaxies at low redshift. Furthermore,
Zabludoff et al. (1996) found that 75\% of their selected galaxies
reside in the field, rather than the cores of rich clusters. This conclusion
was confirmed by Balogh et al. (1999), who also performed a search for
\hd--strong galaxies in the redshift surveys of the Canadian Network for
Observational Cosmology (CNOC; Yee, Ellingson, \& Carlberg 1996), and found
 that the fraction of such galaxies in clusters was similar to
that in the field. Alternatively, the study of Dressler et al. (1999) found an
order--of--magnitude increase in the abundance of E+A galaxies in
distant clusters compared to the field (see also Castander et al. 2001).
Taken together, these studies suggest that the physical interpretation of
\hd--strong galaxies is more complicated than originally envisaged, with
the possibility that different physical mechanisms are important in different
environment, e.g., 5 of the 21 E+A galaxies discovered by
Zabludoff et al. (1996) show signs of tidal features, indicative of
galaxy--galaxy interactions or mergers. Furthermore, redshift evolution might
be an important factor in the differences seen between these surveys.

In addition to studying the environment of \hd--strong galaxies, several
authors have focused on understanding the morphology and dust content of these
galaxies. This has been driven by the fact that on--going star formation in
post--starburst galaxies could be hidden by dust obscuration (See Poggianti \&
Barbaro 1997; Poggianti et al. 1999; Bekki et al. 2001 for more
discussion). In fact, Smail et al (1999) discovered examples of such galaxies
using infrared (IR) and radio observations of galaxies in distant
clusters. They discovered five post--starburst galaxies (based on their
optical spectra) that showed evidence for dust--lanes in their IR
 morphology 
 as well as radio emission consistent with on--going star formation. However,
radio observations of the Zabludoff et al. (1996) sample of nearby
 E+A galaxies indicates that the majority of these galaxies are not
dust--enshrouded starburst galaxies. For example, Miller \& Owen (2001) only
detected radio emission from 2 of the 15 E+A galaxies they
observed, and the derived star--formation rates (SFRs) were consistent with
quiescent star formation and thus much lower than those observed for the
dust--enshrouded starburst galaxies of Smail et al (1999). Chang et al. (2001)
also did not detect radio emission from any of the 5 E+A galaxies they
observed from the Zabludoff et al. (1996) sample and concluded that these
galaxies were not dust--enshrouded starbursts. In summary, these studies
demonstrate that some E+A galaxies have dust--enshrouded star
formation, but the fraction remains ill--determined. Furthermore, it is
unclear how the different sub--classes discussed in the literature are related, and if there
are any environmental and evolutionary processes in play.

The interpretation of \hd--strong galaxies (E+A) suffers from
small number statistics and systematic differences in the selection and
definition of such galaxies among the different surveys constructed to date.
Therefore, many of the difficulties associated with understanding the physical
nature of these galaxies could be solved through the study of a large,
homogeneous sample of \hd\, galaxies. In this Chapter, we present such a sample
derived from the Sloan Digital Sky Survey (SDSS; York et al. 2000). The
advantage of this sample, over previous work, is the quality and quantity of
both the photometric and spectroscopic data, as well as the homogeneous
selection of SDSS galaxies which covers a wide range of local environments.

 We present in this Chapter a sample of galaxies that have been selected based
 solely on the observed strength of their \hd\, absorption line. Our selection
 is thus inclusive, containing many of the sub--classes of galaxies discussed
 in the literature until now, e.g., ``E+A'' galaxies
 (Zabludoff et al. 1996; Dressler et al. 1999), post--starburst galaxies,
 dust--enshrouded starburst galaxies (Smail et al. 1999), \hd--strong galaxies
 (Couch \& Sharples 1987) and the different subsamples of galaxies, i.e.,
 e(a) and  A+em), discussed by Poggianti et al. (1999) and Balogh et
 al. (1999). Following Couch \& Sharples (1987) and Balogh et
 al. (1999), we call our sample of SDSS galaxies as  ``\hd--strong''
 (HDS) galaxies.

In this Chapter, we present the details of our selection and leave the
investigation and interpretation of these HDS galaxies to subsequent papers.
 We publish our sample of HDS galaxies to help the community construct larger
 samples of such galaxies, which are critically needed to advance our
 understanding of these galaxies, as well as to promote the planning of
 follow--up observations and comparisons with studies of such
 galaxies at higher redshifts.

In Section \ref{ea1_data}, we present a brief discussion of the SDSS and the data
used in this Chapter. In Section \ref{ea1_line}, we discuss our techniques for
measuring the \hd\, absorption line and present comparisons between the
different methodologies used to measure this line. In Section \ref{ea1_catalog},
we discuss the criteria used to select of our HDS sample of galaxies and
present data on 3340 such galaxies in our catalog.  In Section
\ref{ea1_discussion}, we compare our sample of galaxies with those in the
literature. A more detailed analysis of the properties of our HDS galaxies
will be discussed in subsequent papers. The cosmological parameters used
throughout this Chapter are ${\rm H_0}$=75 km s$^{-1}$ Mpc$^{-1}$,
$\Omega_m=0.3$ and $\Omega_{\Lambda}=0.7$.

\section{The SDSS Data}
\label{ea1_data}

In this Section, we briefly describe the spectroscopic part of the SDSS. As
discussed in York et al. (2000), the SDSS plans to obtain spectra for
$\simeq10^6$ galaxies to a magnitude limit of $r^*=17.7$ (the ``Main'' galaxy
sample; Strauss et al. 2002), $\simeq10^5$ Luminous Red Galaxies (LRG;
Eisenstein et al. 2001) and $\simeq10^5$ quasars (Richards et al. 2002).  The
reader is referred to Fukugita et al. (1996), Gunn et al. (1998), Lupton et
al. (1999, 2001), York et al. (2000), Hogg et al. (2001), Pier et al. (2002),
Stoughton et al. (2002), Smith et al. (2002) and Blanton et al. (2002a) for
more details of the SDSS and its data.

The SDSS spectra are obtained using two fiber-fed spectrographs (each with 320
fibers), with each fiber subtending 3 arcseconds on the sky. The wavelength
coverage of the spectrographs is 3800\AA\ to 9200\AA, with a spectral
resolution of 1800. The data from these spectrographs is automatically reduced
to produce flux and wavelength--calibrated spectra (SPECTRO2D data analysis
pipeline).

The SDSS spectra are then analyzed via the SDSS SPECTRO1D data processing
pipeline to obtain a host of measured quantities for each spectrum (see
Stoughton et al. 2002; Frieman et al., in prep, for further details). For
example, SPECTRO1D determines the redshift of the spectrum both from
absorption lines (via cross-correlation; Heavens 1993), and emission lines
(via a wavelet--based peak--finding algorithm; see Frieman et al., in prep).
Once the redshift is known, SPECTRO1D estimates the continuum emission at each
pixel using the median value seen in a sliding box of 100 pixels centered on
that pixel.  Emission and absorption lines are then measured automatically by
fitting of a Gaussian, above the best--fit continuum, at the redshifted
rest--wavelength of expected lines. Multiple Gaussians are fit simultaneously
for potential blends of lines, i.e., \ha\, and \nii\, lines). SPECTRO1D
therefore provides an estimate of the equivalent width (EW), continuum, rest
wavelength, identification, goodness--of--fit ($\chi^2$), height and sigma
(and the associated statistical errors on these quantities) for all the major
emission/absorption lines in these spectra. These measurements are done
regardless of whether the line has been detected or not.  For this work, we
have used data from rerun 15 of the SPECTRO1D analysis pipeline, which is
based on version 4.9 of the SPECTRO2D analysis pipeline (see Frieman et al. in
prep for details of these pipelines).

 In order to construct the sample of HDS galaxies presented in this Chapter, we begin with a sample
of SDSS galaxies that satisfy the following selection criteria:
\begin{enumerate}
\item{Spectroscopically--confirmed by SPECTRO1D to be a galaxy;}
\item{Possess a redshift confidence of $\geq0.7$;}
\item{An average spectroscopic signal--to--noise ratio of $>5$ per pixel in
     the SDSS photometric $g$ passband;}
 \item {$z \geq$0.05, to minimize aperture effects as discussed in Zaritsky, Zabludoff \& Willick (1995) and Gomez et al. (2003).}
\end{enumerate}

The reader is referred to Stoughton et al. (2002) for further details on all
these SDSS quantities and how they are determined.  After removing duplicate
observations of the same galaxy (11538 galaxies in total; see Section
\ref{ea1_error}), 106682 galaxies satisfy these criteria, up to and including
spectroscopic plate 804 (observed on a Modified Julian Date of 52266 or
12/23/01; see Stoughton et al. 2002). Of these 106682 galaxies, it was 
 possible to measure the \hd\, line for only 95479 galaxies (see Section
\ref{ea1_Hdelta} below) due to masked pixels at or near the \hd\, line. In Figure
\ref{fig_ea1:sng}, we present the distribution of signal--to--noise ratios for all
106682 spectra (the median value of this distribution is 8.3).

Throughout this analysis, we use the ``smeared'' SDSS spectra, which
improves the overall spectrophotometric calibration of these data by
accounting for light missed from the 3'' fibers due to seeing and
atmospheric refraction (see Gomez et al. 2003; Stoughton et al. 2002 for
observational detail). Unfortunately, this smearing correction can
systematically bias the observed flux of any emission and absorption lines in
the spectrum, as the correction is only applied to the continuum. As shown by
Hopkins et al. in prep., however, this is only a $\simeq10\%$ effect on the flux of
spectral lines, compared to using spectral data without the smearing
correction applied. Furthermore, the equivalent width of our lines is almost
unaffected by this smearing correction as, by definition, they are computed
relative to the height of the continuum.

\section{Spectral Line Measurements}\label{ea1_line}
\subsection{H$\delta$\ Equivalent Width}\label{ea1_Hdelta}

In this Section, we discuss the measurement of the equivalent width (EW) of
the \hd\, absorption line in the SDSS galaxy spectra described in Section
\ref{ea1_data}. The presence of a strong \hd\, absorption line in a galaxy
spectrum indicates that the stellar population of the galaxy contains a
significant fraction of A stars, which must have formed within the last
Gigayear (see Section \ref{ea1_intro}).  The \hd\, line is preferred to other
Hydrogen Balmer lines (e.g., H$\epsilon$, H$\zeta$, H$\gamma$, H$\beta$)
because the line is isolated from other emission and absorption lines, as well
as strong continuum features in the galaxy spectrum (e.g.,
D4000). Furthermore, the higher order Balmer lines (H$\gamma$ and H$\beta$)
can suffer from significant emission--filling (see Section \ref{ea1_emfilling}),
while the lower order lines (H$\epsilon$ and H$\zeta$) are low
signal--to--noise in the SDSS spectra.

In previous studies, several different methods have been employed to measure
the \hd\, line, or select post--starburst galaxies. For example, Zabludoff et
al. (1996) used the average EW of the \hb\ , H$\gamma$ and H$\delta$ lines to
select E+A galaxies. Alternatively, Dressler et al. (1999) and Poggianti et
al. (1999) interactively fit Gaussian profiles to the H$\delta$ line. Finally,
Abraham et al. (1996b), Balogh et al. (1999) and Miller \& Owen (2002)
performed a non--parametric analysis of their galaxy spectra, which involved
summing the flux in a narrow wavelength window centered on the \hd\ line
to determine the EW of the line. Castander et al. (2001) used an 
innovative PCA and wavelet analysis of spectra to select E+A galaxies.  Each
of these methods have different advantages and disadvantages. For example,
fitting a Gaussian to the \hd\, line is optimal for high signal--to--noise
spectra, but can be prone to erroneous results when fit blindly to low
signal--to--noise data or to weak absorption lines (such problems can be
avoided if Gaussians are fit interactively; see Dressler et al. 1999;
Poggianti et al. 1999).  In light of the potential systematic differences
 among the different methods of measuring the \hd\, line, we have
investigated the relative merits of the two main approaches in the literature
-- fitting a Gaussian and summing the flux in a narrow wavelength window --
for determining the EW of the \hd\, line for the signal--to--noise ratio,
resolution, and size of the SDSS spectral dataset used in this Chapter.

First, we investigate the optimal method for computing the EW of the \hd\,
line from the SDSS spectra using the non--parametric methodology outlined in
Abraham et al. (1996b) and Balogh et al. (1999), i.e., summing the flux
within narrow wavelength windows centered on and off the \hd\ absorption
line. We estimate the continuum flux via linear interpolation between two
wavelength windows placed at either side of the \hd\, line (4030\AA\ to 4082\AA\
and 4122\AA\ to 4170\AA). We used the same wavelength windows as in Abraham et
al. (1996b) and Balogh et al. (1999) for estimating the continuum because they
are devoid of any strong emission and absorption features, and the continuum
is relatively smooth within these wavelength ranges. Also, these
continuum windows are
close to the \hd\, line without being contaminated by the \hd\ line itself.
When fitting the continuum flux level, the flux in each pixel was weighted by
the inverse square of the error on the flux in that pixel.  After the initial
fit to the continuum, we re--iterate the fit once by rejecting $3\sigma$
outliers to the original continuum fit.  This guards against noise spikes in
the surrounding continuum.

The rest--frame EW of the \hd\, line was calculated by summing the ratio of
the flux in each pixel of the spectrum, over the estimated continuum flux in
that pixel based on our linear interpolation.  For this summation, we
investigated two different wavelength windows for the \hd\, line; 4088\AA\ to
4116\AA, which is the same as the wavelength range used by Balogh et
al. (1999)\footnote{We note that Table 1 of Balogh et al. (1999) has a
typographical error. The authors used the wavelength range of 4088\AA\ to
4116\AA\ to measure their \hd\, EWs instead of 4082\AA\ to 4122\AA\ as quoted
in the paper.} and 4082\AA\ to 4122\AA, which is the wider range used by Abraham
et al. (1996b). We summarize the wavelength ranges used to measure the \hd\,
EWs in Table \ref{ea1_tab:wavelength}.

In Figure \ref{fig_ea1:hd_narrow_wide}, we compare the two non--parametric
measurements of \hd\, using the narrow and wide wavelength
windows. In this Chapter, positive EWs are absorption lines and negative
EWs are emission lines. 
 We find, as expected, a strong linear relationship between the two
measurements: The scatter about the best fit linear relationship to these
measurements is Gaussian with $\sigma=0.29$\AA. However, there are systematic
differences between the two measurements which are correlated to the intrinsic
width of the \hd\, line. For example, for large EWs of \hd, we find that the
wide wavelength window has a larger value than the narrow window. This is
because the 28\AA\ window is too small to capture the wings of a strong \hd\,
line and thus a wide window is needed.
 The same systematic trend can be found for \hd\ emission lines (i.e., negative
 EWs), where the wide window captures more flux than the narrow
 window and produces smaller values of EWs.

As a compromise, we have empirically determined that the best methodology for
our analysis is to always select the larger of the two \hd\, EW measurements
(this was discovered by visually inspecting many of the spectra and their
various \hd\, measurements). This is a crude adaptive approach of selecting
the size of the window based on the intrinsic strength of the \hd\, line. In
fact, we find that 20.2\% of our HDS galaxies (see Section
\ref{ea1_catalog}) were selected based on the \hd\, measurement in the large
wavelength window.
  Therefore, for the analysis presented in this
paper, we use H$\delta_{\rm max}$, which is the maximum of the two
non--parametric measurements discussed above.

In Figure \ref{fig_ea1:hd_max_1d}, we now compare the H$\delta_{\rm max}$
measurement discussed above to the automatic Gaussian fits to the \hd\, line
from the SDSS SPECTRO1D analysis of the spectra. As expected, the two methods
give similar results for the EW of the \hd\, line for the largest
EWs. However, there are significant differences, as seen in Figure
\ref{fig_ea1:hd_max_1d}, between these two methodologies.  First, there are many
galaxies with a negative EW (emission) as measured by SPECTRO1D, but possess a
(large) positive EW (absorption) using the non--parametric method. These cases
are caused by emission--filling, i.e., a small amount of \hd\, emission
at the bottom of the \hd\ absorption line (see Section \ref{ea1_emfilling}). This
results in SPECTRO1D fitting the Gaussian to the central emission line, thus
producing a negative EW. On the other hand, the non--parametric method simply
sums all the flux in the region averaging over the emission and still
producing a positive EW. In Figure \ref{fig_ea1:spec-0293-51689-144.ps}, we
present five typical examples of this phenomenon.
 
 Another noticeable difference  between the two methods seen in Figure
 \ref{fig_ea1:hd_max_1d} is the deviation 
 from the one--to--one relation for \hd\, EWs near zero, i.e., as the
 \hd\, line becomes weak, it is buried in the noise of the continuum making it
 difficult to automatically fit a Gaussian to the line. In such cases,
 SPECTRO1D tends to overestimate the EW of the \hd\, line because it
 preferentially fits a broad, shallow Gaussian to the noise in the
 spectrum. Typical examples of this problem are shown in Figure
 \ref{fig_ea1:spec-0415-51810-304.ps}. We conclude from our study of the SDSS
 spectra that the non--parametric techniques of Abraham et al. (1996b)
 and Balogh et al. (1999) are preferred to the automatic Gaussian fits of
 SPECTRO1D, especially for the lower signal--to--noise SDSS spectra which
 are the majority in our sample (see Figure \ref{fig_ea1:sng}). We note that many of the
 problems associated with the automatic Gaussian fitting of SPECTRO1D can be
 avoided by fitting Gaussians interactively. However, this is not practical for
 large datasets such as the SDSS.

\subsection{\oii\, and \ha\, Equivalent Widths}\label{ea1_oiiha}

 In addition to estimating the EW of the \hd\, line, we have used our
 flux--summing technique to estimate the rest frame equivalent widths of both
 the \oii\, and \ha\, emission lines. We perform this analysis on all
 95479 SDSS
 spectra. 
  As these emission lines are the primary diagnostics of on--going star--formation
 in a galaxy and thus, we are interested in detecting any evidence of these
 lines in our HDS galaxies. As discussed in Section \ref{ea1_Hdelta}, the
 flux--summing technique is better for the lower signal--to--noise spectra,
 while the Gaussian--fitting method of SPECTRO1D is optimal for higher
 signal--to--noise detections of these emission lines, especially in the case
 of \ha\, where SPECTRO1D deblends the \ha\, and \nii\, lines.

We use the same flux--summing methodology as discussed above for the \hd\,
line. However, we use only one wavelength window centered on the two emission
lines.  We list in Table \ref{ea1_tab:wavelength} the wavelength windows used in
summing the flux for the \oii\, and \ha\, emission lines and the continuum
regions around these lines. Once again, the continuum flux per pixel for each
emission line was estimated using linear interpolation of the continuum
estimated at either side of the emission lines (weighted by the inverse square of
the errors on the pixel values during a line fitting procedure). We again
iterate the continuum fit once rejecting $3\sigma$ outliers to the original
continuum fit. We do not deblend the \ha\, and \nii\, lines and as a result,
some of our \ha\, EW measurements may be overestimated. However, the
contamination is less than 5\% from \nii\, line at 6648\AA\ and less than 30\%
from \nii\, line at 6583\AA. We present estimates of the external error on our
measurements of \oii\, and \ha\, in Section \ref{ea1_error}.

In Figure \ref{fig_ea1:oii_1d_tomo}, we compare our \oii\, equivalent width
measurements to that from SPECTRO1D for all 95479 SDSS spectra. In this Chapter,
positive EWs are absorption lines and negative EWs are emission lines. There
is a good agreement between the two methods for EW(\oii)$>10$\AA, where the
scatter is $\simless 10\%$. However, at lower EWs, the SPECTRO1D measurement
of \oii\, is systematically larger than our flux--summing method which is the
result of SPECTRO1D fitting a broad Gaussian to the noise in the spectrum.
We are
only concerned with making a robust detection of any \oii\, emission, rather
than trying to accurately quantify the properties of the emission line.
Therefore, we prefer our non--parametric method, especially for the low
signal--to--noise cases.

In Figure \ref{fig_ea1:ha_1d_tomo}, we compare our \ha\, equivalent width
measurements against that of SPECTRO1D for all 95479 SDSS spectra regardless of
their \hd\ EW. The two locii of points seen in this figure are caused by
contamination in our estimates of \ha\, due to strong emission lines in AGNs,
 i.e., the top locus of points have larger EWs in our flux--summing, 
method than measured by SPECTRO1D due to contamination by the \nii\,
lines. This is confirmed by the fact that the top locus of points is dominated
by AGNs. At low \ha\, EWs, we again see a systematic difference between our
measurements and those of SPECTRO1D, with SPECTRO1D again over--estimating the
\ha\, line because it is jointly fitting multiple Gaussians to low
signal--to--noise detections of the \ha\, and \nii\, emission lines. Finally,
we do not make any correction for extinction and stellar absorption on our
flux--summed measurements of \ha (see Section \ref{ea1_emfilling}).

\subsection{Emission--Filling of the \hd\, Line}\label{ea1_emfilling}

As mentioned above, our measurements of the \hd\, absorption line can be
affected by emission--filling, i.e., \hd\, emission at the bottom of
the \hd\, absorption line. This problem could be solved by fitting two
Gaussians to the \hd\, line; one for absorption, one for emission. We found,
 however, that this method is only reliable for spectra with a
signal--to--noise ratio 
of $>20$ and, as shown in Figure \ref{fig_ea1:sng}, this is only viable for a
small fraction of our spectra. Therefore, we must explore an alternative
approach for correcting for this potential systematic bias; however, we stress
that the sense of any systematic bias on our non--parametric summing method
would be to always decrease (less absorption) the observed EW of the \hd\
absorption line and thus our technique gives a lower limit to the amount of
\hd\, absorption in the spectrum.

To help rectify the problem of emission--filling, we have used the \ha\, and
\hb\ emission lines (where available) to jointly constrain the amount of
emission--filling at the \hd\, line as well as estimate the effects of
internal dust extinction in the galaxy. Furthermore, our estimates of the
emission--filling are complicated by the effects of stellar absorption on the
\ha\, and \hb\ emission lines. In this analysis, we have used the SPECTRO1D
measurements of \ha\, and \hb\ lines in preference to our flux--summing
technique discussed in Section \ref{ea1_oiiha}, because the emission--filling
correction is only important in strongly star--forming galaxies where the
\ha\, and \hb\ emission lines are well fit by a Gaussian and, for the \ha\,
line, require careful deblending from the \nii\, lines.

To solve the problem of emission--filling, we have adopted two different
methodologies which we describe in detail below.  The first method is an
iterative procedure that begins with a initial estimate for the amount of
stellar absorption at the \hb\ and \ha\, emission lines, i.e., we
assume H$\beta$
EW (absorption) = 1.5\AA\ and H$\alpha$ EW (absorption) = 1.9\AA\ (see
Poggianti \& Barbaro 1997; Miller \& Owen 2002). Then, using the observed
ratio of the \ha\, and \hb\ emission lines (corrected for stellar absorption), in
conjunction with an attenuation law of $\tau=A\,\lambda^{-0.7}$ (Charlot \&
Fall 2000) for galactic extinction and a theoretical \ha\, to \hb\ ratio of
2.87 (case B recombination; Osterbrock 1989), we solve for the
 parameter, $A$, in the
 attenuation law and thus gain extinction--corrected values for both the \hb\
 and \ha\, emission lines.  Next, using the theoretical ratio of \hb\ emission
 to \hd\, emission, we obtain an estimate for the amount of emission--filling
 (extinction--corrected) in the \hd\, absorption line.  We then correct the
 observed \hd\, absorption EW for this emission--filling. Further,
 assuming that the
 EW(H$\delta$) absorption is equal to EW(\hb) absorption and EW(\ha) absorption
 is equal to 1.3 + $0.4\times$ EW(\hb) absorption (Keel 1983), we obtain new estimates for
 the stellar absorption at the \ha\, and \hb\ emission lines, i.e., where
 we begun the iteration.  We iterate this calculation five times, but on
 average, a stable solution converges after only one iteration. 

 Our second method uses the D4000 break to estimate the amount of
 stellar absorption at H$\beta$, using EW(H$\beta$) $= -5.5\times\,{\rm
 D4000} + 11.6$ (Poggianti \& Barbaro 1997; Miller \& Owen 2002). Then,
 assuming that EW(\ha) absorption is equal to 1.3 + $0.4\times$
 EW(H$\beta$) absorption, we obtain an measurement for the amount of
 stellar absorption at both the \ha\, and H$\beta$ absorption lines.  As
 in the first method above, we use the Charlot \& Fall (2000)
 attenuation law, and the theoretical \ha\, to H$\beta$ ratio, to solve
 for the amount of extinction at \ha\, and H$\beta$, and then use these
 extinction--corrected emission lines to estimate the amount of
 emission--filling at \hd. We do not iterate this method, as we have
 used the measured D4000 break to independently estimate the amount
 of stellar absorption at \hb\ and \ha.
 
We have applied these two methods to all our 95479 SDSS spectra, except for any
galaxy that possesses a robust detection of an Active Galactic Nucleus (AGN)
based on the line indices discussed in Kewley et al. (2002) and Gomez et
al. (2003). For these AGN classifications, we have used the SPECTRO1D emission
line measurements. We also stop our emission--filling correction if the ratio
of the \hb\ and \ha\, line becomes unphysical, i.e., greater than
2.87. By definition, the emission--filling correction increases our observed
values of the \hd\, absorption line, with a median correction of 15\% in the
flux of the \hd\, absorption line. In Figure \ref{fig_ea1:hd_emission_hist}, we
show the distributions of H$\delta$ emission EWs calculated using the two
methods described above in a solid (iteration) and a dashed (D4000) line,
respectively. It is reassuring that these two methods broadly give the same
answer and have similar distributions.

\subsection{External Errors on our Measured Equivalent Widths}\label{ea1_error}

 Before we select our HDS sample of galaxies, it is important to accurately
 quantify the errors on our EW measurements.  In our data, there are
 11538 that were
 galaxies spectroscopically observed twice (see Section \ref{ea1_data}). We
 use them to quantify the external error on our EW measurements. In Figures
\ref{fig_ea1:err_from_double_obs_hd},
 \ref{fig_ea1:err_from_double_obs_oii} and 
\ref{fig_ea1:err_from_double_obs_ha}, we present the absolute difference in
equivalent width of the two independent observations of the \hd\,, \ha\, and
\oii\, lines, as a function of signal--to--noise ratio. In these figures, we have
used the lower of the two measured signal--to--noise ratios (in SDSS $g$ band
for \hd\, and \hb, or $r$ band for \ha) because any observed difference in the two
measurements of the EW will be dominated by the error in the noisier (lower
signal--to--noise) of the two spectra.  From this data, we determine the
$1\sigma$ error for each line  as a function of
signal--to--noise ratio, and assign this error to our EW measurements
 for each galaxy. We determine the 
sigma of the distribution by fitting a Gaussian (as a function of
signal--to--noise ratio) as shown in Figures
\ref{fig_ea1:err_from_double_obs_hd_gauss_fit},
\ref{fig_ea1:err_from_double_obs_hd_gauss_fit_oii} and
\ref{fig_ea1:err_from_double_obs_hd_gauss_fit_ha}, and then use a 3rd order
 polynomial as shown in the following equation to interpolate between the four
 signal--to--noise bins, thus obtaining 
 the solid lines shown in Figures \ref{fig_ea1:err_from_double_obs_hd},
 \ref{fig_ea1:err_from_double_obs_oii} and
 \ref{fig_ea1:err_from_double_obs_ha}. 
  \begin{equation} 
 error = a_0 + a_1\times (S/N) + a_2\times(S/N)^2 + a_3\times(S/N)^3 
 \end{equation}
  Using
 these polynomial fits, we can estimate the $1\sigma$ error on our EWs for any
 signal--to--noise ratio.  The coefficients of the fitted 3rd order polynomial for
 each line are given in Table \ref{ea1_equation}.

In addition to quantifying the error on \oii, \ha\, and \hd, we have used the
duplicate observations of SDSS galaxies to determine the error on our
emission--filling corrections. Only 400 (564) of the 11538 duplicate
observations of SDSS galaxies have strong \ha\, and H$\beta$ emission lines
which are required for the iterative (D4000) method of correcting for
emission--filling. The errors on the emission--filling correction are only a
weak function of signal--to--noise ratio, so we have chosen to use a constant value
for their error, rather than varying the error as a function of the galaxy
signal--to--noise ratio as done for \hd, \oii\, and \ha\, emission lines. One sigma
errors on the emission correction of \hd\ for the iterative method (EF1)
and the D4000 method (EF2) are 0.57\AA\ and 0.4\AA\ in EW, respectively.

\section{A Catalog of HDS Galaxies}
\label{ea1_catalog}

\begin{figure}
\includegraphics[scale=0.7]{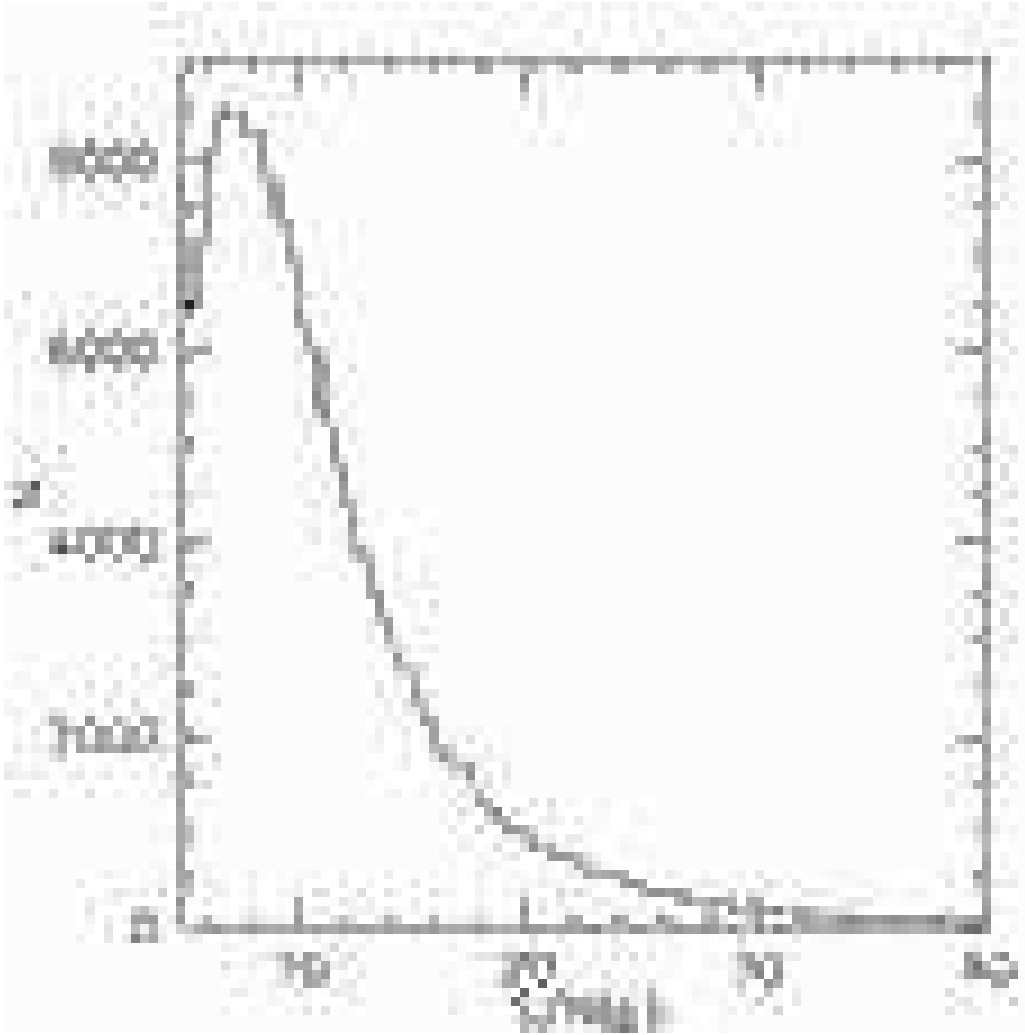}
\caption{
\label{fig_ea1:sng}
The distribution of signal--to--noise ratio for 95479 spectra used in this
analysis (see Section \ref{ea1_data}). The signal--to--noise ratio presented here is the
average signal--to--noise ratio per pixel over the wavelength range defined by the
SDSS photometric $g$ passband. The median signal--to--noise ratio is
8.3. Galaxies with signal--to--noise ratio less than 5 were not used in our study.
 }
\end{figure}

\begin{figure}
\includegraphics[scale=0.7]{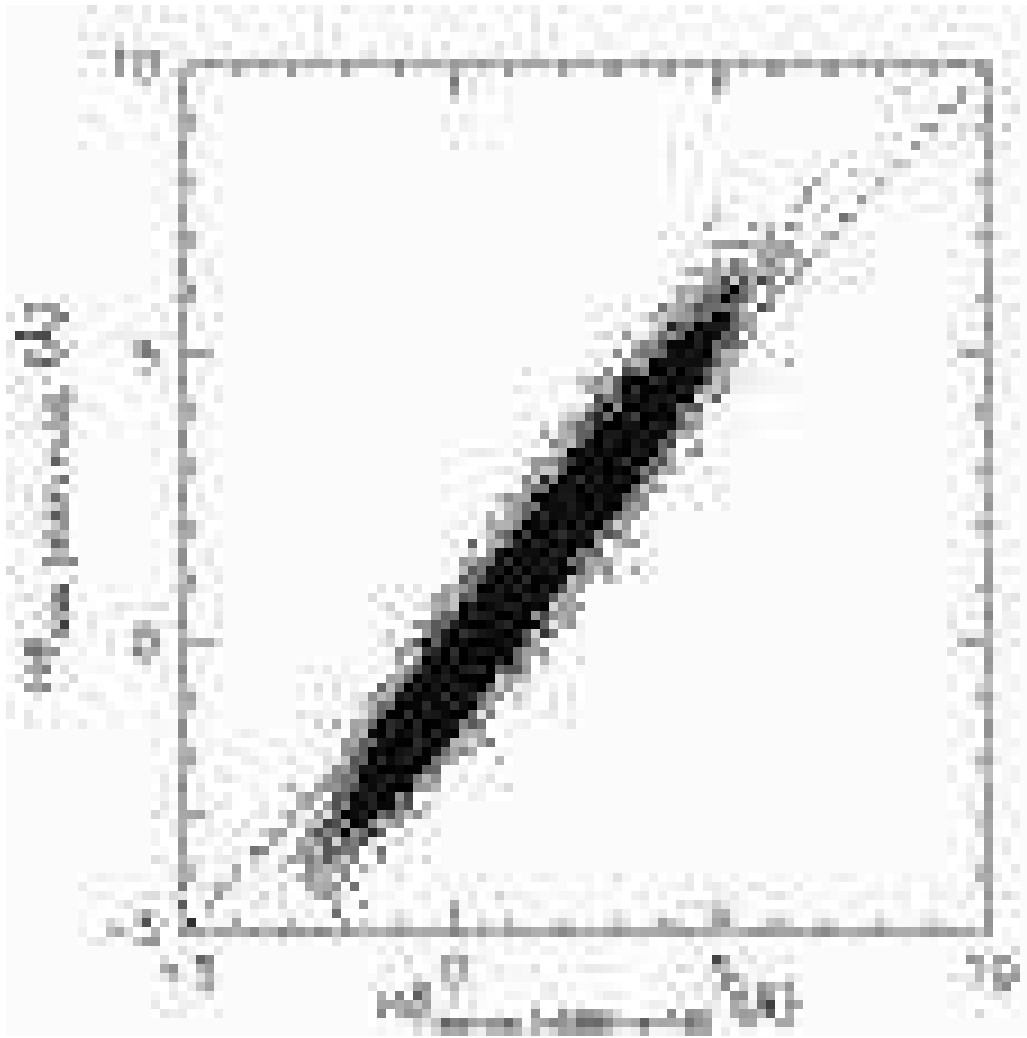}
\caption{
\label{fig_ea1:hd_narrow_wide}
The \hd\, EW (\AA) as measured in two different wavelength windows, i.e.,
the {\it wide} window of Abraham et al. (1996b) and the {\it narrow} window
of Balogh et al. (1999). The expected one--to--one line is plotted to help
guide the eye. For the work presented in this Chapter, absorption lines have a
positive EW values and emission lines have negative EW values. }
\end{figure}

\begin{figure}
\includegraphics[scale=0.7]{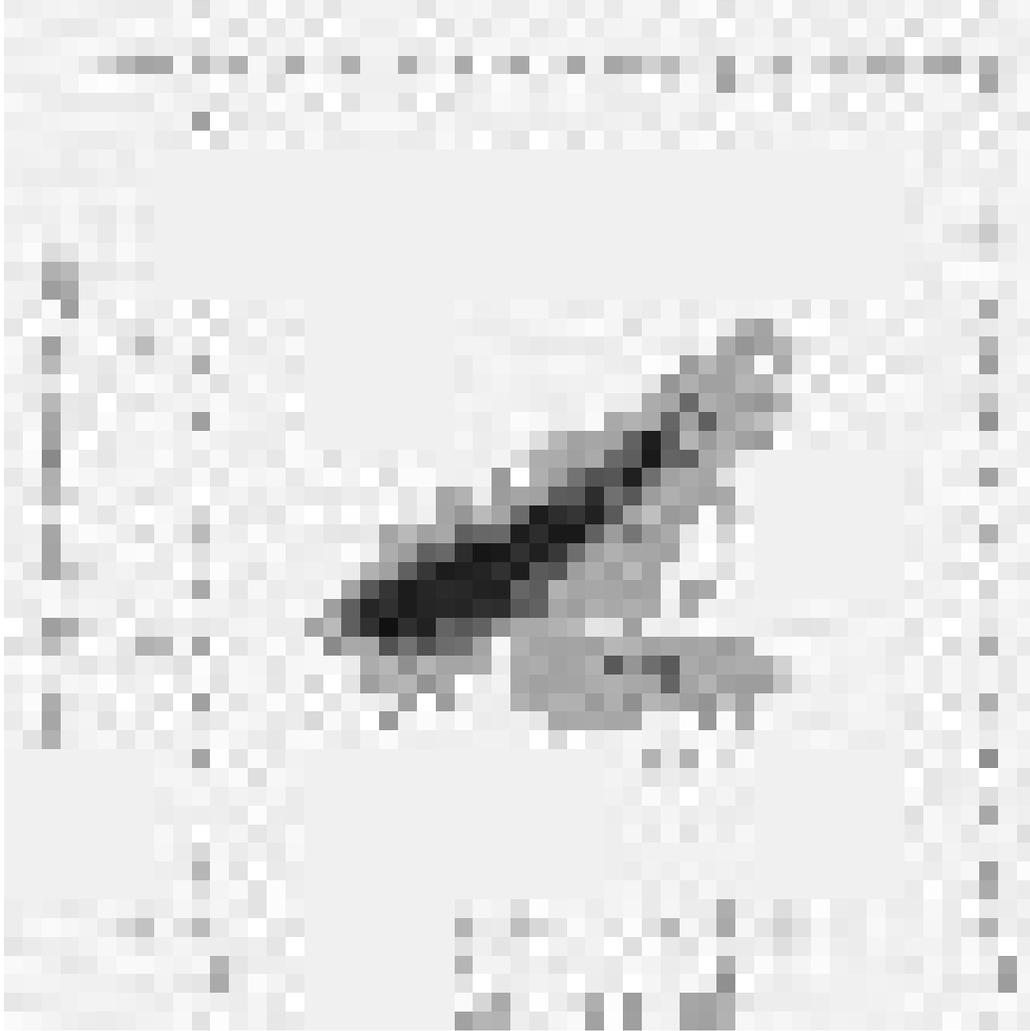}
\caption{
\label{fig_ea1:hd_max_1d} 
A comparison of the \hd\, EW as measured automatically by the SDSS SPECTRO1D
spectroscopic pipeline (a Gaussian fit to the \hd\, line) and the
non--parametric summation technique discussed in this Chapter and presented in
Figure \ref{fig_ea1:hd_narrow_wide}.  The one--to--one line is shown to guide the
eye.  In our work, absorption lines have positive EW values and emission lines
have negative EW values.  }
\end{figure}

\begin{figure}
\includegraphics[scale=0.7]{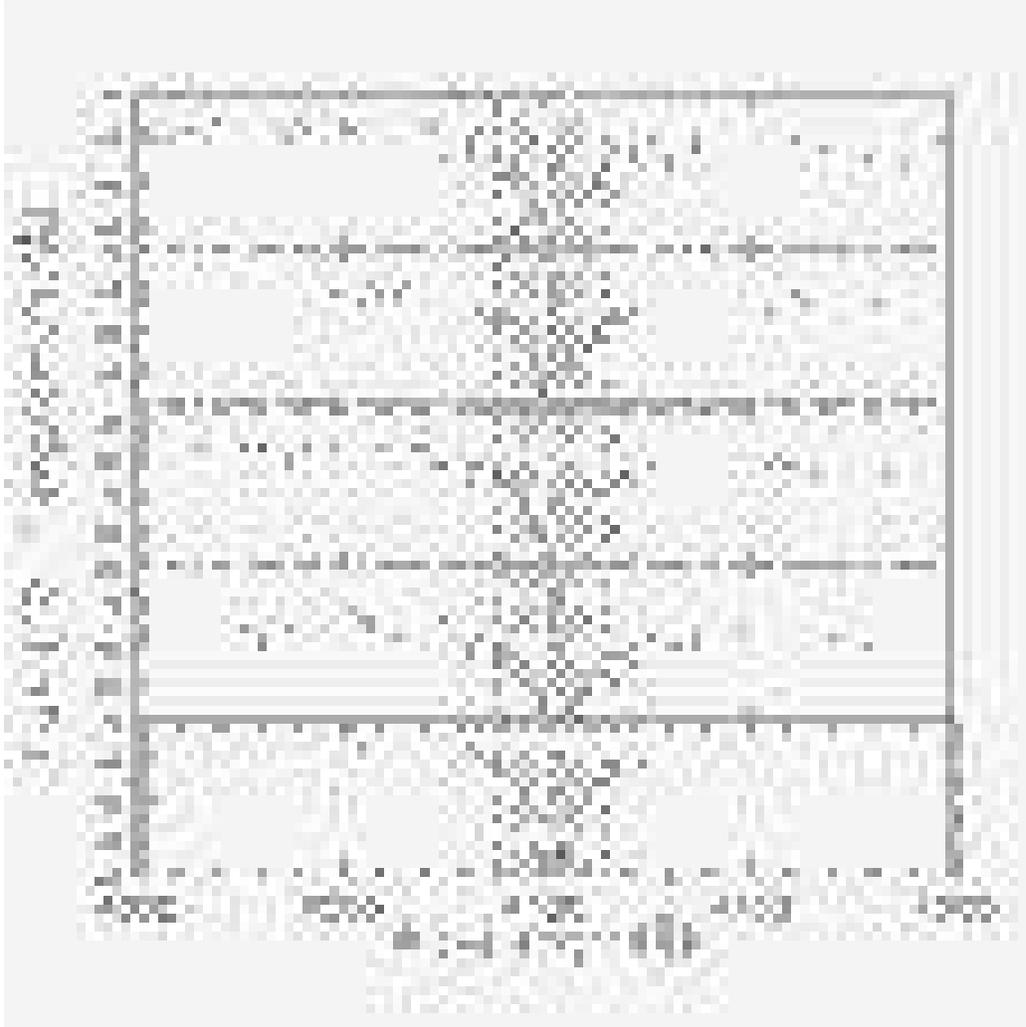}
\caption{
\label{fig_ea1:spec-0293-51689-144.ps}
Five typical examples of SDSS spectra with H$\delta$ emission filling.
In such cases it is difficult to fit the H$\delta$ absorption emission
with a single Gaussian due to a centrally peaked emission. The double
shaded region of this figure, centered on the \hd\, line, represents
the narrow wavelength window used to measure the EW of \hd\, as
explained in Section \ref{ea1_Hdelta}. The slightly wider shaded region,
again centered on the \hd\, line, represents the wide wavelength
window used to measure the \hd\, line (see Section
\ref{ea1_Hdelta}). Finally, the two dashed regions, at each side of the shaded
regions, represent the wavelength regions used to estimate the
continuum flux. See also Table \ref{ea1_tab:wavelength} for details of the wavelength windows
used on measuring the \hd\, line.}
\end{figure}

\begin{figure}
\includegraphics[scale=0.7]{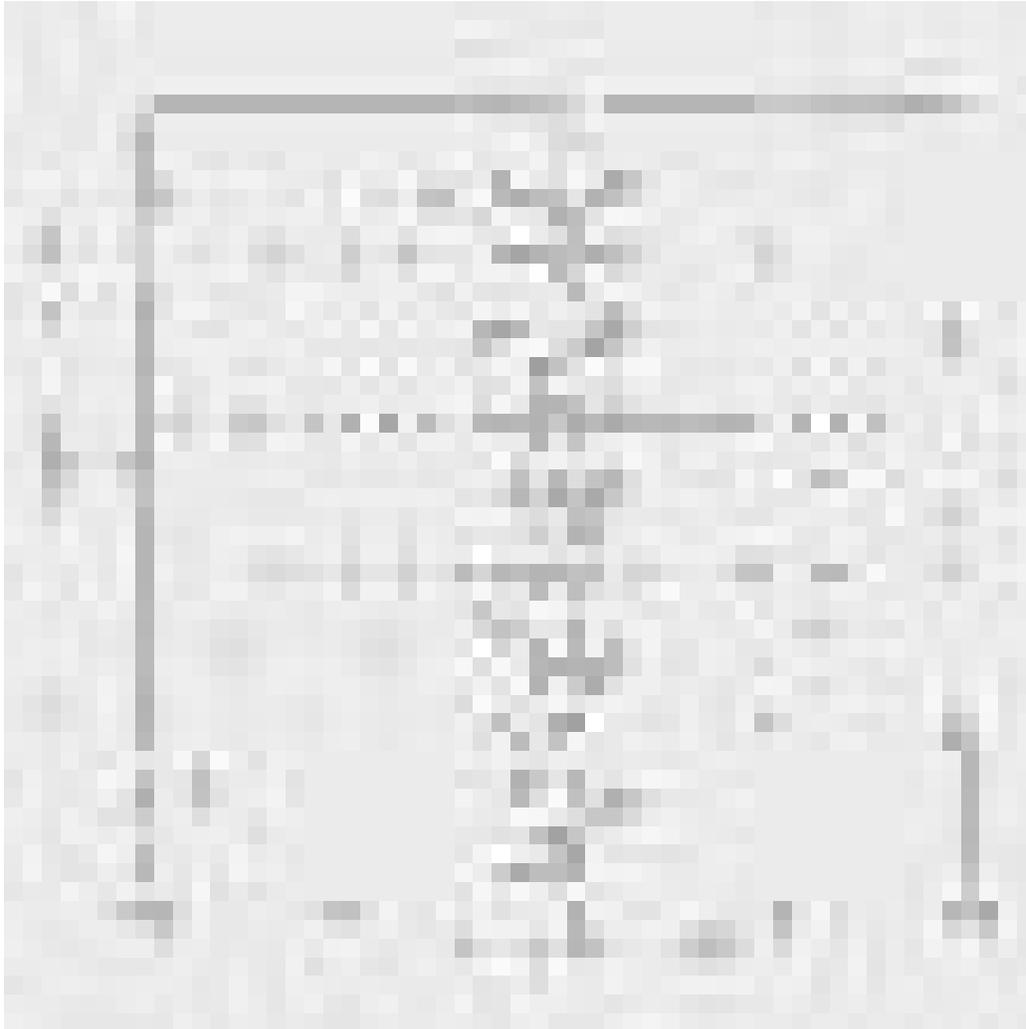}
\caption{
\label{fig_ea1:spec-0415-51810-304.ps}
Five examples of noisy spectra where the SDSS SPECTRO1D pipeline has fit
 a broad absorption line, thus overestimating the \hd\, EW. 
The shaded regions are the same as presented and discussed in Figure 
\ref{fig_ea1:spec-0293-51689-144.ps}.
}
\end{figure}

\begin{figure}
\centering{\includegraphics[scale=0.39]{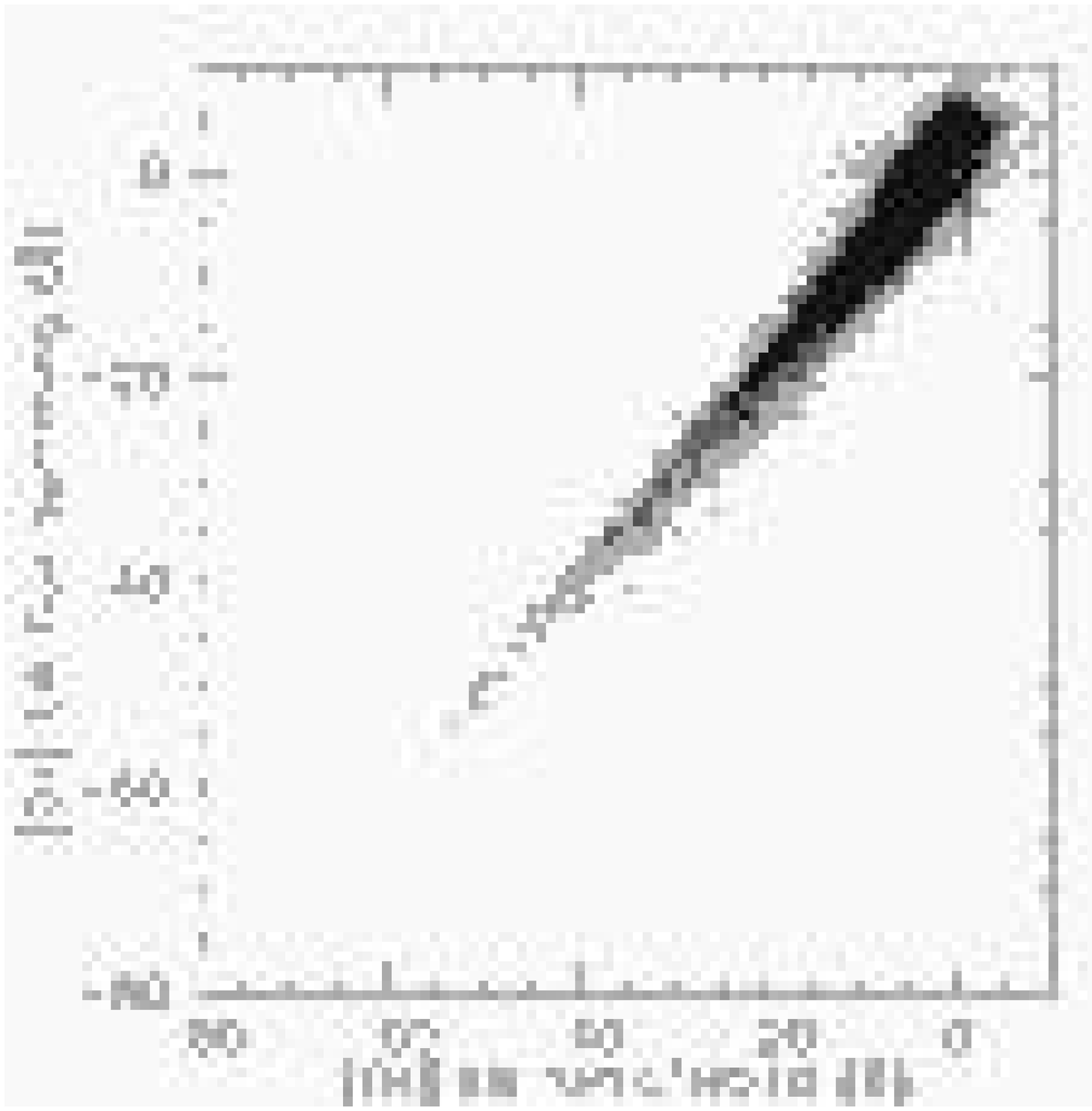}
\includegraphics[scale=0.39]{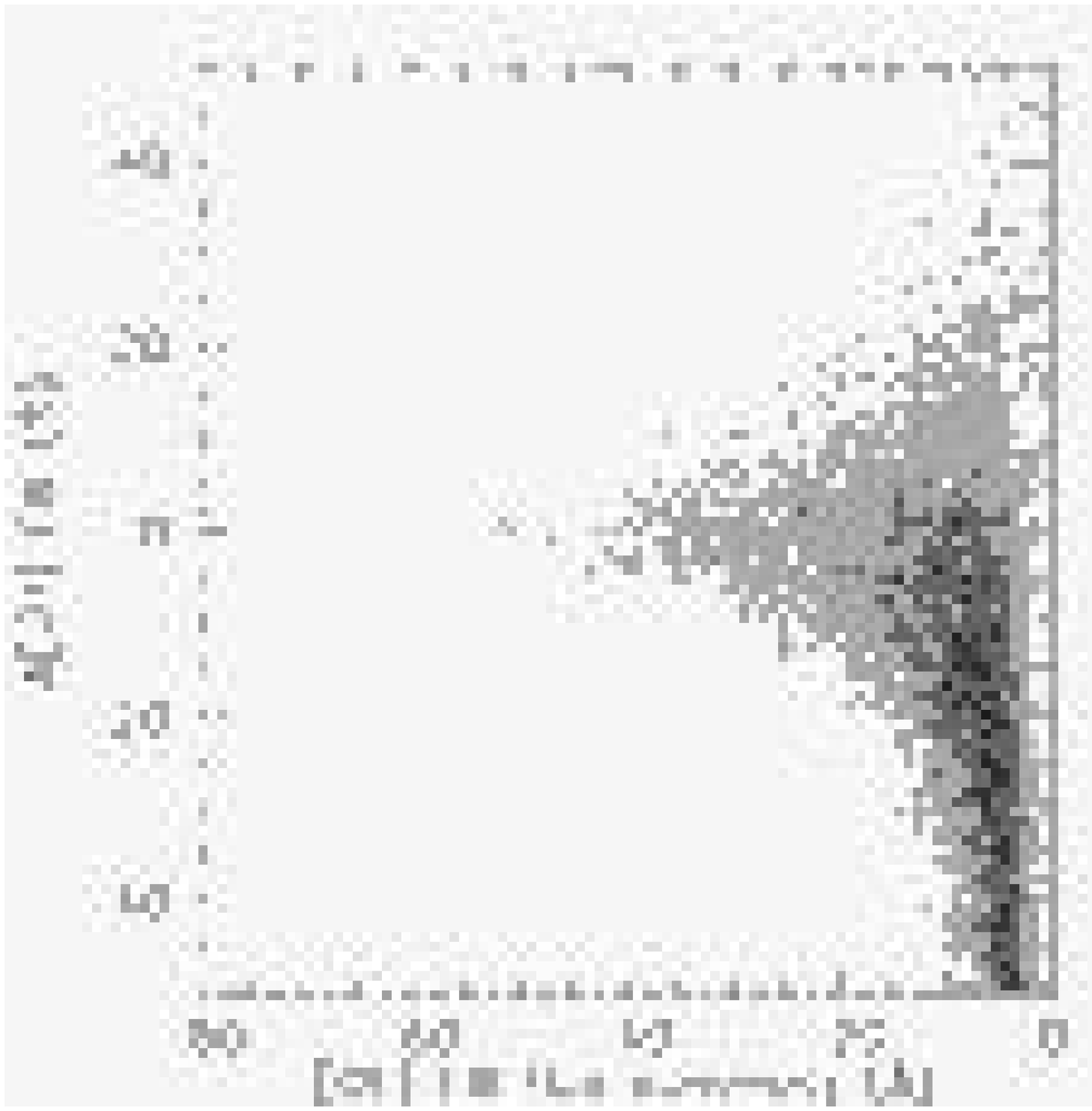}}
\caption{
In the left panel, we present the comparison of our \oii\ EW measurements (flux
summing) and those of SPECTRO1D (Gaussian fitting) for all 94770 SDSS spectra regardless of their \hd\ EWs. In the right panel, we
plot the percentage difference between these two measurements. Positive
percentages mean our flux summing method has a larger value.  }\label{fig_ea1:oii_1d_tomo}
\end{figure}

\begin{figure}
\centering{\includegraphics[scale=0.39]{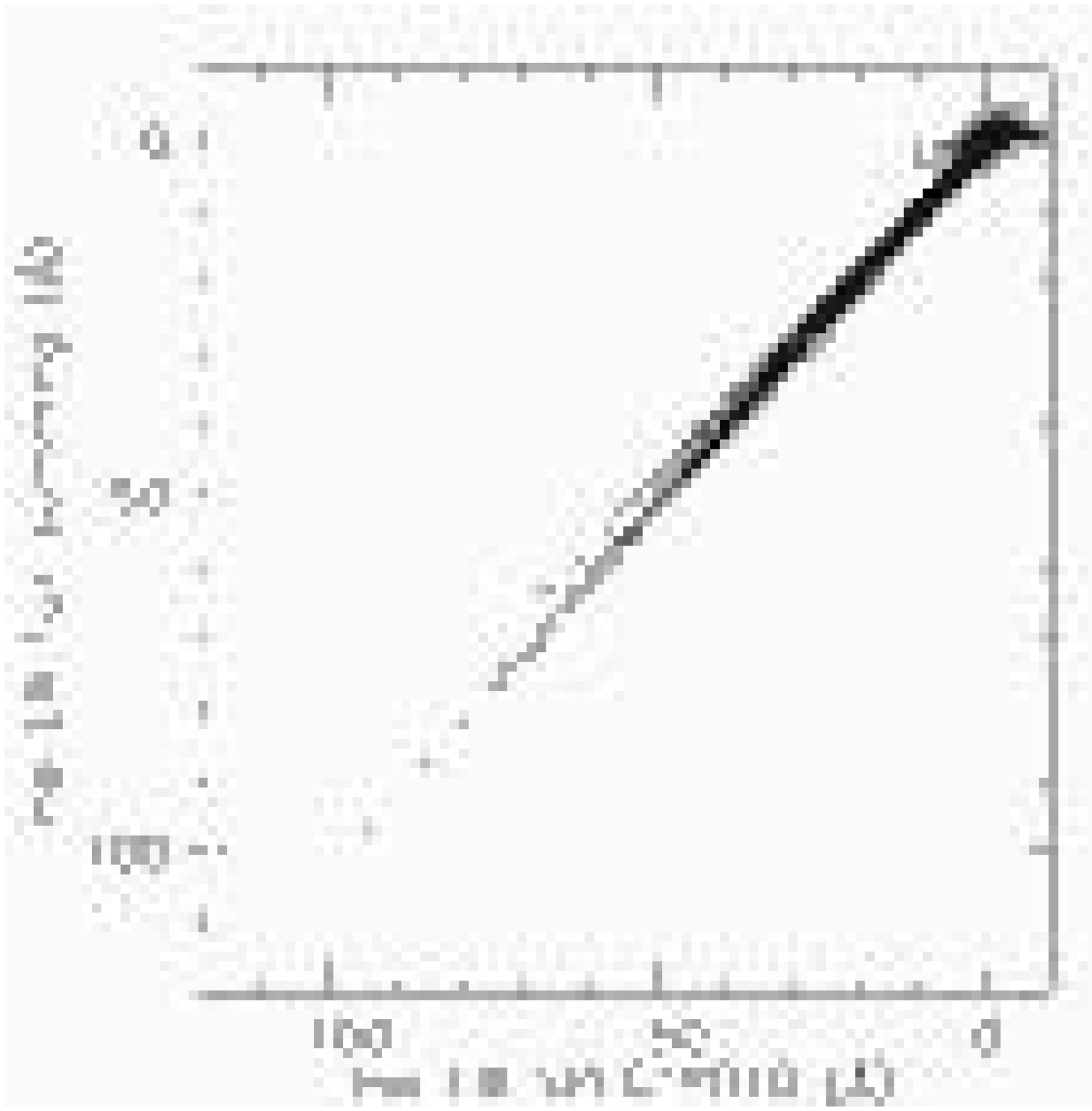}
\includegraphics[scale=0.39]{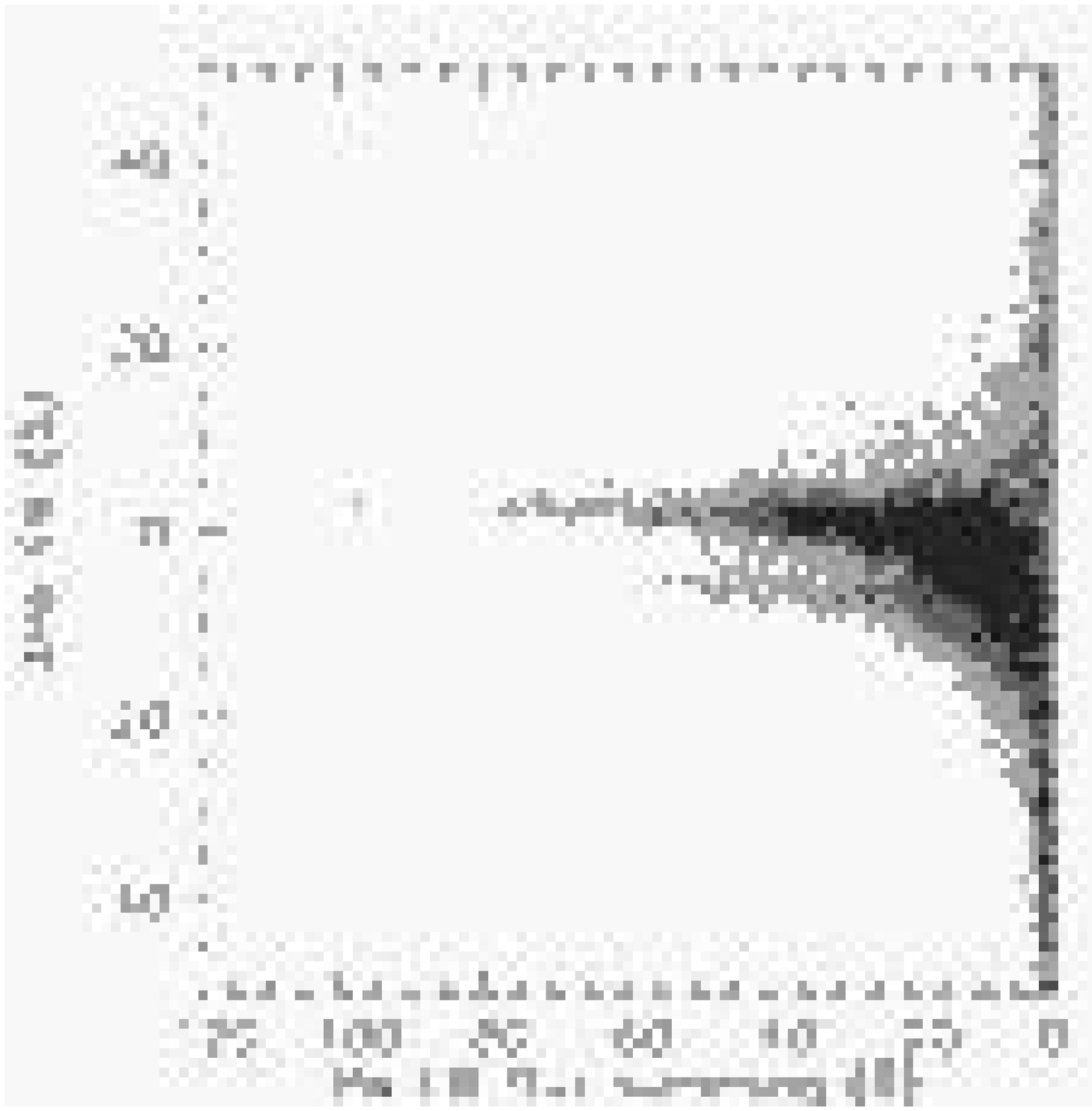}}
\caption{
In the left panel, we present the comparison of our \ha\ EW measurements (flux
summing) and those of SPECTRO1D (Gaussian fitting) for all SDSS 94770 spectra regardless of their \hd\ EWs. In the right panel, we
plot the percentage difference between these two measurements. Positive
percentages mean our flux summing method has a larger value.  }\label{fig_ea1:ha_1d_tomo}
\end{figure}

\begin{figure}
\includegraphics[scale=0.7]{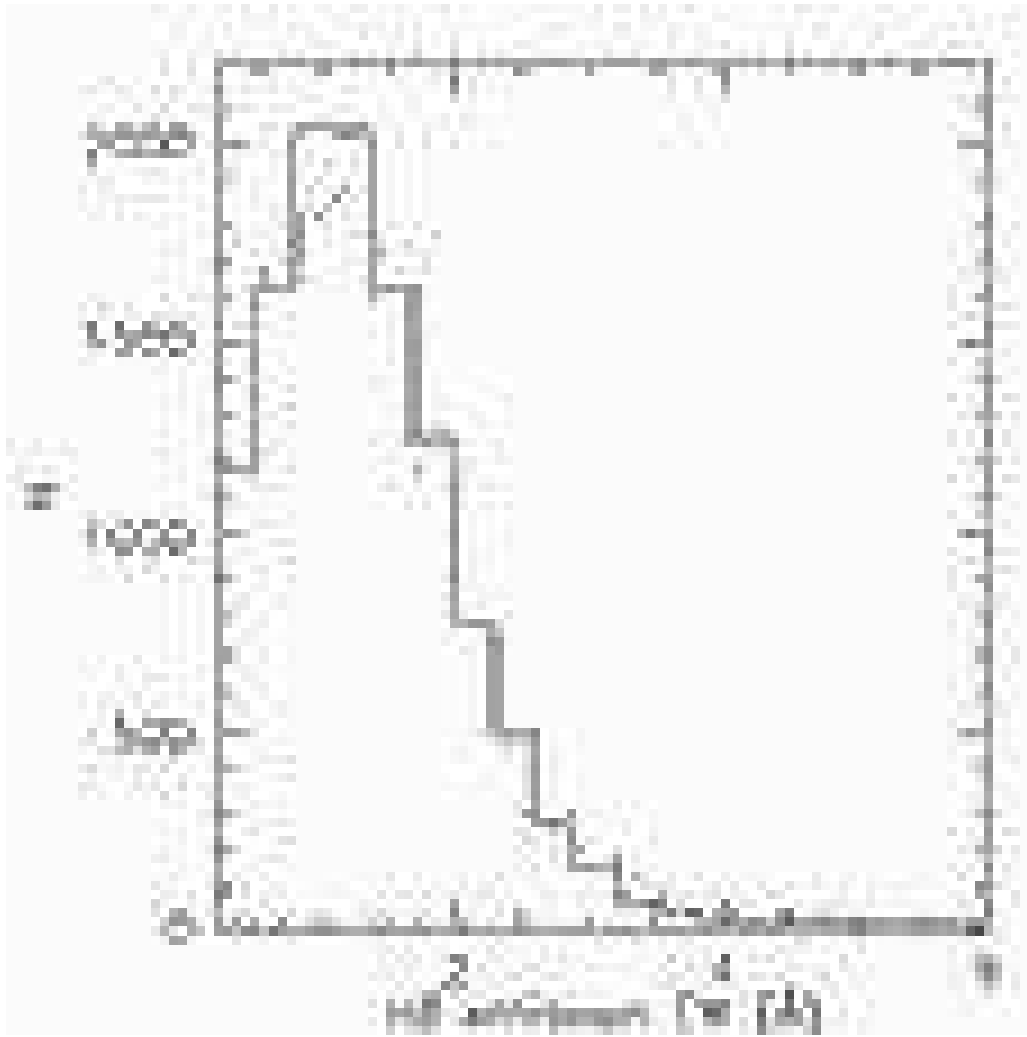}
\caption{
The amount of emission filling correction of H$\delta$ EW. The solid line is
 for the iteration method (EF1) and the shaded histogram uses the D4000 method (EF2).
}\label{fig_ea1:hd_emission_hist}
\end{figure}

\begin{figure}
\includegraphics[scale=0.7]{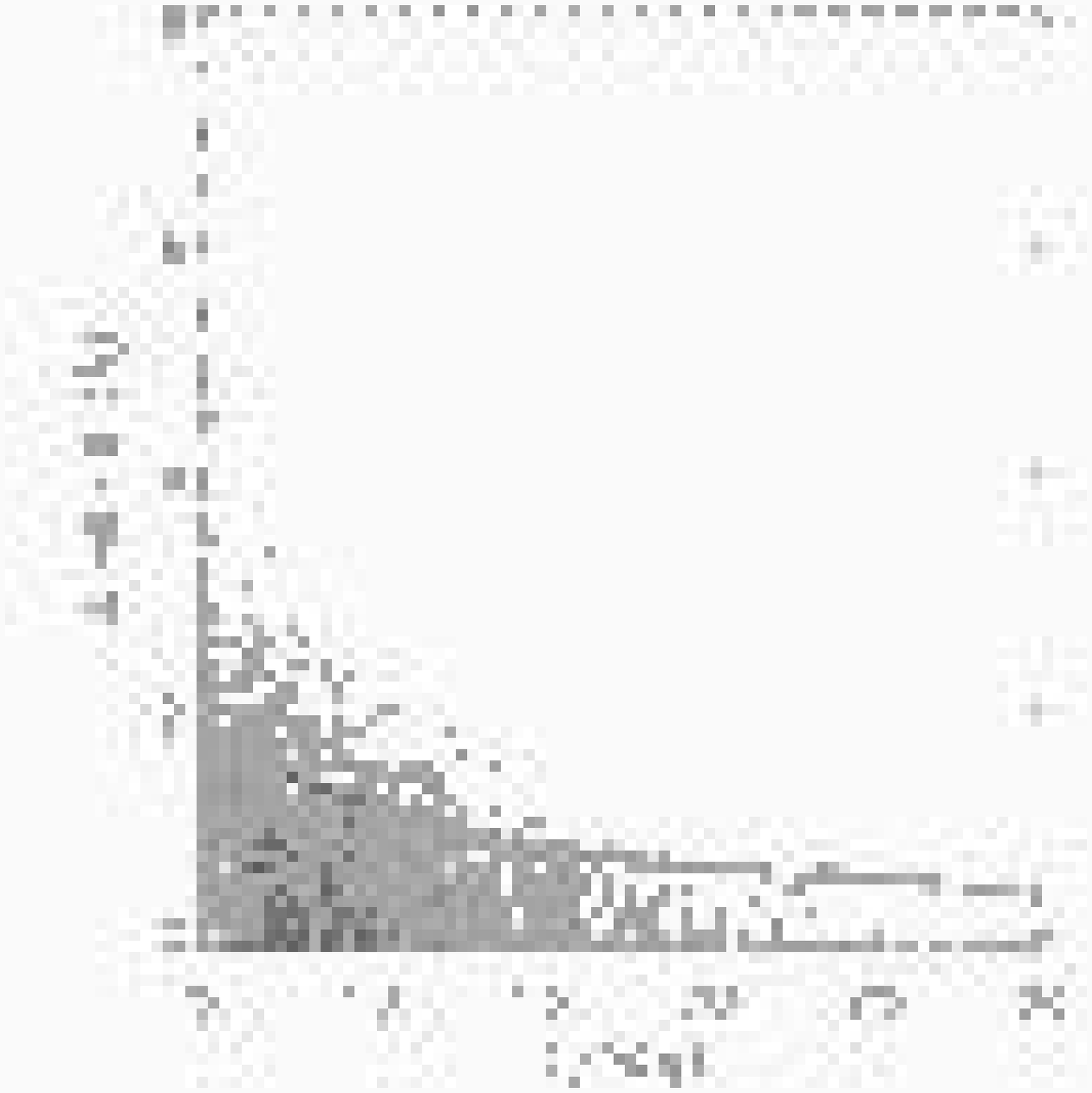}
\caption{
\label{fig_ea1:err_from_double_obs_hd}
The absolute difference in the measured \hd\, EW (\AA) for duplicate
observations of 11538 SDSS galaxies as a function of signal--to--noise ratio (the
lower of the two signal--to--noise ratios has been plotted here). The
solid line shows the $1\sigma$ polynomial line fitted to the
distribution of errors (as a function of signal--to--noise ratio).  }
\end{figure}

\begin{figure}
\includegraphics[scale=0.7]{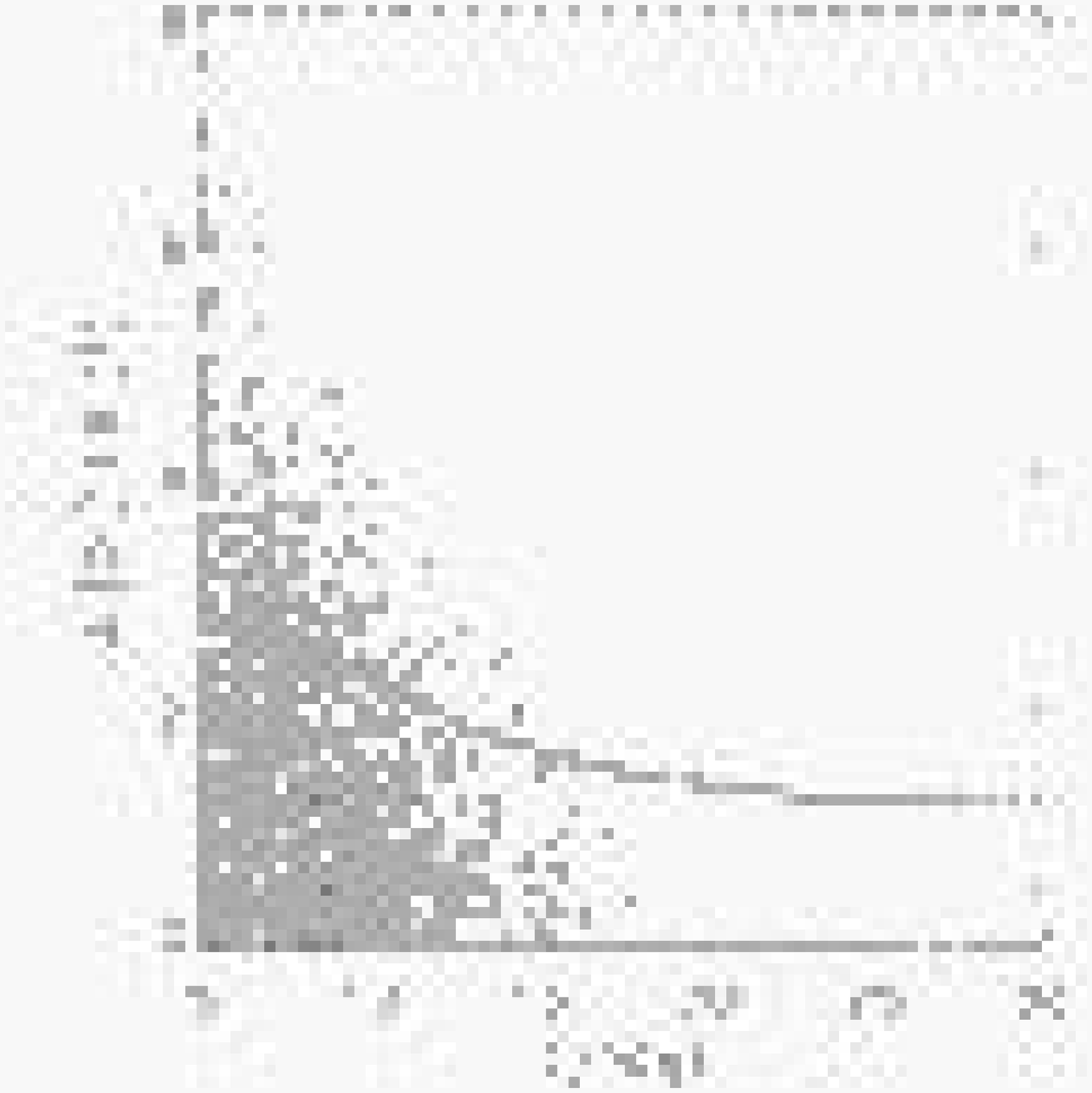}
\caption{
\label{fig_ea1:err_from_double_obs_oii}
The absolute difference in the \oii\, EW (\AA) for duplicate
observations of 11538 SDSS galaxies as a function of signal--to--noise ratio (the
lower of the two signal--to--noise ratios has been plotted here). The
solid line shows the $1\sigma$ polynomial line fitted to the
distribution of errors (as a function of signal--to--noise ratio).}
\end{figure}

\begin{figure}
\includegraphics[scale=0.7]{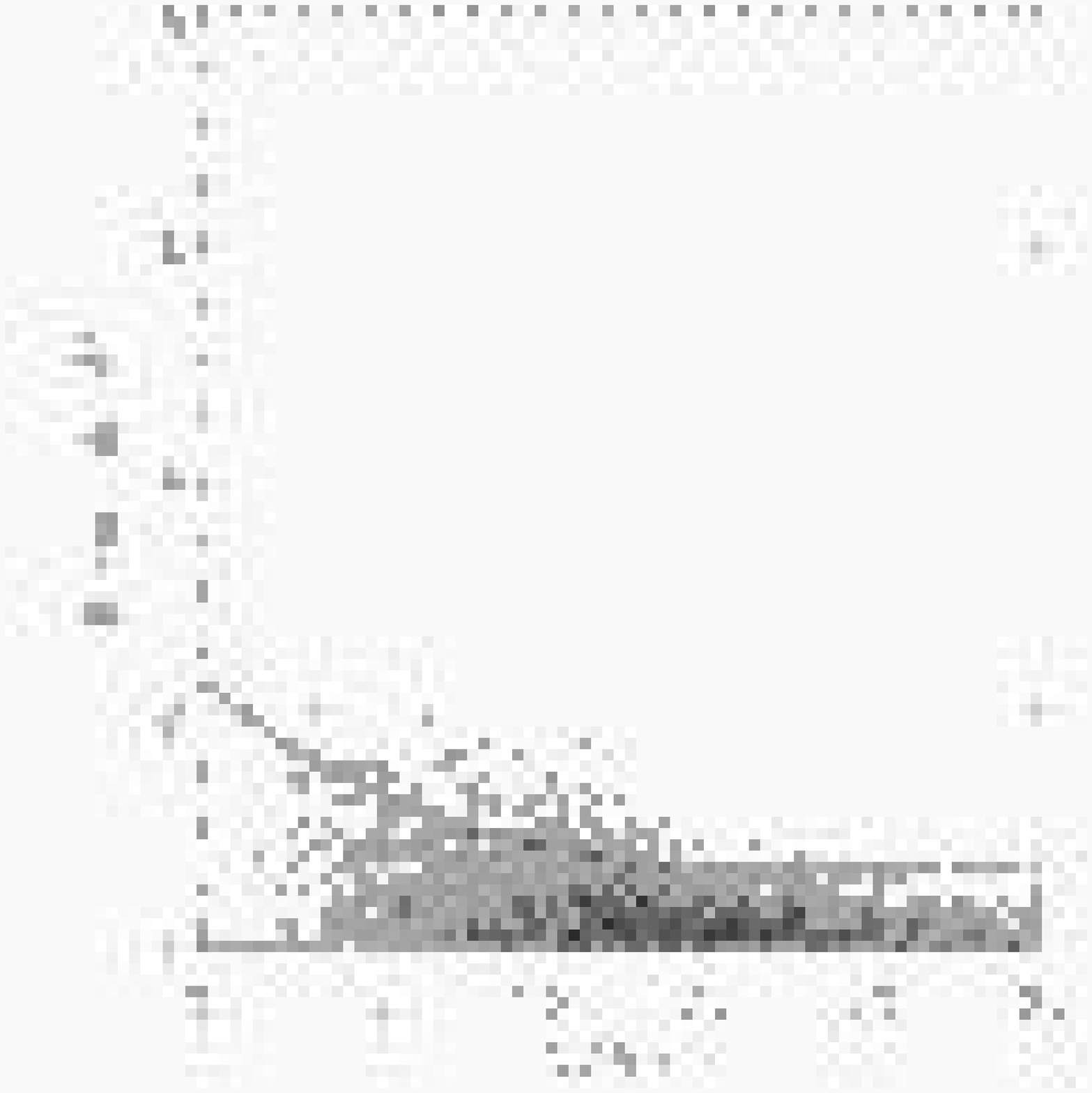}
\caption{
\label{fig_ea1:err_from_double_obs_ha}
The absolute difference in the \ha\, EW (\AA) for duplicate observations
of 11538 SDSS galaxies as a function of signal--to--noise ratio (the lower of the
two signal--to--noise ratios has been plotted here).  The solid line
shows the $1\sigma$ polynomial line fitted to the distribution of
errors (as a function of signal--to--noise ratio).  }
\end{figure}

\begin{figure}
\centering{\includegraphics[scale=0.4]{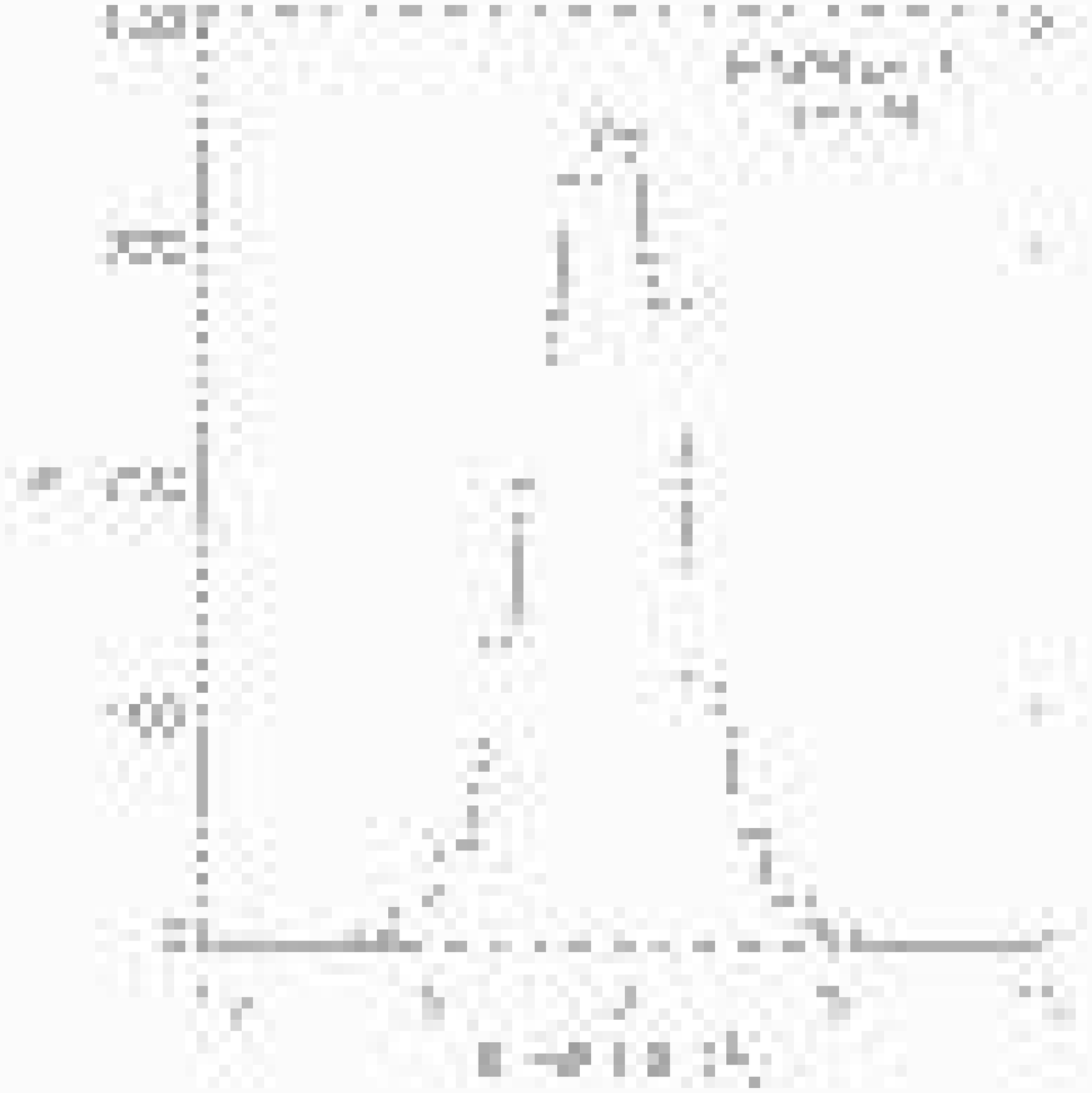}
\includegraphics[scale=0.4]{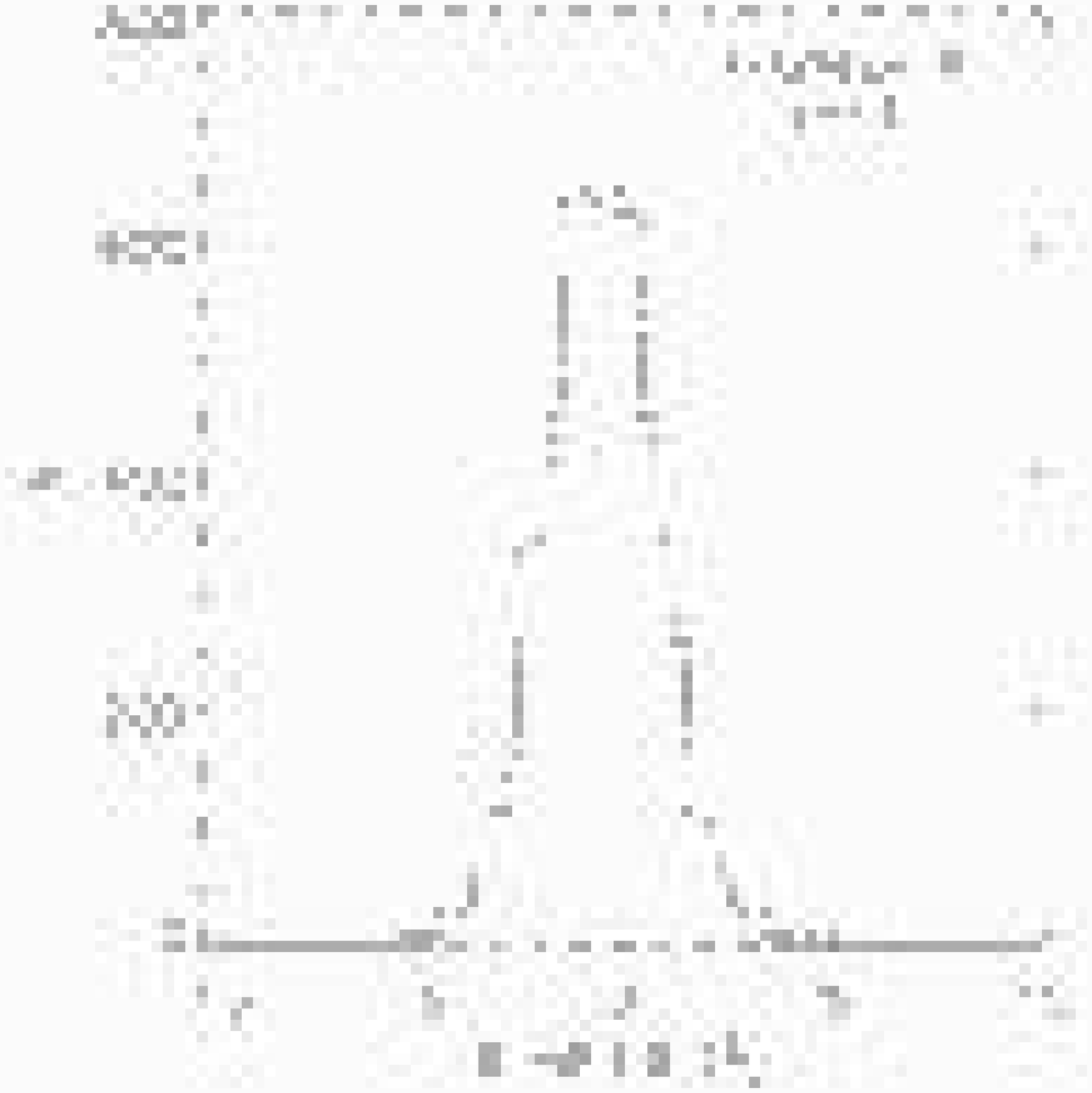}}
\centering{\includegraphics[scale=0.4]{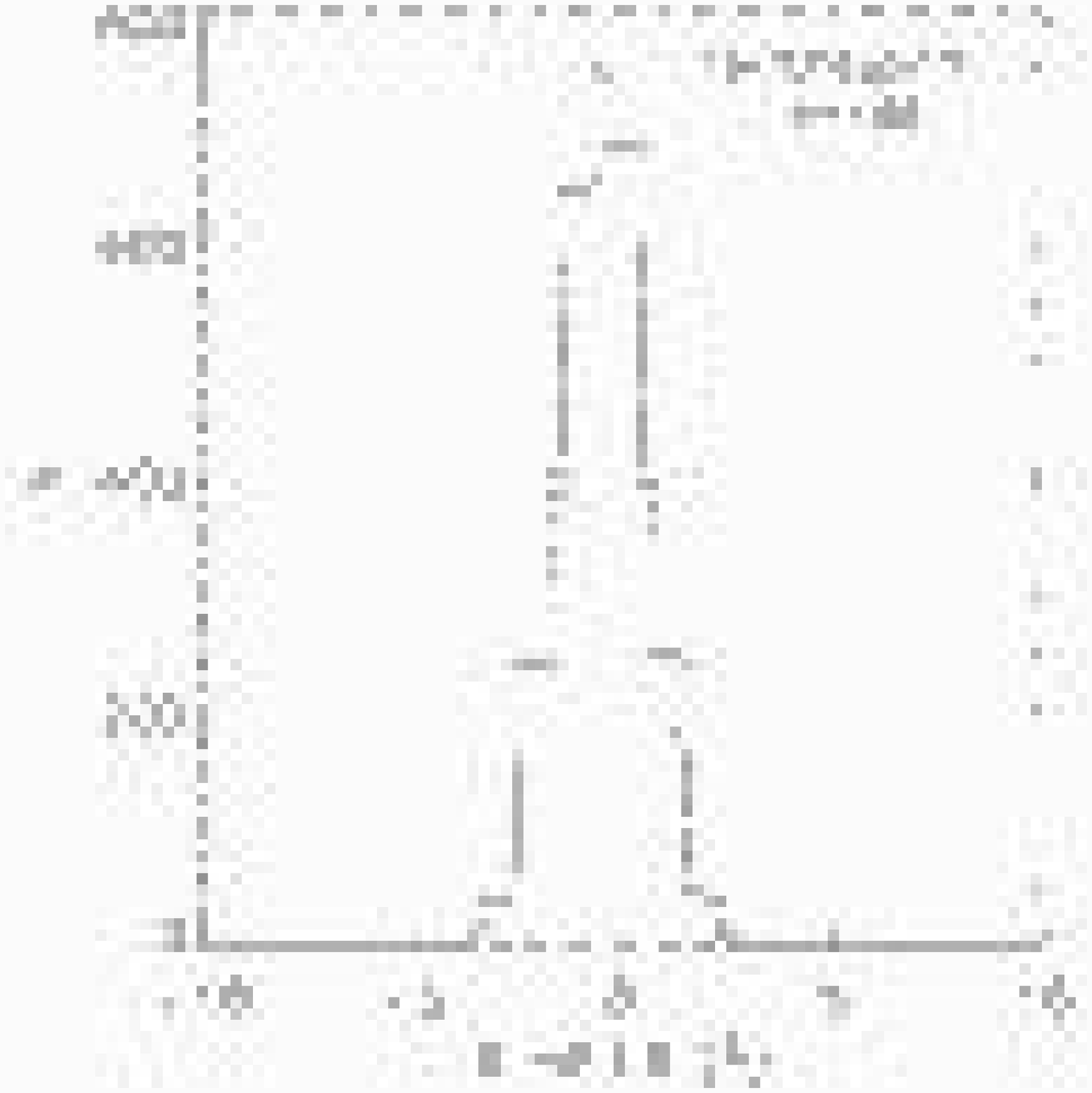}
\includegraphics[scale=0.4]{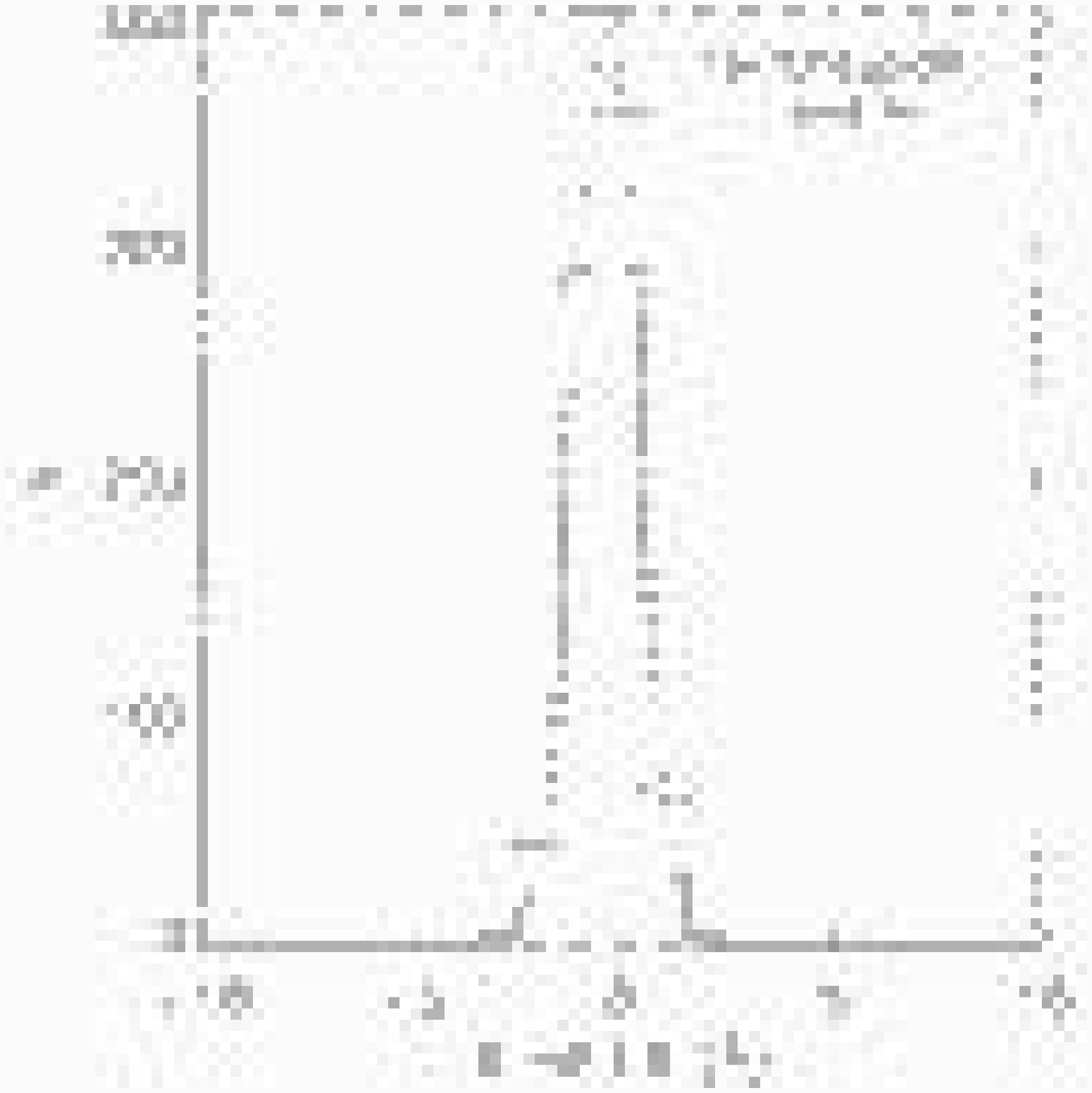}}
\caption{
\label{fig_ea1:err_from_double_obs_hd_gauss_fit}
We present the distribution of differences for the \hd\, line from our
duplicate observations. The four panels denote four different bins in
signal--to--noise ratio, i.e., clockwise from the top--left panel, we have s/n
$<7$, $7<{\rm S/N} <10$, $15<{\rm S/N} <20$ and $10<{\rm S/N} <15$.  We show
in the dotted line the best fit Gaussian to these distributions, which was then
used to determine the $1\sigma$ error (shown for each panel) on \hd\, EW as a
function of signal--to--noise ratio.}
\end{figure}

\begin{figure}
\centering{\includegraphics[scale=0.4]{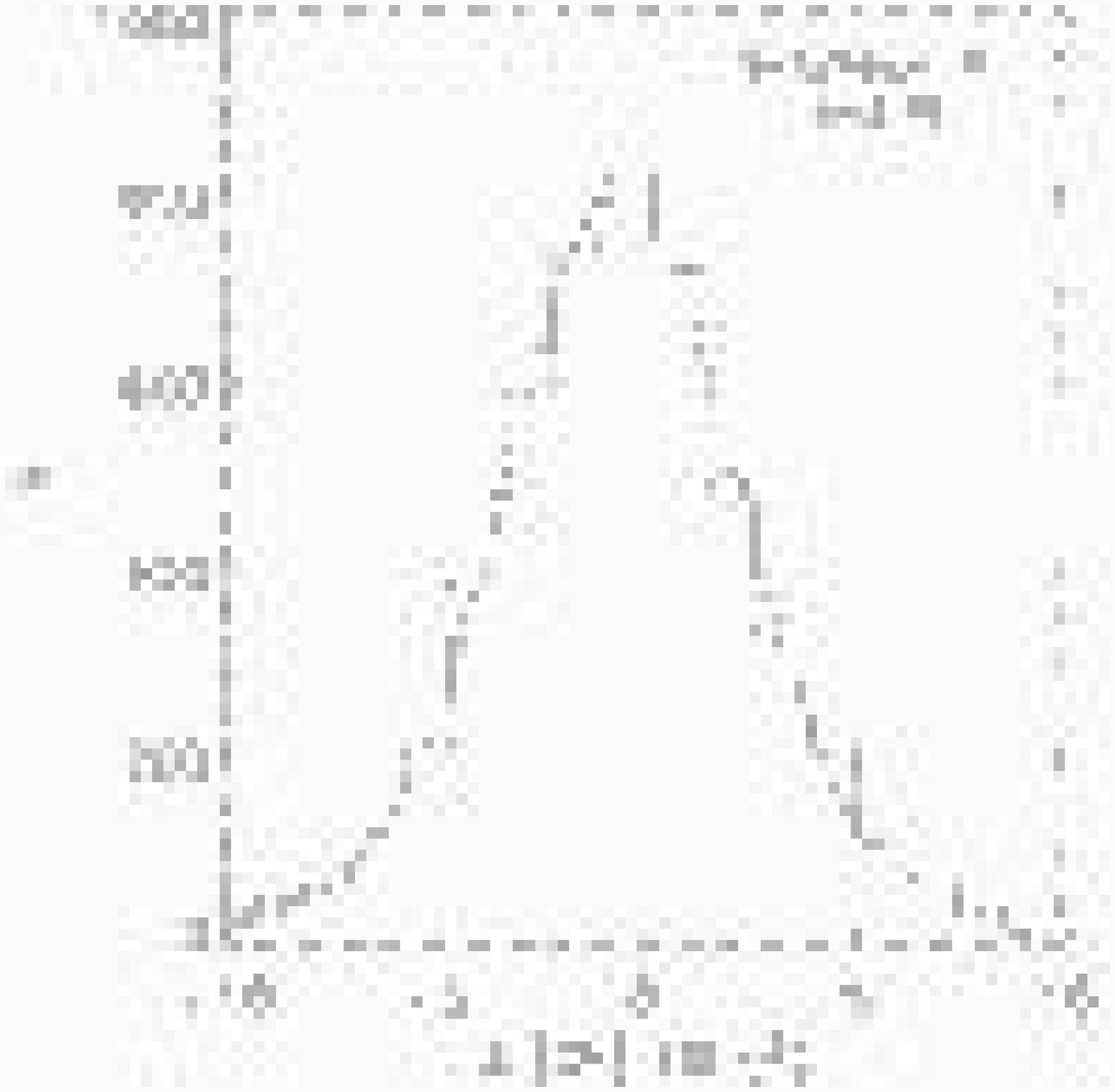}
\includegraphics[scale=0.4]{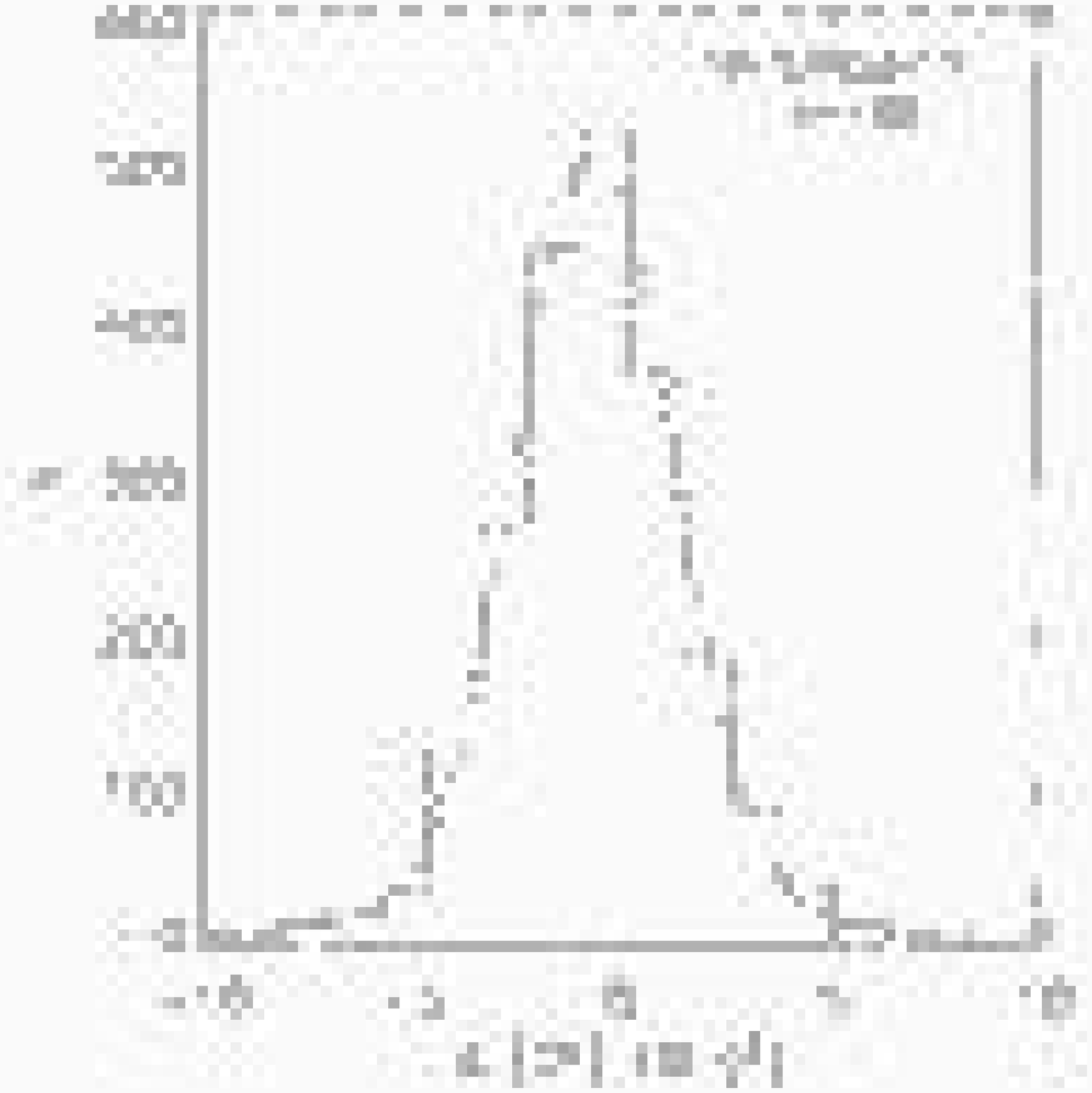}}
\centering{\includegraphics[scale=0.4]{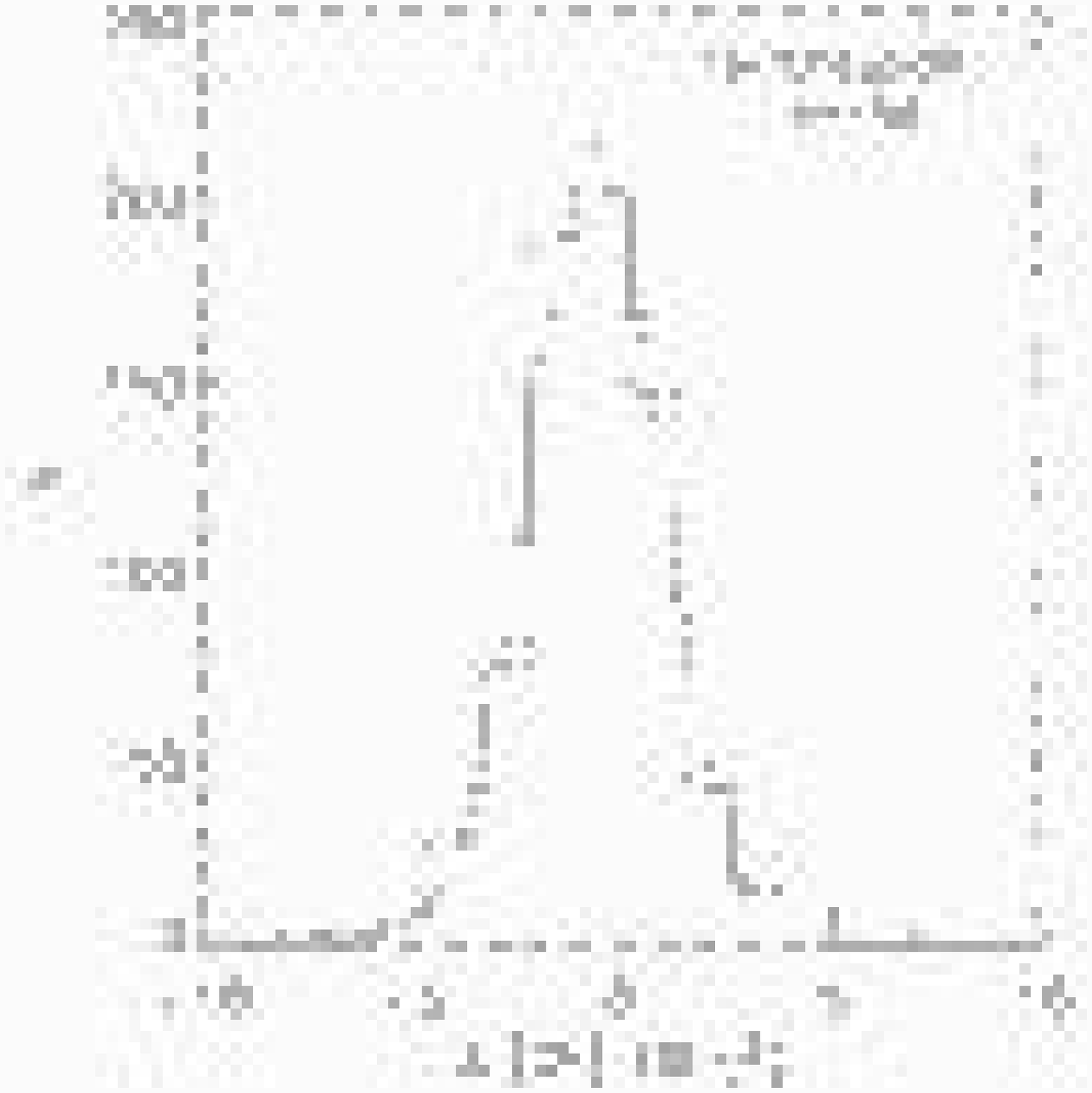}
\includegraphics[scale=0.4]{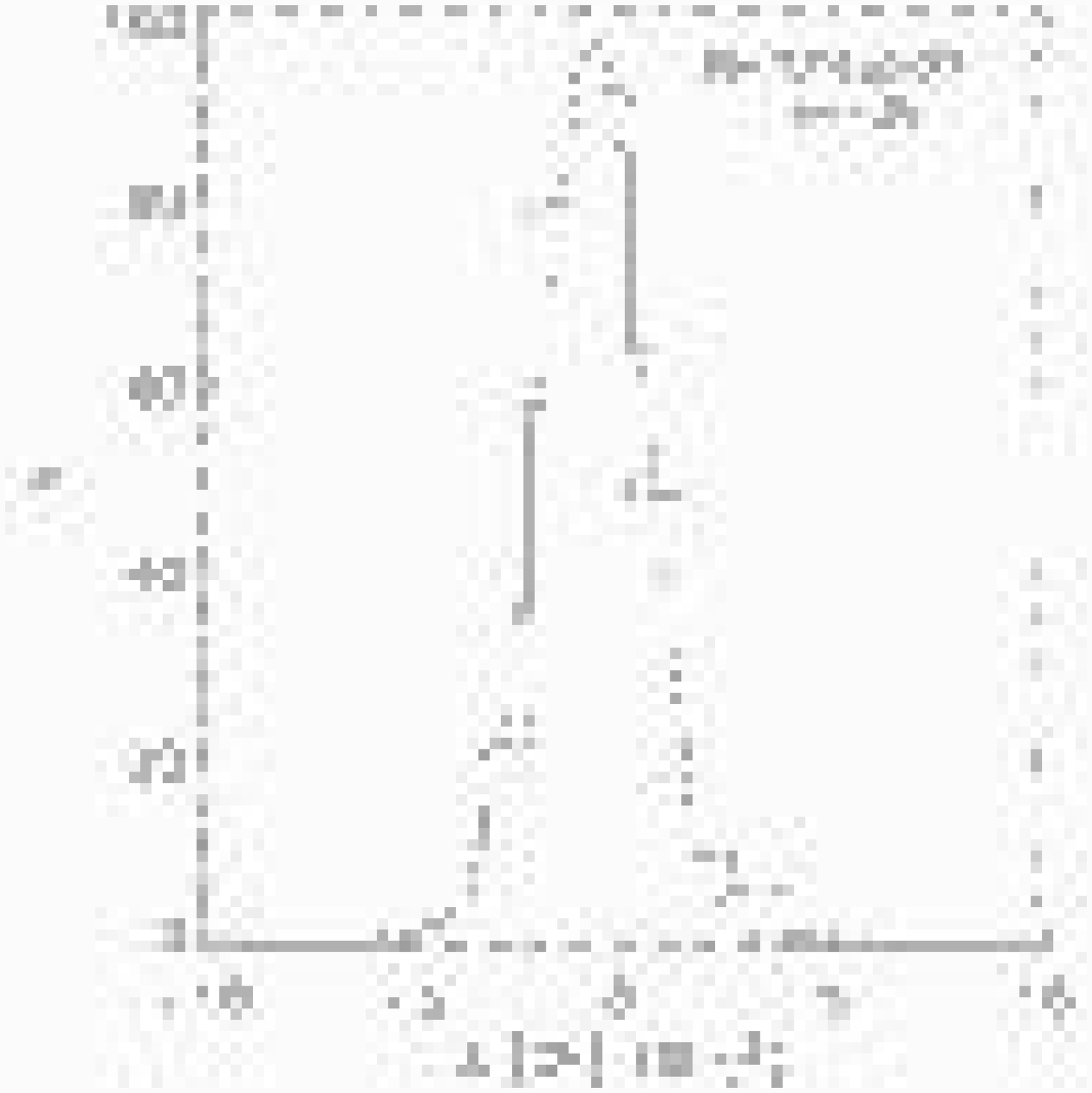}}
\caption{
\label{fig_ea1:err_from_double_obs_hd_gauss_fit_oii}
 The distribution of differences for the \oii\, line
from our duplicate observations. The four panels denote four different
bins in signal--to--noise ratio, i.e., clockwise from the top--left
panel, we have $5<{\rm S/N} <10$, $10<{\rm S/N} <15$, $20<{\rm S/N}
<25$ and $15<{\rm S/N} <20$.  We show in the dotted line the best fit
Gaussian to these distributions, which was then used to determine the
$1\sigma$ error on \oii\, EW as a function of signal--to--noise ratio.
}\end{figure}

\begin{figure}
\centering{\includegraphics[scale=0.4]{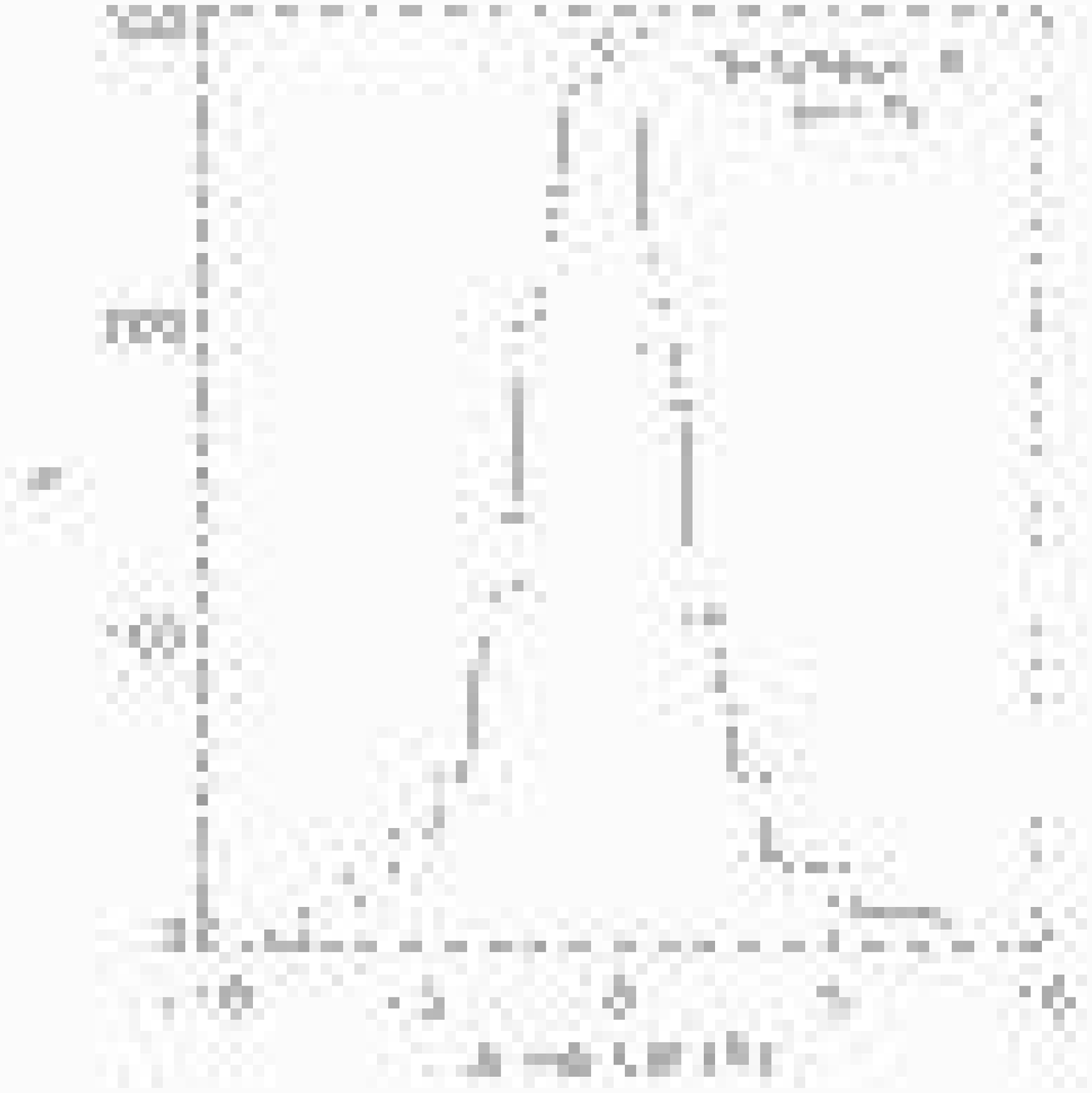}
\includegraphics[scale=0.4]{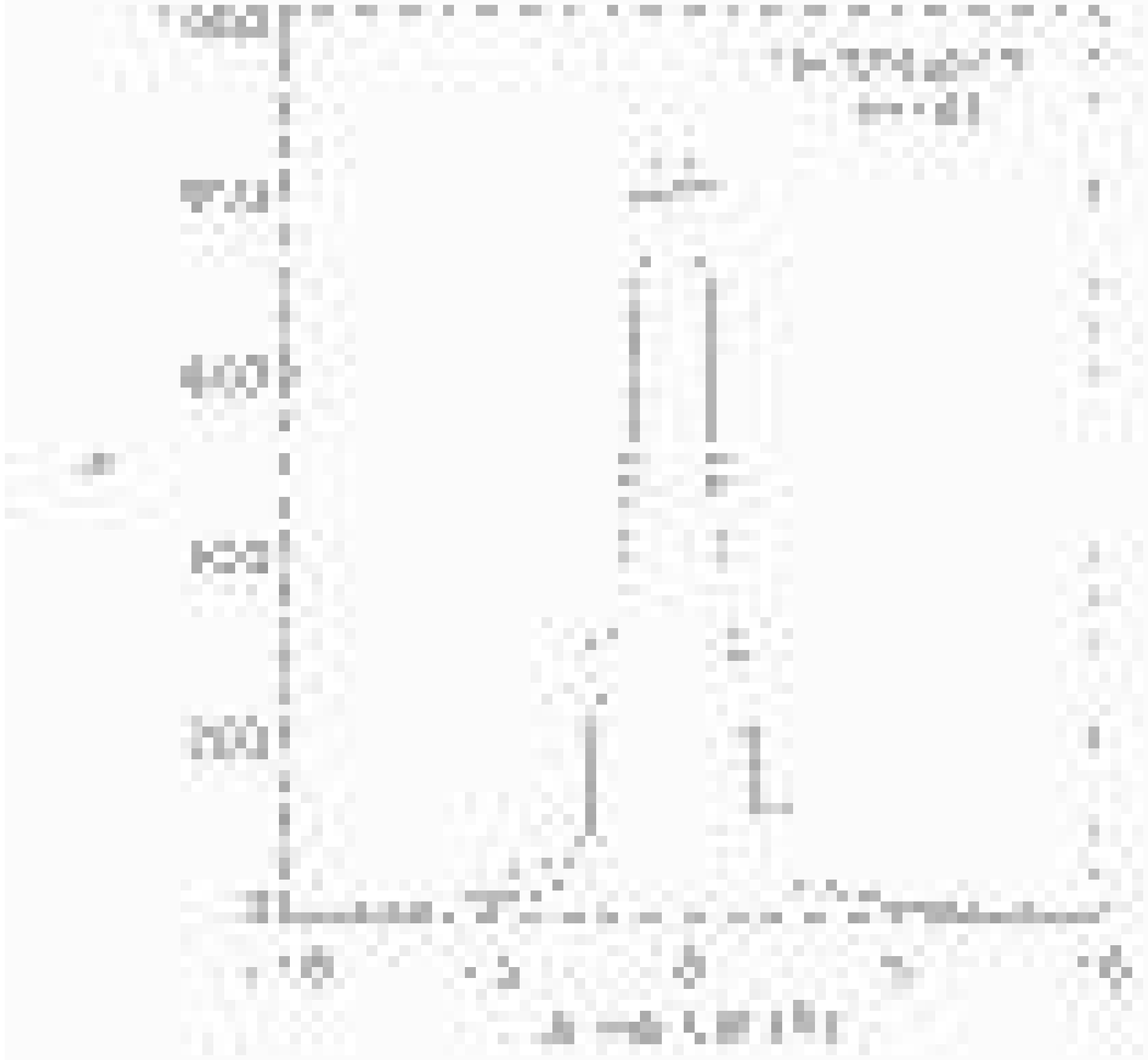}}
\centering{\includegraphics[scale=0.4]{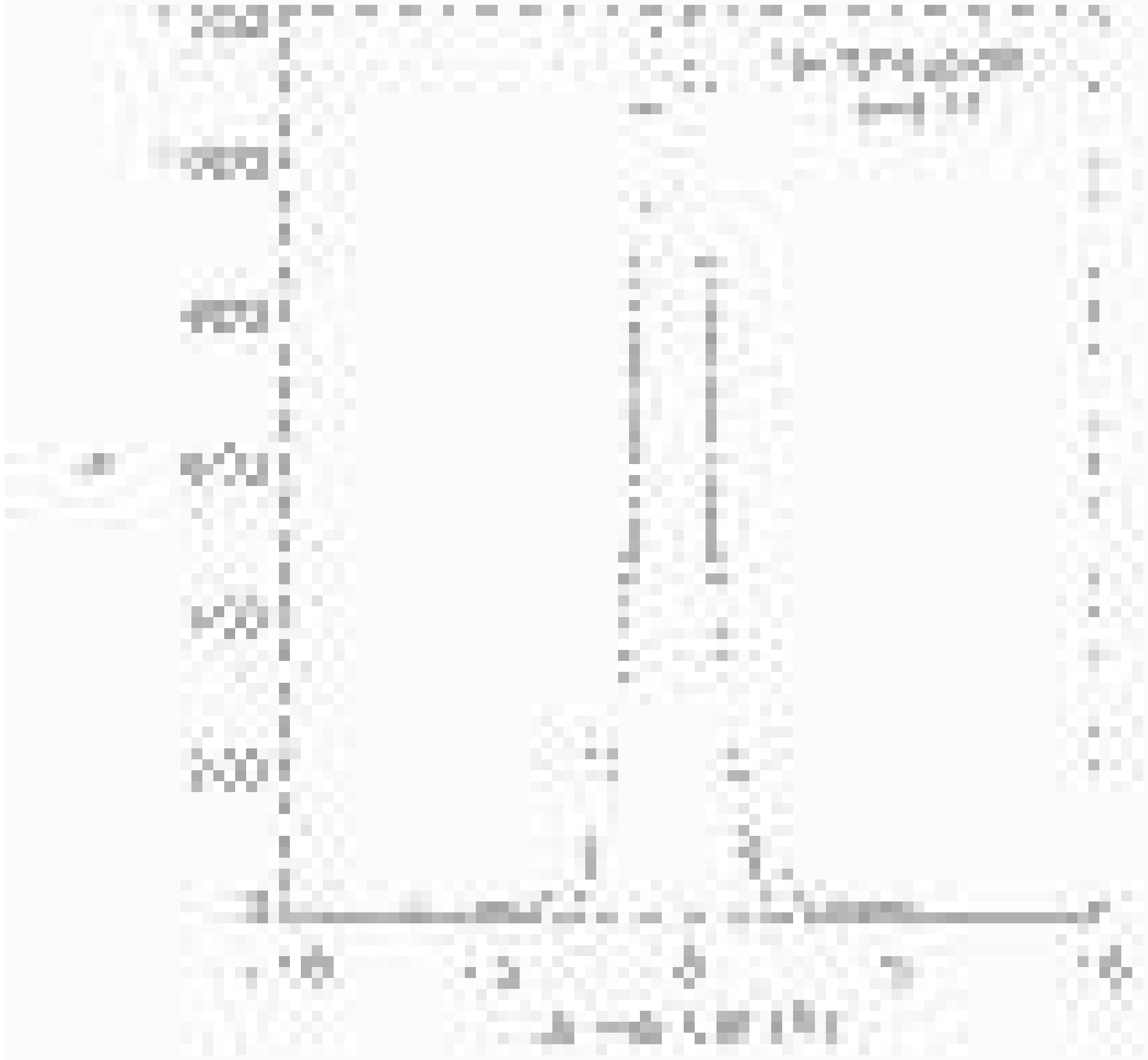}
\includegraphics[scale=0.4]{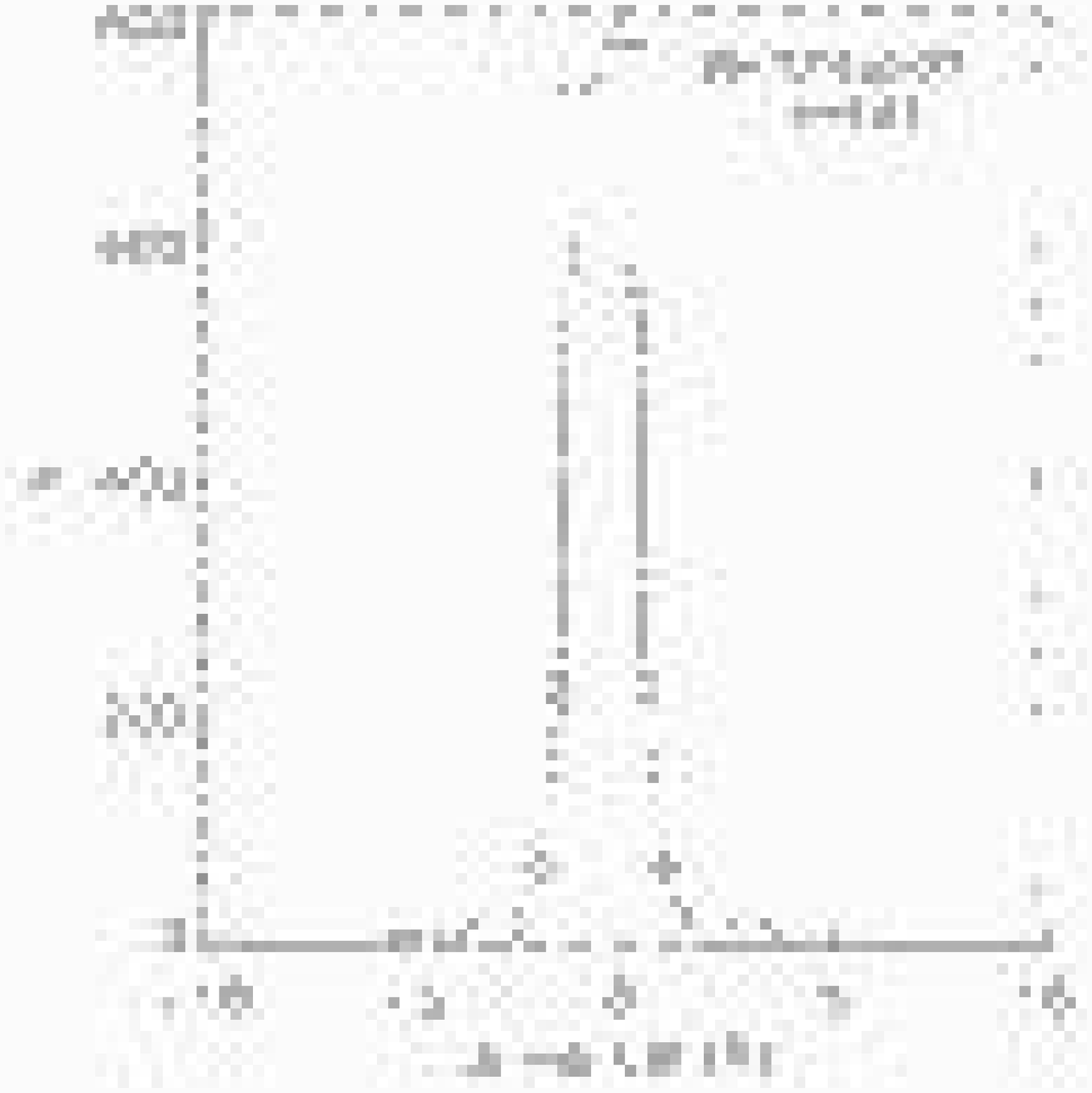}}
\caption{
\label{fig_ea1:err_from_double_obs_hd_gauss_fit_ha}
 The distribution of differences for the \ha\, line from our
duplicate observations. The four panels denote four different bins in
signal--to--noise ratio, i.e., clockwise from the top--left panel, we have
$5<{\rm s/n} <10$, $10<{\rm s/n} <15$, $20<{\rm s/n} <25$ and $15<{\rm s/n}
<20$.  We show in the dotted line the best fit Gaussian to these distributions,
which was then used to determine the $1\sigma$ error on \ha\, as a function of
signal--to--noise ratio.  }\end{figure}

We are now ready to select our sample of HDS galaxies using the non-parametric
measurements of the \hd\, EW (i.e., {\rm EW(H$\delta_{\rm max}$)}).  We
begin by imposing the following threshold on EW(H$\delta_{\rm max}$);

\begin{equation}
{\rm EW(H\delta_{\rm max})} - \Delta {\rm EW(H\delta_{\rm max})} > 4\AA,
\label{mainthres}
\end{equation}

 \noindent where $\Delta{\rm EW(H\delta_{\rm max})}$ is the error
 estimated from the $1\sigma$
 difference in H$\delta$ EWs between two observations of the same galaxy
 (see Figure \ref{fig_ea1:err_from_double_obs_hd}). 
 We have chosen this threshold ($4$\AA)
based on visual inspections of the data and our desire to select galaxies
similar to those selected by other authors (Zabludoff et al. 1996; Balogh et
al. 1999; Poggianti et al. 1999), i.e., galaxies with strong recent star
formation as defined by the \hd\, line. This threshold (Eqn. \ref{mainthres})
is applied without any emission--filling correction. For the signal--to--noise
ratios of our spectra (Figure \ref{fig_ea1:sng}), only galaxies with an observed
\hd\, of $\sim$5\AA\ satisfy Eqn. \ref{mainthres}, which is close to the
$5$\AA\ threshold used by Balogh et al. (1999) to separate normal
star--formating galaxies from post--starburst galaxies (see Figures 8 \& 9 in
their paper). Therefore, our HDS sample should be similar to those already in
the literature, but is still conservative enough to be inclusive of the many
different subsamples of \hd--strong galaxies, like k+a, a+k, A+em and e(a), as
discussed in Pogginati et al. (1999) and Balogh et al. (1999). We will present
a detailed comparison of our HDS sample with models of galaxy evolution in
future papers.

We call the sample of galaxies that satisfy Eqn. \ref{mainthres}
``Sample 1''.
 Sample 1 contains 2760 galaxies. Among these, 2526 galaxies come from
 the main SDSS galaxy sample and 234 were
 galaxies targeted for spectroscopy for other reasons, e.g., mostly
 because they were LRG galaxies (see Eisenstein et al. 2002), or 
 some were targeted as ``stars'' or ``quasars'' (see Richards et al. 2002). For comparison,
 if we remove the  $\Delta{\rm EW(H\delta_{\rm max})}$ term from the
equation, the number of HDS galaxies increases to 10788. Instead of
H$\delta_{\rm max}$, if we only use the narrow or wide window, the
number reduces to 811 and 2273 galaxies, respectively.

 We then apply the emission--filling correction to ${\rm
EW(H\delta_{\rm max})}$ for each galaxy  and select an additional sample of galaxies, which were
not already selected in Sample 1 via Eqn \ref{mainthres}, but now satisfy both
the following criteria;

\begin{equation}
\begin{array}{l}
{\rm EW(H\delta_{\rm max})}-\Delta{\rm EW(H\delta_{\rm max})} - \Delta{\rm EW(EF1)}  > 4\AA, \\
{\rm EW(H\delta_{\rm max})}-\Delta{\rm EW(H\delta_{\rm max})} - \Delta{\rm EW(EF2)}  > 4\AA,
\label{sample2thres}
\end{array}
\end{equation}

\noindent where $\Delta{\rm EW(EF1)}=0.57$\AA\ and $\Delta{\rm
 EW(EF2)}=0.4$\AA\ are the $1\sigma$ errors of the iterative method (EF1) and
 D4000 method (EF2) of the emission--filling correction discussed in Section
 \ref{ea1_error}. Therefore, this additional sample of galaxies represents systems
 that would only satisfy the threshold in Eqn \ref{mainthres} because of the
 emission--filling correction (in addition to galaxies already selected
 as Sample 1).  We call this sample of galaxies ``Sample 2'' and it
 contains 580 galaxies. Among them,
 483 galaxies come from the main SDSS galaxy sample and 97 galaxies were again
 targeted for spectroscopy for other reasons, e.g., LRG galaxies, ``stars''
 or ``quasars''. On average, Sample 2 galaxies have strong emission lines because,
 by definition, they have the largest emission--filling correction at the \hd\,
 line. We have imposed the above two criteria to control the number of extra
 galaxies scattered into the sample. If we relax these criteria (i.e.,
 remove both $\Delta{\rm EW(EF1)}$ and $\Delta{\rm EW(EF2)}$), then the sample
 would increase from 580 galaxies (in total) to 1171. 
 If we only use one of the
 emission filling corrections, the resulting number of HDS galaxies are
 1151 and 1467 for iteration method and D4000 method, respectively. 

\begin{figure}
\includegraphics[scale=0.7]{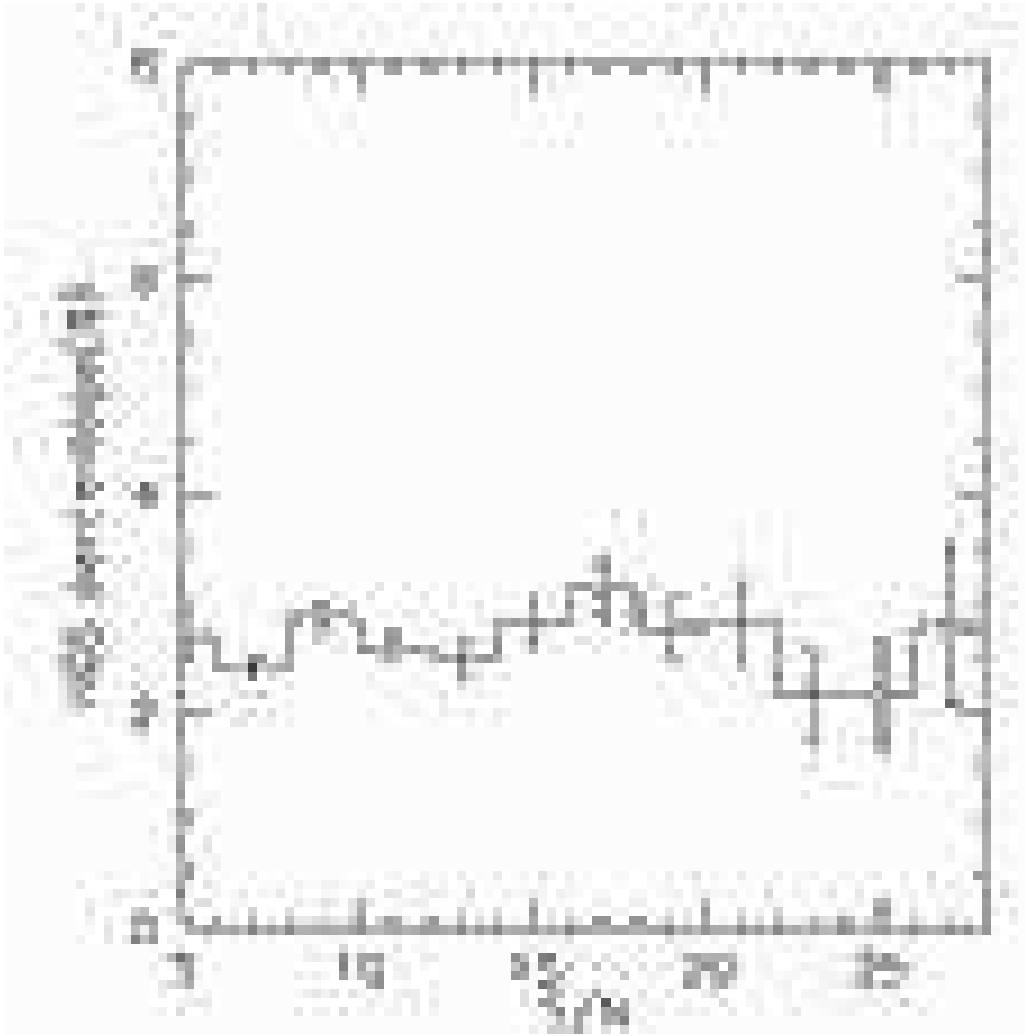}
\caption{
\label{fig_ea1:hds_sn}
The fraction of HDS galaxies as a function signal--to--noise ratio in the $g$
band. The error bars are $\sqrt{N}$, where $N$ is the number of
galaxies in each bin.}
\end{figure}

 In total, 3340 SDSS galaxies satisfy these criteria (Sample 1 plus Sample 2),
 and we present these galaxies as our catalog of HDS galaxies. We note that
 only 131 of these galaxies are securely identified as AGNs using the
 prescription of Kewley et al. (2002) and Gomez et al. (2003). In Figure
 \ref{fig_ea1:hds_sn}, we present the fraction of HDS galaxies selected as a
 function of their signal--to--noise ratio in SDSS $g$ band. It is reassuring that
 there is no observed correlation, which indicates that our selection technique
 is not biased by the signal--to--noise ratio of the original spectra.

For each galaxy in Samples 1 and 2, we present the unique SDSS Name (col.  1),
heliocentric redshift (col. 2), spectroscopic signal--to--noise ratio in the SDSS
photometric $g$ band (col. 3), Right Ascension (J2000; col. 4) and Declination
(J2000; col. 5) in degrees, Right Ascension (J2000; col. 6) and Declination
(J2000; col. 7) in hours, minutes and seconds, the rest--frame EW(\hd) (\AA,
col. 8), the rest--frame $\Delta$EW(\hd) (\AA, col. 9), the rest--frame
EW(\oii) (\AA, col. 10), the rest--frame $\Delta$EW(\oii) (\AA, col. 11), the
rest--frame EW(\ha) (\AA, col. 12), the rest--frame $\Delta$EW(\ha) (\AA,
col. 13), the SDSS Petrosian $g$ band magnitude (col. 14), the SDSS Petrosian
$r$ band magnitude (col. 15), the SDSS Petrosian $i$ band magnitude (col. 16),
the SDSS Petrosian $z$ band magnitude (col. 17; all magnitudes are extinction
corrected), the k--corrected absolute magnitude in the SDSS $r$ band
(col. 18), SDSS measured seeing in $r$ band (col. 19), concentration index
(col. 20, see Shimasaku et al. 2001 and Strateva et al. 2001 for definition).
In Column 21, we present the AGN classification based on the line indices of
Kewley et al. (2002), and in Column 22, we present our E+A classification
flag, which is defined in Section \ref{ea1_previous}.  An electronic version of
our catalog can be obtained at http://sdss2.icrr.u-tokyo.ac.jp/$\sim$yohnis/ea{\tt
}.\footnote{Mirror sites
are available at http://kokki.phys.cmu.edu/$\sim$tomo/ea, and \\
http://astrophysics.phys.cmu.edu/$\sim$tomo/ea}

In addition to presenting Samples 1 and 2, we also present a volume--limited
sample selected from these two samples but within the redshift range of
$0.05<z<0.1$ and with M$(r^*)<-20.5$ (which corresponds to $r=17.7$ at
$z=0.1$, see Gomez et al. 2003).  We use Schlegel, Finkbeiner \& Davis (1998)
to correct for galactic extinction and Blanton et al. (2002b; v1\_11) to
calculate the k--corrections.  In Table \ref{ea1_tab:frequency}, we present the
percentage of HDS galaxies that satisfy our criteria. In this table, the
number of galaxies in the whole sample (shown in the denominator) changes
based on the number of galaxies that could have had their \hd, \oii\, and
\ha\, lines measured because of masked pixels in the spectra.

We note here that we have not corrected our sample for possible aperture
effects, except restricting the sample to $z\geq0.05$: A 3 arcsec fiber
corresponds to $2.7h_{75}^{-1}$ kpc at this redshift, which is comparable to
the half--light radius of most our galaxies (see also Gomez et al. 2003).  We
see an increase of $0.33$\AA\, ($\simless10\%$) in the median observed \hd\ EW
for the whole HDS sample over the redshift range of our volume--limited sample
($0.05<z<0.1$). This is probably caused by more light from the disks of
galaxies coming into the fiber at higher redshifts. We see no such trend for
the subsample of true E+A galaxies (see Section \ref{ea1_discussion}) in our HDS
sample.

\section{Discussion}
\label{ea1_discussion}

In this Section, we compare our sample of HDS galaxies against previous
samples of such galaxies in the literature. Further analysis of the global
properties (luminosity, environment, morphology, {\it etc.}) of these galaxies
will be presented in future papers.

\subsection{Comparison with Previous Work}
\label{ea1_previous}

In this Section, we present a preliminary comparison of our SDSS HDS galaxies
with other samples of post--starburst galaxies in the literature.  We have
attempted to replicate the selection criteria of these previous studies as
closely as possible, but in some cases, this is impossible. Furthermore, we
have not attempted to account for systematic differences in the distribution
of signal--to--noise ratios, differences in the spectral resolutions, and
differences in the selection techniques used by different authors, e.g.,
we are unable to fully replicate the criteria of Zabludoff et al. (1996) as we
do not yet possess accurate measurements of the H$\gamma$ and H$\beta$ lines.
Therefore, these are only crude comparisons and any small differences seen
between the samples should not be over--interpreted until a more detailed
analysis can be carried out. We summarize our comparison in Table
\ref{ea1_comparison} and discuss the details of these comparisons below: Table
\ref{ea1_comparison} does, however, demonstrate once again the rarity of
\hd--strong galaxies, especially at low redshift, as well as illustrating the
sensitivity of their detection to the selection criteria used.

We first compare our sample to that of Zabludoff et al. (1996), which is the
most similar to our work, especially as the magnitude limits of the LCRS (used
by Zabludoff et al. 1996) and the SDSS are close, thus minimizing possible
bias.  Zabludoff et al. (1996) selected E+A galaxies from the LCRS using the
following criteria; a redshift range of $15,000 < cz < 40,000\, {\rm km\,
s^{-1}}$, a signal--to--noise ratio of $>8$ per pixel, an EW of \oii\, of $>-2.5$\AA,
and an average EW for the three Balmer lines (\hd, H$\gamma$ and H$\beta$) of
$>5.5$\AA. Using these four criteria, Zabludoff et al. (1996) selected 21 LCRS
galaxies as E+A galaxies, which corresponds to 0.2\% of all LCRS that satisfy
the same signal--to--noise ratio and redshift limits.  We find 80 SDSS galaxies
(from the whole 95479 SDSS galaxies  analyzed here) which lie in the same redshift range
and satisfy \oii\ EW $>-2.5$\AA  as used by Zabludoff et al. (1996) as well as
having H$\delta$ EW of $>$5.5\AA, which should be close to the average of the
EW of the three Balmer lines (\hd, H$\gamma$ and H$\beta$) used by Zabludoff
et al. (1996). Of these 80 galaxies, 71 are in our HDS sample.  Given all
the caveats discussed above, it is reassuring that we have found HDS
galaxies at a similar frequency (see Table \ref{ea1_comparison}) as Zabludoff et
al. (1996) and it suggests that our criteria are consistent with theirs.

Several other authors have used similar criteria to Zabludoff et al. (1996) to
search for post--starburst galaxies in higher redshift samples of galaxies.
For example, Fisher et al. (1998) used an average EW of the H$\delta$,
H$\gamma$ and H$\beta$ lines of $>4$\AA\ and an EW of \oii\, of $<5$\AA. With
these criteria, they found 4.7\% of their galaxies were E+A galaxies.
Similarly, Hammer et al. (1997) used an average of H$\delta$, H$\gamma$ and
H$\beta$ of $>5.5$\AA, an EW of \oii\, of $<$5--10 \AA\ and $M_B<$-20 to
select E+A galaxies from the CFRS field galaxy sample. They found that 5\% of
their galaxies satisfied these criteria.

We have also attempted to replicate the selection criteria of Dressler et
al. (1999) and Balogh et al. (1999) as closely as possible, using 26863 SDSS
galaxies in the volume--limited sample with measured \hd, \oii\,
and\ha. 
 For example, the MORPHS collaboration of 
 Dressler et al. (1999) and Poggianti et al. (1999) selected E+A galaxies using
 \hd\, EW of $>3$ \AA\ and \oii\, EW of $<5$\AA\ . Using these criteria,
 they found a significant excess of E+A galaxies in their 10 high redshift
 clusters (21\%), compared with the field region (6\%). On the other hand, Balogh
 et al. (1999) selected E+A galaxies using \hd\, EW of $>5$\AA\ and \oii\,
 EW of $<5$\AA\ . They found instead $1.5\pm0.8$\% of cluster galaxies were
 classified as E+A galaxies compared to $1.2\pm0.8$\% for the field (brighter
 than M$(r)=-18.8+5{\rm log}h$ after correcting for several systematic
 effects).

 In Table \ref{ea1_comparison}, we present a qualitative comparison of our HDS
 sample with these two higher redshift studies and, within the quoted errors, the
 frequencies of HDS galaxies we observe are consistent with their values.
 However, we caution the reader not to overinterpret these numbers for
 several reasons. First, we are comparing a low redshift sample ($z<0.1$) of
 HDS galaxies to high redshift studies ($z\simeq0.5$) of such galaxies, and we
 have not accounted for possible evolutionary effects or observational
 biases. In particular, we are comparing our sample against the corrected
 numbers presented by Balogh et al. (1999), which attempt to account for
 scatter in the tail of the \hd\, distribution due to the large intrinsic
 errors on \hd\, measurements, while Dressler et al. (1999) do not make such a
 correction.  Secondly, we are comparing a field sample of HDS galaxies to
 field samples of galaxies selected in the cluster field.
  Finally,
 the luminosity limit of our volume--limited sample is brighter than the high
 redshift studies, which may account for some of the discrepancies.

Finally, we note that the original E+A phenomenon in galaxies, as discussed by
Dressler \& Gunn (1983, 1992), was defined to be a galaxy that possesses
strong Balmer absorption lines, but with no emission lines, i.e., a
galaxy with the signature of recent star--formation activity (A stars), but no
indication of on--going star--formation (e.g., nebular emission
lines). Given the quality of the SDSS spectra, we can re--visit this specific
definition and select galaxies from our sample that possess less than
$1\sigma$ detections of both the \ha\, and \oii\, emission lines, i.e.,
${\rm EW([OII]) + \Delta EW([OII]) \geq0}$\AA\, and ${\rm EW(H\alpha) + \Delta
EW(H\alpha)\geq0}$\AA).  We find that only $3.5\pm0.7$\% (25/717) of galaxies
in the volume limited HDS sample satisfy such a strict criteria (see Table
\ref{ea1_tab:frequency}). We show example spectra of these galaxies in Figure
\ref{fig_ea1:true_ea} and highlight them in the catalog using the E+A
classification flag. This exercise demonstrates that true E+A galaxies -- with
no, or little, evidence for on--going star--formation -- are extremely rare at
low redshift in the field, i.e., $0.09\pm0.02$\% of all SDSS galaxies in our volume--limited sample.

\subsection{HDS Galaxies with Emission--lines}

In this Section, we examine the frequency of nebular emission lines (\oii,
\ha) in the spectra of our HDS galaxies. This is possible because of the large
spectral coverage of the SDSS spectrographs which allow us to study both the
\oii\, and \ha\, emission lines for all galaxies out to a redshift of
$\simeq0.35$.  We begin by looking at HDS galaxies that possess both the \ha\,
and \oii\, emission lines.  Using the criteria that both \oii\, and \ha\,
lines must be detected at $>1\sigma$ significance (i.e., ${\rm EW([OII])
+ \Delta EW([OII]) <0}$\AA\, and ${\rm EW(H\alpha) + \Delta EW(H\alpha)<0}$\AA),
we find that $89\pm5$\% (643/717) of HDS galaxies in our volume--limited
sample are selected. Of these, 131 HDS galaxies possess a robust detection of
an AGN, based on the line indices of Kewley et al. (2001), similar to the AGN
plus post--starburst galaxy found recently in the 2dFGRS (see Sadler, Jackson
\& Cannon 2002). Therefore, a majority of these emission line \hd\, galaxies
may have on--going star formation and are similar to the e(a) and A+em
subsample of galaxies discussed by Poggianti et al. (1999) and Balogh et
al. (1999). We show in Figure \ref{fig_ea1:ea_em} examples of these HDS galaxies
that possess both the \oii\, and \ha\, emission lines.  The median SFR of
these galaxies (calculated from \ha\, flux, see Kennicutt 1998) is
$\simeq0.5{\rm M_{\odot}/yr}$, with a maximum observed SFR of $50{\rm
M_{\odot}/yr}$.  We note that these SFRs have not been corrected for dust
 extinction or aperture effects. The true SFR for the whole galaxy could
 be larger than these values by a factor of 5-10  (see Hopkins et al. in
 prep for more detail). 

We next examine the frequency of HDS galaxies with \oii\, emission, but no
detectable \ha\, emission. Using the criteria of EW(\ha) $<$1 $\sigma$
detection and EW(\oii) $>$ 1$\sigma$ detection (i.e., ${\rm
EW([OII]) + \Delta EW([OII]) <0}$\AA\, and ${\rm EW(H\alpha) + \Delta EW(H\alpha)
\geq0}$\AA), we find that $2.9\pm0.65$\% (21/717) of our HDS galaxies in the
volume--limited sample satisfy these criteria. The presence of \oii\, demonstrates
that the galaxy may possess on--going star formation activity, yet the lack of
the \ha\, emission is curious. Possible explanations for this phenomena are
strong self-absorption of \ha\, by the many A stars in the galaxy and/or
metallicity effect which could increase the \oii\, emission relative to \ha\,
emission. We show several examples of these galaxies in Figure \ref{fig_ea1:oii},
and the lack of \ha\, emission is clearly visible. The fact that many of these
galaxies possess strong \nii\, lines (flanking the \ha\, line) indicates
strong self--absorption is a likely explanation for the missing \ha\, emission
line. Median [OII] EW of these galaxies is 1.3 \AA. Compared with 11.5 
\AA\ of HDS galaxies with both [OII] and H$\alpha$ emission, these
galaxies have much lower amount of [OII] in emission.

 Finally, we find that $3.8\pm 0.7$\% (27/717) of our HDS galaxies in our
 volume--limited sample satisfy the criteria of  ${\rm EW([OII]) +
 \Delta EW([OII]) \geq0}$\AA\, and ${\rm EW(H\alpha) + \Delta EW(H\alpha)
 <0}$\AA; i.e., HDS with no $[OII]$ emission but with H$\alpha$ emission.
  Only 52 of our HDS galaxies in  the volume limited sample have just no
 \oii\ emission detected (only EW([OII]) + $\Delta$ 
 EW([OII]) $\geq$0\AA). Therefore, 52$\pm$12\% (27/52) of the HDS galaxies with no
 detected \oii have detected \ha\ emission.  The existence of such galaxies
 has ramifications on high redshift studies of post--starburst galaxies, as
 such studies use the \oii\, line to constrain the amount of on--going
 star--formation within the galaxies (e.g., Balogh et al. 1999,
 Poggianti et al. 1999).  Therefore, if the \ha\, emission comes from 
star--formation activity, then these previous high redshift studies of post
starburst galaxies may be contaminated by such galaxies.  A possible
explanation for the lack of \oii\, emission is dust
extinction. Miller \& Owen (2002) recently
found dusty star--forming galaxies which do not possess \oii\, in emission, but
have radio fluxes consistent with on--going star formation activity. This
explanation would also be consistent with the lower signal--to--noise
ratio we observe
in the blue--end of the SDSS spectra of these galaxies, relative to the
signal--to--noise ratio seen in the red--end of their spectra. Median H$\alpha
$ EW of these galaxies are 1.5\AA, whereas that of HDS galaxies with
both [OII] and H$\alpha$ emission is 25.9\AA.

\section{Conclusions}
\label{ea1_conclusion}

We present in this Chapter the largest, most homogeneous, search to date for
\hd--strong galaxies (i.e., post--starburst galaxies, E+A's, k+a's,
a+k's {\it etc.}) in the local universe. We provide the astronomical community
with a new catalog of such galaxies selected from the Sloan Digital Sky Survey
(SDSS) based solely on the observed strength of the \hd\, hydrogen Balmer
absorption line within the spectrum of the galaxy. We have carefully studied
different methodologies of measuring this weak absorption line and conclude
that a non--parametric flux--summing technique is most suited for an automated
application to large datasets like the SDSS, and that it is more robust for the observed
signal--to--noise ratios available in these SDSS spectra.  We have studied the effects of
dust extinction, emission--filling and stellar absorption upon the measurements
 of our \hd\, lines and have determined the external error on our measurements
 as a function of signal--to--noise ratio, using duplicate observations of 11358
 galaxies in the SDSS. In total, our catalog of \hd--strong (HDS) galaxies
 contains 3340 galaxies selected from the 95479 galaxies in the Sloan Digital Sky Survey (at the
 time of writing). Our catalog will be a useful basis for the future
 studies to understand the nature of such galaxies and the comparison
 with studies of such systems at higher redshifts.

The measured abundance of our \hd--strong (HDS) galaxies is $2.6\pm$0.1\% of
all galaxies within a volume--limited sample of $0.05<z<0.1$ and
M($r^*$)$<-20.5$, which is consistent with previous studies of post--starburst
galaxies in the literature. We find that only 25 galaxies ($3.5\pm0.7$\%) of
HDS galaxies in this volume limited sample show no, or little, evidence for
\oii\, and \ha\, emission lines. This indicates that true E+A galaxies (as
originally defined by Dressler \& Gunn) are extremely rare objects,
i.e., only $0.09\pm0.02$\% of all galaxies in our volume--limited
sample. In 
contrast, $89\pm5$\% of our HDS galaxies have significant detections of the
\oii\, and \ha\, emission lines. Of these, 131 galaxies are robustly
classified as Active Galactic Nuclei (AGNs) and therefore, the majority of these
emission line HDS galaxies are star--forming galaxies, similar to the e(a) and
A+em galaxies discussed by Poggianti et al. (1999) and Balogh et al. (1999).
We study the global properties of our HDS galaxies in further detail in
 Appendix \ref{EA2}.

\clearpage

\clearpage

\begin{figure}
\includegraphics[scale=0.7]{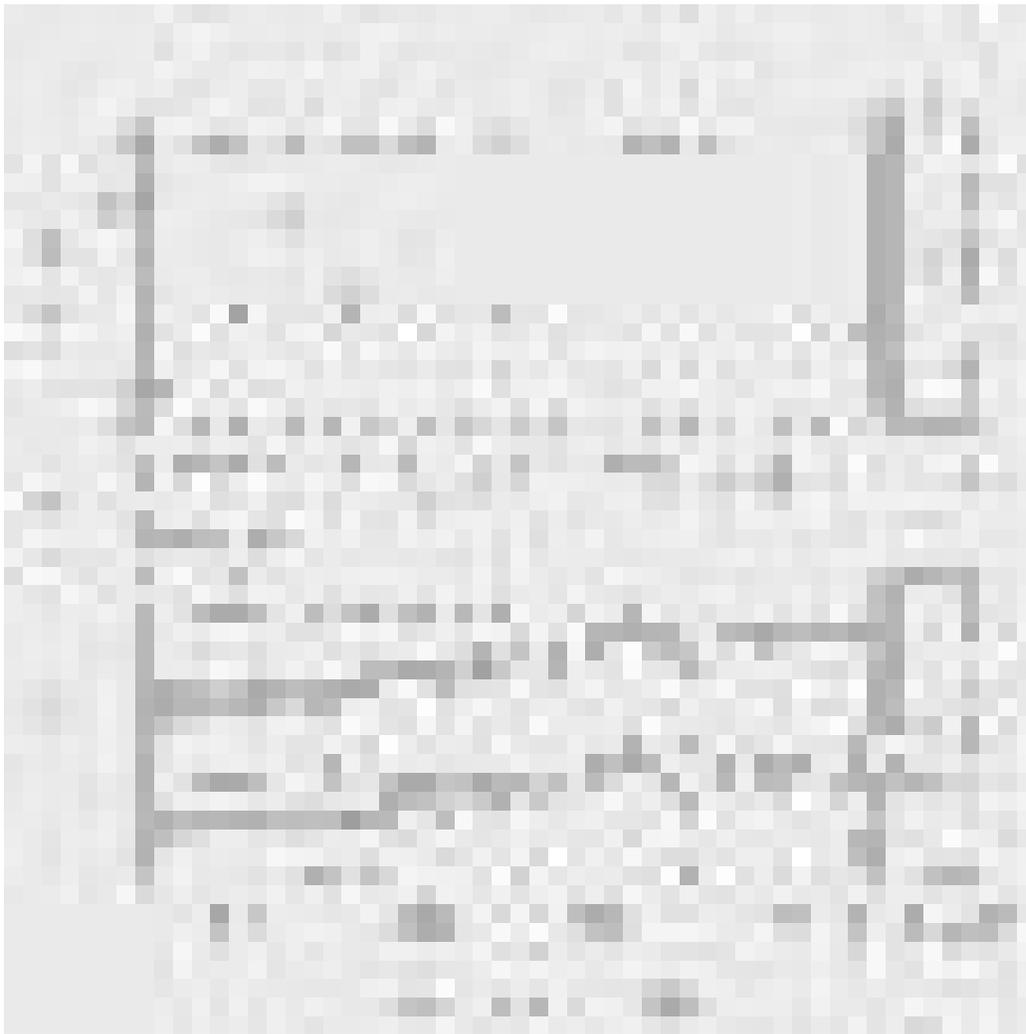}
\caption{
\label{fig_ea1:true_ea}
 Five examples of spectra for the ``true E+A'' subsample
of HDS galaxies discussed in Section \ref{ea1_discussion}.  These galaxies
possess strong Balmer absorption lines, but have no, or little,
detected \oii\, or \ha\, emission. 
 The label shows signal--to--noise ratio,   \hd\, EW, and measured redshift.
}
\end{figure}

\begin{figure}
\includegraphics[scale=0.7]{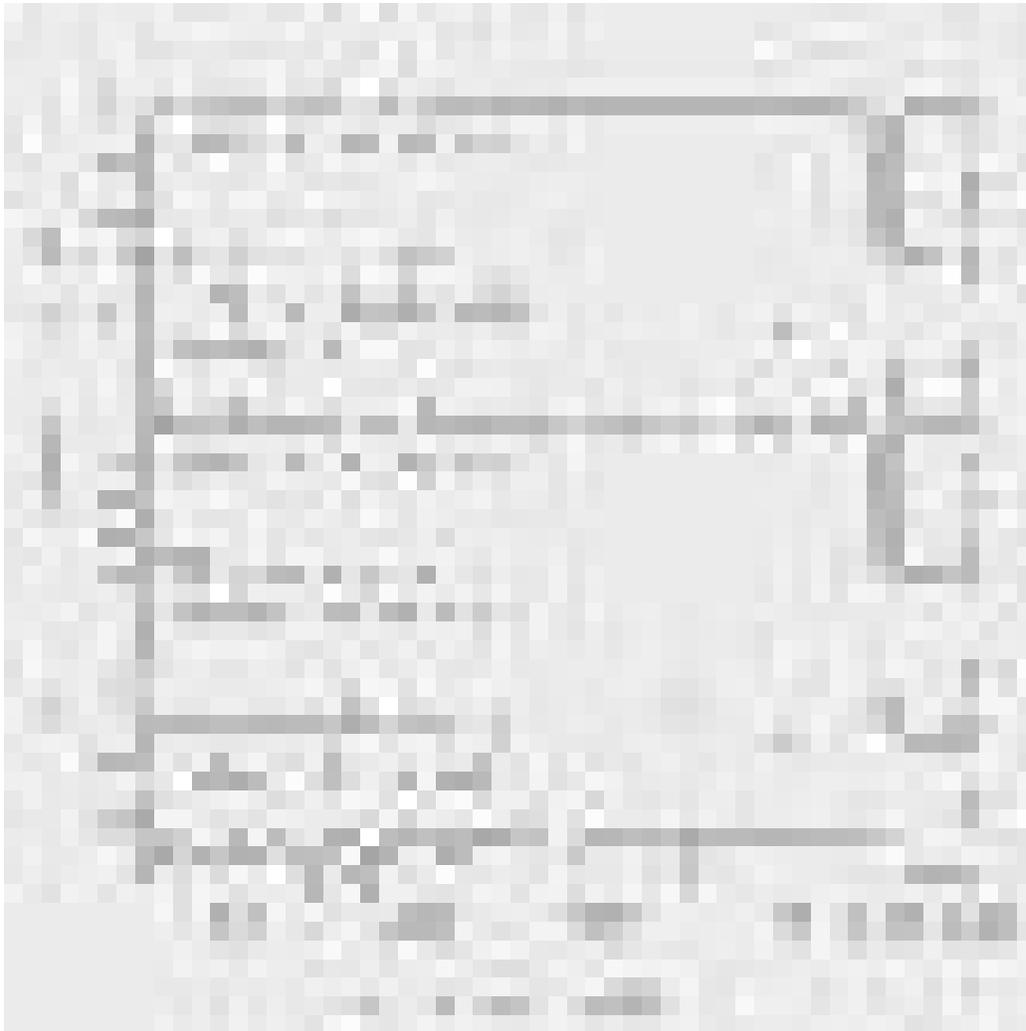}
\caption{
\label{fig_ea1:ea_em}
 Five example spectra of our HDS galaxies that
possess detected \oii\, and \ha\, emission lines. 
 The label shows signal--to--noise ratio,   \hd\, EW, and measured redshift.
}
\end{figure}

\begin{figure}
\includegraphics[scale=0.7]{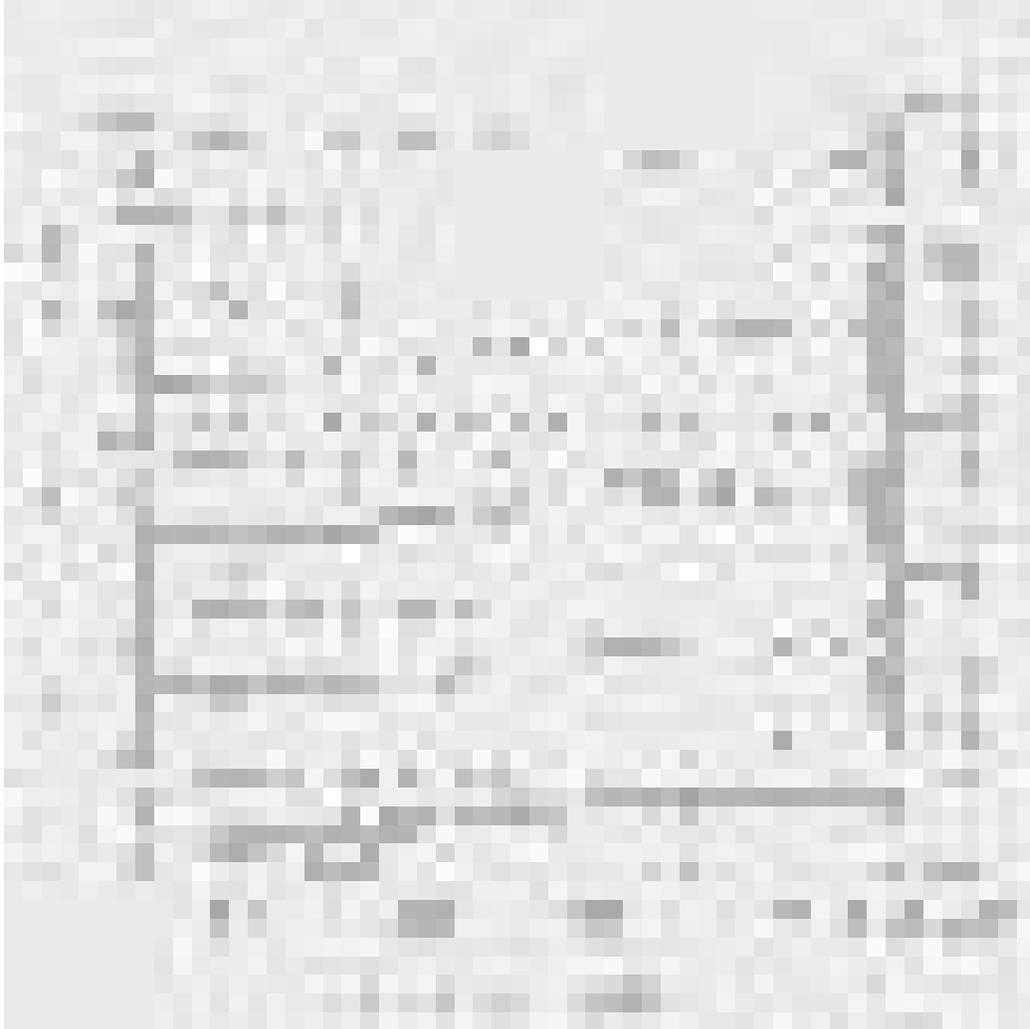}
\caption{
\label{fig_ea1:oii}
 Five example spectra of our HDS galaxies that
possess detected \oii\, emission lines, but no detected \ha. 
  The label shows signal--to--noise ratio,   \hd\, EW, and measured redshift.
}
\end{figure}

\clearpage

\begin{table}[h]
\begin{center}
\caption{
\label{ea1_tab:wavelength}
The wavelength ranges used to measure our \hd, \oii\, and \ha\, EWs.
}
\begin{tabular}{llll}
\hline
  & Blue continuum  & Line & Red continuum \\
\hline
\hline
H$\delta$ (narrow) &  4030-4082\AA &    4088-4116\AA & 4122-4170\AA  \\
H$\delta$ (wide)   &  4030-4082\AA &    4082-4122\AA & 4122-4170\AA  \\ 
 $[OII]$           &  3653-3713\AA &    3713-3741\AA & 3741-3801\AA  \\ 
H$\alpha$          &  6490-6537\AA &    6555-6575\AA & 6594-6640\AA  \\ 
\hline
\end{tabular}
\end{center}
\end{table}

\begin{table}[h]
\begin{center}
\caption{
\label{ea1_equation} Coefficients of third order polynomial fits to the error distributions shown in Figures \ref{fig_ea1:err_from_double_obs_hd},
\ref{fig_ea1:err_from_double_obs_oii} and \ref{fig_ea1:err_from_double_obs_ha}.
}
\begin{tabular}{lllll}
\hline Line & a$_0$ & a$_1$ & a$_2$ & a$_3$  \\ \hline \hline
 H$\delta$ &  2.98 & $-$0.28 & 0.012 &   $-$0.00018\\
 \oii &    4.96 &  $-$0.39 &     0.014 & $-$0.00016\\
 H$\alpha$ &  3.74 & $-$0.36 &  0.014 &  $-$0.00017\\
\hline \hline
\end{tabular}
\end{center}
\end{table}

\begin{table}[h]
\begin{center}
\caption{
\label{ea1_tab:frequency}
The frequency of finding HDS galaxies.}
\begin{tabular}{llll}
\hline
 Category & \% (All galaxies) & \% (Volume Limited) \\
\hline
\hline
Whole HDS sample  & 3340/95479 (3.50$\pm$0.06\%)  & 717/27014 (2.6$\pm$0.1\%) \\
True ``E+A''      & 140/94770 (0.15$\pm$0.01\%)   & 25/26863 (0.09$\pm$0.02\%) \\ 
\hline
\end{tabular}
\end{center}
\end{table}

\begin{table}[h]
\begin{center}
\caption{
\label{ea1_comparison}
A comparison of our HDS sample of galaxies to previous work in the literature.
 }
\begin{tabular}{lllll}
\hline Author & Balmer lines & Emission & Their \% (field) & Our \% \\ \hline \hline
Zabludoff et al. & H$\delta>$5.5\AA & [OII]$>-$2.5\AA
& 0.19$\pm$0.04\% & 0.16$\pm$0.02\% (80/49994)  \\ 
Poggianti et al. & H$\delta>$3 \AA & [OII]$>-$5 \AA & 6$\pm$3\% & 5.79$\pm$0.15\% (1565/27014) \\ 
Balogh et al. & H$\delta>$5 \AA & [OII]$>-$5 \AA & $1.2\pm0.8$\%  & 0.74$\pm$0.05\% (200/27014) \\
\hline \hline
\end{tabular}
\end{center}
\end{table}

\newpage

\clearpage

\chapter{Merger/Interaction Origin of E+A Galaxies}\label{EA2}

\section{Introduction}

  Dressler \& Gunn (1983; 1992) found galaxies with mysterious spectra while
 investigating high redshift cluster galaxies.
 The galaxies had strong Balmer absorption lines with no
 emission in either [OII] or H$\alpha$. These galaxies are called
 ``E+A''
 galaxies since their spectra looked like a superposition of that of
 elliptical galaxies (Mg$_{5175}$, Fe$_{5270}$ and Ca$_{3934,3468}$
 absorption lines) and that of A-type stars (strong
 Balmer absorption)\footnote{Since some of E+A galaxies are
 found to have disk-like morphology (Franx 1993; Couch et al. 1994;
 Dressler et al. 1994; Caldwell et al. 1997; Dressler et al. 1999),
 these galaxies are sometimes called 
 ``k+A'' galaxies. However, considering our findings in Section \ref{Jun
 1 12:00:59 2003}, we call them ``E+A''
 throughout this work.}.  The
 existence of strong Balmer absorption lines shows 
 that these galaxies have experienced starburst recently (within a
 Gyr).  However, these galaxies do not show any sign of on-going star
 formation as non-detection in [OII] or H$\alpha$ emission lines
 indicates.  
   Therefore E+A galaxies are interpreted as a post-starburst galaxy,
 that is,  a galaxy which truncated starburst suddenly (Dressler \& Gunn 1983,
 1992; Couch \& Sharples 1987; MacLaren, Ellis, \& Couch 1988; Newberry
 Boroson \& Kirshner 1990; 
 Fabricant, McClintock, \& Bautz 1991; Abraham et al. 1996).
  The reason why they
 started star burst, and why they abruptly stopped starburst remains one
 of the mysteries in galaxy evolution. 

  At present, there seem to be three popular explanations to the E+A phenomenon. 

\begin{itemize}
 \item Cluster-galaxy interaction
 \item Galaxy-galaxy interaction (in the field)
 \item Dust enshrouded star formation
\end{itemize}

    {\bfseries (i)\ Cluster-Galaxy\ Interaction.}

  At first, E+A galaxies are found in cluster regions, both in low
  redshift clusters (Dressler 1987;  Franx  1993; Caldwell et al.  
  1993, 1996;  Caldwell \& Rose 1997; Castander et al. 2001; Rose et
  al. 2001) and high redshift 
  clusters (Sharples et   al. 1985; Lavery \& Henry 1986; Couch \&
  Sharples 1987; Broadhurst,   Ellis, \& Shanks 1988; Fabricant,
  McClintock, \& Bautz 1991; Belloni   et al. 1995; Barger et al. 1996; Fisher et
  al. 1998; Morris et al. 1998; Couch et al. 1998;   Dressler et al. 1999). Therefore a
  cluster specific phenomenon was thought to be responsible for the
  violent star formation history of E+A galaxies. A ram-pressure
  stripping model (Spitzer \& Baade 1951, Gunn \& Gott 1972, Farouki \&
  Shapiro  1980; Kent 1981; Abadi, Moore \& Bower 1999, Fujita \& Nagashima 1999,
 Quilis, Moore \& Bower 2000; Fujita 2003)
  can first accelerate star formation of cluster galaxies and later turn it
  off. Galaxy-galaxy merger/interaction in cluster regions is also a good candidate to
  explain E+A phenomena (Lonsdale, Persson, \& Matthews 1984; Kennicutt
  et al. 1987; Sanders et al. 1988; Thompson 1988; Larvery \& Henry
  1988,1994; Lavery et al. 1992) although several authors pointed out
  that merging/interaction is difficult to 
  happen in cluster core regions since relative velocities of galaxies are too
  high (Ostriker 1980; Binney  \& Tremaine 1987; Mamon
 1992; Makino \& Hut 1997).
   Other candidate mechanisms include galaxy
  harassment (Moore et al. 1996, 1999), cluster tidal forces (Byrd 
 \& Valtonen 1990, Valluri 1993; Fujita 1998), removal and consumption of the
  gas  (strangulation; Larson, Tinsley \& Caldwell 1980, Balogh et
  al. 1997,1998,1999,2000; Bekki et al. 2001; Mo \& Mao 2002; Oh \& Benson
  2003), and evaporation of the cold gas in disk galaxies via heat
  conduction from the surrounding hot ICM (Cowie \& Songaila 1977;
  Fujita 2003). Fujita et al. (1999b) discussed that a cluster-cluster
  merger can increase the ram-pressure, and thus, can increase the
  fraction of post-starburst galaxies such as E+As. 

    Observationally strong
  evolution in colors of cluster galaxies has been discovered, in a
  sense that there are more 
  blue galaxies in higher redshift clusters (the so-called 
  Butcher-Oemler effect; Butcher \& Oemler 1978,
  1984; Couch \& Sharples 1987; Rakos \& Schombert 1995; Couch et
  al. 1994,1998; Margoniner \& De 
  Carvalho 2000; Margoniner et al. 2001; Ellingson  et al. 2001; Kodama
  \& Bower 2001; Goto et al. 2003a). E+A galaxies appeared 
  to fit in the Butcher-Oemler effect naturally, as blue, star-forming
  galaxies at high redshifts turn off their star formation abruptly,
  changing their spectra to E+A type, and they become red,
  non-star-forming galaxies at low redshifts. In addition, if E+As are cluster
  related, it is 
  of considerable interest to elucidate the connection to the spiral-to-S0 morphological
  evolution of cluster galaxies recently observed (Dresser et al. 1997;
  Fasano et al. 2000; Goto et al. 2003a). 
   E+A galaxies might be a critical link in a galaxy evolution model in
  which a blue, star-forming disk galaxy evolves into a red S0 or
  elliptical galaxy.

   {\bfseries (ii)\ Galaxy-Galaxy\ Interaction\ (in\ the\ field).}

      Zabludoff et al. (1996) found E+A galaxies in the
  field region as well. They selected 21 E+A galaxies in the Las
  Campanas Redshift Survey data (Shectman et al. 1996) and showed that
 they are not preferentially located in cluster regions.
  Balogh et al. (1999) also found
  a comparable amount of E+A galaxies in the field region as in cluster
  regions.   Zabludoff et al. (1996) proposed that galaxy-galaxy interaction
      (Schweizer 1982; Lavery \& Henry 1988; Liu \& Kennicutt 1995a,b;
      Schweizer 1996)
      might be responsible for E+A phenomena since 5 of their 21 E+A
      galaxies had a tidal feature.   Oegerle, Hill, \& Hoessel(1991)
      also found a  nearby E+A galaxy with a tidal feature.   High
      resolution imaging of Hubble Space Telescope supported
      the galaxy-galaxy interaction scenario by identifying that some
      of post-starburst (E+A) galaxies in high redshift clusters show
      disturbed or interacting signatures (Couch et al. 1994,1998;
      Dressler et al 1994; Oemler, Dressler, \& Butcher 1997).
           Liu \& Kennicutt (1995a,b) observed 40 merging/interacting
      systems and found that some of their spectra resemble E+A
      galaxies. 
         Bekki, Shioya, \& Couch (2001) modeled galaxy-galaxy mergers with
    dust extinction, confirming that such systems can produce 
    spectra which evolve into E+A spectra.

 
  {\bfseries (iii)\ Dust\ Enshrouded\ Star\ Formation. }

    Another possible
  explanation for E+A phenomena is dust enshrouded star formation, where
    E+A galaxies are actually star-forming, but emission 
  lines are invisible in optical wavelengths due to the heavy obscuration 
  by dust. 
   Poggianti et
  al. (1999) pointed out that galaxies with strong H$\delta$ and weak
  [OII] emissions (e(a) galaxies in their classification) might be a
  dust-obscured starbursting galaxy. 
   A straightforward test for this scenario is to observe in radio wavelengths
  where dust obscuration is negligible (or in far-infrared, sub-millimeter to detect
  emission from dust). At radio wavelengths, the synchrotron radiation
  from electrons accelerated by supernovae can be observed. Therefore,
  in the absence of a radio-loud active nucleus, the radio flux of a
  star-forming galaxy can be used to estimate its current massive star
  formation rate (Condon 1992; See Hopkins et al. 2003 for comparison between radio
  estimated and optically estimated star formation
  rate). 
    Smail et al. (1999) performed such a radio observation and found that
  among 8 galaxies detected in radio, 5 galaxies have strong Balmer
  absorption with no detection in [OII].
  They concluded that massive stars are
  currently forming in these 5 galaxies. 
    Owen et al. (1999) investigated the radio properties of galaxies in
  a rich cluster at $z\sim$0.25 (A2125) and found that optical line
  luminosities (e.g., H$\alpha$+[NII]) were often weaker than one would
  expect for the SFRs implied by the radio emission.
  On the other hand, 
  Chang et al. (2001) detected none of 5 nearby E+A galaxies in radio
  continuum using VLA and excluded the possibility that their E+As are
  dust-enshrouded massive starburst galaxies. 
    Miller \& Owen (2002) observed radio continua of 15 E+A galaxies found in LCRS
  (Zabludoff et al. 1996) and detected moderate levels of star formation
  in 2 of them. The star formation rates (SFRs) of these two galaxies, however, are
  5.9 and 2.2 M$_{\odot}$ yr$^{-1}$, which are an order of magnitude
  smaller than those in Smail et al. (1999), consistent with normal to
  low SFR instead of starburst. 

   These studies warn us that at least some E+A galaxies might have
  on-going star formation.   
    Therefore, when we discuss E+A galaxies, it is important to acquire
  data in a broad wavelength range. 
  We try to address this problem using H$\alpha$ emission
  (less affected by dust extinction than [OII] since it is at longer
  wavelength), infrared photometry and radio flux.
     As a variation of the dust enshrouded star-forming scenario,
   Poggianti \& Wu (2000) presented the selective dust extinction
   hypothesis, where dust extinction is dependent on stellar age since 
    youngest stars inhabit very dusty star-forming HII regions while older
    stars have had time to migrate out of such dusty regions (also see
    Calzetti, Kinney, \& Storchi-Bergmann 1994; Poggianti et al. 2001). 
     If O, B-type stars in E+A galaxies are embedded in dusty regions and
    only A-type stars have long enough lifetimes (10$^7\sim 1.5\times
    10^9$ yr) to move out from such regions,  
    this scenario can naturally explain E+A phenomena. However, this
    scenario has to explain why the selective extinction happens only in
    E+A galaxies since we can observe many star-forming galaxies with
    detectable [OII] and H$\alpha$ emissions.

 Although these three scenarios are all plausible, the definitive
 conclusion has not been drawn yet. Part of the difficulty stems from
 the extreme rarity of E+A galaxies.  
 Since E+A phase is very short (less than a Gyr), they are very rare. In the Las
  Campanas Redshift survey, there were only 21 E+A galaxies in 11113
 spectra. Its rarity made it more difficult to study a statistically large
 number of E+A galaxies. The largest sample of E+A galaxies to date is
  presented by Galaz (2000), which, however, is a heterogeneous sample
  of only 50 E+As.
  The Sloan Digital Sky Survey (York et al. 2000)
 which is both imaging and redshift surveys of a quarter of the whole
 sky, provides us with the first opportunity to study E+A galaxies in
 a much large number (see also a parallel work by Quintero et al. 2003, which uses a
  sophisticated K/A ratio method to study by far a large number of
  $\sim$1000 k+A galaxies from
  the same SDSS data). 
  Furthermore, the availability of [OII] and 
 H$\alpha$ lines allows us to divide strong Balmer absorption galaxies
 into four categories including E+As, and to study properties of
 galaxies in each category in
  detail. 
 As shown in the following sections, we stress the importance in
  separating the strong Balmer absorption galaxies into  these four
 categories.
 In previous  work, authors might have dealt with heterogeneous samples of
  these four sub-samples of H$\delta$-strong (HDS) galaxies, resulting in wide varieties in
  the properties of H$\delta$-strong galaxies, partly due to the lack of H$\alpha$
  information, and partly due to the low signal-to-noise ratio of spectra.
   In Chapter \ref{EA1}, we described a robust method to
 select H$\delta$-strong galaxies  and presented a catalog of
 such galaxies. In this chapter, we study spectral and photometric
 properties of the whole population of H$\delta$-strong galaxies in
 detail to reveal the origin of these galaxies.
  The cosmological parameters adopted throughout this work are $H_0$=75 km
 s$^{-1}$ Mpc$^{-1}$ and ($\Omega_m$,$\Omega_{\Lambda}$,$\Omega_k$)=(0.3,0.7,0.0).

\section{Data}
  
  To study H$\delta$-strong galaxies as a whole, we created a catalog of
  H$\delta$-strong galaxies in  Chapter \ref{EA1}. Chapter \ref{EA1} presents the details of the selection methodology, which we
  briefly summarize here.  From the Sloan Digital Sky Survey Data
  Release I (see Fukugita et al. 1996, Gunn et al. 1998,  Lupton
  et al. 1999,2001; York et al. 2000, Eisenstein et al. 2001,
  Hogg et al. 2001, Blanton et al. 2003a, Pier et
  al. 2002, Richards et al. 2002, Stoughton et al. 2002, Strauss et
  al. 2002, Smith et al. 2002 and Abazajian et al. 2003 for more detail
  of the SDSS data), we 
  selected galaxies with $z>$0.05 and
  S/N($g$)$>$5. The low redshift cut is applied to exclude the strong
  aperture effect. For these galaxies, we have measured H$\delta$,
  [OII] and H$\alpha$ equivalent widths (EWs) and obtained their errors
  using the flux summing 
  method as described in Section \ref{ea1_line} of Chapter
  \ref{EA1}. This flux summing method 
  is robust for spectra of low S/N ratio and weak lines. We selected
  H$\delta$-strong galaxies as a galaxy with a H$\delta$ equivalent width (EW)
  greater than 4 \AA\ with more than 1 $\sigma$ significance level (Section
  \ref{ea1_catalog}).  Among 94770 galaxies which satisfy the redshift and S/N
  cut with measurable [OII], H$\delta$ and H$\alpha$ lines, 3313 galaxies
  are regarded as H$\delta$-strong galaxies. Among them, we excluded 
  130 H$\delta$-strong galaxies with line ratios consistent with being
  an AGN using the 
  prescription given by Kewley et al. (2001). The final sample consists
  of 3183 H$\delta$-strong galaxies. We classified these H$\delta$-strong galaxies further into
  four sub-categories as described in the next section.

%
%

\section{Defining Four Subsamples of H$\delta$-strong Galaxies}\label{sec:ea2_four_sample}

  Taking the full advantage of the high quality of the SDSS spectra, we divide
 H$\delta$-strong galaxies into four categories. We use [OII] equivalent
 width and H$\alpha$ equivalent width to separate H$\delta$-strong
 galaxies based on their current star formation activity. As seminal
 work by Kennicutt et al. (1992a,b) shows, H$\alpha$ is
 the best star formation indicator in optical wavelength since it is a
 strong line and it has fewer uncertainties (e.g., dust extinction, self
 absorption, metallicity dependence) than the other lines. It thus has been
 used in previous work frequently to study star  
 formation of galaxies when available (e.g., Gomez et al. 2003). 
 We use H$\alpha$ equivalent width measured in Chapter \ref{EA1}.
 The equivalent width of [OII] (3727) line is also used in many studies
 as a star formation indicator (e.g., Zabludoff et al. 1996), especially
 when H$\alpha$ line is not available for high redshift galaxies
 (e.g., Poggianti et al. 1999). [OII] is also a suitable line to study star
 formation of galaxies 
 since the line is strong and the line strength
 does not depend on metallicity very much. However, it is more
 affected by dust extinction than H$\alpha$.   

 Based on these lines, we classify H$\delta$-strong galaxies into four
 categories as follows.

  \begin{equation} 
 E+A:  \left({\rm EW(H\alpha) - \Delta EW(H\alpha)<0}\AA\right) \&  \left( {\rm EW([OII]) - \Delta EW([OII]) <0}\AA\ \right)
 \end{equation}
  \begin{equation} 
 HDS + [OII]: \left({\rm EW(H\alpha) - \Delta EW(H\alpha)<0}\AA \right)\&  \left( {\rm EW([OII]) - \Delta EW([OII]) >0}\AA\ \right)
 \end{equation}
  \begin{equation} 
 HDS + H\alpha: \left({\rm EW(H\alpha) - \Delta EW(H\alpha)>0}\AA\right) \&   \left({\rm EW([OII]) - \Delta EW([OII]) <0}\AA\ \right)
 \end{equation}
  \begin{equation} 
 HDS + em: \left({\rm EW(H\alpha) - \Delta EW(H\alpha)>0}\AA\right) \&   \left({\rm EW([OII]) - \Delta EW([OII]) >0}\AA \right)
 \end{equation}

 All four categories of the H$\delta$-strong galaxies have H$\delta$ EWs greater than
 4 \AA\ with more than 1$\sigma$ significance. Galaxies labeled as a possible
 AGN in Chapter \ref{EA1} are not included in the sample. 
 Table \ref{tab:ea2_hds_sample} shows the number of galaxies in each category. 
   
\begin{figure}
\begin{center}
\includegraphics[scale=0.7]{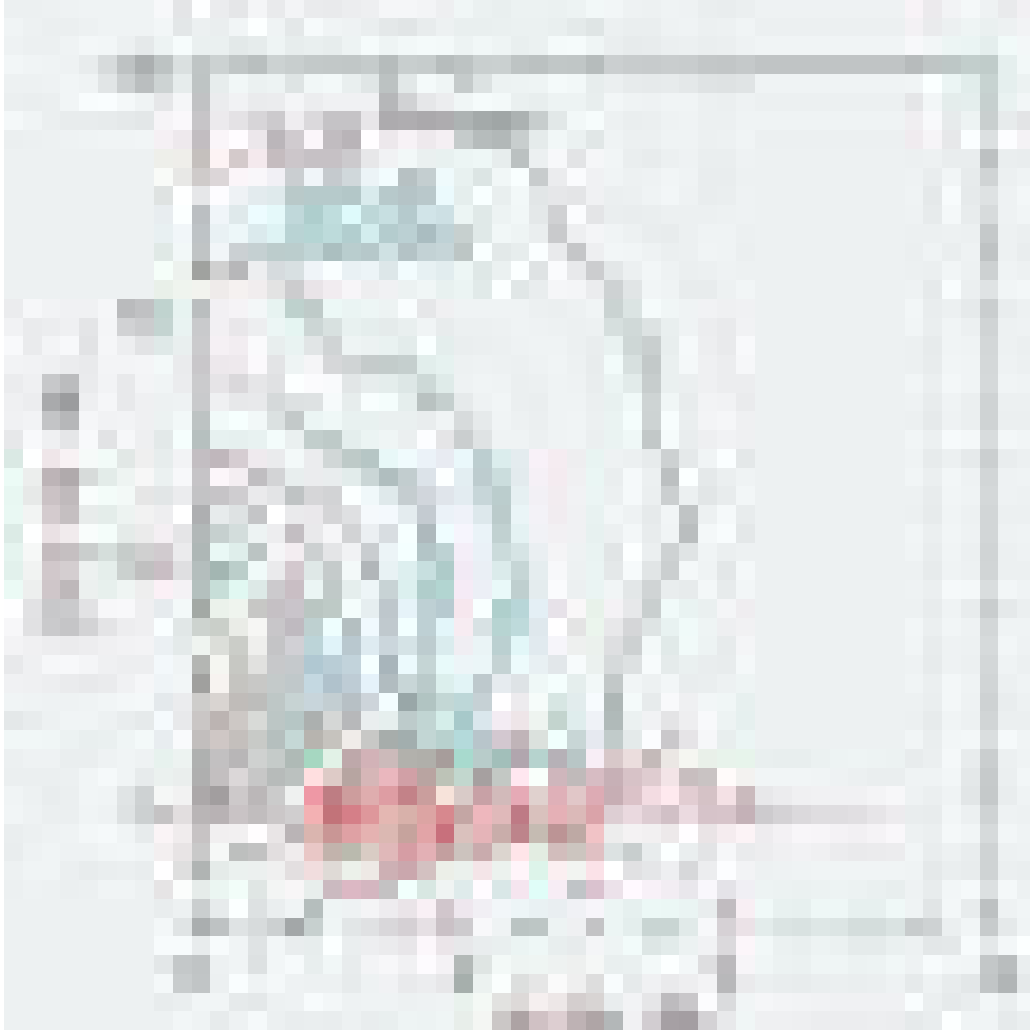}
\end{center}
\caption{  [OII] EWs are plotted against H$\delta$ EWs for four
 sub-samples of H$\delta$-strong galaxies.
 The contours show the distribution of all 94770 galaxies. Large
 open circles, triangles, squares, and small dots represent E+A,
 HDS+[OII], HDS+H$\alpha$ and HDS+em galaxies, 
 respectively. 
}\label{fig:ea2_hd_oii}
\end{figure}

  Figure \ref{fig:ea2_hd_oii} plots [OII] EWs against H$\delta$ EWs.
 Contours show the distribution of all 94770 galaxies. Large 
 open circles, triangles, squares, and small dots represent E+A,
 HDS+[OII], HDS+H$\alpha$ and HDS+em galaxies,  
 respectively.  Generally H$\delta$-strong galaxies are a rare class of galaxies, as
 we see fewer points toward increasing H$\delta$ EWs. Reflecting our
 selection criteria, [OII] EWs are the strongest for HDS+em galaxies and
 become weaker for HDS+[OII], HDS+H$\alpha$ and E+A galaxies. 
  Figure \ref{fig:ea2_hd_ha1}, in turn, plots H$\alpha$ EWs against H$\delta$ EWs.
 Contours and symbols are the same as of Figure \ref{fig:ea2_hd_oii}.
   In this figure, H$\alpha$ EWs are the strongest for HDS+em galaxies, and
 become weaker and weaker toward HDS+H$\alpha$, HDS+[OII] and E+A galaxies. 

    In Figure \ref{fig:ea2_sfr}, we show distributions of star formation rate (SFR)
 calculated using H$\alpha$ luminosity for each subsample of H$\delta$-strong
 galaxies. Solid, long-dashed, dot-dashed, short-dashed and dotted lines
 represent all, E+A, HDS+[OII], HDS+H$\alpha$ and HDS+em galaxies,
 respectively. SFR is calculated using a conversion formula given in Kennicutt
 (1998), assuming constant extinction of 1 magnitude at the wavelength
 of H$\alpha$ (see Hopkins et al. in prep. for more sophisticated SFR
 estimation).  
 Reflecting our selection criteria, HDS+[OII] and E+A
 galaxies have relatively lower SFR. HDS+em galaxies have higher SFR
 compared with all galaxies. HDS+H$\alpha$ have lower SFR than HDS+em,
 showing that H$\delta$-strong galaxies without [OII] emission have in
 average a lower
 amount of H$\alpha$ in emission. This result might be consistent with
 the dusty origin of HDS+H$\alpha$ galaxies.
  In Figure \ref{fig:ea2_absolute}, we show luminosity functions for 
 volume limited subsamples (0.05$<z<$0.1 and $M^*_r<-$20.5) of the H$\delta$-strong galaxies.
 The solid, long-dashed, dot-dashed, short-dashed and dotted lines
 represent all , E+A, HDS+[OII], HDS+H$\alpha$ and HDS+em galaxies,
 respectively. According to a Kolomogorov-Smirnov test, LFs of all the
 four sub-samples do not show any significant 
 difference from that of all galaxies drawn by the solid line except
 HDS+em galaxies. 
  HDS+em galaxies show a
 slightly fainter LF, different from the others with more than 99.99\%
 significance. The fact that we found many
 bright E+As ($Mr^*<-$20.0) is inconsistent with Poggianti, Bridges \& Mobasher et
 al. (2001), where they found no luminous E+A galaxies in Coma
 cluster. Our finding  suggests that E+A phenomena is not restricted to dwarf
 galaxies in the nearby universe.

   Below we briefly comment on these four sub-samples of H$\delta$-strong galaxies.

\subsection{E+A}

  The selection criteria for this category are chosen to match the classical E+A galaxies which have
 been studied in various previous work (e.g., Dresser \& Gunn 1983,1992;
 Zabludoff et al. 1996; Balogh et al. 1999; Poggianti et al. 1999;
 Dressler et al. 1999). 
  Figure \ref{fig_ea1:true_ea} of Chapter \ref{EA1} shows example spectra of true E+A galaxies. 
 The strong H$\delta$ absorption line with no H$\alpha$ and
 [OII] emission lines indicates that these galaxies have had strong star burst in recent 1
 Gyr, but they do not have on-going star formation at all as shown in 
 the lack of emission lines which indicate current star formation. The
 reason why they had experienced starburst, and the reason why they
 stopped star formation is unknown.

\subsection{HDS+[OII]}

  These galaxies are characterized by a strong H$\delta$ absorption line and the lack
 of H$\alpha$ emission line with the existence of detectable emission in
 [OII] (Figure \ref{fig_ea1:oii} of Chapter \ref{EA1}). [OII] emission shows remaining star formation activity.
 However, the lack of H$\alpha$ emission remains a mystery. 
 In Chapter \ref{EA1} we discussed that a possible explanation is self-absorption
 in H$\alpha$ due to many A-type 
 stars and metallicity effect that increases [OII]  emission relative
 to H$\alpha$ emission. As is shown in Figure \ref{fig:ea2_hd_oii}, We note
 that [OII] emission in these systems is 
 not so strong as in normal star-forming galaxies, which might suggest that
 H$\alpha$ emission in these systems is also weak enough to be
 hidden by self-absorption.
  Another possibility includes HI gas cloud in front of these galaxies. Such gas cloud might
 absorb light in H$\alpha$, but not in [OII].

\subsection{HDS+H$\alpha$}

\begin{figure}
\begin{center}
\includegraphics[scale=0.7]{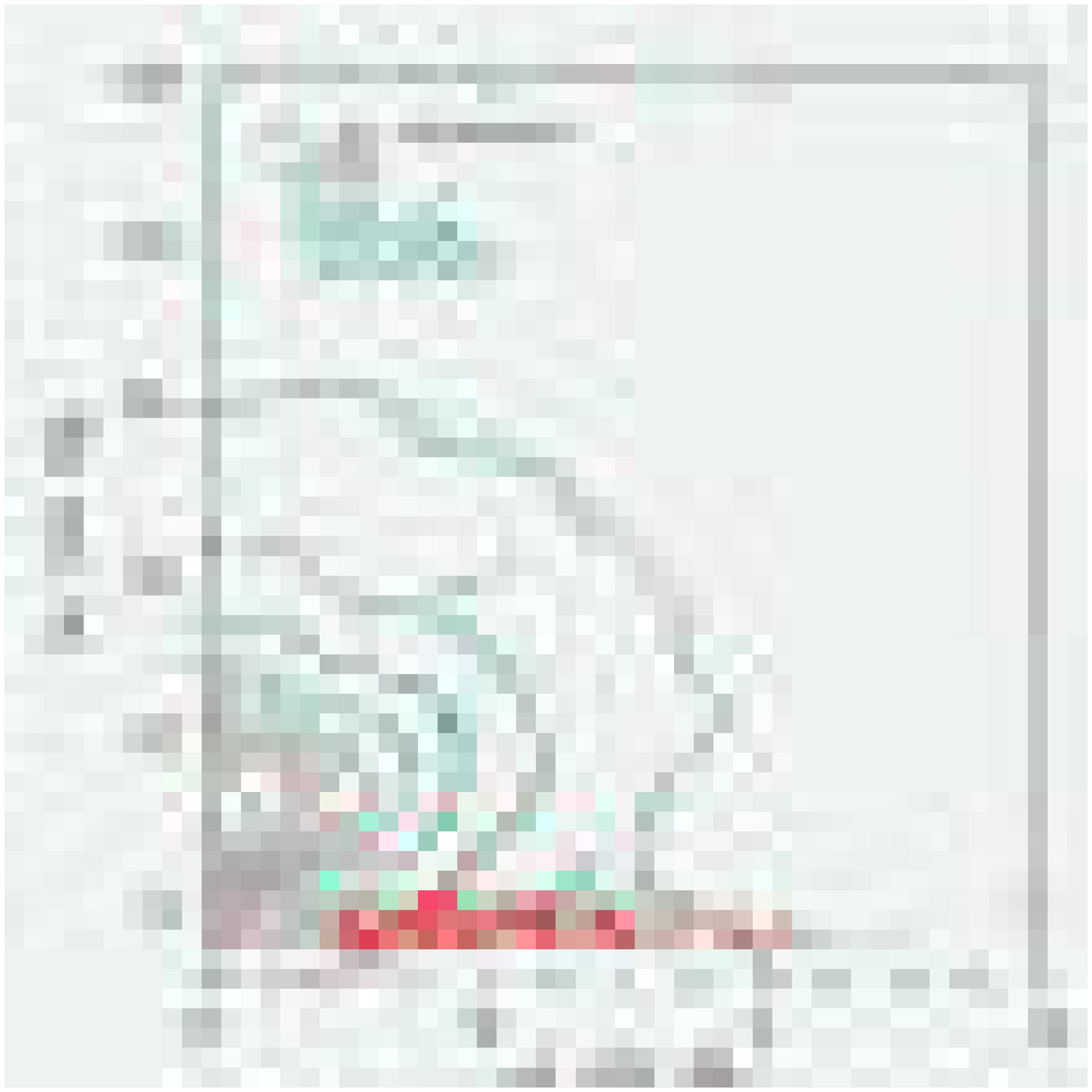}
\end{center}
\caption{ H$\alpha$ EWs are plotted against H$\delta$ EWs for four
 sub-samples of H$\delta$-strong galaxies.
 The contours show the distribution of all 94770 galaxies. Large
 open circles, triangles, squares, and small dots represent E+A,
 HDS+[OII], HDS+H$\alpha$ and HDS+em galaxies, 
 respectively. 
}\label{fig:ea2_hd_ha1}
\end{figure}

\begin{figure}
\begin{center}
\includegraphics[scale=0.7]{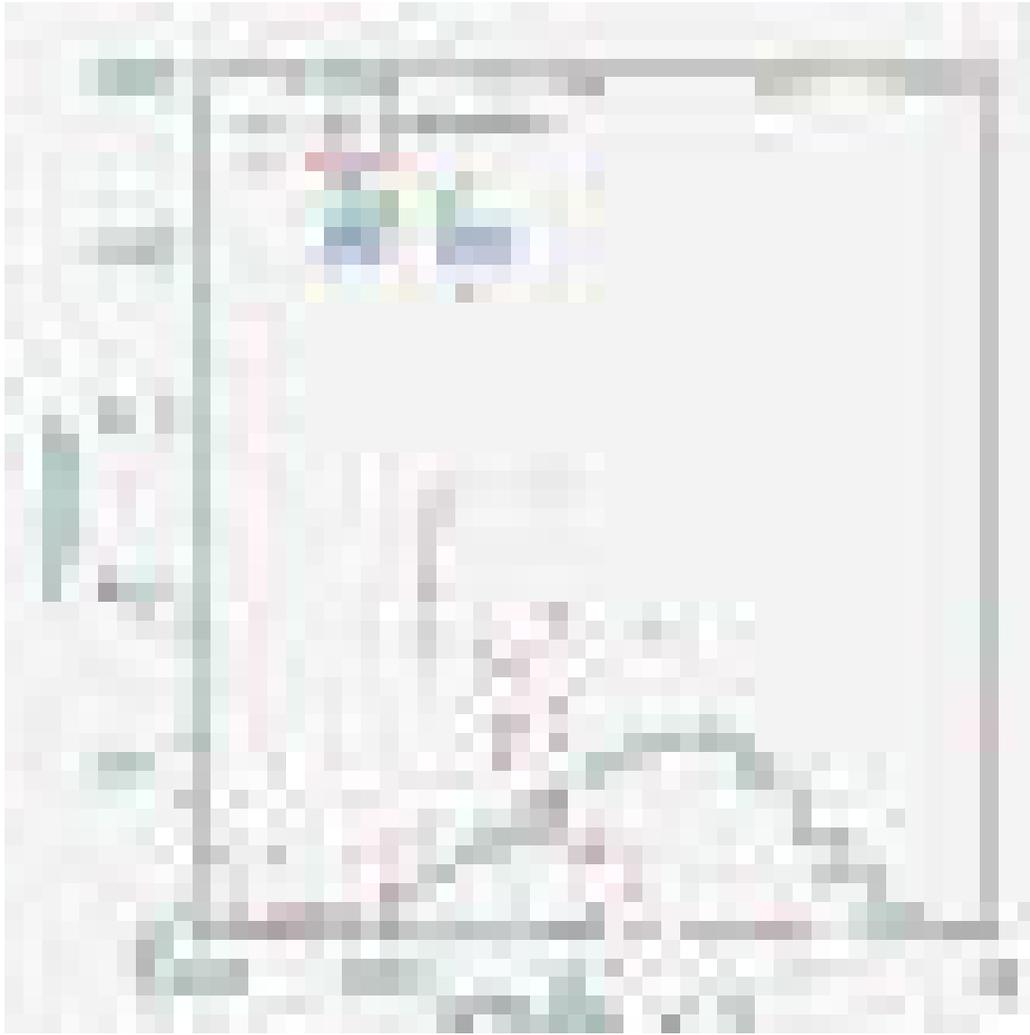}
\end{center}
\caption{
 Star formation rate estimated using H$\alpha$ flux for each class of
 galaxies. The
 solid line is for all 94770 galaxies. The long
 dashed, dotted, short dashed and dotted-dashed lines represent E+A,
 HDS+em, HDS+H$\alpha$ and HDS+[OII] samples, respectively.   }\label{fig:ea2_sfr}
\end{figure}

\begin{figure}
\begin{center}
\includegraphics[scale=0.7]{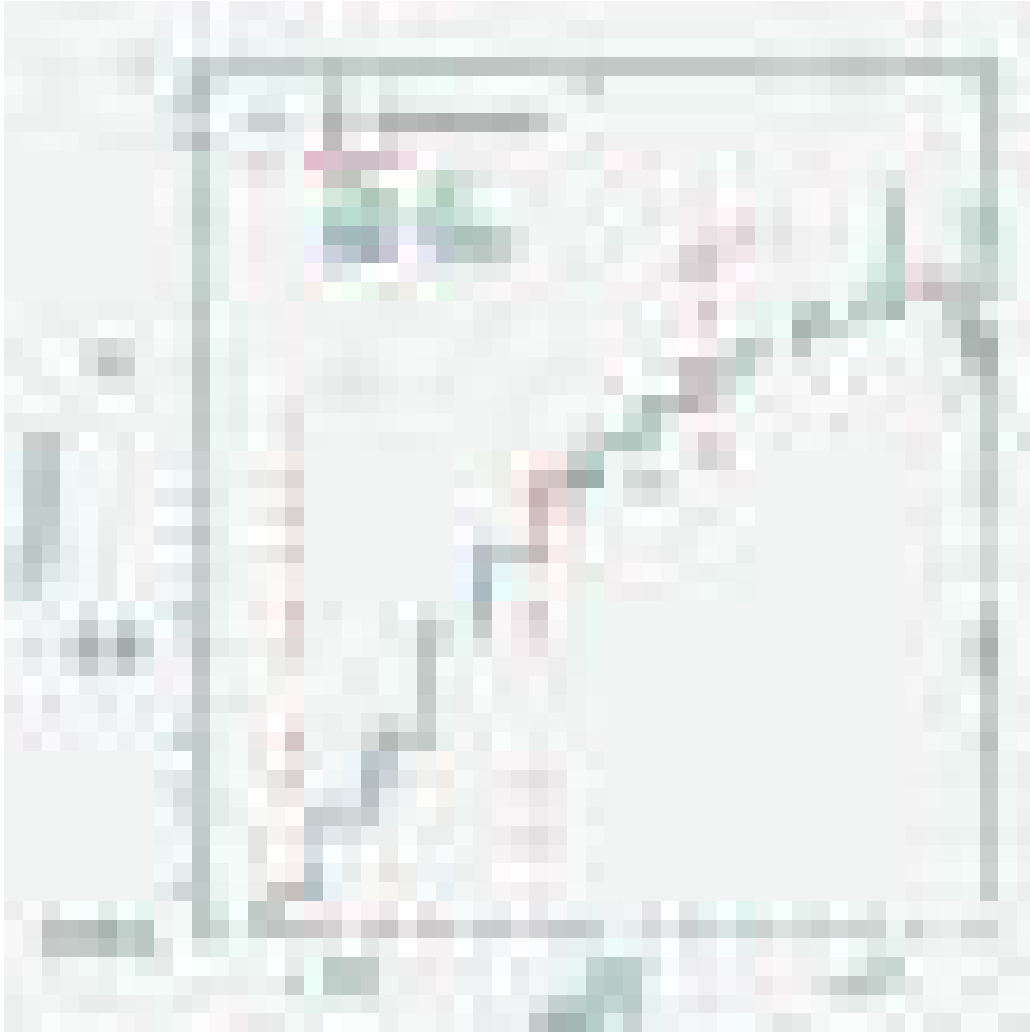}
\end{center}
\caption{
 Luminosity functions in $r$ band for each subclass of H$\delta$-strong galaxies. The
 solid line is for all galaxies in the volume limited sample. The long
 dashed, dotted, short dashed and dotted-dashed lines represent E+A,
 HDS+em, HDS+H$\alpha$ and HDS+[OII] samples, respectively.   }\label{fig:ea2_absolute}
\end{figure}

  These H$\delta$-strong galaxies are characterized by the strong
 H$\delta$ absorption, the lack of [OII] in emission and the 
 existence of H$\alpha$ in emission. The existence of this type of galaxies
 is scary since previous work frequently used the non-existence of [OII] in a
 selection of E+A galaxies due to the non-availability of H$\alpha$
 information. If H$\alpha$ emission comes from star formation activity,
 selecting this type of galaxies as non-star-forming (E+A) galaxies is
 erroneous.  In Chapter \ref{EA1} we discussed that a possible explanation for the
 lack of [OII] emission line 
 is a combination of strong dust extinction and metallicity
 effect. Relatively small H$\alpha$ EWs compared with normal
 star-forming galaxies (Figure \ref{fig:ea2_hd_ha1}) suggest that [OII]
 emissions in these galaxies might be weak enough to be hidden by dust.
   Another possible explanation for this type of galaxies is that
 H$\alpha$ is coming from diffuse ionized gas. In such a case,  diffuse
 ionized gas does not emit in [OII], thus perfectly explains the
 spectral features. In normal galaxies, diffuse ionized gas (not
 associated with star-forming regions) could be responsible for ~50\% of
 the H$\alpha$ emission.  

\subsection{HDS+em} 

 This type of galaxies has an indication of both recent star formation
 (strong H$\delta$ absorption) and currently on-going star formation (H$\alpha$ and
 [OII] in emission).    
 Therefore these galaxies can be understood as a star-forming
 galaxy, possibly reducing its star formation rate  currently.
 Also in terms of statistics, these galaxies are much more numerous than
 the other three types of galaxies (Table
 \ref{tab:ea2_hds_sample}), suggesting more common nature of these galaxies.  
 Figure \ref{fig_ea1:ea_em} of Chapter \ref{EA1} shows five typical spectra of HDS+em galaxies.

\section{The Morphology of H$\delta$-strong Galaxies}
\label{Jun  1 12:00:59 2003}

 In this section we investigate morphologies of H$\delta$-strong galaxies with
 particular attention to E+A
 galaxies. Since E+A galaxies have experienced the truncation of
 starburst fairly recently ($<$1 Gyr), E+As might still hold some traces
 of the truncation of the star-formation in their morphology
 (e.g., dynamically disturbed signs).
 Therefore we might obtain some hint on the
 origins of E+A galaxies by examining their morphology. 
 In Figure \ref{fig:ea2_morphology}, we plot a concentration parameter, $Cin$,
 against $u-r$ color.    
 The concentration parameter, $Cin$, is defined as the  ratio of
 Petrosian 50\% light radius to Petrosian 90\% light radius 
 in $r$ band  (radii which contain 50\% and 90\% of Petrosian fluxes,
 respectively; Shimasaku et al. 2001; Strateva et al. 2001). The border line
 between spiral galaxies and elliptical galaxies is around
 $Cin$=0.4. Since this concentration parameter is an inverse of a
 general definition of concentration parameter, concentrated
 (elliptical) galaxies have a smaller value of $Cin$ than late-type galaxies.   Strateva et
 al. (2001) showed that $u-r$=2.2 also separates early and late type
 galaxies well. 
 In Figure \ref{fig:ea2_morphology}, contours show the distribution of
 all 94770 galaxies in our
 sample.  The distribution shows two peaks, one for elliptical galaxies
 at around ($u-r,Cin$)$\cong$(2.8, 0.35), and one for spiral galaxies at around
 ($u-r,Cin$)$\cong$(1.7, 0.45).
 In the same figure, large open circles, triangles, squares, and small
 dots represent E+A, HDS+[OII], HDS+H$\alpha$ and HDS+em galaxies,
 respectively. 
 This figure shows that E+A galaxies and HDS+[OII] galaxies have more
 elliptical-like morphology with lower $Cin$ values and redder $u-r$
 color. Their $Cin$s are almost as small as those of elliptical galaxies.  Their
 $u-r$ colors, however, are
 not as red as those of elliptical galaxies. HDS+em galaxies have more
 spiral-like morphology with higher $Cin$ values and bluer $u-r$ 
 color. HDS+H$\alpha$ galaxies have intermediate morphology. Some of the
 HDS+H$\alpha$ galaxies have
 elliptical-like morphology and others have spiral-like morphology. The
 center of the distribution is between the elliptical peak and the
 spiral peak. 

  It is interesting to note the following.  We have selected these four
 subsamples of H$\delta$-strong galaxies solely based on 
 spectral line properties. Nevertheless, each sample
 shows clear morphological difference on the $Cin$ vs $u-r$
 plane. In particular, the difference between 
 HDS+em and E+A is clear, suggesting the existence of  a physical phenomenon
 governing morphological properties of these two types of
 H$\delta$-strong galaxies. Our results are consistent with those by Quintero et al. (2003), who
 also report bulge-dominated morphologies of their E+A galaxies.
  Also
 the discovery of elliptical-like morphologies of E+A galaxies is
 interesting since it is against the previous observations where E+As
 were found to have disk-like morphologies (Couch et al. 1994, 1998;
 Dressler et al. 1994,1999; 
 Oemler et al. 1997;  Smail et al. 1997; Chang et al. 2001) and it leaves
 an interesting question why elliptical-like galaxies experienced
 a starburst and truncation of the starburst.  

   These differences in morphologies of E+As in the literature might be understood as a
 result of a significant variety in definitions of  E+A galaxies. Most
 of the 
 previous samples did not have information on H$\alpha$ emission line, and thus,
 used only [OII] emission and Balmer absorption lines to define E+A galaxies. In addition,
 the previous observations did not have as high signal-to-noise spectra as
 the present study.
 In Chapter \ref{EA1}, we showed that there are as many HDS+H$\alpha$ galaxies  as
 E+A galaxies.  These HDS+H$\alpha$ galaxies 
 could be mis-classified as E+A galaxies if H$\alpha$ line is not
 available. In Figure \ref{fig:ea2_morphology},
 HDS+H$\alpha$ galaxies have disk-like morphology. If previous E+A samples are
 contaminated by HDS+H$\alpha$ galaxies, the morphological difference
 between our E+A sample and the previous E+A samples can be naturally
 explained by the disk-like appearance of HDS+H$\alpha$ galaxies.

\section{The Environment of H$\delta$-strong Galaxies}\label{sec:ea2_density}

   The environments of H$\delta$-strong galaxies have been actively
 debated over the past few years. Originally H$\delta$-strong galaxies are found in
 cluster regions, and therefore, thought to be related to cluster induced
 phenomena (Dressler \& Gunn 1983,1992; Sharples et al. 1985; Lavery \&
 Henry 1986; Couch \&  Sharples 1987;Dressler 1987; Broadhurst, Ellis,
 \& Shanks 1988; Fabricant,  McClintock, \& Bautz 1991; Franx  1993; Caldwell et al.  
  1993,1996,1997; Belloni   et al. 1995; Barger et al. 1996; Fisher et
  al. 1998; Morris et al. 1998; Couch et al. 1998; Castander et al. 2001,
  Rose et  al. 2001).
 However, Zabludoff et al. (1996) showed
 that such galaxies were found outside clusters and groups of galaxies. 
 Therefore, it remains controversial if the post-starburst phenomenon is common to
 the whole galaxy population, or specific (or more frequent) in dense
 environments. Dressler et al. (1999) claims that there are an order of magnitude
 more post-starburst galaxies in distant clusters compared to the
 distant  field. On the other hand,
 Balogh et al. (1999) find that the frequency of post-starburst galaxies is the
 same in distant clusters as in the field.

 In this section, we investigate the environments of H$\delta$-strong galaxies by
 measuring local galaxy density in the following way. 
 For each galaxy, we measure a projected distance to the 5th nearest
 galaxy in angular direction using galaxies within $\pm$1000 km s$^{-1}$ in
 redshift space among the volume limited 
 sample ($Mr^*<-$20.5, 0.05$<z<$0.1).
 The  criterion for redshift space ($\pm$1000 km s$^{-1}$) is set to be generous to
 avoid galaxies with a large peculiar velocity slipping out of the
 density measurement, in other words, not to underestimate the density
 in cluster cores. Then, the number of galaxies ($N$=5) within the distance
 is divided by 
 the circular surface area with the radius of the distance to the 5th
 nearest galaxy. 
 When the projected area touches
  the boundary of the data, we corrected the 
 density by correcting the area
 to divide. Since we have redshift information for all of the sample
 galaxies, our density measurement is a
 pseudo-three dimensional density measurement and free from the
 uncertainty in background subtraction.  
  In Figure \ref{fig:ea2_density_cluster}, we present distributions of
 this local galaxy density for cluster galaxies. Galaxies within 0.5 Mpc
 from the nearest cluster, galaxies between 1 and 2 Mpc from the
 nearest  cluster, and all galaxies are plotted  in the  long-dashed,
 short-dashed and solid lines, respectively.    
 In measuring distance from a cluster, we use the C4 cluster catalog
 (Miller et al. in prep.; Gomez et al. 2003). 
  For each galaxy, the distance from the nearest cluster
 center is measured on the 
 projected sky for galaxies within $\pm$1000 km s$^{-1}$ from a cluster
 redshift. The figure shows that the typical local galaxy density is 10, 4, 1
 Mpc$^{-2}$ for environment with a radius of 0.5  and 1-2 Mpc
 from the nearest cluster center.

 In Figure \ref{fig:ea2_hd_density}, we plot H$\delta$ EW against the local
 galaxy density for all galaxies in our volume limited sample
 (0.05$<z<$0.1, $Mr^*<-$20.5). For H$\delta$ line, negative EWs are absorption lines.
  As medians of the distribution (the solid
 line) show, H$\delta$ EW becomes weaker and weaker with increasing
 local galaxy density. 
  In Figure \ref{fig:ea2_density_ea}, we present distributions of local
 galaxy density 
 for each subclass of H$\delta$-strong galaxies. Solid, long-dashed, dot-dashed,
 short-dashed and dotted lines 
 represent all , E+A, HDS+[OII], HDS+H$\alpha$ and HDS+em galaxies,
 respectively. The density distributions of all the four subclasses of
 H$\delta$-strong galaxies are quite similar to that of all galaxies
 (solid line). In contrast, the density distributions of all the four subclasses of
 H$\delta$-strong galaxies are quite different from that of cluster
 galaxies shown in Figure \ref{fig:ea2_density_cluster}.
 In fact, a Kolomogorov-Smirnov test shows that all the four density
 distributions of H$\delta$-strong galaxies are different from that of
 cluster galaxies (within 0.5 Mpc from the cluster center)
 with more than 99.99\% significance level. 
  These results have significant implications for the origins of H$\delta$-strong galaxies.  
  H$\delta$-strong galaxies in local galaxy density $\ll$1 Mpc$^{-2}$ can not be
 explained with the cluster related phenomena. Therefore,
  these results suggest that origins of H$\delta$-strong galaxies are not
 cluster related, and that they are more common phenomena in general field regions. 
  Also, among the four subclasses of H$\delta$-strong galaxies, there are small
 differences between distributions. The distributions of HDS+em and
 HDS+[OII] galaxies
 are slightly shifted to low density regions compared with that of all
 galaxies, while E+A and HDS+H$\alpha$ are not. To clarify these
 differences, we plot the ratio of each subclass of H$\delta$-strong galaxies to all
 galaxies in Figure \ref{fig:ea2_density_ratio}. The left panel shows the
 ratio of HDS+em galaxies to all galaxies in the volume limited
 sample. As indicated in the previous figure, the ratio declines
 continuously as a function of local galaxy density. The relation
 is very similar to the decline of the fraction of spiral galaxies seen in
 the morphology-density relation (Goto et al. 2003c; Chapter \ref{chap:MD}). It is also similar
 to the decline of star formation rate as a function of local galaxy
 density (Lewis et al. 2002; Gomez et al. 2003).
  In the right panel of Figure \ref{fig:ea2_density_ratio}, we show the
 ratios of the rest of H$\delta$-strong subsamples to all galaxies.  The long
 dashed, short dashed and dotted-dashed lines represent E+A,
  HDS+H$\alpha$ and HDS+[OII] galaxies, respectively. Although the
 distributions are dominated by statistical errors, two
 characteristics are found in the panel. The HDS+[OII] fraction shows a
 monotonic decline as found for HDS+em galaxies. On the other hand,
 the HDS+H$\alpha$ and E+A fractions do not show much dependence on the
 local galaxy density.  
 
 When interpreting these results, we have to keep in mind that galaxies
 have the morphology-density relation and the SFR-density relation. From
 Figure \ref{fig:ea2_morphology}, we know that HDS+em galaxies have
 disk-like morphology and HDS+em are the most numerous among four
 sub-samples of H$\delta$-strong galaxies. Therefore it is more likely that Figure
 \ref{fig:ea2_hd_density} reflects the well known morphology-density relation
 in terms of H$\delta$ EWs, and therefore can not be interpreted that
 cluster galaxies have weaker H$\delta$ EWs. In the right panel of
 Figure \ref{fig:ea2_density_ratio}, it is interesting to note that the distribution of
 E+A (and HDS+H$\alpha$) galaxies does not depend much on the local
 galaxy density while the morphologies of these galaxies are more
 elliptical-like. Therefore it is suggested that although E+A galaxies have
 elliptical-like morphologies, the origin of E+As are perhaps different from
 that of bright cluster elliptical galaxies.

\section{Possible Star Formation Histories}
  
 In this section, we compare SED models and observational quantities to
 search for possible star formation histories of H$\delta$-strong galaxies.  Using the
 GISSEL model by Bruzual A.~\& Charlot(1993), we simulated three
  representative star formation histories as follows; (i) $Burst$ model, which has an
 instantaneous starburst at the beginning and no star formation
 thereafter. (ii) $Constant$ star formation. (iii) $Exponentially$
 decaying star formation. In all the three models, we use the Salpeter
 initial mass function (Salpeter 1955) with a single stellar population.
 Figure \ref{fig:ea2_time_hd} plots H$\delta$ EWs against time (or galaxy
 age) for the three models. Dashed, solid and dotted
 lines show the models with instantaneous burst, constant star formation and
 exponentially decaying star formation rate. The burst model has a
 strong H$\delta$ EW right after its burst at 1 Gyr. However, its
 H$\delta$ EW declines rapidly, and becomes less than 3\AA\ at 1 Gyr after
 the burst. The exponentially decaying model maintains strong H$\delta$
 for a longer time. Its H$\delta$ EW becomes 3\AA\ in 5 Gyrs.
 The constantly star-forming model
 maintains a large H$\delta$ EW ($>$6\AA) beyond 14 Gyr.

 In Figure \ref{fig:ea2_gri}, we plot restframe $g-r$ color against $r-i$
 color.    We first plot the models in this figure to compare later with the observational
 data. The dashed, solid and dotted
 lines are for the models with an instantaneous burst, constant star formation and
 exponentially decaying star formation rate. We show the models for two
 different metallicity in this figure ($Z$=0.02 and $Z=0.1$). While a
 galaxy is star-forming, its colors stay around
 ($g-r$,$r-i$)$\cong$(0.4, 0.2), where the constant star formation model
 stays. As soon as the galaxy stops star formation, its colors become
 redder and redder, approaching to the peak colors of elliptical
 galaxies at around  ($g-r$,$r-i$)$\cong$(0.8,0.4). An increase in metallicity
 results in  redder $r-i$ color. 
  
  We now add the observational data to compare with these models.
  Observed colors are shifted to the restframe using the
 $k$-correction code by Blanton et al. (2003; v1\_11).  Compared with contours showing
 distribution of all galaxies in our sample, the SED models reproduce
  observed colors of galaxies reasonably well. 
  Open circles,
 small dots, triangles and squares represent E+A, HDS+em, HDS+H$\alpha$
 and HDS+[OII], respectively.  HDS+em galaxies (small dots) have bluer
 colors in both $g-r$ and $r-i$, which are consistent with the colors of
 normal star-forming galaxies. All the three models also show blue
  colors consistent with  HDS+em galaxies, when the models have star
  formation in their early stage.
   E+A galaxies (open circles) have somewhat
 redder color in $g-r$, but not as red as elliptical galaxies
 ($g-r\sim$0.8) with some exceptions, suggesting that they are a still
  evolving population of galaxies. The $r-i$ colors of E+As are
 widely spread. Some of them are as blue as star-forming galaxies in
 $r-i$.  These bluer colors of E+As are consistent with the continuum
 dominated by many A-type stars.
  Compared with the models, colors of E+As are consistent with
  the exponentially decaying model and the burst models at the stage
  where the models start
  stopping their star formation. 
 HDS+H$\alpha$ galaxies have a wide spread
 color distribution in both $g-r$ and $r-i$, with its center between
 elliptical galaxies and star-forming galaxies. In the exponentially
  decaying and the burst model, these colors are at the
  stage where star formation is declining. 
  The distribution of
 HDS+[OII] galaxies is similar to that of E+A galaxies except that there
 are no HDS+[OII] galaxies as blue in $r-i$ as E+A galaxies, perhaps representing
 a  somewhat flatter continuum shape than E+As (see Figure \ref{fig_ea1:oii} of Chapter \ref{EA1}). 
 
  In Figure \ref{fig:ea2_jk_rk}, we plot $J-K$ vs $r-K$, where infrared
 magnitudes are derived from the Two Micron All Sky Survey (2MASS;
 Jarrett et al. 2000) and $k$-corrected to restframe using  Mannucci et
 al. (2001). Note that infrared colors are in AB system in this figure.
 Optical-infrared color is 
 sensitive to the amount of dust since infrared color is
 less obscured by dust than optical. Dusty galaxies should be redder by
 $\sim$1 mag in optical-infrared color (e.g., see Figure 2 of Smail et al. 1999).
 In Figure \ref{fig:ea2_jk_rk}, contours show the distribution of all galaxies
 in our sample.  Open circles,  small dots, triangles and squares
 represent E+A, HDS+em, HDS+H$\alpha$  and HDS+[OII], respectively. 
 E+A galaxies (open circles), which are often suspected to be dust enshrouded star
 forming galaxies, do not show any redder colors than the normal galaxies
 (contours). Therefore our data suggest that E+A galaxies are not likely
 to be a  dust enshrouded star-forming galaxy. 
  Some of the HDS+em galaxies are about 0.5 mag redder in $r-K$ than the normal
 galaxies. These galaxies might have a significant amount of dust. 
  In the figure, the models are plotted in dashed, solid and dotted
 lines for the instantaneous burst, constant star formation and
 exponentially decaying models. Three sets of models are plotted for
 different metallicities. All three models with solar metallicity ($Z$=0.02) show a good
 agreement with the behavior of the observational data. 

                We provide further evidence suggesting that E+As are not dusty
 star-forming galaxies in Figure \ref{fig:ea2_radio_sfr}, where radio
 derived SFR is plotted against redshift. Under the assumption that the
 radio emission is due to star formation, we calculated radio
 estimated SFR using the following conversion (Yun, Reddy \& Condon 2001).
  \begin{equation} 
 SFR (M_{\odot}\ yr^{-1})= 5.9 \times 10^{-22} L_{1.4 Ghz} (W Hz^{-1})
\end{equation}
  This conversion assumes a Salpeter initial mass function integrated
  over all stars ranging from 0.1 to 100 M$_{\odot}$ and hence
  represents the total SFR of a galaxy. We also have applied
  $k$-correction in the shape of (1.0+$z$)$^{-0.8}$.
  The radio data are taken from
  the Faint Images of the Radio Sky at Twenty-cm Survey (FIRST; Becker
  et al. 1995), whose detection limit is regarded to be 1 mJy. In Figure
  \ref{fig:ea2_radio_sfr},  contours show the distribution of all galaxies
 in our sample.  Open circles,  small dots, triangles and squares
 represent E+A, HDS+em, HDS+H$\alpha$  and HDS+[OII], respectively.
 We have excluded galaxies with a possible AGN sign (section \ref{sec:ea2_four_sample}).
  When a galaxy is within the FIRST survey area, and is not detected, we
  assigned 1mJy to the galaxy as an upper limit of the radio flux. In Figure
  \ref{fig:ea2_radio_sfr}, points along with the line around 10 M$_{\odot}$ yr$^{-1}$ are
  thus those with no detection, showing an upper limit of the
  radio SFR. Only one E+A has a moderate SFR of $\sim$10
  M$_{\odot}$ yr$^{-1}$. The other E+As are not detected in the FIRST and thus
  only shown as an upper limit. In this redshift range, none of the E+As
  has radio SFR greater than 100 M$_{\odot}$ yr$^{-1}$. Although it is
  difficult to exclude the possibility that E+As have a moderate rate of
  dust hidden SFR, we can safely conclude that E+As do not hold
  dust hidden starburst. The same is true for HDS+H$\alpha$ and
  HDS+[OII] galaxies except one  HDS+H$\alpha$ galaxy with SFR of
  $\sim$40  M$_{\odot}$ yr$^{-1}$. On the other hand, some of the HDS+em galaxies
  have strong star formation of $\sim$100 M$_{\odot}$ yr$^{-1}$, consistent
  with their emission lines in optical.

 Next we compare the models and the data on the H$\delta$ EW vs D4000 plane
 (strength of the 4000\AA\ break) in Figure \ref{fig:ea2_hd_d4000}. 
 Since D4000 is sensitive to old stellar populations and
 H$\delta$ EW traces the amount of young A-type stars, the H$\delta$ EW vs D4000
 plane is a suitable space to study the star formation history of H$\delta$-strong
 galaxies. A caveat, however, is that models become less accurate on the
 plane since both  H$\delta$ EWs and D4000 is more difficult quantities to
 reproduce than broad band colors. For models, we use H$\delta$ EWs
 given in the GISSEL model, which were measured using the flux between
 4083.50 and 4122.25 \AA. This is essentially the same window as used to
 measure H$\delta$ EWs from the observational data (4082-4122\AA; Chapter \ref{EA1}). For D4000, the
 SED model uses the flux ratio of the 3750-3950\AA\ window to the 4050-4250
 \AA\ window (Bruzual 1983). Observationally D4000 is measured using the
 ratio of the flux in
 the 3751-3951\AA\ window to that in the 4051-4251 \AA\ (Stoughton et
 al. 2002). We regard these two D4000 
 measurements as essentially the same.   
   
   In all the panels in Figure \ref{fig:ea2_hd_d4000}, we plot the three
 models. The dashed, solid and dotted
 lines are for the models with an instantaneous burst, constant star formation and
 exponentially decaying star formation rate. We subtracted 1\AA\ from the
 model H$\delta$ EWs to compensate for possible stellar
 absorption. Different panels are for three different metallicities
 ($Z$=0.0001, 0.02 and 0.1). Compared with the distribution of all
 the observed galaxies represented by the solid contours, the models might have a
 slight shift toward 
 larger D4000 and H$\delta$ directions. However, the behavior of the
 models on this plane well reproduce expected behavior of galaxies, in
 a sense that star-forming galaxies evolve into large D4000 and small
 H$\delta$ EWs. Therefore, we regard that qualitative interpretation
 based on the models as valid. The latest version of the GISSEL model will solve
 these problems (Kauffmann et al. 2003a,b; Charlot et al. in prep.).   
    Observational data are plotted using open circles,
 small dots, triangles and squares for E+A, HDS+em, HDS+H$\alpha$
 and HDS+[OII], respectively. In the plane, E+A galaxies occupy an upper
 right part of the panel, having high H$\delta$ EWs ($>$5\AA) $and$ large
 D4000 ($\sim$1.5) at the same time. 
 This part of the plane can
 be reached only with the burst models, which is consistent with the
 previous interpretation of E+A galaxies as a post-starburst galaxy. 
  According to the instantaneous burst model with the solar metallicity ($Z$=0.02), E+A
 phase is found to be 130-800 Myrs after the burst. 
   On the other hand, HDS+em galaxies are at the tip of the contours,
 and can be reached using both the
 exponentially decaying model and the constant star formation model,
 which reflects that these galaxies are more common galaxies. These star
 formation histories are also consistent with HDS+em galaxies having
 both [OII] and H$\alpha$ in emission. 
   Note that in the lowest metallicity model ($Z$=0.0001), the
 exponentially decaying model can reach the region occupied by E+A
 galaxies. Therefore, some of E+As might be extremely metal poor galaxies
 with an exponentially decaying star formation history.
  HDS+[OII] galaxies occupy the similar place on the plane to E+A
 galaxies. HDS+H$\alpha$ galaxies show a somewhat wider
 distribution. However, their distribution is closer to that of E+As than
 HDS+em. (See Kauffmann et al. 2003a,b for a more detailed discussion of
 the  behavior of various types of galaxies on the D4000 vs H$\delta$ plane
 and its dependence on stellar mass).

 Finally, we compare the models and the data on the $u-g$ vs H$\delta$
 EW plane. We aim to study the relation between the on-going and
 previous (recent) star
 formation activities, using $u-g$ and H$\delta$ EWs as indicators of
 on-going and previous star formation activities, respectively.  Although the emission lines such as
 [OII] and H$\alpha$ are more direct indicators of on-going star
 formation, we found it more difficult to reproduce them with the current
 version of the model, and therefore, more difficult to perform
 quantitative comparisons with the observational data (However, see
 Charlot et al. 2001; Shioya et al. 2001,2002; Bekki et al. 2001 for
 such an attempt). Instead of 
 H$\alpha$ and [OII], we use $u-g$ color as an indicator of on-going
 star formation since it shows better agreement between the SED models and  the
 observational data.  
  In Figure \ref{fig:ea2_ug_hd}, we plot $u-g$ against H$\delta$ EW for the
 three models and the observational data. Three panels represent
 different metallicities. Symbols are the same as previous figures. 
 The model $u-g$ color depends on metallicity to some extent, showing bluer $u-g$ color
 with decreasing metallicity. The solar metallicity model in the upper
 right panel shows a good agreement with the observational data. In the
 panel, the exponentially decaying model and the constant star formation
 model well explain the HDS+em galaxies. The instantaneous burst model,
 on the other hand, explains E+A galaxies well. The agreement again
 indicates that E+A galaxies are in a post-starburst phase, and cannot be
 explained by more normal star formation histories. If the burst model with
 solar metallicity is assumed, somewhat tighter constraints on the time
 scale can be obtained from this figure than that obtained by Figure \ref{fig:ea2_hd_d4000}.
 E+A galaxies are found to be in 270-800 Myrs after the burst. 

 More interestingly, E+A galaxies are beautifully aligned on the
 time sequence of the instantaneous burst model from 270 Myr to 800
 Myr. Therefore, according to this model, E+A galaxies with larger
 H$\delta$ EWs are younger than E+A galaxies with smaller  H$\delta$
 EWs. In the next section, we investigate those E+A galaxies with large
 H$\delta$ EWs in more detail, as young E+A galaxies.


\section{Properties of young E+A Galaxies}\label{sec:ea2_progenitor}
 
 In the last section, we found that E+A galaxies with larger H$\delta$
 EWs are younger E+A galaxies, assuming the instantaneous burst model.
  We select possible young E+A galaxies from our E+A sample using 
 the following criteria. 

 \begin{equation}
  E+A\ and\ H\delta\ EW > 7\AA 
 \end{equation}
 Among our 133 E+A galaxies, 28 galaxies satisfy these criteria.  
   In Figure \ref{fig:ea2_progenitor_spectra_individual}, we show example
   spectra of these young E+A galaxies. All the spectra show very strong H$\delta$
   absorption with the lack of  [OII] and H$\alpha$ emission lines.
   According to the instantaneous burst model, these young E+As are
   found to be around 270-430 Myrs after the burst. The entire E+A phase
   continues until $\sim$800 Myrs after the burst (Figures
   \ref{fig:ea2_time_hd} and \ref{fig:ea2_ug_hd}). 
   Table \ref{tab:ea2_progenitor_median} shows median
    properties of these young E+As and all the E+As in our
   sample. Errors are quoted using 75 
 and 25 percentiles. Both samples show quite similar properties
   except strong H$\delta$ EWs and absolute magnitude.  The brighter
   absolute magnitude for the young E+As is consistent with the
   hypothesis that these are younger galaxies which evolve into normal E+A
   galaxies as their stellar populations become older and fainter.

   If these galaxies are truly in younger phase of E+As,  they are also 
 closer to the initial burst, and therefore closer to the epoch of physical
 change (only 270-430 Myr after the burst and truncation of star
 formation). Therefore, by examining
 the images of the young E+As, we might obtain some hints on the
 physical mechanisms causing the burst and truncation.  
 Figure \ref{fig:ea2_progenitor_image_individual} shows randomly sampled
 example images of these young E+A galaxies. Each panel in the figure
 corresponds to that in Figure  \ref{fig:ea2_progenitor_spectra_individual}.
  Interestingly, more than half of these galaxies have their
 companions/tidal features within 60'' of the image, suggesting
 recent merger/interaction with accompanying galaxies.
  These results indicate that the origin of E+A galaxies might be
 merger/interaction, which can dynamically disturb galaxies and cause starburst and sudden
 truncation of it (Lonsdale, Persson, \& Matthews 1984; Kennicutt
  et al. 1987; Sanders et al. 1988).  
   Elliptical-like morphology of E+As
 (Figures \ref{fig:ea2_morphology}, \ref{fig:ea2_progenitor_image_individual})
 might stem from the merger/interaction already in progress. Indeed,
 numerical simulations have shown that merger/interaction is able to
 produce elliptical morphology of galaxies (Ostriker \&
 Hausman 1977; Hausman \& Ostriker 1978; Miller 1983; Merritt 1984;
 Malumuth \& Richstone 1984; Bode et al. 1994; Athanassoula, Garijo, \& Garc{\'{\i}}a G{\' 
o}mez 2001). 

  We calculated the number of accompanying galaxies within 50 or 75 kpc 
 from the young E+As using galaxies from the SDSS
 imaging data, and
 compared it with that of 1000 randomly selected SDSS galaxies with a
 similar redshift distribution. 
 We calculated
 $Mr^*$ assuming that all galaxies within 50 or 75 kpc in the imaging data are at the same
 redshift as the central galaxy, and using the $k$-correction by Blanton et
 al. (2003; v1\_11). Then, we limit galaxies
 between $-23<Mr^*<-19.5$ and subtract global galaxy number
 count adjusted to the angular area that 50 or 75 kpc subtends at the
 redshift. The faint end of the magnitude limit
 ($Mr^*=-$19.5) corresponds to $r^*$=22.2 at the 
 highest redshift ($z$=0.3) of the young E+A sample for galaxies with
 large $k$-correction. 
 Although star/galaxy separation become unstable at $r^*>21.5$, the object
 detection of the SDSS imaging data is 95\%
 complete at  $r^*$=22.2 (Stoughton et al. 2002). 
 Table \ref{tab:ea2_nearby_galaxies} summarizes the number of accompanying
 galaxies within 50 and 75 kpc for the young E+A, all E+A and 1000 randomly
 selected galaxy samples.
 The young E+A galaxies
 have 0.40$\pm$0.12 accompanying galaxies per galaxy on average within
 75 kpc, while the 1000 randomly 
 selected galaxies with a similar redshift distribution have 0.16$\pm$0.01 
 accompanying galaxies per galaxy. 
 Thus, the young E+A galaxies have 2.5 times more
 accompanying galaxies than the randomly selected galaxies with more than
 two   $\sigma$ significance.
  A similar result can be obtained when we use 50 kpc radius, where 
 young E+As have  accompanying galaxies of 0.24$\pm$0.09, which is eight
 times more than the random sample of galaxies with 0.03$\pm$0.01 accompanying galaxies.
   In addition, we calculated the number of accompanying galaxies for all
 the E+A galaxies to be  0.12$\pm$0.03 and 0.26$\pm$0.04
 within 50 and 75 kpc, respectively. 
  It is interesting that the number of accompanying galaxies for all E+A
  sample is  between the young E+A sample and the random
 sample for both 50 and 75 kpc radii.
 These results might suggest that
 accompanying galaxies of the young E+As will be merged into the central
 E+A galaxy in a few hundred Myrs when it is seen as a normal (older) E+A galaxy.
   In Figure \ref{fig:ea2_accompany_hist}, we show a preliminary absolute magnitude
 distribution of accompanying galaxies  within 75 kpc radius after fore/background
 subtraction for young E+As. A slight peak
 may be found at the faintest magnitude bin 
 ($Mr^*\sim-$19.5), suggesting that accompanying galaxies are
 fainter than the central E+A galaxies. Thus, the possible origin of E+A
 galaxies can be
 called minor (different mass) merger/interaction rather than major
 (equal mass) merger/interaction. However, since star/galaxy separation and
 $k$-correction become uncertain at the faintest magnitude of
 $r^*\sim$22.2 ($Mr^*=-$19.5 at $z=0.3$), Figure
 \ref{fig:ea2_accompany_hist} should not be over-interpreted. Follow-up
 observations are urgently needed.

\section{Discussion : Origins of E+A Galaxies}\label{discussion}

\subsection{Are E+As Cluster Related Phenomena?}
 
 In section \ref{sec:ea2_density}, we showed that the spatial 
 distribution of E+A galaxies does not depend much on the local galaxy density.
 There are many E+A galaxies at a local galaxy density of $\ll$1
 Mpc$^{-2}$.   Our result is consistent with Zabludoff et al. (1996) and Balogh et
 al. (1999), where they also found E+A galaxies in the field.  Recently Quintero
 et al. (2003) reported that E+A galaxies do not lie in high-density
 regions using a sophisticated K/A ratio method to select $\sim$1000 E+A
 galaxies from the same SDSS data.
   There is, however, other work which reported that E+A galaxies live
 preferentially in cluster regions (e.g., Poggianti et al. 1999; Dressler
 et al. 1999). 
  We would like to stress the high quality of our spectra data. 
  As we showed in Section \ref{ea1_discussion} of Chapter \ref{EA1}, we found a comparable number of
 HDS+[OII] galaxies to E+A galaxies (52\%). Therefore high redshift E+A
 samples often selected without H$\alpha$ information could be
 contaminated with HDS+[OII] galaxies up to 52\%. 
   Also if errors in measuring EWs of lines were larger in the previous work, 
 the errors could affect the resulting ratio of E+A
 galaxies. Since E+A galaxies are rare, there are more galaxies
 scattering into the E+A sample than E+A galaxies slipping out of E+A
 criteria. 
  We tried to avoid these problems using high quality spectra from the
 SDSS and a much larger number of 133 E+As than used in previous work.
 In addition, most of the work reporting an excess of E+A galaxies in 
  cluster regions usually observed only cluster regions,
 and did not have good field data. All the field survey data (LCRS,
  CNOC and the SDSS) found E+A galaxies in field regions.
  Considering all of these, our result indicates that E+A phenomena are common to
 various environments including the general field region, 
 rather than cluster specific phenomena.
  At the very least, it is clear that the field E+A galaxies we found
 can not be explained by cluster 
 related phenomena since there exists no ram-pressure or cluster tidal
 effects in the field. Additional mechanisms which work in the field is
 needed.  

   Historically many people connected the existence of E+A
  galaxies to the evolution of cluster galaxies such as the
  Butcher-Oemler effect or the spiral-to-S0 transition. However, our
  results indicate that E+A galaxies has little to do with these
  cluster-related phenomena. It seems that we have to search for the physical mechanism
  responsible for cluster galaxy evolution elsewhere
  (e.g., Goto et al. 2003d; Chapter \ref{chap:PS}).

\subsection{Are E+As Dusty Star-forming Galaxies?}

  Among the 8 radio detected galaxies of Smail et al. (1999), 5 have
 strong Balmer absorption and  no detectable [OII] emission. 
   Miller \& Owen (2002) detected 2 galaxies in radio out of 15 E+A galaxies
 defined in Zabludoff et al. (1996). These two studies indicate that E+A
 galaxies might have an on-going star formation, but their star formation might be
 hidden by dust.   
  However, we would like to point out that E+A galaxies
 defined in these two samples do not have H$\alpha$ information. 
 In Chapter \ref{EA1}, we found that 52\% of H$\delta$-strong galaxies
 with no [OII] emission  do have H$\alpha$ in emission (HDS+H$\alpha$). Therefore, it is
 no wonder that
 5/8 or 2/15 of their post-starburst galaxies had star formation, and
 thus radio emission. These galaxies may be HDS+H$\alpha$ galaxies in
 our category. 
   Strictly speaking on E+A galaxies in our criteria (HDS with no [OII]
 or H$\alpha$ in emission), E+A galaxies are not redder
 in $r-K$ than other galaxies as shown in Figure \ref{fig:ea2_jk_rk}. For example, radio detected
 galaxies in Smail et al. (1999) are redder than other galaxies by
 $\sim$1 mag in $r-H$. Therefore, Figure \ref{fig:ea2_jk_rk} suggests that our E+A
 galaxies are not dust-enshrouded star-forming galaxies. 
  In addition, in Figure \ref{fig:ea2_radio_sfr}, we showed that the upper
 limit of the radio SFR of the E+A galaxies is around 10 M$_{\odot}$ yr$^{-1}$.
 We can not rule out the possibility that E+As might have moderate
 star formation rates. However,  we can conclude that
 E+As are not dusty starbursting ($\sim$100 M$_{\odot}$ yr$^{-1}$) galaxies. 
   In the literature, Poggianti \& Wu (2000) detected HDS+[OII] galaxies (e(a)
 galaxies in their criteria) in far-infrared (FIR). However, they did not
 detect E+A galaxies (k+a/a+k in their criteria) in FIR,
 suggesting that E+A galaxies do not have dust hidden star
 formation. Duc et al. (2002) studied mid-infrared emission in the A1689
 region and found that none of the galaxies with post-starburst optical
 spectra is  detected in 15 $\mu$m down to its flux limit, which
 corresponds to SFR$_{IR}$ of 1.4 M$_{\odot}$ yr$^{-1}$. This does not support
 the dust-enshrouded origin of E+A galaxies, either. 
  Furthermore, Quillen et al. (1999) observed 7 E+A galaxies at 12 $\mu$m
 using ISO to find that E+A's 12$\mu$m flux is consistent with that of
 old stellar populations. They concluded that the E+A phase appears to be
 truly a post-starburst phase with little on-going star formation. 
  We still can not reject the possibility that some of E+A galaxies
 might have moderate rate of star formation hidden by dust. 
 However, it is also difficult to
 explain all of the E+A galaxies as dusty star-forming  
 galaxies. Therefore, the majority of E+A galaxies
 seems to originate from other physical mechanisms than dust
 enshrouded star formation. 


 A selective extinction (Poggianti \& Wu 2000) is also capable of explaining the E+A
 phenomena. In this scenario, O- and  B-type stars are embedded in heavily dust obscured star-forming
 HII regions, and therefore, [OII] and H$\alpha$ emission lines are
 invisible in optical. However, A-type stars responsible for strong
 H$\delta$ absorption have a longer life time
 (10$^7 - 1.5\times 10^9$ yr) to drift away from (or disperse) such
 dusty regions where they were born. In this scenario, the effect of
 dust is maximum for the youngest generation of stars that provides the
 ionizing flux responsible for the emission lines, and decreases for
 older stars. This scenarios is equivalent to a clumpy distribution of
 dust where the location and thickness of the patches are not random but
 dependent on the age of the embedded stellar populations.  
   Observationally this scenario is very difficult to verify unless
 high spatial resolution Mid-/Far-IR observation which can resolve each
 HII region becomes
 available. However, we would like to point out that we observed many
 HDS+em galaxies in which [OII] and H$\alpha$ emission
 lines are visible in addition to strong H$\delta$ absorption.
   If the selective extinction is the
 explanation for E+A galaxies, it needs to explain why selective
 extinction only happens in a certain sub-sample of H$\delta$-strong galaxies, instead
 of all H$\delta$-strong galaxies.

\subsection{Merger/Interaction Origin of  E+A Galaxies} 

 In Section \ref{sec:ea2_progenitor}, we found that the instantaneous burst
 model can explain the observational properties of E+A galaxies very
 well, including optical colors (Figure \ref{fig:ea2_gri}), infrared
 colors (Figure \ref{fig:ea2_jk_rk}), H$\delta$ strength and stellar
 population indicators (D4000 in Figure \ref{fig:ea2_hd_d4000}; $u-g$ in
 Figure \ref{fig:ea2_ug_hd}). Assuming that this  instantaneous burst
 model is correct, we selected young E+A galaxies
  (E+A galaxy between 270 and 430 Myrs after the burst; Section
 \ref{sec:ea2_progenitor}). Since these young E+As are closer to the
 burst epoch than the other E+As, they have  stronger H$\delta$
 absorption and brighter absolute magnitude (Table \ref{tab:ea2_progenitor_median}).
 In a few hundred Myrs, when H$\delta$ absorption becomes a little weaker,
 these galaxies will be seen as a typical E+A galaxy. 
 In Figure \ref{fig:ea2_progenitor_image_individual}, we examined the
 image of young E+As and found that these galaxies
 frequently have accompanying galaxies.
 In fact, these
 galaxies have $\sim$2.5 times more accompanying galaxies within 75 kpc than 1000 randomly selected
 galaxies with two $\sigma$ significance. This statistical result
 suggests that 
 merger/interaction is the origin of these 
 galaxies (Table \ref{tab:ea2_nearby_galaxies}). 

 In the literature, there have been some evidence on a few nearby E+A
 galaxies, suggesting that interaction/merger between galaxies can trigger starburst.
 For example,
 Oegerle, Hill, \& Hoessel(1991)  observationally found a
 tidal feature in a nearby E+A galaxy G515 (see also Carter et
 al. 1988). Belloni et al. (1995) reported the merging origin of E+A
 galaxies based on the HST image. 
  Schweizer (1996) found several nearby E+As that have highly disturbed
 morphologies consistent with the products of galaxy-galaxy mergers.
  Also 5 out of 21 E+A galaxies in LCRS have a tidal feature (Zabludoff et
 al. 1996).  
  One of two E+A galaxies observed with the HST has an elliptical
 morphology with extended tidal tails (Zabludoff 1999).
  One E+A galaxy observed with the VLA has clear tidal tails (Chang
 et al. 2001), which support the galaxy-galaxy interaction
 picture for E+A formation.
   Poggianti \& Wu (2000) reported that  proportion of close mergers are very
 high among their e(a) sample (HDS+em in our definition).  Liu \&
 Kennicutt (1995a,b) also found E+A galaxies among 40
 merging/interacting systems they observed.
   Bartholomew et al. (2001) and Norton et al. (2001) found that
 star formation in E+A galaxies is centrally concentrated. They argued that
 their observational results are consistent with the tidal interaction
 origin, considering the fact that Moss \& Whittle (1993,2000) reported
 that early-type tidally distorted spiral galaxies are often found with
 compact nuclear H$\alpha$ emission.
   Theoretically,  Bekki et al. (2001) showed that galaxy-galaxy mergers with
    high infrared luminosity  can produce e(a) spectra which evolve
 into E+A spectra (See also Shioya et al. 2001,2002).

 Since these previous results were based on only a few nearby E+A
 galaxies, and thus lacked statistical significance, previous authors did not
 conclude on  origins of  E+A galaxies. 
 However, considering our findings of excess number of
 accompanying galaxies around E+As using a much larger sample of 133
 E+As, it is likely that these previous results are truly observing a
  merger/interaction remnant of E+A galaxies.
 In the last two subsections, the dusty origin and cluster origin are not all plausible.
 Therefore, we conclude that the E+A phenomena are most
 likely to be the results of merger/interaction with accompanying galaxies.

%
%
%
%
%
%
%

\section{Summary and Conclusions}\label{conclusion}

 E+A spectra (strong H$\delta$ absorption with no [OII] nor H$\alpha$ emission) can
 be only understood with an instantaneous starburst and its
 truncation. However, the cause of the starburst and its truncation has
 been unknown for more than 20 years since the discovery of E+As (Dressler \& Gunn 1983).
 We have pursued the origin of this interesting population
 of galaxies, using the four statistically large subsamples of 3183
 H$\delta$-strong (HDS; H$\delta$ EW $>$4 \AA) 
 galaxies presented in Chapter \ref{EA1}; 133 E+A (H$\delta$-strong with no [OII]
 nor H$\alpha$ emission), 42
 HDS+[OII] (HDS with [OII] and without H$\alpha$), 108 HDS+H$\alpha$
 (H$\delta$-strong with H$\alpha$ and without [OII]), and 2900 HDS+em (H$\delta$-strong with both
 of [OII] and H$\alpha$) galaxies.
  Our findings can be summarized as follows.

\begin{figure}
\begin{center}
\includegraphics[scale=0.7]{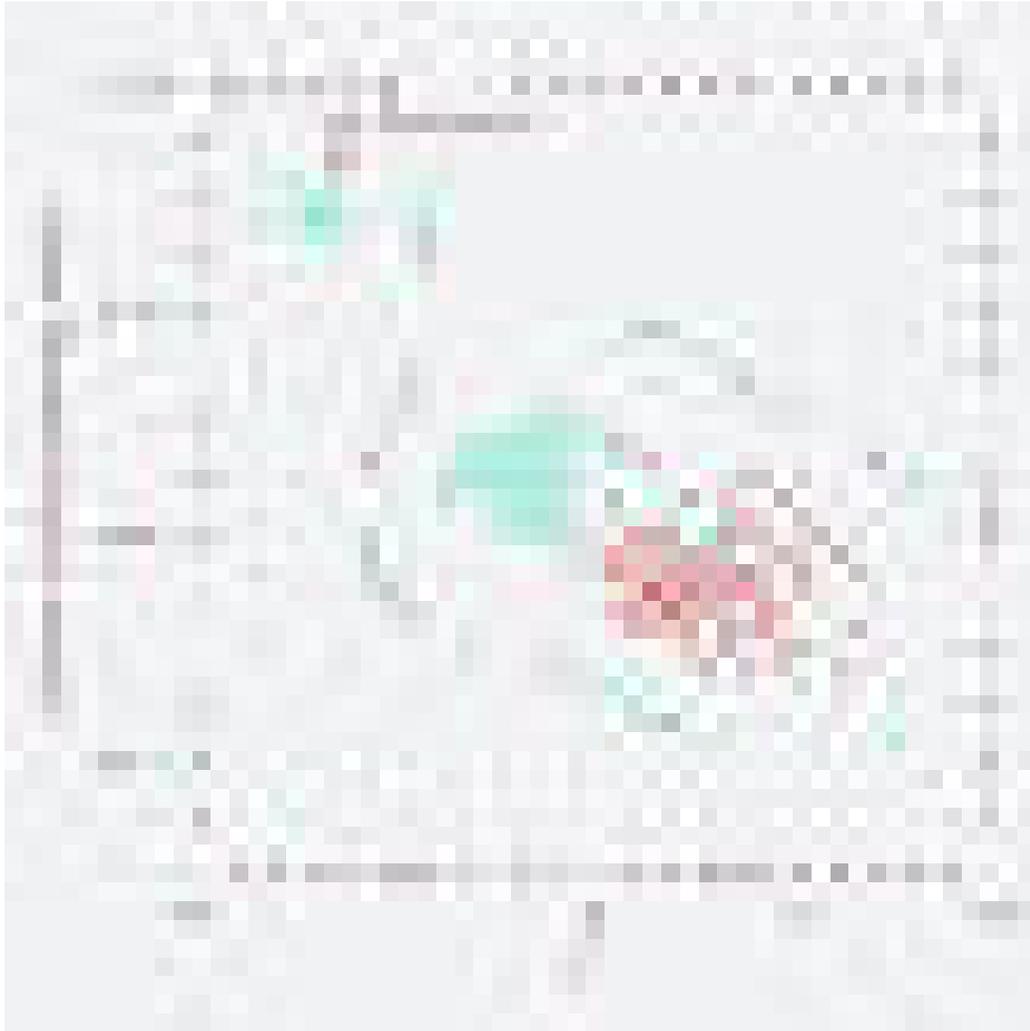}
\end{center}
\caption{
 Distributions of each subclass of galaxies in $Cin$ v.s. $u-r$
 plane. The contours show the distribution of all 94770 galaxies. The large
 open circles, triangles, squares, and small dots represent E+A,
 HDS+[OII], HDS+H$\alpha$ and HDS+em galaxies, 
 respectively.     }\label{fig:ea2_morphology}
\end{figure}

\begin{figure}
\begin{center}
\includegraphics[scale=0.7]{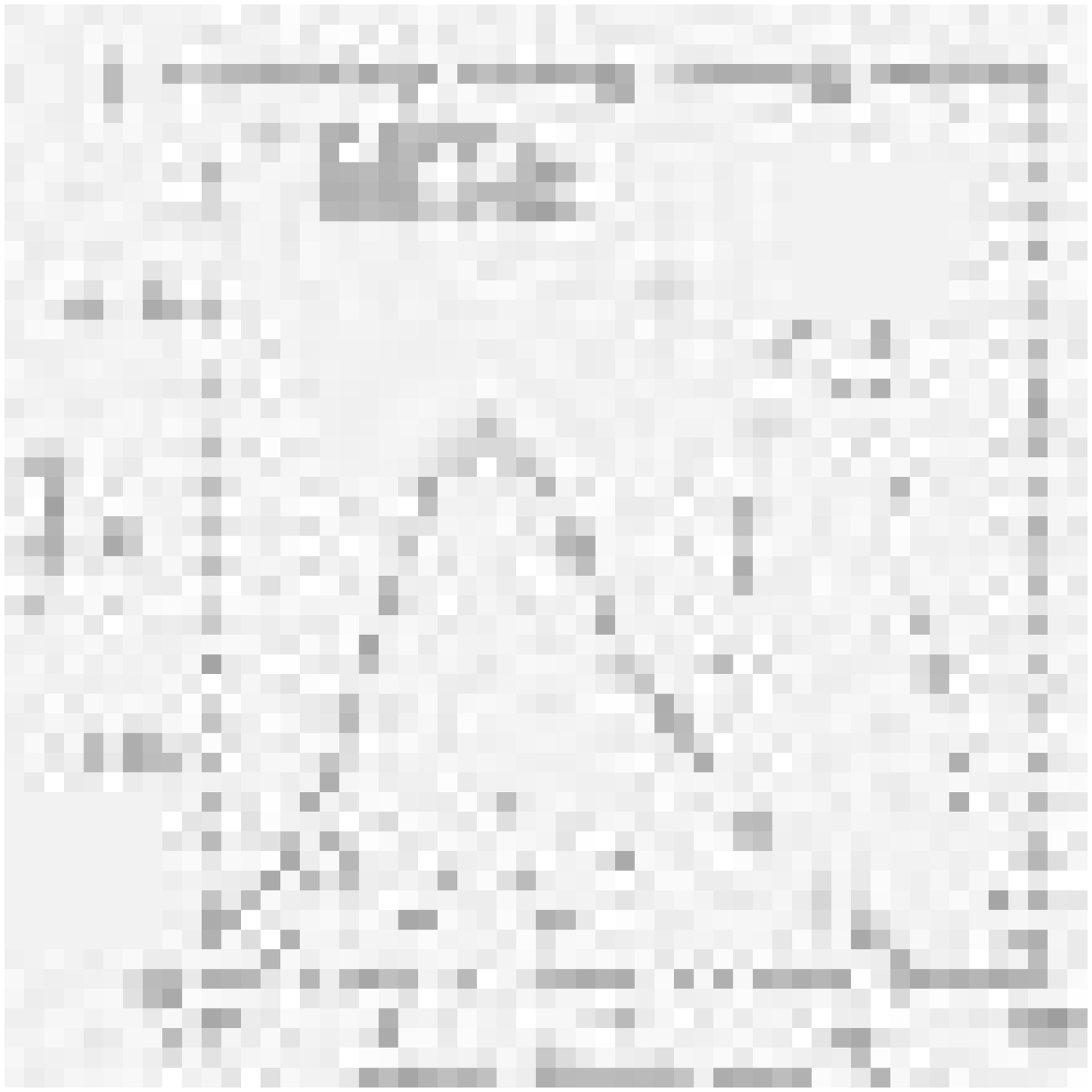}
\end{center}
\caption{
 Distribution of local galaxy density. The solid, dashed and dotted lines
 show distributions for all 94770 galaxies, galaxies within 0.5 Mpc from the nearest
 cluster and  galaxies between 1 and 2 Mpc from the nearest cluster, respectively.
}\label{fig:ea2_density_cluster}
\end{figure}

\begin{figure}
\begin{center}
\includegraphics[scale=0.7]{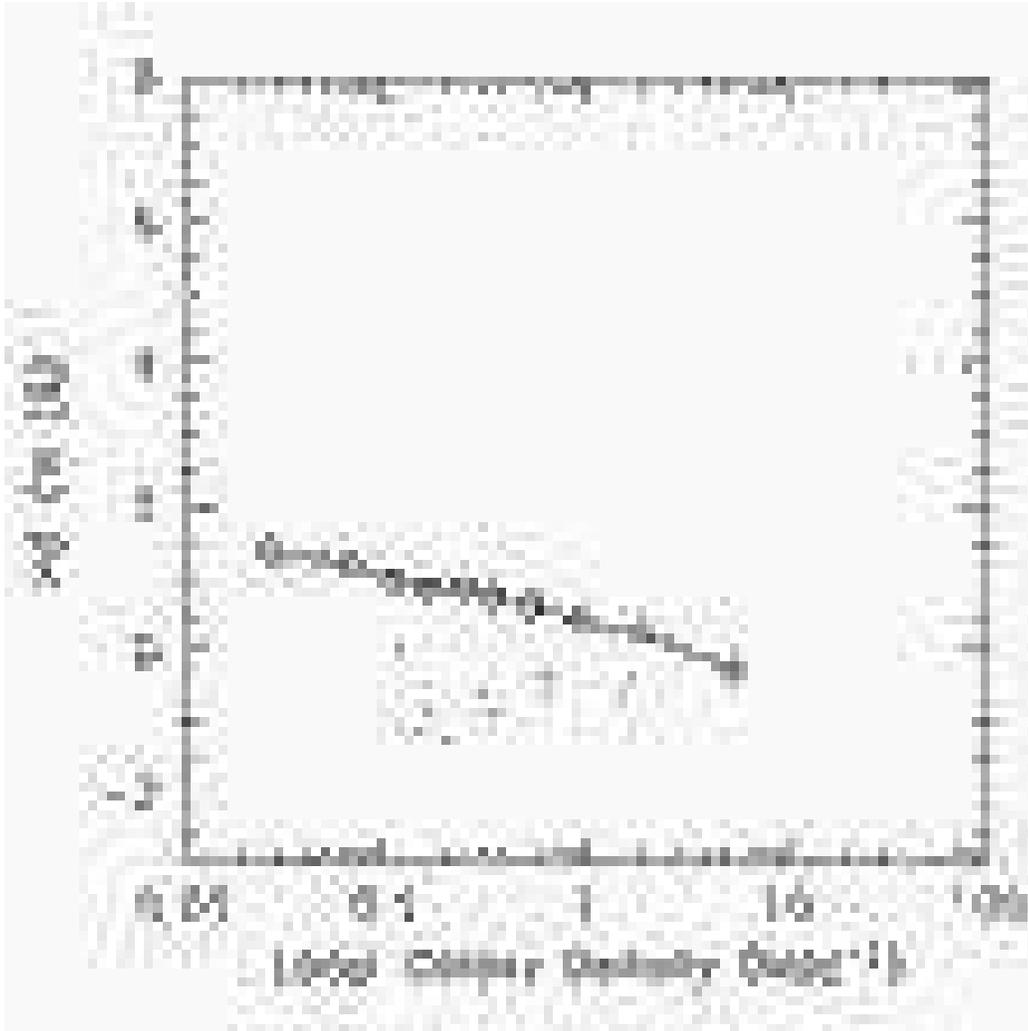}
\end{center}
\caption{
 H$\delta$ EW is plotted against local galaxy density. Negative EWs are
 absorption lines. The solid line shows the median of the distribution.}\label{fig:ea2_hd_density}
\end{figure}

\begin{figure}
\begin{center}
\includegraphics[scale=0.39]{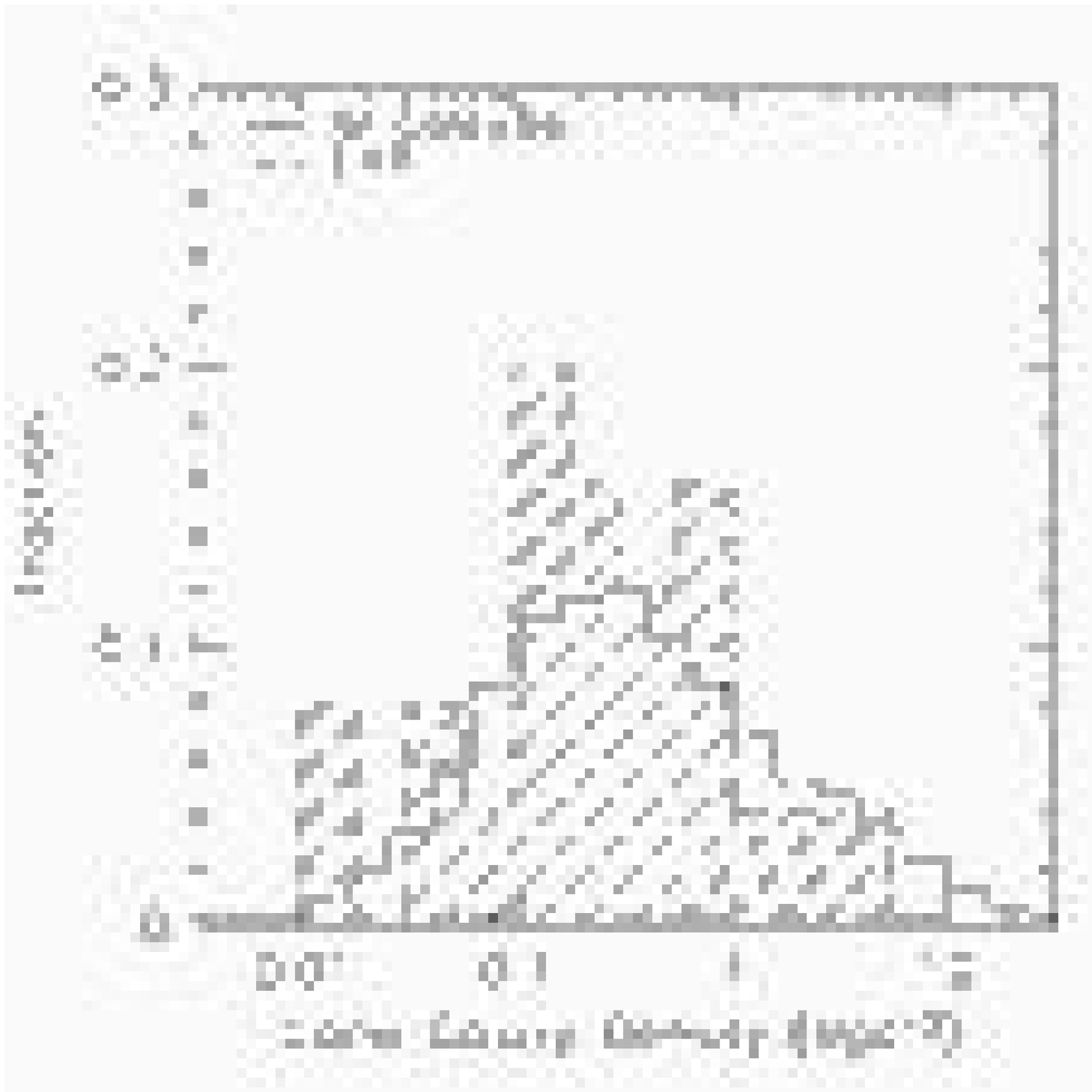}
\includegraphics[scale=0.39]{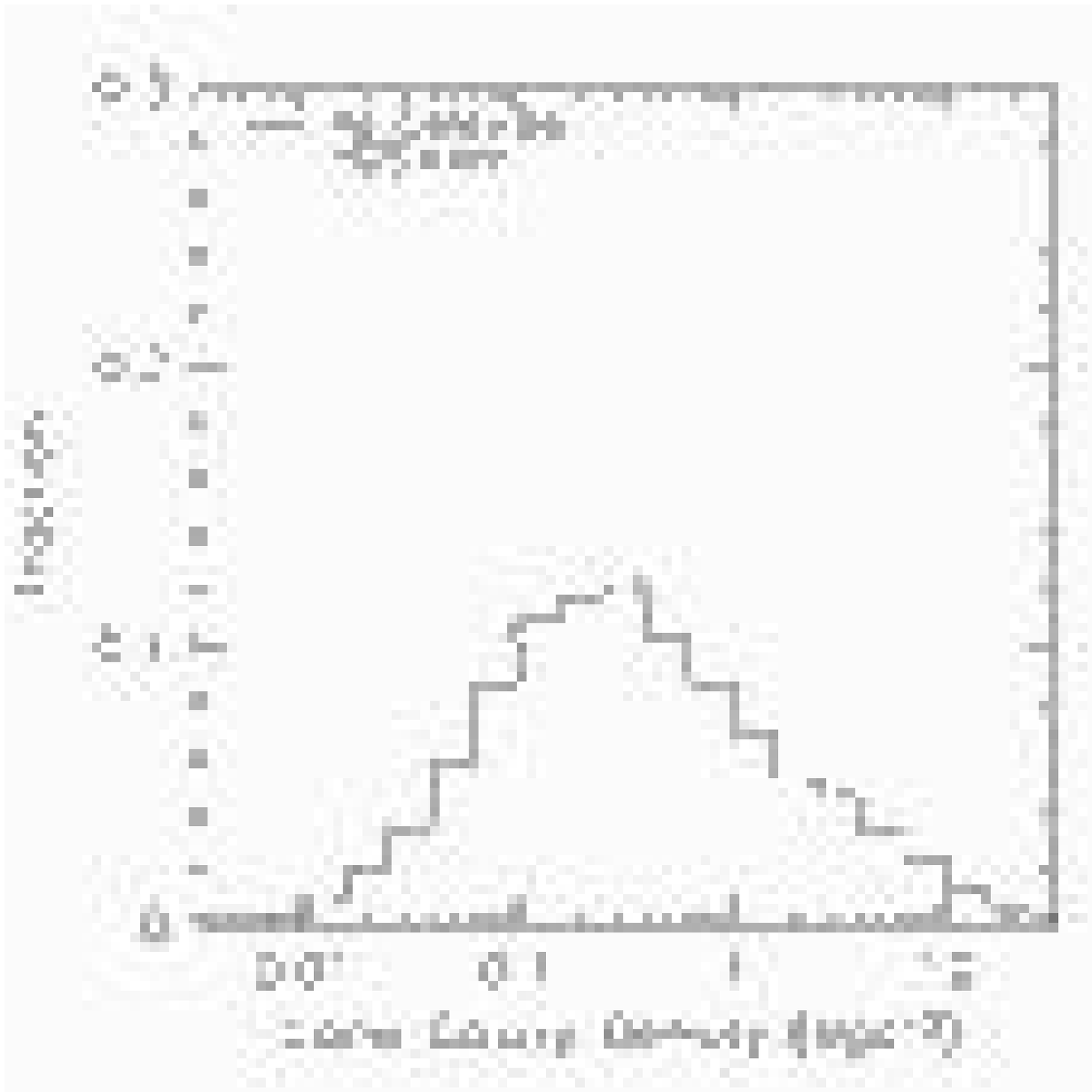}
\includegraphics[scale=0.39]{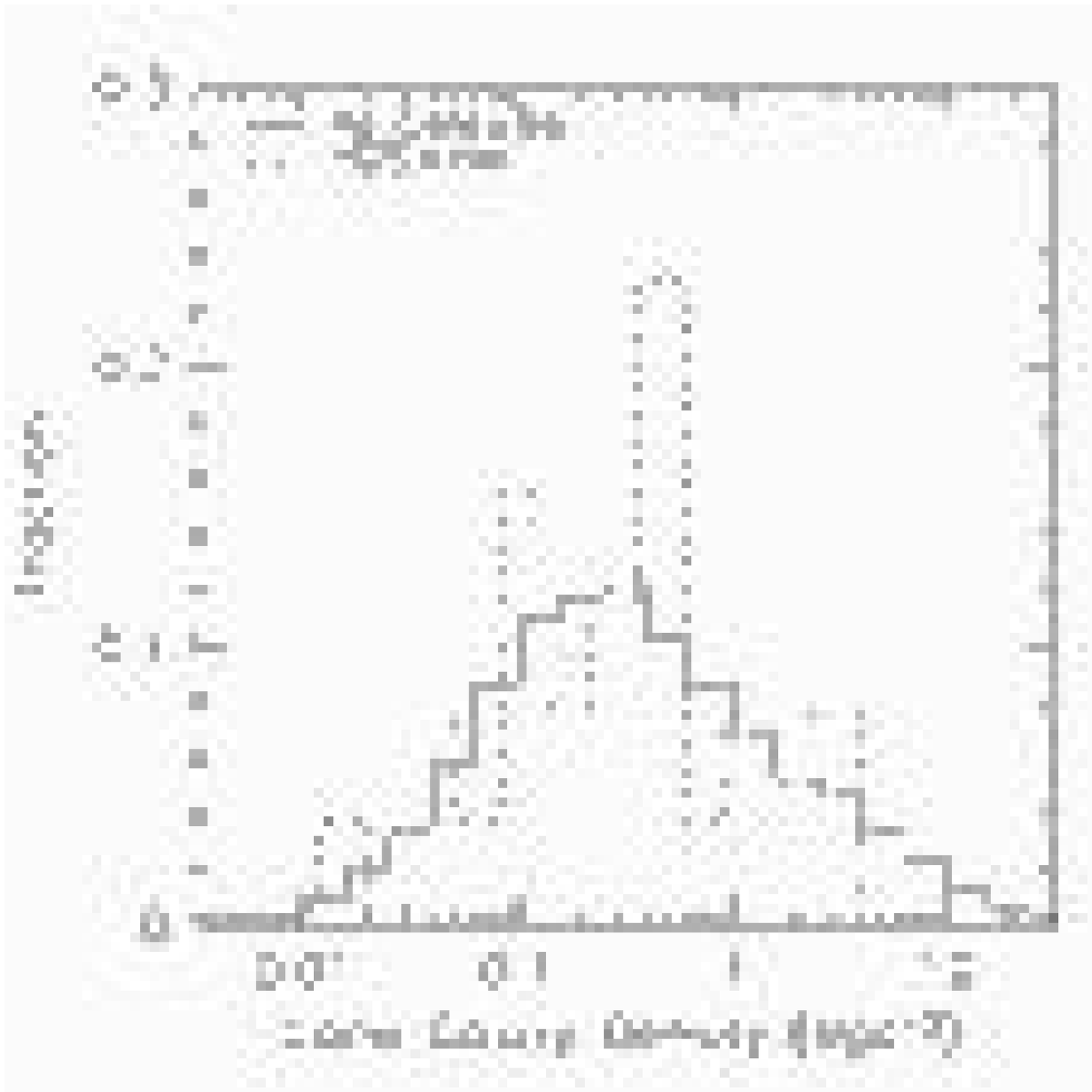}
\includegraphics[scale=0.39]{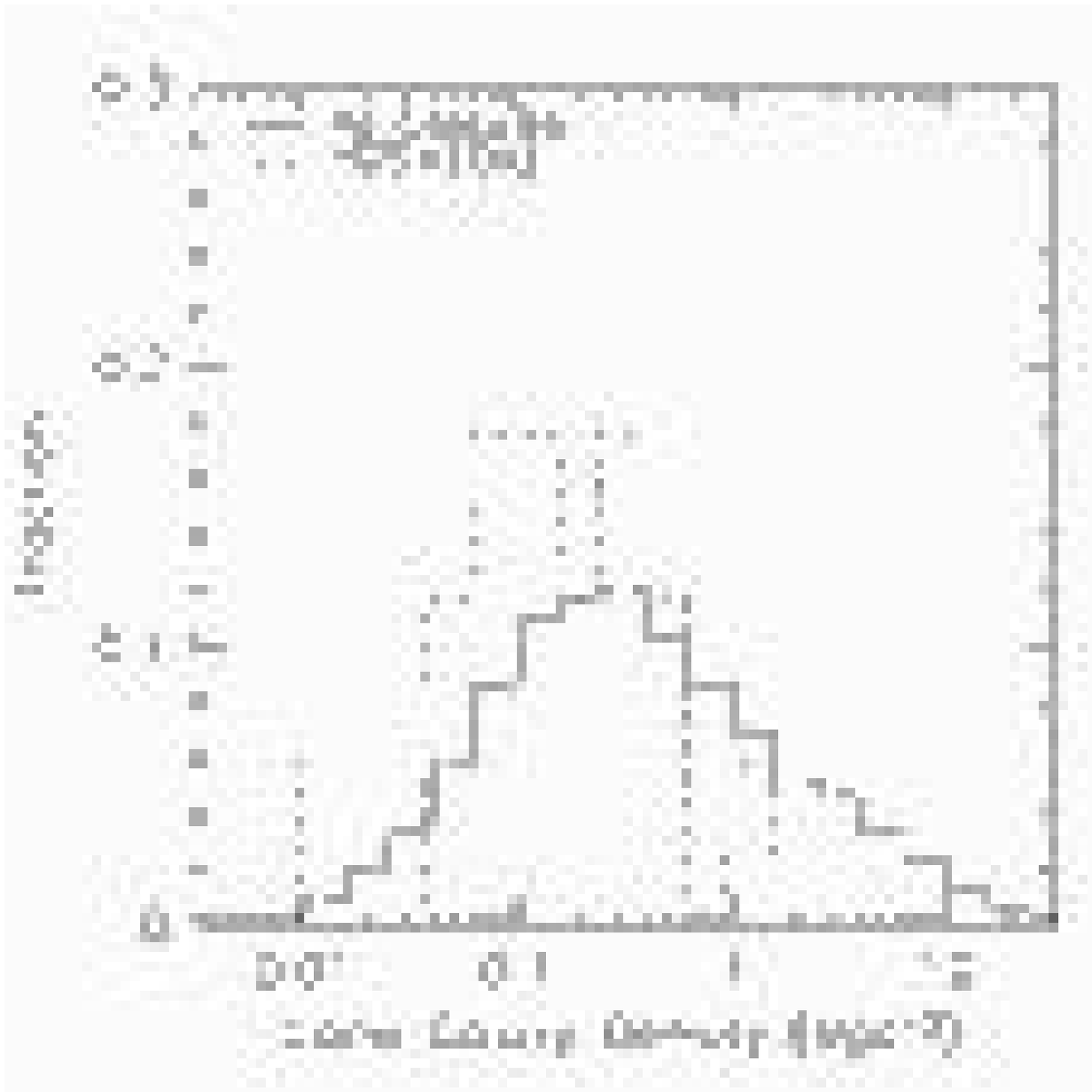}
\end{center}
\caption{
\label{fig:ea2_density_ea}
 Distributions of local galaxy density for each subsample of H$\delta$-strong
 galaxies and all  94770 galaxies in the volume limited sample. The 
 solid line is for all galaxies. The long
 dashed, dotted, short dashed and dot-dashed lines represent E+A,
 HDS+em, HDS+H$\alpha$ and HDS+[OII] samples, respectively. 
}
\end{figure}

\begin{figure}
\begin{center}
\centering{
\includegraphics[scale=0.39]{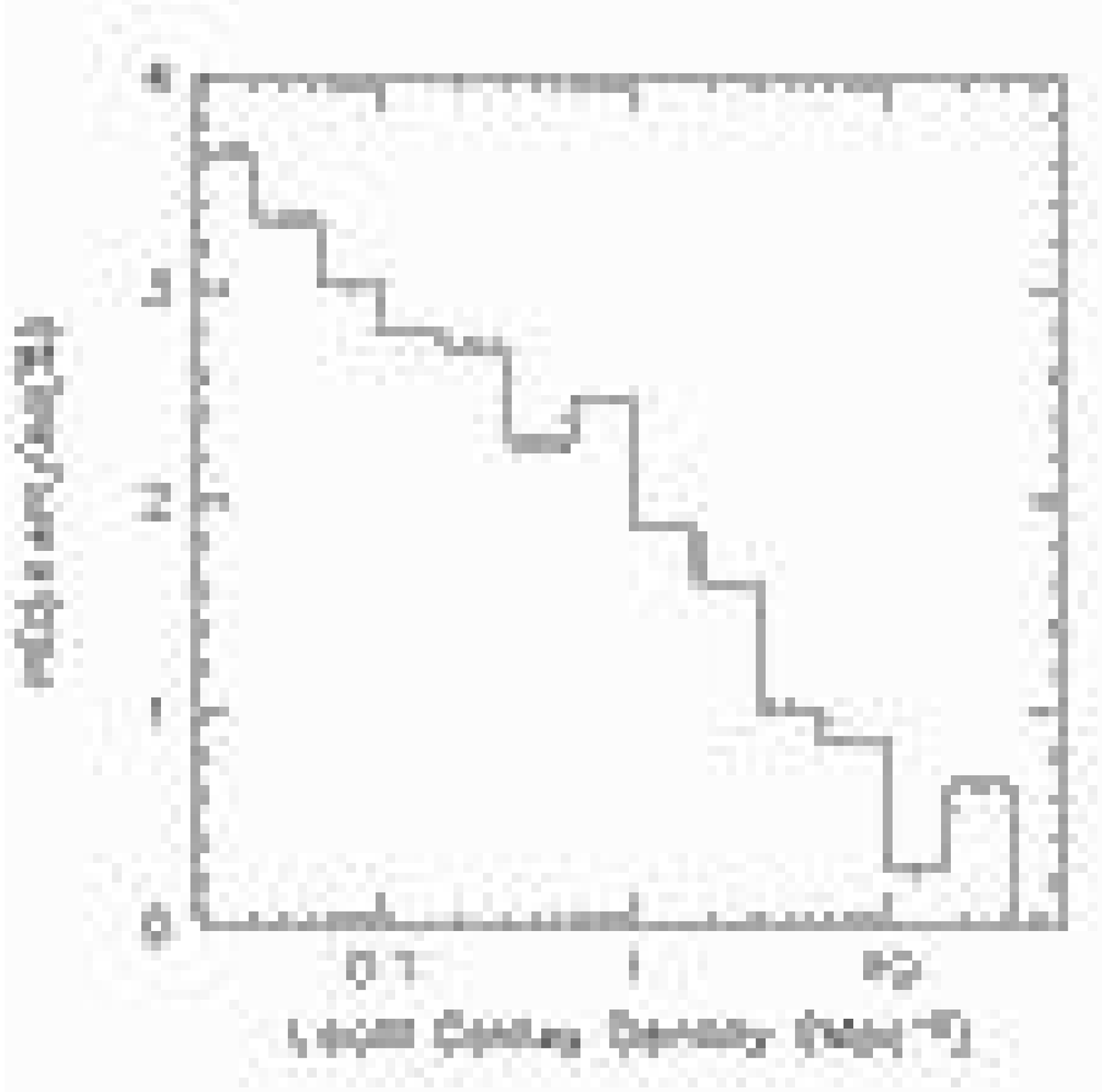}
\includegraphics[scale=0.39]{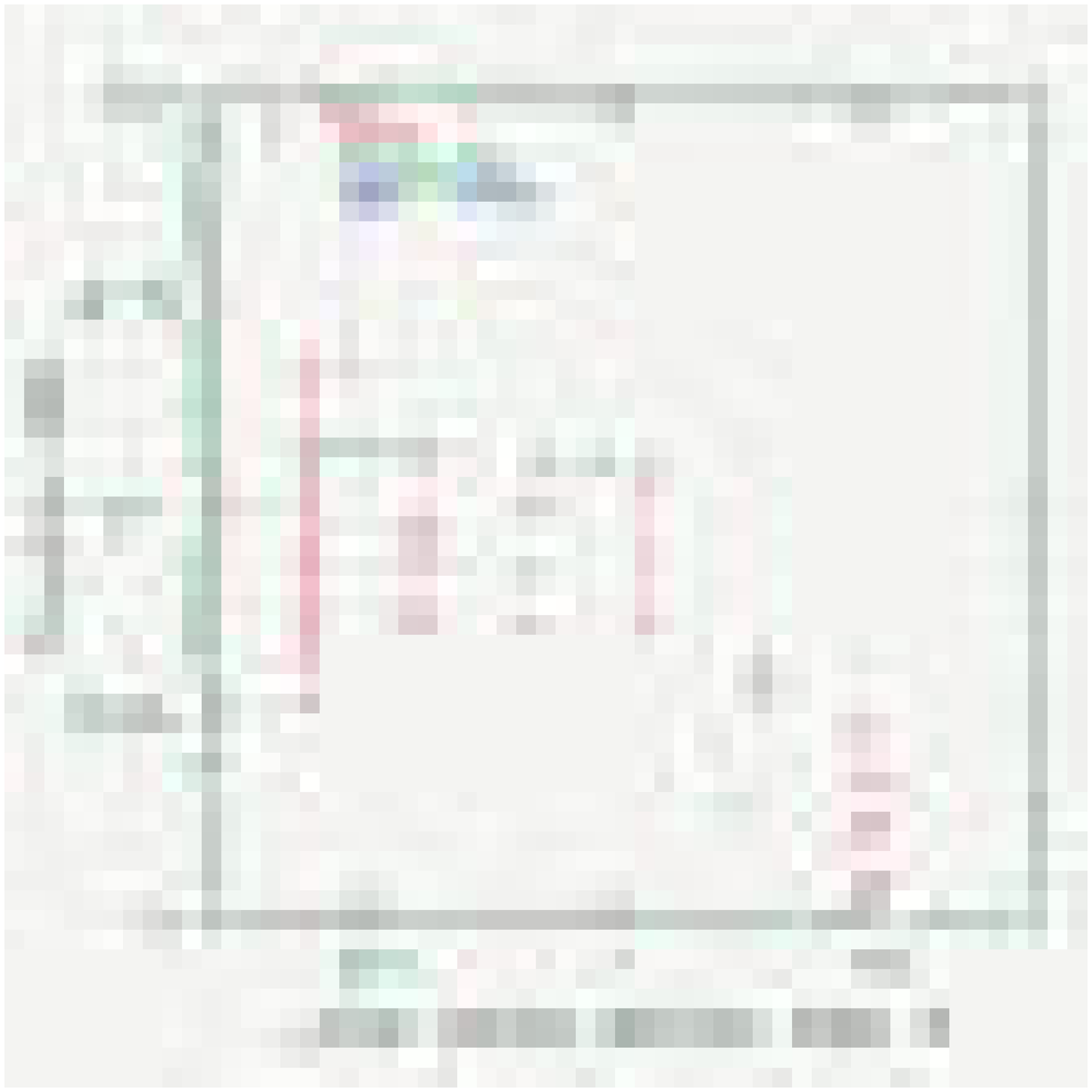}
}
\end{center}
\caption{
\label{fig:ea2_density_ratio}
 Ratio of each subclass of H$\delta$-strong galaxies to all galaxies as a function of
 local galaxy density. The left panel shows ratio for HDS+em galaxies to
 all galaxies in the volume limited sample.
 In the right panel, a long
 dashed, short dashed and dotted-dashed lines represent E+A,
  HDS+H$\alpha$ and HDS+[OII] galaxies, respectively. 
    }
\end{figure}

\begin{figure}
\begin{center}
\includegraphics[scale=0.7]{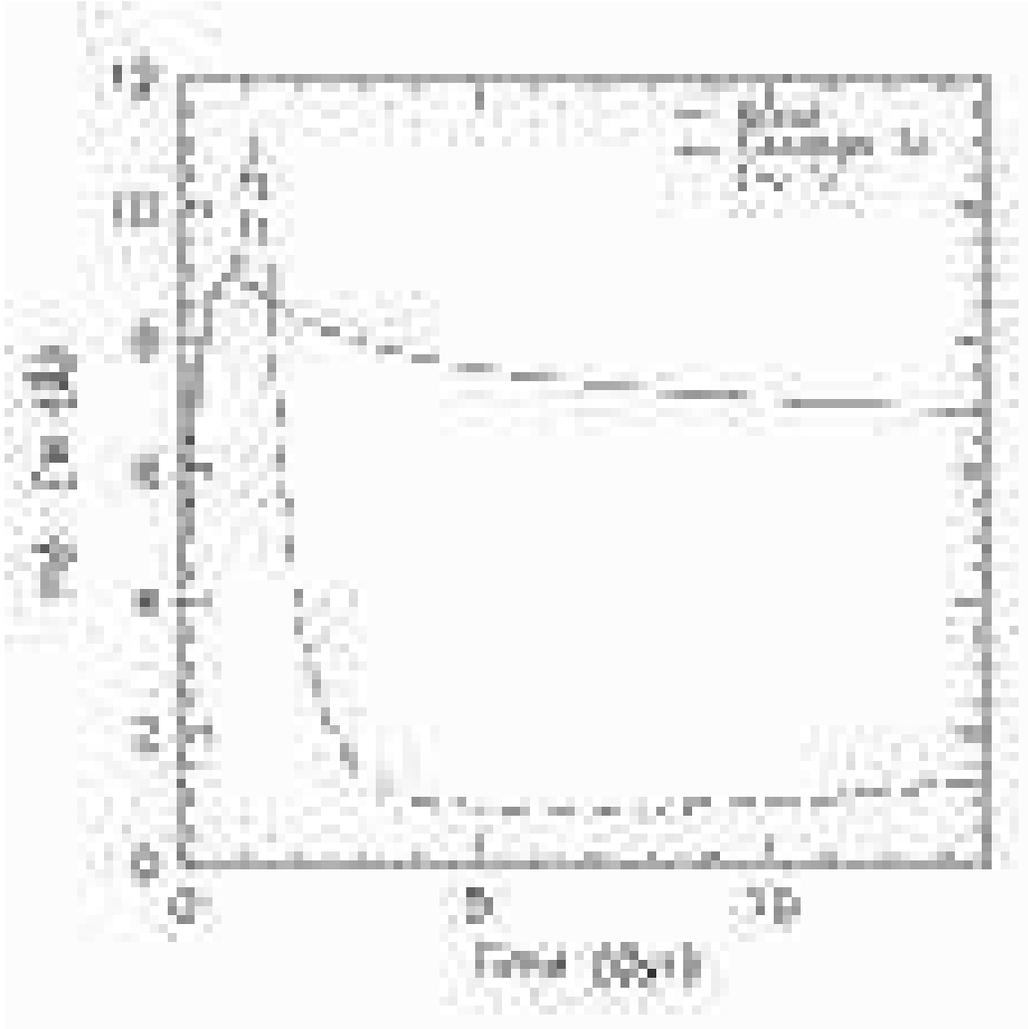}
\end{center}
\caption{
 H$\delta$ EWs are plotted against time (age) for three star
 formation histories with the GISSEL model.  The dashed, solid and dotted
 lines show the models with instantaneous burst, constant star formation and
 exponentially decaying star formation rate. The models in this figure
 assume Salpeter IMF and solar metallicity. }\label{fig:ea2_time_hd}
\end{figure}

\begin{figure}
\begin{center}
\includegraphics[scale=0.7]{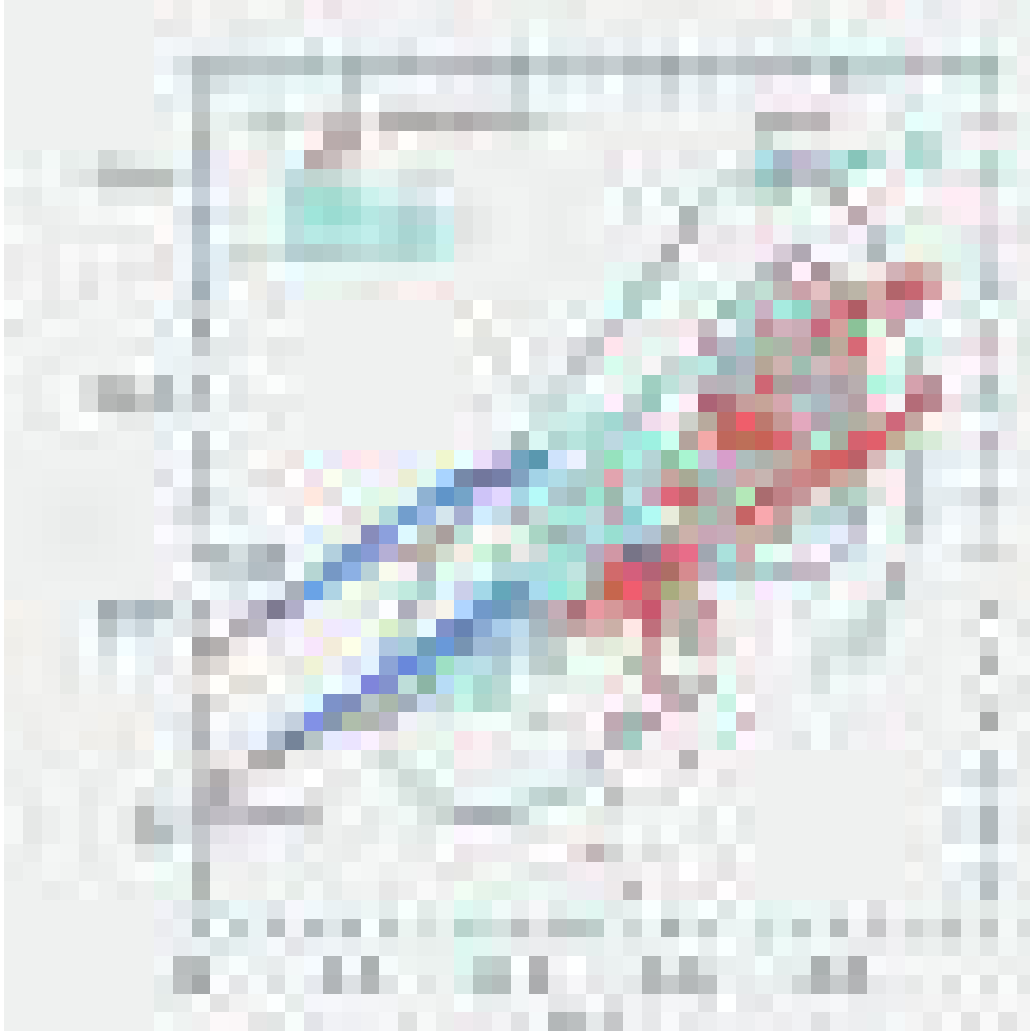}
\end{center}
\caption{
 Restframe $g-r$ color is plotted against $r-i$ color.  The dashed, solid and dotted
 lines show the models with instantaneous burst, constant star formation and
 exponentially decaying star formation rate.  Two sets of the models are
 present for solar metallicity ($Z$=0.02) and 5 times solar metallicity
 ($Z$=0.1). Open circles,
 small dots, triangles and squares represent E+A, HDS+em, HDS+H$\alpha$
 and HDS+[OII], respectively.}
\label{fig:ea2_gri}
\end{figure}

\begin{figure}
\begin{center}
\includegraphics[scale=0.7]{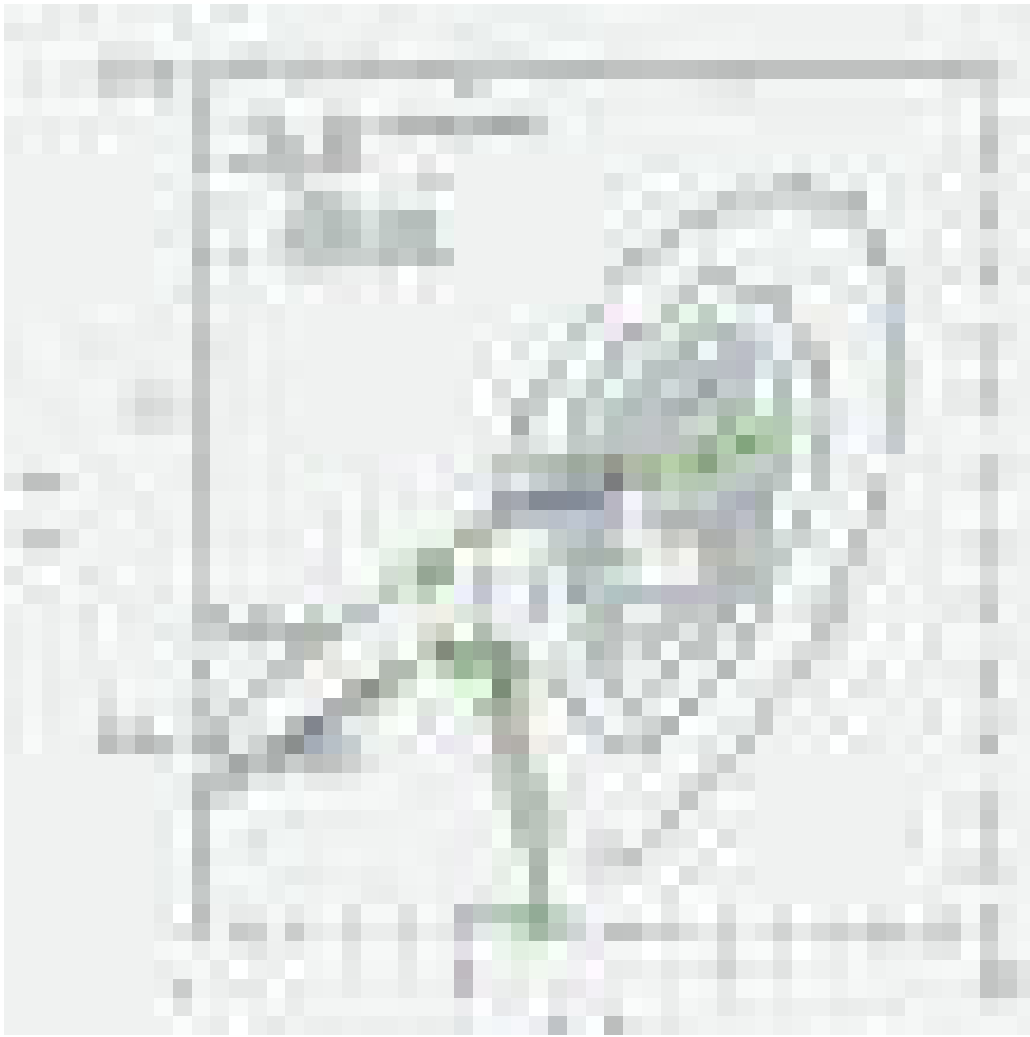}
\end{center}
\caption{$J-K$ is plotted against $r-K$. All magnitudes are in restframe
 AB system.
 The contours show distribution of all galaxies
 in our sample.  Open circles,  small dots, triangles and squares
 represent E+A, HDS+em, HDS+H$\alpha$  and HDS+[OII], respectively.
  The dashed, solid and dotted
 lines show the models with instantaneous burst, constant star formation and
 exponentially decaying star formation rate. Three sets of the models
 are plotted for different metallicities.  
}\label{fig:ea2_jk_rk}
\end{figure}

\begin{figure}
\begin{center}
\includegraphics[scale=0.7]{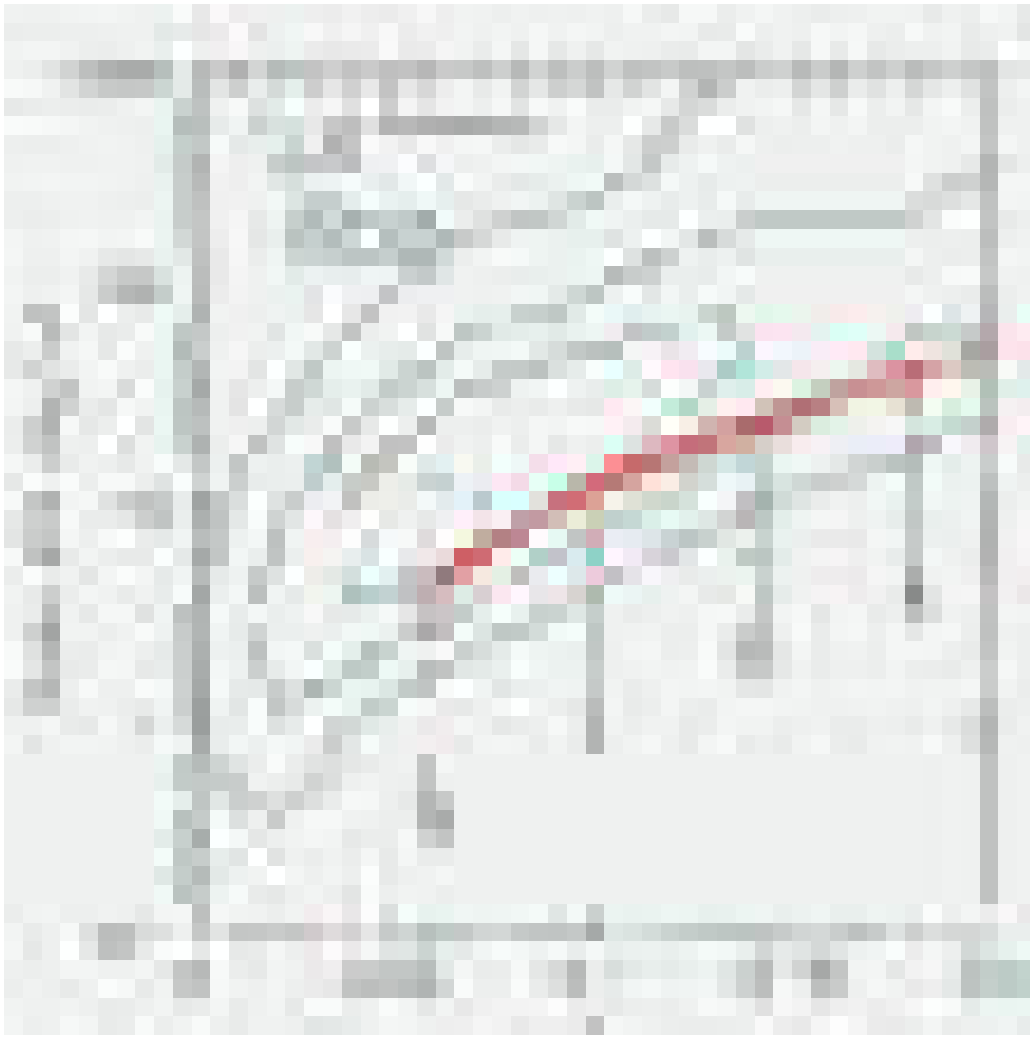}
\end{center}
\caption{ 
 Radio SFR calculated using the FIRST data is plotted against redshift.
 The contours show the distribution of all galaxies
 in our sample.  Open circles,  small dots, triangles and squares
 represent E+A, HDS+em, HDS+H$\alpha$  and HDS+[OII], respectively.
  When a H$\delta$-strong galaxy is not detected in the FIRST data, we assigned 1 mJy as
 an upper limit of radio flux. Those galaxies with no radio detection
 appear in the plot as a line around 
 10 M$_{\odot}$ yr$^{-1}$ showing the upper limit of radio SFR.
}\label{fig:ea2_radio_sfr}
\end{figure}

\begin{figure}
\begin{center}
\includegraphics[scale=0.39]{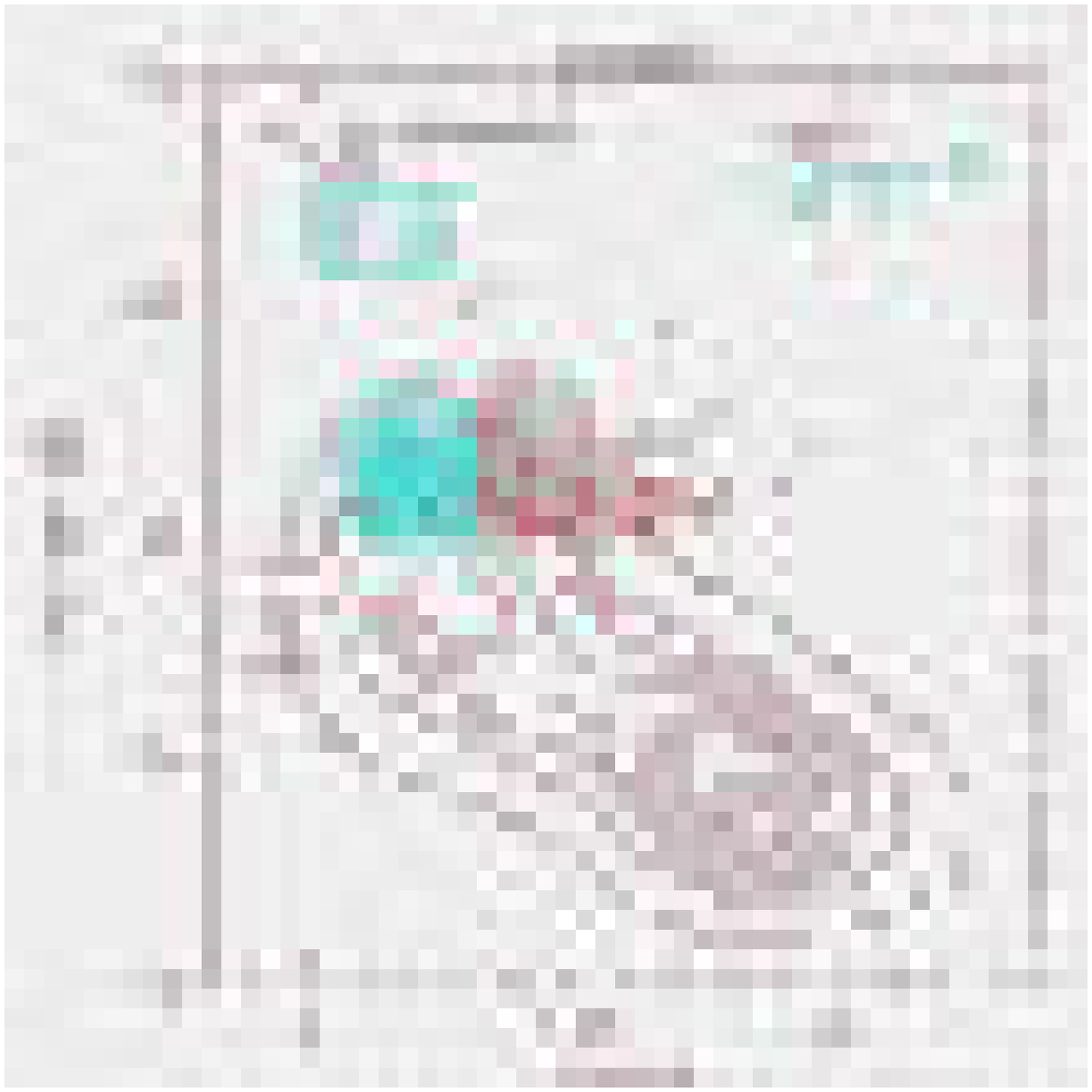}
\includegraphics[scale=0.39]{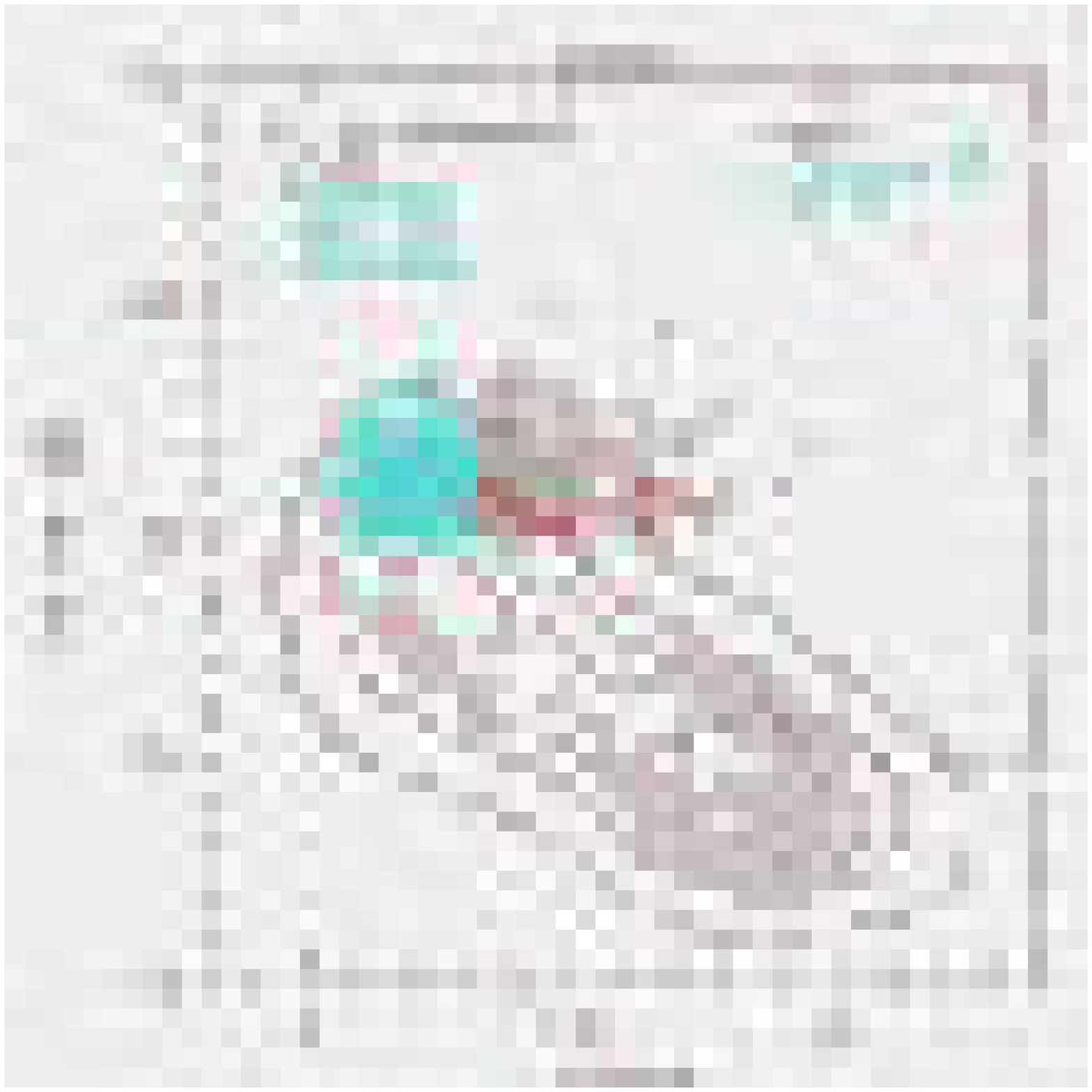}
\includegraphics[scale=0.39]{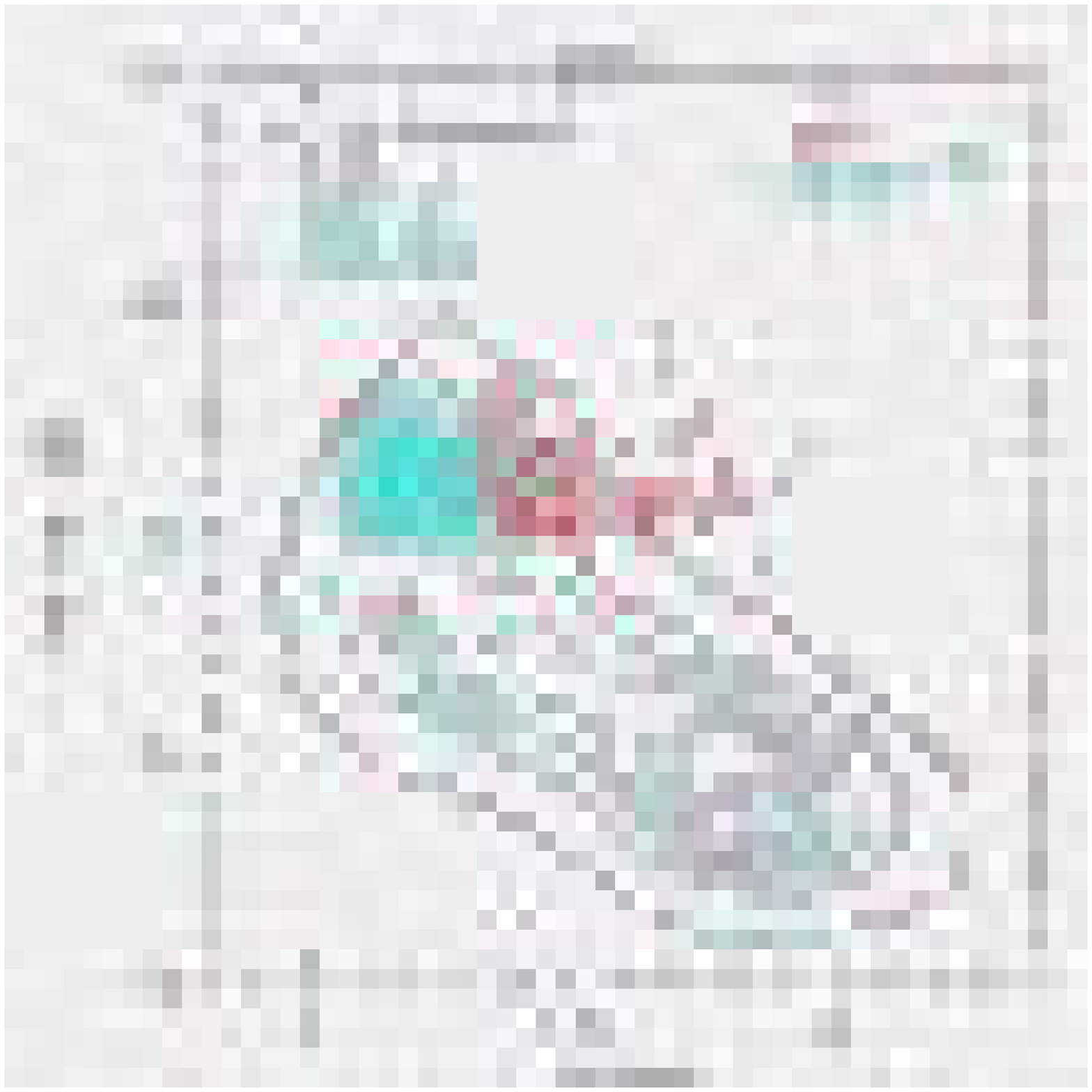}
\end{center}
\caption{
 H$\delta$ EWs are plotted against D4000 for the models with $Z$=0.0001(0.5\%
 solar), 0.02(solar) and 0.1 (5
 times solar) from top to bottom.
  The dashed, solid and dotted
 lines are for the models with instantaneous burst, constant star formation and
 exponentially decaying star formation rate. 
    Observational data are plotted using open circles,
 small dots, triangles and squares for E+A, HDS+em, HDS+H$\alpha$
 and HDS+[OII], respectively.
}\label{fig:ea2_hd_d4000}
\end{figure}

\begin{figure}
\begin{center}
\includegraphics[scale=0.39]{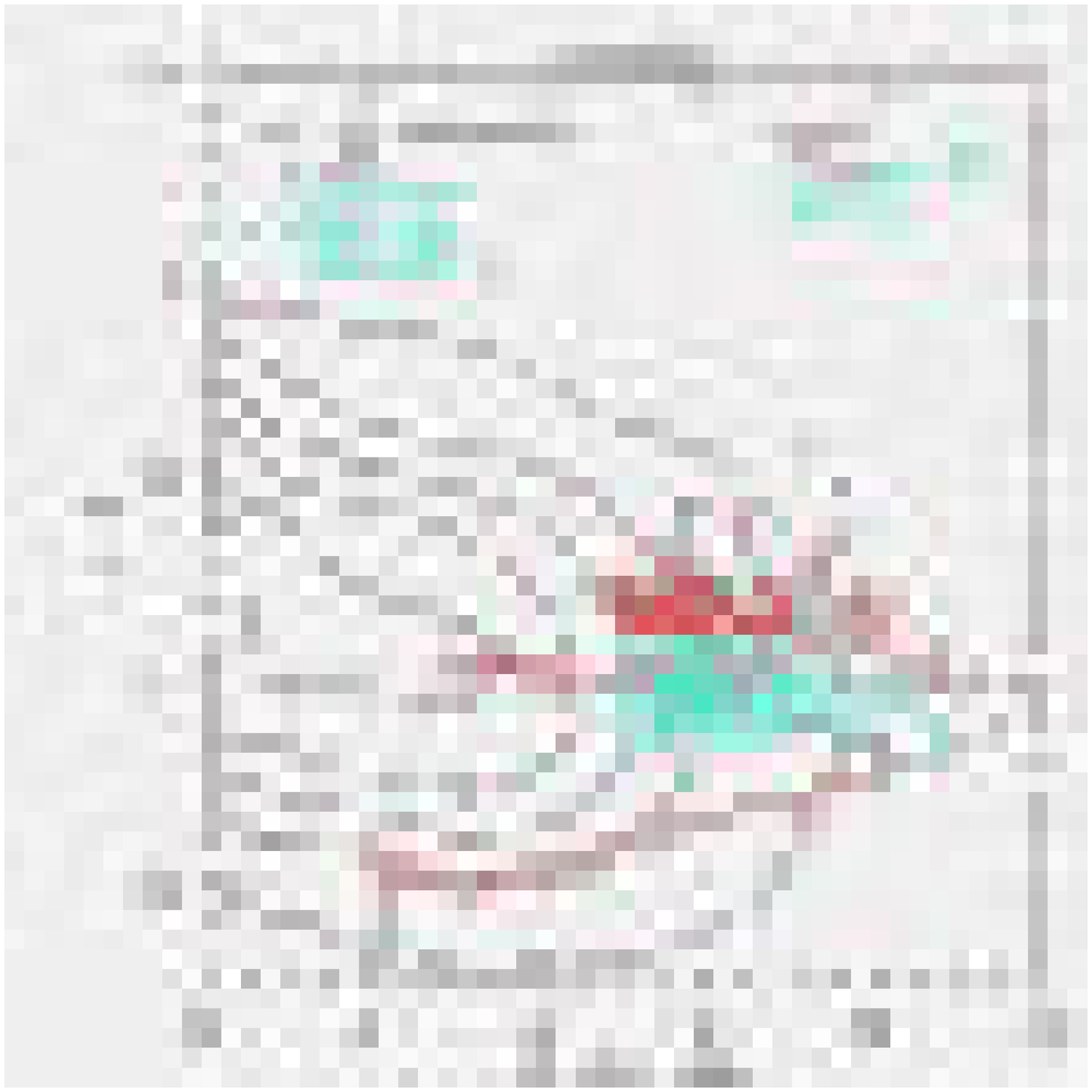}
\includegraphics[scale=0.39]{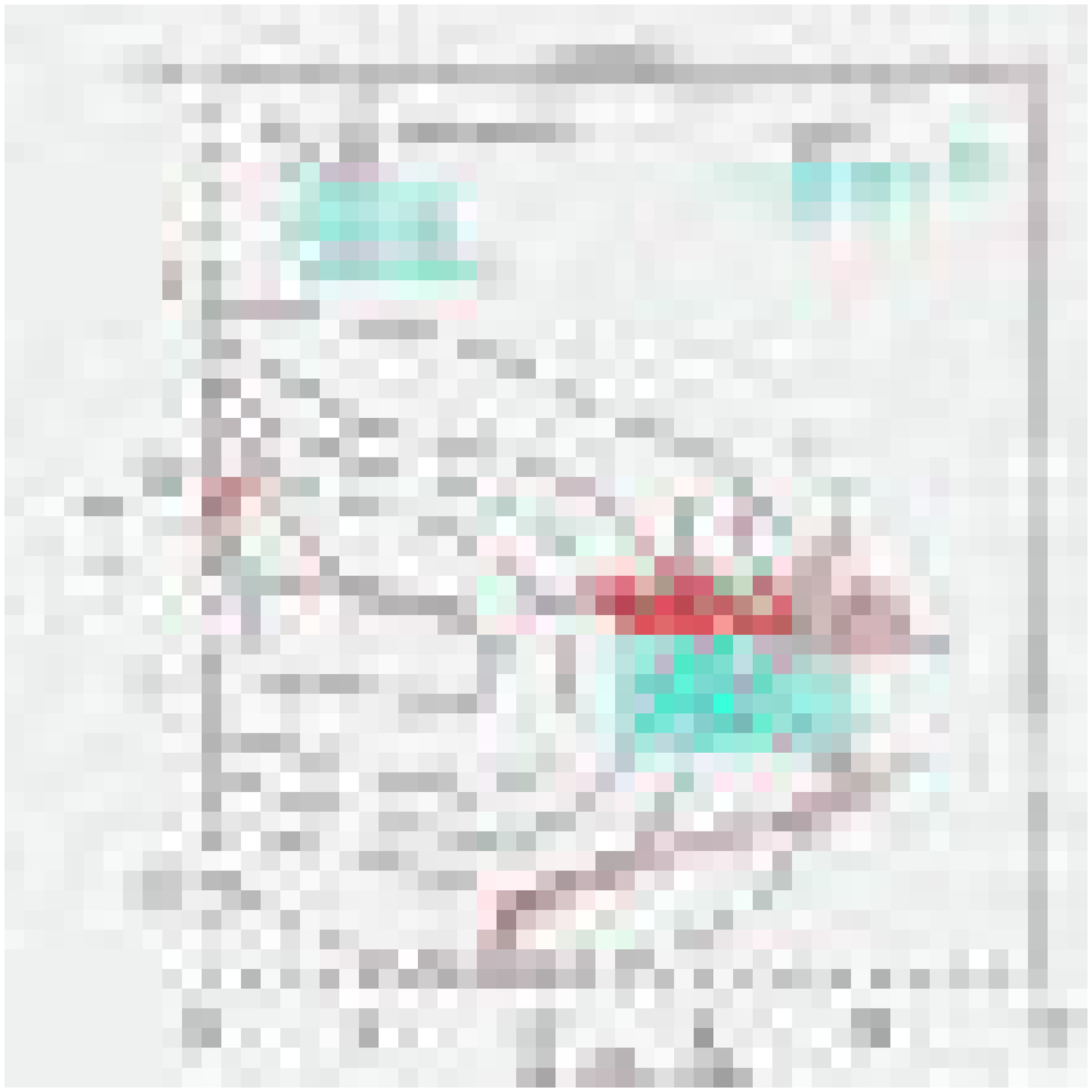}
\includegraphics[scale=0.39]{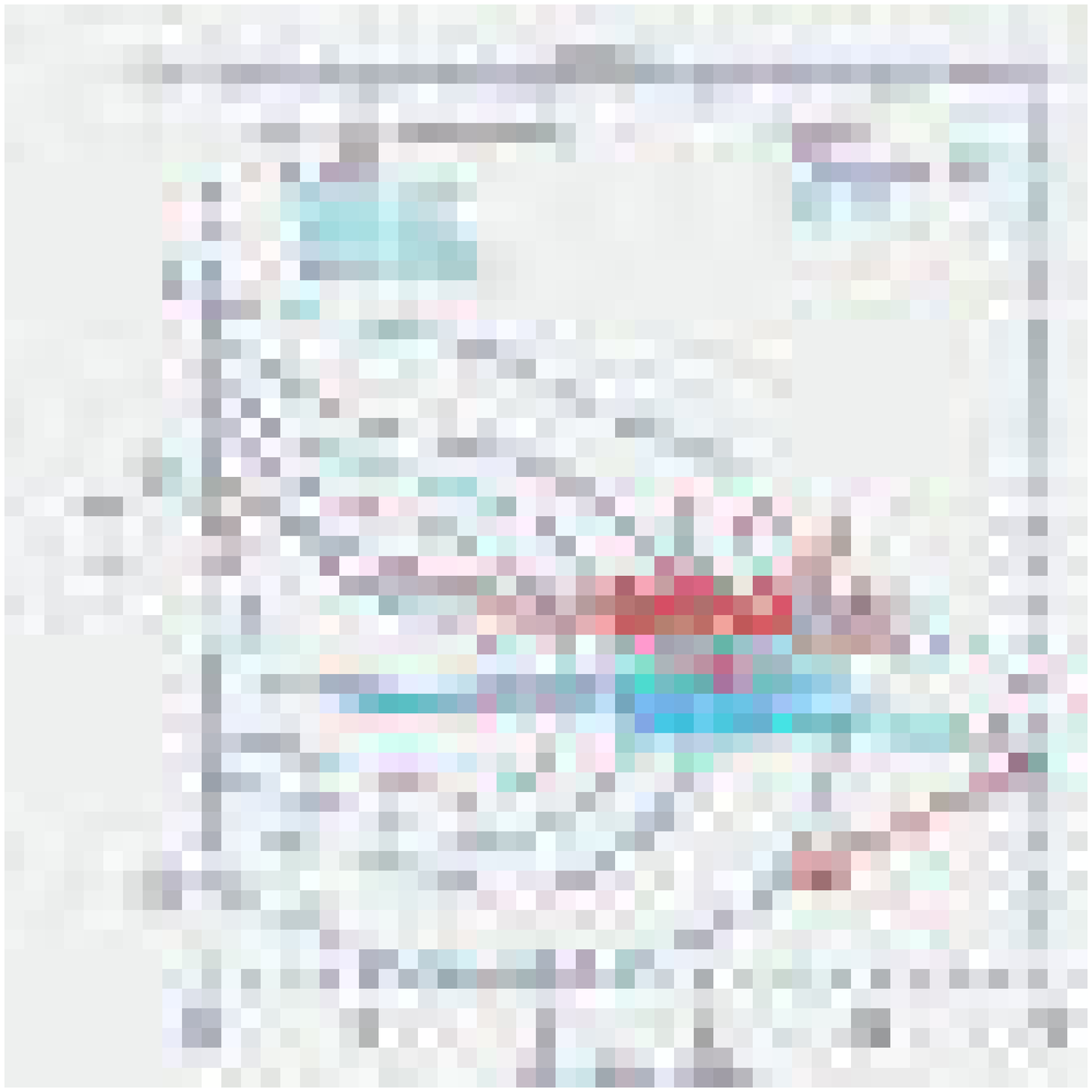}
\end{center}
\caption{
 The $u-g$ color is plotted against  H$\delta$ EWs for three metallicity models
 and three star formation histories. Metallicities are $Z$=0.0001(0.5\%
 solar), 0.0004(2\% solar), 0.04(20\% solar), 0.02(solar), 0.10(5 times
 solar) from top to bottom. Star formation histories are the burst,
 constant and exponentially decreasing, shown by 
  the dashed, solid and dotted lines, respectively. 
    Observational data are plotted using open circles,
 small dots, triangles and squares for E+A, HDS+em, HDS+H$\alpha$
 and HDS+[OII], respectively.
}\label{fig:ea2_ug_hd}
\end{figure}

\begin{figure}
\begin{center}
\includegraphics[scale=0.25]{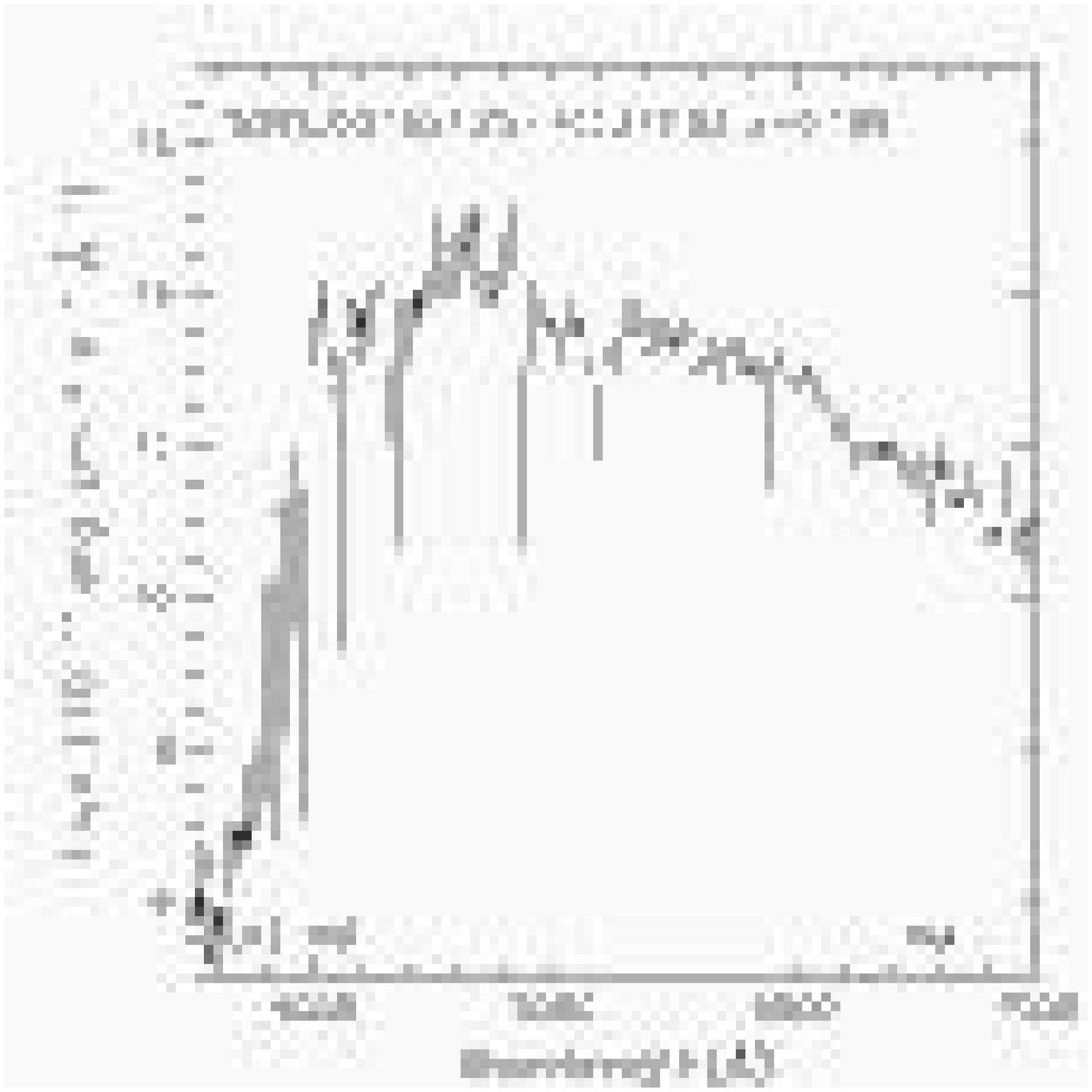}
\includegraphics[scale=0.25]{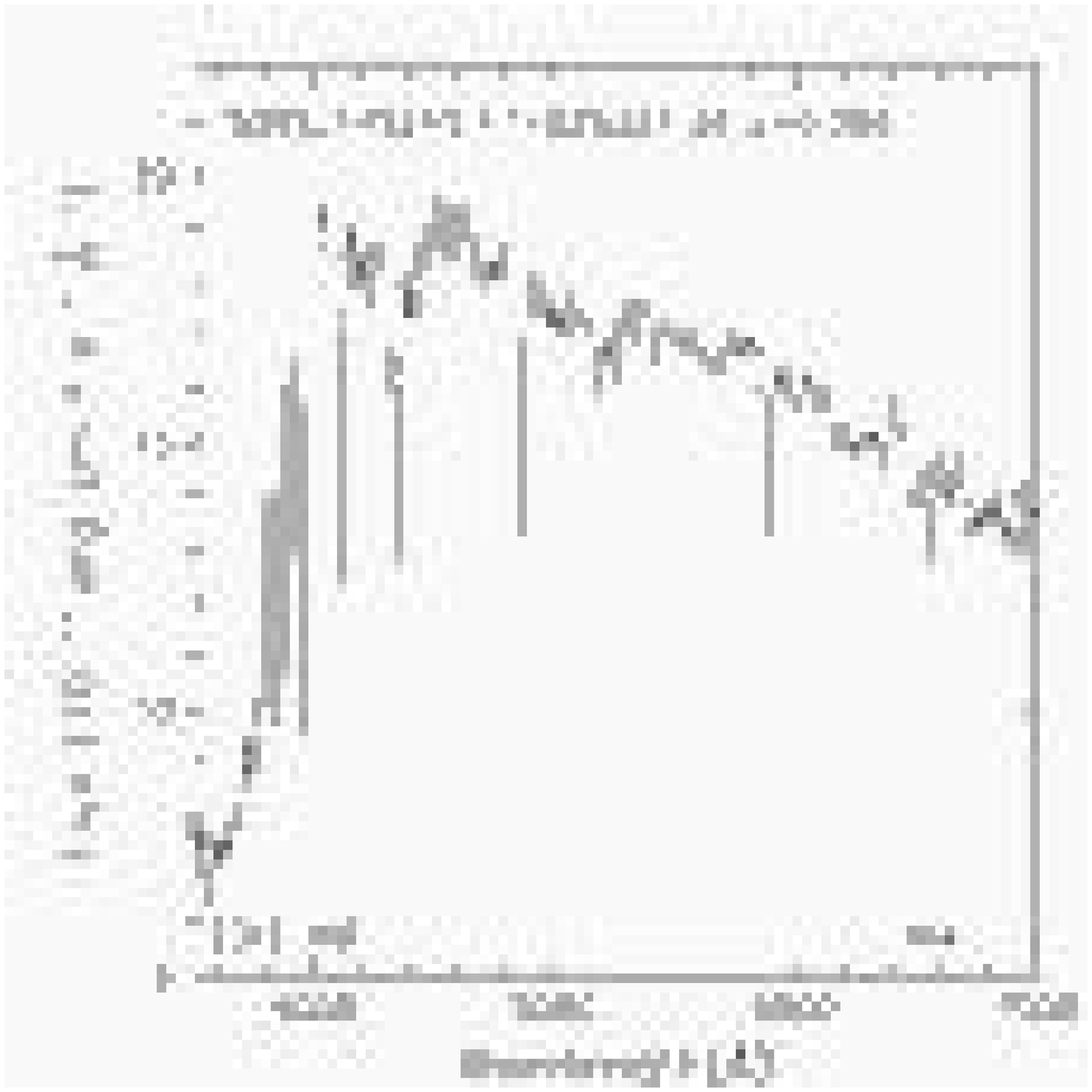}
\includegraphics[scale=0.25]{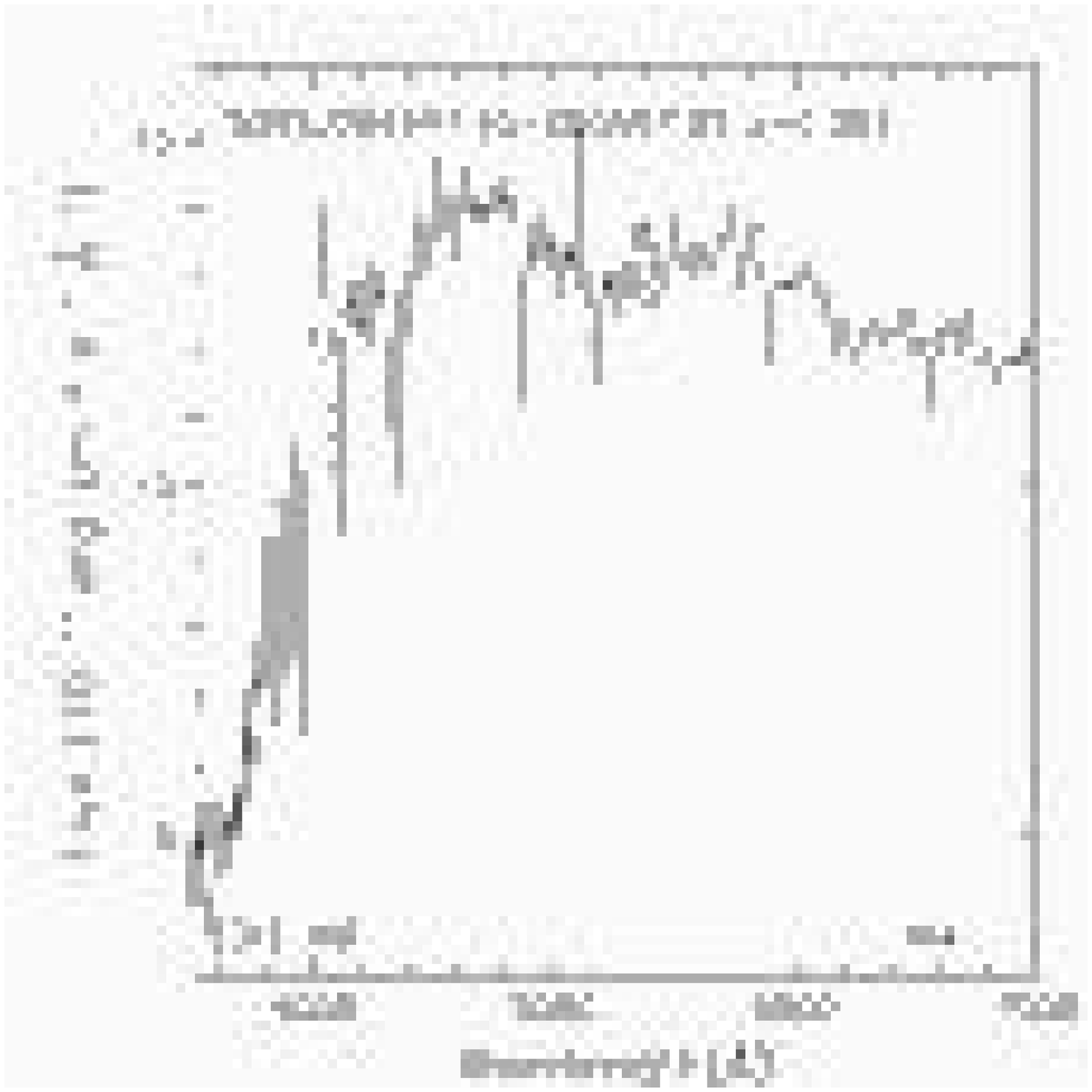}
\includegraphics[scale=0.25]{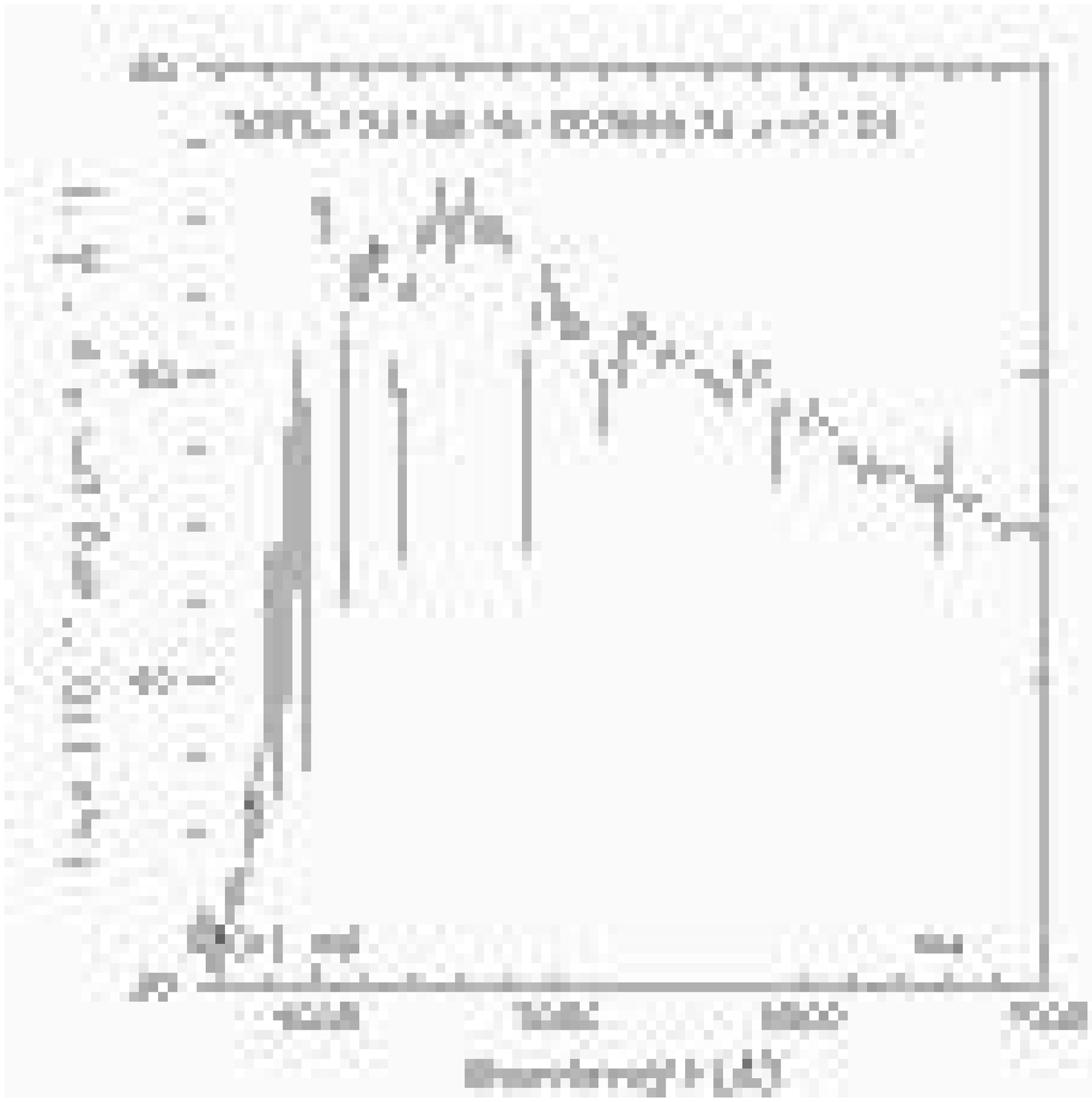}
\includegraphics[scale=0.25]{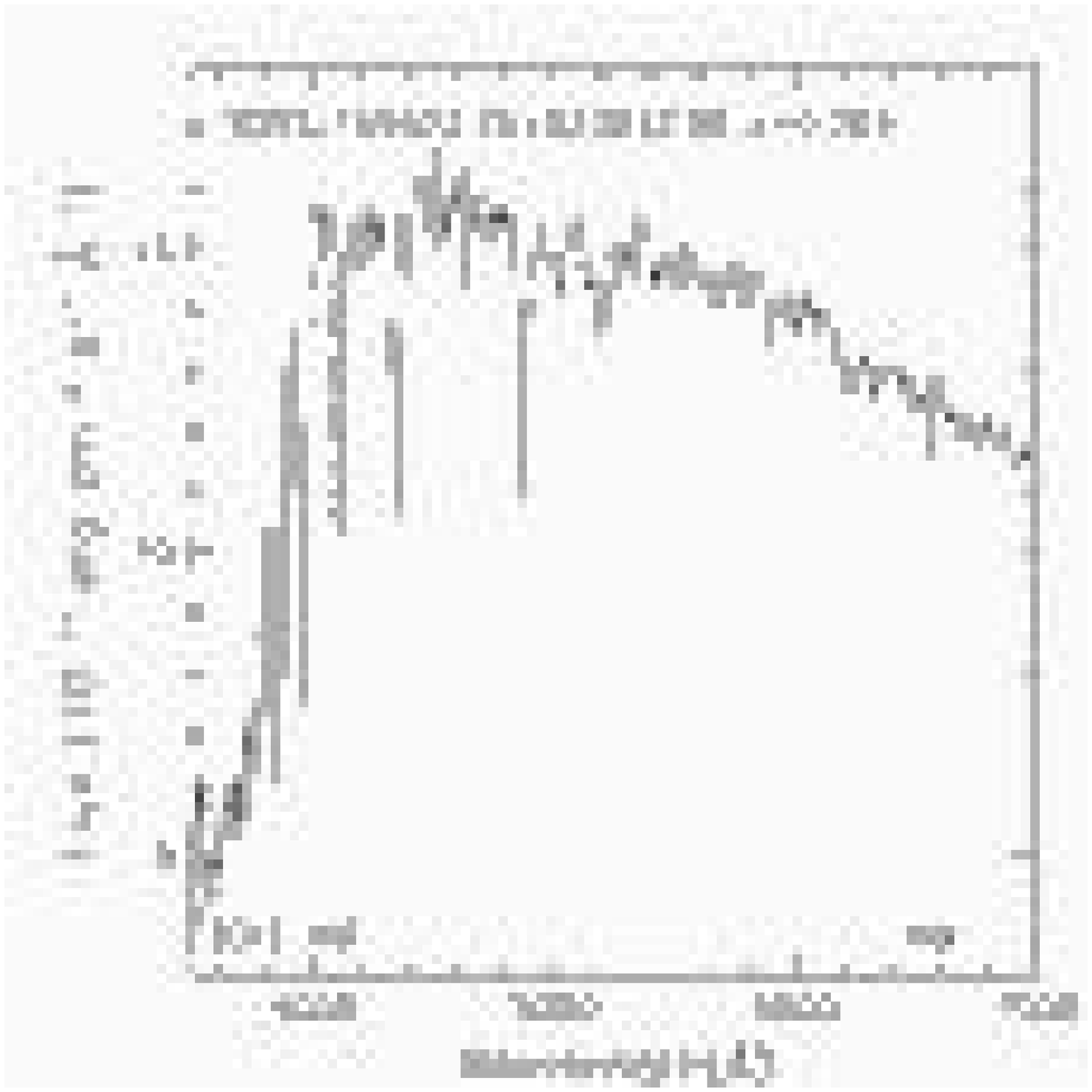}
\includegraphics[scale=0.25]{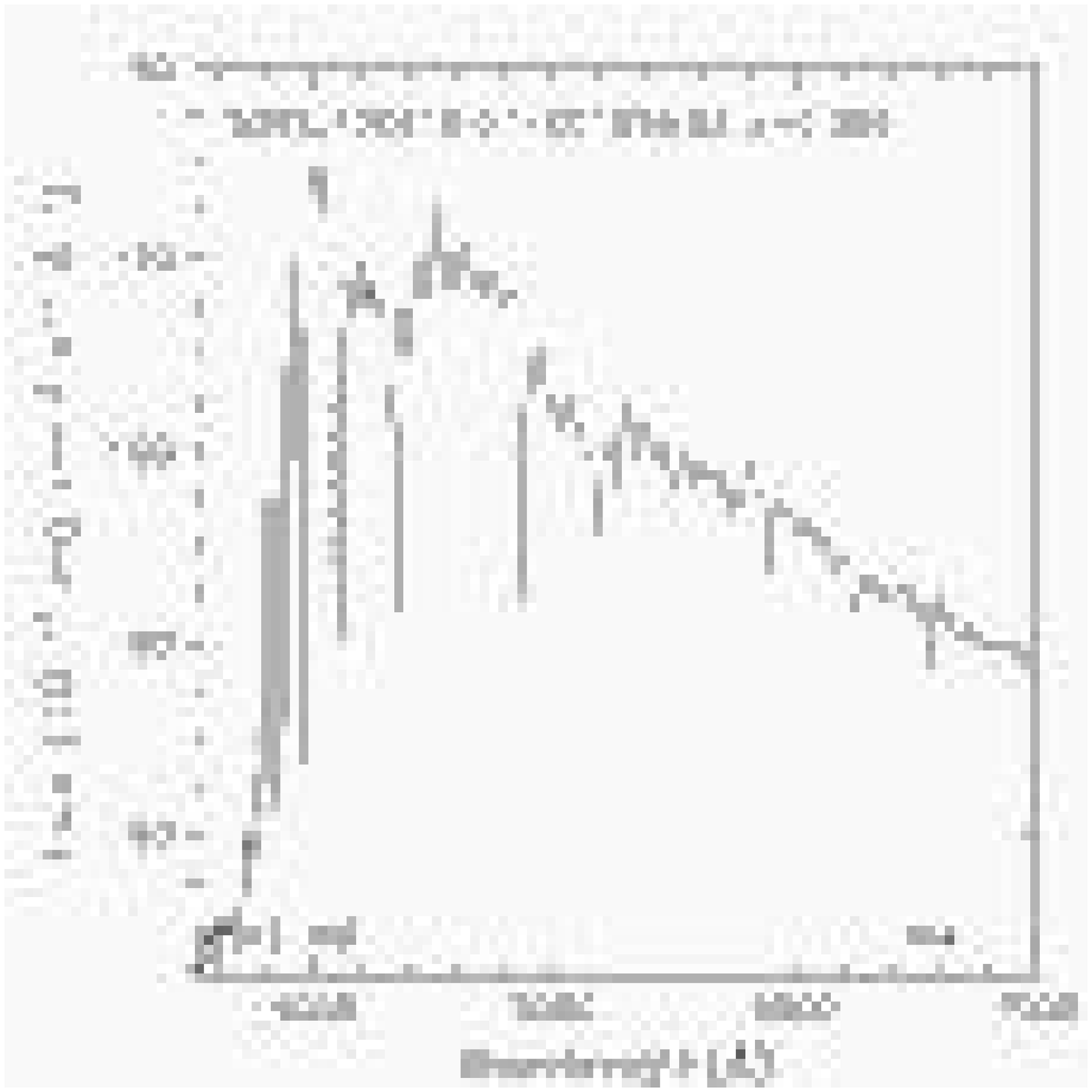}
\includegraphics[scale=0.25]{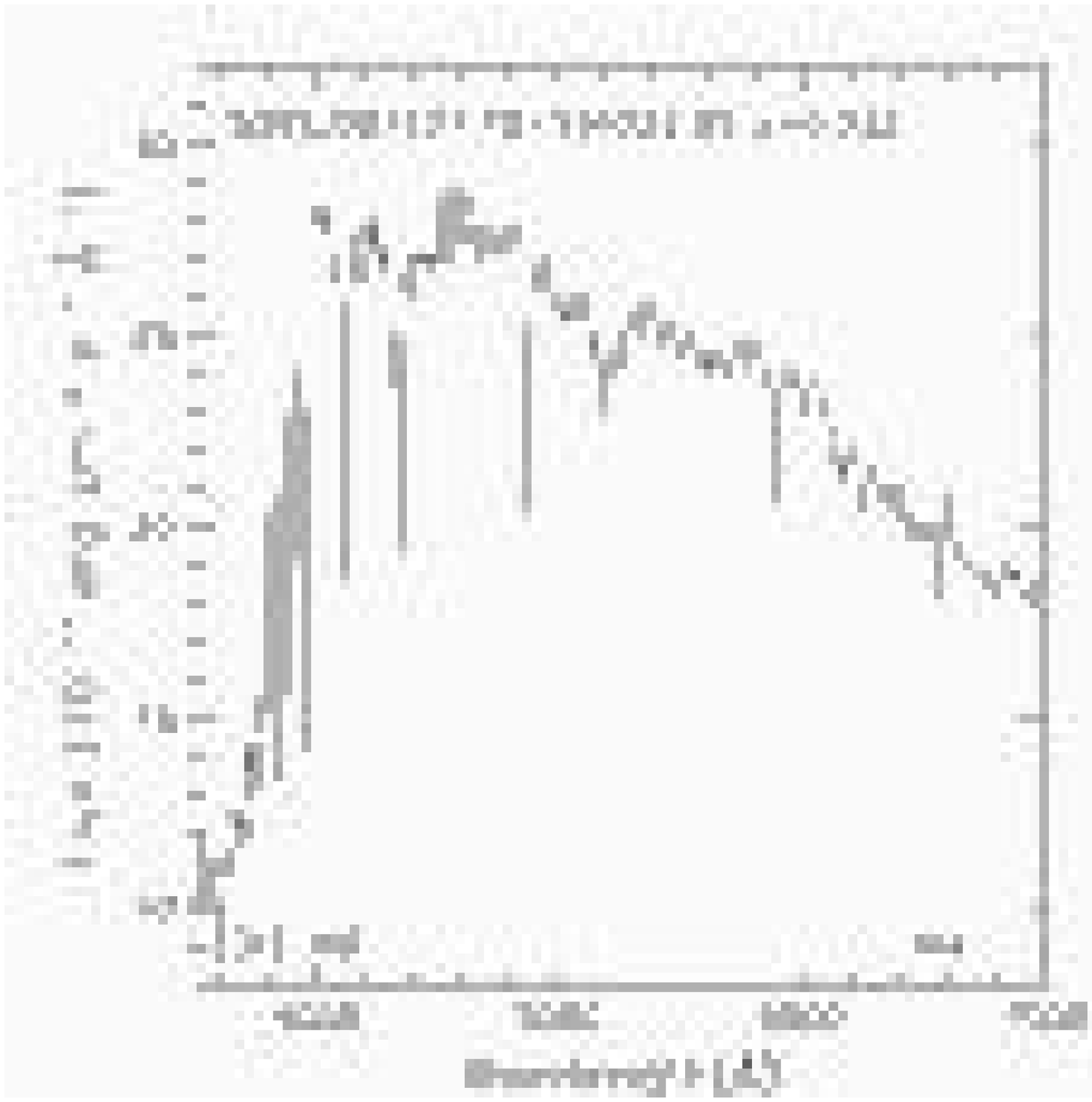}
\includegraphics[scale=0.25]{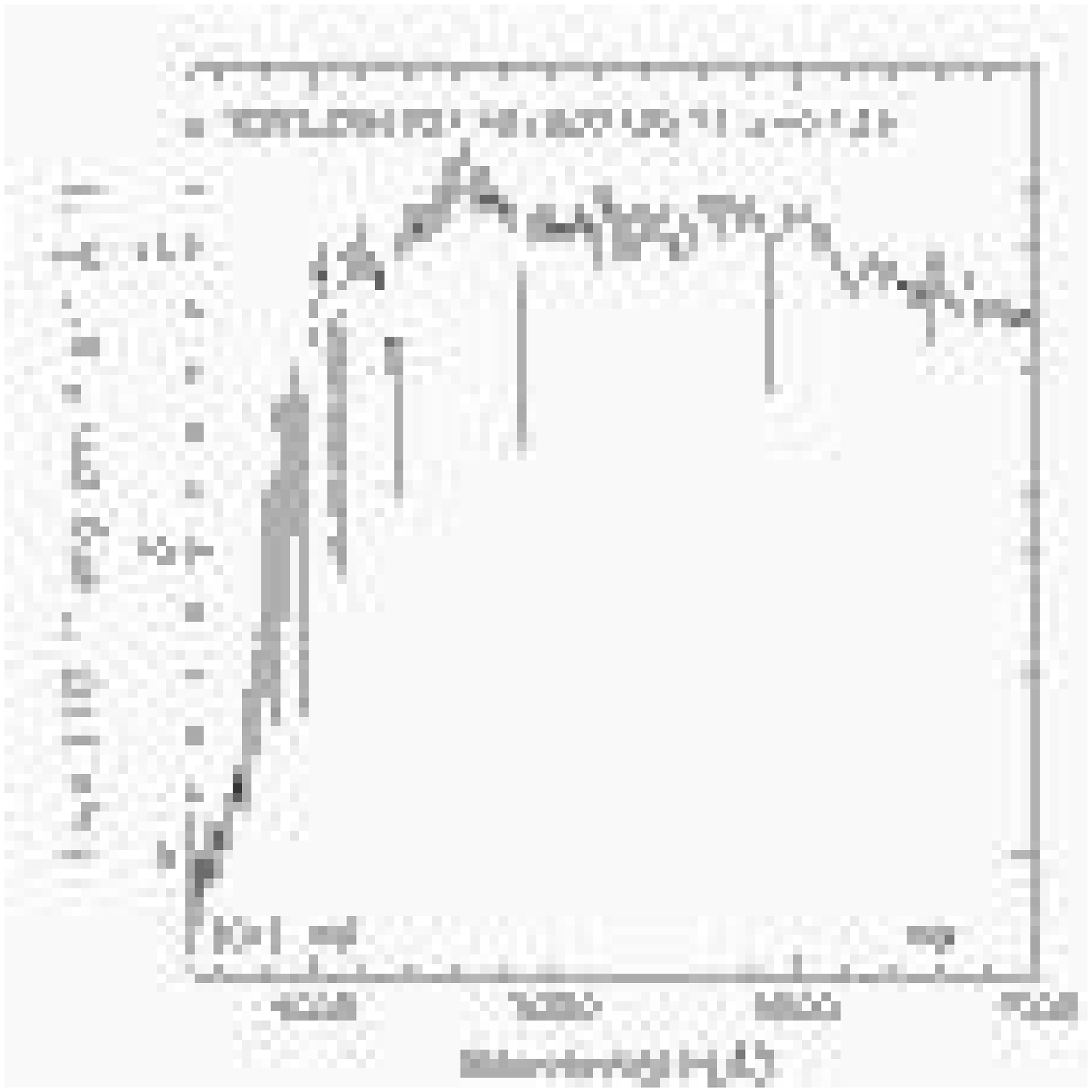}
\includegraphics[scale=0.25]{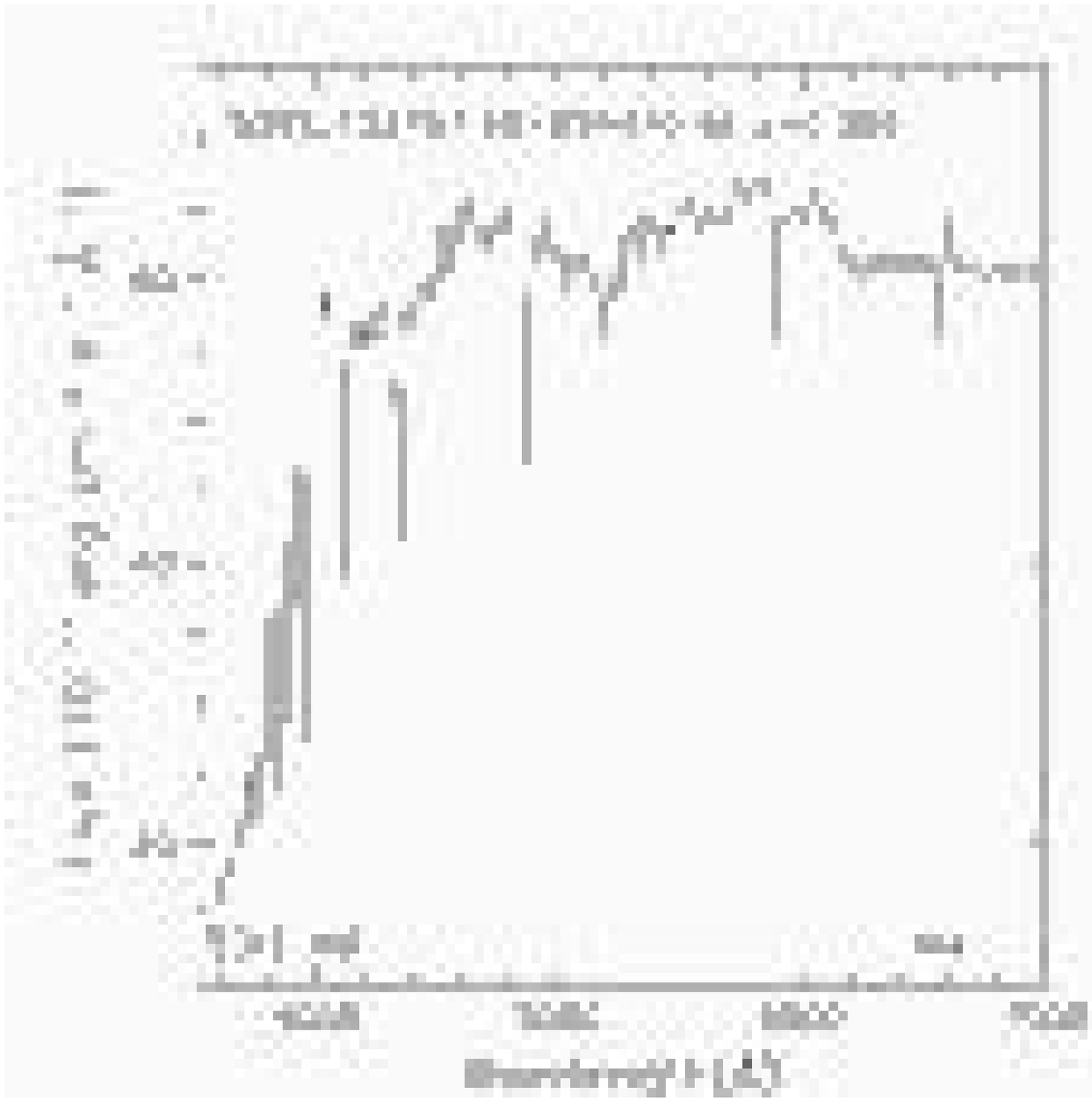}
\end{center}
\caption{
 Nine example spectra of young E+A galaxies (E+As with H$\delta$ EW $>$7 \AA). Spectra are shifted
 to restframe and smoothed using a 20\AA box. 
}\label{fig:ea2_progenitor_spectra_individual}
\end{figure}

\begin{figure}
\begin{center}
\includegraphics[scale=0.2]{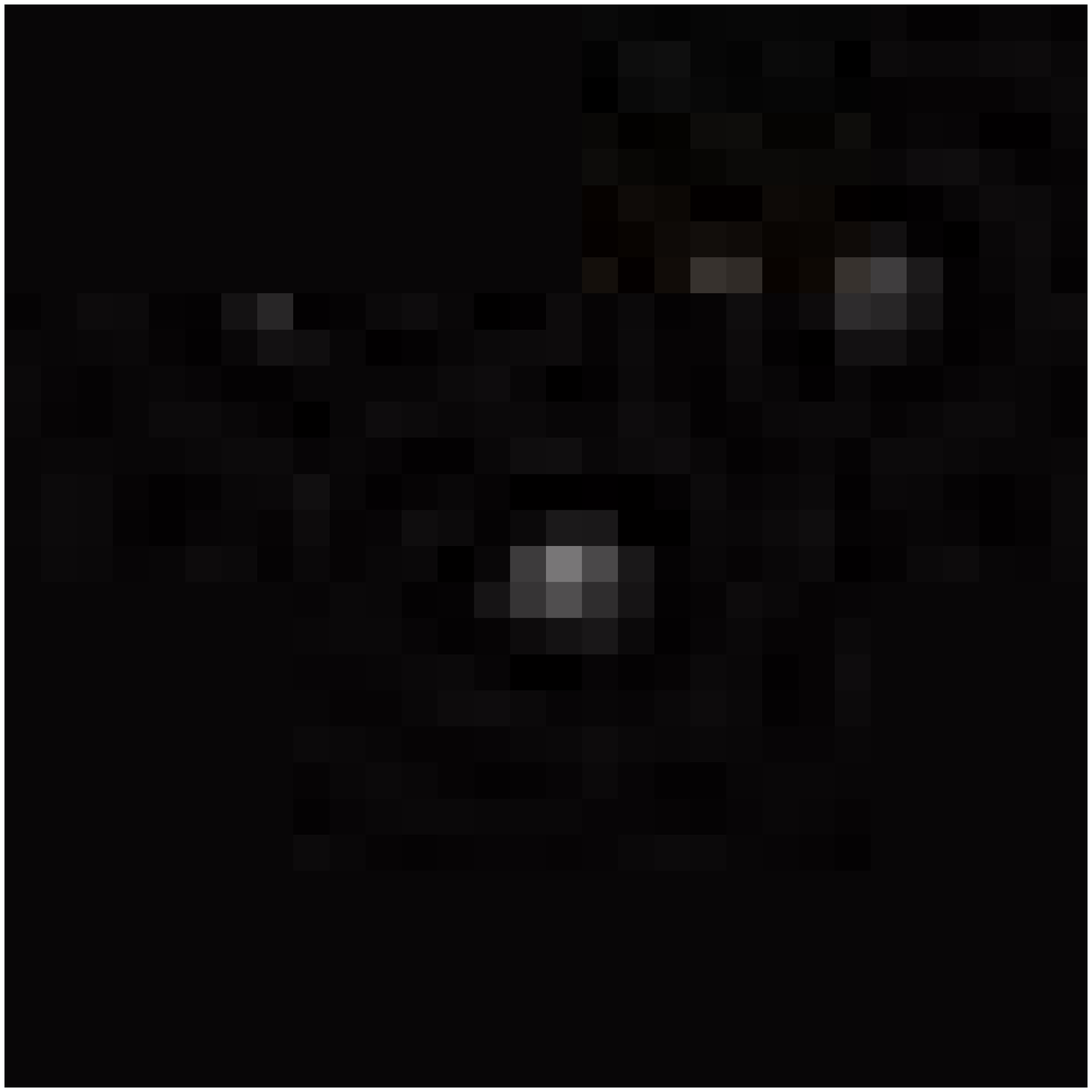}
\includegraphics[scale=0.2]{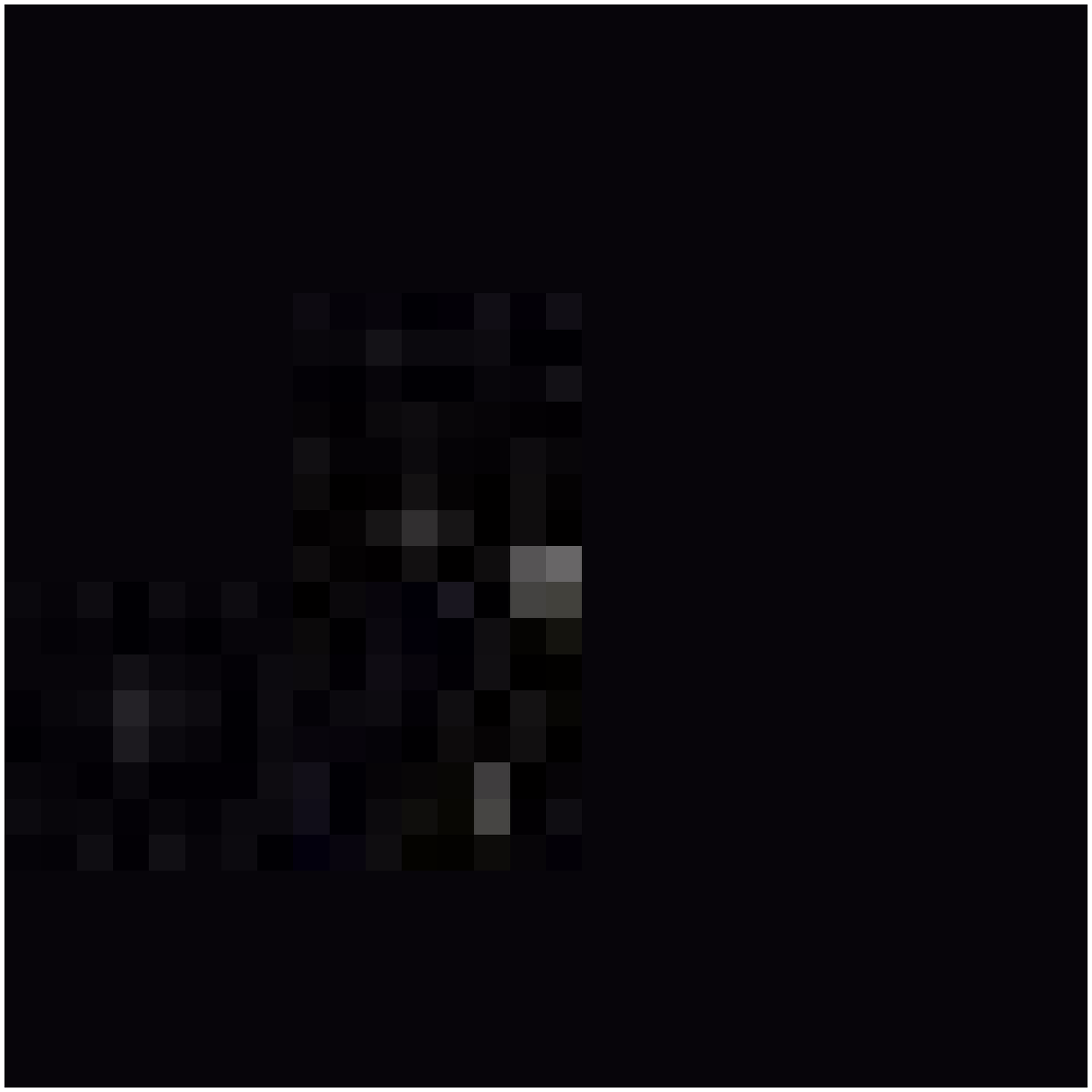}
\includegraphics[scale=0.2]{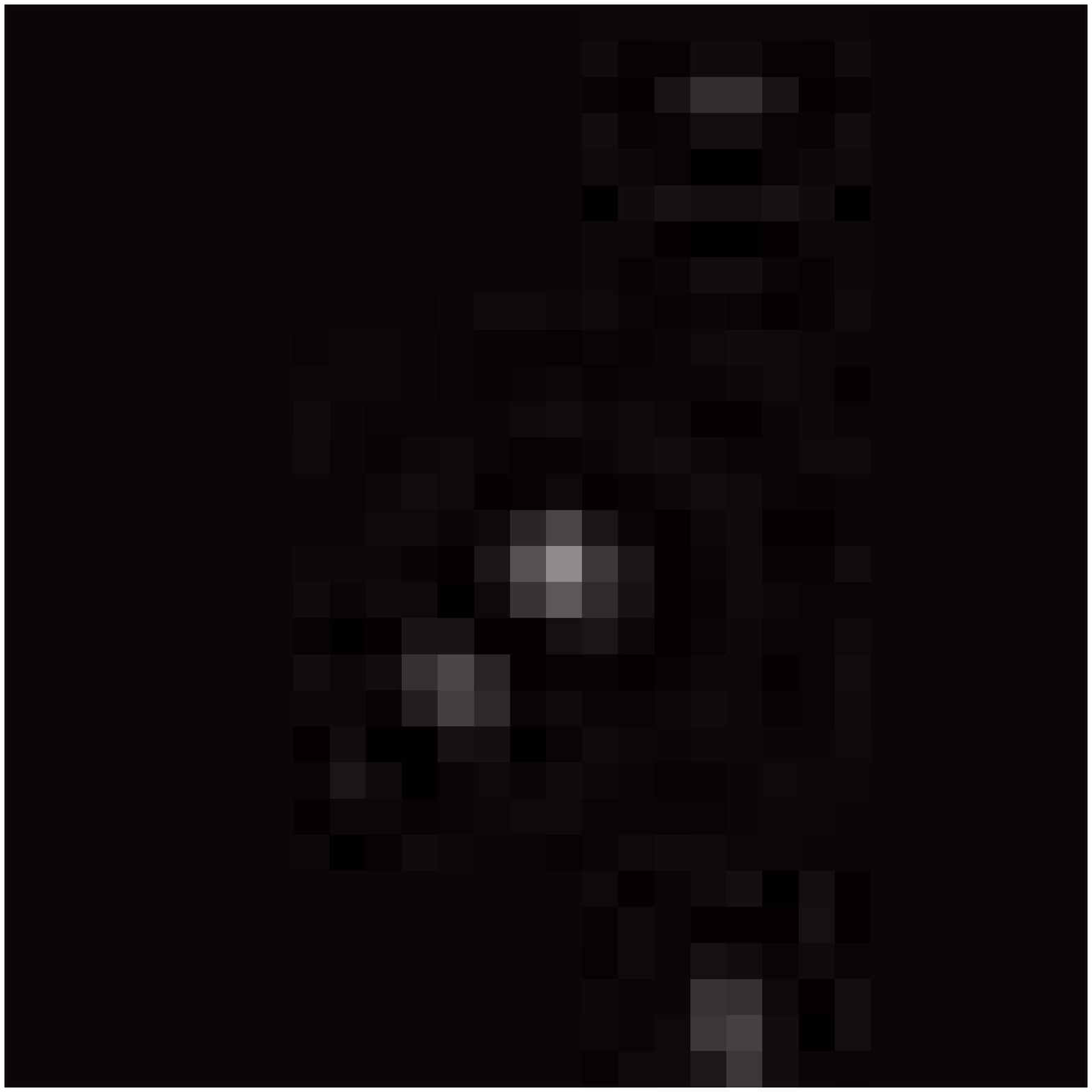}
\includegraphics[scale=0.2]{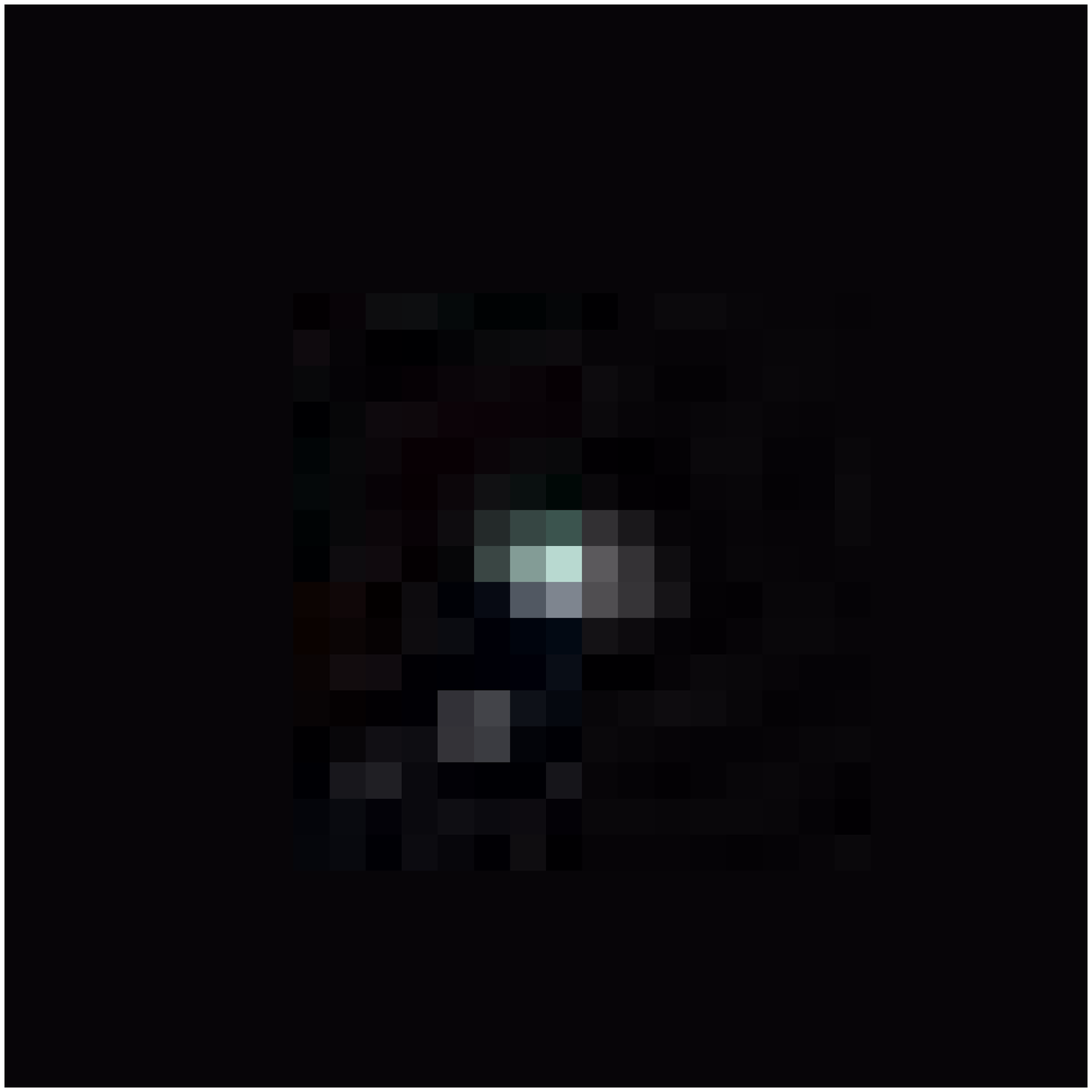}
\includegraphics[scale=0.2]{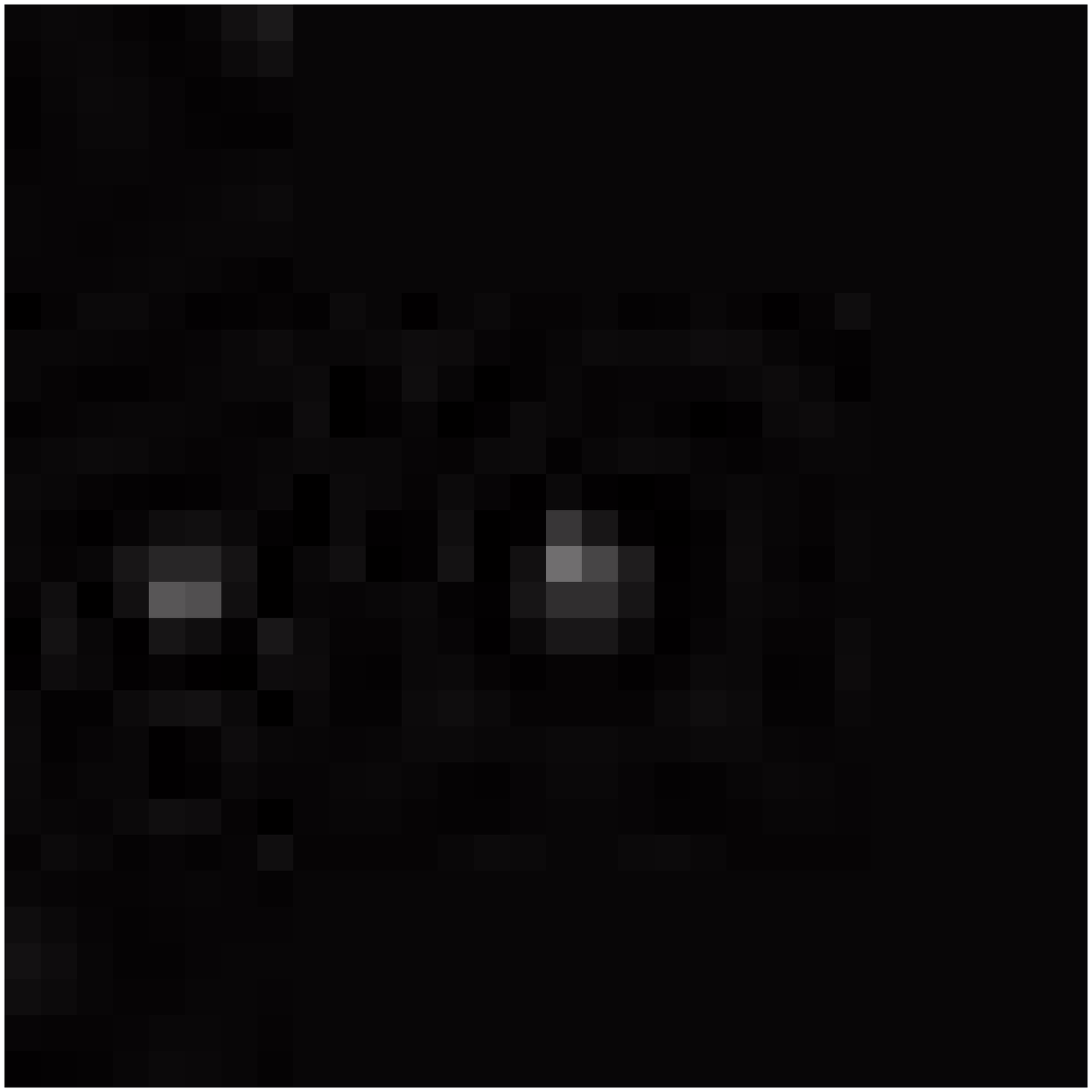}
\includegraphics[scale=0.2]{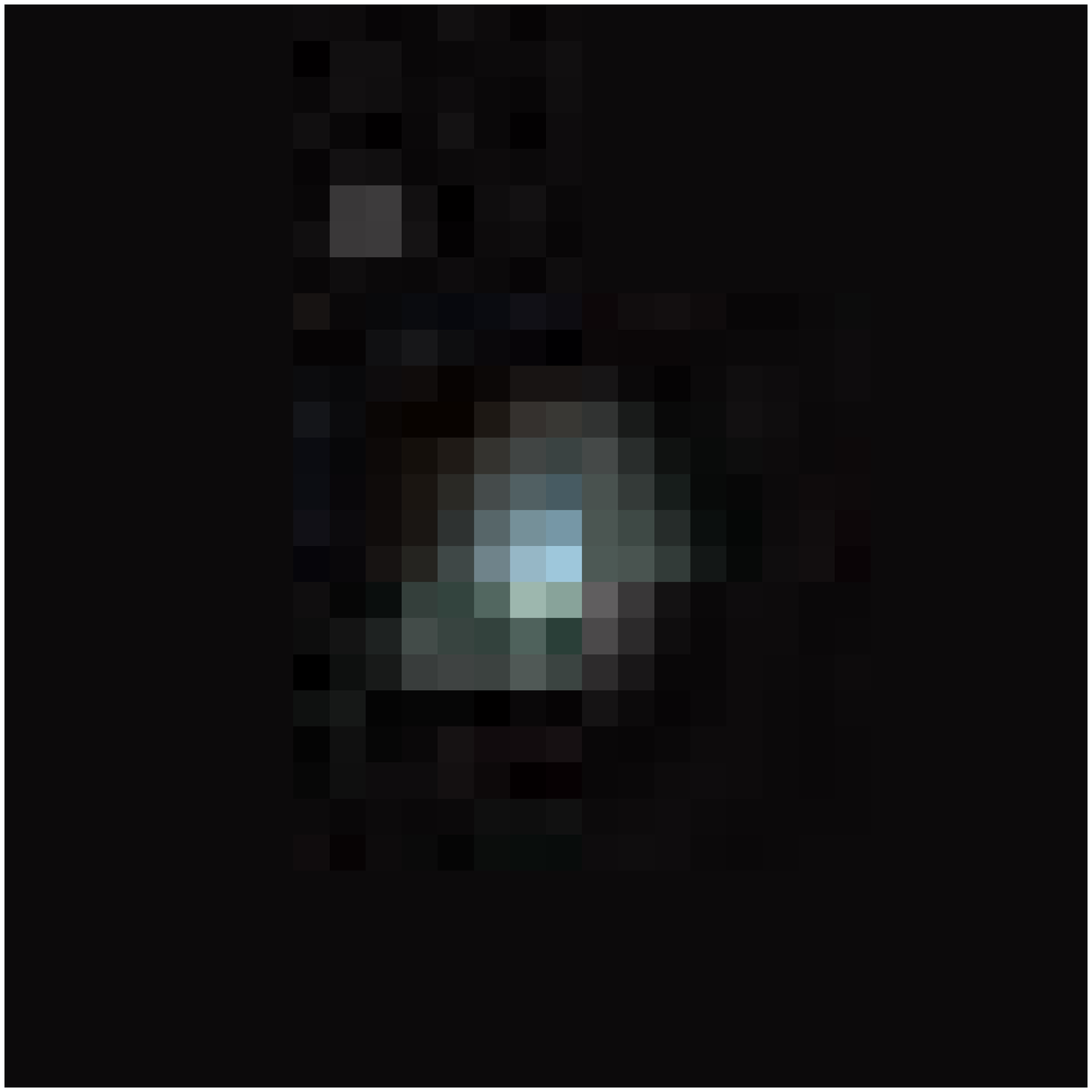}
\includegraphics[scale=0.2]{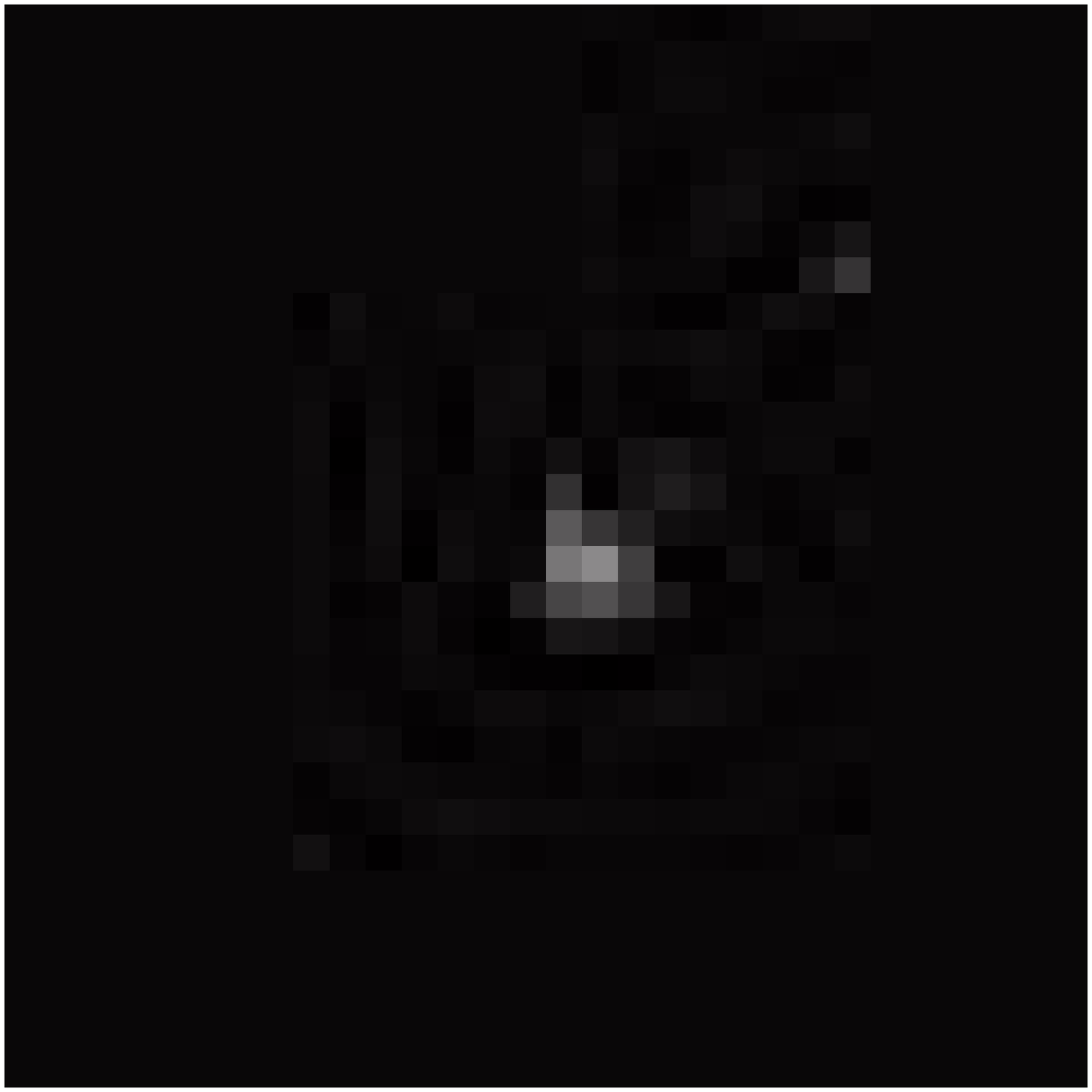}
\includegraphics[scale=0.2]{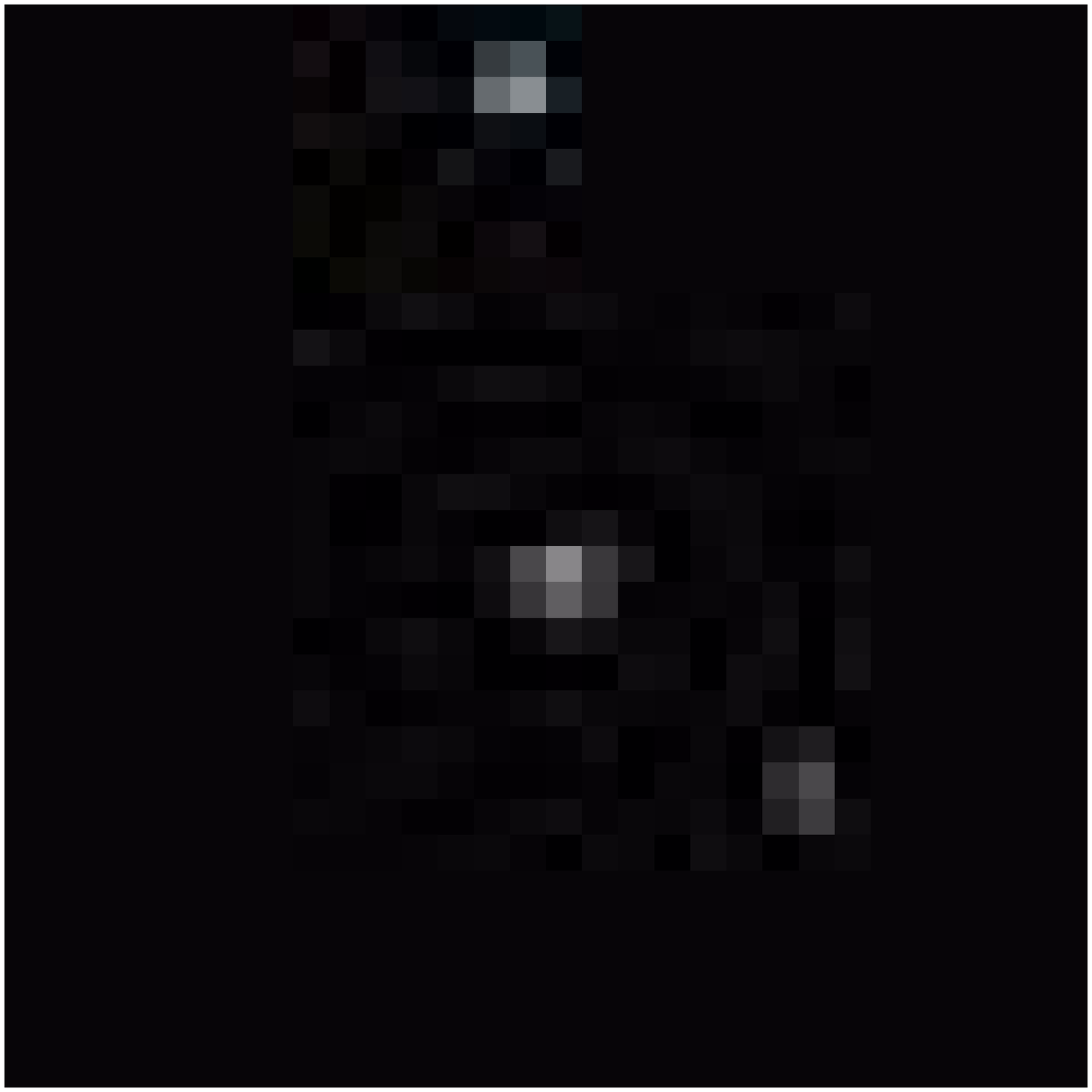}
\includegraphics[scale=0.2]{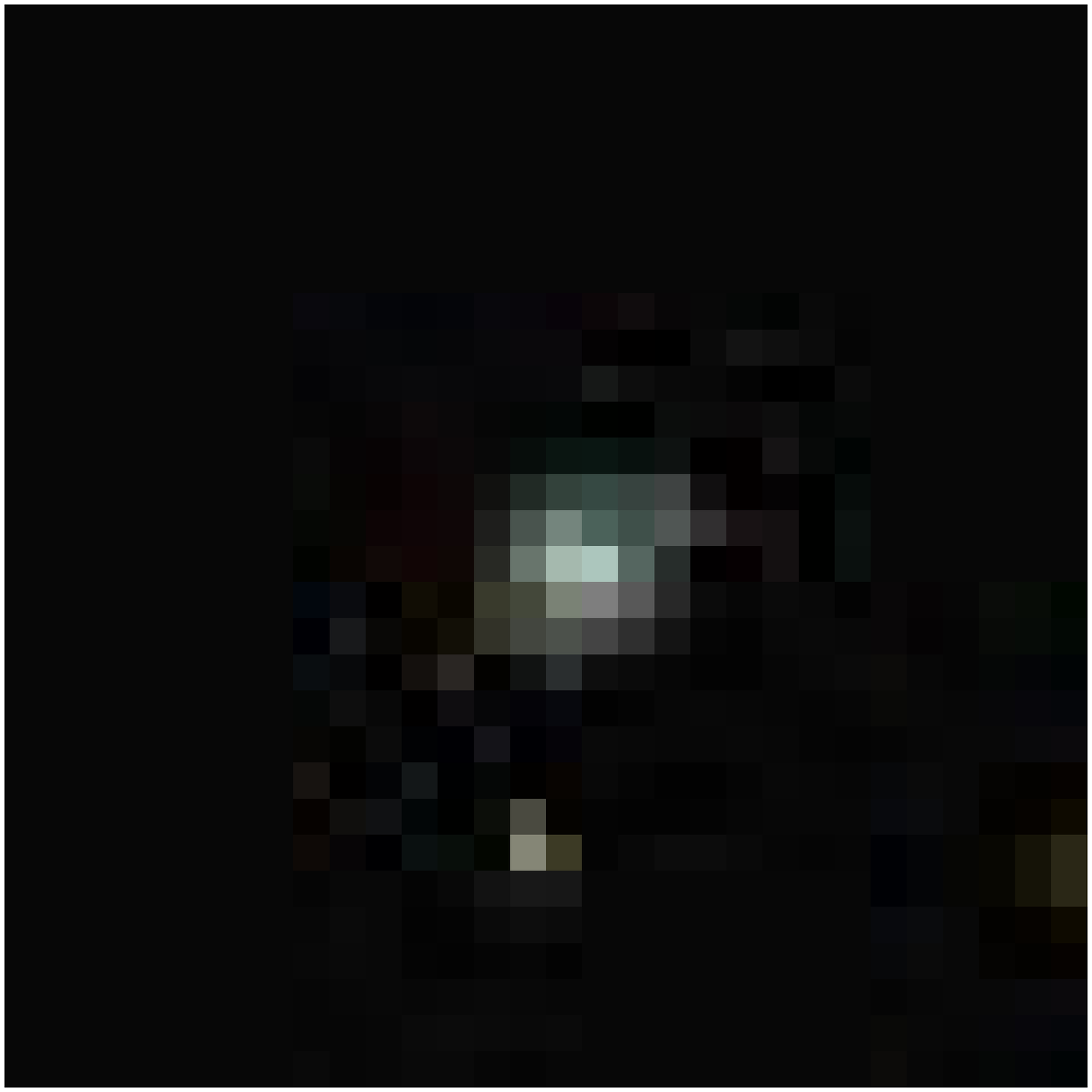}
\end{center}
\caption{Nine example images of  young E+A galaxies (E+As with H$\delta$
 EW $>$7 \AA). Image size is
 60''$\times$60'' and north is up. Each panel corresponds to that in
 Figure \ref{fig:ea2_progenitor_spectra_individual}.
}\label{fig:ea2_progenitor_image_individual}
\end{figure}

\begin{figure}
\begin{center}
\includegraphics[scale=0.7]{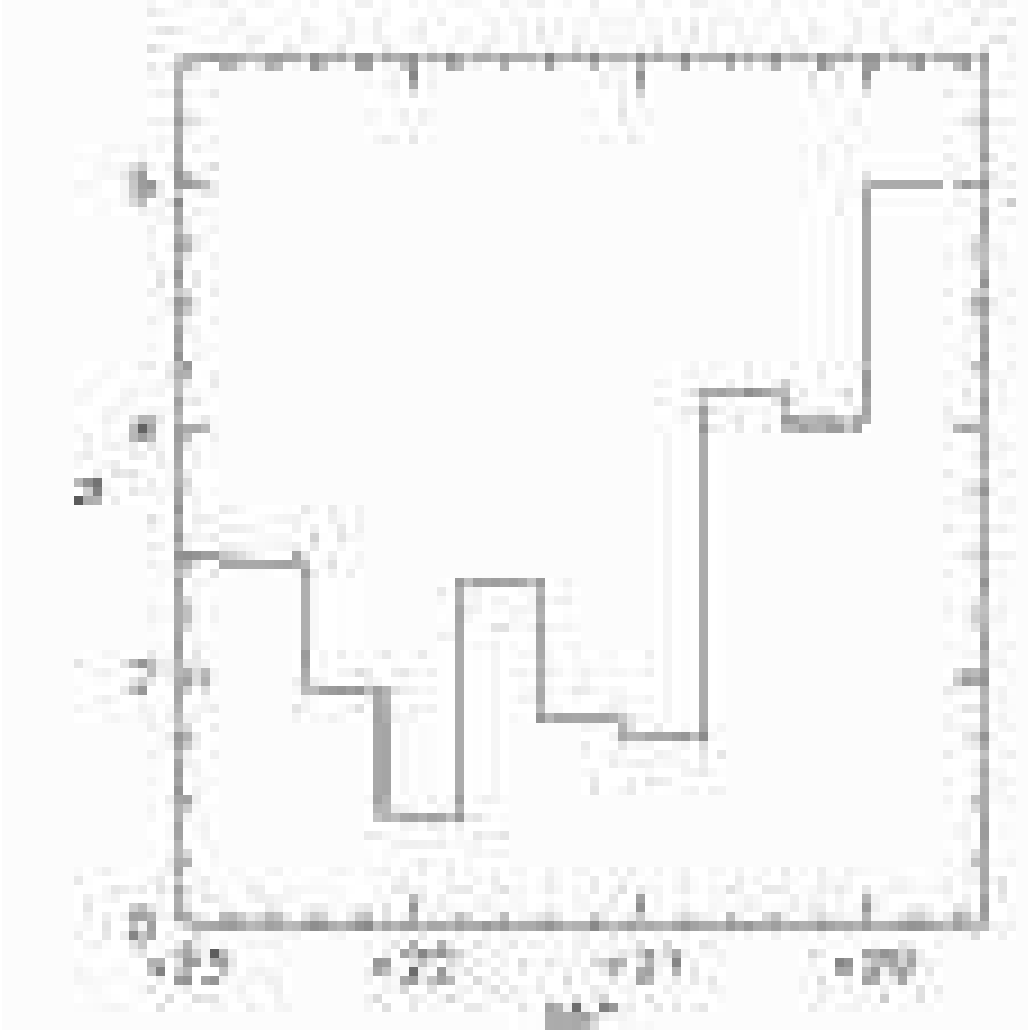}
\end{center}
\caption{
 Absolute magnitude distribution of accompanying galaxies within 75 kpc
 of young E+A galaxies. Since uncertainty in $k$-correction and star/galaxy separation
 of the SDSS increase at $r^*\sim$22.2 ($Mr^*=-$19.5 at $z=0.3$), this figure should not be over-interpreted.
}\label{fig:ea2_accompany_hist}
\end{figure}

\begin{itemize}
 \item Morphologies of E+A galaxies are elliptical-like, while those of HDS+em galaxies
       are disk-like (Figure \ref{fig:ea2_morphology}).
 \item The local galaxy density distribution of E+A galaxies is
       consistent with that of the field galaxies.
       And thus, many E+A galaxies are found in field regions. These field E+A galaxies
       can not be explained with cluster-related phenomena such as
       ram-pressure stripping or gravitational interaction with cluster
       potential. Therefore, the origin of E+A galaxies is not likely to
       be cluster-related.
 \item Dusty star-forming galaxies are expected to have redder
       color in $r-K$ by $\sim$1 mag. However,
       $r-K$ colors of E+As (Figure \ref{fig:ea2_jk_rk}) are not much redder
       than those  of normal galaxies. 
        We derived an upper limit on the dust
       enshrouded star formation using radio data from the FIRST survey,
       and found that dust enshrouded star formation rate of E+A galaxies is
       well below $\sim$10 M$_{\odot}$ yr$^{-1}$. Therefore, it is not likely
       that E+A galaxies are dusty starburst galaxies.
 \item  We compared three typical star formation histories in the GISSEL
       model with the observational quantities including H$\delta$ EW,
       D4000 and $u-g$ color. As is found in previous work, only the
       instantaneous burst model can explain unusual
       properties of E+A galaxies, assuring the post-starburst nature of
       these galaxies. 
 \item  Assuming the instantaneous burst model, 
       we selected young E+A galaxies of an age of 270-430 Myrs
       after the burst. These young E+As have a stronger H$\delta$
       absorption and a brighter absolute magnitude than normal (all) E+As.
       In Figure
       \ref{fig:ea2_progenitor_image_individual}, we investigated images of
       the young E+As and found that they frequently have close
       accompanying galaxies.
         Statistically, these galaxies have accompanying
       galaxies within 75 kpc $\sim$2.5 times more frequent  than randomly selected
       galaxies at two $\sigma$ significance level.
          Considering that cluster related origins and dusty star 
       formation origins are not all plausible in terms of  our data, we conclude
       that a merger/interaction with closely accompanying galaxies
       is the most likely mechanism to be responsible for the violent
       star formation history of E+A galaxies.
\end{itemize}

\clearpage

\clearpage

\clearpage

\clearpage

\clearpage

\clearpage

\clearpage

\clearpage

\clearpage

\clearpage

%

%

%
%

%

\clearpage

\clearpage

\clearpage

\clearpage

\clearpage

\clearpage

\clearpage

\begin{table}[h]
\caption{
 Number of galaxies in each subsample of H$\delta$-strong galaxies. Galaxies with
 the line ratios consistent with an AGN are not included in the sample
 of H$\delta$-strong galaxies .
}\label{tab:ea2_hds_sample}
\begin{tabular}{ll}
\hline
 Category  & Number \\
\hline
\hline
E+A                &    133  \\ 
HDS+em             &   2900  \\ 
HDS+H$\alpha$      &    108  \\ 
HDS+[OII]          &     42  \\ 
All H$\delta$-strong            &   3183  \\ 
All                &  94770  \\ 
\hline
\end{tabular}
\end{table}

\begin{table}[h]
\caption{
 Median properties of all E+A galaxies and young E+As (with  H$\delta$ EW$>$7 \AA). Errors are
 quoted using 75 and 25 percentile. 
}\label{tab:ea2_progenitor_median}
\begin{tabular}{lll}
\hline
 Parameter  & E+As with H$\delta$ EW$>$7 $\AA$ &  All E+As\\
\hline
\hline
 H$\delta$ EW (\AA)   &   7.67$^{+0.40}_{-0.34}$  &   5.94$^{+0.92}_{-0.48}$    \\ 
 $u-g$             &   1.35$^{+0.28}_{-0.15}$     &   1.37$^{+0.09}_{-0.07}$ \\ 
 $u-r$             &   2.45$^{+0.27}_{-0.12}$     &   2.43$^{+0.27}_{-0.19}$ \\ 
 $g-r$             &   0.57$^{+0.05}_{-0.04}$     &   0.59$^{+0.07}_{-0.04}$\\ 
 $r-i$             &   0.26$^{+0.08}_{-0.12}$     &   0.25$^{+0.07}_{-0.05}$\\ 
 D4000          &      1.41$^{+0.13}_{-0.04}$     &    1.48$^{+0.11}_{-0.09}$\\ 
 $[OII]$ EW (\AA) &  $-$0.01$^{+0.72}_{-0.41}$      &   $-$0.11$^{+0.92}_{-0.79}$\\ 
 H$\alpha$ EW (\AA) &    $-$1.75$^{+1.10}_{-0.35}$  &   $-$1.61$^{+0.46}_{-0.34}$\\ 
 Local Galaxy Density (Mpc$^{-2}$) &     0.05$^{+0.32}_{-0.04}$  &   0.08$^{+0.23}_{-0.06}$ \\ 
 $Cin$             &   0.36$^{+0.02}_{-0.02}$     &   0.35$^{+0.01}_{-0.02}$ \\ 
 $Mr^*$            &    $-$22.23$^{+0.85}_{-0.64}$  &   $-$21.70$^{+0.45}_{-0.65}$\\ 
\hline
\end{tabular}
\end{table}

\begin{table}[h]
\caption{
 Number of accompanying galaxies around all the E+A galaxies and young E+A
 galaxies (with H$\delta$ EW$>$7) are compared with that of 1000
 randomly picked
 galaxies with similar redshift distribution. Numbers of  accompanying
 galaxies are counted using galaxies 
 with -23.0$<Mr^*<$-19.5 after $k$-correction within the radius of 50 and 75
 kpc. Fore/backgournd galaxy number counts are 
 statistically subtracted. 
}\label{tab:ea2_nearby_galaxies}
\begin{tabular}{lll}
\hline
 Sample  & $N_{nearby\ galaxy}$ within 50 kpc  & $N_{nearby\ galaxy}$ within 75 kpc \\
\hline
\hline
 Random                         &   0.03$\pm$0.01 & 0.16$\pm$0.01  \\ 
 All E+As                       &   0.12$\pm$0.03 & 0.26$\pm$0.04  \\ 
 E+A with H$\delta$ EW$>$7 \AA  &   0.24$\pm$0.09 & 0.40$\pm$0.12  \\ 
\hline
\end{tabular}
\end{table}


\listoffigures 
\listoftables

\end{document}